\documentclass[english,a4paper]{book}
\usepackage[T1]{fontenc}
\usepackage[numbers]{natbib}
\usepackage{titling}
\usepackage{tikz}
\usetikzlibrary{patterns}
\usepackage{array}
\usepackage{amsmath, amssymb}
\usepackage{pgf}
\usepackage{amsfonts}
\usepackage{dsfont}
\usepackage{mathrsfs}
\usepackage{mathabx}
\usepackage{stmaryrd}
\usepackage{makeidx}
\usepackage{amsbsy}
\usepackage{amsthm}
\usepackage{color}
\usepackage{fullpage}
\usepackage{enumitem}
\usepackage{float}
\usepackage[bookmarks=false, colorlinks, citecolor=red, linkcolor=blue, urlcolor=purple]{hyperref}

\newsavebox{\accentbox}
\newcommand{\compositeaccents}[2]{\sbox\accentbox{$#2$}#1{\usebox\accentbox}}

\newtheorem{defi}{Definition} 
\newtheorem{rem}{Remark}
\newtheorem{prop}[defi]{Proposition}
\newtheorem{theo}[defi]{Theorem}

\newtheorem{lemma}[defi]{Lemma}

\usepackage[nopostdot]{glossaries}
\renewcommand*{\glstextformat}[1]{\textcolor{black}{#1}}

\makeatletter
\@ifundefined{glossentry}{%
    \newcommand{\glossentry}[1]{%
      \xdef\glscurrententrylabel{\glsdetoklabel{##1}}%
      \gls@org@glossaryentryfield{##1}%
    }%
}{}
\makeatother

\def\sC{\mathscr{C}} 
\def\sD{\mathscr{D}} 
\def\sE{\mathscr{E}} 
\def\sF{\mathscr{F}} 
\def\sG{\mathscr{G}}

\def\sO{\mathscr{O}}

\def\sR{\mathscr{R}}

\def\sU{\mathscr{U}}

\def\bC{{\mathbb C}}

\def\bN{{\mathbb N}} 

\def\bP{{\mathbb P}}

\def\bR{{\mathbb R}}

\newcommand{\ca}[1]{{\cal #1}}         

\def\cC{{\ca C}}
\def\cD{{\ca D}}
\def\cE{{\ca E}}
\def\cF{{\ca F}}

\def\cH{{\ca H}}

\def\cL{{\ca L}}

\def\cO{{\ca O}}
\def\cP{{\ca P}}

\def\cU{{\ca U}}
\def\cV{{\ca V}}
\def\cW{{\ca W}}

\def\id{{\mathrm{id}}}  
\def\1{{\mathds{1}}}

\def\Rea{{\, \mathrm{Re} \,}}
\def\Imm{{\, \mathrm{Im} \,}}

\def\supp{{\mathrm{supp} \,}}
\def\WF{{\mathrm{WF}}}

\def\vol{{\mathrm{vol}}}

\def\const{{\mathrm{const}}}

\def\Deg{{\mathrm{Deg}}}

\def\ad{{ \, \mathrm{ad}}}

\def\cl{{\mathrm{cl}}}

\def\loc{{\mathrm{loc}}}

\def\hami{{\mathfrak{h}}}

\def\wedgevee{{{\setlength{\medmuskip}{0mu} \wedge\vee}}}

\makeglossaries

\newglossaryentry{sigma_c}%
{%
  name={\ensuremath{\sigma_c}},
  description={Distributional kernel for the standard symplectic form associated to linear Klein-Gordon-type equations. Page},
  sort={Cd}
}

\newglossaryentry{E_phi}%
{%
  name={\ensuremath{E_\phi}},
  description={Causal propagator associated to the Klein-Gordon type operator $P_\phi$. Page},
  sort={Cc}
}

\newglossaryentry{sigma_phi}%
{%
  name={\ensuremath{\sigma_\phi}},
  description={On-shell $W$-smooth symplectic structure on $S$. Page},
  sort={Ce}
}

\newglossaryentry{hatF}%
{%
  name={\ensuremath{\hat{F}}},
  description={Quantum observable corresponding to $F$ defined via Haag's formula. Page},
  sort={Di}
}

\newglossaryentry{nablaR}%
{%
  name={\ensuremath{\nabla^R}},
  description={Retarded derivative. Page},
  sort={Dh}
}

\newglossaryentry{omegaR_phi}%
{%
  name={\ensuremath{\omega^R_\phi}},
  description={Retarded $2$-point function. Page},
  sort={Cb}
}

\newglossaryentry{omega_phi}%
{%
  name={\ensuremath{\omega_\phi}},
  description={Pure Hadamard $2$-point function associated to the Klein-Gordon type operator $P_\phi$. Page},
  sort={Ca}
}

\newglossaryentry{G_phi}%
{%
  name={\ensuremath{G_\phi}},
  description={Symmetric part of the pure Hadamard $2$-point function $\omega_\phi$. Page},
  sort={Cf}
}

\newglossaryentry{alpha}%
{%
  name={\ensuremath{\alpha}},
  description={Isomorphism $C^\infty_W(S,\cW') \to C^\infty_W(S,\cW)$. Page},
  sort={Dl}
}

\newglossaryentry{omegaflat_phi}%
{%
name={\ensuremath{\omega^\flat_\phi}},
  description={On-shell $W$-smooth K\"{a}hler structure on $S$ corresponding to an admissible assignment $\phi \mapsto \omega_\phi$ of $2$-point function. Page},
  sort={Ce}
}

\newglossaryentry{H_phi}%
{%
  name={\ensuremath{H_\phi}},
  description={Hadamard parametrix with respect to the Klein-Gorndon type operator $P_\phi$. Page},
  sort={Cg}
}

\newglossaryentry{cE_W'_n}%
{%
  name={\ensuremath{\cE_W'(M^n)}},
  description={Compactly supported distributions on $M^n$ with wave-front sets contained in the set $W_n$. Page},
  sort={Ad}
}

\newglossaryentry{partial}%
{%
  name={\ensuremath{\partial}},
  description={Derivative operator acting on on-shell $W$-smooth covariant sections. Page},
  sort={Da}
}

\newglossaryentry{mathring_nabla_W}%
{%
  name={\ensuremath{\mathring{\nabla}^W}},
  description={$W$-covariant derivative (in the sense of def.~\ref{def_W_cov_der}) corresponding to the Levi-Civita connection. Page},
  sort={Dc}
}

\newglossaryentry{nabla_W}%
{%
  name={\ensuremath{\nabla^W}},
  description={$W$-covariant derivative (in the sense of def.~\ref{def_W_cov_der}) corresponding to the Yano connection. Page},
  sort={Dd}
}

\newglossaryentry{delta}%
{%
  name={\ensuremath{\delta}},
  description={Fedosov operator on on-shell $W$-smooth forms. Page},
  sort={De}
}

\newglossaryentry{delta-1}%
{%
  name={\ensuremath{\delta^{-1}}},
  description={``Inverse'' Fedosov operator on on-shell $W$-smooth forms. Page},
  sort={Df}
}

\newglossaryentry{DW}%
{%
  name={\ensuremath{D^W}},
  description={Fedosov connection on on-shell $W$-smooth forms. Page},
  sort={Dg}
}

\newglossaryentry{C_infty_W_S}%
{%
  name={\ensuremath{C^\infty_W(S)}},
  description={On-shell $W$-smooth functionals. Page},
  sort={Ba}
}

\newglossaryentry{C_infty_W_S_n}%
{%
  name={\ensuremath{C^\infty_W(S, \boxtimes^n_W T^* S)}},
  description={On-shell $W$-smooth covariant sections of rank $n$. Page},
  sort={Bb}
}

\newglossaryentry{Omega_W_k}%
{%
  name={\ensuremath{\Omega_W^k(S)}},
  description={On-shell $W$-smooth $k$-forms. Page},
  sort={Bc}
}

\newglossaryentry{Omega_W_S_cW}%
{%
  name={\ensuremath{\Omega_W(S, \cW)}},
  description={On-shell $W$-smooth form with values in the formal Wick algebra $\cW$. Page},
  sort={Be}
}

\newglossaryentry{C_W_infty_S_cW}%
{%
  name={\ensuremath{C_W^\infty(S,\cW)}},
  description={On-shell $W$-smooth sections in the formal Wick algebra bundle $\cW$. Page},
  sort={Bd}
}

\newglossaryentry{S}%
{%
  name={\ensuremath{S}},
  description={Smooth solutions to the non-linear Klein-Gordon equation. Page},
  sort={Aa}
}

\newglossaryentry{T_S}%
{%
  name={\ensuremath{T_\phi S}},
  description={Tangent space of $S$ at $\phi \in S$, i.e. the space smooth solutions to the linearised equation around $\phi$. Page},
  sort={Ab}
}

\newglossaryentry{boxtimes_W_n}%
{%
  name={\ensuremath{\boxtimes_W^n T_\phi^* S}},
  description={The (completion of the) $n$-fold tensor product of the cotangent space $T^*_\phi S$ at the solution $\phi \in S$. Page},
  sort={Ac}
}

\newglossaryentry{cW_phi}%
{%
  name={\ensuremath{\cW_\phi}},
  description={The formal Wick algebra at the solution $\phi \in S$. Page},
  sort={Ae}
}

\newglossaryentry{t_bullet_s_phi}%
{%
  name={\ensuremath{(t \bullet s)_\phi}},
  description={Wick product on sections in $C^\infty_W(S,\cW)$. Page},
  sort={Db}
}

\newglossaryentry{W}%
{%
  name={\ensuremath{W_n}},
  description={Set related to wave-front set estimates. Page},
  sort={Ao}
}

\newglossaryentry{X}%
{%
  name={\ensuremath{X_{2+\nu}}},
  description={Set related to wave-front set estimates. Page},
  sort={Ap}
}

\newglossaryentry{Z_nu}%
{%
  name={\ensuremath{Z_{2+\nu}}},
  description={Set related to wave-front set estimate. Page},
  sort={Aq}
}

\newglossaryentry{P_phi}%
{%
  name={\ensuremath{P_\phi}},
  description={Klein-Gordon type operator for the linearised equation around $\phi$. Page},
  sort={Aba}
}

\newglossaryentry{P_pm}%
{%
  name={\ensuremath{\bP^\pm}},
  description={Symmetrization/Antisymmetrization operator. Pages},
 	user1={\ensuremath{\bP^+}},
 	user2={\ensuremath{\bP^-}},
  sort={E}
}

\begin{document}
	
	\frontmatter
	\begin{titlepage}
   \centering
	{}
	\vspace*{2cm}
	\noindent\rule[0.5ex]{\linewidth}{2pt}
	{}\\
	\vspace*{0.5cm}
	{\Huge \em Fedosov Quantization and Perturbative Quantum Field Theory}\\
	\vspace{1.5cm}
	{\Large {Giovanni Collini\footnote{\href{mailto:collini@itp.uni-leipzig.de}{\texttt{collini@itp.uni-leipzig.de}}}}}\\
	Institut f\"{u}r Theoretische Physik, Universit\"{a}t Leipzig, Br\"{u}derstrasse 16, D-04103 Leipzig, Germany\\
	\vspace*{0.5cm}
	\noindent\rule[0.5ex]{\linewidth}{2pt}
	{}\\
	\vspace*{1cm}
	\noindent
	{\bf Abstract}\\
	\begin{flushleft}
	Fedosov has described a geometro-algebraic method to  construct in a canonical way a deformation of the Poisson algebra  associated with a finite-dimensional symplectic manifold (``phase  space''). His algorithm gives a non-commutative, but associative,  product (a so-called ``star-product'') between smooth phase space  functions parameterized by Planck's constant $\hbar$, which is treated  as a deformation parameter. In the limit as $\hbar$ goes to zero, the  star product commutator goes to $\hbar$ times the Poisson bracket, so  in this sense his method provides a quantization of the algebra of  classical observables. In this work, we develop a generalization of Fedosov's  method which applies to the infinite-dimensional  symplectic ``manifolds'' that occur in Lagrangian field theories. We show that the procedure remains mathematically well-defined, and we  explain the relationship of this method to more standard perturbative  quantization schemes in quantum field theory.
	\end{flushleft}
	\vfill
	{31.03.2016}

 	\end{titlepage}

	\tableofcontents

	\mainmatter

\chapter*{Introduction}
\addcontentsline{toc}{chapter}{Introduction}
It is well known that the quantization of classical systems is, in many cases, not a straightforward procedure, and, furthermore, usually suffers from certain ambiguities. In the case of a mechanical system defined on a finite-dimensional  phase space spanned by coordinates $(q,p)$, one usually proceeds by  representing the quantum operators corresponding to $q$ and $p$ by the multiplication and the differential operators $Q = q$ and $P = i\hbar \partial/\partial q$. It arises the problem of how to consistently assign a quantum operator to a general phase space function $f(q,p)$ because the quantum operators no longer commute. In practice, one typically deals only with a restricted class of phase space functions such as the Hamiltonian (often of the form $H(q,p) = \frac{p^2}{2m} + V(q)$) suffering only from mild --if any-- ordering ambiguities. However, this is an issue for general phase space functions.\\

It is, of course, legitimate to take the viewpoint that the quantum observables are simply the (self-adjoint) elements of the algebra  generated by $(Q,P)$, (the CCR, Weyl, or resolvent~\citep{buchholz2008resolvent} $*$-algebras, depending on the precise framework) and that one can worry about the classical limit later. However, this procedure only works for simple (linear) phase spaces, and problems also occur when one passes to infinite-dimensional phase spaces, such as for Klein-Gordon, or gauge theory, especially in versions of such theories exhibiting self-interactions.\\
At some level, these problems can be ascribed  to ``Haag's theorem''~\citep{haag1955quantum}, which states that the representation of the canonical commutation relations (CCR) is no longer unique in the infinite-dimensional setting\footnote{The Stone-von Neumann theorem~\citep{reed1972methods, W94} no longer holds.}. Instead, the determination of the representation is, in a sense, a dynamical problem which must be solved as part of the construction of the quantum theory. One manifestation of these issues is the appearance of the well-known ``renormalization'' procedures in quantum field theory,  which seem unavoidable if one wants to give proper mathematical sense  to the naive quantization procedures extrapolated from finite-dimensional quantum mechanical systems (with  linear phase space).\\

An alternative approach to quantization which is both somewhat more general than that sketched above and well-adapted to quantum field theory --as we shall see-- is ``deformation quantization''. As before, the input is a finite-dimensional phase space $S$ equipped with a Poisson bracket $\{f, g\}$ between phase space functions $f,g$. However,  rather than trying to promote these to operators in some way or other,  one tries to ``deform'' the algebraic structure on space of phase  space functions $C^\infty(S)$ in such a way that the Poisson bracket is recovered in  the limit $\hbar \to 0$. More precisely, one looks for an associative  product $\star_\hbar$ (called ``star product'' in this context) on  $C^\infty(S)$ depending on Planck's constant, which is now considered as a  deformation parameter\footnote{Often $\hbar$ is treated as a ``formal parameter'' in the sense that the objects considered are formal series in $\hbar$, i.e. $\hbar$ does not take any numerical value. Questions of the convergence of the series are thereby avoided/ignored.}. To have a correspondence with the Poisson  bracket, one postulates that $(f \star_\hbar g - g \star_\hbar  f)/\hbar \to \{f, g\}$ and that $f \star_\hbar g \to fg$ as $\hbar \to 0$. Since the  underlying space of functions, $C^\infty(S)$, is unchanged, the  ordering problem seems to have disappeared at first glance. On the  other hand, the precise definition of the product $\star_\hbar$ is now no longer evident, and one has in fact  many possible ways to define $\star_\hbar$ consistent with these  requirements. Thus, one can say that the ambiguities have simply been  shifted into the precise definition of the associative structure on  $C^\infty(S)$, and one might be tempted to conclude that not much has  been gained after all.\\

This impression is, however, incorrect. Firstly, the framework of  deformation quantization is more general than the usual one, since one  does not assume, even, the existence of an underlying symplectic structure $\sigma$ on $S$ (i.e., closed, non-degenerate 2-form), but only a Poisson structure, and one certainly does not have to assume that $S$  is a linear space with constant symplectic structure $\sigma = dq^i  \wedge dp_i$ as for the CCR algebra.\\
Also, the framework is naturally embedded into the algebraic framework  of deformations of algebras, for which natural notions of equivalence  are available. Indeed, considering --as seems perfectly natural-- different  deformations to be equivalent if they lead to isomorphic algebras, one  gets a classification of non-equivalent star products in terms of  certain cohomological data on $(S,\sigma)$~\citep{deligne1995deformations, nest1995algebraic, nest1995algebraicb, bertelson1997equivalence, weinstein1998hochschild, gutt1999equivalence, neumaier2002local, neumaier2003universality}. Furthermore, as we shall review, there  exist very natural geometrical constructions of star products that are  not only very appealing from the mathematical viewpoint, but also give  new insights into the nature of the quantization problem. Finally, and most importantly  for us, deformation quantization seems also to be very well-adapted to  the field theoretic setting, i.e. to the quantization of field theories.\\

The connection between quantum field theory and deformation quantization was investigated for the first time by Dito~\citep{dito1990star, dito1992star} and, in the algebraic approach to quantum field theory, has been made transparent in the paper~\citep{DF01a} by D\" utsch and Fredenhagen. The essence of their paper is the observation that the Wick-product in free quantum field theory (e.g. Klein-Gordon theory) can be viewed as  a certain special kind of star product on the space of classical  functionals on phase space. More precisely, the authors suggest to  view the Klein-Gordon field $\varphi(x), x \in \bR^4$, as an ``evaluation functional'' on phase space $S = \{ \mbox{classical smooth solutions to } (\boxempty -  m^2)u=0 \}$, defined by $\varphi(x)[u] = u(x)$. They observe that  $S$ carries a natural Poisson structure. This structure is inherited from the Lagrangian  formulation of the theory and is sometimes also called  ``Peierls-bracket''~\citep{peierls1952commutation}. They then proceed by defining a star product, setting
\begin{equation}\label{star_prod_intro}
\varphi(x_1) \star_{\hbar} \varphi(x_2) = \varphi(x_1) \varphi(x_2) + \hbar  \omega(x_1,x_2) 1,
\end{equation}
where the product $\varphi(x_1) \varphi(x_2)$ is the usual product of  evaluation functionals, i.e. $\varphi(x_1)\varphi(x_2)[u] = u(x_1)  u(x_2)$, and where $\omega$ is the so-called ``Wightman function'' (vacuum $2$-point function) of  the Klein-Gordon field, i.e.
\begin{equation*}
\omega(x,y) = \frac{1}{(2\pi)^2}\int_{\overline{V}^+} \delta^4( p^2 - m^2) \exp(i p_\mu (x-y)^\mu) dp
\end{equation*}
As they continue to show, this defines consistently a star product on  the space of (polynomial) evaluation functionals on $S$, which are  more precisely functions $F: S \to \bC$ of the form
\begin{equation}\label{functional_intro}
F = \int_{M^n} f(x_1, \dots, x_n) \varphi(x_1) \dots \varphi(x_n) dx_1 \dots  dx_n,
\end{equation}
where $f$ can even be a distribution with certain well-described  singularities such as a delta distribution. The above procedure looks  unfamiliar to a field theorist at first sight, but becomes natural if  we observe that the product rule is precisely ``Wick's theorem'' if we formally identify
\begin{equation*}
\varphi(x_1)  \dots \varphi(x_n) \leftrightarrow :\hat{\varphi}(x_1) \dots \hat{\varphi}(x_n):,
\end{equation*}
where the hat denotes the usual field operator on  Fock-space, and where the double dots $: \cdots :$ mean normal ordering. In fact, this correspondence precisely defines a  Hilbert-space representation of the associative algebra $\cW$,  generated by these $F$'s under the star product.\\

Apart from clarifying the connection between ``ordinary'' quantization  using Fock-space methods and deformation quantization, the  construction of~\citep{DF01a} has several advantages. First of all, the  resulting algebra $\cW$ is, as an abstract algebra, independent of any choices such as a vacuum state. Indeed, the only  datum entering the construction is $\omega$, and it is shown that passing to a new $\omega'$ within a certain natural class (``Hadamard  states'', see def.~\ref{def_hadamard_2-point} below), yields an isomorphic algebra $\cW'$. This is a strong conceptual advantage if one wants to  consider a Klein-Gordon field on a general Lorentzian manifold $M$, where no  preferred structures such as a vacuum state are available~\citep{dewitt1975quantum,kay1978linear,D80,fulling1989aspects,W94}.\\
Another advantage of the formalism is that, within $\cW$, there are contained not only observables such as $\varphi(x)$, but also the Wick powers $\varphi(x)^k$ and their ``time-ordered  products''. These in turn are the building blocks of the usual perturbative series for a corresponding interacting quantum field  theory.\\
The authors of~\citep{DF01a} indeed go on to explain in detail how such series are constructed within $\cW$ using the methods of ``causal  perturbation theory''~\citep{BFK96, brunetti2000microlocal, DF01b, DF04} (based on earlier ideas by Epstein and Glaser~\citep{epstein1973role}) on Minkowski space. It turns out that  these constructions can also be generalized to a general globally  hyperbolic curved Lorentzian manifold $M$~\citep{HW01,HW02,HW03, brunetti2003generally, HW05}.\\

Even though the constructions of~\citep{BFK96, brunetti2000microlocal, DF01b, DF01a, HW01, HW02, HW03, brunetti2003generally, DF04, HW05} (for reviews see~\citep{fredenhagen2015perturbative, fredenhagen2015perturbative2, hollands2015quantum}) are mathematically clear and rigorous, a conceptually unsatisfactory aspect remains. The point is that, although the construction precisely follows the philosophy of deformation quantization in the case of linear field theories, one deviates from it in the case of interacting theories. Indeed, what one constructs are perturbative series in $\cW$ for the observables in the interacting theory, but the star product $\star_\hbar$ {\em remains}  that given by eq.~\eqref{star_prod_intro} for the underlying {\em free} theory. On the other hand, according to the philosophy of deformation quantization, it would be more natural to keep the observables unchanged, but rather deform the underlying star product now taking also into account the self-interaction of the field. According to this approach, one would  hence start more naturally with the ``phase space'' $S = \{  \text{classical smooth solutions to } (\boxempty- m^2)\phi - \frac{\lambda}{3!}  \phi^3=0 \}$ of the theory {\em with interactions} encoded in the non-linear term $\frac{\lambda}{3!} \phi^3$. As for $\lambda = 0$, this space carries a natural symplectic structure,  hence Poisson bracket, inherited from the underlying Lagrangian formulation. The task would then be to deform this Poisson bracket according to the general rules for deformation quantization. This star product would certainly not be the same as for the free theory~\eqref{star_prod_intro}, but how to construct it? Also, once it has been constructed, what is the relation to the construction of~\citep{BFK96, brunetti2000microlocal, DF01b,DF01a, HW01, HW02, HW03, brunetti2003generally, DF04, HW05}?\\

In this work, we address and answer these two questions. In order to do so, we go back to the case of a finite-dimensional symplectic manifold $(S,\sigma)$ and review how one can construct a deformation quantization there. The method which we will follow is that pioneered by Fedosov~\citep{F94,F96} and elaborated upon by many other people~\citep{bordemann1997fedosov,bordemann1998homogeneous,KS01,grigoriev2001fedosov}.  Our main result will be that a variant of his method can also be  applied in the infinite-dimensional setting of field theory, i.e. to  $S = \{ \text{classical smooth solutions to }(\boxempty - m^2)\phi -  \frac{\lambda}{3!} \phi^3=0 \}$, and we will be able to say how this construction relates to that via causal perturbation theory of~\citep{BFK96, brunetti2000microlocal, DF01b,DF01a, HW01,HW02,HW03, brunetti2003generally, DF04, HW05}. In order to explain our methods and results in more detail, we must however first outline the essential  ideas of Fedosov's method in finite dimensions (a more detailed  outline is given for the convenience of the reader in chapter~\ref{sec_Fedosov_fin}).\\

We start considering as phase space a finite-dimensional symplectic manifold $(S, \sigma)$. We denote phase space points by $x$. Fedosov's method can be explained as follows. Choose an arbitrary but fixed $x$. The cotangent space  $T^*_x S$ with symplectic form $\sigma_x$  clearly is a {\em linear} phase space (of dimension $n= \dim(S)$)  with {\em constant} symplectic form. For polynomial functions $F(y), H(y)$ on this linear phase space $T^*_x S$ --{\em not} on $S$--, one defines a star product by
\begin{equation}\label{fiberwise_prod}
F \bullet_x H = m \left( \exp( \hbar \omega^{ij}_x \partial_{y^i} \otimes \partial_{y^j}) (F \otimes H) \right),
\end{equation}
or equivalently by
\begin{equation*}
y^i \bullet_x y^j = y^i y^j + \hbar \omega^{ij}_x,
\end{equation*}
where $y^i, i=1, \dots, n$ are coordinates on $T^*_x S$ --{\em not}  $S$-- and where $\omega^{ij}_x$ is a complex tensor such that its imaginary part is the symplectic form $\sigma^{ij}_x$ on $T^*_x S$ and its real part $G^{ij}_x$ is a positive definite (real) inner product on  $T^*_x S$. The choice is made in such a way that $(J_x)^i{}_j = G_x^{ik}(\sigma_x)_{kj} $ is a complex structure on $T^*_x S$. The product $\bullet_x$ can be extended to formal power series $F(y), H(y)$, i.e. roughly speaking we allow polynomials in $y^1, \dots, y^n$ of infinite degree\footnote{More rigorously, for $n=1$ the ring of formal power series $\bC[[y^1]]$ is the direct product $\bC^{\bN}$, i.e. the sequences $(a_0, a_1, \dots)$ with possibly infinitely many non-vanishing elements (conventionally written also as $\sum_{n \leq 0} a_n y^n$), equipped with the ring structure $(a_n)_{n \in \bN} + (b_n)_{n \in \bN} = (a_n + b_n)_{n \in \bN}$ and $(a_n)_{n \in \bN} (b_n)_{n \in \bN} = (\sum_{k \leq n} a_{n-k}b_k)_{n \in \bN}$. See eg.~\citep{zariski1960commutative}.}. The algebra of formal power series $F(y)$ on $V^*_x$ with product $\bullet_x$ is denoted by $\cW_x$. We may repeat this construction for any other point $x$ if we provide such an $\omega_x$ at each point of $S$, i.e. if $S$ is equipped with an {\em almost-K\"{a}hler structure} (the section $x \mapsto \omega_x$ is called almost-K\"{a}hler section), and thereby get an algebra $\cW_x$ for all $x \in S$. The union of these algebras defines a bundle over $S$, and the product in each fibre evidently gives a  product between the sections of this bundle.\\
The next step in the  scheme is to define a {\em flat} derivative operator $D$ in this bundle. One starts from a natural connection $\nabla$ associated with the almost-K\"{a}hler structure $\omega$ (the Yano connection). Since $\nabla$ is not flat, one modifies it proceeding recursively in the polynomial order and in the $\hbar$-order. The corrections depend on the $y^i$, the symplectic form $\sigma_{ij}$, the curvature tensor $R^i{}_{jkl}$, the torsion tensor  $T^i{}_{jk}$ and an increasing number of covariant derivatives of theses two tensors.\\
The {\em flat} sections in $\cW$ relative to $D$ form a sub algebra of all sections. Furthermore, it is seen that for any smooth phase space function $f$ on $S$, there exists a  corresponding smooth flat section $F$ in $\cW$ such that $F_x = f(x) 1  + O(\hbar)$ for each $x \in S$. If we denote by $\tau$ the projection of $F$ onto its part proportional to the section $1$, then we can also characterize the relation between $F$ and $f$ by $\tau F = f$. The correspondence $f \leftrightarrow F$ is in fact one-to-one.\\
The desired star product on $C^\infty(S)$ is now defined as follows. For $f,h \in C^\infty(S)$, first find the flat sections $F,H$ of $\cW$  under this correspondence, then form $F \bullet H$ via the fiberwise  product~\eqref{fiberwise_prod},  then project $F \bullet H$ onto the part proportional to the section $1$, i.e. acting with the map $\tau$. The projection is a formal power series in $\hbar$ with functions on $S$ as coefficients. This provides the star product $f \star h:= \tau (F \bullet H)$.\\

We have presented Fedosov's method in such a way that the analogies to field theory suggest themselves: in field theory, $S$ is the space of  solutions of the theory, i.e. a ``point'' $\phi \in S$ is a (smooth) solution to $(\boxempty - m^2) \phi - \frac{\lambda}{3!} \phi^3 = 0$. $T_\phi S$ is the space of solutions of the linearized equations around $\phi$, i.e. solutions $u$ to $(\boxempty - m^2 - \frac{\lambda}{2}  \phi^2) u = 0$. $T^*_\phi S$ is the space of linear functionals from linearized  solutions to $\bR$, e.g. functionals $\varphi(x)$ of the form  $\varphi(x)[u] = u(x)$, or more generally functions $F$ of the form~\eqref{functional_intro}.
The fiberwise product $\bullet$ corresponds to~\eqref{fiberwise_prod}, where $\omega_\phi$ is now a chosen $2$-point function in for each $\phi$ for the linear Klein-Gordon theory described by the equation $(\boxempty - m^2 - \frac{\lambda}{2} \phi^2) u = 0$. The functions $F$ of the form~\eqref{functional_intro} together with the product $\bullet_\phi$ define an algebra $\cW_\phi$, and the union of these fibres forms a bundle $\cW$ over $S$, just as in the finite-dimensional case. Thus, we are in  principle set to start Fedosov's construction in the field theory setting.\\

However, it is far from evident that this will make any actual mathematical sense. Obvious potential problems that come to mind are:
	\begin{enumerate}
		\item The ``manifold'' $S$ is clearly infinite-dimensional in the  field theory setting. Thus, we need to first define a suitable  manifold structure on $S$, which will depend on the behaviour of solutions to the non-linear Klein-Gordon equation. As is well-known,  even if this can be achieved, we are left with the task of giving a  precise meaning to bundles like $(T^*S)^{\otimes n}$ entering the recursion procedure in Fedosov's method. In the infinite-dimensional  case, such tensor products could have a priori different meanings. For  instance, they could mean various distribution spaces in $n$ space-time variables. Experience from ordinary perturbative quantum field theory suggests  that very singular distributions should be expected to occur.
		\item We need to show that the recursion procedure in Fedosov's  method, which involves taking an increasing number of ``derivatives''  along $S$, can be carried through. In the infinite-dimensional  setting, index contractions such as in $\sigma_{si} R^i{}_{jkl} R^j{}_{mnr} G^{jm}  G^{kn} y^l y^r \ dx^s$ (a typical example of a $\cW$-valued 1-form  appearing in the construction of $D$) would formally become ``integrals''  over space-time (``continuous index summation''). Such integrals of distributions  have no a priori reason to make any sense.
		\item Even if the above problems can be solved, it is a priori highly unclear what would be the relation of the quantization scheme to more standard methods.
		\item In quantum field theory, there is a well-defined notion of  space-time locality (``Einstein causality'') meaning that quantum field  observables localized at space-like related regions should commute. For the free field theory with product (1), this property is evident  because $\omega(x,y) = \omega(y,x)$ if $x,y$ are space-like to each other. However, if we follow Fedosov's  algorithm in the context of interacting quantum field theories --  assuming even this can be done -- there is no guarantee that space-time  locality will hold. Indeed, there is no analogue of space-time locality  in finite-dimensional systems, hence this property is very far from  being manifest in the quantum field theory generalization of Fedosov's  method.
	\end{enumerate}

In this work, we propose a possible solution to these issues. We proceed in the following manner:
	\begin{enumerate}
		\item We propose a notion of smoothness  for functions on $S$ (or more  generally, for sections in the bundle $\cW \to S$), which we call  ``on-shell $W$-smoothness''.
This notion encapsulates the following ideas. First of all, a function on $S$ (or section in $\cW$) should be extendible to a function (or section) on the space $C^\infty(M)$, i.e.
it should not only be defined for smooth solutions $\phi$ of the non-linear Klein-Gordon equation, but also for arbitrary smooth functions $\phi$. Of course, there could be many extensions of a given function (or section) on $S$. We require that there exists an extension such that for any $\nu \in \bN$ the $\nu$-th Gateaux derivative (i.e. the variational derivative in $\phi$) of the extension not only exists, but defines a distribution on $M^\nu$ with a certain restricted ``wave-front set''. In the case of sections in $\cW$, we also require the extension to have a specific form. The  terminology ``on-shell'' refers to the fact that it is made for functions (or sections) on the solution space $S$, and the  letter `$W$' is used throughout this work for a sequence of sets  $\{W_n\}_{n \in \bN}$, where $W_n$ is in the cotangent bundle $T^*M^n$, appearing in the wave front condition.
		\item It turns out that the notion of on-shell $W$ smoothness has the  desired properties for our purposes: We can show that it behaves well under the products $\bullet$, and derivatives. In particular, we can  show that the infinite-dimensional analogues of the curvature tensors $R^i{}_{jkl}$, the torsion tensor $T^i{}_{jk}$, and their covariant derivatives are on-shell $W$-smooth (for a suitable choice of the ``tensor'' $\omega$  on $S$). Using such results, we can show furthermore that these properties suffice to construct the infinite-dimensional analogue of the flat Fedosov connection $D^W$, and that this connection maps on-shell $W$-smooth
sections into on-shell $W$-smooth sections. The methods of microlocal analysis give a convenient calculus for the wave front set of the various  distributions that come up in this construction and are instrumental in demonstrating these  results.
		\item We then show that different (suitable) choices $\omega,\omega'$  give ``gauge equivalent'' Fedosov  connections $D^W, D^{\prime W}$ on the bundles $\cW, \cW'$. This result will enable  us to see how Fedosov's method in quantum field theory is related to more standard  methods of quantization.  The method which can be compared most easily with Fedosov's method is that of ``causal perturbation theory''~\citep{DF01b, DF01a, HW01, HW02, HW03}. In this method, one constructs, for each classical (local, polynomial function) $F(\phi)$ on $S$ a corresponding quantum observable. This observable is defined  separately for each $\phi \in S$ and is denoted by $\hat{F}_\phi$. Here, the notation reflects a splitting of the quantum field into a ``classical background'' $\phi  \in S$ and a ``quantum fluctuation'' $\varphi$. The quantity $\hat{F}_\phi$ is an element of $\cW_\phi$, and is constructed by a perturbation series involving retarded products of the interaction. This interaction is obtained by expanding the Lagrangian $\cL(\phi + \varphi)$ in $\varphi$  keeping only the part that is higher than quadratic in the ``fluctuation'', $\varphi$, which is treated ``as an operator''. In $\frac{\lambda}{4!} \phi^4$-theory, this would be $\cV_\phi(\varphi) = ( \frac{\lambda}{3!} \phi \varphi^3 + \frac{\lambda}{4!}\varphi^4)$. In the simplest case when the classical observable is the field itself, $F(\phi) = \phi$, the corresponding quantum observable would be given by
\begin{equation}\label{haag_series_intro}
	\hat{\phi}(x) = \phi(x) 1 + \varphi(x) + \underbrace{\sum_{n \geq 1}  \frac{(i\lambda)^n}{\hbar^n n!} \int dy_1 \dots dy_n  R_{n,\phi}(\varphi(x), \cV_\phi(y_1) \otimes \dots \otimes \cV_\phi(y_n))}_{\mbox{perturbation series}}.
\end{equation}
The precise meaning of the terms in the perturbation series (``retarded products'' $R_{n,\phi}$) is recalled in sec.~\ref{subsec_int_QFT_per}.\\
It is possible to show with our methods that the map $\phi \mapsto \hat{F}_\phi$ is, in fact, an on-shell $W$-smooth section in $\cW$. The first guess might be that
this section is flat with respect to the Fedosov connection $D^W$, but this turns out to be not the case. However, we shall show that it is ``gauge  equivalent'' in a natural sense to a flat section in $\cW$. We therefore obtain two different algebras of flat sections with respect to the Fedosov connection $D$: one consists of the flat sections obtained via the correspondence $\tau$ in Fedosov's method, while the other is generated by the flat sections obtained acting with the ``gauge equivalence'' on any possible $\hat{F}$.
	\item The flat sections of the type~\eqref{haag_series_intro} appearing in causal perturbation theory are known to satisfy Einstein causality~\citep{DF04}. It follows that the corresponding flat sections relative to a general Fedosov operator $D$ related via ``gauge equivalence'' also respect Einstein causality, as gauge equivalence respects the product structure $\bullet$.
\end{enumerate}

The approaches to the quantization of field theories described and compared in this work are not the only possible ones. Another  possibility is to take as the fundamental input the so-called operator  product expansion (OPE)~\citep{wilson1969non, zimmermann1973normal, keller1992perturbative, keller1993perturbative, hollands2007operator}. (This framework seems to work best in the  context of Euclidean quantum field theories, i.e. versions of the  theory on a Riemannian manifold). In this approach, the ``product'' is  encoded in a set of ``structure functions''. More precisely, the OPE  is an expansion of the form
\begin{equation*}
O_{A_1}(x_1) \cdots O_{A_n}(x_n) = \sum_C C^C_{A_1, \dots, A_n}(x_1,  \dots, x_n) O_C(x_n),
\end{equation*}
where the $O_A$ for a ``basis'' of local functionals of the basic  field (i.e. they are monomials in $\phi, \partial \phi, \partial^2 \phi, \dots$ in the present  setting). The meaning of the sum and the equality is that both sides  should be equal when inserted into a suitable correlation function of  the Euclidean theory, see~\citep{hollands2008quantum, hollands2012operator, holland2014operator} for more details and explanations. ``Associativity'' is encoded in a set of highly non-trivial  consistency conditions between the coefficient functions $C_{A_1,  \dots, A_n}^C$, see~\citep{hollands2008quantum, holland2015associativity}.\\
The OPE framework looks rather different at first sight from that  presented here, where the ultimate goal is to construct a star-product $\star_\hbar$ of the interacting theory, rather than a corresponding  set of coefficient functions. Nevertheless, the two approaches are closely related. This becomes  more evident if one expands the OPE coefficients out in a deformation  parameter, which can be $\hbar$ (or also other parameters, such as the  coupling, $\lambda$, or even $1/N$ in a theory with $N$-component  fields transforming e.g. under $O(N)$). The different expansion orders in $\hbar$ of the OPE coefficient functions then correspond to the different expansion orders of the product $\star_\hbar$, and so, in a  sense, the coefficients $C$ are to be seen as the ``structure constants'' of the  product $\star_\hbar$, i.e. both approaches are complementary. The OPE  approach seems to be more geared towards Euclidean quantum field  theory and its advantage is that the algebraic structure is directly  linked to the short distance properties of correlation functions. The  methods described in this thesis (Fedosov's method, causal  perturbation theory) are more naturally geared to the Lorentzian  quantum field theories and nicely emphasize the dependence on any  classical backgrounds. Thus, the two methods are, in a sense,  complementary. A formal proof of their equivalence would be highly  desirable. This should certainly be possible, since also Euclidean  quantum field theories have a Hamiltonian formulation, which is  underlying Fedosov's strategy.\\

For the convenience of the reader, we now summarize the contents of  this work.\\

\noindent
{\bf Chapter~\ref{sec_Fedosov_fin}:} We begin this chapter by restating the programme of deformation quantization in sec.~\ref{subsec_intro_deformation}.\\
Then, in sec.~\ref{subsec_Fedosov_fin}, we outline the variant of Fedosov's method applicable to almost-K\"{a}hler manifolds, i.e. symplectic manifolds $(S,\sigma)$ with an additional almost-complex structure $J$ compatible with the symplectic form (equivalently with a compatible almost-K\"{a}hler section $\omega$). In particular, we present the two fundamental results in this context: there is a unique Fedosov connection $D$ determined by $\sigma$, $J$ (or equivalently, the corresponding almost-K\"{a}hler section $\omega$) and certain auxiliary data (thm.~\ref{theo_Fedosov_1_fin}), and how the associated star product is constructed using the one-to-one correspondence between smooth functions on $S$ and smooth flat sections with respect to $D$ (thm.~\ref{theo_Fedosov_2_fin}).\\
In the last section, sec.~\ref{subsec_equivalence_fin}, we consider for a given symplectic manifold $(S,\sigma)$ two different almost-K\"{a}hler sections $\omega$, $\omega'$ both compatible with the same $\sigma$. We give a proof of the equivalence (in the sense of def.~\ref{def_equiv}) of the star products corresponding to $\omega$ and $\omega'$ (for the same choice of auxiliary data). In particular, we give an explicit construction of the gauge equivalence between the Fedosov connections corresponding to $\omega$ and $\omega'$ (thm.~\ref{thm1_bis}).\\

\noindent
{\bf Chapter~\ref{sec_pQFT}:} In sec.~\ref{subsec_free_QFT}, we present the deformation quantization of free Klein-Gordon field theory, following the approach of~\citep{DF01a}. We recall the notion of Hadamard $2$-point function (def.~\ref{def_hadamard_2-point}). Then, we define the algebra $\cW$ (def.~\ref{defi_free_wick_per}) as the space~\eqref{formal_Wick_algebra_space_QFT} of sequences of ($\bC[[\hbar]]$-valued) compactly supported symmetric distributions with wave-front set bounded by the sets $\{ W_n \}$ defined by~\eqref{W_set_def} (modulo distributions obtained by acting with the Klein-Gordon operator) equipped with the product $\bullet$ given by eq.~\eqref{wick_prod_per} in terms of a (pure) Hadamard $2$-point function $\omega$. This product $\bullet$ can be viewed as a star product for the Poisson structure given by the Peierls bracket~\eqref{Peierls_brkt}. We summarize the similarities to Fedosov's method in Table~\ref{table:fin_linQFT}.\\
In sec.~\ref{subsec_int_QFT_per}, we discuss the interacting Klein-Gordon field. We will not directly extend Fedosv's method to the interacting Klein-Gordon theory yet, but rather we present the approach based on causal perturbation theory. The idea is to fix a background $\phi$ and expand the classical action of the interacting Klein-Gordon theory around $\phi$. For each $\phi$, the quadratic part of the expansion gives a linear theory and, similarly as in the free case, we can construct the algebra $\cW_\phi$ once a (pure) Hadamard $2$-point function $\omega_\phi$ is provided. The higher than quadratic part is treated as the interaction. Then, for any local functional (def.~\ref{def_funct_class}), its corresponding quantum interacting field in $\phi$ is the element $\hat{F}_\phi \in \cW_\phi$ defined by the Haag series (eq.~\ref{Haag_series}), given in terms of the retarded-products. (This gives the precise structure of the perturbation series~\eqref{haag_series_intro}.) We recall the axiomatic definition of the retarded products~\ref{R0}-\ref{R12} and the characterization of their ``renormalization ambiguities'' (thm.~\ref{theo_R12}). We conclude this section proving that the map $\phi \mapsto \hat{F}_\phi$ which assigns to a background $\phi$ the quantum interacting field in $\phi$ corresponding to a local functional satisfies a functional equation (eq.~\eqref{Fedosov_per}), which has a striking similarity with the flatness condition $D\hat{F} = 0$ for the Fedosov connection $D$ in finite dimensions. This gives the first hint of the relation between Fedosov's method and the causal perturbative approach to interacting quantum field theories.\\

\noindent
{\bf Chapter~\ref{sec_Fedosov_inf}:} In this chapter, we show that Fedosov's procedure can be directly implemented in the infinite-dimensional framework of a Klein-Gordon quantum field theory with a non-linear equation of motion. In sec.~\ref{subsec_manifold_inf}, we define rigorously the infinite-dimensional manifold $S$ of the smooth solutions to the $\frac{\lambda}{4!} \phi^4$-interacting Klein-Gordon equation on ultra-static space-times exploiting the global well-posedness of the initial value problem of the corresponding non-linear equation of motions (see app.~\ref{app_nnlin}). The tangent space $T_\phi S$ is defined in terms of smooth solutions of the linearized equation around $\phi$. The definitions of the cotangent space $T^*_\phi S$ and its tensor powers $\boxtimes_W T^*_\phi S$ are given in terms of compactly supported distributions with wave-front set bounded by a set $W_n$ modulo distributions obtained by acting with the Klein-Gordon operator for the linearized equation of motion around $\phi$ (see~\eqref{cov_tens_S}). We provide the definition of on-shell $W$-smoothness for functionals on $S$ (see def.~\ref{def_smooth_on_f}) or more general sections on $\boxtimes_W^n T^*S$ (see def.~\ref{def_smooth_on_tens}). This notion of smoothness is tailored to the choice of the sets $\{W_n\}_{n \in \bN}$ and it will be sufficient to guarantee the well-definiteness of the tensor product (prop.~\ref{prop_tensor_prod_W_sec}), the differential (prop.~\ref{prop_der_smooth_on}) and the Poisson bracket (prop.~\ref{prop_W-poisson}) as maps acting on on-shell $W$-smooth sections which preserve the on-shell $W$-smoothness property. We conclude this section defining the bundle $\cW$, its on-shell $W$-smooth sections, $C^\infty_W(S,\cW)$, and the on-shell $W$-smooth forms with values in $\cW$, $\Omega(S,\cW)$.\\
In sec.~\ref{subsec_W_smooth_symp_metric}, we provide this geometrical framework with two important on-shell $W$-smooth sections: the symplectic form $\sigma$ (thm.~\ref{theo_W_symp}) and an almost-K\"{a}hler structure $\omega^\flat = -\frac{1}{2} \mu + \frac{i}{2}\sigma$ (thm.~\ref{theo_var_controll_state}). For the latter, it is required a tight control on the dependence of the pure Hadamard $2$-point function $\omega_\phi$ with respect to the background $\phi$, which is formalized in the notion of ``admissible'' assignment $\phi \mapsto \omega_\phi$ (see def.~\ref{def_suit_omega}). It is also proved that there exists a non-trivial $\phi \mapsto \omega_\phi$ satisfying such requirements.\\
In sec.~\ref{subsec_algebra_W_smooth_sec}, we prove that the product $\bullet_\phi$ in each fibre $\cW_\phi$ defined in terms of an admissible $\omega_\phi$ preserves the on-shell $W$-smoothness and therefore provides an algebra structure $\bullet$ for $C^\infty_W(S,\cW)$ (prop.~\ref{prop_prod_smooth_on_W}) and $\Omega_W(S, \cW)$ (prop.~\ref{prop_prod_smooth_on_W_form}).\\
In sec.~\ref{subsec_W_covariant_der}, we define the infinite-dimensional analogue $\nabla^W$ of the Yano connection $\nabla$ in finite dimension. $\nabla^W$ is shown to be a $W$-smooth covariant derivative (in the sense of def.~\ref{def_W_cov_der}) which preserves the sections $\sigma$ and $\mu$ (prop.~\ref{prop_Yano_W-smooth_on}). Then, we extend $\nabla^W$ to a derivative operator on $\Omega(S,\cW)$ (prop.~\ref{prop_Yano_W-smooth_on_form}).\\
In sec.~\ref{subsec_Fedosov_inf}, after defining the Fedosov operators, we prove the infinite-dimensional analogue of Fedosov's theorems (thm.~\ref{theo_Fedosov_inf}). In particular, we show that there is a unique Fedosov $W$-smooth connection $D^W$ corresponding to $\sigma$, $\omega^\flat$ and the $W$-smooth connection $\nabla^W$.\\

\noindent
{\bf Chapter~\ref{sec_rel_pQFT_Fedosov}:} The last chapter is devoted to  the relation between the perturbative approach to quantum field theory of sec.~\ref{subsec_int_QFT_per} and Fedosov's method in infinite-dimensions constructed in chapter~\ref{sec_Fedosov_inf}. In particular, the aim is to understand eq.~\eqref{Fedosov_per} in the light of Fedosov quantization.  In sec.~\ref{subsec_gaugeeq}, we prove that the operator $\nabla^R - \langle \cdot, \frac{\delta}{\delta \varphi} \rangle$ appearing in~\eqref{Fedosov_per} is precisely the Fedosov connection with respect to the ``retarded $2$-point function'', or ``in-state'',  (thm.~\ref{theo_ret_conn_fedosov_conn}). We then extend the results obtained in sec.~\ref{subsec_equivalence_fin} for the finite-dimensional context to our infinite-dimensional case. In particular, we prove that the two Fedosov $W$-connections corresponding to different choices of $2$-point functions as in lemma~\ref{lemma_state} are gauge equivalent (thm.~\ref{theo_gauge_inf}). We then conclude that the map $\phi \mapsto \hat{F}_\phi$, which is an on-shell $W$-smooth section (as proved in sec.~\ref{subsec_W_smooth_ret_prod}), is gauge equivalent to a flat section with respect the Fedosov connection corresponding to any admissible assignment of (pure Hadamard) $2$-point functions (in the sense of def.~\ref{def_suit_omega}). Finally, we check that flat sections obtained by acting with the gauge equivalence on $\hat{F}$ satisfy Einstein causality (prop.~\ref{prop_einst_loc}). In this sense, Fedosov's method respects Einstein Causality.\\

\newpage
\noindent
{\bf Acknowledgements:}\\
I would like to thank my Ph.D. supervisor Professor S. Hollands  for suggesting the research program carried out in this work and  for providing guidance and advice along the way. The research leading to these results has received funding from the European Research Council under the European Union's Seventh Framework Programme (FP7 2007-2013) ERC grant agreement no. QC \& C 259562.\\
I would also like to thank Professor S. Waldmann for discussions about deformation quantization and his hospitality in W\"{u}rzburg in February 2015. I thank Dr. K. Sanders, Dr. J. Zahn, Dr. J. Holland, Dr. M. Fr\"{o}b for discussions and proofreading.

\chapter{Fedosov deformation quantization of finite-dimensional manifolds}\label{sec_Fedosov_fin}

\section{Deformation quantization}\label{subsec_intro_deformation}

		 There are several possible approaches to the quantization of a classical system. Among these, we focus in this work on deformation quantization. This approach was introduced in the form used here by Bayen, Flato, Fronsdal, Lichnerowicz and Sternheimer~\citep{BFFLS77, bayen1978deformation, bayern1978deformation2}, although antecedents can be found in earlier investigations e.g. by Weyl~\citep{weyl1927quantenmechanik} and Moyal~\citep{moyal1949quantum}. In these papers, quantization is considered as a deformation of the structure of the algebra of classical observables, rather than as a change in the nature of the observables themselves. Thus, mathematically, the approach has a close relationship to the theory of deformations of algebraic structures as described e.g. by Gerstenhaber in~\citep{gerstenhaber1964deformation}. For a summary of the general approach, we refer to~\citep{S98} and the references therein. The basic set-up of deformation quantization is as follows.\\
		 
		 The input is a finite-dimensional {\em Poisson manifold} $(S,\{ \cdot, \cdot \})$, which is a manifold $S$ equipped with a bilinear, skew-symmetric map $\{ \cdot, \cdot \}: C^\infty(S) \times C^\infty(S) \to C^\infty(S)$ satisfying the Jacobi identity and the Leibniz rule with respect to the pointwise multiplication of functions, i.e.
		 \begin{equation*}
		 	\left\{ \{f,g\}, h \right\} + \mbox{ cyclic permutations } = 0, \qquad \{ f, g h\} = \{f,g\} h + \{f , h \} g.
		 \end{equation*}
		By $C^\infty(S)[[\hbar]]$ one denotes the space of {\em formal power series} in $\hbar$ whose coefficients are smooth complex valued functions, i.e. each element can be written as
			\begin{equation*}
				f(x, \hbar) = \sum_{k \geq 0} \hbar^k f_k(x),
			\end{equation*}						
		with coefficients $f_k(x) \in C^\infty(S)$. The formal power series form a ring: such series are added and multiplied in the usual way (as if they were converging power series) but we ignore questions of convergence not assuming that $\hbar$ takes any numerical value. Deformation quantization consists in providing an associative algebra structure on $C^\infty(S)[[\hbar]]$, a so-called {\em star-product} $\star$, which satisfies the following conditions: 
			\begin{enumerate}
				\item For any $f,h \in C^\infty(S)$, it holds
						\begin{equation}\label{star}
							f \star h = \sum_{k \geq 0} \hbar^k C_k(f,h),
						\end{equation}
					where $C_k$ are $\bC$-bilinear (differential) operators on $C^\infty(S)$. Eq.~\eqref{star} extends $\bC[[\hbar]]$-linearly to $C^\infty(S)[[\hbar]]$.
				\item The algebra $C^\infty(S)[[\hbar]]$ equipped with the star-product $\star$ is a formal deformation of the commutative algebra of functions $C^\infty(S)$ equipped with the pointwise multiplication, i.e. $C_0(f,h)=fh$.
				\item The product $\star$ satisfies the correspondence principle, i.e. $C_1(f,h) - C_1(h,f) = i \{f,h\}$.
			\end{enumerate}
		In the this chapter we consider a special case of Poisson manifolds. We focus on {\em symplectic manifolds}. A symplectic manifold $(S, \sigma)$ is an even dimensional manifold $S$ with a $2$-form $\sigma = \sigma_{ij} dx^i \wedge dx^j$, called {\em symplectic form}, that is non-degenerate, i.e. $v=0$ if and only if $\sigma(v,v') = 0$ for any vector field $v'$, and closed, i.e. $d\sigma = 0$. Such a $2$-form induces a Poisson bracket according to the usual rules of Hamiltonian mechanics: if $E := \sigma^{ij} \partial_i \wedge \partial_j$ is the inverse of the symplectic form, one sets $\{ f,h \} := E(df,dh)$.\\	
		The existence of a deformation quantization for finite-dimensional symplectic manifolds was established by De Wilde and Lecomte~\citep{dWL83}. Later these results were conceptualized by Omori, Maeda, Yoshioka~\citep{OMY91} and, in particular, by Fedosov~\citep{F94, F96}. In this chapter, we will follow a variant of Fedosov's construction which has the advantage that it can be generalized to quantum field theories as will be discussed in later sections. As an aside, it is worth mentioning that the existence of a deformation quantization for the more general case of finite-dimensional Poisson manifolds was proved by Kontsevich~\citep{K03}, but this work is not relevant for us here, since we will always be given a symplectic form.

\section{Fedosov's method}\label{subsec_Fedosov_fin}
		The Fedosov's method, as described in his original paper~\citep{F94}, only requires as input a symplectic structure $\sigma$. In our application of the method to quantum field theory, it will be necessary to consider a variant of his method, described by Karabegov and Schlichenmaier in~\citep{KS01}. This variant uses as input a positive semidefinite section $\omega= \omega^{ij} \partial_i \otimes \partial_j$ of $\bC \otimes TS \otimes TS$ such that
			\begin{equation}\label{Im_and_Re_fin} 
				\Imm \omega^{ij} = \frac{1}{2} \sigma^{ij}, \qquad \Rea \omega^{ij} = \frac{1}{2}G^{ij},
			\end{equation}
		where $G= G^{ij} \partial_i \otimes \partial_j$ is the inverse of a Riemannian metric $\mu=G_{ij} dx^i \otimes dx^j$ on $S$. In other words,
			\begin{equation*}
				\omega = \frac{1}{2} G + \frac{i}{2} E,
			\end{equation*}
		where $E= \sigma^{ij} \partial_i \wedge \partial_j$ is the inverse of the symplectic form $\sigma$. Positive semidefinite means that for an arbitrary section $t$ of $\bC \otimes T^* S$ we have 
			\begin{equation}\label{pos_omega_fin}
				\omega(\bar{t}, t) \geq 0.
			\end{equation}	
		This condition is equivalent to a Cauchy-Schwarz-type inequality for $G$ and $E$, i.e.
			\begin{equation}\label{pos_fin}
				|E(t_1,t_2)| \leq \left( G(t_1,t_1) G(t_2,t_2) \right)^{1/2},
			\end{equation}
		for any pair $t_1,t_2$ of real valued sections of $T^*S$. As we will discuss extensively later, the tensor field $\omega$ can be interpreted as the finite-dimensional analogue of a $2$-point function of a quasi-free state in the quantum field theory setting. We assume also that $\omega$ defines an almost-complex structure compatible with the symplectic form $\sigma$, i.e. there exists a section $J:TS \to TS$ such that 
			\begin{equation*}
					J^2=-\id, \qquad \sigma(Jv, Jw) = \sigma(v,w), \qquad \mu(v,w)=\sigma(Jv,w).
			\end{equation*}
		In local coordinates, the almost-complex structure is given by
			\begin{equation*}
				J^i{}_j = G^{i \ell} \sigma_{\ell j} = -\sigma^{i \ell} G_{\ell j}.
			\end{equation*}
		In other words, we assume that $S$ is an {\em almost-K\"{a}hler manifold} and we call such $\omega$ an {\em almost-K\"{a}hler-section}. As we will see, this condition will correspond to $\omega$ being pure in quantum field theory setting. Of course, for a given $\sigma$, there are many such corresponding $\omega$'s. These ambiguities are discussed in sec.~\ref{subsec_equivalence_fin} below.\\
		Note that $\omega$ corresponds to the hermitian form $\omega^\flat$ on $\bC \otimes T^*S \otimes T^*S$ given by
			\begin{equation}\label{hermitian_fin}
				\omega^\flat := - \sigma_{ik} \omega^{k \ell} \sigma_{\ell j} dx^i \otimes dx^j = \frac{1}{2} \mu - \frac{i}{2} \sigma,
			\end{equation}
		and conditions~\eqref{pos_omega_fin} and~\eqref{pos_fin} imply
			\begin{equation*}
				\omega^\flat(\bar{u},u) \geq 0, \qquad |\sigma(u_1,u_2)| \leq \left( \mu(u_1,u_1) \mu(u_2,u_2) \right)^{1/2},
			\end{equation*}
		for any section $u$ of $\bC \otimes T S$, and for any pair $u_1,u_2$ of real valued sections of $TS$.\\
		
		The basic example is $S = \bR^{2d}$ with constant almost-K\"{a}hler structure. In this case, the desired deformation quantization, denoted by $\star=\bullet$ from now on, is elementary to describe.
		\begin{rem}[\bf basic example]\label{rem_key_ex}
			Assume that $S=\bR^{2d}$ and that $\sigma$ is the standard constant symplectic form, i.e.
				\begin{equation*}
					(\sigma^{ij}) = \left(
						\begin{matrix}
							0 & 1 & \hdots & 0 & 0 \\
							-1 & 0 & \hdots & 0 & 0 \\
							\vdots & \vdots & & \vdots & \vdots\\
							0 & 0 & \hdots &0 & 1 \\
							0 & 0 & \hdots &-1 & 0
						\end{matrix}
					\right)
				\end{equation*}
			in a suitable basis. Let $\omega^{ij}$ be any constant complex hermitian matrix with the properties just described, for instance $\omega^{ij}= \frac{1}{2} G^{ij} + \frac{i}{2} \sigma^{ij}$ with
				\begin{equation*}
					(G^{ij}) = \left(
						\begin{matrix}
							1 & 0 & \hdots & 0 & 0 \\
							0 & 1 & \hdots & 0 & 0 \\
							\vdots & \vdots & & \vdots & \vdots\\
							0 & 0 & \hdots &1 & 0 \\
							0 & 0 & \hdots &0 & 1
						\end{matrix}
					\right).
				\end{equation*}
			We define the star product $\star = \bullet$ by
				\begin{equation}\label{star_free}
					f \bullet h := m \left( \exp( \hbar \omega^{ij} \partial_i \otimes \partial_j) (f \otimes h) \right),
				\end{equation}
			where $m$ is the pointwise multiplication, $m(f,h)=fh$, and where the exponential is understood in the sense of formal power series.
		\end{rem}			
		
		The construction above will serve as a model for the case of general almost-K\"{a}hler manifolds. For this purpose, we first reformulate the construction. For Fedosov's method, we actually need the above star product not for general smooth functions on $S=\bR^{2d}$, but in fact only for formal power series in $2d$ coordinates,
			\begin{equation*}
				\bC[[y^1, \dots, y^{2d}]] \cong \bC \otimes \bigoplus_{n\geq0} \vee^n \bR^{2d},
			\end{equation*}
		where $\bigoplus$ denotes the direct product, and where $\vee^n$ denote the symmetrized $n$-fold tensor product. 
		As usual in formal deformation quantization, the complex coefficients are then further promoted to power series in the formal parameter $\hbar$.
			\begin{defi}\label{def_Wick_alg_const}
				The {\em formal Wick algebra} $\cW = \cW(\bR^{2d}, \omega)$ is the vector space 
					\begin{equation}\label{Wick_algebra_free}
						\bC[[y^1, \dots, y^{2d}]][[\hbar]] \cong \bC[[\hbar]] \otimes \bigoplus_{n\geq0} \vee^n \bR^{2d}
					\end{equation}
				equipped with the star-product $\bullet$ defined by eq.~\eqref{star_free}. More explicitly, 
				for the monomials $t = t_{i_1 \dots i_n} y^{i_1} \cdots y^{i_n}$ and $s = s_{i_1...i_m} y^{i_1} \cdots y^{i_m}$, the component of $t \bullet s$ in $\bC[[\hbar]] \otimes \vee^j  \bR^{2d}$ is given by
					\begin{equation}\label{star_co}
						\begin{split}
							&(t \bullet s)^j =\\
							&= \hbar^k \frac{n!m!}{k!(n-k)!(m-k)!} t_{\ell_1 \dots \ell_k i_1, \dots i_{n-k}}  s_{\ell'_1 \dots \ell'_k, i_{n-k+1}, \dots i_{j}} \omega^{\ell_1 \ell'_1} \cdots \omega^{\ell_k \ell'_k} y^{i_1} \cdots y^{i_{n-k}} y^{i_{n-k+1}} \cdots y^{i_j}
						\end{split}
					\end{equation}
				if $j= n +m -2k$ for $k \leq n,m$, otherwise $(t \bullet s)^j = 0$.
			\end{defi}
		\noindent
		It is useful and natural in the context of def.~\ref{def_Wick_alg_const} to introduce two gradings, called the {\em symmetric degree} $\deg_s$ and the {\em formal degree} $\deg_\hbar$. They are defined by
			\begin{equation*}
				\deg_s(y^i) := 1, \qquad \deg_\hbar(\hbar) := 1,
			\end{equation*}
		and extended to $\cW$ in the natural way. We define also the {\em total degree} $\Deg := 2 \deg_\hbar + \deg_s$.
			\begin{rem}\label{rem_filtration}
				The product $\bullet$ preserves the total degree $\Deg$ and therefore we can filtrate the formal Wick algebra $\cW$ with respect to the total degree $\Deg$. It follows that if we decompose $t \bullet s$ in terms homogeneous in $\Deg$, then each of these terms is a finite sum of products of components of $t, s$ homogeneous in $\Deg$ with degree not greater than $k$. Note that each element in the formal Wick algebra $\cW$ which is homogeneous in $\Deg$ is a finite sum of elements homogeneous in $\deg_s$ and $\deg_\hbar$. Therefore, each of these products of components of $t, s$ homogeneous in $\Deg$ further decomposes into a finite sum of terms as in eq.~\eqref{star_co}, for some appropriate components of $t, s$ homogeneous in $\deg_s$, $\deg_\hbar$.
			\end{rem}
		In addition to the product $\bullet$, the algebra $\cW$ has a natural involutive, anti-linear, $*$-operation, which we denote by $\dagger$\footnote{The star symbol is already over-used.}. 
		For the monomial $t=t_{i_1 \dots i_\ell} y^{i_1} \cdots y^{i_\ell} \in \bC[[y^1, \dots, y^{2d}]][[\hbar]]$, it is defined by
			\begin{equation*}
				t^\dagger := \overline{t_{i_1 \dots i_\ell}} y^{i_1} \cdots y^{i_\ell},
			\end{equation*}					
		 where the overbar denotes complex conjugation. Note that each $y^i$ is by definition hermitian with respect to $\dagger$. It is easy to check that the operation $\dagger$ satisfies 
			\begin{equation*}
				(t \bullet s)^\dagger = s^\dagger \bullet t^\dagger,
			\end{equation*}
		as required.\\
		
		We now review Fedosov's method, as adapted to almost-K\"{a}hler manifolds in the work of Karabegov and Schlichenmaier~\citep{KS01}. As in the original approach by Fedosov, there are three main steps:
		\begin{enumerate}
			\item For each $x \in S$, define the so-called formal Wick polynomial $*$-algebra $(\cW_x,\bullet_x)$ (associated with $\omega_x$) for the cotangent space $T^*_xS$. This defines a bundle of associative algebras, called $\cW$. The space of smooth sections of $\cW$ is denoted by $C^\infty(S,\cW)$. It is an associative algebra with respect to a product, called $\bullet$, which is naturally induced by the product $\bullet_x$ on each fiber.
			\item In the bundle $\cW$, construct a {\em flat} covariant derivative $D$, called {\em ``Fedosov connection''}, which is compatible with the product $\bullet$ in the sense that the Leibniz rule holds:
					\begin{equation*}
						D(t \bullet s) = (D t) \bullet s + t \bullet (Ds),
					\end{equation*}
				where $t, s$ are smooth sections on $\cW$. Furthermore, $D$ is compatible with the hermitian conjugation operation $\dagger$ on $\cW$ in the sense that 
					\begin{equation*}
						(Dt)^\dagger = D(t^\dagger).
					\end{equation*}	
				This condition is usually not emphasized, but it is necessary to provide the space of flat sections in $C^\infty(S,\cW)$, denoted by $\ker D$, with the natural structure of a $*$-sub-algebra of $C^\infty(S,\cW)$.
			\item The last step consists in defining an isomorphism $\tau$ between $\ker D$ and $C^\infty(S)[[\hbar]]$, and, finally, proving that a deformation quantization is given by the star-product defined by
					\begin{equation*}
						f \star h := \tau\left( \tau^{-1}(f) \bullet \tau^{-1}(h) \right),
					\end{equation*}
			Since the inverse $\tau^{-1}$ can be shown to be compatible with the $*$-operation on $\cW$ in the sense that
					\begin{equation*}
						(\tau^{-1}(f))^\dagger = \tau^{-1}(f^\dagger),
					\end{equation*}
				the algebra $(C^\infty(S)[[\hbar]], \star)$ equipped with the complex conjugation $f \mapsto f^\dagger := \sum_k \hbar^k \overline{f_k}$ is indeed a $*$-algebra. 
		\end{enumerate}

		We now explain in more detail how the above steps are carried out. We begin by defining the formal Wick algebra at $x \in S$, which is given by our local model in def.~\ref{def_Wick_alg_const} replacing $\bR^{2d}$ with the cotangent space $T^*_x S$, and $\omega$ with $\omega_x$ (the value of the almost-K\"{a}hler section at $x$) in eq.~\eqref{star_co} i.e.
			\begin{equation*}
				\cW_x := \cW(T^*_x S, \omega_x). 
			\end{equation*}	
		Thus, as a vector space, $\cW_x$ is the formal symmetric algebra over $T^\ast_x S$ with values in $\bC[[\hbar]]$, i.e.
			\begin{equation}\label{formal_wick_alg_fin}
				\cW_x =\bC[[\hbar]]\otimes \bigoplus_{n\geq0} \vee^n T^\ast_xS,
			\end{equation}
		(compare with~\eqref{Wick_algebra_free}). To simplify the notation, we introduce the symmetric tensor fields $y^{i_1}\cdots y^{i_n} := dx^{i_1} \vee \cdots \vee dx^{i_n}$. In other words, 
		$y^i_x$ are commuting variables similarly as before. Hence, an element of $\cW_x$ can be again identified with $t= (t^0, t^1, \dots)$, where $t^0 \in \bC[[\hbar]]$, and where, for $n >0$, 
			\begin{equation*}
				t^n=t_{i_1 \dots i_n} y^{i_1}_x \cdots y^{i_n}_x,
			\end{equation*}
		with $t_{i_1 \dots i_n} \in \bC[[\hbar]]$ symmetric. Similarly as before we can introduce the symmetric degree $\deg_s$, the formal degree $\deg_\hbar$ and the total degree $\Deg$.\\
		The Wick product in $\cW_x$, denoted by $\bullet_x$, is defined as in eq.~\eqref{star_co} using $\omega_x$, i.e. the value at $x$ of the complex tensor field $\omega$. Again the product is interpreted as given with respect to the $\Deg$-filtration (see remark~\ref{rem_filtration}). The bundle of formal Wick algebras is defined as the disjoint union of all the fibers $\cW_x$, i.e.
			\begin{equation*}
				\cW := \bigsqcup_{x \in S} \{x\} \times \cW_x.
			\end{equation*}
		Because $\cW$ is just given by tensor products of the cotangent bundle, it has the structure of a smooth vector bundle. The product defined in each fiber induces naturally an associative product on the space of smooth sections of $\cW$\footnote{To be precise, one should consider elements in the space $\left(\bigoplus_{n\geq 0} C^\infty(S,\vee^n T^*S) \right)[[\hbar]]$, rather than in the space $C^\infty(S,\cW)$. Abusing the notations, we identify these two notions.} denoted by $C^\infty(S,\cW)$, namely for any $t,s$ smooth sections on $\cW$ we set
			\begin{equation}\label{fiber_prod_fin}
				(t \bullet s)(x) := t(x) \bullet_x s(x).
			\end{equation}
		The product $\bullet$ is smooth in the sense that the product of two smooth sections gives another smooth section. This follows from the smoothness of $\omega$. Finally, each algebra $\cW_x$ is also a $*$-algebra with hermitian conjugation operation $\dagger$ and, therefore, the prescription
			\begin{equation*}
				t^\dagger(x) := (t(x))^\dagger
			\end{equation*}					
		provides the structure of a $*$-algebra for the space of smooth sections of $\cW$.\\
		
		We are still far from completing the deformation quantization of $C^\infty(S)$. In fact, instead of defining a star product on $C^\infty(S)[[\hbar]]$, we have given an algebra structure on the much larger space of sections $C^\infty(S,\cW)$ in the algebra bundle $\cW$. The key idea of Fedosov is to get around this problem by defining a special flat covariant derivative and restricting to the corresponding flat sections. These flat sections are then put into correspondence with functions on $S$. We now outline the procedure.\\
		Since we assume that $S$ is an almost-K\"{a}hler manifold, we have the Riemannian metric $\mu = G_{ij} dx^i \vee dx^j$. Let $\mathring{\nabla}$ be the Levi-Civita connection with respect to $\mu$, i.e. $\mathring{\nabla}$ is the unique torsion-free connection such that $\mathring{\nabla} \mu = 0$. In local coordinates, the Christoffel symbols of this connection take the well-known form
			\begin{equation*}
				\mathring{\Gamma}^k {}_{ij} = \frac{1}{2} G^{k\ell} \left(\partial_i G_{\ell j} + \partial_j G_{i\ell} - \partial_\ell G_{ij} \right).
			\end{equation*}
		Because the bundle $\cW$ is a formal series of tensor products of $T^\ast S$ (with values in $\bC[[\hbar]]$), the Levi-Civita connection extends ($\bC[[\hbar]]$-linearly) to a torsion-less covariant derivative on $\cW$, which is denoted again by $\mathring{\nabla}$.\\
		In general this connection is not flat and it does not satisfy the Leibniz rule with respect to the product $\bullet$ unless $\mathring{\nabla} \sigma = 0$, i.e. unless $S$ is a K\"{a}hler manifold. The second issue can be solved by passing to another natural connection defined by Yano in~\citep{Y65}. Introduce the Nijenhuis tensor by
			\begin{equation*}
				N(v,w) := [v,w] + J[Jv,w] + J[v,Jw] - [Jv,Jw], 
			\end{equation*}
		for $v,w$ vector fields. In local coordinates takes the form
			\begin{equation*}
				N^k{}_{ij} = (\partial_\ell J^k{}_i) J^\ell{}_j + J^k{}_\ell (\partial_i J^\ell{}_j) - (\partial_\ell J^k{}_j) J^\ell{}_i - J^k{}_\ell (\partial_j J^\ell{}_i).
			\end{equation*}					
		The following proposition is proved in \citep{Y65}:
			\begin{prop}[Yano Connection]\label{prop_Yano}
				Let $(S,\sigma,J)$ be an almost-K\"{a}hler manifold. There is a unique connection $\nabla$, called {\em Yano connection}, such that
					\begin{equation}\label{Yano_res_fin}
						\nabla \mu = 0, \quad \nabla \sigma =0, \quad \mbox{and} \quad T(v,w) =- \frac{1}{4}N(v,w)
					\end{equation}
				for $v,w$ vector fields on $S$, where $T(v,w) = \nabla_v w - \nabla_w v - [v,w]$ is the torsion tensor. In local coordinates the Christoffel symbols of $\nabla$ are
					\begin{equation}\label{Christoffel}
						\Gamma^k_{ij} := \mathring{\Gamma}^k{}_{ij} - \frac{1}{8}\left(N^k{}_{ij} + G^{k s}(N^r{}_{si} G_{rj} + N^r{}_{sj}G_{ri})\right).
					\end{equation}									
			\end{prop}
		\noindent		
		In general, the torsion of the Yano connection does not vanish (nor does the curvature). In fact, it vanishes precisely when $N=0$, i.e. when $S$ is a K\"{a}hler manifold. In this case, the Yano connection coincides with the Levi-Civita connection. We can naturally extend $\nabla$ to a covariant derivative on $\cW$. Because $\nabla$ annihilates both $\sigma$ and $\mu$, and, therefore, $\omega$, it follows that $\nabla$ satisfies the Leibniz rule with respect to the product $\bullet$. Furthermore, $\nabla$ is compatible with the conjugation $\dagger$ because by construction the Christoffel symbols of $\nabla$ are real. In other words, if $t,s$ are smooth sections of $\cW$, then it holds
			\begin{equation*}
				\nabla(t \bullet s) = (\nabla t) \bullet s + t \bullet (\nabla s), \quad \nabla(t^\dagger) = (\nabla t)^\dagger. 
			\end{equation*}
			
		To implement Fedosov's idea, we would like to have a {\em flat} connection satisfying the Leibniz rule which is compatible with the conjugation $\dagger$. The Yano connection is not generally flat and, consequently, we need to consider yet another connection. The construction of a flat connection becomes more natural if we consider the algebra of {\em $\cW$-valued forms on $S$}. The $\cW$-valued $k$-forms are smooth sections of $(\wedge^k T^\ast S) \otimes \cW$\footnote{To be precise, one should consider elements in the space $\left(\bigoplus_{n\geq 0} C^\infty(S,\wedge^k T^*S \otimes \vee^n T^*S) \right)[[\hbar]]$, rather that in the space $\Omega^k(S,\cW)$. Abusing the notations, we identify these two notions.} and they form a vector space denoted by $\Omega^k(S,\cW)$. Hence, a $k$-form $t$ with values in $\cW$ consists of a sequence $(t^{k,0}, t^{k,1}, \dots)$, where
			\begin{equation}\label{k_form_W_valued}
				t^{k,n}=t_{i_1 \cdots i_k; j_1 \cdots j_n} dx^{i_1} \wedge \cdots \wedge dx^{i_k} \otimes y^{j_1} \cdots y^{j_n},
			\end{equation}
		and where $t_{i_1 \cdots i_k; j_1 \cdots j_n}: S \to \bC[[\hbar]]$ are smooth functions anti-symmetric in the first $k$ indices and symmetric in the remaining $n$ indices.\\
		We can extend canonically the degrees $\deg_s$, $\deg_\hbar$ and $\Deg$ to forms with values in $\cW$. In addition, we define the anti-symmetric degree $\deg_a$ as $\deg_a dx^i := 1$. The space of $\cW$-valued forms of arbitrary anti-symmetric degree is denoted by
			\begin{equation*}
				\Omega(S,\cW) := \bigoplus_{k = 0}^{\dim S} \Omega^k(S,\cW).
			\end{equation*}	
		An element $t$ in $\Omega(S,\cW)$ is a collection $(t^{k,n})_{k=0, \dots, \dim S; n \in \bN}$ where $t^{k,n}$ is the same as~\eqref{k_form_W_valued}. It is clear that the anti-symmetric degree does not exceed the dimension of the manifold $S$. The product $\bullet$ can be extended to a product on $\Omega(S,\cW)$ in the following way. Consider two $\deg_a$-homogeneous elements in $\Omega(S,\cW)$. Without loss of generality, they can be written as $t \otimes \lambda, s \otimes \lambda'$, where $t,s \in C^\infty(S,\cW)$, and $\lambda \in \Omega^k(S)$, $\lambda' \in \Omega^{k'}(S)$, i.e. $\lambda$ and $\lambda'$ are two ordinary forms (with values in $\bC$) of rank $k$ and $k'$ respectively. The product is then defined as
			\begin{equation}\label{graded_prod_fin}
				(t \otimes \lambda) \bullet (s \otimes \lambda') := (t \bullet s) \otimes (\lambda \wedge \lambda') \in \Omega^{k+k'}(S,\cW).
			\end{equation}	
		The definition of the product $\bullet$ extends to forms with values in the formal Wick algebra $\cW$ using the $\deg_a$-filtration. The algebra $(\Omega(S,\cW),\bullet)$ inherits the structure of an associative algebra. Furthermore, this product $\bullet$ is bi-graded with respect to the gradings $\deg_a$ and $\Deg$.\\
		We can extend the Yano connection to $\Omega(S,\cW)$ in a natural way by defining for $t \in C^\infty(S,\cW)$, $\lambda \in \Omega^k(S)$
			\begin{equation}\label{graded_Yano_conn}
				\nabla( t \otimes \lambda) := (\nabla_i t) \otimes (dx^i \wedge \lambda) + t \otimes d\lambda,
			\end{equation}						
		where $d$ is the ordinary exterior differential acting on differential forms. Following further the procedure outlined by Fedosov, we introduce the operators $\delta$ and $\delta^{-1}$ on $\Omega(S,\cW)$, called {\em ``Fedosov operators''}. Let $t \in \Omega(S,\cW)$ with $\deg_a t=k$ and $\deg_s t =n$, then we define $\delta t, \delta^{-1}t$ by
			\begin{equation}\label{Fedosov_op}
		 		\delta t := dx^i \wedge \partial_{y^i} t,
		 	\end{equation}
		and
		 	\begin{equation}\label{Fedosov_op_-1}
		 		\delta^{-1} t := \left\{ \begin{array}{ll}
				\frac{1}{n+k} y^j i_a(\partial_{x^j}) t & k \neq 0 \\
				0 & \mbox{otherwise} \end{array} \right.
		 	\end{equation}
		In the previous formula, $i_a(\partial_{x^j})$ means the contraction of the vector field $\partial_{x^j}$ with the first anti-symmetric index of $t$. It is clear that $\delta$ increases by one the anti-symmetric degree while reducing by one the symmetric degree of a given element homogeneous in $\deg_a,\deg_s$. The operator $\delta^{-1}$ is doing the opposite. More explicitly, we can write
			\begin{equation}\label{Fedosov_op_alt}
				(\delta t)^{k+1,n-1} :=\left\{ \begin{array}{ll}
				n t^{k,n}_{i_2 \dots i_{k+1} ; (i_{1} j_1 \dots j_{n-1})} dx^{i_1} \wedge \cdots \wedge dx^{i_{k+1}} \otimes y^{j_1} \cdots y^{j_{n-1}} & n \neq 0\\
				 0 & \mbox{otherwise} \end{array} \right.
			\end{equation}
		and
			\begin{equation}\label{Fedosov_op_-1_alt}
				(\delta^{-1} t)^{k-1,n+1} := \left\{ \begin{array}{ll}
				\frac{k}{n+k} t^{k,n}_{[i_1 \dots i_{k-1} j_1] ; j_2 \dots j_{n+1}} dx^{i_1} \wedge \cdots \wedge dx^{i_{k-1}} \otimes y^{j_1} \cdots y^{j_{n+1}} & k \neq 0 \\
				0 & \mbox{otherwise} \end{array} \right.
			\end{equation}
		The following identities involving $\delta$, $\delta^{-1}$ and $\nabla$ are essential for the construction of the Fedosov connection and can be proved by direct computation, see~\citep{KS01}.
			\begin{lemma}\label{lemma_various_res_fin}
				Let $\hat{T}$ and $\hat{R}$ be the elements in $\Omega(S,\cW)$ respectively constructed from the torsion tensor $T$ and the Riemann tensor $R$ of the Yano connection $\nabla$ and defined by
					\begin{equation}\label{R_hat_T_hat}
						\hat{T} := \frac{1}{2} \sigma_{j_1 \ell} T^\ell {}_{i_1 i_2} dx^{i_1} \wedge dx^{i_2} \otimes y^{j_1}, \quad \hat{R} := \frac{1}{4} \sigma_{j_1 \ell} R^\ell {}_{j_2 i_1 i_2} dx^{i_1} \wedge dx^{i_2} \otimes y^{j_1} y^{j_2}.
					\end{equation}
				The following relations hold:
				\begin{enumerate}[label=(\roman*)]
					\item\label{adj_action_Fedosov_op_fin} $\delta = \frac{2i}{\hbar} \ad_\bullet(\delta^{-1} \sigma)$,
					\item\label{flat_Fedosov_fin} $\delta^2= (\delta^{-1})^2=0$,
					\item\label{Fedosov_op_vs_Yano_fin} $\delta \nabla + \nabla \delta = \frac{i}{\hbar} \ad_\bullet \hat{T}$,
					\item\label{square_Yano_fin} $\nabla^2 = - \frac{i}{\hbar} \ad_\bullet \hat{R}$,
					\item\label{extra_fin} $\delta \hat{T} = 0$, $\nabla \hat{T} = \delta \hat{R}$, and $\nabla \hat{R} = 0$,
				\end{enumerate}
				where $\ad_\bullet(t) := [t, \cdot]_\bullet$ is the adjoint action defined via the $\deg_a$-graded commutator in $\Omega(S,\cW)$.\\
				Moreover, the Fedosov operators satisfy a Hodge-type decomposition
				\begin{enumerate}[label=(\roman*)]
					\setcounter{enumi}{5}
					\item\label{Hodge_fin} $\delta\delta^{-1} + \delta^{-1} \delta + \tau = \id$,
				\end{enumerate}
				where $\tau$ is the projection on the $\deg_s,\deg_a=0$ part of $\Omega(S,\cW)$.\\
			\end{lemma}
			
		Following Fedosov, one makes the following ansatz for our desired flat connection operator, called $D$:
			\begin{equation}\label{Fedosov_conn_fin}
				D := \nabla - \delta + \frac{i}{\hbar} \ad_\bullet(r),
			\end{equation}
		where $r=r^\dagger$ is a suitable $1$-form with values in $\cW$ that we need to construct. A sufficient condition on $r$ to ensure the flatness of $D$ is
			\begin{equation}\label{r_eq}
				\nabla r - \delta r + \frac{i}{\hbar} r \bullet r -\hat{R} - \hat{T} = \Omega,
			\end{equation}
		where $\Omega$ is a closed 2-form valued in $\bC[[\hbar]]$, i.e. $\Omega=\sum_{k \geq 1} \hbar^k \Omega_k$ and each $\Omega_k$ is a real valued closed $2$-form on $S$.\\
		The following theorems are modifications of the original results of Fedosov~\citep[thm. 3.2 and thm. 3.3]{F94} to connections with non-vanishing torsion (as discussed in~\citep{KS01}) and to non-trivial $\Omega$ and $s$ (as detailed in~\citep{neumaier2003universality}).
			\begin{theo}[Fedosov's 1st Theorem]\label{theo_Fedosov_1_fin}
				There is a unique element $r \in \Omega^1(S,\cW)$ satisfying the equation~\eqref{r_eq} for any closed $\bC[[\hbar]]$-valued 2-form $\Omega$ under the requirements
					\begin{equation}\label{r_reqs}
						r = r^\dagger, \quad r^{(0)}=r^{(1)}=0, \quad (\delta^{-1} r)^{(k)} = s^{(k)},
					\end{equation}
				where $r^{(k)}$ denotes the component homogeneous in $\Deg$ of degree $k$, and where $s \in C^\infty(S,\cW)$ is some arbitrary self-adjoint element with $\Deg s \geq 3$.\\
				Consequently, the Fedosov connection $D$ defined via eq.~\eqref{Fedosov_conn_fin} is flat, satisfies the Leibniz rule with respect to the product $\bullet$, and is compatible with the hermitian conjugation operation $\dagger$.
			\end{theo}
		\noindent
		Note that the Fedosov connection $D$ depends only on the following input: the (non-flat) Yano connection, which in turn depends on $J$ and $G$ or, equivalently, $\sigma$ and $G$, the closed form $\Omega$ on $S$ taking values in $\bC[[\hbar]]$, and the datum $s$ (subject only to the constraint $\Deg s \geq 3$). We refer to $\Omega$ and $s$ as {\em ``auxiliary data''}. We will mostly use the Fedosov's First Theorem for the case of $\Omega = s = 0$, but it will occasionally be necessary to have the more general form with non-vanishing auxiliary data.\\
		\noindent
		Once we have defined the Fedosov connection, we can perform the last step in the construction of the deformation quantization, which is encoded in the following theorem (see~\citep{F94}).
			\begin{theo}[Fedosov's 2nd Theorem]\label{theo_Fedosov_2_fin}
				Let $\tau$ be the projection of a smooth section on $\cW$ onto its component with $\deg_s=0$.  For each $f \in C^\infty(S)[[\hbar]]$, there exists a unique $t \in C^\infty(S,\cW)$ such that
					\begin{equation*}
						D t = 0, \mbox{ i.e. } t \in \ker D, \qquad \tau t= f.
					\end{equation*}
				In other words, the restriction to $C^\infty(S,\cW) \cap \ker D$ of the projection $\tau$ is a bijection. Let us denote its inverse by
					\begin{equation*}
						\tau^{-1}: C^\infty(S)[[\hbar]] \to C^\infty(S,\cW) \cap \ker D.
					\end{equation*}
				Then
					\begin{equation*}
						f \star h := \tau \left( \tau^{-1}(f) \bullet \tau^{-1}( h) \right)
					\end{equation*}
				is a star-product, and the standard conjugation map $f \mapsto \overline{f}$ gives $(C^\infty(S)[[\hbar]], \star)$ the structure of a $*$-algebra, i.e. $\overline{(f \star h)} = \overline{h} \star \overline{f}$.
			\end{theo}
			\begin{proof}
				The proof of this theorem is given in~\citep{F94} with the exception of the statement concerning the hermitian conjugation. This can be seen as follows. First of all, we notice that $\tau( t^\dagger) = \overline{\tau(t)}$, simply because $t^\dagger = (\overline{t^0}, \dots)$ for any $t = (t^0, \dots) \in C^\infty(S,\cW)$. Then, for any $f \in C^\infty(S)[[\hbar]]$ the sections $\tau^{-1}(\overline{f})$ and $(\tau^{-1} (f))^\dagger$ satisfy
					\begin{equation*}
						D (\tau^{-1}(f))^\dagger = (D \tau^{-1}( f))^\dagger = 0 = D \tau^{-1}(\overline{f}), \qquad \tau (\tau^{-1} (f))^\dagger = \overline{\tau \tau^{-1} (f)} = \overline{f} = \tau \tau^{-1} (\overline{f}).
					\end{equation*}	
				As consequence of the first part of the theorem, there is a unique $D$-flat section in $C^\infty(S,\cW)$ such that $\overline{f}$ is its component with $\deg_s=0$. Therefore, $\tau^{-1}(\overline{f}) = (\tau^{-1} (f))^\dagger$, which implies straightforwardly the statement about the $*$-algebra structure.
			\end{proof}
			
		We conclude this section by giving some details concerning the construction, see~\citep{F94, KS01, neumaier2003universality}. 
			\begin{rem}\label{rem_Fedosov}
				\begin{enumerate}
					\item The $\cW$-valued $1$-form $r$ is constructed iteratively. For the case $\Omega=0=s$, it is defined by
						\begin{equation*}
							\begin{split}						
								&r^{(2)}=\delta^{-1}\hat{T}, \quad r^{(3)}=\delta^{-1}\left(\hat{R} + \nabla r^{(2)} - \frac{i}{\hbar}r^{(2)} \bullet r^{(2)} \right),\\
								&r^{(3+\ell)} = \delta^{-1} \left( \nabla r^{(\ell+2)} - \frac{i}{\hbar}\sum_{\ell' \leq \ell} r^{(\ell' +2)} \bullet r^{(\ell-\ell'+2)}\right).
							\end{split}
						\end{equation*}
					\item The map $\tau^{-1}$ is a formal quantization map, in the sense that it takes a classical observable $f \in C^\infty(S)$ to an element in the non-commutative algebra $C^\infty(S,\cW) \cap \ker D$. Moreover, $\tau^{-1}(f)$ is constructed iteratively
						\begin{equation*}
							\begin{split}						
								&(\tau^{-1} f)^{(0)}=f,\\
								&(\tau^{-1} f)^{(\ell+1)} = \delta^{-1} \left( \nabla (\tau^{-1} f)^{(\ell)} - \frac{i}{\hbar}\sum_{\ell' \leq \ell} [r^{(\ell'+2)}, (\tau^{-1} f)^{\ell-\ell'}]_{\bullet}\right).
							\end{split}
						\end{equation*}	
					\item Fedosov's theorems, even in their generalized versions with non-vanishing data $\Omega$ and $s$, are valid if instead of the Wick product $\bullet$, we consider the Weyl-Moyal product $\diamond$, defined on $\cW_x$ as
							\begin{equation}\label{weyl_prod_fin}
								t \diamond t' := m \left( \exp \left(i \frac{\hbar}{2} \sigma_x^{ij} \partial_{y^i} \otimes \partial_{y^j} \right) t \otimes t' \right).
							\end{equation}	
						The product $\diamond$ is given (fiberwisely) by the same formula as~\eqref{star_free}, except that $\omega^{ij}$ is now replaced by $i/2 \sigma^{ij}$. The definition is extended to forms with values in $\cW$ similarly as done in~\eqref{graded_prod_fin} for the product $\bullet$ (trivial action on the anti-symmetric part). One can directly check that relations~\ref{adj_action_Fedosov_op_fin}-\ref{extra_fin} in lemma~\ref{lemma_various_res_fin} still hold for the product $\diamond$. The reason why we do not use $\diamond$ throughout this work is that it is not suitable for generalization to quantum field theories.
				\end{enumerate}
			\end{rem}

\section{Equivalence of the Fedosov quantization of two different almost-K\"{a}hler structures}\label{subsec_equivalence_fin}
		The construction of the star product on $(S, \sigma)$ we outlined in the previous section depends on a choice of almost-K\"{a}hler structure $J$, or equivalently of the almost-K\"{a}hler section $\omega$. Consider a given symplectic manifold $(S, \sigma)$ endowed with two different almost-K\"{a}hler sections $\omega$ and $\omega'$ compatible with the {\em same} $\sigma$. It is then natural to ask how the corresponding star-products (quantizations) are related. We will answer this question in the present section. Our analysis is based on a construction of Neumaier~\citep{neumaier2003universality}. The author was concerned with the case that $J, J'$ define two K\"{a}hler structures, whereas we need to consider the almost-K\"{a}hler situation.\\
		
		We first make a general definition. 
			\begin{defi}\label{def_equiv}
				Consider two deformations $(C^\infty(S)[[\hbar]],\star)$ and $(C^\infty(S)[[\hbar]],\star')$ of the classical algebra $(C^\infty(S),\{ \cdot, \cdot \})$. The star-products are called {\em equivalent} if there is an isomorphism of algebras $B:(C^\infty(S)[[\hbar]],\star) \to (C^\infty(S)[[\hbar]],\star')$ such that
					\begin{equation*}
						B=\id + \hbar B_1 + \hbar^2 B_2 + \dots,
					\end{equation*}								
				where each $B_\ell$ is given by a map $C^\infty(S) \to C^\infty(S)$ which vanishes on constant functions. 
			\end{defi}
			
		We would like to decide whether two star-products on two almost-K\"{a}hler manifolds, obtained using Fedosov's method, are equivalent, and we would also 
like to give explicitly the corresponding isomorphisms. For this, we must look at Fedosov's construction associated with the two given almost-K\"{a}hler sections $\omega,\omega'$. Following the previous subsection, we refer to $G$, $\bullet$, $\nabla$, $T$, and $R$ as the Riemannian metric, the Wick product, the Yano connection, its torsion, and its Riemann tensor corresponding to $\omega$. Similarly, $G'$, $\bullet'$, $T'$ and $R'$ are the corresponding quantities associated with $\omega'$. As vector spaces, the algebras $\Omega(S,\cW)$ and $\Omega(S,\cW')$ coincide. The difference is in the choice of the product, respectively $\bullet$ and $\bullet'$. Fedosov's construction, in particular the Fedosov's First Theorem (thm.~\ref{theo_Fedosov_1_fin}), provides for the algebra $\Omega(S,\cW)$ a flat connection $D= -\delta + \nabla -i/\hbar \ad_{\bullet} (r)$ corresponding to the first almost-K\"{a}hler section $\omega$, and similarly for $\Omega(S,\cW')$ a flat connection $D':= -\delta + \nabla' -i/\hbar \ad_{\bullet'} (r')$ corresponding to the second almost-K\"{a}hler section $\omega'$. The connections $D$ and $D'$ are uniquely determined by $\omega$ and respectively $\omega'$ if we assume, as we will, that the associated auxiliary data $\Omega, s$ and respectively $\Omega', s'$ are zero. We first observe:
			\begin{lemma}\label{lemma_Wick_iso_fin}
				$\Omega(S,\cW)$ and $\Omega(S,\cW')$ are isomorphic as algebras.
			\end{lemma}
		\begin{proof}
			We first consider $\cW(\bR^{2n}, \omega)$ with constant $\omega$, i.e. with constant K\"{a}hler structure $J$ with respect to a fixed constant symplectic form $\sigma$, and $\cW(\bR^{2n}, \omega')$, where $\omega'$ corresponds to another constant K\"{a}hler structure $J'$ with respect to the same symplectic form $\sigma$. These two algebras are isomorphic and the isomorphism $\alpha: \cW(\bR^{2n}, \omega) \to \cW(\bR^{2n}, \omega')$ is explicitly given by
			\begin{equation}\label{iso_fin}
				\alpha = \exp\left( \frac{\hbar}{2} (\omega - \omega')^{ij} \partial_{y^i} \partial_{y^j} \right).
			\end{equation}
		For a monomial $t_{i_1 \dots i_n} y^{i_1} \cdots y^{i_n}$ we have
			\begin{equation}\label{iso_fin_alt}
				\alpha(t_{i_1 \dots i_n} y^{i_1} \cdots y^{i_n}) = \sum_{k=0}^{[n/2]} \frac{\hbar^k n!}{(2k)! (n-2k)!} t_{i_1 \dots i_n} (\omega' - \omega)^{(i_1 i_2} \cdots (\omega' - \omega)^{i_{2k-1} i_{2k}} y^{i_{2k+1}} \cdots y^{i_n)}.
			\end{equation}
		The same construction then gives an isomorphism $\alpha_x : \cW_x \to \cW'_x$ for any fiber. Since both $\omega$ and $\omega'$ are smooth sections, and since the algebra structures $\bullet$ and $\bullet'$ are defined fiberwise, we obtain an isomorphism $\alpha$ for $C^\infty(S,\cW)$ and $C^\infty(S,\cW')$.\\
		 We can extend naturally the map defined in~\eqref{iso_fin},~\eqref{iso_fin_alt} as an isomorphism $\Omega(S,\cW) \to \Omega(S,\cW')$, i.e. we allow non-trivial anti-symmetric degree. This concludes the proof.
		\end{proof}
		
		The Fedosov connection $D'$ on $\cW'$ can be pulled back to a connection on $\cW$ via the bundle map $\alpha$. We denote this pull-back by $D^\alpha := \alpha^{-1} D' \alpha$. Concerning this connection $D^\alpha$, we have the following result.
			\begin{lemma}\label{lemma_pullback_Fedosov}
				$D^\alpha$ is a Fedosov connection. More precisely, $D^\alpha$ coincides with the covariant derivative obtained from Fedosov's first theorem (thm.~\ref{theo_Fedosov_1_fin}) with respect to the product $\bullet$ and characterized by the following input data: the Yano connection $\nabla$, $\Omega^\alpha = 0$ and $s^\alpha \in C^\infty(S,\cW)$ with $\Deg s^\alpha \geq 3$ given by~\eqref{s_alpha}.
			\end{lemma}
		\begin{proof}
			First of all, $D^\alpha$ as a map $C^\infty(S,\cW) \to \Omega^1(S, \cW)$ is linear by definition. Since $D'$ is flat, it follows immediately that $D^\alpha$ is also flat, $(D^\alpha)^2 = \alpha^{-1} (D')^2 \alpha = 0$.\\
			By definition, $\alpha$ is an algebra isomorphism $\cW \to \cW'$, therefore $t \bullet s = \alpha^{-1} ((\alpha t) \bullet' (\alpha s))$ for any $t, s \in C^\infty(S,\cW)$. The Leibniz rule follows from this consideration and the properties of $D'$, i.e.
				\begin{align}
					D^\alpha (t \bullet s) &= (\alpha^{-1} D') ((\alpha t) \bullet' (\alpha s)) = \alpha^{-1} \left\{  (D' \alpha t) \bullet' (\alpha s) + (\alpha t) \bullet' (D'\alpha s) \right\} \nonumber \\
					&= (D^\alpha t) \bullet s + t \bullet (D^\alpha s).
				\end{align}
			By definition $\alpha$, (and also $\alpha^{-1}$) acts as the identity on the elements of $\Omega(S,\cW)$ with $\deg_s = 0$. Therefore, $D^\alpha(f) = df$ for any $f \in C^\infty(S)$. Keeping this in mind, it follows as a particular case of the Leibniz rule that $D^\alpha(f \cdot t) = f \cdot D^\alpha (t) + df \otimes t$, i.e. the linear map $D^\alpha$ is indeed a connection.\\
			To prove that $D^\alpha$ is a Fedosov connection, first we note that $\delta \alpha = \alpha \delta$ as follows from~\ref{adj_action_Fedosov_op_fin} and the definitions involved. It is clear that we can rewrite the derivative $D^\alpha = \alpha^{-1} D' \alpha$ in the following form
				\begin{equation}\label{D^alpha_fin}
					D^\alpha = - \delta + \alpha^{-1} \nabla' \alpha + \frac{i}{\hbar} \ad_{\bullet} (\alpha^{-1} r').
				\end{equation}
			We express the difference between $\nabla$ and $\alpha^{-1} \nabla' \alpha$ as
				\begin{equation}\label{alpha_on_Yano}
					\alpha^{-1} \nabla' \alpha = \nabla - \frac{i}{\hbar} \ad_{\bullet} (C),  
				\end{equation}
			where 
				\begin{equation}\label{c_fin}
					C := \frac{1}{2} \sigma_{j_1 \ell} (\Gamma' - \Gamma)^\ell_{i j_2} dx^i \otimes y^{j_1} y^{j_2}, 
				\end{equation}
			and where $\Gamma, \Gamma'$ are the Christoffel symbols~\eqref{Christoffel} for the Yano connections corresponding to $\omega$ and $\omega'$ respectively. Then, one finds that $D^\alpha$ can be written as:
				\begin{equation}\label{fedprimetilde_bis}
					D^\alpha = -\delta + \nabla + \frac{i}{\hbar}\ad_{\bullet} (r^\alpha),
				\end{equation}
			where $r^\alpha =\alpha^{-1} r' - C$. The map $\alpha^{-1}$ changes neither the total degree $\Deg$ nor the antisymmetric degree $\deg_a$, then $r^\alpha$ is $\Deg \geq 2$ and ${r^\alpha}^{(0)}={r^\alpha}^{(1)}=0$. A direct computation shows that
				\begin{equation}\label{technical_proper_c}
					\delta C = \hat{T} - \hat{T}', \qquad \alpha^{-1}\hat{R}' =  \hat{R}  + \nabla C - \frac{i}{\hbar} C \bullet C,
				\end{equation}
			and then we straightforwardly obtain
				\begin{equation}\label{r^alpha_fin}
					\begin{split}
						\delta r^\alpha &= \alpha^{-1} \delta r' - \delta C = \alpha^{-1} ( \nabla' r' - \hat{R}' - \hat{T}' + \frac{i}{\hbar} r' \bullet' r')  - (\hat{T} - \hat{T}') \\
						&= (\alpha^{-1} \nabla' \alpha)(\alpha^{-1} r') - \hat{R} - \hat{T} + \frac{i}{\hbar} (\alpha^{-1} r') \bullet (\alpha^{-1} r') + (\hat{R} - \alpha^{-1} \hat{R}') \\
						&= \nabla r^\alpha - \hat{R} - \hat{T} + \frac{i}{\hbar} r^\alpha \bullet r^\alpha.
					\end{split}
				\end{equation}
			Therefore, the derivative $D^\alpha$ coincides with the flat covariant derivative obtained from Fedosov's first theorem with respect to the product $\bullet$ and is uniquely defined by the input data $\nabla$, $\Omega^\alpha := 0$, and
				\begin{equation}\label{s_alpha}
					s^\alpha:=\delta^{-1} \alpha^{-1} r' - \delta^{-1} C,
				\end{equation}
			as we wanted to prove.
		\end{proof}

		Since $D$ and $D^\alpha=\alpha^{-1} D' \alpha$ are derivative operators for the same algebra $\Omega(S,\cW)$ (both satisfy the Leibniz rule with respect to the product $\bullet$), we can compare them, unlike $D$ and $D'$. Our claim is that they are gauge equivalent, in the sense explained below, and this implies that the star-products $\star$, $\star'$ on $C^\infty(S)[[\hbar]]$ are equivalent. 
			\begin{theo}\label{thm1_bis}
				There exists a smooth section $H \in C^\infty(S,\cW)$ such that $\Deg H \geq 3$, $\tau H =0$, $H^\dagger = H$ and
					\begin{equation}\label{interpolation_bis}
						D = \exp \left( \frac{i}{\hbar} \ad_\bullet(H) \right)  \alpha^{-1} D' \alpha  \exp \left( - \frac{i}{\hbar} \ad_\bullet (H) \right).
					\end{equation}
				In particular, a solution $H$ for \eqref{interpolation_bis} is uniquely determined by a closed $1$-form $\theta \in \Omega^1(S)[[\hbar]]$. For $\theta=0$ the solution $H = \sum H^{(k)}$ (where $\Deg H^{(k)} = k$) is given by  
						\begin{equation}\label{low_H_bis}
								\begin{split}
									&H^{(0)} = H^{(1)} = H^{(2)} = 0\\
									&H^{(3)} = \frac{1}{2} \sigma_{j_1\ell} (\Gamma' - \Gamma)^\ell_{j_2 j_3} y^{j_1} y^{j_2} y^{j_3} - \frac{\hbar}{4} J^\ell_k (\Gamma' - \Gamma)^k_{j\ell} y^j = \delta^{-1} C, 
								\end{split}
							\end{equation}
						and by the following recursive definition for $H^{(k)}$ with $k >3$
							\begin{equation}\label{iter_H_bis}
								\begin{split}
									&H^{(k+1)} = \delta^{-1} \left(\nabla H^{(k)} - \frac{3 i}{2 \hbar} \left[ C , H^{(k)} \right]_\bullet + \frac{i}{2\hbar} \sum_{\ell=0}^{k-3} \left[ 3 \alpha^{-1} (r')^{(2+\ell)} - r^{(2 + \ell)}, H^{(k-\ell)} \right]_\bullet + \right. \\
									&\qquad + \alpha^{-1} (r')^{(k)} - r^{(k)}  - \sum_{\lambda = 2}^{k-2} n_\lambda \left(\frac{i}{\hbar} \right)^\lambda \sum_{\ell_1+ \dots +\ell_\lambda = k + 2\lambda - 2} [H^{(\ell_1)}, \dots [H^{(\ell_\lambda)}, C]_\bullet \dots ]_\bullet + \\
									&\qquad \left. + \sum_{\lambda = 2}^{k-2} n_\lambda \left(\frac{i}{\hbar} \right)^\lambda \sum_{\ell=0}^{k-2-\lambda} \sum_{\ell_1+ \dots +\ell_\lambda = k - \ell + 2\lambda - 2} [H^{(\ell_1)}, \dots [H^{(\ell_\lambda)}, \alpha^{-1} (r')^{(2+\ell)} - r^{(2+\ell)}]_\bullet \dots ]_\bullet \right),
								\end{split}
							\end{equation}
						where $C$ is given by eq.~\eqref{c_fin}, and where the numbers $\ell_i$ are all taken $\geq 3$. The numbers $n_\lambda$ are defined recursively through eq.~\eqref{n_coefficient_bis}.\\
				Furthermore, the Fedosov star-products $\star$ and $\star'$ are equivalent and the isomorphism $B: (C^\infty(S)[[\hbar]], \star) \to (C^\infty(S)[[\hbar]], \star')$ is explicitly given by
					\begin{equation}\label{isomorphism_bis}
						B(f) := \tau \alpha \exp \left( \frac{i}{\hbar} \ad_\bullet(H) \right) \tau^{-1}(f),
					\end{equation}
				Here $\tau^{-1}$ is the ``quantization map'' as defined in the Fedosov's Second Theorem (thm.~\ref{theo_Fedosov_2_fin}). 
			\end{theo}
			\begin{proof}
				The proof is very similar to the one presented in~\citep[prop. 3.2 ii]{neumaier2003universality},~\citep[prop. 3.5.3, 3.5.4]{neumaier2001klassifikationsergebnisse}. The set-up considered in these reference differs to our case mainly for two points: (1) it is assumed that K\"{a}hler manifold $S$ and the connection $\nabla$ is torsion free (while in our case we cannot exclude a non-vanishing torsion), and (2) an equivalence of the type~\eqref{interpolation_bis} is derived not for $D, D^\alpha$, but for a more general pair of Fedosov connections $D_1, D_2$ corresponding to auxiliary data $(\Omega_1, s_1), (\Omega_2, s_2)$ such that $\Omega_1 - \Omega_2 = d \theta$ for a general $\theta \in \Omega^1(S)[[\hbar]]$ (in our case $\theta$ is necessarily closed). The argument exploited in~\citep{neumaier2001klassifikationsergebnisse, neumaier2003universality} consists in rewriting eq.~\eqref{interpolation_bis} in a form suitable for applying the fixed-point theorem with respect to the total degree.\\
				We proved in lemma~\ref{lemma_pullback_Fedosov} that $D^\alpha: \Omega^k(S,\cW) \to \Omega^{k+1}(S,\cW)$ respects the product $\bullet$. As an algebraic consequence of this fact\footnote{In the terminology of~\citep{neumaier2001klassifikationsergebnisse}, $D^\alpha$ is a ``$\bullet$-superderivative'' and then eq.~\eqref{rewrite_e_D_alpha_e_-1} is  given by~\citep[lemma 1.3.20]{neumaier2001klassifikationsergebnisse}.} and the assumptions on $H$, it holds that
					\begin{equation}\label{rewrite_e_D_alpha_e_-1}
						\begin{split}
							&\exp \left( \frac{i}{\hbar} \ad_\bullet(H) \right)  \alpha D' \alpha^{-1}  \exp \left( - \frac{i}{\hbar} \ad_\bullet (H) \right) =\\
							&\quad = D^\alpha - \frac{i}{\hbar} \ad_\bullet \left( \sum_{k \geq 0} \frac{1}{(k+1)!} \left(\frac{i}{\hbar} \ad_\bullet(H) \right)^k (D^\alpha H) \right)\\
							&\quad =D - \frac{i}{\hbar} \ad_\bullet \left( r - r^\alpha + \sum_{k \geq 0} \frac{1}{(k+1)!} \left(\frac{i}{\hbar} \ad_\bullet(H) \right)^k (D^\alpha H) \right).
						\end{split}
					\end{equation}
				Therefore, eq.~\eqref{interpolation_bis} holds for the section $H$ if and only if there is an element $\theta \in \Omega^1(S)[[\hbar]]$ such that
					\begin{equation}\label{equiv_H-formula_bis}
						r - r^\alpha + \sum_{k \geq 0} \frac{1}{(k+1)!} \left( \frac{i}{\hbar} \ad_\bullet(H) \right)^k (D^\alpha H) = \theta,
					\end{equation}
				i.e. the section in $\Omega^1(S,\cW)$ given by the left-hand side of~\eqref{equiv_H-formula_bis} must be in the center of the algebra.\\
				Although in~\citep[lemma 3.5.1]{neumaier2001klassifikationsergebnisse} the author considered the case of $\nabla$ torsionless, an inspection of his proof shows that the presence of a non-vanishing torsion can at most affect the following equation used in the aforementioned lemma
					\begin{equation*}
						D^\alpha (r^\alpha - r) = \frac{i}{\hbar} (r^\alpha -r) \bullet (r^\alpha -r).
					\end{equation*}						
				This equation is still valid in our case as can be seen directly from the definitions of $r$ and $r^\alpha$ and, consequently, the results of~\citep[lemma 3.5.1]{neumaier2001klassifikationsergebnisse} are valid also in our case. In particular, it is necessary for eq.~\eqref{equiv_H-formula_bis} to be solvable that $\theta$ is closed because $d\theta = \Omega - \Omega^\alpha = 0$.\\
				Making use of the Hodge-type decomposition~\ref{Hodge_fin} and the assumption $\tau H = 0$, we get that $H$ is a solution to eq.~\eqref{equiv_H-formula_bis} if $H$ solves the following equation
					\begin{equation}\label{req_H-formula_bis}
						H=\delta^{-1} \left( \theta + \nabla H+ \frac{i}{\hbar}\ad_{\bullet} (r^\alpha)(H) + \sum_{\lambda \geq 0} n_\lambda \left( \frac{i}{\hbar} \right)^\lambda \left( \ad_{\bullet} (H)\right)^{\lambda}(r^\alpha - r) \right),
					\end{equation}
				where $n_\lambda$ are real numbers such that $\sum_\lambda n_\lambda x^\lambda$ is the inverse of the formal power series $\sum_k 1/(k+1)! x^k$, i.e. $n_\lambda$ are defined recursively by
					\begin{equation}\label{n_coefficient_bis}
						n_0 = 1, \qquad n_{\lambda >0} = - \sum_{\lambda' = 1}^\lambda \frac{1}{(\lambda' +1)!} n_{\lambda - \lambda'}.
					\end{equation}
				Now proceeding as in the proof of~\citep[prop. 3.2 ii]{neumaier2003universality},~\citep[prop. 3.5.3, 3.5.4]{neumaier2001klassifikationsergebnisse}. We first notice that the right-hand side of eq.~\eqref{req_H-formula_bis} is in the form $L(H)$, where $L$ is a contracting map with respect to the total degree. Therefore, the fixed point theorem guarantees the existence and the uniqueness of the solution $H$ to eq.~\eqref{req_H-formula_bis}. Arguing as in the aforementioned references, we verify that such $H$ solves eq.~\eqref{equiv_H-formula_bis} and necessarily also eq.~\eqref{interpolation_bis}.\\
				Finally, the map $B$ defined via~\eqref{isomorphism_bis} is indeed a star-isomorphism as follows immediately from~\eqref{interpolation_bis} and the definitions of $\alpha$, $\star$ and $\star'$.
		\end{proof}
		\begin{rem}
			There are already several results in the literature relating the gauge equivalence of certain star products based on cohomological considerations. One associates to a star-product $\star$ on a symplectic manifold its Deligne's characteristic class $\cl(\star) \in i/\hbar [\sigma] + H_{\mathrm{dR}}^2(S)[[\hbar]]$, where $H^2_{\mathrm{dR}}(S)[[\hbar]]$ are formal power series in $\hbar$ with values in the de Rham cohomology of $S$. According to~\citep{gutt1999equivalence} this class consists of two different parts
				\begin{equation*}
					\cl(\star) = c_0(\star) + \frac{i}{\hbar} d(\star),
				\end{equation*}
			namely, the zero-th order term $c_0(\star)$, and the Deligne's intrinsic class
				\begin{equation*}
					d(\star)=\sum_{k=0}^\infty \frac{1}{k!} \left( \frac{\hbar}{i} \right)^k d_{k}(\star),
				\end{equation*}
			which can be constructed in terms of local derivation (see e.g.~\citep{neumaier2002local, gutt1999equivalence}). By construction the zero-th order term $d_0(\star)$ in the formal power series is $[\sigma]$, while the first order term $d_1(\star)$ vanishes. As proven in~\citep{gutt1999equivalence}, the Deligne's characteristic class specifies uniquely the equivalence class of a given star-product. Said differently, two star-products are equivalent if and only if their Deligne's characteristic classes coincide. For the star-products $\star$, $\star'$ we are considering, computing the corresponding Deligne's characteristic classes and checking that they coincide\footnote{In a nutshell it is just needed to adapt \citep{neumaier2002local, neumaier2003universality} to connections with non-vanishing torsion and to combine with the direct computation of $c_0$ presented in~\citep{KS01}.} is another possible line of argument leading to thm.~\ref{thm1_bis}, but without explicit formula for $H$.
	\end{rem}

\chapter{Reformulation of perturbative quantum field theory}\label{sec_pQFT}
		Quantum field theories are often based on classical field theories described by a Lagrangian or Hamiltonian. Such theories, thus, have a symplectic structure at the classical level. Therefore, it is conceivable that Fedosov's method of quantization could be applied to such systems. The difference to the symplectic manifolds discussed so far is, of course, that in a field theoretic setting, the manifold $S$ is infinite-dimensional as it corresponds to the space of classical solutions of the equations of motion, or their initial data. However, even ignoring this point, if one looks at standard presentations of the quantization of field theories, the connection to Fedosov's method is absolutely not evident even at a purely formal level. The purpose of this work is to explain this connection. This is  straightforward for {\em free} quantum fields theories--all we need to do is to properly interpret~\citep{DF01a}. We are going to present the details in sec.~\ref{subsec_free_QFT}. The situation is much more involved for {\em interacting} quantum field theories. For those, we will first present the method of causal perturbation theory (see~\citep{BFK96, brunetti2000microlocal, DF01a, DF01b, HW01, HW02, HW03, brunetti2003generally, DF04, HW05}) in sec.~\ref{subsec_int_QFT_per}. At the end of this section, we present an interesting consequence of the principle of perturbative agreement which gives a first hint to a possible connection to Fedosov's method. This connection will then be established step by step in the remaining sections.

\section{Free scalar field in curved spacetime}\label{subsec_free_QFT}
		We present our formalism first for the {\em free, real, scalar, Klein-Gordon field} on a {\em Lorentzian manifold} $(M,g)$. This case should be thought of as the ``model case'' in the same sense as the basic example of remark~\ref{rem_key_ex} is the ``model case'' in the finite-dimensional framework. The Klein-Gordon field is denoted by $\varphi$ in the following. The field equation (with source) is
			\begin{equation}\label{K-G_w_source}
				j = (\boxempty - m^2 - v) \varphi, 
			\end{equation}
		where $\boxempty$ is the d'Alembertian operator (wave operator) associated with the metric $g$, where $j \in C^\infty(M)$ is some fixed source, and $v \in C^\infty(M)$ is some smooth external potential. In order for this theory to behave reasonably, we need to assume that the underlying space-time $(M,g)$ is a {\em globally hyperbolic} manifold (see e.g.~\citep{wald1984general}). This means that $M$ has a {\em smooth Cauchy surface}, i.e. a surface $\Sigma$ such that every causal, inextendible curve intersects $\Sigma$ precisely once. Under this assumption, as proved e.g. in~\citep{BGP07}, the Klein-Gordon equation has a unique solution $\varphi$ for any choice of initial data $(q,p) \in C^\infty_0(\Sigma) \times C^\infty_0(\Sigma)$ satisfying
			\begin{equation*}
				\varphi|_\Sigma = q, \qquad \partial_n \varphi|_\Sigma = p. 
			\end{equation*}
		Here $\partial_n$ is the normal derivative to $\Sigma$.  The hallmark of the Klein-Gordon equation is the causal propagation of disturbances: if $(q,p)=0$ and if the support of $j$ is contained in some subset $O \subset M$, then the support of the corresponding solution $\varphi$ is contained in $J^+(O) \cup J^-(O)$, where $J^\pm(O)$ denote the causal future/past of $O$. The solutions of the Klein-Gordon equation can be obtained in terms of the {\em advanced/retarded fundamental solutions} $E^{A/R}$. The action $E^{A/R}(j)$ on a compactly supported smooth source $j \in C^\infty_0(M)$ is defined by demanding that $\varphi = E^{A/R}(j)$ is the unique solution to the Klein-Gordon equation~\eqref{K-G_w_source} having initial data $(q,p) =0$ on some Cauchy surface in the future/past of the support of $j$. The advanced and retarded fundamental solutions are continuous functions $E^{A/R}:C^\infty_0(M) \to C^\infty(M)$, and so, as a consequence of the Schwartz Kernel theorem (see e.g.~\citep{H83}), $E^{A/R}$ may be viewed alternatively as distributions on $M \times M$. In the distributional sense, we have
			\begin{equation*}
				(\boxempty - m^2 - v)_{x_1} E^{A/R}(x_1,x_2) = \delta(x_1,x_2), \quad \supp(E^{A/R}(x_1,x_2)) \subset \{ (x_1,x_2) \in M^2 | x_1 \in J^\mp(x_2) \},
			\end{equation*}
		and
			\begin{equation*}
				E^{A/R}(x_1,x_2) = E^{R/A}(x_2,x_1).
			\end{equation*}
		Furthermore, the wave-front sets of $E^{A/R}$ are well-known (see~\citep{duistermaat1972fourier}) and they take the form
			\begin{equation}\label{A/R_WF}
				\WF(E^{A/R}) = \cC^{A/R}(M),
		\end{equation}
	where the sets $\cC^{A/R}(M)$ are defined by
		\begin{equation}\label{C^A/R}
			\cC^{A/R}(M) :=\left\{ (x_1, x_2; k_1, k_2) \in  \dot{T}^\ast M^2 : x_1 \in J^\mp (x_2), \; (x_1, k_1) \sim (x_2, -k_2) \mbox{ or } x_1=x_2, k_1=-k_2 \right\}.
		\end{equation}
		The {\em causal propagator}, also called ``commutator function'' in some references, is the quantity
			\begin{equation*}
				E= E^A - E^R.
			\end{equation*}
		In the distributional sense, $E$ is a bi-solution for the homogeneous ($j=0$) Klein-Gordon equation~\eqref{K-G_w_source} and it is anti-symmetric, $E(x_1,x_2) = -E(x_2,x_1)$. A well-known computation (see~\citep{duistermaat1972fourier}) implies that
			\begin{equation}\label{WF_causal}
				\WF(E(x_1,x_2)) = \left\{ (x_1,x_2;k_1,k_2) \in \dot{T}^\ast M^2: (x_1,k_1) \sim (x_2,-k_2) \right\}.
			\end{equation}

		Let $S$ be the space of smooth, spatially compact solutions to the Klein-Gordon equation~\eqref{K-G_w_source}. For $j=0$, this space is obviously linear. Let us focus, for the moment, on this case. We will show that $S$ carries a natural symplectic form and we will describe a way to realize the construction of the finite-dimensional ``model case'', i.e. the basic example of remark~\ref{rem_key_ex}, for case of this homogeneous Klein-Gordon equation.  For each $f \in C^\infty_0(M)$, let $\varphi(f)$ be the map $S \to \bR$ which assigns to a smooth solution $u \in S$ of the homogeneous Klein-Gordon equation, i.e. $(\boxempty - m^2 - v)u = 0$, its $f$-weighted average, namely
			\begin{equation*}
				\varphi(f)[u] := \int_M f(x) u(x) dx. 
			\end{equation*}
		Thus, $\varphi(f)$ defines an element in the dual $S^*$. More generally, we may consider ``observables'' of the form $\varphi^{\otimes n}(f^{(n)})$ defined by
			\begin{equation*}
				\varphi^{\otimes n}(f^{(n)})[u] := \int_{M^n} f^{(n)}(x_1, \dots, x_n) u(x_1) \cdots u(x_n) dx_1 \dots dx_n,
			\end{equation*}
		where $u \in S$, and where $f^{(n)}$ is a smooth, complex-valued, symmetric function of compact support on $M^n$. In a moment we will even allow certain distributional $f^{(n)}$'s. Note that as a functional on solutions $\varphi^{\otimes n}(f^{(n)}) = 0$ if the $f^{(n)}$'s satisfy the relation 
			\begin{equation}\label{equiv_rel_lin}
				f^{(n)}(x_1, \dots, x_n) = (\boxempty - m^2 - v)_{x_i} h^{(n)}(x_1, \dots, x_n)
			\end{equation}
		for some smooth $h^{(n)}$ of compact support. It follows, in particular, that $f^{(1)} \mapsto \varphi(f^{(1)})$ defines a mapping $C^\infty_0(M)/(\boxempty - m^2-v) C^\infty_0(M) \to S^*$. Thus, in this sense, we can say that
			\begin{equation}\label{identification}
				C^\infty_0(M)/(\boxempty - m^2 - v) C^\infty_0(M) \subset S^*.
			\end{equation}
		We will often use this relation in the following. We may alternatively write the observable $\varphi^{\otimes n}(f^{(n)})$ in the form
			\begin{equation}\label{obs}
				\varphi^{\otimes n}(f^{(n)}) = \int_{M^n} f^{(n)}(x_1, \dots, x_n) \varphi(x_1) \cdots \varphi(x_n) dx_1 \dots dx_n,
			\end{equation}
		where $\varphi(x): S \to \bR$ is viewed as the {\em evaluation functional} $\varphi(x)[u] = u(x)$ for $u \in S$. Thus, $\varphi^{\otimes n}(f^{(n)})$ is a function $S \to \bC$, or alternatively, can be viewed as an element of a suitable closure of the (complex) symmetric algebra of $S^*$. 
		For any pair of observables $t, s$ of this form, we can define a Poisson bracket $\{ t, s\}$, the {\em Peierls bracket}, by demanding that 
			\begin{equation}\label{Peierls_brkt}
				\{ \varphi(x), \varphi(y) \} := E(x,y), 
			\end{equation}
		and extending $\{ \cdot , \cdot \}$ to elements of the form~\eqref{obs} by the Leibniz rule. This Poisson bracket comes from a symplectic structure, explicitly  the symplectic structure $\sigma: S \times S \to \bR$ defined by 
			\begin{equation}\label{symp_form}
				\sigma(u_1, u_2) := \int_\Sigma u_1(z) \overleftrightarrow{\partial_n} u_2(z) d\Sigma(z),
			\end{equation}
		 for two solutions $u_1, u_2 \in S$. Formula~\eqref{symp_form} does not depend on the choice of Cauchy surface $\Sigma$ in $M$ as can be proved using the Stokes theorem (see e.g.~\citep{W94}). The commutator function $E$ is a bisolution and so, by the identification~\eqref{identification}, it can formally be viewed as a map $E: S^* \times S^* \to \bR$, by setting $E(\varphi(x), \varphi(y)) = E(x,y)$. Moreover, the causal propagator $E$ can be interpreted as the inverse of $\sigma: S \times S \to \bR$ in the following sense: as proved in~\citep[lemma 3.2.1 part (3)]{W94}\footnote{Note that in the reference the symplectic structure has the opposite sign.}, for any $u \in S$ and any $f \in C^\infty_0(M)$ we have
			\begin{equation}\label{symp_kernel/causal_prop_simple}
				\sigma(u, E(f)) = \int_{M} f(x) u(x) dx. 
			\end{equation}
		These relations will be elaborated more and will be reinterpreted in terms of the infinite-dimensional geometry on $S$ in sec.~\ref{subsec_manifold_inf}.\\
		
		Next, we define a star-product for this symplectic structure. Our construction is going to follow the one given in the basic example (remark~\ref{rem_key_ex}). In particular, we want to produce an analogue of the formal Wick algebra~\eqref{Wick_algebra_free}. For this, it is necessary to introduce the concept of a {\em pure Hadamard $2$-point function} $\omega$, which will play the same role as the constant complex hermitian matrix $\omega^{ij}$ in the finite-dimensional context of the basic example (remark~\ref{rem_key_ex}). Because this concept is of vital importance for the entire rest of this paper, we formalize the definition of a Hadamard $2$-point function following~\citep{radzikowski1996micro}.
			\begin{defi}\label{def_hadamard_2-point}
				A Hadamard 2-point function $\omega$ is a $\bC$-valued distribution on $M \times M$ satisfying the following properties:
					\begin{enumerate}
						\item $\omega$ is a bi-solution for the homogeneous Klein-Gordon equation, i.e. 
								\begin{equation*}
									(\boxempty - m^2 - v)_{x} \omega(x,y) = (\boxempty - m^2 - v)_{y} \omega(x,y)=0.
								\end{equation*}
						\item The anti-symmetric part of $\omega$ is $i/2$ times the causal propagator, i.e. 
								\begin{equation*}
									\omega(x,y) - \omega(y,x) = iE(x,y).
								\end{equation*}
						\item $\omega$ is positive (semi-definite) in the sense that it holds
								\begin{equation*}
									\int_{M^2} \omega(x,y) \overline{f}(x) f(y) dx dy\geq 0,
								\end{equation*}
							where $\overline{f}$ is the complex conjugate of $f$.
						\item $\omega$ satisfies the ``Hadamard condition'', namely
							\begin{equation}\label{WF_Hadamard}
								\WF(\omega) = \cC^\triangleright[g] := \left\{ (x,y;k_x,k_y) \in \dot{T}^* M^2 : (x, k_x) \sim (y, - k_y) \mbox{ and } k_x \in V_x^+\right\}.
							\end{equation}
					\end{enumerate}
			\end{defi}
		\noindent
		We can decompose the $2$-point function (c.f.~\eqref{Im_and_Re_fin}) as
			\begin{equation}\label{decomp_quasi-free_state}
				\Imm \omega(x,y) =\frac{1}{2} E(x,y), \qquad \Rea \omega(x,y)=\frac{1}{2} G(x,y),  
			\end{equation}
		where $G$ is a real-valued and symmetric distribution on $M \times M$. We should think of $\omega(x,y)$ as analogous to $\omega^{ij}$, $E(x,y)$ as analogous to $\sigma^{ij}$ and $G(x,y)$ as analogous to $G^{ij}$ in the basic example of remark~\ref{rem_key_ex}. Positivity of $\omega$ (cf.~\eqref{pos_omega_fin}) is now equivalent to the condition (cf.~\eqref{pos_fin})
			\begin{equation}\label{pos_2-point}
				|E(f,h)| \leq \left( G(f,f) G(h,h) \right)^{1/2}, 
			\end{equation}
		for all $f,h \in C^\infty_0(M)/(\boxempty - m^2-v) C^\infty_0(M) \subset S^*$. In the finite-dimensional case, we assumed in addition that $\omega^{ij}$ defines a constant almost-K\"{a}hler structure. The analogue of this condition in the infinite-dimensional context consists in requiring $\omega$ to be {\em pure}, see e.g.~\citep{ashtekar1975quantum,KW91,W94}. As discussed in these references, the $2$-point function $\omega$ is pure if and only if eq.~\eqref{pos_2-point} is saturated, i.e.
			\begin{equation*}
				G(f,f) = \sup_{h \neq 0} \frac{|E(f,h)|^2}{G(h,h)}.
			\end{equation*}
		We next define the analogue of the algebra $\cW$. 
		This algebra will include the $\varphi^{\otimes n}(f^{(n)})$ defined in eq.~\eqref{obs}, but for later purposes, we need to extend the class of allowed smearing $f^{(n)}$ beyond smooth symmetric functions of compact support. The extended class of $f^{(n)}$'s is defined in terms of wave front sets. For any $n$, we define
			\begin{equation}\label{W_set_def}
				\gls{W} := \dot{T}^\ast M^n \backslash (C^+_n \cup C^-_n),
			\end{equation}
		where $C^\pm_n$ are the subsets of $T^\ast M^n$ defined by
			\begin{equation}\label{C_set_def}
				C^\pm_n := \{ (x_1, \dots, x_n ; k_1, \dots, k_n) \in \dot{T}^\ast M^n : k_i \in \overline{V}^\pm  \, \forall i \mbox{ or } \exists ! k_j \notin \overline{V}, k_{i\neq j} \in \overline{V}^\pm\}.
			\end{equation}	 
		We define the corresponding spaces of distributions as
			\begin{equation}\label{W_dist_compact}
				\gls{cE_W'_n} = \{ f^{(n)} \mbox{ distributions of compact support with } \WF(f^{(n)}) \subset W_n \}.
			\end{equation}
		Arguing as in~\eqref{identification}, we view
			\begin{equation}\label{n_box_spaces}
				\boxtimes_W^n S^* := \cE_W'(M^n)/(\boxempty - m^2-v)\cE_W'(M^n)
			\end{equation}
		as a completion of the algebraic tensor product $\otimes^n S^*$. In the above formula, we mean that we quotient out distributions in the form~\eqref{equiv_rel_lin}. Now we can define the formal Wick algebra $\cW(S,\omega)$ for this infinite-dimensional context imitating the finite-dimensional case. More precisely, as vector space, $\cW(S,\omega)$ is defined by
			\begin{equation}\label{formal_Wick_algebra_space_QFT}
				\cW(S, \omega) := \bC[[\hbar]] \otimes \bigoplus_{n = 0}^\infty \vee^n_W S^* = \bC[[\hbar]] \otimes \bigoplus_{n = 0}^\infty \bP^+ \cE_W'(M^n)/(\boxempty - m^2-v) \bP^+ \cE_W'(M^n),
			\end{equation}
		where $\vee^n_W S^*$ denotes the totally symmetric elements in~\eqref{n_box_spaces}. By $\glsuseri{P_pm}$ we mean the symmetrization. In other words, $\cW(S, \omega)$ is the vector space\footnote{Addition and scalar multiplication are defined componentwise.} of sequences $t=(t^0,t^1, \dots)$ where $t^0 \in \bC[[\hbar]]$ and $t^n \in \bC[[\hbar]] \otimes \vee^n_W S^\ast$ (cf.~\eqref{formal_wick_alg_fin}). We stress that it is not required that only finitely many elements of the sequence $(t^0,t^1, \dots)$ are non-zero. Based on the analogy with the finite-dimensional case, there is an obvious way to define the gradings $\deg_s$, $\deg_\hbar$, $\Deg$ in the field theory context, namely
			\begin{equation}\label{gradings_QFT}
				\deg_\hbar \hbar := 1, \quad \deg_s t^n := n \mbox{ for } t^n \in \vee^n_W S^\ast, \quad \Deg:=2\deg_\hbar + \deg_s.
			\end{equation}
		In order to avoid heavy notation, we will often identify an equivalence class in $\boxtimes^n_W S^\ast$ with a representative, i.e. we will identify $t^n$ with a distribution $t^n \in \cE'_W(M^n)$ in its class. Any other representative then differs by a distribution in $(\boxempty - m^2-v)\cE_W'(M^n)$, and we must be careful that our subsequent constructions do not depend on the given choice of the representative in the equivalence class.\\
		We summarize our construction:
			\begin{defi}\label{defi_free_wick_per}
				As a vector space, the algebra $\cW(S,\omega)$ is defined to be the vector space~\eqref{formal_Wick_algebra_space_QFT} equipped with the gradings~\eqref{gradings_QFT}. The product $\bullet$ is defined by analogy with eq.~\eqref{star_free} in the finite-dimensional case: let $t$, $s$ be two elements in $\cW$ homogeneous in $\deg_s$, respectively $\deg_s t = n$ and $\deg_s s = m$, we define the $\deg_s = j$ part of $t \bullet s$ by
					\begin{equation}\label{wick_prod_per}
						\begin{split}
							(t \bullet s)^j (x_1, \dots, x_j) &:= \hbar^k \sC_{n,m,k} \bP^+ \int_{M^{2k}} t^n(z_1, \dots, z_k, x_1, \dots, x_{n-k}) \times \\
							&\quad \times s^m(z'_1, \dots, z'_k, x_{n-k+1}, \dots, x_j) \prod_{\ell=1}^{k} \omega(z_\ell, z'_\ell) dz_\ell dz'_\ell,
						\end{split}
					\end{equation}
				if $j = n +m -2k$ for $k \leq n,m$, and $(t \bullet s)^j =0$ otherwise. By $\bP^+$ we mean the symmetrization operator acting on the free variables $x_1, \dots, x_j$. The combinatorial factor $\sC_{n,m,k} = \frac{n!m!}{k! (n-k)! (m-k)!}$ is the same appearing in~\eqref{star_co}.  Making use of the $\Deg$-filtration, the product $\bullet$ extends to all $\cW$.
			\end{defi}
		\noindent
		Despite the strong analogy with the finite-dimensional case, there is one significant difference: we have to prove that formula~\eqref{wick_prod_per} actually makes mathematical sense, because the right-hand side involves products and compositions of distributions. Products of distributions are generally not automatically well-defined. However, if the wave-front sets of the factors satisfy a certain relative condition, the product makes sense in a canonical way due to~\citep[thm. 8.2.10]{H83}. A similar result, \citep[thm. 8.2.14]{H83}, ensures the compositions of distributions are well-defined provided that the distributions involved satisfy an additional condition. Because these results are used extensively in this work, we summarize them for the convenience of the reader.
			\begin{theo}\label{theo_WF_horma}
				Let $\theta\in \cD'(X \times Z)$ and  $\theta' \in \cD'(Z \times Y)$. If $\theta$, $\theta'$ satisfy the {\em multiplication condition}
					\begin{equation}\label{mult_cond}
						\WF(\theta(x,z))_z \cap \WF'(\theta'(z,y))_z = \emptyset\footnote{We used the notation $\WF'(\theta) = \{(x,z;-k,-p) \in \WF(\theta)\}$ and $\WF(\theta(x,z))_z = \{ (z,p) \in T^\ast Z: (x,z; 0, p) \in \WF(\theta)\}$.},
					\end{equation}
				then the product $\theta(x,z)\theta'(z,y)$ (or simply $\theta \cdot \theta'$) can be defined as a distribution $\cD'(X \times Z \times Y)$ and
					\begin{equation*}
						\begin{split}
							\WF(\theta \cdot \theta') \subset \left\{ (x,z,y;k_x, p'+p'',k_y) \in \dot{T}^\ast X :\right. &(x,z;k_x,p') \in WF(\theta) \mbox{ or } k_x,p'=0,\\
							 &\left. (z,y;p'',k_y) \in WF(\theta') \mbox{ or } p'',k_y=0 \right\}.
						\end{split}
					\end{equation*}
			Moreover, if $\theta$, $\theta'$ satisfy the {\em integration condition} 
				\begin{equation}\label{int_cond}
					\supp (\theta \cdot \theta') \ni (x,z,y) \mapsto (x,y) \in X \times Y \mbox{ is a proper map}\footnote{The inverse image of any compact set is compact.},
				\end{equation}
			 then the composition $\int_{Z} \theta(x,z) \theta'(z,y) dz$ (or simply $\theta \circ \theta'$) can be defined as distribution in $\cD'(X \times Y)$ and 
					\begin{equation*}
						\begin{split}
							\WF(\theta \circ \theta') \subset \left\{ (x,y; k_x, k_y) \in \dot{T}^\ast (X \times Y) :\right. &(x,z; - k_x, -p) \in \WF(\theta) \mbox{ or } k_x,p=0,\\
							&\left. (z,y; p, k_y) \in \WF(\theta) \mbox{ or } k_y,p=0, \mbox{ for } (z, p) \in T^\ast Z \right\}.
						\end{split}
					\end{equation*}
			\end{theo}
		\noindent
		As a corollary of the previous theorem, we get the following result. 	
			\begin{lemma}\label{lemma_W_comp}
				Consider a distribution $\theta \in \cD'(M^n)$ and a distribution $\theta' \in \cD'(M^m)$ such that their wave-front sets are contained in $W_n$, respectively $W_m$. Then, the product $\theta(x_1, \dots, z, \dots, x_n)\theta(y_1, \dots, z, \dots, y_m)$ is well-defined and it satisfies the following wave-front set estimate
					\begin{equation*}
						\WF(\theta \cdot \theta') \subset W_{n+m-1}.
					\end{equation*}
				If in addition the integration condition holds, for example if one distribution has compact support, then the composition $\int_M \theta(x_1, \dots, z, \dots, x_n)\theta(y_1, \dots, z, \dots, y_m) dz$ is well-defined and it satisfies the following wave-front set estimate
					\begin{equation}\label{W_WF_est}
						\WF(\theta \circ \theta') \subset W_{n+m-2}.
					\end{equation}
			\end{lemma}
		\begin{proof}[Proof that eq.~\eqref{wick_prod_per} is well-defined]
			The argument is similar to that presented in \citep[thm. 2.1]{HW01}, but is adapted to our more stringent wave-front set restrictions compared to the one considered in~\citep{HW01}. For sake of completeness, we provide the full proof. Let $t^n$ and $s^m$ be two representatives, i.e. two compactly supported distributions with $\WF (t^n) \subset W_n$ and $\WF(s^m) \subset W_m$. By definition $\WF(\omega) \subset \cC^\triangleright \subset W_2$. For any $k \leq m$, the composition
				\begin{equation}\label{half_wick_prod_free}
					 (\omega^{\otimes k} \circ s^m)(z_1, \dots, z_k, \{x_{i > n-k}\}):= \int_{M^k} s^m(z'_1, \dots, z'_k, \{ x_{i > n-k} \}) \prod_{\ell=1}^k \omega(z_\ell, z'_\ell) dz'_\ell
				\end{equation}
			is a well-defined distribution with wave-front set contained in $W_m$ as a consequence of lemma~\ref{lemma_W_comp}. As a consequence of the Hadamard condition, the wave-front set of distribution~\eqref{half_wick_prod_free} can contain only elements of the type $(z_1, \dots, z_k, (x_{i>n-k}); p_1, \dots, p_k, (k_{i>n-k})) \in W_m$ with $p_\ell \in V^+$ or $p_\ell=0$. Clearly $\WF(\omega^{\otimes k} \circ s^m)_{z_1, \dots, z_k} \subset (V^+)^k$ and then it does not intersect $\WF(t^n)_{z_1, \dots, z_k}$ by definition of $W_n$. It follows that we can apply thm.~\ref{theo_WF_horma} and we get that for any $k \leq m,n$, the composition in $z_1, \dots, z_k$ of $t^n$ with $\omega^{\otimes k} \circ s^m$ is well-defined as a (compactly supported) distribution in $x_1, \dots, x_{n+m-2k}$ and $\WF(t^n \circ (\omega^{\otimes k} \circ s^m)) \subset W_{n+m-2k}$.\\
		Because $\omega$ is a bi-solution, the equivalence class $[(t \bullet s)^j]$ in $\vee_W^j S^*$ defined by eq.~\eqref{wick_prod_per} does not depend on the choice of the representative of $t^n$ and $s^m$. This concludes the proof that the product $\bullet$ in def.~\ref{defi_free_wick_per} is well-defined.
		\end{proof}
	
In the algebraic approach to quantum field theory, one defines the algebraic states as positive, normalized linear functionals on the algebra of observables\footnote{In the finite-dimensional context, the concept of algebraic states is discussed in~\citep{bordemann1998formal, waldmann2005states}.}. In our context, the underlying field $\bC$ is replaced by the ring $\bC[[\hbar]]$, so a {\em state} on $\cW$ is a $\bC[[\hbar]]$-linear functional $\omega: \cW_\phi \to \bC[[\hbar]]$\footnote{Any attempt to consider $\bC$-linear positive functionals $\omega: \cW \to \bC$ is affected by serious convergence problems.} which is normalized to $1$, in the sense that $\omega(1) = 1$. In this context, by positive we mean that $\omega( t^\dagger \bullet t)$ is a positive element in $\bR[[\hbar]]$ for any $t \in \cW_\phi$, where an element $\sum_{k \geq 0} \hbar^k a_k \in \bR[[\hbar]]$ is positive if the first non-vanishing real coefficient $a_k$ is positive. A state $\omega$ is equivalently defined in terms of all its {\em $n$-point functions}, i.e. the distributions $\omega_n$ such that
			\begin{equation*}
				\int_{M^n} f_1(x_1) \dots f_n(x_n) \omega_n (x_1, \dots, x_n) dx_1 \dots dx_n := \omega( \varphi(f_1) \bullet \cdots \bullet \varphi(f_n)),
			\end{equation*}
		where $f_1, \dots, f_n$ are test functions on $M$. The positivity condition of the state $\omega$ becomes a complicated hierarchy of conditions on $\omega_n$, the simplest of which is the positivity of the $2$-point function, i.e. $\omega_2 (\overline{f}, f) \geq 0$ for any test function $f$.\\
		Once a Hadamard $2$-point function $\omega(x,y)$, in the sense of def.~\ref{def_hadamard_2-point}, is provided, we can define a state requiring that all the $n$-point functions with $n$ odd vanish and all the $n$-point functions with $n$ even are given by appropriate combinations of tensor products of $\omega(x,y)$, i.e.
			\begin{equation*}
				\omega_2(x,y) = \omega(x,y), \quad \omega_n(x_1, \dots, x_{2n}) = \sum_{I} \prod_{(i,j) \in I} \omega(x_{i}, x_{j}),
			\end{equation*}
		where $I$ is any possible arrangement of $\{1, \dots, 2n\}$ into a collection disjoint pairs $(i,j)$ such that $i <j$. Such a state is called a {\em quasi-free Hadamard} state. By abuse of notation, we will identify a quasi-free Hadamard state $\omega$ with its $2$-point function $\omega(x,y)$ in the following.\\
		
		As in the finite-dimensional case, the product $\bullet$ can be viewed as a star-product for the Peierls bracket $\{ \cdot , \cdot \}$ defined by $E$, see eq.~\eqref{Peierls_brkt} (for further details, see~\citep{DF01a}). The analogies between the finite and infinite-dimensional cases are summarized in the following table.

		\begin{table}[H]
		\begin{center}
			\renewcommand{\arraystretch}{1.5}
			\begin{tabular}{ | l | l |}
				\hline
				 	{\bf finite-dim} & {\bf linear QFT} \\
  				\hline                       
 					$S \ni y$ vector in $\bR^{2d}$	& $S \ni u$ smooth sol. of K-G eq.~\eqref{K-G_w_source} with $j=0$	\\
 				\hline
 					$\sigma(u,v) = \sigma_{ij} u^i v^j$ constant symp. form 	&	$\sigma(u,v) = \int_{\Sigma} u \overleftrightarrow{\partial_n} v d \Sigma$\\
 				\hline
 					$t \in \bR^{2d}$,	& $t \in C_0^\infty(M)/ (\boxempty - m^2 -v) C_0^\infty(M) \subset  S^\ast$, \\
 					$t(y)=t_i y^i$ where $t_i \in \bR^{2d}$	& $t(u)= \int_M t(x) u(x)$, where $t(x) \in t \subset C^\infty_0(M)$ \\
 				\hline
 					$\sigma^{ij}$	& $E(x_1, x_2)$ \\
  				\hline
  					$\omega^{ij}$ almost-K\"{a}hler pos. Hermitian form,	& $\omega(x_1,x_2)$ pure Hadamard quasi-free state\\
  				 	$2 \Imm \omega^{ij} = \sigma^{ij}$,	& $2 \Imm \omega(x_1,x_2) = E(x_1,x_2)$, \\
  				 	$2 \Rea \omega^{ij} = G^{ij}$ inv. metric & $2 \Rea \omega(x_1,x_2) = G(x_1,x_2)$ symm. distr.\\
  				\hline
  					$\sum_n t_{i_1 \dots i_n} y^{i_1} \cdots y^{i_n}$ observables 	&	$\sum_n \varphi^{\otimes n} (t^{n})$ observables\\
  					$\{y^i,y^j\}=\sigma^{ij}$	& $\{\varphi(x_1),\varphi(x_2)\} = E(x_1,x_2)$\\
  				\hline
  					$\cW(S,\omega) = \bC[[\hbar]] \otimes \bigoplus_{n \geq 0} \vee^n \bR^{2d}$, 	& $\cW(S,\omega) =\bC[[\hbar]] \otimes \bigoplus_{n \geq 0} \vee^n_W S^\ast$ \\
  					$t = (t^0,t^1, \dots)$	& $t = (t^0,t^1, \dots)$ \\
  					$t^n$ symmetric covariant tensor	& $t^n \in \bP^+ \cE'_W(M^n)/(\boxempty - m^2 -v)\bP^+\cE'_W(M^n),$\\
  				 	with coefficients in $\bC[[\hbar]],$	& with coefficients in $\bC[[\hbar]]$, \\
  					$\bullet$ given by \eqref{star_co}	& $\bullet$ given by \eqref{wick_prod_per}\\
  				\hline
			\end{tabular}
			\caption{Analogies between the finite-dimensional framework and the linear quantum field theory setting.}\label{table:fin_linQFT}
		\end{center}
		\end{table}
		A difference to the finite-dimensional case is that in the infinite-dimensional setting, we need to discuss the topological structure of the formal Wick algebra $\cW$. In finite dimensions there is only one reasonable topology, while many inequivalent definitions are a priori available if $S$ is a space of smooth functions. The formal Wick algebra $\cW$ has a natural notion of convergence (not actually a topology) that is inherited from the wave-front set condition satisfied by the distributions in $\cE'_W$ (see~\eqref{W_dist_compact}) . This topological structure is defined as follows. Firstly, we introduce on the distribution spaces $\cE'_W(M^n)$ the notion of convergence by saying that a sequence $\{ t_{\ell} \}_{\ell \in \bN} \subset \cE'_W(M^n)$ converges to $t \in \cE'_W(M^n)$ in the {\em H\"{o}rmander pseudo-topology}, written $t_{\ell} \stackrel{W_n}{\to} t$, if for any pseudo-differential operator $A$ on $M^n$ with $\mathrm{char}(A) \subset T^* M^n \setminus W_n$, it holds $At_{\ell} \to At$ in the sense of $C^\infty_0(M)$. Because $\cW$ is basically the direct product of spaces $\cE'_W(M^n)$, the topological structure on $\cE'_W(M^n)$ we have just introduced naturally leads to a notion of sequential convergence also for $\cW$. We denote it by $\stackrel{W}{\to}$. One can show that:
			\begin{enumerate}
				\item The algebra $\cW$ is closed under taking the sequential completion with respect to the notion of convergence $\stackrel{W}{\to}$.
				\item The product $\bullet$ and the $\dagger$-conjugation are continuous. 
			\end{enumerate}
	
		We emphasize that, in the infinite-dimensional context, our choice of distribution spaces (representing suitable closures) is, a priori, only one among many possibilities to get a well defined analogue of the algebra $\cW$ in the finite-dimensional case. Our choice is guided by experience from perturbative quantum field theory and will turn out to be suitable for our purposes in the following sections. In the literature, see e.g.~\citep{BFK96,brunetti2000microlocal, DF01a, DF01b, HW01, HW02, HW03, DF04, HW05}, a less restrictive wave-front set condition is imposed, namely in formula~\eqref{W_dist_compact} the set $W_n$ is replaced with $T^\ast M^n \backslash ((\overline{V}^+)^n \cup (\overline{V}^-)^n)$. Our choice will be motivated in sec.~\ref{subsec_Fedosov_inf} when we generalize Fedosov's method to an infinite-dimensional geometry based on such restrictive constraints on the wave-front sets.\\
		
		So far we have discussed a {\em linear} quantum field theory. This framework suffices to treat the linearised theory around a classical ``background'' solution $\phi$ to a non-linear equation, as we explain in more detail in the next section. In the following, we need to consider also the more general situation where the ``background'' is not a solution. This more general framework requires a slight generalization of the formal Wick algebra. Namely, we allow a non-vanishing smooth source $j$ in~\eqref{K-G_w_source}, i.e. we want to implement on the quantum algebra $\cW$ the condition that $\varphi$ is a solution to the {\em inhomogeneous} equation $(\boxempty - m^2 -v)_x \varphi(x) = j(x)$. To do so, we consider $\bC[[\hbar]] \otimes \oplus_{n \geq 0} \cE'_W(M^n)$ and we quotient out the elements $t=(t^0, t^1, \dots )$ in form
			\begin{equation}\label{equiv_rel_inhomo}
				\begin{split}
					&t^{0}= - \int_M j(z) h^{1}_1(z) dz, \\
					&t^{n}(x_1, \dots, x_n) = \sum_{i=1}^n (\boxempty - m^2 - v)_{x_i} h^n_i(x_1, \dots, x_n) -\\
					&\qquad \qquad \qquad \qquad - \sum_{k=0}^{n} \int_M j(z) h^{n+1}_k(x_1, \dots, x_k,z, x_{k+1}, \dots, x_n) dz
				\end{split}
			\end{equation}
		for a collection $\{h^{n}_i\}_{n,i \in \bN}$ of smooth compactly supported functions. Note that for the inhomogeneous case the equivalence relation compares distributions with different symmetric degrees, i.e. it cannot be written as a relation on each $\cE'_W(M^n)$ (c.f.~\eqref{equiv_rel_lin}). The modified algebra $\cW$ (cf.~\eqref{formal_Wick_algebra_space_QFT}) is then defined as a vector space by
			\begin{equation}\label{formal_Wick_algebra_space_QFT_inhomo}
				\cW(S, \omega) = \bC[[\hbar]] \otimes \left( \bigoplus_{n=0}^\infty \bP^+ \cE_W'(M^n)\right) /(\boxempty - m^2-v - j \circ \id),
			\end{equation}
		where we mean that we now quotient out the new relations eq.~\eqref{equiv_rel_inhomo}. The product on the algebra is defined as before by formula~\eqref{wick_prod_per}, which is seen to give a consistent definition.

\section{Interacting Klein-Gordon equation, quantization \`{a} la ``causal perturbation theory''}\label{subsec_int_QFT_per}
		In the previous section we have described the ``deformation quantization'' of a linear, scalar Klein-Gordon field. If there is no source, the solution space has a linear structure, while it has an affine structure if there is a source. As we discussed, under these circumstances the deformation quantization procedure is a precise analogue of the quantization of the finite-dimensional classical symplectic manifold  described as our ``model case'' in the basic example (remark~\ref{rem_key_ex}). The situation changes drastically if we want to apply deformation quantization to a {\em non-linear} Klein-Gordon equation of the type
			\begin{equation}\label{eom} 
				(\boxempty - m^2)\phi - V'(\phi) = 0, 
			\end{equation}
		which is the Euler-Lagrange equation of the action
			\begin{equation}\label{action}
				I(\phi) =  \int_M \left( \frac{1}{2}|\nabla \phi(x)|_g^2 + \frac{1}{2}m^2 \phi(x)^2 + V(x, \phi(x))  \right) dx.				
			\end{equation}
		Here, $V(\phi)$ is a potential, which we will typically take to be of the form $V = \frac{\lambda}{4!} \phi^4$, where $\lambda$ could be a smooth function of $x$ or just a constant. It seems natural to try to apply Fedosov method to get a deformation quantization of this system by proceeding along the lines described in the classical case in sec.~\ref{subsec_Fedosov_fin}, where $S$ would now be the space of solution to the non-linear Klein-Gordon theory. We will indeed do this below in sec.~\ref{subsec_Fedosov_inf}, after the necessary concepts in infinite-dimensional geometry will have been introduced in sec.~\ref{subsec_manifold_inf}-\ref{subsec_W_covariant_der}. However, to get a better perspective of the construction, and to relate it to more conventional constructions in quantum field theory, we will present here first a different approach which is based, roughly speaking, on the ideas of ``causal perturbation theory'' in the sense of Epstein-Glaser~\citep{epstein1973role} following \citep{brunetti2000microlocal, DF01a, HW01, HW02, HW03, DF04, HW05}.\\

		The starting point of this type of perturbation theory is to fix some ``background'' $\phi \in C^\infty(M)$, and to expand the classical action around $\phi$. It is not assumed at this stage that $\phi$ is a solution to the non-linear Klein-Gordon equation, although we will be interested in that case later on. We first consider the action $I(\phi+\varphi)$ up to quadratic order in the ``perturbation'' $\varphi$. Thus, letting
			\begin{equation}\label{action_exp}
				I^{(p)}_\phi(\varphi) := \frac{1}{p!} \left. \frac{d^p}{d\epsilon^p} I(\phi +\epsilon\varphi)\right|_{\epsilon=0},
			\end{equation}
		we consider the ``free'' action $I^{(0)}_\phi + I^{(1)}_\phi + I^{(2)}_\phi$, i.e. up to quadratic terms in $\varphi$. The zeroth order term evidently does not depend on $\varphi$ at all, and so does not contribute to the equations of motion for $\varphi$. The variation of the first term with respect to $\varphi$ of the first order term vanishes if $\phi$ itself is a solution to the background Klein-Gordon equation, and otherwise gives a source in the equation of motion of $\varphi$. Thus, the equation of motion for the theory corresponding to the truncated action $I^{(0)}_\phi + I^{(1)}_\phi + I^{(2)}_\phi$ is:
			\begin{equation}\label{leom+in}
				[\boxempty - m^2 - v_\phi]\varphi=j_\phi, \qquad v_\phi(x) = V''(\phi(x)), \qquad j_\phi(x) = (\boxempty - m^2) \phi(x) - V'(\phi(x)).
			\end{equation}
		Note that in general $j_\phi$ is not compactly supported. However, we are interested in the case where the background $\phi$ is a smooth solution to the non-linear equation and in this case $j_\phi $ simply vanishes.
		We have already explained how to quantize this theory for fixed background $\phi$ in the previous section. These constructions give an algebra $\cW_\phi$, the formal Wick algebra for the background $\phi$:
			\begin{defi}
				For an arbitrary background $\phi$, consider a pure Hadamard $2$-point function $\gls{omega_phi}$ with respect to the linearised KG-equation $[\boxempty -m^2 - v_\phi] \varphi = 0$. The algebra $\cW_\phi$ is defined as in Def.~\ref{defi_free_wick_per} of Sec.~\ref{subsec_free_QFT} for the $2$-point function $\omega_\phi$ with the modification given by formula~\eqref{formal_Wick_algebra_space_QFT_inhomo}.
			\end{defi}

		Of course we need to say how to incorporate the corrections arising from the higher-than-quadratic parts in the action, $I^{(p)}_\phi$ for $p>2$. These corrections are organized in certain (formal) series, which are valued in the algebra $\cW_\phi$. In order to describe these series in more detail, we first make some definitions.
			\begin{defi}\label{def_funct_class}
				Let $F$ be a functional $C^\infty(M) \to \bC$.
				\begin{itemize}
					\item $F$ is called {\em $W$-smooth} if the following two conditions hold:
						\begin{enumerate}
							\item All its Gateaux derivatives exist in the sense of distributions on the appropriate Cartesian power of $M$, i.e. for any $\phi \in C^\infty(M)$ and any $\nu \in \bN$, it holds
									\begin{equation*}
										\begin{split}
										&\left. \frac{\partial^\nu}{\partial\epsilon_1 \dots \partial\epsilon_\nu} F\left(\phi + \sum_i \epsilon_i h_i\right) \right|_{\epsilon_1, \dots, \epsilon_\nu=0} =\\
										 &\quad = \int_{M^\nu} \frac{\delta^\nu F(\phi)}{\delta \phi(y_1) \cdots \delta \phi(y_\nu)} h_1(y_1) \cdots h_\nu(y_\nu) dy_1 \dots dy_\nu,
										\end{split}
									\end{equation*}
								where $\delta^\nu F(\phi) / \delta^\nu \phi$ is a symmetric distribution in $\cE'_W(M^\nu)$ (see~\eqref{W_dist_compact}), and where $h_1, \dots, h_\nu$ are arbitrary smooth functions.
							\item The Gateaux derivatives depend smoothly on $\phi$ in the following sense: consider a smooth $1$-parameter family of backgrounds $\bR \ni \epsilon \mapsto \phi(\epsilon)$ (in the topology on $C^\infty(M)$), and view $\delta^\nu F(\phi(\epsilon)) / \delta \phi(y_1) \cdots \delta \phi(y_\nu)$ as a distribution in the variables $\epsilon, y_1, \dots, y_\nu$, i.e. as a distribution on $\bR \times M^\nu$. It is required that its wave-front set satisfies
									\begin{equation*}
										\WF\left(\frac{\delta^\nu F(\phi(\epsilon))}{\delta \phi(y_1) \cdots \delta \phi(y_\nu)}\right) \subset \bR \times \{ 0 \} \times W_\nu,
									\end{equation*}	
								where $W_\nu$ is the set defined by~\eqref{W_set_def}.
						\end{enumerate}	
					\item $F$ is said to be {\em polynomial} if all the Gateux derivatives of sufficiently high degree vanish.
					\item $F$ is said to be {\em compactly supported} if its support, defined as the closed set
							\begin{equation*}
								\supp F : = \left\{ p \in M \vert \forall U \ni p, \exists \phi, \psi \neq \phi \in C^\infty(M), \supp \psi \subset U, F(\phi + \psi) \neq F(\phi) \right\},
							\end{equation*}
						is compact.
					\item $F$ is said to be {\em additive} if for any $\phi_1, \phi_2, \phi_3 \in C^\infty(M)$ such that $\supp \phi_1 \cap \supp \phi_3 = \emptyset$ and $\phi_1, \phi_3, \phi_1+ \phi_3 \neq \phi_2$ it holds
							\begin{equation*}
								F(\phi_1 + \phi_2 + \phi_3) = F(\phi_1 + \phi_2) - F(\phi_2) + F(\phi_2 + \phi_3).
							\end{equation*}
				\end{itemize}
				The set of $W$-smooth, additive, polynomial functionals of compact support is denoted $\cF_{\loc}(M)$ and the elements are called {\em local functionals}\footnote{Smooth, additive, polynomial functionals with compact support  are indeed local in the sense of~\citep{DF04,brunetti2009perturbative}), i.e. $\supp (\delta^\nu F(\phi) /\delta \phi^\nu) \subset \Delta_\nu$ and $\WF(\delta^\nu F(\phi) /\delta \phi^\nu) \perp T\Delta_\nu$, where $\Delta_\nu = \{ (x, \dots, x) \in M^\nu\}$ is the diagonal in $M$. The first condition follows form~\citep[prop. 2.3.11]{brunetti2012algebraic}, while the second one is a consequence of the fact that any smooth, additive, polynomial functional with compact support is in the form~\eqref{F_loc_general form}.}. 
			\end{defi}
		\noindent
		It can be shown\footnote{Since $W_1 = \emptyset$, local functionals in $\cF_{\loc}(M)$ are indeed microlocal functional in the sense of~\citep{brunetti2012algebraic} and, hence, the statement is a consequence of~\citep[prop. 2.3.12]{brunetti2012algebraic}.} that every local functional in $\cF_{\loc}(M)$ must have the form
			\begin{equation}\label{F_loc_general form}
					F(\phi) = \int_M \cP(x, \phi(x), \nabla \phi(x), \dots, \nabla^n \phi(x), \dots) dx,
			\end{equation}
		where $\cP$ is a polynomial with smooth compactly supported coefficients, and with degree locally bounded on compact sets. Among such functionals, a prime example is a local self-interaction of the form
			\begin{equation}\label{int_funct}
				\int_M V(x,\phi(x)) dx= \frac{1}{4!}\int_M \lambda(x) \phi(x)^4 dx, 
			\end{equation}
		where $\lambda$ is smooth and of compact support on $M$. Here, $\lambda$ plays the role of a ``coupling function'' that can, for example, be smoothly switched on and off.\\
		
		In the ``causal approach'' to the quantization of the theory described classically by the action $I(\phi)$ (cf.~\eqref{action}), one proceeds as follows. First, one fixes an arbitrary  smooth background $\phi$ which solves eq.~\eqref{eom}. For such $\phi$, one considers the {\em free} theory described by the quadratic part of the action $I^{(2)}_\phi(\varphi)$, see~\eqref{action_exp}. For the corresponding equation of motion 
			\begin{equation}\label{leom}
				[\boxempty - m^2 - v_\phi]\varphi=0
			\end{equation}
		(cf.~\eqref{leom+in}), one picks a pure Hadamard $2$-point function $\omega_\phi$ and defines the corresponding algebra
			\begin{equation}\label{wick_alg_per_int}
				\cW_\phi := \cW\left( \{ \mbox{ smooth solutions of eq.~\eqref{leom} } \},\omega_\phi \right),
			\end{equation}
		as explained in the preceding section.\\
		
		For any local functional $F \in \cF_\loc(M)$ one next wishes to define a corresponding ``interacting field observable'' associated to the full action $I(\phi +\varphi)$~\eqref{action}. One denotes by
			\begin{equation}\label{local_int}
				\cV_\phi (\varphi) := \sum_{p >2} I_\phi^{(p)}(\varphi) \equiv \sum_{p >2} \frac{1}{p!} \int_M \cL^{(p)}_\phi (x) dx,
			\end{equation}
		 the part of $I(\phi + \varphi)$ higher than quadratic in $\varphi$\footnote{So that $I(\phi + \varphi) = I^{(0)}_\phi(\varphi) + I^{(1)}_\phi(\varphi) + I^{(2)}_\phi(\varphi)+ \cV_\phi(\varphi)$.}. Then, one writes the quantum field observable $\hat{F}_\phi (\varphi)$ associated to $F(\phi)$ (cf.~\eqref{F_loc_general form}) in the interacting theory as the series
			\begin{equation}\label{naive_Haag_formula}
				\begin{split}
					\mbox{``} \hat{F}_\phi (\varphi) := \sum_{n=0}^{\infty} \frac{i^n}{\hbar^n n!} \sum_{p_i >2} &\int_{M^{n+1}} \left[ \dots \left[ \cP_\phi(x), \cL^{(p_1)}_\phi(y_1) \right]_{\bullet_\phi}, \dots, \cL^{(p_n)}_\phi (y_n) \right]_{\bullet_\phi} \times\\
					&\qquad \times \theta(x^0, y^0_1, \dots, y^0_n) dx dy_1 \dots dy_n \mbox{''},
				\end{split}
			\end{equation}
		where $\cP_\phi(x, \varphi(x), \nabla \varphi(x), \dots)$ is the density for $F(\phi + \varphi)$, i.e. $\int_M \cP_\phi(x) dx = F(\phi + \varphi)$, and where $\theta(x^0_1, \dots, x^0_n)$ is the product of the Heaviside step-functions $\prod_{i=1}^n \theta(x^0_i - x^0_{i+1})$. For example if $x \in M$ and $F(\phi) = \phi(x)$, then $\hat{\phi}(x)$ formally satisfies the interacting Klein-Gordon equation~\eqref{eom}.\\
		
		There are several problems with eq.~\eqref{naive_Haag_formula}:
		\begin{enumerate}
			\item The integrand is not a well-defined distribution because the commutators are too singular to be multiplied by $\theta(x^0, y^0_1, \dots, y^0_n)$ (this follows e.g. from the wave-front set calculus). This is a manifestation of the usual UV-divergences in perturbative quantum field theories.
			\item The $dy_i$-integrals can suffer from IR-divergences, e.g. if $m=0$ and $M = (\bR^4, \eta)$ the Minkowski space-time.
			\item The series $\sum_n$ cannot be expected to converge.
		\end{enumerate}
		Note that the third problem does not affect us (or rather, is ignored), because we only work with formal series in $\hbar$, so we only need to make sense of the individual terms appearing in eq.~\eqref{naive_Haag_formula}. Dealing with the second problem in general requires further analysis and depends on the choices of $(M,g)$, $V$ and $\phi$. We sidestep this issue by choosing compactly supported interactions, e.g.~\eqref{local_int} with $\lambda$ compactly supported. The first problem needs to be dealt with by some form of ``renormalization''.\\
		
		Our approach to the renormalization problem is to characterize the integrand in eq.~\eqref{naive_Haag_formula} axiomatically, keeping as many formal properties as possible. In a second step, we will then prove that there exists a non-trivial solutions to these prescribed axioms. This program is called ``causal perturbation theory''~\citep{brunetti2000microlocal, HW01, HW02, HW03, HW05}. It turns out that one has to formulate quite a few axioms to characterize the integrand on the right-hand side of eq.~\eqref{naive_Haag_formula} with sufficient precision. The objects to be characterized by these axioms are called {\em retarded products} and formally correspond to
			\begin{equation}\label{naive_retarder_product}
				\begin{split}
					&\mbox{``}R_{n,\phi} \left( F(\phi + \varphi); I^{(p_1)}_\phi(\varphi) \otimes \cdots \otimes I^{(p_n)}_\phi(\varphi) \right)  =\\
					&\quad =  \int_{M^{n+1}} \left[ \dots \left[ \cP_\phi(x), \cL^{(p_1)}_\phi(y_1) \right]_{\bullet_\phi}, \dots, \cL^{(p_n)}_\phi (y_n) \right]_{\bullet_\phi} \theta(x^0, y^0_1, \dots, y^0_n) dx dy_1 \dots dy_n \mbox{''}.
				\end{split}
			\end{equation}
		The terminology is due to the support property of~\eqref{naive_retarder_product}: the retarded product~\eqref{naive_retarder_product} vanishes if none of the terms $\cL^{(p)}_\phi$ has the support in the causal past of the support of $\cP_\phi(x)$. This support property is encoded in the causality axiom~\ref{R8}. The other axioms~\ref{R1}-\ref{R12} are described in detail below and similarly encode other properties that formally hold for~\eqref{naive_retarder_product}.\\
		
		We now present the abstract properties of the retarded products. A prescription for retarded products is a collection of maps
			\begin{equation*}
				R_{n,\phi}: \cF_{\loc}(M) \otimes \cF_{\loc}(M)^{\otimes n} \to \cW_\phi(M,g)
			\end{equation*}
		given for every value of $n \geq 0$, for every $\phi \in C^\infty(M)$, and for every globally hyperbolic manifold $(M,g)$. For $\phi, \lambda \in C^\infty(M)$ we first define the formal Wick algebra $\cW[g, m, \phi, \lambda]$ as in def.~\ref{defi_free_wick_per}, where the underlying Klein-Gordon equation~\eqref{leom} is characterized by~\eqref{local_int}, i.e. $v_\phi = \frac{\lambda}{2} \phi^2$. In principle, for an arbitrary smooth $\phi$, we should construct the formal Wick algebra with respect to the inhomogeneous Klein-Gordon equation~\eqref{leom+in}, but ultimately we want to consider $\phi$ a smooth solution to eq.~\eqref{eom}), so we restrict to the formal Wick algebra corresponding to the homogeneous Klein-Gordon equation~\eqref{leom} also when $\phi$ is not a solution.\\
		The desired properties that the retarded products are supposed to satisfy are:		
			\begin{enumerate}[label=(R\arabic*), start=0]
				\item\label{R0} {\em Initial conditions}:\\
					If $F \in \cF_{\loc}(M)$ is independent of $\varphi$ then $R_{n,\phi}(F; \dots ) = F \delta_{n,0} \1$.\\
					If $f\in C^\infty_0(M)$ then $R_{0,\phi}(\int_M f(x) \varphi(x)) = \int_M f(x) \varphi(x) dx \in \cW_\phi$.
				\item\label{R1} {\em Locality/covariance}:\\
					Consider an isometric embedding $\iota:(M',g') \hookrightarrow (M,g)$, i.e. $g' = \iota^* g$ and a background $\phi'$ on $M'$ such that $\phi' = \iota^* \phi$. It can be proved that $\omega'_{\phi'} = \iota^* \omega_\phi$ is a Hadamard $2$-point function for the linearised Klein-Gordon equation in $(M',g')$ around $\phi'$, where the  ``mass'' and the coupling are now $m' = \iota^* m$ and $\lambda' = \iota^* \lambda$. One defines the corresponding algebra $\cW'_{\phi'}=\cW[g', m', \phi', \lambda']$. Let $\alpha_\iota : \cW'_{\phi'}  \to \cW_\phi$ be the natural injective $*$-homomorphism corresponding to $\iota$ (see~\citep{HW01}). Then, it should hold
						\begin{equation*}
							\alpha_\iota \left( R_{n,\phi'} \left( F; \bigotimes_{j=1}^m H_j \right) \right) = R_{n,\phi} \left( \iota_* F; \bigotimes_{j=1}^n \iota_* H_j \right).
						\end{equation*}
				\item\label{R2} {\em Scaling}:\\
					The retarded products scale almost homogeneously (in the sense of~\citep[def. 4.2]{HW01}) under a rescaling $g \mapsto \Lambda^{-2} g$ where $\Lambda \in \bR$ and under the corresponding rescaling of $m$, $V$, $\varphi$ and $\phi$ chosen in such a way that the truncated action $I_\phi^{(0)} +I_\phi^{(1)} + I_\phi^{(2)}$ is invariant, i.e.  $m \mapsto \Lambda m$, $\varphi \mapsto \Lambda \varphi$, $\phi \mapsto \Lambda \phi$ and $\lambda$ does not scale\footnote{The last is a consequence of the choice $V(\phi) = \frac{\lambda}{4!}\phi^4$. Different choices for the interaction could require a rescaling of $\lambda$.}. More precisely, let $\sigma_\Lambda: (\cW[\Lambda^{-2}g, \Lambda m, \Lambda \phi, \lambda], \Lambda^2 \omega_\phi) \to (\cW[g, m, \phi, \lambda], \omega_\phi)$ be the canonical homomorphism between two formal Wick algebras at different scales introduced in~\citep[lemma 4.2]{HW01}, then there exists some $N \geq 0$ such that
						\begin{equation*}
							\frac{\partial^N}{\partial^N \log \Lambda} \Lambda^{-d_R - 4(n+1)} \sigma_\Lambda R_{n,[\Lambda^{-2}g, \Lambda m, \Lambda \phi, \lambda]}\left( F ; \bigotimes_{j=1}^n H_j \right) = 0.
						\end{equation*}
					In the formula above $d_R$ is the engineering dimension of the retarded product, which is defined as follows (see~\citep{HW02, HW05}). The functionals $F, H_1, \dots, H_n$ can be written as $F=\int_M f(x) \Phi_{0,\phi}(x) dx$ and $H_j = \int_M h_j(x) \Phi_{j,\phi}(x) dx$ for any $j=1, \dots, n$, where $f, h_1, \dots, h_n$ are compactly supported tensor fields, and where $\Phi_\phi$ are monomials in the classical field $\varphi$, its symmetrized covariant derivatives, the metric, arbitrary curvature tensors, the functions $m$ and $\phi$ and their symmetric covariant derivatives. We assign to each $\Phi_\phi$ an engineering dimension
						\begin{equation*}
							\begin{split}
							d_{\Phi_\phi} &= \#(\mbox{factors of } \varphi) + \#(\mbox{factors of m}) + \#(\mbox{factors of } \phi) + \#(\mbox{derivatives})  +\\
							&\quad + 2 \times \#(\mbox{factors of curvature}) + \#(\mbox{``up'' indices}) - \#(\mbox{``down'' indices}).
							\end{split}
						\end{equation*}
					The engineering dimension $d_R$ is just the sum $d_{\Phi_{0,\phi}} + d_{\Phi_{1,\phi}} + \dots + d_{\Phi_{n,\phi}}$.
				\item\label{R3} {\em Microlocal spectrum condition}:\\
					Let $\omega_\phi$ be any quasi-free Hadamard state on $\cW_\phi$\footnote{We are free to use the $2$-point function of $\omega_\phi$ to define the Wick product $\bullet_\phi$, exploiting the fact that two $2$-point functions satisfying the Hadamard condition differ only for a smooth function, and such smooth function induces an isomorphism of the formal Wick algebra (see~\citep{HW01}).}, i.e. $\WF(\omega_{2,\phi})=\cC^\triangleright$. Let
						\begin{equation}\label{omega_ret_m}
							\omega_{R,n,\phi}(y;x_1, \dots, x_n) := \omega_\phi \left( R_{n,\phi}\left( \Phi_{\phi}(y); \bigotimes_{j=1}^n \Phi_{j, \phi}(x_j) \right) \right),
						\end{equation}
					for any $\Phi_\phi$ monomials in the classical field $\varphi$, its symmetrized covariant derivatives, the metric, arbitrary curvature tensors and the background $\phi$. Then, we require that
						\begin{equation*}
							\WF(\omega_{R,n,\phi}) \subset \cC^R_{1+n}[g],
						\end{equation*}
					where $\cC^R_n[g]$ is the set defined by
						\begin{equation}\label{C^R_m}
							\begin{split}
								\cC^R_{1+n}[g] := &\left\{ (y,x_1, \dots, x_n ; q, k_0, \dots, k_n) \in \dot{T}^\ast M^{m+1}: \exists \mbox{ decorated graph } \sG \right.\\
								&\left. \mbox{ with external vertex } y \mbox{ and with internal vertices } x_1, \dots, x_n \right. \\
								&\left. \mbox{ such that } x_j \in J^-(y) \forall j \right.\\
								&\left. q = \sum_{e: s(e)=y} p_e(y) - \sum_{e: t(e) =y} p_e(y), \quad k_j = \sum_{e: s(e)=x_j} p_e(x_j) - \sum_{e: t(e) =x_j} p_e(x_j) \right\}.
							\end{split}
						\end{equation}
					Following~\citep{BFK96,brunetti2000microlocal, HW01}, a decorated graph $\sG$ is an embedded graph in $M$ with vertices $y_1, \dots, y_\ell$, $x_1 \dots, x_n$, where $y_1, \dots, y_\ell$ are ``external vertices'' and $x_1, \dots, x_n$ are ``internal vertices'', and with edges connecting the vertices given by oriented null-geodesic curves. The valence of a vertex in the graph is here restricted to be less or equal to the number of field factors appearing in the corresponding classical functionals $\Phi$. An abstract ordering $<$ of the vertices is chosen (not related to the causal structure of $M$). It is required that the ordering satisfies $x_1 < \dots < x_n$ for the internal vertices, while no restrictions are imposed for the external vertices. For each edge $e$ we call source (denoted by $s(e)$) the smaller endpoint with respect to $<$ and we call target (denoted by $t(e)$) the bigger endpoint with respect to $<$. We consistently impose an orientation for the null-geodesic corresponding to $e$ in such a way that the curve starts at $s(e)$. Each edge is equipped with a future-directed covector field $p_e$ which is cotangent and coparallel to the geodesic curve associated to the edge $e$.
				\item\label{R4} {\em Smoothness}:\\
					The retarded products have a smooth functional dependence on $g, m,\phi,\lambda$ in the following sense. Consider the smooth $1$-parameter families $\{g^{(s)}\}_{s \in \Omega}$, $\{ m^{(s)} \}_{s \in \Omega}$, $\{\phi^{(s)}\}_{s \in \Omega}$ and $\{\lambda^{(s)}\}_{s \in \Omega}$, where $\Omega$ an domain in $\bR^p$. Furthermore, let $\{ \omega^{(s)}\}_{s \in \Omega}$ be a collection of quasi-free Hadamard states $\omega^{(s)}$ for the algebras $\cW_{\phi^{(s)}} = \cW[g^{(s)},m^{(s)},\phi^{(s)}, \lambda^{(s)}]$ such that the $2$-point functions $\omega^{(s)}(x_1,x_2)$, seen as a distribution in $\Omega \times M^2$, satisfies
						\begin{equation*}
							\WF(\omega^{(s)}(x_1,x_2)) \subset \left\{ (s,x_1,x_2; \rho, k_1, k_2) \in \dot{T}^\ast (\Omega \times M^2) : (x_1,x_2;k_1,k_2) \in \cC^\triangleright [g^{(s)}] \right\}.
						\end{equation*}
					For any $n$, the collection $\{ \omega^{(s)}_{R,n}(y;x_1, \dots x_n) \}_{s \in \Omega}$, where $\omega^{(s)}_{R,n}$ is defined as in~\eqref{omega_ret_m}, can be interpreted as a distribution in the variables $(s,y,x_1, \dots x_n)$. We require that this distribution satisfies
						\begin{equation*}
							\begin{split}
								\WF( \omega^{(s)}_{R,n}) \subset &\left\{ (s,y,x_1, \dots, x_n; \rho, p, k_1, \dots, k_n) \in \dot{T}^\ast(\Lambda \times M^n): \right. \\
								&\qquad \left. (y,x_1, \dots, x_n; p, k_1, \dots, k_n) \in \cC^R_{1+n}[g^{(s)}] \right\}.
							\end{split}
						\end{equation*}
					Furthermore, if we consider variations only of the background $\phi$, i.e. $g^{(s)} = g$, $m^{(s)}=m$ and $\lambda^{(s)} = \lambda$, and if $\{ \omega^{(s)} \}_{s \in \Omega}$ is a collection of quasi-free Hadamard states such that $\WF( \omega^{(s)}(x_1,x_2) ) \subset \Omega \times \{0\} \times \cC^\triangleright$, then we have $\WF( \omega^{(s)}_{R,n}) \subset \Omega \times \{0\} \times \cC^R_{1+n}$.
				\item\label{R5} {\em Analyticity}\footnote{It is worth mentioning that a recent result \citep{khavkine2014analytic} suggests that the analyticity condition can be dropped.}:\\
					Similarly, we require that for analytic families of analytic metric, masses, backgrounds and couplings (and analytic $V$), the expectation value of the retarded products in an analytic family of states varies analytically in the same sense as~\ref{R4}, replacing the smooth wave-front set with the analytic wave-front set (see \citep{H83}).
				\item\label{R6} {\em Symmetry}:\\
					The map $R_{n,\phi}$ is symmetric in the last $n$ entries.
				\item\label{R7} {\em Unitarity}:\\
					We require that
						\begin{equation*}
							R_{n,\phi} \left( F ; \bigotimes_{j=1}^n H_j \right)^\dagger = (-1)^n R_{n,\phi} \left( \overline{F} ; \bigotimes_{j=1}^n \overline{H}_j \right),
						\end{equation*}
					where $\overline{F}$ denotes complex conjugation, i.e. if the local functional $F$ is $F(\phi)= \int_M \cP(x) dx$ for a polynomial $\cP$ as in~\eqref{F_loc_general form}, then $\overline{F}$ is defined by $F(\phi):=\int_M \overline{\cP(x)} dx$.
				\item\label{R8} {\em Causality}:\\
					If $(\cup_j \supp H_j) \cap J^-(\supp F )= \emptyset$, then  $R_{n,\phi} ( F ; \otimes_{j=1}^n H_j)=0$.
				\item\label{R9} {\em Field independence}:\\
					Let $u$ be a (space-like compact) smooth solution of $P_\phi u =0$, then
						\begin{equation*}
							\begin{split}
								\langle u , \frac{\delta}{\delta \varphi} \rangle R_{n,\phi} \left( F ; \bigotimes_{j=1}^n H_j \right) &= R_{n,\phi} \left( \langle u , \frac{\delta}{\delta \varphi} \rangle F ; \otimes_{j=1}^n H_j \right) +\\
								&\quad + \sum_{j=1}^n R_{n,\phi} \left( F ; H_1 \otimes \cdots \langle u , \frac{\delta}{\delta \varphi} \rangle H_j \cdots \otimes H_n \right),
							\end{split}
						\end{equation*}
					where $\langle u , \delta / \delta \varphi \rangle$ acts on $\cW_\phi$ as\footnote{Equivalently $\langle u , \frac{\delta}{\delta \varphi} \rangle = [\cdot, \varphi(f_\phi)]_\bullet$ for any $f_\phi \in C^\infty_0(M)$ such that $u= E_\phi(f_\phi)$.}
						\begin{equation}\label{variation_varphi}
							\langle u , \frac{\delta}{\delta \varphi} \rangle \varphi^{\otimes n}(f^{(n)}) := n \int_{M^n} f^{(n)}(x_1, \dots, x_n) u(x_1) \varphi(x_2) \cdots \varphi(x_n) dx_1 \dots dx_n,
						\end{equation}	
					whereas $\langle u, \delta / \delta \varphi \rangle$ acts on $\cF_\loc(M)$ as the Gateaux derivative along the direction of $u$.							
				\item\label{R10} {\em Leibniz rule/Action Ward Identities}:\\
					$R_{n,\phi}$ commutes with the derivatives, i.e.
							\begin{equation*}
								\nabla_{x_i} R_{n,\phi}(\Phi_{1,\phi}(x_1) \otimes \cdots \otimes \Phi_{n,\phi}(x_n)) = R_{n,\phi}(\Phi_{1,\phi}(x_1) \otimes \cdots \otimes \nabla_{x_i}\Phi_{i,\phi}(x_i) \otimes \cdots \otimes \Phi_{n,\phi}(x_n)).
							\end{equation*}
				\item\label{R11} {\em GLZ (Glaser-Lehmann-Zimmermann) formula}:\\
					For $n \geq 2$
						\begin{equation}\label{GLZ}
							\begin{split}
								&R_{n,\phi} \left( F ; F' \otimes \bigotimes_{j=1}^{n-1} H_j \right) - R_{n,\phi} \left( F' ; F \otimes \bigotimes_{j=1}^{n-1} H_j \right) = \\
								&\qquad = \sum_{I \subset \{1, \dots, n-2 \}} \left[ R_{n,\phi} \left( F ; \bigotimes_{i \in I} H_i \right), R_{n,\phi} \left( F' ; \bigotimes_{j \in I^c} H_j \right)\right]_{\bullet_\phi}.
							\end{split}
						\end{equation}
			\end{enumerate}
		\noindent
		The final key property of retarded products is the {\em principle of perturbative agreement} discussed in~\citep{HW05}. This principle can be invoked to relate the quantum field defined by the retarded products $\{R_{m,\phi}\}$ and $\{R_{m,\phi'}\}$ for different backgrounds $\phi, \phi'$ as follows. The quantum field theories corresponding to the quadratic actions $I^{(2)}_\phi, I^{(2)}_{\phi'}$ are both exactly solvable and trivially
			\begin{equation}\label{split_quadratic_quadratic}
				 I^{(2)}_{\phi'}(\varphi) = I^{(2)}_\phi(\varphi) + \int_M \frac{1}{2} (v_{\phi'} - v_\phi)(x) \varphi(x)^2 dx \equiv I^{(2)}_\phi(\varphi) + \cV_{\phi,\phi'}(\varphi)
			\end{equation}
		If $V$ is compactly supported as in eq.~\eqref{leom}, the second term in the right-hand side is a local functional $\cV_{\phi,\phi'}(\varphi)$, which we may choose to treat perturbatively via the series~\eqref{naive_Haag_formula}, where $\cL^{(p)}$ is now $\frac{1}{2} (v_{\phi'} - v_\phi) \varphi^2$. Of course, there is no need to do this really, because the theory can be defined ``exactly'' proceeding with $I^{(2)}_{\phi'}$ in the first place. If we demand that the two procedures gives the ``same'' result, then we get non-trivial relations between $\{R_{\phi,n}\}$ and $\{ R_{\phi', n} \}$.\\
		We now state these relations in a precise manner. First, we note that the retarded products $\{R_{\phi,n}\}$ and $\{ R_{\phi', n} \}$ take values in {\em different} algebras, $\cW_\phi$ and $\cW_{\phi'}$. So before we can compare them, we must first define a suitable isomorphism between these algebras. This isomorphism is constructed following~\citep{HW05}. First, we fix a Hadamard $2$-point function $\omega_0$ with respect to the Klein-Gordon operator $\boxempty - m^2$ (in the case of a static space-time $\omega_0$ can be chosen as the ground state). Then, we construct the so-called {\em retarded state} (or ``in state'') with respect to $\omega_0$, denoted by $\gls{omegaR_phi}$. This state is uniquely characterized by the fact that for all $x,y \notin J^+(\supp \lambda)$ the corresponding $2$-point function satisfies 
			\begin{equation}\label{in-state}
				\omega_\phi^R(x, y) = \omega_0 (x,y).
			\end{equation}
		This requirement is consistent because $\omega_\phi^R(x, y)$ is a $2$-point function with respect to $(\boxempty - m^2-v_\phi)$ and $v_\phi(x) = V''(\phi(x))=\frac{1}{2}\lambda(x) \phi^2(x)$ vanishes for $x \notin J^+(\supp \lambda)$.  The complement of the region $J^+(\supp \lambda)$ contains a Cauchy surface. Therefore, the requirement~\eqref{in-state} uniquely defines the quasi-free state $\omega^R_\phi$ because its $2$-point function obeys a hyperbolic equation in both the entries (see~\citep{FSW78, FNW81}). We can similarly construct the retarded state $\omega^R_{\phi'}$ (with respect to $\omega_0$) for the background $\phi'$.\\
		We next define our algebras $\cW_\phi = \cW(M, \omega^R_\phi)$, $\cW_{\phi'} = \cW(M, \omega^R_{\phi'})$ by constructing the product via the Hadamard $2$-point functions of the retarded states just described. The desired isomorphism $\alpha_{\phi,\phi'}^R : \cW_{\phi} \to \cW_{\phi'}$ is then constructed as follows. Let $t=(t^n)_{n \in \bN}$ be an element of $\cW_\phi$ with $\deg_\hbar = 0$. Let us identify each $t^n$, which is an equivalence class in $\cE'_W(M^n) / (\boxempty - m^2 v)\cE'_W(M^n)$ (see~\eqref{formal_Wick_algebra_space_QFT}), with one of its representatives in $\cE'_W(M^n)$. Then, we define $\alpha_{\phi,\phi'}^R$ by
			\begin{equation}\label{alpha_R_wick_per}
				\alpha_{\phi,\phi'}^R (t^n) := \left[ \left(A^R_{\phi,\phi'} \right)^{\otimes n}t^n \right],
			\end{equation}
		and then we extend $\alpha_{\phi,\phi'}^R$ by $\bC[[\hbar]]$-linearity to the whole algebra $\cW_\phi$. The distribution $A^R_{\phi,\phi'} \in \cD(M^2)$ is uniquely characterized by demanding that:
			\begin{enumerate}
				\item $\alpha_{\phi,\phi'}^R$ is a homomorphism of algebras.
				\item $\alpha^R$ satisfies the ``cocycle condition''
					\begin{equation}\label{cocycle}
						\alpha_{\phi',\phi''}^R \alpha_{\phi,\phi'}^R = \alpha_{\phi, \phi''}^R.
					\end{equation}
				\item Outside the future of the support of $v_\phi, v_{\phi'}$ the map $\alpha^R_{\phi, \phi'}$ is the identity.
			\end{enumerate}
		It turns out that $A^R_{\phi, \phi'}$ must be given explicitly by\footnote{We make use of the Schwartz kernel theorem (see e.g.~\citep{H83}) to identify distributions $\cD'(X \times Y)$ with continuous functionals $C^\infty_0(X) \to \cD'(Y)$.}
			\begin{equation}\label{A_R}
				A^R_{\phi,\phi'}(f) := - (\boxempty - m^2 - v_{\phi'}) \left(c S^R_{\phi,\phi'}(f) \right),
			\end{equation}
		where $S^R$ is defined by
			\begin{equation}\label{S_R}
				S^R_{\phi,\phi'}(f_1,f_2) := \int_{\Sigma_-} E_{\phi'}[f_1](z) \overleftrightarrow{\partial_n} E_{\phi}[f_2](z) d\Sigma(z),
			\end{equation}
		where $\Sigma_-$ is a Cauchy surface in the complement of $J^+(\supp \lambda)$, and where $c$ is a ``(retarded) regularized step function'' with respect to $\Sigma_-$, i.e. a smooth function with values in $[0,1]$ such that $c = 1$ in $J^-(\Sigma_-)$ and $c = 0$ in $J^+(\Sigma_+)$ for a Cauchy surface $\Sigma_+$ in the complement of $J^-(\supp \lambda)$. The situation is sketched in fig.~\ref{fig:ret_cutoff}.
		\begin{figure}
		\centering
		\begin{tikzpicture}
		\node (c=0) at (4,3.9) {$c=0$};
		\path [pattern=north west lines, pattern color = black!20] (0,2) -- (8,2) -- (8,3.4)-- (0,3.4) -- (0,2);
		\path [fill = black!20] (0,1) -- (8,1) -- (8,2)-- (0,2) -- (0,1);
		\node (c=1) at (4,1.5) {$c=1$};
		\draw [thick] (0,3.4) -- (8,3.4);
		\node (a1) at (1, 3.35) {};
		\node (a2) at (0,4) {$\Sigma_+$};
		\draw [-latex, bend left] (a2) to (a1);
		\draw [thick] (0,2) -- (8,2);
		\node (b1) at (1, 1.95) {};
		\node (b2) at (0,2.6) {$\Sigma_-$};
		\draw [-latex, bend left] (b2) to (b1);
		\draw [draw=red, ultra thick]  (4,2.7) ellipse (1 and 0.5);
		\node [red] (lambda) at (4,2.7) {$\supp \lambda$};
		\end{tikzpicture}
		\caption{Choice of $\Sigma_\pm$ and $c$ adapted to $\supp V$.}\label{fig:ret_cutoff}
		\end{figure}
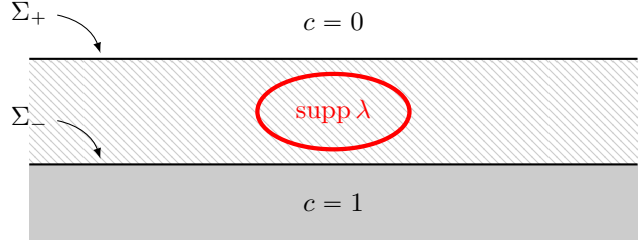\\
		After these preparations, we can now formulate the principle of perturbative agreement.
		\begin{enumerate}[label=(R\arabic*), start=12]
			\item\label{R12} {\em Principle of perturbative agreement for variations of the background $\phi$}:\\
				Let $\{ \phi^{(s)} \}_{s \in \Omega}$ be a smooth $1$-parameter family of backgrounds, where $\Omega$ is an open interval of $0$ and where $\phi := \phi^{(0)}$. Let $F_\phi, G_\phi$ be local functionals depending ($W$-smoothly) on the background $\phi$\footnote{In particular, we are interested in $F_\phi = I^{(p)}_\phi$ defined above in \eqref{action_exp}.}. We require that
				\begin{equation*}
					\begin{split}
						&\frac{\partial}{\partial s} \alpha^R_{\phi^{(s)}, \phi} \left( R_{n,\phi^{(s)}} \left( F_{\phi^{(s)}}, \bigotimes_{j=1}^n G_{j,\phi^{(s)}} \right) \right) =\\
						&\quad = \frac{i}{\hbar} R_{n+1, \phi} \left( F_{\phi} ; \bigotimes_j G_{j,\phi} \otimes \frac{\partial I^{(2)}_{\phi^{(s)}}}{\partial s} \right) + R_{n,\phi} \left( \frac{\partial F_{\phi^{(s)}}}{\partial s} ; \bigotimes_j G_{j,\phi} \right) + \\
						&\qquad +\sum_{\ell=1}^n R_{n,\phi} \left( F_{\phi} ; \frac{\partial G_{\ell,\phi^{(s)}}}{\partial s}\otimes  \bigotimes_{j \neq \ell} G_{j,\phi} \right).
					\end{split}
				\end{equation*}
			To simplify the notation (here and in appendix~\ref{app_proof_R12}) we always consider the derivative $\partial/ \partial s$ as evaluated at $0$, unless stated otherwise.
		\end{enumerate}
		We refer to~\citep{HW05} for an explanation why this encodes the heuristic idea discussed around eq.~\eqref{split_quadratic_quadratic}.\\
		
		The fundamental result is that there exists a prescription for retarded products satisfying~\ref{R0}-\ref{R12}, and that these axioms uniquely define the maps $\{R_{n,\phi}\}$ up to well-characterized ``finite renormalization ambiguities'':
			\begin{theo}\label{theo_R12}
				There exists a prescription $R_\phi=\{R_{n,\phi}\}_{n \in \bN}$ for retarded products which satisfies axioms~\ref{R0}-\ref{R11} and~\ref{R12}.\\
				Moreover, if $R'_\phi=\{R'_{n,\phi}\}_{n \in \bN}$ is another prescriptions for the retarded products which satisfies the axioms~\ref{R0}-\ref{R12}, then there exists a hierarchy $D_\phi=\{D_{n,\phi}\}_{n \in \bN}$ of maps
					\begin{equation*}
						D_{n,\phi}: \cF_{\loc}(M)^{\otimes n} \to \cF_{\loc}(M)[[\hbar]],
					\end{equation*}
				which satisfies
					\begin{equation*}
						R_\phi' \left( F ; \exp_\otimes \left( \frac{i}{\hbar} H \right) \right) = R_\phi \left( F + D_\phi \left( F \otimes \exp_\otimes H \right) ; \exp_\otimes \left( \frac{i}{\hbar} \left( H + D (\exp_\otimes H) \right) \right) \right),
					\end{equation*}
				and the following properties:
					\begin{itemize}
						\item $D_{n,\phi}(F_{1} \otimes \cdots \otimes F_{n})$ is of order $O(\hbar)$.
						\item $D_{1,\phi}(1)=0$, $D_{1,\phi}(\varphi(f))=0$.
						\item $D_{n,\phi}$ are local/covariant functionals in the following sense: let $\iota: M \to M'$ be any causality and orientation preserving isometric embedding between two space-times, i.e. $\iota^* g' = g$, then $\iota^* \circ D'_{n,\iota^*\phi}=D_{n,\phi} \circ (\iota^*)^{\otimes n}$.
						\item For any monomials $\Phi_{1,\phi}, \dots, \Phi_{n,\phi}$ as in~\ref{R2}, the distribution $D_{n,\phi}(\Phi_{1,\phi}(x_1) \otimes \cdots \otimes \Phi_{n,\phi}(x_n))$ is supported on the diagonal and it satisfies the wave-front set condition
									\begin{equation*}
										\left. \WF(D_{n,\phi}) \right|_{\Delta_{n+1}} \perp T\Delta_{n+1}.
									\end{equation*}
								Furthermore, $D_{n,\phi}$ depends smoothly (even analytically) on the background $\phi$, i.e. for a smooth (repsectively analytic) family $\bR \ni s \mapsto \phi_s$ of backgrounds it holds
									\begin{equation*}
										\left. \WF(D_{n,\phi_s}) \right|_{\bR \times \Delta_{n+1}} \perp T(\bR \times \Delta_{n+1}),
									\end{equation*}
								(where the smooth wave-front set $\WF$ must be replaced with the analytic wave-front set $\WF_A$ in the analytic case).
						\item $D_{n,\phi}$ depends only polynomially on the Riemann curvature tensor, and the functions $m$, $\phi$ and $\varphi$ (as well as their covariant derivatives).\\
						Moreover, $D_{n,\phi}$ satisfies the following scaling constraint: there exists $N$ such that
								\begin{equation*}
									\frac{\partial^N}{\partial^N \log \Lambda} \Lambda^{-d_D} D_{n,\phi}(\Phi_{1,\phi}(x_1) \otimes \cdots \otimes \Phi_{n,\phi}(x_n)) = 0,
								\end{equation*}
							where $d_D = \sum_j d_{\Phi_j}$ denotes the engineering dimension.
						\item $D_{n,\phi}$ is symmetric.
						\item For $u \in C^\infty(M)$, it holds
								\begin{equation*}
									\langle u , \frac{\delta}{\delta \varphi} \rangle D_{n,\phi} \left(\bigotimes_{j=1}^n F_{j} \right) =\sum_{\ell=1}^n D_{n,\phi} \left( \langle u , \frac{\delta}{\delta \varphi} \rangle F_\ell \otimes \bigotimes_{j \neq \ell} F_j \right).
								\end{equation*}
							Remember that $\langle u ,\delta /\delta \varphi \rangle$ acts on $\cF_{\loc}$ as the Gateaux derivative in the direction $u$.
						\item $D_{n,\phi}$ commutes with the derivative, i.e.
							\begin{equation*}
								\nabla_{x_i} D_{n,\phi}(\Phi_{1,\phi}(x_1) \otimes \cdots \otimes \Phi_{n,\phi}(x_n)) = D_{n,\phi}(\Phi_{1,\phi}(x_1) \otimes \cdots \otimes \nabla_{x_i}\Phi_{i,\phi}(x_i) \otimes \cdots \otimes \Phi_{n,\phi}(x_n)).
							\end{equation*}
					\end{itemize}
			\end{theo}
		\noindent
		A proof of this theorem can be be given following the methods of~\citep{HW05}. Compared to the existing constructions in the literature, a non-trivial extra point is that the axiom~\ref{R12}, i.e. the principle of perturbative agreement, can be consistently imposed with~\ref{R0}-\ref{R11}. This is proved in appendix~\ref{app_proof_R12}.\\
		 
		We now construct the interacting fields, referring to the literature~\citep{DF01a, DF01b, HW03, DF04, HW05} for more details. In our perturbative setting, interacting quantum fields are given by formal power series in the algebra $\cW_\phi$ involving retarded products as we already anticipated in~\eqref{naive_Haag_formula}. The precise definition is as follows:
			\begin{defi}
				Let $F \in \cF_{\loc}$ be a local observable, and let $V$ be a potential with compact support such as $V(x, \phi(x)) = \frac{1}{4!}\lambda(x) \phi^4(x)$, where $\lambda$ has compact support. For each background configuration $\phi \in C^\infty(M)$, the corresponding interacting quantum field observable (with respect to the action $I_\phi^{(2)} + \sum_{p >2} I_\phi^{(p)}$) is an element $\hat{F}_\phi \in \cW_\phi$ defined by the {\em Haag series}~\citep{haag1955quantum, glaser1957field}
					\begin{equation}\label{Haag_series}
						\gls{hatF}_\phi := \sum_{n \geq 0} \frac{i^n}{\hbar^n n!} \sum_{p_1, \dots, p_n >2} R_{n,\phi} \left( F(\phi + \varphi); I^{(p_1)}_\phi(\varphi) \otimes \cdots \otimes I^{(p_n)}_\phi(\varphi) \right),
					\end{equation}
				where $\{R_{n,\phi}\}$ denote the retarded products in the background $\phi$. 
			\end{defi}
		\noindent
		The definition of $\hat{F}_\phi$ just presented makes precise formula~\eqref{naive_Haag_formula}.\\
		
		We now want to investigate how the interacting field changes under a change of background $\phi$. We can understand this in the light of the principle of perturbative agreement~\ref{R12}. Let $\phi$ be a smooth solution to the background equations of motion eq.~\eqref{eom} and let $u$ be a smooth solution to the linearized equation~\eqref{leom} around $\phi$. We consider a map $S \ni \phi \mapsto t_\phi \in \cW_\phi$ which satisfies the following smoothness properties: for any test functions $f_1, \dots, f_n$, and for any smooth map $\epsilon \mapsto \phi(\epsilon) \in S$, $\epsilon \mapsto t^n_{\phi(\epsilon)}(E_{\phi(\epsilon)}(f_1) \otimes \cdots \otimes E_{\phi(\epsilon)}(f_n))$ is smooth, where $t^n_\phi$ is the $\deg_s =n$ part of $t_\phi$. We may define its ``retarded directional derivative in the direction of $u$'' as 
			\begin{equation}\label{R/A_der_per}
				\left( \gls{nablaR}_u t \right)_\phi := \left. \frac{d}{d\epsilon} \alpha^R_{\phi(\epsilon), \phi} (t_{\phi(\epsilon)}) \right|_{\epsilon = 0}.
			\end{equation}
		where $\alpha^R$ is the same map as that defined in the principle of perturbative agreement~\ref{R12}, and where $\bR \ni \epsilon \mapsto \phi(\epsilon) \in S$ is a smooth map such that $\phi(0)=\phi$ and $d\phi(\epsilon) / d \epsilon |_{\epsilon=0} = u$. We call $\nabla^R$ the ``retarded connection''.  From the cocycle condition~\eqref{cocycle} it follows immediately that the retarded connection is {\em flat}.\\

		We focus now on the case $V(\phi)=\frac{\lambda}{4!} \phi^4$ with $\lambda \in C^\infty_0(M)$ and we consider a smooth global solution $\phi$ to the corresponding background equation~\eqref{eom}, i.e.
			\begin{equation}\label{eom_phi4}
				(\boxempty - m^2)\phi - \frac{1}{3!}\lambda \phi^3 = 0.
			\end{equation}
		Regarding the existence of smooth global solutions of eq.~\eqref{eom_phi4}, at least for ultra-static space-times with compact Cauchy surfaces, we refer to appendix~\ref{app_nnlin}. Let $u$ be a smooth solution to the linearised equations at $\phi$, i.e.
			\begin{equation}\label{leom_phi4}
				(\boxempty - m^2 - \frac{1}{2} \lambda \phi^2)u = 0.
			\end{equation}
		Let $F$ be a local observable, and let $\hat{F}_\phi \in \cW_\phi$ be the corresponding quantum observable defined by the series (\ref{Haag_series}). We think of $u$ as a ``tangent vector'' at $\phi$ to the ``manifold'' of smooth non-linear solutions (the rigorous definitions of the infinite-dimensional geometry will be provided in chapter~\ref{sec_Fedosov_inf}). We compute:
			\begin{equation*}
				\begin{split}
				\left( \nabla_u^R \hat{F} \right)_\phi &= \sum_{k \geq 0} \frac{i^k}{\hbar^k k!} \sum_{p_1, \dots, p_k >2} \left\{  \frac{i}{\hbar} R_\phi \left( F(\phi + \varphi);  \langle u, \frac{\delta}{\delta \phi} \rangle I^{(2)}_\phi (\varphi) \otimes \bigotimes_{i=1}^k I^{(p_i)}_\phi(\varphi)\right) + \right. \\
				&\quad + R_\phi \left( \langle u, \frac{\delta}{\delta \phi} \rangle F(\phi + \varphi); \bigotimes_{i=1}^k I^{(p_i)}_\phi(\varphi) \right) + \\
				&\quad  + \left. \sum_{j\leq k} R_\phi \left( F(\phi + \varphi); \langle u, \frac{\delta}{\delta \phi} \rangle I^{(p_j)}_\phi(\varphi) \otimes \bigotimes_{i \neq j} I^{(p_i)}_\phi(\varphi) \right) \right\},
				\end{split}
			\end{equation*}
			where $\langle u, \delta / \delta \phi \rangle$ is the Gateaux derivative in $\phi$ along the direction $u$. This formula can be simplified as follows. Using \eqref{action_exp} and \eqref{variation_varphi}, it holds
			\begin{equation*}
				\langle u, \frac{\delta}{\delta \phi} \rangle I^{(p)}_\phi(\varphi) = \langle u, \frac{\delta}{\delta \varphi} \rangle I^{(p +1)}_\phi (\varphi), \qquad \langle u, \frac{\delta}{\delta \phi} \rangle F(\phi +\varphi) = \langle u, \frac{\delta}{\delta \varphi} \rangle F(\phi +\varphi).
			\end{equation*}
		Next, we apply the field-independence axiom~\ref{R9} to pull the operator $\langle u, \frac{\delta}{\delta \varphi} \rangle$ in front of everything. We summarize the above computation by the following result.
			\begin{theo}\label{theo_Fedosov_per}
				Let $F$ be a local observable in $\cF_{\loc}(M)$ and $\phi \mapsto \hat{F}_\phi$ as in eq.~\eqref{Haag_series}. Then, for any smooth solution $u$ to the linearised equation~\eqref{leom_phi4} we have
				\begin{equation}\label{Fedosov_per}
					\left( \nabla_u^R - \langle u, \frac{\delta}{\delta \varphi} \rangle \right) \hat{F} = 0.
				\end{equation}
			\end{theo}
		\noindent		
		The operator $\nabla^R - \langle \cdot , \delta / \delta \varphi \rangle$ clearly has a striking similarity with Fedosov connection~\eqref{Fedosov_conn_fin}, noting that $\langle \cdot, \delta / \delta \varphi \rangle$ is equal to the Fedosov operator $\delta$ (see~\eqref{Fedosov_op}) in the present context. Furthermore, from this point of view, the condition~\eqref{Fedosov_per} simply means that the interacting observables $\hat{F}$ are, as functions of the background solution $\phi$, flat sections in the ``algebra bundle'' $\sqcup_{\phi} \cW_\phi$ (more carefully defined below in sec.~\ref{subsec_manifold_inf}). We thereby get a first hint that the ``standard'' method of quantization based on retarded products -- while looking completely different at first sight-- might have something to do with Fedosov quantization. In the following sections, we will describe a version of Fedosov's method appropriate for the setting of field theory. Then we will investigate the relation of these methods in chapter~\ref{sec_rel_pQFT_Fedosov}.

\chapter{Fedosov quantization for quantum field theory}\label{sec_Fedosov_inf}
 In this chapter, we prove that it is indeed possible to implement Fedosov's procedure in the infinite-dimensional framework of a quantum field theory for non-linear equations of motion, but many new ideas, which are going to be extensively explained, are required. In sec.~\ref{subsec_manifold_inf}, we first characterize rigorously the infinite-dimensional symplectic manifold of the smooth solutions of the non-linear (more precisely, $\lambda \phi^4$-interacting) Klein-Gordon equation on an ultra-static space-time with compact Cauchy surfaces. We then define the geometric set-up to discuss Fedosov's scheme in infinite dimensions. In particular, we provide the appropriate notion of smoothness, called ``on-shell $W$-smoothness'', and the definitions of the corresponding covariant tensor bundles necessary for constructing the vector space structure of the formal Wick algebra. In sec.~\ref{subsec_W_smooth_symp_metric}, we provide two concrete on-shell $W$-smooth tensor fields corresponding respectively to the symplectic structure and the almost-K\"{a}hler structure. In sec.~\ref{subsec_algebra_W_smooth_sec}, we discuss the algebra structure of the formal Wick algebra. In particular, we define the product of on-shell $W$-smooth sections on the formal Wick algebra and, more generally, of on-shell $W$-smooth forms with values in the formal Wick algebra. The appropriate notion of covariant derivative is presented in sec.~\ref{subsec_W_covariant_der}. We define two concrete covariant derivatives, corresponding to the Levi-Civita connection and the Yano connection in the finite-dimensional case. The non-trivial results proved in sec.~\ref{subsec_algebra_W_smooth_sec} and~\ref{subsec_W_covariant_der} concern the consistency of the product and, respectively, the covariant derivatives with the notion of on-shell $W$-smoothness. With the description of setting completed, we state and prove the infinite-dimensional version of Fedosov's theorems in sec.~\ref{subsec_Fedosov_inf}.\\
 The remaining task will be then to explain the precise relationship between this construction and the construction based on the ``causal perturbation theory'' described in sec.~\ref{subsec_int_QFT_per}. This question will be addressed in chapter~\ref{sec_rel_pQFT_Fedosov}.
 
\section{The manifold structure of $S$}\label{subsec_manifold_inf}
	We have already highlighted in sec.~\ref{subsec_free_QFT} the formal similarities between Fedosov quantization for finite-dimensional almost-K\"{a}hler manifolds and perturbative quantization in the case of the free field (see table~\ref{table:fin_linQFT}). At the end of sec.~\ref{subsec_int_QFT_per}, we have seen a hint that these formal analogies can be extended to interacting models. Throughout the rest of the work we substantiate this.\\
	For technical reasons, we consider only the case of the interaction $V(\phi) = \frac{\lambda}{4!} \phi^4$, where $\lambda \in C^\infty_0(M)$. The analogue of the classical underlying almost-K\"{a}hler manifold, $S$, is the topological space of smooth solutions to the non-linear Klein-Gordon equation
		\begin{equation*}
			\gls{S} = \left\{ \phi \in C^\infty(M) : (\boxempty - m^2) \phi +  \frac{\lambda}{3!} \phi^3 = 0 \right\}.
		\end{equation*}
	Smooth solutions $u$ to the linearised equation around a background $\phi \in S$ are naturally viewed as tangent vectors to $S$, i.e. elements $u \in T_\phi S$. The algebra $\cW_\phi$ can next be defined for all $\phi \in S$ as the corresponding algebra in the finite-dimensional situation. It is modelled over the symmetrized tensor powers of $T^*_\phi S$, and the product is given in terms of a suitable smooth assignment $S \ni \phi \mapsto \omega_\phi$, where each $\omega_\phi$ is a pure Hadamard $2$-point function. This provides the analogue of the almost-K\"{a}hler structure for $S$. In particular, we may choose $\omega_\phi$ as the retarded state $\omega^R_\phi$ with respect to the unique ground state $\omega_0$ (cf.~\eqref{in-state}). In this case, we will see in sec.~\ref{subsec_gaugeeq} then, that the operator $\nabla^R - \langle \cdot, \delta / \delta \varphi \rangle$, defined at the end of sec.~\ref{subsec_int_QFT_per}, is roughly speaking the Fedosov connection associated with this particular almost-K\"{a}hler structure. However, in order to turn these formal analogies into precise mathematical ones, we must be careful about the infinite-dimensional nature of $S$. Thus, we will begin by equipping $S$ with the structure of an infinite-dimensional {\em Fr\'{e}chet manifold}, and then we will define precisely the bundles over $S$ needed in Fedosov's method, namely $TS$, $T^*S$, and $\cW$, and their differentiable structures. This will be done in the rest of the present section.\\
	
	First of all, we recall the definition of ``Fr\'{e}chet spaces'' and of ``Fr\'{e}chet manifolds''. A {\em Fr\'{e}chet space} is a locally convex vector space, i.e. a vector space equipped with a family of countably many seminorms such that the topology is induced by this family of seminorms. One can define naturally a metric for a Fr\'{e}chet space\footnote{Once chosen a family of seminorms $\{ p_n \}_{n \in \bN}$ generating the topology of the Fr\'{e}chet space $\sF$, the metric is defined by
		\begin{equation*}
			d(f,h) := \sum_{n\in \bN} 2^{-n} \frac{p_n(f-h)}{1 + p_n(f -h)} . 
		\end{equation*}
	for $f,h \in \sF$.}. A Fr\'{e}chet space is required to be complete with respect to this metric.  The space of smooth functions over a finite-dimensional manifold endowed with the compact-open topology\footnote{In appendix~\ref{app_nnlin}, we review explicitly the definition of the compact-open topology.}, also called ``topology of uniform convergence on compact set of $M$'', is the prime example of a Fr\'{e}chet space, see e.g.~\citep{brunetti2012algebraic}.\\
	A {\em Fr\'{e}chet manifold} is a topological space $F$ modelled upon Fr\'{e}chet spaces, in the same way as a smooth\footnote{Transition maps between two overlapping charts are smooth.} $n$-dimensional manifold is a topological space modelled upon $\bR^n$. More precisely, in the Fr\'{e}chet manifold context, an atlas is a collection of charts $\{(U_\alpha, \rho_\alpha)\}_{\alpha \in A}$, where $\{U_\alpha\}_{\alpha \in A}$ is an open covering the topological space $F$ and each $\rho_\alpha$ is an homeomorphism from $U_\alpha$ into an open subset of a Fr\'{e}chet space $\sE_\alpha$.\\
	We would now like to equip the space $S$ of smooth solutions to~\eqref{eom_phi4} with the structure of a Fr\'{e}chet manifold. To define the manifold structure of $S$, we use a description of $S$ in terms of initial data. To avoid excessive technicalities, we shall restrict attention, from now on, to space-times $(M,g)$ that are {\em spatially compact}, i.e. have a compact Cauchy surface $\Sigma$, and carry an {\em ultra-static} metric
		\begin{equation}\label{ultra_stat_metric}
			g = -dt^2 + h_{ij} dx^i dx^j,
		\end{equation}
	where the spatial part $h$ does not depend on the global time coordinate $t$. As is shown in appendix~\ref{app_nnlin} (prop.~\ref{prop_nn-lin_Cauchy_prob}), in this situation (and probably more generally, too) for each set of smooth initial data $(q,p) \in C^\infty(\Sigma) \times C^\infty(\Sigma)$ there exists a unique, globally defined, smooth solution $\phi \in S$ such that
		\begin{equation*}
			q = \phi |_{\Sigma}, \qquad p = \partial_n \phi |_{\Sigma} . 
		\end{equation*}
	This correspondence naturally establishes an isomorphism between $S$ and the space $\sE := C^\infty(\Sigma) \oplus C^\infty(\Sigma)$. The linear space $\sE$ has the structure of a Fr\'{e}chet space when equipped with the canonical topology defined by the direct sum of the Fr\'{e}chet seminorm of each copy of $C^\infty(\Sigma)$, see e.g.~\citep[Chapter 10]{treves2006topological}. The isomorphism between solutions and initial data thereby induces a Fr\'{e}chet manifold structure on $S$. By the continuous dependence of the solution $\phi$ on its initial data, proved in appendix~\ref{app_nnlin} (prop.~\ref{prop_U_cont}), it follows that the topology on $S$ induced by the compact open topology on $C^\infty(M)$ is compatible with that manifold structure. In detail, let $\rho: S \to \sE$ be defined  as the ``restriction map'', i.e.
		\begin{equation}\label{restriction_map}
			\phi \mapsto (q,p) := (\rho_1(\phi), \rho_0(\phi)).
		\end{equation}
	By existence and uniqueness of the initial value problem for~\eqref{eom_phi4}, this map has an inverse, $U:\sE \mapsto S$, the ``time evolution map''. If we endow $S \subset C^\infty(M)$ with the relative topology, and $\sE$ with the canonical Fr\'{e}chet topology we discussed, then the map $U$ is continuous (prop.~\ref{prop_U_cont}). By definition, it therefore provides a global chart of $S$.\\

	Since $S$ is a Fr\'{e}chet manifold, it comes with a natural notion of smoothness\footnote{In the context of manifolds modelled on locally convex vector spaces, there are in general many inequivalent notions of smoothness. Let $\sF_1, \sF_2$ be two locally convex spaces and let $\sO \subset \sF_1$ be an open subset. The most common definition of smoothness, in some references called Michal-Bastiani smoothness, states that a continuous map $F:\sO \to \sF_2$ is smooth if for any $\nu$ the $\nu$-th Gateaux derivative exists as a continuous maps $\sO \times \sF_1^\nu \to \sF_2$. If $\sU \subset \sF_1$ is a subset, not necessarily open, then a continuous map $F: \sU \to \sF_2$ is said to be smooth if there is $\sO \supset \sU$ open and a smooth map $\tilde{F}: \sO \to \sF_2$ extending $F$. A problem with this definition is that, for completely arbitrary $\cU$, the directional derivatives of $F$ depend on the extension chosen. More details are presented in~\citep{michal1938differential, bastiani1964applications}, and also discussed in~\citep{hamilton1982inverse, neeb2005monastir, brunetti2012algebraic}. For general locally convex spaces, this notion of smoothness is not equivalent to the notion $C^\infty$-open smoothness of~\citep{KM97}. However, in the context of Fr\'{e}chet spaces, which is our setting, these coincide.}. However, for our purposes below, we will require a stronger notion. Recall that a function $F: C^\infty(M) \to \bC$ was called $W$-smooth (def.~\ref{def_funct_class}) if all its Gateaux derivatives exist in the sense of distributions on the appropriate Cartesian power of $M$, if their wave-front sets are contained in corresponding $W$ sets~\eqref{W_set_def}, and if they depend continuously on $\phi$.  The appropriate strengthened notion of smoothness for functionals on $S$ is to require the existence of a $W$-smooth {\em extension} on $C^\infty(M)$:
		\begin{defi}\label{def_smooth_on_f}
			A functional $F:S \to \bC$ is called {\em on-shell $W$-smooth} if there is an extension $\tilde{F}:C^\infty(M) \to \bC$ of $F$, i.e. $\tilde{F}(\phi) = F(\phi)$ for all $\phi \in S$, which is $W$-smooth, i.e. the following conditions holds:
				\begin{enumerate}[label=(W\arabic*)]
					\item\label{W1_f} For all $\nu \in \bN$, the $\nu$-th Gateaux derivative $\delta^\nu \tilde{F}_\phi / \delta \phi(y_1) \dots \delta \phi(y_\nu)$ exists as compactly supported symmetric distribution in $M^\nu$ and 
							\begin{equation}\label{WF_gateaux_f}
								\WF\left( \frac{\delta^\nu \tilde{F}_\phi}{\delta \phi(y_1) \dots \phi(y_\nu)} \right) \subset W_{\nu}. 
							\end{equation}
						\item\label{W2_f} Let $\bR \ni \epsilon \mapsto \phi(\epsilon) \in C^\infty(M)$ be smooth and view $\delta^\nu \tilde{F}_{\phi(\epsilon)} / \delta \phi(y_1) \dots \delta \phi(y_\nu)$ as a distribution in $\bR \times M^{\nu}$, i.e. in the variables $\epsilon, y_1, \dots, y_\nu$. For all $\nu \in \bN$, it is required to satisfy
							\begin{equation}\label{WF_cont_gateaux_f}
								\WF\left( \frac{\delta^\nu \tilde{F}_{\phi(\epsilon)}}{\delta \phi(y_1) \dots \phi(y_\nu)} \right) \subset \bR \times \{0\} \times W_{\nu}.
							\end{equation}
				\end{enumerate}
			The space of on-shell $W$-smooth functionals is denoted by $\gls{C_infty_W_S}$.
		\end{defi}
	\noindent
	For a $W$-smooth extension $\tilde{F}$, all Gateaux derivatives exist as compactly supported distributions on suitable Cartesian powers of $M$ by definition. Therefore, if we feed $\delta F_\phi / \delta \phi$ with a smooth solution $u$ to the linearised wave equation~\eqref{leom_phi4} at $\phi$, we expect to be able to define a covariant derivative along $S$. We will provide this construction in more generality for covariant tensor fields below in def.~\ref{def_smooth_on_tens}.\\
			
	We next introduce the tangent bundle of $S$, denoted by $TS$, in the standard way. Let $\phi \in S$ and consider the set of all smooth curves $\gamma: I \to S$ such that $\gamma(0) = \phi$, where $I$ is an open interval around $0$ in $\bR$. A tangent vector at $\phi$ is identified with one of the equivalence classes of such curves, where two curves $\gamma, \tilde{\gamma}$ are defined to be equivalent if it holds
		\begin{equation*}
			 \left. \frac{d}{d \epsilon} F(\gamma(\epsilon)) \right|_{\epsilon = 0}  = \left. \frac{d}{d \epsilon} F(\tilde \gamma(\epsilon)) \right|_{\epsilon = 0},
		\end{equation*}
	for any on-shell $W$-smooth function $F: S \to \bC$. The tangent space $\gls{T_S}$ is defined as the collection of all such tangent vectors. Note that this definition coincides with the usual ``kinematic'' definition of the tangent space for finite-dimensional manifolds~\footnote{In infinite dimensions one has to be careful and in general has to distinguish between ``kinematic tangent vector'', i.e. given as velocity of curves,  and ``operational tangent vector'', i.e. given as bounded derivations of local smooth functions. These concepts do not coincide unless the locally convex space which models the infinite-dimensional manifold is reflexive~\citep[thm. 28.7]{KM97}. In our situation, reflexivity does {\em not} hold.}. Since the elements of $S$ are the smooth solutions of equation~\eqref{eom_phi4}, we can alternatively characterize the tangent space $T_\phi S$ as the space of smooth solutions to the linearised equation at $\phi$, namely
		\begin{equation*}
			T_\phi S \simeq \{ u \in C^\infty(M): P_\phi u =0 \},
		\end{equation*}
	where we used the notation $\gls{P_phi} := \boxempty - m^2 - \frac{\lambda}{2}\phi^2$. The tangent bundle is defined as the disjoint union of its fibers, 
		\begin{equation}\label{tan_S}
			TS = \bigsqcup_{\phi \in S} T_\phi S.
		\end{equation}
	For any Fr\'{e}chet manifold, the kinematic tangent bundle inherits a natural topology and a natural manifold structure. In our case, these structures are particularly easy to describe, because $S$ can be covered by a single chart via initial data. More precisely, we define the map $\dot{U}:\sE \oplus \sE \to TS$ by
		\begin{equation*}
			\dot{U}((q_1,p_1),(q_2,p_2)):= (U(q_1,p_1), u_{U(q_1,p_1)} (q_2,p_2)),
		\end{equation*}
	where $u_\phi: \sE \ni (q,p) \mapsto u_\phi(q,p) \in T_\phi S$ maps the Cauchy data $q,p$ into the unique smooth solution $u_\phi(q,p)$ of the linear equation~\eqref{leom_phi4} such that $\rho_0 u_\phi(q,p) = q$ and $\rho_1 u_\phi(q,p) = p$\footnote{As in~\citep{D80}, the $u_\phi$ can be expressed in terms of the causal propagator as $u_\phi(q,p) = E_\phi[ \rho_0' p - \rho_1' q]$, where $\rho'_{0,1}:\cE'(\Sigma) \to \cE'(M)$ are the ``adjoints'' of $\rho_{0,1} : C^\infty(M) \to C^\infty(\Sigma)$, i.e. they are defined by $\langle \rho'_{0,1} t, f\rangle = \langle t, \rho_{0,1} f \rangle$ for any $t \in \cE'(\Sigma)$ and $f \in C^\infty(M)$.}.\\
	The topology of $TS$ is defined as the topology induced by the Fr\`{e}chet topology of $\sE \oplus \sE$ and the map $\dot{U}$\footnote{I.e. a subset $\cO \subset TS$ is open if and only if $\dot{U}^{-1}(\cO) \subset \sE \oplus \sE$ is open.}. It is easy to see that $(\sE \oplus \sE, \dot{U})$ is a global chart for $TS$. In fact, $\dot{U}$ is bijective and its inverse is $\dot{U}^{-1} (\phi,u):=(\rho(\phi), \rho(u))$ because the Cauchy problems for both eq.~\eqref{eom_phi4} and its linearisation eq.~\eqref{leom_phi4} are well-posed. By construction the maps $\dot{U}$ and $\dot{U}^{-1}$ are continuous. Note that the continuity properties of $\phi \mapsto u_\phi(q,p)$ proved in appendix~\ref{app_nnlin} (prop.~\ref{prop_cont_causal}) imply that $\dot{U}$ is continuous also if we endow $TS \subset C^\infty(M) \oplus C^\infty(M)$ with the relative topology. With respect to this topology $\dot{U}^{-1}$ is continuous too. Thus, the relative topology on $TS$ is compatible with the natural topology induced by the global chart.\\

	We would next like to define the cotangent bundle $T^*S$ and its tensor powers. A well-known issue in infinite dimensions is that there is no natural manifold structure for the cotangent bundle. For instance, if we define the cotangent space as the topological dual of the tangent space, then we can endow the cotangent bundle with a vector bundle structure, but generally not with a smooth manifold structure. If we consider the stronger category of manifolds modelled on Banach spaces, i.e. complete normed vector spaces, the issue can be resolved, as discussed in~\citep[Remark II.3.5]{neeb2005monastir}. But for the case of manifolds modelled on a Fr\'{e}chet space there is no natural definition. A similar problem arises for the tensor powers of the cotangent bundle. The key point is that, to define the tensor product of locally convex spaces, we need to take the completion of the algebraic tensor product of these spaces with respect to some topology. In~\citep{KM97}, the authors proved that choosing the bornological completion, i.e. the finest locally convex topology such that the canonical tensor map is bounded, allows one  to construct the full theory of calculus for locally convex spaces . For the purposes of this work, a more direct approach, based on the specific infinite-dimensional structure we are considering, is preferable.\\
	 In our concrete case, $S$ is a set of smooth functions and $T_\phi S$ for any background $\phi$ is a linear space of smooth functions. The topological dual space of $T_\phi S$ is a space of distributions, and similarly for the dual space of $\otimes^n T_\phi S$. For our constructions below, we cannot consider arbitrary distributions, because we would like to define on these spaces a product structure in order to define the algebras $\cW_\phi$. Actually, we had already encountered this problem when we defined the algebra $\cW(S,\omega)$~\eqref{Wick_algebra_free} in sec.~\ref{subsec_free_QFT} that serves as a model for $\cW_\phi$. We shall proceed in the exact same way. For each fixed $\phi \in S$, we note that any compactly supported distribution $t$ on $M$, modulo compactly supported distributions of the form $P_\phi t'$ gives rise to a well-defined linear form on $T_\phi S$, i.e. on smooth solutions of associated with the operator $P_\phi$. Similar statements hold true for distributions of more variables. By analogy with our discussion in sec.~\ref{subsec_int_QFT_per}, we therefore define
		\begin{equation}\label{cov_tens_S}
			\gls{boxtimes_W_n} := \cE_W'(M^n)/P_\phi \cE_W'(M^n),
		\end{equation}
	which we can interpret as a completion of the algebraic tensor product $\otimes^n T^*_\phi S$ viewed as tensor product of smooth functions. In the above quotient, we mean that the Klein-Gordon operator $P_\phi$ can act on any argument as in~\eqref{n_box_spaces}.\\
	The bundle corresponding to~\eqref{cov_tens_S} is defined as the set-theoretic union of its fibers, i.e.
		\begin{equation}\label{cov_tens_S_bundle}
			\boxtimes^n_W T^*S := \bigsqcup_{\phi \in S} \boxtimes^n_W T_\phi^* S.
		\end{equation}
	We need to equip $T^*S$ and more generally the bundles $\boxtimes^n T^*S$ with a smooth structure, i.e. we need to define the notion of smooth sections on these bundles.\\
	
\begin{figure}
		\centering
		\begin{tikzpicture}
		\node (c=0) at (4,3.9) {$c=0$};
		\path [pattern=north west lines, pattern color = black!20] (0,2) -- (8,2) -- (8,3.4)-- (0,3.4) -- (0,2);
		\path [fill = black!20] (0,1) -- (8,1) -- (8,2)-- (0,2) -- (0,1);
		\node (c=1) at (4,1.5) {$c=1$};
		\draw [thick] (0,3.4) -- (8,3.4);
		\node (a1) at (1, 3.35) {};
		\node (a2) at (0,4) {$\Sigma_+$};
		\draw [-latex, bend left] (a2) to (a1);
		\draw [thick] (0,2) -- (8,2);
		\node (b1) at (1, 1.95) {};
		\node (b2) at (0,2.6) {$\Sigma_-$};
		\draw [-latex, bend left] (b2) to (b1);
		\end{tikzpicture}
		\caption{Choice of $c$.}\label{fig:cutoff}
		\end{figure}
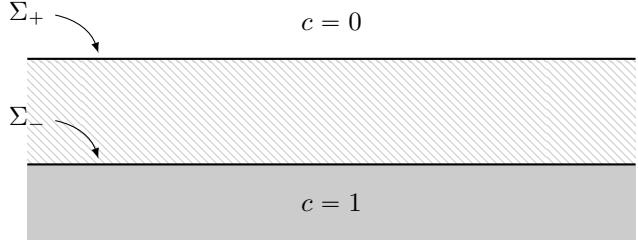

	For the purpose of discussing the smooth structure, it will be convenient to have an alternative characterization of the fibers of the covariant tensor bundles. In order to set up this characterization, we first introduce a special distributional integral kernel $\sigma_c$ that will appear throughout the following sections. We begin by choosing two disjoint Cauchy surfaces $\Sigma_+, \Sigma_-$, such that $\Sigma_+$ is in the future of $\Sigma_-$. Then, consider a function $c \in C^\infty(M)$ such that $c(M) \subset [0,1]$, $c= 0$ in $J^+ (\Sigma_+)$ and $c=1$ in $J^- (\Sigma_-)$. Roughly speaking, $c$ is a smoothed out version of the step function that jumps from $1$ to $0$ across the Cauchy surface $\Sigma_-$. See fig.~\ref{fig:cutoff} for a sketch of the situation. 
		Then, we put
		\begin{equation}\label{kernel_symp}
			\gls{sigma_c}(x_1, x_2) := -(\boxempty c(x_1) )\delta(x_1,x_2) - 2 ( \nabla c(x_1), \nabla \delta (x_1, x_2) )_g,
		\end{equation}
	where $\nabla$ is the Levi-Civita connection with respect to the space-time metric $g$, and where $( \cdot, \cdot )_g$ is the contraction with $g$. For later use, we notice that the wave-front set of $\sigma_c$ satisfies the following bound:
		\begin{equation}\label{WF_kernel_symp}
			\WF (\sigma_c) \subset \left\{ (x_1,x_2;k_1,k_2) \in \dot{T}^\ast M^2 : x_1 = x_2 \in J^-(\Sigma_+) \cap J^+(\Sigma_-), \, k_1 = - k_2 \right\} \subset W_2.
		\end{equation}
	Next, we consider the composition
		\begin{equation*}
			(\sigma_c \circ E_\phi)(x,y) = \int_M \sigma_c(x,z) E_\phi(z,y) dz,
		\end{equation*}
	for any cut-off function $c$ as in eq.~\eqref{kernel_symp}, where $\gls{E_phi}$ is the causal propagator of $P_\phi$. This distribution will be extensively used in the construction of our infinite-dimensional setting.
	\begin{lemma}\label{lemma_project}
		For any cut-off function $c$ as in eq.~\eqref{kernel_symp}, $(\sigma_c \circ E_\phi)(x,y)$ is a well-defined distribution which has wave-front set in $W_2$ and which is compactly supported in $x$.
	\end{lemma}
	\begin{proof}
		We first notice that the wave-front set of $\sigma_c$ is estimated by~\eqref{WF_kernel_symp}, the wave-front set of $E_\phi$ is given by~\eqref{WF_causal}, and they are both contained in $W_2$. Furthermore, $\sigma_c$ is compactly supported and so the integration condition~\eqref{int_cond} is fulfilled. All the hypotheses of lemma~\ref{lemma_W_comp} are satisfied and thus $(\sigma_c \circ E_\phi)(x,y)$ is a well-defined distribution which has wave-front set in $W_2$. The support property of $(\sigma_c \circ E_\phi)(x,y)$ is a straightforward consequence of the fact that $\sigma_c$ is compactly supported.
	\end{proof}
	\noindent
	The following lemma establishes that~\eqref{kernel_symp} is basically an equivalent way of writing the standard symplectic form on $T_\phi S$ and that $E_{\phi} \circ \sigma_c$ is the identity on each $T_\phi S$, or, in other words, that $E_{\phi}$ is the left inverse of $\sigma_c$ on $T_\phi S$. We also show some useful consequences of these two important facts. We present the results for a more general situation than $T_\phi S$, namely for the space of smooth solutions $u$ of $P_\phi u = (\boxempty - m^2 - \frac{\lambda}{2}\phi^2)u=0$ for $\phi \in C^\infty(M)$ and not just for $\phi \in S$.
		\begin{lemma}\label{lemma_kernel_symp}
			Let $\phi \in C^\infty(M)$ and $c$ be a cut-off function as in eq.~\eqref{kernel_symp}. Then, for any two smooth solutions $u_1, u_2$ to $P_\phi u_{1,2}=0$ and for any Cauchy surface $\Sigma$, it holds
				\begin{equation}\label{sympl_form/symp_kernel_sol}
					\int_{\Sigma} u_1(x) \overleftrightarrow{\partial_n} u_2(x) d\Sigma(x) = \int_{M^2} u_1(x_1) \sigma_c(x_1,x_2) u_2(x_2) dx_1 dx_2.
				\end{equation}
			Furthermore, for any smooth $u$ such that $P_\phi u = 0$ we have
				\begin{equation}\label{causal_prop_inv_symp_kernel_sol}
					u(x) = \int_{M^2} E_\phi(x,y_1) \sigma_c(y_1, y_2) u(y_2) dy_1 dy_2. 
				\end{equation}
			It also holds
				\begin{equation}\label{causal_prop/symp_kernel}
					E_\phi(x_1,x_2) = \int_{M^2} E_\phi(x_1,y_1) \sigma_c(y_1, y_2) E_\phi(y_2,x_2) dy_1 dy_2,
				\end{equation}
			for any cut-off function $c$ as in eq.~\eqref{kernel_symp}, and
				\begin{equation}\label{project_E_sigma}
					(E_\phi \circ \sigma_{c'}) \circ (E_\phi \circ \sigma_c) = E_\phi \circ \sigma_c,
				\end{equation}
			for any cut-off functions $c, c'$ as in eq.~\eqref{kernel_symp}.
		\end{lemma}
		\begin{proof}
			The proof of~\eqref{sympl_form/symp_kernel_sol} follows easily from the Stokes theorem.\\
			As proved in~\citep[lemma 3.2.1 part (3)]{W94}\footnote{Note that in the reference the symplectic structure has the opposite sign.}, we have 
				\begin{equation*}
					\int_M f(x) u(x) dx = - \sigma(E(f), u) = \int_{M} \int_{\Sigma} f(x) E_\phi(x,z) \overleftrightarrow{\partial_n} u(z) dx d\Sigma(z),
				\end{equation*}
			for any test function $f$ and any Cauchy surface $\Sigma$. Using eq.~\eqref{sympl_form/symp_kernel_sol}, then eq.~\eqref{causal_prop_inv_symp_kernel_sol} follows.\\
			Since $E_\phi(f)$ is a smooth $P_\phi$-solution for any $f \in C^\infty(M)$, eq.~\eqref{causal_prop/symp_kernel} is a straightforward consequence of eq.~\eqref{causal_prop_inv_symp_kernel_sol}.\\
			Finally, eq.~\eqref{project_E_sigma} is just a corollary of~\eqref{causal_prop/symp_kernel}.
		\end{proof}
		\noindent
		For later, we note that in the more general situation where $u_1, u_2$ are arbitrary smooth functions, not necessarily solutions to $P_\phi u_{1,2}=0$, eq.~\eqref{sympl_form/symp_kernel_sol} becomes
			\begin{equation}\label{kernel_symp_not_sol}
				\begin{split}
					&\int_{M^2} u_1(x_1) \sigma_c(x_1,x_2) u_2(x_2) dx_1 dx_2 = \\
					&\quad =\int_{\Sigma_-} u_1(x) \overleftrightarrow{\partial_n} u_2(x) d\Sigma(x) - \int_B (P_\phi u_1)(x) c(x) u_2(x) dx + \int_B u_1(x) c(x) (P_\phi u_2)(x) dx,
				\end{split}
			\end{equation}
		where $c$ is a cut-off function as in  eq.~\eqref{kernel_symp}, i.e. $c$ is identically $1$ in the past of a Cauchy surface $\Sigma_-$ and is identically $0$ in the future of a Cauchy surface $\Sigma_+$, and where $B$ is the compact region $J^-(\Sigma_+) \cap J^+(\Sigma_-)$.\\
		If $u_1, u_2$ are arbitrary smooth functions, not necessarily solutions to $P_\phi u_{1,2}=0$, then it follows that the integral $\int_{\Sigma} u_1(x) \overleftrightarrow{\partial_n} u_2(x) d\Sigma(x)$ is not any more independent to the choice of the Cauchy surface $\Sigma$. In particular, if $\Sigma'$ is another Cauchy surface which is in the (strict) past of the Cauchy surface $\Sigma$, then it holds
				\begin{equation}\label{symp_form_not_sol}
					\begin{split}
						\int_{\Sigma'} u_1(x) \overleftrightarrow{\partial_n} u_2(x) dx = \int_{\Sigma} u_1(x) \overleftrightarrow{\partial_n} u_2(x) d\Sigma(x) + \int_{B'} (P_\phi u_1)(x) u_2(x) dx - \int_{B'} u_1(x) (P_\phi u_2)(x) dx,
					\end{split}
				\end{equation}
		where $B'$ is the compact region $J^+( \Sigma') \cap J^- (\Sigma)$.\\
		
		The desired alternative description of the cotangent bundle and its tensor powers is given by the following proposition, which is presented again (as in lemma~\ref{lemma_kernel_symp}) for a generic smooth function $\phi$.
		\begin{prop}\label{prop_cov_tens_S_var}
			Let $c$ be any cut-off function as in~\eqref{kernel_symp}. For any $\phi \in C^\infty(M)$ and any $n \in \bN$ the spaces $(\sigma_c \circ E_\phi)^{\otimes n} \cE^\prime_{W}(M^n)$ and $\cE_W'(M^n)/P_\phi \cE_W'(M^n)$ are naturally isomorphic. The isomorphism is given by~\eqref{map_iso}.\\
			In case $\phi \in S$, we thus have $\boxtimes_W^n T^*_\phi S \simeq (\sigma_c \circ E_\phi)^{\otimes n} \cE_W' (M^n)$.
		\end{prop}
		\begin{proof}
			First, we show that the space of distributions $(\sigma_c \circ E_\phi)^{\otimes n} \circ \cE^\prime_{W}(M^n)$ is actually well-defined. We already established in lemma~\ref{lemma_project} that the distribution $(\sigma_c \circ E_\phi)(x,y)$ has wave-front set contained in $W_2$ and is compactly supported in $x$. Thus, using lemma~\ref{lemma_W_comp} we see that the composition of $(\sigma_c \circ E_\phi)^{\otimes n}$ with a distribution in $\cE'_W(M^n)$ is well-defined and $(\sigma_c \circ E_\phi)^{\otimes n} \cE'_W(M^n) \subset \cE'_W(M^n)$.\\
			The desired isomorphism between $(\sigma_c \circ E_\phi)^{\otimes n} \cE^\prime_{W}(M^n)$ and $\cE_W'(M^n)/P_\phi \cE_W'(M^n)$ is given by
				\begin{equation}\label{map_iso}
					(\sigma_c \circ E_\phi)^{\otimes n} \cE^\prime_{W}(M^n) \ni (\sigma_c \circ E_\phi)^{\otimes n} t \mapsto [t] \in \cE_W'(M^n)/P_\phi \cE_W'(M^n).
				\end{equation}
			First, we need to check that the proposed definition is actually consistent. It is sufficient for this purpose to prove that if $(\sigma_c \circ E_\phi)^{\otimes n} t = 0$, then $[t] = [0]$, i.e. $t \in P_\phi \cE^\prime_{W}(M^n)$.\\
			For $n=1$, the distributional space $\cE^\prime_{W}(M)$ is just $C^\infty_0(M)$. If we assume that $(\sigma_c \circ E_\phi)f=0$ for a certain $f \in C^\infty_0(M)$, then using eq.~\eqref{causal_prop/symp_kernel}, we have necessarily
				\begin{equation*}
					0 = (E_\phi \circ \sigma_c \circ E_\phi)f = E_\phi(f).
				\end{equation*}
			As proven in~\citep[lemma 3.2.1 part (2)]{W94}, $E _\phi (f) = 0$ if and only if $f \in P_\phi C^\infty_0(M)$ as we desired to show.\\
			For $n>1$, the proof is a bit more involved. We present explicitly only the case $n=2$, but exactly the same argument can be adapted to the general case. Using the hypothesis $(\sigma_c \circ E_\phi)^{\otimes 2} t = 0$, we write $t$ as
				\begin{equation}\label{iso_tech}
					\begin{split}
						t &= t - (\sigma_c \circ E_\phi) \circ t + (\sigma_c \circ E_\phi) \circ t - (\sigma_c \circ E_\phi) \circ t \circ  (E_\phi \circ \sigma_c)\\
						&=(\id - \sigma_c \circ E_\phi) \circ t  + (\sigma_c \circ E_\phi) \circ t \circ (\id - E_\phi \circ \sigma_c).
					\end{split}
				\end{equation}
			Let us focus on the distribution $(\id - \sigma_c \circ E_\phi) \circ t$. By construction, it is compactly supported. We show that the composition $E^{A/R}_\phi \circ (\id - \sigma_c \circ E_\phi) \circ t$ is a well-defined distribution with wave-front set contained in $W_2$. In fact, the wave-front sets of $E^{A/R}_\phi$ are estimated, respectively, by the sets $\cC^{A/R}$ defined in~\eqref{C^A/R}. By definition, $\cC^{A/R}$ are subsets of $W_2$. As already show in lemma~\eqref{lemma_project}, the distribution $(\sigma_c \circ E_\phi)(x,y)$ has wave-front set contained in $W_2$ and it is compactly supported in $x$. By hypothesis, $t$ is a distribution in $\cE'_W(M^n)$. Therefore, all the compositions $E^{A/R}_\phi \circ t$ and $E^{A/R}_\phi \circ (\sigma_c \circ E_\phi) \circ t$ satisfy the integration condition~\eqref{int_cond} and involve distributions with wave-front sets contained in $W_2$. Thus, we can apply lemma~\ref{lemma_W_comp} and we conclude that $E^{A/R}_\phi \circ (\id - \sigma_c \circ E_\phi) \circ t$ is indeed a well-defined distribution with wave-front set contained in $W_2$.\\
			As a consequence of eq.~\eqref{causal_prop_inv_symp_kernel_sol}, it holds 
				\begin{equation}\label{tech_formula}
					(E^A_\phi - E^R_\phi) \circ (\id - \sigma_c \circ E_\phi) \circ t=0.
				\end{equation}
			By the support properties of the advanced/retarded propagators $E^{A/R}_\phi$, it follows from~\eqref{tech_formula} that $E^A_\phi \circ (\id - \sigma_c \circ E_\phi) \circ t$ must be a compactly supported distribution.\\
			Summing up, we obtained that $E^A_\phi \circ (\id - \sigma_c \circ E_\phi) \circ t \in \cE'_W(M^2)$. With a similar argument, we can prove that $(\sigma_c \circ E_\phi) \circ t \circ (\id - E_\phi \circ \sigma_c) \circ E^A_\phi \in \cE'_W(M^2)$.\\
			Finally, if we set $h_1:=E^A_\phi \circ (\sigma_c \circ E_\phi - \id) \circ t$ and $h_2:=(\sigma_c \circ E_\phi) \circ t  \circ (E_\phi \circ \sigma_c - \id) \circ E^A_\phi$, then it follows that
				\begin{equation*}
					t=(P_\phi \otimes 1) h_1 + (1 \otimes P_\phi) h_2,
				\end{equation*}
			so $t \in P_\phi \cE'_W(M^2)$ and the map~\eqref{map_iso} is consistently defined.\\
			To conclude the proof, we notice that the map~\eqref{map_iso} is clearly surjective and it is injective because $(\sigma_c \circ E_\phi)^{\otimes n}$ vanishes when acting on $P_\phi \cE_W'(M^n)$.
		\end{proof}
	\noindent
	The following lemma clarifies the dependence on the cut-off $c$ of the alternative description of the cotangent space and its tensor powers we have just presented in prop.~\ref{prop_cov_tens_S_var}.
	 	\begin{lemma}\label{lemma_approx}
	 		Let $c, c'$ be two cut-off functions satisfying the properties required by eq.~\eqref{kernel_symp}, then for any $t \in \cE'_W(M^n)$ and any $\phi \in C^\infty(M)$ it holds
				\begin{equation}\label{approx}
					(\sigma_c \circ E_\phi)^{\otimes n} t \approx (\sigma_{c'} \circ E_\phi)^{\otimes n} t,
				\end{equation}
			where $\approx$ means that the distributions differ by an element in $P_\phi \cE'_W(M^n)$.
		\end{lemma}
		\begin{proof}
			By eq.~\eqref{project_E_sigma}, we have that
				\begin{equation*}
					(\sigma_c \circ E_\phi)^{\otimes n} ((\sigma_c \circ E_\phi)^{\otimes n} t - (\sigma_{c'} \circ E_\phi)^{\otimes n} t)= (\sigma_c \circ E_\phi)^{\otimes n} t - (\sigma_c \circ E_\phi)^{\otimes n} t = 0.
				\end{equation*}
			Arguing as in the proof of prop.~\ref{prop_cov_tens_S_var}, we have $(\sigma_c \circ E_\phi)^{\otimes n} t - (\sigma_{c'} \circ E_\phi)^{\otimes n} t \in P_\phi \cE'_W(M^n)$ as we desired to show.
		\end{proof}
	
	We next wish to define the notion of smooth sections for the cotangent bundle $T^*S$ and, more generally, for the bundles $\boxtimes^n T^*S$. It turns out that the best way to define this smooth structure for our purposes is again via the notion of ``on-shell $W$-smoothness''. Above in def.~\ref{def_smooth_on_f}, we had already defined the notion of an on-shell $W$-smooth function $F: S\to \bC$, and we now essentially repeat this definition for sections on $\boxtimes^n T^*S$, which are called ``covariant sections of rank $n$''.  First of all, given a covariant section $t: S \to \boxtimes_W^n T^*S$, we say that $\tilde{t}: C^\infty(M) \to \cE'_W(M^n)$ is an extension of $t$ if for all $\phi \in S$ and $u_1, \dots, u_n \in T_\phi S$ it holds
		\begin{equation}\label{extension_on_W_tens}
			t_\phi(u_1, \dots, u_n) = \tilde{t}_\phi(u_1, \dots, u_n),
		\end{equation}
	where $t_\phi$ is understood as a distributional representative of the equivalence class in $\boxtimes_W^n T_\phi^*S$.
		\begin{defi}\label{def_smooth_on_tens}
			A covariant section $t: S \to \boxtimes_W^n T^*S$ is called on-shell $W$-smooth if there is a $W$-smooth extension $\tilde{t}: C^\infty(M) \ni \phi \mapsto \tilde{t}_\phi \in (\sigma_c \circ E_\phi)^{\otimes n} \cE'_W(M^n)$ for some cut-off function $c$ as in eq.~\eqref{kernel_symp}, by which we mean an extension such that:
				\begin{enumerate}[label=(W\arabic*)]
					\item\label{W1} For all $\nu \in \bN$, the $\nu$-th Gateaux derivative $\delta^\nu \tilde{t}_\phi(x_1, \dots, x_n) / \delta \phi(y_1) \dots \delta \phi(y_\nu)$ exists as distribution of compact support in $M^{n+\nu}$ and it holds
							\begin{equation}\label{WF_gateaux_t}
								\WF\left( \frac{\delta^\nu \tilde{t}_\phi(x_1, \dots, x_n)}{\delta \phi(y_1) \dots \phi(y_\nu)} \right) \subset W_{n+\nu}; 
							\end{equation}
						\item\label{W2} Let $\bR \ni \epsilon \mapsto \phi(\epsilon) \in C^\infty(M)$ be smooth and we view $\delta^\nu \tilde{t}_{\phi(\epsilon)}(x_1, \dots, x_n) / \delta \phi(y_1) \dots \delta \phi(y_\nu)$ as a distribution in $\bR \times M^{n + \nu}$, i.e. with respect to the variables $\epsilon, x_1, \dots, x_n, y_1, \dots, y_\nu$. For all $\nu \in \bN$, it required to satisfy
							\begin{equation}\label{WF_cont_gateaux_t}
								\WF\left( \frac{\delta^\nu \tilde{t}_{\phi(\epsilon)}(x_1, \dots, x_n)}{\delta \phi(y_1) \dots \phi(y_\nu)} \right) \subset \bR \times \{0\} \times W_{n+\nu}; 
							\end{equation}
				\end{enumerate}
			\noindent
			We denote the space of on-shell $W$-smooth covariant sections of rank $n$ by $\gls{C_infty_W_S_n}$.\\
			A on-shell $W$-smooth $k$-form is a totally anti-symmetric element in $C_{W}^\infty(S, \boxtimes^k T^* S)$. The space of on-shell $W$-smooth $k$-forms is denoted by $\gls{Omega_W_k}$.
		\end{defi}
		\noindent
		We want to prove that the notion of on-shell $W$-smoothness is independent of the choice of the cut-off function, in the sense that if an extension satisfying~\ref{W1},~\ref{W2} can be found for a specific $c$ as in eq.~\eqref{kernel_symp}, then it can be found also for any other cut-off function $c'$ of the same kind. For this purpose, we to investigate the variational derivatives of the causal propagator.
			\begin{prop}\label{prop_var_ders_causal}
				For any $\phi \in C^\infty(M)$, and for any $\nu \in \bN$, the $\nu$-th Gateaux derivative 
					\begin{equation*}
						\frac{\delta^\nu E_\phi (x_1,x_2)}{\delta \phi(y_1) \dots \delta \phi(y_\nu)}
					\end{equation*}
				is a well-defined distribution which satisfies the following properties:
				\begin{enumerate}
					\item The distribution $\delta^\nu E_\phi (x_1,x_2) / \delta \phi(y_1) \dots \delta \phi(y_\nu)$ is compactly supported in $y_1, \dots, y_\nu$, more precisely $y_1, \dots, y_\nu$ must belong to $\supp \lambda$, where $\lambda$ enters via $P_\phi = \boxempty - m^2 - \frac{\lambda}{2} \phi^2$.
					\item It holds
							\begin{equation}\label{var_ders_causal_WF}
								\WF\left( \frac{\delta^\nu E_{\phi}(x_1,x_2)}{\delta \phi(y_1) \cdots \delta \phi(y_\nu)} \right) \subset X_{2+\nu},
							\end{equation}
						where the set $X_{2 + \nu}$ is defined as
							\begin{equation}\label{X}
								\begin{split}
									\gls{X} &:=\left\{ (x_1,x_2,y_1,\dots, y_\nu; k_1, k_2, p_1, \dots,  p_\nu) \in \dot{T}^\ast M^{\nu+2} : \right. \\
									&\quad \left. p_1=p_1' + p_1'', \dots, p_\nu=p_\nu' + p_\nu'' \mbox{ and } \exists \mbox{ a permutation } \pi \mbox{ of } \{1, \dots, \nu\} \mbox{ such that} \right. \\
									&\quad \left. (x_1; k_1) \sim (y_{\pi(1)}; -p_{\pi(1)}') \mbox{ or } (x_1, k_1)=(y_{\pi(1)}, -p_{\pi(1)}') \mbox{ or } k_1,p_{\pi(1)}'=0 \right.\\
									&\quad \left. (y_{\pi(i)}; p_{\pi(i)}'') \sim (y_{\pi(i+1)}; -p_{\pi(i+1)}') \mbox{ or } (y_{\pi(i)}, p_{\pi(i)}'') =(y_{\pi(i+1)}, -p_{\pi(i+1)}') \right. \\
									&\qquad \qquad \qquad \qquad \qquad \qquad \qquad \quad \mbox{ or } p_{\pi(i)}'',p_{\pi(i+1)}'=0 \\
									&\quad \left. (y_{\pi(\nu)}; p_{\pi(\nu)}'') \sim (x_2; -k_2) \mbox{ or } y_{\pi(\nu)}=x_2, p_{\pi(\nu)}''= -k_2 \mbox{ or } p_{\pi(\nu)}'', k_2 = 0 \right\}.
								\end{split}
							\end{equation}
					\item Let $\bR \ni \epsilon \mapsto \phi(\epsilon) \in C^\infty(M)$ be smooth and view $\delta^\nu E_{\phi(\epsilon)}(x_1, x_2)/\delta \phi(y_1) \cdots \delta \phi(y_\nu)$ as a distribution in $\bR \times M^{2+\nu}$. It holds
							\begin{equation}\label{cont_var_ders_causal_WF}
								\WF\left( \frac{\delta^\nu E_{\phi(\epsilon)}(x_1,x_2)}{\delta \phi(y_1) \cdots \delta \phi(y_\nu)} \right) \subset \bR \times \{0\} \times X_{2+\nu}.
							\end{equation}
				\end{enumerate}
				
			\end{prop}
			\begin{proof}
					The causal propagator is by definition $E_\phi = E^A_\phi - E^R_\phi$ and, consequently, the properties 1,2,3 follow from the corresponding properties 1,2,3 of $\delta^\nu E^{A/R}_\phi / \delta \phi^\nu$ proved in prop.~\ref{prop_var_ders_A/R} of appendix~\ref{app_bkgr_dep_propa}.
			\end{proof}
			\noindent
			The following technical lemma clarifies the relation between the sets $W_\nu$ we defined by~\eqref{W_set_def} and the sets $X_{2+\nu}$ described in~\eqref{X}.
			\begin{lemma}\label{lemma_tech_X_W}
					For any $\nu$
						\begin{equation*}
							X_{2 + \nu} \subset W_{2 +\nu}.
						\end{equation*}
				\end{lemma}
				\begin{proof}
					We proceed by induction in $\nu$.\\
					The induction starts at $\nu =1$. Let $(x_1, x_2, y; k_1, k_2, p)$ be an element of $X_{2+1}$. If we assume that two of the covectors $k_1, k_2,p$ belong to $\overline{V}^+$ (respectively $\overline{V}^-$), then the third is necessarily contained in $\overline{V}^-$ (respectively $\overline{V}^+$) by the definition of $X_{2+1}$. Thus, $X_{2+1} \subset W_{2+1}$ as we needed to prove.\\
					We then prove the induction step: suppose that $X_{2+\nu'}\subset W_{2+ \nu'}$ holds for any $\nu' < \nu$, then we show that $X_{2 +\nu} \subset W_{2+\nu}$. Let $(x_1, x_2, y_1, \dots,  y_{\nu}; k_1, k_2, p_1, \dots,  p_{\nu})$ be an element in $X_{2+\nu}$. By the definition of $X_{2 +\nu}$, it follows that there exists a permutation $\pi$ of $\{1, \dots, \nu\}$ and decompositions $p_1=p_1' + p_1'', \dots, p_\nu=p_\nu' + p_\nu''$ such that the relations in the right-hand side of~\eqref{X} are satisfied. This means that
						\begin{equation}\label{conf_X_W}
							\begin{split}
								&(x_1, y_{\pi(\nu)}, y_{\pi(1)}, \dots, y_{\pi(\nu)} ; k_1, p_{\pi(\nu)}', p_{\pi(1)}, \dots, p_{\pi(\nu-1)}) \in X_{2+\nu} \\
								&(y_{\pi(\nu)}; p_{\pi(\nu)}'') \sim (x_2; -k_2) \mbox{ or } y_{\pi(\nu)}=x_2, p_{\pi(\nu)}''= -k_2 \mbox{ or } p_{\pi(\nu)}''=k_2=0.
							\end{split}
						\end{equation}
					We prove by reductio ad absurdum that $(x_1, x_2, y_1, \dots,  y_{\nu}; k_1, k_2, p_1, \dots,  p_{\nu}) \in W_{2+\nu}$, i.e. if we assume that all covectors $k_1, k_2, p_1, \dots, p_{\nu+1}$ belong to $\overline{V}^{+}$ (or all belong to $\overline{V}^{-}$), except at most one which can be space-like, then we get a contradiction. We present the argument for $\overline{V}^{+}$, the other situation can be treated similarly.\\
					We consider three cases separately: (a) $k_1, k_2, p_{\pi(1)}, \dots p_{\pi(\nu)} \in \overline{V}^+$ except at most one covector among $k_1, p_{\pi(1)}, \dots p_{\pi(\nu-1)}$ which can be space-like, (b) $k_2$ is space-like and $k_1, p_{\pi(1)}, \dots p_{\pi(\nu)} \in \overline{V}^+$, and (c) $p_{\pi(\nu)}$ is space-like and $k_1, k_2, p_{\pi(1)}, \dots p_{\pi(\nu-1)} \in \overline{V}^+$.
					\begin{enumerate}[label=(\alph*), start=1]
						\item As a consequence of the assumptions and the inductive hypothesis $X_{2+\nu -1}\subset W_{2+ \nu -1}$, it must necessarily hold $p_{\pi(\nu)}' \notin \overline{V}^+$. Since $p_{\pi(\nu)}= p_{\pi(\nu)}'+p_{\pi(\nu)}'' \in \overline{V}^+$ by assumption, we obtain $p_{\pi(\nu)}'' \in \overline{V}^+$ and $p_{\pi(\nu)}'' \neq 0$. Moreover, we also assume $k_2 \in \overline{V}^+$. We clearly get a contradiction with the second requirement of~\eqref{conf_X_W}.
						\item By assumption all $k_1, p_{\pi(1)}, \dots p_{\pi(\nu)} \in \overline{V}^+$, and so the inductive hypothesis $X_{2+\nu -1}\subset W_{2+ \nu -1}$ implies that $p_{\pi(\nu)}' \in \overline{V}^-$. Since $p_{\pi(\nu)}= p_{\pi(\nu)}'+p_{\pi(\nu)}'' \in \overline{V}^+$ by assumption, we have again that $p_{\pi(\nu)}'' \in \overline{V}^+$ and $p_{\pi(\nu)}'' \neq 0$. Since $k_2$ is assumed to be space-like, we obtain again a contradiction with the second requirement of~\eqref{conf_X_W}.
						\item As for the case (b), it follows from the assumption and the inductive hypothesis that $p_{\pi(\nu)}' \in \overline{V}^-$. Since $p_{\pi(\nu)}= p_{\pi(\nu)}'+p_{\pi(\nu)}''$ is space-like by assumption, we obtain $p_{\pi(\nu)}'' \notin \overline{V}^-$. Since $k_2 \in \overline{V}^+$ by assumption, also for the case (c) we get a contradiction with the second requirement of~\eqref{conf_X_W}.
					\end{enumerate}
					This concludes the proof.
				\end{proof}
				
		With these two results at our disposal, we show that the notion of on-shell $W$-smoothness is independent of the choice of the cut-off function $c$.
		\begin{lemma}\label{lemma_free_c_ext}
			Let $c$ be a cut-off function as in eq.~\eqref{kernel_symp} and let $\tilde{t}: C^\infty(M) \ni \phi \mapsto \tilde{t}_\phi \in (\sigma_c \circ E_\phi)^{\otimes n} \cE'_W(M^n)$ be a $W$-smooth extension of an on-shell $W$-smooth section $t: S \to \boxtimes_W^n T^*S$. For any other cut-off function $c'$ as in eq.~\eqref{kernel_symp}, the map
			\begin{equation}\label{change_c_rep}
				C^\infty(M) \ni \phi \mapsto (\sigma_{c'} \circ E_\phi)^{\otimes n} \tilde{t}_\phi \in (\sigma_{c'} \circ E_\phi)^{\otimes n} \cE'_W(M^n)
			\end{equation}
		is a $W$-smooth extension of $t$ in the sense of def.~\ref{def_smooth_on_tens}.
		\end{lemma}
		\begin{proof}
			We first show that the map~\eqref{change_c_rep} is consistently defined. In fact, by the properties of $(\sigma_c \circ E_\phi)$ given in lemma~\eqref{lemma_project}, it holds $\tilde{t}_\phi \in (\sigma_c \circ E_\phi)^{\otimes n} \cE'_W(M^n) \subset \cE'_W(M^n)$ and thus $(\sigma_{c'} \circ E_\phi)^{\otimes n} \tilde{t}_\phi$ is indeed an element of $(\sigma_{c'} \circ E_\phi)^{\otimes n} \cE'_W(M^n)$.\\
			Next, we prove that the map~\eqref{change_c_rep} is an extension of $t$, i.e. for any $\phi \in S$ and $u_1, \dots, u_n \in T_\phi S$ it holds
				\begin{equation}\label{exten_change_c}
					t_\phi(u_1, \dots, u_n) =\left( (\sigma_{c'} \circ E_\phi)^{\otimes n}\tilde{t}_\phi \right) (u_1, \dots, u_n).
				\end{equation}
			The right-hand side of the equation~\eqref{exten_change_c} above can be rewritten as
				\begin{equation*}
					\left( (\sigma_{c'} \circ E_\phi)^{\otimes n}\tilde{t}_\phi \right) (u_1, \dots, u_n) = \tilde{t}_\phi \left( (E_\phi \circ \sigma_{c'})u_1, \dots,  (E_\phi \circ \sigma_{c'})u_n \right) = \tilde{t}_\phi(u_1, \dots, u_n),
				\end{equation*}
			where we used the fact that $(E_\phi \circ \sigma_{c'})u_i = u_i$ for any $i=1, \dots, n$, see eq.~\eqref{causal_prop_inv_symp_kernel_sol}. Since $\tilde{t}_\phi$ is by hypothesis an extension of $t$, it follows that eq.~\eqref{exten_change_c} is satisfied as we needed to prove.\\
			To conclude the proof, we need to show that $(\sigma_{c'} \circ E_\phi)^{\otimes n} \tilde{t}_\phi$ satisfies the conditions~\ref{W1},~\ref{W2} of def.~\ref{def_smooth_on_tens}.\\
			In order to show~\ref{W1}, we compute $\delta^\nu (\sigma_{c'} \circ E_\phi)^{\otimes n} \tilde{t}_\phi / \delta \phi^\nu$ by distributing the variational derivatives among the factors of $ (\sigma_{c'} \circ E_\phi)^{\otimes n} \tilde{t}_\phi$. It follows that $\delta^\nu (\sigma_{c'} \circ E_\phi)^{\otimes n} \tilde{t}_\phi / \delta \phi^\nu$ is a finite sum of terms in the form
				\begin{equation}\label{term_var_chang_c}
					 	\int_{M^{2n}} \prod_{i=1}^{n} \sigma_{c'}(x_i, x'_i) \frac{\delta^{|N_i|} E_\phi (x'_i, x''_i)}{\delta \phi^{|N_i|} (\{y_{r} \}_{r \in N_i})} \frac{\delta^{|N_{t}|} \tilde{t}_\phi (x''_1, \dots, x''_{n})}{\delta \phi^{|N_t|} (\{y_{r} \}_{r \in N_t})} dx'_1 \dots dx'_{n} dx''_1 \dots dx''_n,
				\end{equation}
			where $N_1, \dots, N_n, N_t$ is a partition of $\{1, \dots, \nu\}$. To establish that $(\sigma_{c'} \circ E_\phi)^{\otimes n} \tilde{t}_\phi$ satisfies~\ref{W1} it is sufficient to show that each term~\eqref{term_var_chang_c} is a well-defined distribution in $\cE'_W(M^n)$.\\
			By construction, $\sigma_{c'}$ is a compactly supported distribution and its wave-front set is contained in $W_2$. As a consequence of the estimate~\eqref{var_ders_causal_WF} and lemma~\ref{lemma_tech_X_W}, we have that $\WF(\delta^{|N_i|} E_\phi / \delta \phi^{|N_i|})$ is contained in $W_{2 + |N_i|}$ for any $i$. By hypothesis, $\tilde{t}_\phi$ is a $W$-smooth extension and so it satisfies condition~\ref{W1} of def.~\ref{def_smooth_on_tens}, i.e. $\delta^{|N_t|} \tilde{t}_\phi / \delta \phi^{|N_t|}$ is a compactly supported distribution and its wave-front set is contained in $W_{n + |N_t|}$. These considerations imply that we can apply lemma~\ref{lemma_W_comp} and so the distribution~\eqref{term_var_chang_c} is a well-defined distribution with wave-front set in $W_{n + \nu}$. To verify the condition~\ref{W1} we still need to prove that the distribution~\eqref{term_var_chang_c} is of compact support. This follows from the fact that $\sigma_{c'}$ and $\delta^{|N_t|} \tilde{t}_\phi / \delta \phi^{|N_t|}$ are compactly supported and the fact that $\WF(\delta^{|N_i|} E_\phi / \delta \phi^{|N_i|})$ is compactly supported in the variables $(y_{r})_{r \in N_i}$ (see (2) of prop.~\ref{prop_var_ders_causal}).\\
			In order to prove~\ref{W2}, let $\bR \ni \epsilon \mapsto \phi(\epsilon) \in C^\infty(M)$ be smooth and consider $\delta^\nu (\sigma_{c'} \circ E_{\phi(\epsilon)})^{\otimes n} \tilde{t}_{\phi(\epsilon)} / \delta \phi^\nu$ as a distribution in $\bR \times M^{n +\nu}$. To prove~\ref{W2}, we need to show that the wave-front set of $\delta^\nu (\sigma_{c'} \circ E_{\phi(\epsilon)})^{\otimes n} \tilde{t}_{\phi(\epsilon)} / \delta \phi^\nu$ is contained in $\bR \times \{ 0 \} \times W_{n+\nu}$. We use a similar argument as the one presented for the proof of~\ref{W1}. More precisely, we notice that $\delta^\nu (\sigma_{c'} \circ E_{\phi(\epsilon)})^{\otimes n} \tilde{t}_{\phi(\epsilon)} / \delta \phi^\nu$ is again a finite sum of terms in the form~\eqref{term_var_chang_c}, with the only difference that $\phi$ is replaced by $\phi(\epsilon)$ in any occurrence. Then,  estimates~\eqref{cont_var_ders_causal_WF} and lemma~\ref{lemma_tech_X_W} imply that $\WF(\delta^{|N_i|} E_{\phi(\epsilon)} / \delta \phi^{|N_i|})$ is contained in $\bR \times \{0 \} \times W_{2 + |N_i|}$, and, by hypothesis, the wave-front set of $\delta^{|N_t|} \tilde{t}_{\phi(\epsilon)} / \delta \phi^{|N_t|}$ is bounded by $\bR \times \{0 \} \times W_{2 + |N_t|}$. To conclude that~\ref{W2} holds for $(\sigma_{c'} \circ E_{\phi(\epsilon)})^{\otimes n} \tilde{t}_{\phi(\epsilon)}$, we just need to use the wave-front set calculus (thm.~\ref{theo_WF_horma}). This concludes the proof.
		\end{proof}
		
	The first operation we introduce on on-shell $W$-smooth covariant sections is the tensor product.
			\begin{prop}\label{prop_tensor_prod_W_sec}
				Let $t,s$ be two on-shell $W$-smooth covariant field of rank respectively $n$ and $m$. For any $\phi \in S$, we define $(t \otimes s)_\phi \in \boxtimes^{n+m} T^*_\phi S$ as
					\begin{equation}\label{tensor_prod_W_sec}
						(t \otimes s)_\phi := t_\phi \otimes s_\phi.
					\end{equation}
				Note that, by abuse of notation, we identify an equivalence class in $\boxtimes^{\bullet} T^*_\phi S = \cE'_W(M^{\bullet})/ P_\phi \cE'_W(M^{\bullet})$ with one of its representatives.\\
				The map $S \ni \phi \mapsto (t \otimes s)_\phi$ is an on-shell $W$-smooth covariant section of rank $n+m$.\\
				Furthermore, $\otimes$ is a bilinear map $C^\infty_W(S, \boxtimes^n_W T^* S) \times C^\infty_W(S, \boxtimes^m_W T^* S) \to C^\infty_W(S, \boxtimes^{n+m}_W T^* S)$.
			\end{prop}
			\begin{proof}
				Let $c$ be a cut-off function as in eq.~\eqref{kernel_symp}. By lemma~\ref{lemma_free_c_ext}, we can choose two $W$-smooth extensions $\tilde{t}, \tilde{s}$ of $t, s$ in the sense of def.~\ref{def_smooth_on_tens} such that for any $\phi \in C^\infty(M)$ we have $\tilde{t}_\phi \in (\sigma_c \circ E_\phi)^{\otimes n} \cE'_W(M^n)$ and $\tilde{s}_\phi \in (\sigma_c \circ E_\phi)^{\otimes m} \cE'_W(M^m)$ for the same fixed $c$. The desired extension of $(t \otimes s)$ is defined by
					\begin{equation}\label{tilde_prod_tensor_W_sec}
						\widetilde{(t \otimes s)}_\phi := \tilde{t}_\phi \otimes \tilde{s}_\phi,
					\end{equation}
				for any $\phi \in C^\infty(M)$. Since $(W_n \times W_m) \cup (\{0\} \times W_m) \cup (W_n \times \{0\}) \subset W_{n+m}$, the estimate of the wave-front set of the tensor product of two distributions~\citep[thm. 8.2.9]{H83} implies that $\tilde{t}_\phi \otimes \tilde{s}_\phi \in (\sigma_c \circ E_\phi)^{\otimes n+m} \cE'_W(M^{n+m})$.\\
				By hypothesis, $\tilde{t}_\phi$ and $\tilde{s}_\phi$ satisfy conditions~\ref{W1},~\ref{W2}. Then, by distributing the variational derivatives onto the factors in $\tilde{t}_\phi \otimes \tilde{s}_\phi$, it follows again from~\citep[thm. 8.2.9]{H83} that we have $\widetilde{(t \otimes s)}_\phi$ satisfies conditions~\ref{W1},~\ref{W2}.\\
				Finally, $\otimes$ is linear by definition.
			\end{proof}
			
		Based on def.~\ref{def_smooth_on_f} and def.~\ref{def_smooth_on_tens}, we next define a natural derivative operator $\partial$ acting on on-shell $W$-smooth functions or covariant sections via the Gateaux derivative of a corresponding extension. Looking at these definitions, it is clear that the extensions depend on a choice of the cut-off function $c$ satisfying the properties required by eq.~\eqref{kernel_symp}. This choice will also be reflected in the definition of $\partial$.
		\begin{rem}\label{rem_lin_sol_var_der}
			The situation is simpler for functions $F \in C^\infty_W(S)$ (rather than covariant sections). In this case, we can show that along directions in $T_\phi S$ the first Gateaux derivative of all possible extensions of $F$ coincide. More precisely, consider two extensions $\tilde{F}_1, \tilde{F}_2$ of the same on-shell $W$-smooth function $F$. Obviously, $(\tilde{F}_1 - \tilde{F}_2)(\phi)=0$ for any $\phi \in S$. Let $u$ be an element of $T_\phi S$ and consider the smooth non-linear solution $\phi_\epsilon := U(\rho(\phi) + \epsilon \rho(u))$, i.e. the unique smooth solution of the non-linear eq.~\eqref{eom_phi4} corresponding to the Cauchy data $\rho(\phi) + \epsilon \rho(u)$, where $\rho$ is the restriction map~\eqref{restriction_map}. Because $u$ is a smooth solution of the linearised eq.~\eqref{leom_phi4} around $\phi \in S$, it holds that $\phi_\epsilon = \phi + \epsilon u + o(\epsilon^2)$ and, therefore, we can conclude that
		 \begin{equation}\label{vanising}
		 	\int_M \left( \frac{\delta \tilde{F}_1(\phi)}{\delta \phi (y)} - \frac{\delta \tilde{F}_2(\phi)}{\delta \phi (y)} \right) u(y) dy = \left. \frac{d}{d \epsilon} (\tilde{F}_1 - \tilde{F}_2)(\phi + \epsilon u) \right|_{\epsilon =0} = \left. \frac{d}{d \epsilon} (\tilde{F}_1 - \tilde{F}_2)(\phi_\epsilon) \right|_{\epsilon =0} = 0.
		 \end{equation}
		\end{rem}
	\noindent
	Before stating the definition of $\partial$ for on-shell $W$-smooth covariant sections, we need to prove that for any $\phi \in C^\infty(M)$ and for any $f_1, f_2 \in C^\infty_0(M)$ the map 
					\begin{equation}\label{inf_comm}
						M \ni x \mapsto \int_{M^3} \left( E_\phi (y,x_2) \frac{\delta E_\phi (x,x_1)}{\delta \phi(y)} - E_\phi (y,x_1) \frac{\delta E_\phi(x,x_2)}{\delta \phi(y)}\right) f_1(x_1)f_2(x_2) dx_1 dx_2 dy
					\end{equation}
				is a smooth solution with respect to $P_\phi = \boxempty - m^2 - V''(\phi)$. Actually, we prove the following more general result, which will be needed later on.
		\begin{lemma}\label{lemma_inf_comm}
		For any $\phi \in C^\infty(M)$ let $A_\phi, B_\phi$ be two distributions in $\cD'(M^2)$ such that:
			\begin{itemize}
				\item $A_\phi(x_1,x_2), B_\phi(x_1,x_2)$ are bi-solutions with respect to $P_\phi$.
				\item $\WF(A_\phi), \WF(B_\phi) \subset W_2$.
				\item The Gateaux derivatives $\delta A(x_1,x_2) / \delta \phi(y)$, $\delta B(x_1,x_2) / \delta \phi(y)$ are compactly supported distributions in $y$ and they satisfy
						\begin{equation*}
							\WF \left( \frac{\delta A_\phi(x_1,x_2)}{\delta \phi(y)} \right), \WF \left( \frac{\delta B_\phi(x_1,x_2)}{\delta \phi(y)} \right) \subset W_3.
						\end{equation*}
			\end{itemize}
		Then, for any $f_1, f_2 \in C^\infty_0(M)$, the map 
					\begin{equation}\label{inf_comm_gen}
						M \ni x \mapsto \int_{M^3} \left( A_\phi (y,x_2) \frac{\delta B_\phi (x,x_1)}{\delta \phi(y)} - B_\phi (y,x_1) \frac{\delta A_\phi(x,x_2)}{\delta \phi(y)}\right) f_1(x_1)f_2(x_2) dx_1 dx_2 dy
					\end{equation}
				is a smooth solution with respect to $P_\phi$.
	\end{lemma}
	\begin{proof}
		We first prove that the map~\eqref{inf_comm_gen} is actually a well-defined smooth function. By hypothesis, we can apply lemma~\ref{lemma_W_comp} and conclude that the composition of distributions 
			\begin{equation*}
				\int_M \left( A_\phi (y,x_2) \frac{\delta B_\phi (x,x_1)}{\delta \phi(y)} - B_\phi (y,x_1) \frac{\delta A_\phi (x,x_2)}{\delta \phi(y)} \right)dy
			\end{equation*}
		is a well-defined distribution with wave-front set contained in $W_3$. By smearing this distribution in $x_1, x_2$ with the two compactly supported functions $f_1, f_2$, we obtain the map~\eqref{inf_comm_gen}. We get a distribution in $x$ with wave-front set in $\{ (x,x_1,x_2; k, 0, 0) \in W_3 \}$. However, by definition, $W_3$ cannot contain any elements in the form $(x,x_1,x_2; k, 0, 0)$ and so the map~\eqref{inf_comm_gen} is a smooth function.\\
		Since $P_\phi^{(x_1)} A_\phi(x_1,x_2)=0$, it follows that
			\begin{equation*}
				P_\phi^{(x_1)} \frac{\delta A_\phi(x_1,x_2)}{\delta \phi(y)} = \frac{\delta P^{(x_1)}_\phi A_\phi(x_1,x_2)}{\delta \phi(y)} - \frac{\delta P_\phi^{(x_1)}}{\delta \phi(y)} A_\phi(x_1,x_2) = \lambda(y)\phi(y) \delta(y,x_1) A_\phi(x_1,x_2).
			\end{equation*}
		A similar result holds for $B_\phi$. It follows that if we act with the operator $P_\phi$ on the function~\eqref{inf_comm_gen}, then we obtain
				\begin{equation*}
					\int_M \left( A_\phi(f_2)(y) \delta(x,y) \lambda(y)\phi(y) B_\phi (f_1)(y) - B_\phi(f_1)(y) \delta(x,y) \lambda(y)\phi(y) A_\phi (f_2)(y) \right) dy =0,
				\end{equation*}
		and this concludes the proof.
	\end{proof}
	\noindent
	According to prop.~\ref{prop_var_ders_causal}, the causal propagator $E_\phi$ satisfies the conditions required by lemma~\ref{lemma_inf_comm}, and, therefore, the map~\eqref{inf_comm} is indeed a  smooth solution respect to $P_\phi$.\\
	
	We now consider on-shell $W$-smooth covariant sections in the sense of def.~\ref{def_smooth_on_tens} and we define the derivative operator $\partial$ acting on these sections.
		\begin{prop}\label{prop_der_smooth_on}
			Let $c$ be a fixed cut-off function as in eq.~\eqref{kernel_symp} and $t$ be an on-shell $W$-smooth covariant section of rank $n$. For $\phi \in C^\infty(M)$, let $\tilde{t}_\phi \in (\sigma_c \circ E_\phi)^{\otimes n} \cE'_W(M^n)$ be an extension of $t$ as in def.~\ref{def_smooth_on_tens}. Define
				\begin{equation}\label{tilde_diff}
					\begin{split}
						\widetilde{(\partial t)}_\phi(x_1, \dots, x_{n+1}) &:=\int_{M^{n+1}} \prod_{i=1}^{n+1} (\sigma_c \circ E_\phi)(x_i, x_i') \frac{\delta \tilde{t}_\phi (x'_2, \dots, x'_{n+1})}{\delta \phi(x'_1)} dx'_1 \dots dx'_{n+1},
					\end{split}
				\end{equation}
			If $\phi \in S$, then the distribution $\widetilde{(\partial t)}_\phi$ does not depend on the choice of the extension $\tilde{t}_\phi$ in $(\sigma_c \circ E_\phi)^{\otimes n} \cE'_W(M^n)$.\\
			Moreover, it defines an on-shell $W$-smooth covariant section $\partial t$ with rank $(n+1)$ by restriction to $(TS)^{\otimes n+1}$,  i.e.
				\begin{equation}\label{diff_on-shell}
					(\partial t)_\phi (u_1, \dots, u_{n+1}) := \widetilde{(\partial t)}_\phi(u_1, \dots, u_{n+1}) \quad \forall \phi \in S, \, u_i \in T_\phi S.
				\end{equation}
			Note that $\partial t$ depends on the choice of $c$. $\gls{partial}$ is a linear map $C^\infty_W(S, \boxtimes^n_W T^*S) \to C^\infty_W(S, \boxtimes^{n+1}_W T^*S)$.\\
			We define the map $d: \Omega_{W}^k(S) \to \Omega_{W}^{k+1} (S)$ acting via anti-symmetrisation $\glsuserii{P_pm}$ on $\partial$, i.e.
				\begin{equation*}
					\widetilde{(d t)}_\phi := \bP^- \widetilde{(\partial t)}_\phi
				\end{equation*}
			It satisfies the following properties:
			\begin{enumerate}[label=(\arabic*), start=1]
				\item The section $dt$ does not depend on the choice of the cut-off $c$, unlike $\partial$.
				\item For any $F \in C_{W}^\infty(S)$, $\phi \in S$ and $u \in T_\phi S$, $dF_\phi (u)$ coincides with the directional derivative of $F$ along $u$.
				\item For $t, s$ on-shell $W$-smooth forms with rank respectively $k$ and $k'$ it holds that $d(t \wedge s) = dt \wedge s + (-1)^{k} t \wedge ds$,
					where $\wedge$ is the anti-symmetrisation of the map $\otimes$ defined in prop.~\ref{prop_tensor_prod_W_sec}.
				\item $d$ is flat, i.e. $d^2=0$.
			\end{enumerate}
		\end{prop}
		\begin{proof}
			First of all, we verify that formula~\eqref{tilde_diff} provides a well-defined distribution which belongs to $\cE'_W(M^{n+1})$. The wave-front set of $(\sigma_c \circ E_\phi)$ is bounded by $W_2$ as shown in lemma~\ref{lemma_project}. By hypothesis, we have $\WF (\delta \tilde{t}_\phi / \delta \phi) \subset W_{n+1}$. By construction, $\sigma_c$ is compactly supported. Therefore, the claim is a consequence of lemma~\ref{lemma_W_comp}.\\
			We next show that any two $W$-smooth extensions $\tilde{t}_{1,\phi}, \tilde{t}_{2,\phi} \in (\sigma_c \circ E_\phi)^{\otimes n} \cE'_W(M^n)$ of the same on-shell $W$-smooth covariant section $t$ for our fixed choice of $c$ give the same $\widetilde{(\partial t)}_\phi$ for $\phi \in S$. After smearing with arbitrary test functions $f_1, \dots, f_{n+1}$, the difference between the distributions~\eqref{tilde_diff} corresponding to the extensions $\tilde{t}_1, \tilde{t}_2$ is
				\begin{equation}\label{difference}
					\begin{split}
						&\left[ \widetilde{(\partial t_1)}_\phi- \widetilde{(\partial t_2)}_\phi \right] (f_1, \dots, f_{n+1}) = \\
						&\quad = \int_{M} (E_\phi \circ \sigma_c \circ f_1)(x_1) \frac{\delta}{\delta \phi(x_1)} \left\{ (\tilde{t}_{1,\phi} - \tilde{t}_{2,\phi})\left( (E_\phi \circ \sigma_c \circ f_2), \dots, (E_\phi \circ \sigma_c \circ f_{n+1}) \right) \right\} dx_1- \\
						&\qquad - \sum_{i=2}^{n+1}  \int_{M^{n+1}} (E_\phi \circ \sigma_c \circ f_1)(x_1) \frac{\delta (E_\phi \circ \sigma_c \circ f_i)(x_i)}{\delta \phi(x_1)} \prod_{j \neq i} (E_\phi \circ \sigma_c\circ f_j)(x_j) \times \\
						&\qquad \qquad \qquad  \times (\tilde{t}_{1,\phi} - \tilde{t}_{2,\phi})(x_2, \dots, x_{n+1})  dx_1 \dots dx_{n+1},
					\end{split}
				\end{equation}
			where we applied the Leibniz rule for the variational derivative. We need to prove that the $(\widetilde{\partial t_1} - \widetilde{\partial t_2})_\phi(f_1, \dots, f_{n+1})$ vanishes if $\phi$ belongs to $S$.\\
			Due to $(\sigma_c \circ E_\phi)(\sigma_c \circ E_\phi) = \sigma_c \circ E_\phi$ (see~\eqref{project_E_sigma}), we have $ (\tilde{t}_{1,\phi} - \tilde{t}_{2,\phi}) = (\sigma_c \circ E_\phi)^{\otimes n} (\tilde{t}_{1,\phi} - \tilde{t}_{2,\phi})$ since, by hypothesis, we assume $\tilde{t}_{1,\phi}, \tilde{t}_{2,\phi} \in (\sigma_c \circ E_\phi)^{\otimes n} \cE'_W(M^n)$. Then, the distribution $ (\sigma_c \circ E_\phi)^{\otimes n} (\tilde{t}_{1,\phi} - \tilde{t}_{2,\phi})$ is identically zero for $\phi \in S$ because $\tilde{t}_1, \tilde{t}_2$ are extensions of the same on-shell $W$-smooth section (see~\eqref{extension_on_W_tens}). Thus, the second term on the right-hand side of~\eqref{difference} must vanish if $\phi \in S$. The first term in~\eqref{difference} is the Gateaux derivative along the smooth $P_\phi$-solution $E_\phi \circ \sigma_c \circ f_1$ of the $W$-smooth function $C^\infty(M) \ni \phi \mapsto (\tilde{t}_{1,\phi} - \tilde{t}_{2,\phi})((E_\phi \circ \sigma_c \circ f_2), \dots, (E_\phi \circ \sigma_c \circ f_{n+1}))$ which is identically zero whenever $\phi \in S$. Arguing as in remark~\ref{rem_lin_sol_var_der}, we conclude that for $\phi \in S$ also the first term in~\eqref{difference} vanishes. Thus, we have verified the independence with respect to the choice of the extension.\\
			To prove that $\partial t$ is an on-shell $W$-smooth section, we need to show that $\widetilde{(\partial t)}_\phi$ satisfies the conditions~\ref{W1},~\ref{W2} of def.~\ref{def_smooth_on_tens} for any $\phi \in C^\infty(M)$.\\
			To show~\ref{W1}, we need to compute the $\nu$-th Gateaux derivative by distributing the functional derivative $\delta / \delta \phi(y)$ over the various factors on the right-hand side of~\eqref{tilde_diff}. It follows that $\delta^\nu \widetilde{(\partial t)}_\phi / \delta \phi(y_1) \cdots \delta \phi(y_\nu)$ is a linear combination of terms in the form
				\begin{equation}\label{term_var_tilde_diff}
					 	\int_{M^{n+1}} \prod_{i=1}^{n+1} \frac{\delta^{|N_i|} (\sigma_c \circ E_\phi)(x_i, x'_i)}{\delta \phi^{|N_i|} (\{y_{r} \}_{r \in N_i})} \frac{\delta^{|N_{t}|+1} \tilde{t}_\phi (x'_2, \dots, x'_{n+1})}{\delta \phi(x'_1) \delta \phi^{|N_t|} (\{y_{r} \}_{r \in N_t})} dx'_1 \dots dx'_{n+1},
				\end{equation}
			where $N_t, N_1, \dots, N_{n+1}$ is a partition of $\{1, \dots, \nu\}$. The wave-front set of $\delta^{|N_i|} E_\phi / \delta \phi^{|N_i|}$ is estimated in~\eqref{var_ders_causal_WF}. The wave-front set of $\sigma_c$ is estimated in~\eqref{WF_kernel_symp}. The wave-front set of $\delta^{|N_t| +1} \tilde{t}_\phi / \delta \phi^{|N_t| +1}$ is contained in $W_{n + {|N_t|} +1}$ since, by hypothesis, $\tilde{t}_\phi$ satisfies~\ref{W1}. We can then apply lemma~\ref{lemma_W_comp} and lemma~\ref{lemma_tech_X_W} and thereby we find that the distribution~\eqref{term_var_tilde_diff} is well-defined and its wave-front set is contained in $W_{n +1 + \nu}$. Thus, the requirement~\ref{W2} holds.\\
			In order to show~\ref{W2}, let $\bR \ni \epsilon \mapsto \phi(\epsilon) \in C^\infty(M)$ be continuous, and consider $\delta^\nu \widetilde{\partial t}_{\phi(\epsilon)} / \delta \phi^\nu$ as a distribution in $\bR \times M^{n+1+\nu}$. This distribution is again a linear combination of terms in the form~\eqref{term_var_tilde_diff} with the only difference that $\phi$ is replaced by $\phi(\epsilon)$ everywhere. The wave-front set of $\delta^{|N_t| +1} \tilde{t}_{\phi(\epsilon)} / \delta \phi^{|N_t| +1}$ is contained in $\bR \times \{0\} \times W_{n + |N_t| +1}$ because, by hypothesis, $\tilde{t}_\phi$ satisfies~\ref{W2}.  Formula~\eqref{cont_var_ders_causal_WF} implies that $\delta^{|N_i|} E_{\phi(\epsilon)} / \delta \phi^{|N_i|}$ is contained in $\bR \times \{0\} \times W_{2 + |N_i|}$. Thus, using the wave-front set calculus (thm.\ref{theo_WF_horma}), it follow that $\delta^\nu \widetilde{\partial t}_{\phi(\epsilon)} / \delta \phi^\nu$ has wave-front set contained in $\bR \times \{0\} \times W_{n +1 + \nu}$ which is precisely condition~\ref{W2}.\\
		Since the map $d$ is defined simply by acting with $\partial$ followed by an anti-symmetrization, it is clearly well-defined.\\
		We prove that $d$ satisfies (1)-(4).
		\begin{enumerate}[label=(\arabic*), start=1]
			\item To prove that $d$ does not depend on the choice of the cut-off function $c$, we first need to investigate the dependence of $\partial$ on $c$. Let be $c, c'$ are two cut-off functions satisfying the properties required by eq.~\eqref{kernel_symp}, i.e. there exist Cauchy surfaces $\Sigma_{\pm}$ and $\Sigma'_{\pm}$ such that $c = 0$ on $J^+(\Sigma_+)$, $c = 1$ on $J^-(\Sigma_-)$, $c' = 0$ on $J^+( \Sigma'_+)$, and $c' = 1$ on $J^- (\Sigma'_-)$. We write $\widetilde{(\partial t)}$ for the distribution defined by eq.~\eqref{tilde_diff} for the cut-off $c$ and for a $W$-smooth extension $\tilde{t}_\phi \in (\sigma_c \circ E_\phi)^{\otimes n} \cE'_W(M^n)$ of $t$. Due to the fact that $(\sigma_c \circ E_\phi)(\sigma_c \circ E_\phi)= \sigma_c \circ E_\phi$ (see eq.~\eqref{project_E_sigma}), we have that
			\begin{equation}\label{particular_ext}
				\begin{split}
					\widetilde{(\partial t)}_\phi(x_1, \dots, x_{n+1}) &= \int_{M^{n+1}} \prod_{i=1}^{n+1} (\sigma_{c} \circ E_\phi)(x_i, x_i') \frac{\delta ((\sigma_{c} \circ E_\phi)^{\otimes n} \tilde{t}_\phi) (x'_2, \dots, x'_{n+1})}{\delta \phi(x'_1)} \prod_i dx'_i.
				\end{split}
			\end{equation}
		We write $\widetilde{(\partial t)'}$ for the distribution defined by eq.~\eqref{tilde_diff} using the cut-off $c'$ and a $W$-smooth extension of $\tilde{t}'_\phi \in (\sigma_{c'} \circ E_\phi)^{\otimes n} \cE'_W(M^n)$ of $t$. As proved in lemma~\ref{lemma_free_c_ext}, the distribution $(\sigma_{c'} \circ E_\phi)^{\otimes n} \tilde{t}_\phi$ is a $W$-smooth extension for $t$ with respect to the cut-off $c'$. We have shown at the beginning of this proof that $\widetilde{(\partial t)'}$ does not depend on the choice of the $W$-smooth extension in $\tilde{t}'_\phi$ of $t$, therefore we can use $(\sigma_{c'} \circ E_\phi)^{\otimes n} \tilde{t}_\phi$ as extension for $t$ where $\tilde{t}_\phi$ is the same extension as in~\eqref{particular_ext}, i.e.
			\begin{equation*}
				\begin{split}
					\widetilde{(\partial t)'}_\phi(x_1, \dots, x_{n+1}) = &\int_{M^{n+1}} \prod_{i=1}^{n+1} (\sigma_{c'} \circ E_\phi)(x_i, x_i') \frac{\delta ((\sigma_{c'} \circ E_\phi)^{\otimes n} \tilde{t}_\phi) (x'_2, \dots, x'_{n+1})}{\delta \phi(x'_1)} \prod_i dx'_i.
				\end{split}
			\end{equation*}
		It now follows from lemma~\ref{lemma_approx} that
			\begin{equation}\label{rel_c_c'_partial_1}
				\begin{split}
					&\left( \widetilde{(\partial t)}_\phi - \widetilde{(\partial t)'}_\phi \right)(x_1, \dots, x_{n+1}) \approx \\
					&\quad \approx \int_{M^{n+1}} \prod_{i=1}^{n+1} (\sigma_c \circ E_\phi) (x_i,x'_i) \frac{\delta \left( (\sigma_{c} \circ E_\phi)^{\otimes n} - (\sigma_{c'} \circ E_\phi)^{\otimes n} \right) \tilde{t}_\phi}{\delta \phi (x'_1)} (x'_2, \dots,  x'_{n+1}) \prod_i dx'_i,
				\end{split}
			\end{equation}
		where $\approx$ means ``equal up to distributions in $P_\phi \cE'_W(M^{n+1})$''. We can express $(\sigma_{c} \circ E_\phi)^{\otimes n} - (\sigma_{c'} \circ E_\phi)^{\otimes n}$ as
			\begin{equation*}
				\begin{split}
					&\left( (\sigma_{c} \circ E_\phi)^{\otimes n} - (\sigma_{c'} \circ E_\phi)^{\otimes n} \right)(x_1, \dots, x_n, y_1, \dots, y_n) = \\
					&\quad = \sum_{j=1}^{n}  \left( \prod_{i <j} (\sigma_c' \circ E_\phi) (x_i, y_i) \right) ( (\sigma_c -\sigma_{c'}) \circ E_\phi) (x_j, x'_j) \left( \prod_{\ell > i} (\sigma_{c} \circ E_\phi)(x_{\ell}, y_{\ell}) \right).
				\end{split}
			\end{equation*}
		Using eq.~\eqref{project_E_sigma}, it holds $(\sigma_c \circ E_\phi)( (\sigma_c -\sigma_{c'}) \circ E_\phi) =0$. Eq.~\eqref{rel_c_c'_partial_1} thereby becomes
			\begin{equation}\label{rel_c_c'_partial_2}
				\begin{split}
					&\left( \widetilde{(\partial t)}_\phi - \widetilde{(\partial t)'}_\phi \right)(x_1, \dots, x_{n+1}) \approx \\
					&\quad \approx \sum_{j =2}^{n+1} \int_{M^{n+2}} \prod_{i =1}^{n+1} (\sigma_c \circ E_\phi)(x_i, x'_i) \frac{\delta (\sigma_c -\sigma_{c'}) \circ E_\phi}{\delta \phi (x'_1)} (x'_j,z_j) \times\\
					&\qquad \qquad \qquad \qquad \times \tilde{t}_\phi(x'_2, \dots, z_j, \dots, x'_{n+1})  dz_j \prod_i dx'_i\\
					&\quad \approx \sum_{j =2}^{n+1} \int_{M^2} (\sigma_c \circ E_\phi)(x_1, x'_1) \left( \sigma_c \circ E_\phi \circ (\sigma_c -\sigma_{c'}) \circ \frac{\delta E_\phi}{\delta \phi (x'_1)} \right)(x_j,z_j) \times \\
					&\qquad \qquad \qquad \qquad \times \tilde{t}_\phi(x_2, \dots, z_j, \dots, x_{n+1}) dx'_1 dz_j,
				\end{split}
			\end{equation}
		where we used $(\sigma_c \circ E_\phi)(\sigma_c \circ E_\phi)= \sigma_c \circ E_\phi$ (see eq.~\eqref{project_E_sigma}) and the fact that $\tilde{t} \in (\sigma_c \circ E_\phi)^{\otimes n} \cE'_W(M^n)$ by hypothesis. We now substitute
			\begin{equation*}
				\begin{split}
					\frac{\delta E_\phi (x,z)}{\delta \phi(y)} &=\frac{\delta E^A_\phi (x,z)}{\delta \phi(y)} - \frac{\delta E^R_\phi (x,z)}{\delta \phi(y)} \\
					&= E^A_\phi(x,y) \lambda(y) \phi(y) E^A_\phi(y,z) -  E^R_\phi(x,y) \lambda(y) \phi(y) E^R_\phi(y,z).
				\end{split}
			\end{equation*}
		for the variational derivative of $E_\phi$. Then, we need to analyse $E_\phi \circ (\sigma_c -\sigma_{c'}) \circ E^{A/R}_\phi$ in order to simplify further eq.~\eqref{rel_c_c'_partial_2}. It follows from eq.~\eqref{kernel_symp_not_sol} that for any test functions $f_1, f_2$ we have
			\begin{equation}\label{cut-off_difference}
				\begin{split}
					&(E_\phi \circ (\sigma_c - \sigma_{c'}) \circ E^{A/R}_\phi)(f_1, f_2) =\\
					&\quad = - \int_{\Sigma_-} E_\phi(f_1)(x) \overleftrightarrow{\partial_n} E^{A/R}_\phi(f_2)(x) d\Sigma(x) + \int_{\Sigma'_-} E_\phi(f_1)(x) \overleftrightarrow{\partial_n} E^{A/R}_\phi(f_2)(x) d\Sigma(x) + \\
					&\qquad + \int_B (E_\phi(f_1))(x) c(x) f_2(x) dx - \int_{B'} (E_\phi(f_1))(x) c'(x) f_2(x) dx
				\end{split}
			\end{equation}
		where $B = J^- (\Sigma_+) \cap J^+ (\Sigma_-)$ and $B' = J^-( \Sigma'_+) \cap J^+ (\Sigma'_-)$. Let us choose another pair of Cauchy surfaces $\Sigma''_\pm$ such that $\Sigma''_+$ is in the (strict) future of both $\Sigma_+,\Sigma'_+$ and $\Sigma''_-$, is in the (strict) past of both $\Sigma_-,\Sigma'_-$. We identify by $B''$ the space-time region $J^- (\Sigma''_+) \cap J^+ (\Sigma''_-)$. We can use eq.~\eqref{symp_form_not_sol} to rewrite the first two integrals in the right-hand side of eq.~\eqref{cut-off_difference} in terms of the Cauchy surface $\Sigma''_+$. We obtain
			\begin{equation}\label{cut-off_difference_term}
					(E_\phi \circ (\sigma_c - \sigma_{c'}) \circ E^{A/R}_\phi)(f_1, f_2) =  \int_{B''} (E_\phi(f_1))(x) (c-c')(x) f_2(x) dx.
			\end{equation}
			 Notice that the domain of the integration in the right-hand side of eq.~\eqref{cut-off_difference_term} can be extended to the whole $M$ because $c - c' =0$ outside $B''$. Using eq.~\eqref{cut-off_difference_term}, we conclude that the following relation holds
			\begin{equation}\label{rel_c_c'_partial}
				\begin{split}
					&\left( \widetilde{(\partial t)}_\phi - \widetilde{(\partial t)'}_\phi \right)(x_1, \dots, x_{n+1}) \approx  \\
					&\quad \approx \sum_{j =2}^{n+1} \int_{M^{2}} (\sigma_c \circ E_\phi)(x_1, y)  (\sigma_c \circ E_\phi)(x_j, y) \lambda(y)\phi(y) (c-c')(y) E_\phi(y,z_j) \times \\
					&\qquad \qquad \qquad \qquad \times \tilde{t}_\phi(x_2, \dots, z_j, \dots, x_{n+1}) dy dz_j.
				\end{split}
			\end{equation}
		The right-hand side of eq.~\eqref{rel_c_c'_partial} does not vanish in general. Thus, we see explicitly that $\partial t^{(c)}$ depends on the choice of the cut-off. Nevertheless, the right-hand side of eq.~\eqref{rel_c_c'_partial} is a finite sum of distributions that are symmetric in $x_1, x_j$. Therefore, it follows that
			\begin{equation}\label{vanish_anti}
				\begin{split}
					&\bP^-\left( \widetilde{(\partial t)}_\phi - \widetilde{(\partial t)'}_\phi \right) \approx 0.
				\end{split}
			\end{equation}
		where $\bP^-$ denotes the anti-symmetrization. Because the on-shell $W$-smooth form $dt$ is defined by restriction to $(TS)^{\otimes n+1}$ of $\widetilde{(dt)} = \bP^- \widetilde{(\partial t)}$, the relation~\eqref{vanish_anti} implies that $dt$ does not depend on the choice of the cut-off $c$.
		\item Let $\phi \in S$ and $u \in T_\phi S$. For any on-shell $W$-smooth functional $F$, it holds
			\begin{equation*}
				\begin{split}
					dF_\phi(u) &= \int_M \left(\bP^- \widetilde{(\partial F)}_\phi\right)(x) u(x) dx = \int_{M^2} u(x) (\sigma_c \circ E_\phi)(x,x') \frac{\delta \tilde{F}(\phi)}{\delta \phi(x')} dx dx' \\
					&= \int_{M} u(x) \frac{\delta \tilde{F}(\phi)}{\delta \phi(x')} dx,
				\end{split}
			\end{equation*}
		where we simply apply eq.~\eqref{causal_prop_inv_symp_kernel_sol}. Then, arguing as in remark~\ref{rem_lin_sol_var_der}, we have
			\begin{equation*}
				dF_\phi(u) = \left. \frac{d}{d \epsilon} \tilde{F}(\phi + \epsilon u) \right|_{\epsilon = 0} = \left. \frac{d}{d \epsilon} \tilde{F}(\phi_\epsilon) \right|_{\epsilon = 0} =  \left. \frac{d}{d \epsilon} F(\phi_\epsilon) \right|_{\epsilon = 0} = \left. \frac{d}{d \epsilon} F(\phi + \epsilon u) \right|_{\epsilon = 0},
			\end{equation*}
		where $\phi_\epsilon$ is the unique solution of the non-linear eq.~\eqref{eom_phi4} with Cauchy data $\rho(\phi) + \epsilon \rho(u)$. Thus, we have verified that property (2) holds.
		\item It can be easily seen that (3) is just a straightforward consequence of the definition of $d$, the definitions the tensor product $\otimes$, and the Leibniz rule for the variational derivative.
		\item To show that $d$ is flat, it is sufficient to prove that the off-shell extension $(\widetilde{d(dt)})$ vanishes. This can be shown by direct calculation. In fact, by definition, we have
		\begin{equation}\label{eq_flat_1}
			\begin{split}
				&\widetilde{(d(dt))}_\phi(x_1, \dots, x_{n+2}) =\\
				&\quad = \bP^- \int_{M^{n+2}} \prod_{i=1}^{n+2} (\sigma_c \circ E_\phi)(x_i, x'_i) \frac{\delta}{\delta \phi(x'_1)} \left\{ \widetilde{dt}_\phi(x'_2, \dots, x'_{n+2}) \right\} \prod_i  dx'_i \\
				&\quad = \bP^- \int_{M^{2n +3}} \prod_{i=1}^{n+2} (\sigma_c \circ E_\phi)(x_i, x'_i) \times \\
				&\qquad \qquad \qquad \qquad \times \frac{\delta}{\delta \phi(x'_1)} \left\{ \prod_{j=2}^{n+2} (\sigma_c \circ E_\phi)(x'_j, x''_j) \frac{\delta \tilde{t}_\phi (x''_3, \dots, x''_{n+2})}{\delta \phi(x'_2)}\right\} \prod_i  dx'_i \prod_j  dx''_j,
			\end{split}
		\end{equation}
		where $c$ is a fixed cut-off function as in eq.~\eqref{kernel_symp}, and where $\tilde{t}_\phi \in (\sigma_c \circ E_\phi)^{\otimes n} \cE'_W(M^n)$ is a $W$-smooth extension of the on-shell $W$-smooth section $t$. Next, applying the Leibniz rule for the variational derivative and using eq.~\eqref{project_E_sigma}, it follows that
		\begin{equation}\label{eq_flat}
			\begin{split}
				&\widetilde{(d(dt))}_\phi(x_1, \dots, x_{n+2}) = \\
				&\quad= \bP^- \int_{M^{n+2}} (\sigma_c \circ E_\phi)(x_1, x'_1) \times \\
				&\qquad \qquad \qquad \qquad \times \frac{\delta}{\delta \phi(x'_1)} \left\{ \prod_{j=2}^{n+2} (\sigma_c \circ E_\phi)(x_j, x''_j) \frac{\delta \tilde{t}_\phi (x''_3, \dots, x''_{n+2})}{\delta \phi(x''_2)} \right\} dx'_1 \prod_j  dx''_j - \\
				&\qquad - \bP^- \int_{M^{n+3}} (\sigma_c \circ E_\phi)(x_1, x'_1) \frac{\delta (\sigma_c \circ E_\phi)(x_2, x'_2)}{\delta \phi(x'_1)} (\sigma_c \circ E_\phi) (x'_2, x''_2)  \times \\
				&\quad \qquad \qquad \qquad \qquad \times \prod_{j=3}^{n+2} (\sigma_c \circ E_\phi)(x_j, x''_j)  \frac{\delta \tilde{t}_\phi (x''_3, \dots, x''_{n+2})}{\delta \phi(x''_2)} dx'_1 dx'_2 dx''_2\prod_j  dx''_j- \\
				&\qquad - \bP^- \sum_{\ell = 3}^{n+2} \int_{M^{n+3}} (\sigma_c \circ E_\phi)(x_1, x'_1) \frac{\delta (\sigma_c \circ E_\phi)(x_\ell,x'_\ell)}{\delta \phi(x'_1)} (\sigma_c \circ E_\phi) (x'_\ell, x''_\ell)  \times \\
				&\quad \qquad \qquad \qquad \qquad \times \prod_{j \neq \ell} (\sigma_c \circ E_\phi)(x_j, x''_j) \frac{\delta \tilde{t}_\phi (x''_3, \dots, x''_{n+2})}{\delta \phi(x''_2)} dx'_1 dx'_\ell dx''_\ell \prod_j  dx''_j
			\end{split}
		\end{equation}
		Using the fact that we anti-symmetrize in all variables $x_1, \dots, x_{n+2}$, we rewrite the second term in the equation~\eqref{eq_flat} above as
		\begin{equation}\label{first_term_flat}
			\begin{split}
				 &- \frac{1}{2} \bP^- \int_{M^{n+3}} \left[ (\sigma_c \circ E_\phi)(x_1, y) \frac{\delta (\sigma_c \circ E_\phi)(x_2, z)}{\delta \phi(y)} - (\sigma_c \circ E_\phi)(x_2, y) \frac{\delta (\sigma_c \circ E_\phi)(x_1, z)}{\delta \phi(y)} \right]  \times \\
				 &\qquad \times (\sigma_c \circ E_\phi) (z, x''_2) \prod_{j=3}^{n+2} (\sigma_c \circ E_\phi)(x_j, x''_j) \frac{\delta \tilde{t}_\phi (x''_3, \dots, x''_{n+2})}{\delta \phi(x''_2)} dy dz dx''_2 \prod_j dx''_j
			\end{split}
		\end{equation}
		Note that the term in the bracket $[\dots]$ is a $P_\phi$-solution in $z$ as a consequence of lemma~\ref{lemma_inf_comm}. Since $\sigma_c \circ E_\phi$ is the identity on $P_\phi$-solutions (see eq.~\eqref{project_E_sigma}), then it follows that the distribution~\eqref{first_term_flat} is equal to
			\begin{equation*}
			\begin{split}
				 &- \frac{1}{2} \bP^- \int_{M^{n+2}} \left[ (\sigma_c \circ E_\phi)(x_1, y) \frac{\delta (\sigma_c \circ E_\phi)(x_2, x''_2)}{\delta \phi(y)} - (\sigma_c \circ E_\phi)(x_2, y) \frac{\delta (\sigma_c \circ E_\phi)(x_1, x''_2)}{\delta \phi(y)} \right] \times \\
				 &\qquad \times \prod_{j=3}^{n+2} (\sigma_c \circ E_\phi)(x_j, x''_j) \frac{\delta \tilde{t}_\phi (x''_3, \dots, x''_{n+2})}{\delta \phi(x''_2)} dy dx''_2 \prod_j dx''_j\\
				 &\quad = - \bP^- \int_{M^{n+2}} (\sigma_c \circ E_\phi)(x_1, x'_1) \frac{\delta (\sigma_c \circ E_\phi)(x_2, x''_2)}{\delta \phi(x'_1)}  \prod_{j=3}^{n+2} (\sigma_c \circ E_\phi)(x_j, x''_j)  \times \\
				&\quad \qquad \times \frac{\delta \tilde{t}_\phi (x''_3, \dots, x''_{n+2})}{\delta \phi(x''_2)} dx'_1 dx''_2 \prod_j  dx''_j
			\end{split}
		\end{equation*}
		A similar argument holds also for the last term in eq.~\eqref{eq_flat}. Putting together and applying again the Leibniz rule of the variational derivative, we finally have
		\begin{equation*}
			\begin{split}
				&\widetilde{(d(dt))}_\phi(x_1, \dots, x_{n+2}) = \bP^- \int_{M^{n+2}} \prod_{i=1}^{n+2} (\sigma_c \circ E_\phi)(x_i, x'_i) \frac{\delta^2 \tilde{t} (x'_3, \dots, x'_{n+2})}{\delta \phi(x'_1) \delta\phi(x'_2)} \prod_i  dx'_i = 0.\\
			\end{split}
		\end{equation*}
	\end{enumerate}
		This concludes the proof.
		\end{proof}
		
	Now, we can discuss what is meant by ``deformation quantization'' in the infinite-dimensional context we provided. The notion of smoothness we are considering in this framework is the concept of on-shell $W$-smoothness. Thus, we consider deformations of the commutative algebra $(C_W^\infty(S), \cdot)$, where $\cdot$ is the pointwise product. Note that the pointwise product is a well-defined bilinear map $C_W^\infty(S) \times C_W^\infty(S) \to C_W^\infty(S)$ because it is just a special case of the tensor product of on-shell $W$-smooth covariant section discussed in prop.~\ref{prop_tensor_prod_W_sec}, namely for covariant sections with rank $0$. For this commutative algebra, we define the Poisson structure as follows:
		\begin{prop}\label{prop_W-poisson}
			Let $F_1,F_2 \in C^\infty_W(S)$. For any $\phi \in C^\infty(M)$, we define
				\begin{equation}\label{W-poisson}
					\widetilde{\left\{ F_1, F_2 \right\}}_\phi := E_\phi \left( \widetilde{(\partial F_1)}_\phi \otimes \widetilde{(\partial F_2)}_\phi \right).
				\end{equation}
			If $\phi \in S$, then the distribution $\widetilde{\left\{ F_1, F_2 \right\}}_\phi$ does not depend on the choice for the cut-off function $c$ and the extensions of $\tilde{F}_1, \tilde{F}_2$ implicit in the definition of $\widetilde{(\partial F_1)}, \widetilde{(\partial F_2)}$.\\
			Moreover, \eqref{W-poisson} defines an on-shell $W$-smooth function $\left\{ F_1, F_2 \right\}$ simply by $\left\{ F_1, F_2 \right\}_\phi := \widetilde{\left\{ F_1, F_2 \right\}}_\phi$ for any $\phi \in S$.\\
			Furthermore, $\left\{ \cdot, \cdot \right\}$ is a Poisson bracket $C^\infty_W(S) \times C^\infty_W(S) \to C^\infty_W(S)$ for the commutative algebra $( C^\infty_W(S), \cdot)$.
		\end{prop}
		\begin{proof}
			First of all, we notice that if $\phi \in S$, then $\widetilde{(\partial F_1)}_\phi, \widetilde{(\partial F_2)}_\phi$ does not depend on the choice of the extensions $\tilde{F}_1, \tilde{F}_2$, as we already proved in prop.~\ref{prop_der_smooth_on}.\\
			Next, we show that $\widetilde{\left\{ F_1, F_2 \right\}}_\phi$ does not depend on the choice of the cut-off functions. Let $\tilde{F}_{1}, \tilde{F}_{2}$ be $W$-smooth extensions of $F_1, F_2$ in the sense of def.~\ref{def_smooth_on_f}. For any $\phi \in C^\infty(M)$, consider the distributions $\widetilde{(\partial F_1)}_\phi \in (\sigma_c \circ E_\phi) \cE_W'(M)$ and $\widetilde{(\partial F_2)}_\phi \in (\sigma_{c '} \circ E_\phi) \cE_W'(M)$ as given by eq.~\eqref{tilde_diff} in terms of $\tilde{F}_{1}, \tilde{F}_{2}$ and for two possibly different cut-off functions $c,c'$ as in eq.~\eqref{kernel_symp}. Since $E_\phi \circ \sigma_c \circ E_\phi =E_\phi$ (see eq.~\eqref{causal_prop/symp_kernel}), we have
			\begin{equation}\label{W-poisson_alt}
				\begin{split}
					\widetilde{\left\{ F_1, F_2 \right\}}_\phi &= \int_{M^2} \frac{ \delta \tilde{F_1}_\phi}{\delta \phi(x_1)} (E_\phi \circ \sigma_c \circ E_\phi \circ \sigma_{c'} \circ E_\phi) (x_1,x_2) \frac{ \delta \tilde{F_2}_\phi}{\delta \phi(x_2)} dx_1 dx_2 \\
					&= \int_{M^2} \frac{ \delta \tilde{F_1}_\phi}{\delta \phi(x_1)} E_\phi(x_1,x_2) \frac{ \delta \tilde{F_2}_\phi}{\delta \phi(x_2)} dx_1 dx_2,
				\end{split}
			\end{equation}
		which implies that $\widetilde{\left\{ F_1, F_2 \right\}}_\phi$ does not depend on $c$ even for $\phi \in C^\infty(M)$ and not just for $\phi \in S$.\\
		In order to prove that $S \ni \phi \mapsto \left\{ F_1, F_2 \right\}_\phi$ is an on-shell $W$-smooth functional, we have to verify that $\widetilde{\left\{ F_1, F_2 \right\}}_\phi$ satisfies conditions \ref{W1_f}, \ref{W2_f} of def.~\ref{def_smooth_on_f}.\\
			To show \ref{W1_f}, we need to compute the $\nu$-th Gateaux derivative of $\widetilde{\left\{ F_1, F_2 \right\}}$ by distributing the functional derivative $\delta/ \delta \phi(y)$ over the factors in the right-hand side of eq.~\eqref{W-poisson_alt}. It follows that $\delta^\nu \widetilde{\left\{ F_1, F_2 \right\}}_\phi / \delta \phi^\nu$ is a finite sum of terms in the form
				\begin{equation}\label{term_var_tilde_W-poisson}
					\int_{M^2} \frac{\delta^{|N_1| +1} \widetilde{F}_{1,\phi}}{\delta \phi(x_1) \delta \phi^{|N_1|}(\{y_{r}\}_{r \in N_1})} \frac{ \delta^{|N_2|} E_\phi(x_1,x_2)}{\delta \phi^{|N_2|} (\{y_{r} \}_{r \in N_2})} \frac{\delta^{|N_3| +1} \widetilde{F}_{2,\phi}}{\delta \phi(x_2) \delta \phi^{|N_3|}(\{y_{r}\}_{r \in N_3})} dx_1 dx_2,
				\end{equation}
			where $N_1, N_2, N_3$ form a partition of $\{1, \dots, \nu\}$. The wave-front set of $\delta^{|N_2|} E_\phi / \delta \phi^{|N_2|}$ is contained in $W_{2 + |N_2|}$ as follows from~\eqref{var_ders_causal_WF} and lemma~\ref{lemma_tech_X_W}. By hypothesis, $\widetilde{F}_{1,\phi}, \widetilde{F}_{2,\phi}$ satisfies \ref{W1_f}, and so the wave-front sets of the compactly supported distributions $\delta^{|N_1| +1} \widetilde{F}_{1,\phi} / \delta \phi^{|N_1| +1}$ and $\delta^{|N_3| +1} \widetilde{F}_{2,\phi} / \delta \phi^{|N_3| +1}$ are contained in $W_{|N_1|+1}$ and in $W_{|N_3|+1}$ respectively. Then, we can apply lemma~\ref{lemma_W_comp} to prove that each term~\eqref{term_var_tilde_W-poisson} is a well-defined distribution and its wave-front set is contained in $W_{\nu}$. Furthermore, the distribution~\eqref{term_var_tilde_W-poisson} is compactly supported as follows from the support properties of the distributions involved. Thus, $\delta^\nu \widetilde{\left\{ F_1, F_2 \right\}}_\phi / \delta \phi^\nu \in \cE'_W(M^\nu)$ which is precisely the condition \ref{W1_f}.\\
			To show \ref{W2_f}, let $\bR \ni \epsilon \mapsto \phi(\epsilon) \in C^\infty(M)$ be continuous, and consider $\delta^\nu \widetilde{\left\{ F_1, F_2 \right\}}_{\phi(\epsilon)} / \delta \phi^\nu$ as a distribution in $\bR \times M^{\nu}$. This distribution is again a linear combination of terms in the form~\eqref{term_var_tilde_W-poisson} with the only difference that $\phi$ is replaced by $\phi(\epsilon)$. By hypothesis, $\widetilde{F}_{1,\phi}, \widetilde{F}_{2,\phi}$ satisfies \ref{W2_f}, and so the wave-front sets of $\delta^{|N_1| +1} \widetilde{F}_{1,\phi(\epsilon)} / \delta \phi^{|N_1| +1}$ and $\delta^{|N_3| +1} \widetilde{F}_{2,\phi(\epsilon)} / \delta \phi^{|N_3| +1}$ are contained in $\bR \times \{0\} \times W_{|N_1|+1}$ and in $\bR \times \{0\} \times W_{|N_3|+1}$ respectively. Formula~\eqref{cont_var_ders_causal_WF} implies that $\delta^{|N_2|} E_{\phi(\epsilon)} / \delta \phi^{|N_2|}$ is contained in $\bR \times \{0\} \times W_{2 + |N_i|}$. Thus, by the wave-front set calculus (thm.~\ref{theo_WF_horma}), we have that $\delta^\nu \widetilde{\left\{ F_1, F_2 \right\}}_{\phi(\epsilon)} / \delta \phi^\nu$ has wave-front set contained in $\bR \times \{0\} \times W_{\nu}$ which is precisely condition \ref{W2_f}.\\
			We have verified that $\phi \mapsto \left\{ F, F' \right\}_\phi$ is a well-defined on-shell $W$-smooth function.\\
			We notice that eq.~\eqref{W-poisson_alt} implies that $\widetilde{\left\{ F_1, F_2 \right\}}_\phi$ coincides precisely with the Peierls bracket~\eqref{Peierls_brkt}, which is a Poisson structure as established in~\citep{DF01a}. Therefore, $\left\{ \cdot, \cdot \right\}$ is a Poisson bracket for the algebra $C^\infty_W(S)$. This concludes the proof.
		\end{proof}
	\noindent
	In our infinite-dimensional setting, the notion of deformation quantization is the same as for finite dimensions, sec.~\ref{subsec_intro_deformation}, with the only adjustment that on-shell $W$-smoothness replaces ordinary smoothness everywhere.
		\begin{defi}	
			A deformation quantization on $S$ consists in providing an associative algebra structure, the star-product $\star$, on $C_W^\infty(S)[[\hbar]]$ such that $F \star F' = F \cdot F' + o(\hbar)$, and $[ F, F']_{\star} = i \hbar \{ F, F'\} + o(\hbar)$.
		\end{defi}
		
	As already mentioned, our intention is to define a deformation quantization on $S$ by mimicking Fedosov's construction in finite dimensions reviewed throughout sec.~\ref{subsec_Fedosov_fin}. Following this logic, we define in our infinite-dimensional framework the bundle of formal Wick algebras and its smooth sections.\\
	These notions are natural, basically functorial, generalizations of the definitions of the covariant tensor bundle~\eqref{cov_tens_S_bundle} and on-shell $W$-smooth covariant tensor fields (def.~\ref{def_smooth_on_tens}). The formal Wick algebra $\cW_\phi$ is defined as a vector space (cf.~\eqref{formal_wick_alg_fin}) in terms of the algebraic direct product
		\begin{equation*}
			\gls{cW_phi} = \bC[[\hbar]] \otimes \bigoplus_{n \geq 0} \vee_W^n T^*_\phi S,
		\end{equation*}
	where $\vee_W^n T^*_\phi S$ denotes the totally symmetric elements of $\boxtimes_W^n T^*_\phi S$. Let us consider the bundle
		\begin{equation*}
			\cW = \bigsqcup_{\phi \in S} \cW_\phi.
		\end{equation*}
	An on-shell $W$-smooth section on this bundle is a sequence $(t^0, t^1, \dots)$, where each $t^n$ is a $\bC[[\hbar]]$-valued totally symmetric on-shell $W$-smooth covariant section with rank $n$. We denote the space of such sections by $\gls{C_W_infty_S_cW}$. Similarly as in the finite-dimensional case, we introduce on $\cW$, and then canonically on $C_W^\infty(S,\cW)$, the symmetric degree $\deg_s$, the formal degree $\deg_\hbar$, and the total degree $\Deg$, which are defined by
		\begin{equation}\label{deg_s,deg_h,Deg,inf}
			\deg_s t^n := n, \quad \deg_\hbar \hbar :=1, \quad \Deg:=\deg_s + 2\deg_\hbar,
		\end{equation}
	where $t^n \in C_W^\infty(S,\vee_W^n T^* S)$. Exactly as in finite dimensions, a $\Deg$-homogeneous element $t$ has only finitely many non-zero elements $t^n$ in $(t^0, t^1, \dots)$ and each $t^n$ is a polynomial in $\hbar$.\\
	Other natural definitions from the finite-dimensional setting may then also be generalized. In particular, $\cW$-valued $k$-forms are defined as on-shell $W$-smooth sections in the bundle
		\begin{equation*}
			\bigsqcup_{\phi \in S} \bC[[\hbar]] \otimes \bigoplus_{n \geq 0} \wedgevee^{k,n}_W T^*_\phi S,
		\end{equation*}
	where $\wedgevee^{k,n}_W T^*_\phi S$ denotes the elements of $\boxtimes_W^{k+n} T^*_\phi S$ which are anti-symmetric in the first $k$ entries and symmetric in the remaining $n$. More precisely, a $\cW$-valued on-shell $W$-smooth $k$-form is a sequence $(t^{k,0}, t^{k,1}, \dots)$ where each $t^{k,n}$ is a $\bC[[\hbar]]$-valued on-shell $W$-smooth covariant section with rank $(k+n)$, anti-symmetric in the first $k$ entries and symmetric in the remaining $n$. The space of such $\cW$-valued forms is denoted by $\Omega_W^k(S, \cW)$. The three degrees defined by~\eqref{deg_s,deg_h,Deg,inf} extend to the space of $\cW$-valued $k$-forms. In addition, we can introduce the anti-symmetric degree $\deg_a$. Namely, we have
		\begin{equation}\label{gradings_inf_form}
			\deg_s t^{k,n} := n, \quad \deg_a t^{k,n} := k, \quad \deg_\hbar \hbar := 1, \quad \Deg: =\deg_s + 2\deg_\hbar,
		\end{equation}
	where $t^{k,n} \in C_W^\infty(S,\wedgevee^{k,n} T^* S)$.\\	
	Finally, we introduce the space of $\cW$-valued (on-shell $W$-smooth) forms with arbitrary anti-symmetric degree, i.e. the direct product
		\begin{equation*}
			\gls{Omega_W_S_cW} := \bigoplus_{k \geq 0} \Omega_W^k(S,\cW).
		\end{equation*}
	In contrast to the finite-dimensional setting, where the anti-symmetric degree cannot exceed the dimension of $S$, in the infinite-dimensional $\deg_a$ does not have a maximum value. An element $t$ in $\Omega_W(S,\cW)$ is a collection $(t^{k,n})_{k \in \bN; n \in \bN}$ where $t^{k,n}$ is the same as before. It is clear that a $\cW$-valued on-shell $W$-smooth form $t$ which is homogeneous in both $\Deg$ and $\deg_a$ is a finite collection of covariant sections homogeneous in $\deg_a, \deg_\hbar, \deg_s$, i.e. only finitely many $t^{k,n}$ appearing in the array defining $t$ are non-zero, and those which do not vanish are polynomial in $\hbar$. The $\Deg$-filtration is one cornerstone of Fedosov's method in the finite-dimensional setting, and it will be fundamental also in our infinite-dimensional construction.\\
	
	In the following sections~\ref{subsec_W_smooth_symp_metric}-\ref{subsec_Fedosov_inf} we will provide the notions needed to rigorously translate the construction of the Fedosov connection into the infinite-dimensional framework. This line of argument results in thm.~\ref{theo_Fedosov_inf}, which provides the infinite-dimensional version of Fedosov quantization, thm.~\ref{theo_Fedosov_1_fin} and thm.~\ref{theo_Fedosov_2_fin}.
	
	\section{Examples of on-shell $W$-smooth tensor fields on $S$}\label{subsec_W_smooth_symp_metric}
	In the previous section, we have discussed the manifold structure of $S$, i.e. the space of smooth solutions to $(\boxempty - m^2)\phi - \frac{\lambda}{3!}\phi^3 = 0$, and we have defined various bundles over $S$ and the corresponding notions of smooth (more precisely, on-shell $W$-smooth) sections. We would now like to give concrete examples for such sections which generalize the covariant tensor fields $\sigma_{ij}, G_{ij}$, and $\omega_{ij}=\sigma_{i\ell} \omega^{\ell k} \sigma_{k j} = - \frac{1}{2} G_{ij} + \frac{i}{2} \sigma_{ij}$ in the finite-dimensional case.

	\subsection{Symplectic structure on $S$}\label{subsubsec_sympl_inf}
		For each $\phi \in S$ and each pair $u_1, u_2 \in T_\phi S$ of solutions to the linearized equations around $\phi$, we consider the standard symplectic structure
			\begin{equation}\label{symp_struct}
				\gls{sigma_phi}(u_1,  u_2) := \int_\Sigma u_1(x) \overleftrightarrow{\partial_n} u_2(x) d\Sigma(x),
			\end{equation}
		and its associated distributional kernel $\sigma_c \in \cE'_W(M^2)$ defined by eq.~\eqref{kernel_symp}. Moreover, $\sigma_\phi$ is anti-symmetric and is the analogue of the tensor $(\sigma_x)_{ij}$ in finite dimensions.\\
		As we already proved in lemma~\ref{lemma_kernel_symp}, the distribution $E_\phi \circ \sigma_c$ gives the identity on $T_\phi S$. Thus, the causal propagator $E_\phi$ is the analogue of the tensor $(\sigma_x)^{ij}$ in finite dimensions.\\
		Concerning the dependence of $\sigma_\phi$ on $\phi$, we have:
		\begin{theo}\label{theo_W_symp}
			The map $S\ni \phi \mapsto \sigma_\phi$ is an on-shell $W$-smooth $2$-form, which we denote by $\sigma$.\\
			Furthermore, $\sigma$ is closed as an on-shell $W$-smooth form, i.e. $d\sigma = 0$, where $d$ is defined as in prop.~\ref{prop_der_smooth_on}.
		\end{theo}
		\begin{proof}
			To prove that the map $S \ni \phi \mapsto \sigma_\phi$ is on-shell $W$-smooth, we define for any $\phi \in C^\infty(M)$ and for a fixed cut-off function $c$ as in eq.~\eqref{kernel_symp} a distributional kernel $\tilde{\sigma}_\phi(x,y) \in (\sigma_c \circ E_\phi)^{\otimes 2} \cE'_W(M^2)$ which provides an extension in the sense of eq.~\eqref{extension_on_W_tens} of the symplectic structure $T_\phi S \times T_\phi S \ni (u,v) \mapsto \sigma_\phi(u,v)$ for $\phi \in S$, and which fulfils the requirements of def.~\ref{def_smooth_on_tens}. We set
				\begin{equation}\label{tilde_symp}
					\tilde{\sigma}_\phi (x_1,x_2) := \int_{M^2} \prod_{i=1}^2 (\sigma_c \circ E_\phi)(x_i,x'_i)\sigma_c(x'_1,x'_2) dx'_1 dx'_2 = (\sigma_c \circ E_\phi \circ \sigma_c \circ E_\phi \circ \sigma_c)(x_1,x_2).
				\end{equation}
			For any $\phi \in C^\infty(M)$, the distribution~\eqref{tilde_symp} is in $(\sigma_c \circ E_\phi)^{\otimes 2}\cE'_W(M^2)$ because $\sigma_c$ is compactly supported and its wave-front set is contained in $W_2$, see the estimate~\eqref{WF_kernel_symp}. Making use of eq.~\eqref{causal_prop/symp_kernel}, we can rewrite $\tilde{\sigma}_\phi$ as
				\begin{equation}\label{tilde_symp_alt}
					\tilde{\sigma}_\phi (x_1,x_2) = (\sigma_c \circ E_\phi \circ \sigma_c)(x_1,x_2).
			\end{equation}
			Then, the map $C^\infty(M) \ni \phi \mapsto \tilde{\sigma}_\phi$ is indeed an extension of $S \ni \phi \mapsto \sigma_\phi$ as can be verified directly using eq.~\eqref{causal_prop_inv_symp_kernel_sol} and eq.~\eqref{sympl_form/symp_kernel_sol}. In fact, for any $\phi \in S$ and for $u_1, u_2 \in T_\phi S$, it holds
				\begin{equation*}
					\tilde{\sigma}_\phi(u_1,u_2) =(\sigma_c \circ E_\phi \circ \sigma_c)(u_1,u_2) = \sigma_c(u_1, u_2) = \int_\Sigma u_1(x) \overleftrightarrow{\partial_n} u_2(x) d\Sigma(x) = \sigma_\phi(u_1, u_2).
				\end{equation*}
			To conclude the proof of the on-shell $W$-smoothness of  $S \ni \phi \mapsto \sigma_\phi$, we need to check that $\tilde{\sigma}_\phi$ satisfies the properties~\ref{W1},~\ref{W2} in def.~\ref{def_smooth_on_tens}.\\
			In order to prove~\ref{W1}, we need to compute the $\nu$-th Gateaux derivative of $\tilde{\sigma}_\phi$. By distributing the variational derivative on the factors in the right-hand side of eq.~\eqref{tilde_symp_alt}, it follows 
				\begin{equation}\label{term_var_der_tilde_symp}
					\frac{\delta^\nu \tilde{\sigma}_\phi (x_1,x_2)}{\delta \phi(y_1) \dots \delta \phi(y_\nu)} = \frac{\delta^\nu (\sigma_c \circ E_\phi \circ \sigma_c) (x_1,x_2)}{\delta \phi(y_1) \dots \delta \phi(y_\nu)}  = \int_{M^2} \sigma_c(x_1, x'_1) \frac{\delta^{\nu} E_\phi(x'_1,x'_2)}{\delta \phi(y_1) \dots \delta \phi(y_\nu)} \sigma_c(x'_2, x_2) dx'_1 dx'_2.
				\end{equation}
			By formula~\eqref{var_ders_causal_WF} and lemma~\ref{lemma_tech_X_W}, we know that the wave-front set of $\delta^\nu E_\phi / \delta \phi^\nu$ is contained in $W_{2+\nu}$. Since $\sigma_c$ is compactly supported and its wave-front set is contained in $W_2$, lemma~\ref{lemma_W_comp} implies that $\WF( \delta^\nu \tilde{\sigma}_\phi / \delta \phi^\nu) \subset W_{2+\nu}$. Because $\sigma_c$ is compactly supported and $\delta^\nu E_\phi(x_1,x_2) / \delta \phi^\nu(y_1, \dots, y_\nu)$ is compactly supported in $y_1, \dots, y_\nu$ (see (2) in prop~\ref{prop_var_ders_causal}), we conclude that $\delta^\nu \tilde{\sigma}_\phi / \delta \phi^\nu \in \cE'_W(M^{2+\nu})$, which is precisely the requirement~\ref{W1}.\\
			To prove~\ref{W2}, let $\bR \ni \epsilon \mapsto \phi(\epsilon) \in C^\infty(M)$ be smooth. We argue similarly as just done for~\ref{W1} starting again from eq.~\eqref{term_var_der_tilde_symp} and using in this case formula~\eqref{cont_var_ders_causal_WF} and thm.~\ref{theo_WF_horma}. It follows that $\delta^\nu \tilde{\sigma}_{\phi(\epsilon)} / \delta \phi^\nu$, viewed as a distribution in $\bR \times M^{2+\nu}$, has wave-front set contained in $\bR \times \{0\} \times W_{2+\nu}$, which is precisely the condition~\ref{W2}. This concludes the proof of $S\ni \phi \mapsto \sigma_\phi$ being an on-shell $W$-smooth $2$-form.\\
			We now argue that $\sigma$ is closed, i.e. $d\sigma = 0$. According to prop.~\ref{prop_der_smooth_on}, the exterior derivative $d$ is defined by anti-symmetrization of $\partial$. Thus, it is sufficient to show $\partial \sigma =0$. For our fixed cut-off function $c$, the off-shell extension $\widetilde{\partial \sigma}_\phi$, given by eq.~\eqref{tilde_diff}, does not depend on the choice of the extension $\tilde{\sigma}_\phi$. Therefore, we are free to chose as extension the distribution~\eqref{tilde_symp}. As a consequence of eq.~\eqref{causal_prop/symp_kernel}, it holds $\sigma_c \circ E_\phi \circ \sigma_c \circ E_\phi \circ \sigma_c = \sigma_c \circ E_\phi  \circ \sigma_c$. Then, applying the Leibniz rule, we obtain
				\begin{equation*}
					\begin{split}
						\frac{\delta (\sigma_c \circ E_\phi \circ \sigma_c)(x_1,x_2)}{\delta \phi(y)} &= \left( \sigma_c \circ E_\phi \circ \frac{\delta (\sigma_c \circ E_\phi \circ \sigma_c)}{\delta \phi(y)}\right)(x_1,x_2) +\\
						&\quad + \left( \frac{\delta (\sigma_c \circ E_\phi \circ \sigma_c)}{\delta \phi(y)} \circ E_\phi \circ \sigma_c\right)(x_1,x_2).
					\end{split}
				\end{equation*}
			By repeatedly applying the last equation, it holds
				\begin{equation}\label{partial_tilde_sigma}
					\begin{split}
						\widetilde{\partial \sigma}_\phi (x_1, x_2, x_3) &= \int_{M^3} \prod_{i=1}^3 (\sigma_c \circ E_\phi)(x_i,x'_i) \frac{\delta (\sigma_c \circ E_\phi \circ \sigma_c)(x'_2, x'_3)}{\delta \phi(x'_1)} dx'_1 dx'_2 dx'_3\\
						&= \int_M (\sigma_c \circ E_\phi)(x_1, x'_1) \frac{\delta (\sigma_c \circ E_\phi \circ \sigma_c)(x_2, x_3)}{\delta \phi(x'_1)} dx'_1 - \\
						&\quad - \int_M (\sigma_c \circ E_\phi)(x_1, x'_1) \left( \frac{\delta (\sigma_c \circ E_\phi \circ \sigma_c)}{\delta \phi(x'_1)}  \circ E_\phi \circ \sigma_c \right) (x_2, x_3) dx'_1 - \\
						&\quad - \int_M (\sigma_c \circ E_\phi)(x_1, x'_1) \left( \sigma_c \circ E_\phi \circ \frac{\delta (\sigma_c \circ E_\phi \circ \sigma_c)}{\delta \phi(x'_1)} \right) (x_2, x_3) dx'_1\\
						&=0,
					\end{split}
				\end{equation}
			as we wanted to prove. This concludes the proof.
		\end{proof}
			
 	\subsection{Almost-K\"{a}hler structure on $S$}\label{subsubsec_kahler_inf}
 		We define an almost-K\"{a}hler structure on $S$. This will be provided by a choice of pure Hadamard $2$-point function $\omega_\phi$ for each $\phi \in S$. By definition of a $2$-point function (see def.~\ref{def_hadamard_2-point}), $\omega_\phi$ decomposes into $\frac{1}{2} G_\phi + \frac{i}{2} E_\phi$ (cf.~\eqref{decomp_quasi-free_state}) where $\gls{G_phi}$ is a real-valued positive definite symmetric distribution. Thus, $\omega_\phi$ for $\phi$ smooth non-linear solution in $S$ is the analogue of the tensor given by $\omega^{ij}_x = \frac{1}{2} G^{ij}_x + \frac{i}{2} \sigma^{ij}_x$, where $x$ is a point in a finite-dimensional almost-K\"{a}hler manifold.
   Given any such $2$-point function $\omega_\phi$ for any $\phi \in S$, we define its action on a pair of smooth solutions $u_1, u_2 \in T_\phi S$ for the linearised Klein-Gordon equation by the ``symplectic smearing'', i.e. we set
		\begin{equation}\label{symp_2-point}
				\gls{omegaflat_phi}(u_1, u_2) := \int_{\Sigma \times \Sigma} u_1(z_1) \overleftrightarrow{\partial_{n}} \omega_\phi(z_1,z_2) \overleftrightarrow{\partial_{n}} u_2(z_2) d\Sigma(z_1) d\Sigma(z_2),
		\end{equation}
	where $\Sigma$ is a Cauchy surface.\\
	Formula~\eqref{symp_2-point} is actually well-defined as can be seen from the following argument. Any $u$ in $T_\phi S$ is a by definition a smooth $P_\phi$-solution. Therefore, the restriction to $\Sigma$ of $u$ and its normal derivative $\partial_n u$ are smooth functions on the compact surface $\Sigma$. We must show that the restrictions to $\Sigma \times \Sigma$ of the distribution $\omega_\phi(x_1, x_2)$ and its normal derivatives $\partial_n^{(x_1)} \omega_\phi(x_1, x_2)$, $\partial_n^{(x_2)} \omega_\phi(x_1, x_2)$, $\partial_n^{(x_1)} \partial_n^{(x_2)} \omega_\phi(x_1, x_2)$ are well-defined. This can be shown as follows. Since differential operators do not change the wave-front set of a distribution~\citep[8.1.11]{H83}, the wave-front sets of $\omega_\phi$ and its normal derivatives are contained in the set $\cC^\triangleright$, given by~\eqref{WF_Hadamard}. By definition, $\cC^\triangleright$ does not contain elements $(x_1,x_2; k_1,k_2)$ with $k_1, k_2$ time-like. On the other hand, the normal bundle of the Cauchy surface $\Sigma$ must contain only time-like co-vectors because $\Sigma$ is a space-like surface.  Thus, it follows from~\citep[thm 8.2.4]{H83} that the restrictions are well-defined.
	
	\paragraph{The analogies to the finite-dimensional case.}
	We now explain in detail the analogies to the finite-dimensional case. More precisely, we want to establish that $\omega^\flat_\phi$ is the analogue of the finite-dimensional Hermitian tensor
	\begin{equation*}
		(\omega_x)_{ij} := (\sigma_x)_{i\ell} \omega_x^{\ell k} (\sigma_x)_{kj} = -\frac{1}{2}(G_x)_{ij} + \frac{i}{2} (\sigma_x)_{ij}.
	\end{equation*}
	In the second equality, we used the fact $G_{ij} = - \sigma_{i\ell} G^{\ell k} \sigma_{k j}$ which is a consequence of the almost-K\"{a}hler structure.\\
	By definition of $\sigma_\phi$ (see~\eqref{symp_struct}), $\omega^\flat_\phi$ corresponds to $(\sigma_x)_{i\ell} \omega_x^{\ell k} (\sigma_x)_{k j}$. Furthermore, by the results obtained in sec.~\ref{subsubsec_sympl_inf}, it immediately follows that the imaginary part of $\omega^\flat_\phi$ is $\frac{1}{2}\sigma_\phi$. To establish the claimed analogy, we will show that the real part of $\omega^\flat_\phi$ is $-\frac{1}{2} \mu_\phi$, where $\mu_\phi$ is the inner product on $T_\phi S$ which is the inverse of $G_\phi$. In other words, $\mu_\phi$ corresponds to $(G_x)_{ij}$. This condition is equivalent to the almost-K\"{a}haler condition $J^2 =-1$ for $J^i{}_j = G^{i \ell} \sigma_{\ell j}$. So,
			\begin{equation}\label{state_decomp}
					\omega^\flat_\phi = - \frac{1}{2} \mu_\phi + \frac{i}{2} \sigma_\phi.
				\end{equation}	
	First, we give an equivalent description of $\omega^\flat_\phi$:
	\begin{lemma}\label{lemma_symp_2-pt_alt}
		The quantity $\omega^\flat_\phi$ can be written as
		\begin{equation}\label{symp_2-point_alt}
			\begin{split}
				\omega^\flat_\phi(u_1, u_2) &= (\sigma_c \circ \omega_\phi \circ \sigma_c)(u_1, u_2) \\
				&= \int_{M^4} u_1(x_1) \sigma_c(x_1,x'_1) \omega_\phi(x'_1, x'_2) \sigma_c(x'_2, x_2) u_2(x_2) dx_1 dx'_1 dx_2 dx'_2.
			\end{split}
		\end{equation}
	\end{lemma}
	\begin{proof}
		We begin by showing that 
			\begin{equation}\label{1_symp_smear}
				M \ni x_1 \mapsto \int_\Sigma \omega_\phi(x_1,z_2) \overleftrightarrow{\partial_{n}} u_2(z_2) d\Sigma(z_2)
			\end{equation}
		 is a well-defined smooth $P_\phi$-solution. Note that $P_\phi^{(x_1)} \omega_\phi(x_1,z_2)=0$ by definition. Thus, we need to show that the map~\eqref{1_symp_smear} is a well-defined smooth function. The normal bundle of the Cauchy surface $\Sigma$ contains only time-like covectors. As a consequence of~\citep[thm 8.2.4]{H83}, the distributions $\omega_\phi$, $\partial_n \omega_\phi$, which both have wave-front sets contained in $\cC^\triangleright$, can be restricted to $M \times \Sigma$. The wave-front sets of both the restrictions are bounded by
		\begin{equation*}
			\{ (x_1, x_2; k_1, k_2) \in \dot{T}^* (M \times \Sigma) : \exists t, \eta \in \bR, \mbox{ such that } (x_1, k_1) \sim ((t,x_2), -(\eta, k_2)), k_1 \in V^+ \}.
		\end{equation*}	
		Since $u_2|_\Sigma$ and $\partial_n u_2 |_\Sigma$ are smooth functions on the compact manifold $\Sigma$, it follows form thm.~\ref{theo_WF_horma} that the distribution $\int_\Sigma \omega_\phi(x_1,z_2) \overleftrightarrow{\partial_{n}} u_2(z_2) d\Sigma(z_2)$ is well-defined and that it has empty wave-front set, as we wanted to show.\\
	Next, we apply formula~\eqref{sympl_form/symp_kernel_sol} of lemma~\ref{lemma_kernel_symp} to the right-hand side of eq.~\eqref{symp_2-point} and we obtain
		\begin{equation}\label{symp_2-point_alt_inter}
			\omega^\flat_\phi(u_1, u_2) = \int_{M^2} \int_\Sigma  u_1(x_1) \sigma_c(x_1,x'_1) \omega_\phi(x'_1, z_2) \overleftrightarrow{\partial_{n}} u_2(z_2) dx_1 dx'_1 d\Sigma(z_2).
		\end{equation}
	Now, the map $M \ni x_2 \mapsto \int_{M} u(x_1) (\sigma_c \circ \omega_\phi)(x_1,x_2) dx_1$ is a smooth function as follows using the wave-front set calculus (thm.~\ref{theo_WF_horma}) and the wave-front sets of $\sigma_c$ (given by~\eqref{WF_kernel_symp}) and $\omega_\phi$ (given by the Hadamard condition). This smooth function is a $P_\phi$-solution because $P^{(x_2)} \omega_\phi(x_1, x_2)=0$ by definition. Applying formula~\eqref{sympl_form/symp_kernel_sol} to the right-hand side of eq.~\eqref{symp_2-point_alt_inter}, we obtain eq.~\eqref{symp_2-point_alt}.
	\end{proof}	
	\noindent
	Concerning the claim, it follows from lemma~\ref{lemma_symp_2-pt_alt} that we have
		\begin{equation}\label{mu_phi}
			\begin{split}
				\mu_\phi (u_1, u_2)&= -2 \Rea \omega^\flat (u_1,u_2)= -(\sigma_c \circ G_\phi \circ \sigma_c)(u_1,u_2)\\
				 &= - \int_{M^4} u_1(x_1) \sigma_c(x_1,x'_1) G_\phi(x'_1, x'_2) \sigma_c(x'_2, x_2) u_2(x_2) dx_1 dx'_1 dx_2 dx'_2,
			\end{split}
		\end{equation}
	Next, we prove the following two facts: (a) $\mu_\phi \in \vee^2_W T^*_\phi S$ is a real inner product, i.e. a real symmetric bilinear form on $T_\phi S$ which is positive definite, and (b) the distribution $-(\sigma_c \circ G_\phi \circ \sigma_c)(x,y) \in \cE'_W(M^2)$ is the ``inverse'' of $G_\phi(x,y)$ in the same sense as $\sigma_c(x,y)$ is the ``inverse'' of $E_\phi(x,y)$, namely for any $u \in T_\phi S$ it holds
		\begin{equation}\label{G_mu}
			u(x) =- \int_{M^2} G_\phi(x,y) (\sigma_c \circ G_\phi \circ \sigma_c)(y,z) u(z) dy dz.
		\end{equation}
	Proving (a) is quite straightforward. By construction, $\mu_\phi$ is real, symmetric and bilinear. The positive definite property follows from the fact that the $2$-point function $\omega_\phi$ is positive semidefinite\footnote{In particular, we use $|E_\phi(f,h)| \leq \left( G_\phi(f,f) G_\phi(h,h) \right)^{1/2}$ for any $f,h \in C^\infty_0(M)$ (cf. eq.~\eqref{pos_2-point}).} and the fact that the symplectic structure $\sigma_\phi$ is non-degenerate.\\
	The proof of (b) relies on the fact that $\omega_\phi$ is chosen to be pure. It is known (see e.g.~\citep{ashtekar1975quantum,KW91,W94}) that a pure $2$-point function induces a complex structure on (the completion of) $T_\phi S$ in the following way. The inequality~\eqref{pos_2-point} guarantees the existence and uniqueness of the continuous extension of the symplectic structure on the Hilbert completion of $T_\phi S$ with respect to the real inner product $\mu_\phi$. We denote this real Hilbert space by $\cH_\phi$. The Riesz lemma implies that there exists a unique operator $J_\phi$ on $\cH_\phi$, such that $\sigma_\phi(\hat{u}, \hat{v})= \mu_\phi(\hat{u},J_\phi \hat{v})$ for any $\hat{u}, \hat{v} \in \cH_\phi$. Because the $2$-point function $\omega_\phi$ is pure, it follows that $\ker J_\phi = \emptyset$ as shown e.g. in~\citep[Appendix A]{KW91}. Furthermore, it follows that $J_\phi$ satisfies $J_\phi^2= - \id$, $J_\phi^+ = - J_\phi$, where $(\cdot)^+$ denotes the Hilbert adjoint defined by $\mu_\phi$, and, consequently,
		\begin{equation}\label{J_completion}
			\sigma_\phi(J_\phi \hat{u}, \hat{v})=\mu_\phi(\hat{u},\hat{v}).
		\end{equation}
	Because $\ker J_\phi = \emptyset$, and because $\sigma_\phi$ is non-degenerate, $J_\phi$ is uniquely defined by eq.~\eqref{J_completion}. For any $u \in T_\phi S$, we can write
		\begin{equation}\label{J_phi_on_sol}
			J_\phi (u) = (G_\phi \circ \sigma_c)(u),
		\end{equation}			
	because
		\begin{equation*}
			\begin{split}
				\sigma_\phi((G_\phi \circ \sigma_c)(u), \hat{v}) &=\sigma_\phi((G_\phi \circ \sigma_c)(u), \lim_n v_n) =- \lim_n \sigma_\phi(v_n, (G_\phi \circ \sigma_c)(u)) = \lim_n \mu_\phi(u,v_n)\\
				&= \mu_\phi(u,\hat{v}),
			\end{split}	
		\end{equation*}
	for any smooth $u \in T_\phi S$, for any $\hat{v} \in \cH_\phi$, and for any sequence $\{v_n\}_{n \in \bN} \subset T_\phi S$ such that $\hat{v}=\lim_n v_n$. Since $J_\phi$ is the unique operator which satisfies eq.~\eqref{J_completion}, it follows that eq.~\eqref{J_phi_on_sol} holds for any $u \in T_\phi S$.\\
	Finally, because $J_\phi$ is anti-involutive and it maps $T_\phi S$ into $T_\phi S$, we have
		\begin{equation}\label{inverse_G}
			\id_{T_\phi S} = - (G_\phi \circ \sigma_c)^2,
		\end{equation}
	which implies eq.~\eqref{G_mu} as we wanted to prove.\\
	
	For later use, we state the following remark:
	\begin{rem}
		The result can be generalized replacing $T_\phi S$ by the space of smooth solutions of $P_\phi u =0$, where $\phi$ is now any arbitrary smooth function in $C^\infty(M)$ (not necessarily in $S$), and where $\omega_\phi$ now is a pure Hadamard $2$-point function corresponding to the operator $P_\phi$. In particular, for any $\phi \in C^\infty(M)$, it holds
		\begin{equation}\label{G_tilde_G_flat}
			-G_\phi \circ \sigma_c \circ G_\phi = -G_\phi \circ \sigma_c \circ G_\phi \circ \sigma_c \circ E_\phi =  E_\phi,
		\end{equation}
	where eq.~\eqref{inverse_G} and eq.~\eqref{causal_prop_inv_symp_kernel_sol} were used.
	\end{rem}
	
	\paragraph{The on-shell $W$-smoothness of the almost-K\"{a}hler section.}
	So far, our considerations have been for an arbitrary but fixed $\phi \in S$ and a corresponding $\omega_\phi^\flat$. What we will need is some information about the dependence of $\omega_\phi$ on $\phi$.\\
	According to our general framework, $\phi \mapsto \omega^\flat_\phi$ should be on-shell $W$-smooth. First of all, it is unclear a priori how to get such an on-shell $W$-smooth section. As we will see in a moment, it is sufficient for this purpose to find an assignment $C^\infty(M) \ni \phi \mapsto \omega_\phi$, where $\omega_\phi$ is a pure Hadamard $2$-point function with respect to $P_\phi$, such that for any $\nu$ the Gateaux derivative $\delta^\nu \omega_{\phi} (x_1, x_2) / \delta\phi(y_1) \cdots \delta\phi(y_\nu)$ is a well-defined distribution which is compactly supported in $y_1, \dots, y_\nu$ and satisfies the following conditions:
		\begin{itemize}
			\item It holds
					\begin{equation}\label{WF_o_2}
						\WF \left( \frac{\delta^\nu \omega_{\phi} (x_1, x_2) }{\delta\phi(y_1) \cdots \delta\phi(y_\nu)}\right) \subset W_{2 + \nu}.
					\end{equation}
			\item Let $\bR \ni \epsilon \mapsto \phi(\epsilon) \in C^\infty(M)$ be smooth. We can view $\delta^\nu \omega_{\phi(\epsilon)} (x_1, x_2) / \delta\phi(y_1) \cdots \delta\phi(y_\nu)$ as a distribution in $\epsilon,x_1,x_2, y_1, \dots, y_\nu$ and it holds
				\begin{equation}\label{WF_o_3}
					\WF \left( \frac{\delta^\nu \omega_{\phi(\epsilon)} (x_1, x_2) }{\delta\phi(y_1) \cdots \delta\phi(y_\nu)}\right) \subset \bR \times \{0\} \times W_{2 + \nu}.
				\end{equation}
		\end{itemize}
	It is not obvious that such assignment exists.\\
	
	Also, as we will see later, we need for our subsequent construction (in particular for prop.~\ref{prop_prod_smooth_on_W}) more stringent constraints on the dependence of $\omega_\phi$ on $\phi$. The conditions that will work are collected in the following definition.
	\begin{defi}\label{def_suit_omega}
		An assignment $C^\infty(M) \ni \phi \mapsto \omega_\phi$, where $\omega_\phi$ is a pure Hadamard $2$-point function with respect to $P_\phi$, is called {\em admissible} if for any $\nu$ the Gateaux derivative $\delta^\nu \omega_{\phi} (x_1, x_2) / \delta\phi(y_1) \cdots \delta\phi(y_\nu)$ is a well-defined distribution which is compactly supported in $y_1, \dots, y_\nu$ and satisfies the following conditions:
		\begin{enumerate}[label=($\omega$\arabic*), start=1]
			\item\label{o'_2} It holds
					\begin{equation}\label{WF_better}
							\WF \left( \frac{\delta^\nu \omega_\phi(x_1,x_2)}{\delta \phi(y_1) \cdots \delta \phi(y_\nu)} \right) \subset Z_{2+\nu},
						\end{equation}
					where the sets $Z_{2+\nu}$ are defined by
						\begin{equation}\label{Z}
							\gls{Z_nu} := \dot{T}^\ast M^{2+\nu} \backslash (C^{2;+}_{2+\nu} \cup C^{1;-}_{2+\nu}),
						\end{equation}
					and where $C^{i;\pm}_{2+\nu}$ are the subset of $T^* M^{2 +\nu}$ defined by
						\begin{equation}\label{aux_C_sets}
							\begin{split}
								C^{i;\pm}_{2+\nu} &:=\left\{ (x_1,x_2, y_1, \dots, y_\nu ; k_1, k_2, p_1, \dots, p_\nu) \in \dot{T}^\ast M^{2+\nu} : k_i \notin \overline{V}^\mp, p_r \in \overline{V}^\pm \right. \\
								&\qquad \qquad \qquad \qquad \qquad \qquad  \quad  \left.  \mbox { or } k_i \in \overline{V}^\pm, k_i \neq 0, \exists ! \; p_{r'} \notin \overline{V}, p_{r \neq r'} \in \overline{V}^\pm \right\}.						
							\end{split}
						\end{equation}
			\item\label{o'_3} Let $\bR \ni \epsilon \mapsto \phi(\epsilon) \in C^\infty(M)$ be smooth. We can view $\delta^\nu \omega_{\phi(\epsilon)} (x_1, x_2) / \delta\phi(y_1) \cdots \delta\phi(y_\nu)$ as a distribution in $\epsilon,x_1,x_2, y_1, \dots, y_\nu$ and it holds
				\begin{equation}\label{WF_better_s}
					\WF \left( \frac{\delta^\nu \omega_{\phi(\epsilon)} (x_1, x_2) }{\delta\phi(y_1) \cdots \delta\phi(y_\nu)}\right) \subset \bR \times \{0\} \times Z_{2+\nu}.
				\end{equation}
		\end{enumerate}
	\end{defi}
	\noindent
	These new conditions~\ref{o'_2},~\ref{o'_3} imply the previous ones because
		\begin{equation}\label{upper_bound_W}
			Z_{2+\nu} \subset W_{2 + \nu},
		\end{equation}
	as follows from the definitions of $Z_{2+\nu}$~\eqref{Z} and $W_{2+\nu}$~\eqref{W_set_def}.\\
	
	Of course, it is even less obvious that an admissible assignment $\phi \mapsto \omega_\phi$ exists. We will therefore provide one now. We construct a pure Hadamard $2$-point function $\omega_\phi$ for each $\phi \in C^\infty(M)$ using the well-known procedure of ``space-time deformation'' developed by Fulling, Narcowich, Sweeny and Wald~\citep{FSW78, FNW81}. For this, we pick a reference pure Hadamard $2$-point function $\omega_0$ for the free theory on the background $\phi=0$, i.e. with respect to the Klein-Gordon operator $P_0=\boxempty - m^2$. We could basically choose any pure Hadamard $2$-point function we want, but for the sake of being explicit, we take the ground state. As explained e.g. in~\citep[\S 7]{kay1978linear} (see also~\citep[sec. 3.4]{junker1996hadamard}), the $2$-point function of the ground state in an ultra-static space-time is given by
		\begin{equation}\label{ground_state}
			\omega_0(f_1,f_2) = -\frac{1}{2} \left(  (A^{\frac{1}{2}} - i \partial_{n})E_0(f_1), A^{-\frac{1}{2}} (A^{\frac{1}{2}} - i \partial_{n}) E_0(f_2) \right)_{L^2(\Sigma)},
		\end{equation}
	where $A$ is the square root of the unique self-adjoint extension of the operator $-\Delta^{(h)}+m^2$ on $\Sigma$ (see \citep{chernoff1973essential}), and where $\Delta^{(h)}$ is the Laplacian associated with the metric $h$ on $\Sigma$ as in eq.~\eqref{ultra_stat_metric}. It is well-known that this formula defines a pure Hadamard $2$-point function with respect to $P_0=\boxempty - m^2$, see e.g.~\citep[Corollary 3.16]{junker1996hadamard}. Next, we choose two Cauchy surfaces $\Sigma_-, \Sigma_+$ such that $\Sigma_-$ is in the past of $\Sigma_+$, and a smooth function $\phi_-$ such that $\phi_- = 0$ in the past of $\Sigma_-$ and $\phi_- = \phi$ in the future of $\Sigma_+$. We define a $2$-point function $\omega_-$ with respect to $P_{\phi_-} = \boxempty - m^2 - V''_{\phi_-}$ requiring that $\omega_- (x,y) := \omega_0(x,y)$ for $x,y$ in the past of $\Sigma_-$. Finally, we define a $2$-point function $\omega_\phi$ with respect to $P_\phi =  \boxempty - m^2 - V''_{\phi}$ by demanding that $\omega_\phi(x,y) := \omega_-(x,y)$ for $x,y$ in the future of $\Sigma_+$. Applying the results of~\citep{FSW78, FNW81}, it follows that $\omega_-$ and $\omega_\phi$ are Hadamard $2$-point functions. Furthermore, they are pure because $\omega_0$ is chosen pure. We can clearly perform this construction for any $\phi \in C^\infty(M)$. To make the construction completely canonical, we only need to specify how we choose $\phi_-$ for a given $\phi$. This can be done by introducing an arbitrary smooth cut-off function $\chi$ which is $1$ in the future of $\Sigma_+$ and $0$ in the past of $\Sigma_-$, and then setting $\phi_- := \chi \phi$.\\
	We present a different representation of the $2$-point function $\omega_\phi$ just constructed, which will be more efficient for computing the variational derivatives of $\omega_\phi$.
		\begin{lemma}\label{lemma_state}
			We choose four Cauchy surfaces $\Sigma_{\pm \pm}$ such that 
				\begin{equation}\label{causal_order}
					\Sigma_{--} \prec \Sigma_{-+} \prec \Sigma_- \prec \Sigma_+\prec\Sigma_{+-} \prec \Sigma_{++},
				\end{equation}
			where $\prec$ the ordering is understood in terms of the causal structure. We consider two smooth cut-off functions $c_\pm$ as in eq.~\eqref{kernel_symp} and such that $c_{\pm} =0$ in the future of $\Sigma_{\pm+}$ and $c_{\pm}=1$ in the past of $\Sigma_{\pm-}$. Let $\phi$ be an arbitrary smooth function. The $2$-point function $\omega_\phi(x_1,x_2)$ defined previously can be written in terms of $c_\pm$ as
				\begin{equation}\label{state}
					\omega_\phi(x_1,x_2) = \left( E_\phi \circ \sigma_{c_+}  \circ E_{\phi_-} \circ \sigma_{c_-}  \circ \omega_0 \circ \sigma_{c_-} \circ E_{\phi_-} \circ \sigma_{c_+} \circ E_\phi \right) (x_1,x_2)
				\end{equation}
			where $\sigma_{c_\pm}$ are the distributional kernels defined in eq.~\eqref{kernel_symp} respectively for $c_{\pm}$, and where $E_{\phi_-}$ is the causal propagator for the Klein-Gordon operator $P_{\phi_-}$. 
		\end{lemma}
		\begin{proof}
			We proceed by showing first that the $2$-point function $\omega_-$ can be written as
				\begin{equation}\label{minus_state}
					\omega_-(x_1,x_2) = \left( E_{\phi_-} \circ \sigma_{c_-} \circ \omega_0 \circ \sigma_{c_-} \circ E_{\phi_-} \right)(x_1,x_2).
				\end{equation}
			The support of $\sigma_{c_-}$ is contained in $K_- \times K_-$ where $K_-$ is a compact subset in $J^-(\Sigma_-) \backslash \Sigma_-$. In the past of $\Sigma_-$ it holds $P_{\phi_-} = P_0$ because $\phi_- = \chi \phi$ and because the smooth function $\chi$ vanishes in the past of $\Sigma_-$. Therefore, when the right-hand side of~\eqref{minus_state} is smeared with two test functions $f_1, f_2 \in C^\infty_0(M)$ supported in the past of $\Sigma_-$, we can replace $E_{\phi_-}$ with $E_0$. As a consequence of eq.~\eqref{causal_prop_inv_symp_kernel_sol} and the fact that $\omega_0$ is a bi-solution with respect to $P_0$, it necessarily holds
				\begin{equation*}
					\left( E_{\phi_-} \circ \sigma_{c_-} \circ \omega_0 \circ \sigma_{c_-} \circ E_{\phi_-} \right)(f_1,f_2) = \left( E_{0} \circ \sigma_{c_-} \circ \omega_0 \circ \sigma_{c_-} \circ E_{0} \right)(f_1,f_2)=\omega_0(f_1,f_2),
				\end{equation*}
			which is exactly what we have to show to prove eq.~\eqref{minus_state} since the $2$-point function $\omega_-$ is defined by the requirement $\omega_-(x,y) = \omega_0(x,y)$ for $x,y$ in the past of $\Sigma_-$.\\
			As a consequence of  eq.~\eqref{minus_state}, the right-hand side of eq.~\eqref{state} can be rewritten as
				\begin{equation}\label{state_minus_state}
					\left( E_\phi \circ \sigma_{c_+}  \circ \omega_- \circ \sigma_{c_+} \circ E_\phi \right)(x_1,x_2).
				\end{equation}
			We now proceed by showing that the distribution~\eqref{state_minus_state} coincides with $\omega_\phi$. The support of $\sigma_{c_+}$ is contained in $K_+ \times K_+$ where $K_+$ is a compact subset of $J^+(\Sigma_+)\backslash \Sigma_+$. In the  future of $\Sigma_+$ it holds $P_\phi = P_{\phi_-}$ because $\phi_- = \chi \phi$ and because the smooth function $\chi$ is equal $1$ in the future of $\Sigma_+$. When the distribution~\eqref{state_minus_state} is smeared with two test functions $f_1, f_2 \in C^\infty_0(M)$ supported in the future of $\Sigma_+$, we can replace $E_\phi$ with $E_{\phi_-}$ in~\eqref{state_minus_state}. Using again eq.~\eqref{causal_prop_inv_symp_kernel_sol} and the fact that $\omega_-$ is a bi-solution with respect to $P_{\phi_-}$, it necessarily holds
				\begin{equation*}
					\begin{split}
						&\left( E_\phi \circ \sigma_{c_+}  \circ E_{\phi_-} \circ \sigma_{c_-}  \circ \omega_0 \circ \sigma_{c_-} \circ E_{\phi_-} \circ \sigma_{c_+} \circ E_\phi \right)(f_1,f_2) =\\
						&\quad = \left( E_{\phi_-} \circ \sigma_{c_+}  \circ \omega_- \circ \sigma_{c_+} \circ E_{\phi_-} \right)(f_1,f_2) =\omega_-(f_1,f_2),
					\end{split}
				\end{equation*}
			which is exactly what we have to show to prove eq.~\eqref{state} because the $2$-point function $\omega_\phi$ is defined by the requirement $\omega_\phi(x,y) = \omega_-(x,y)$ for $x,y$ in the future of $\Sigma_+$. This concludes the proof.
		\end{proof}
	\begin{rem}\label{rem_state}
		The definition of $\omega_\phi$ depends only on the choice of the Cauchy surfaces $\Sigma_\pm$ and the cut-off function $\chi$. Therefore, eq.~\eqref{state} holds for any choice of the four Cauchy surface $\Sigma_{\pm \pm}$, as long as the causal ordering~\eqref{causal_order} holds, and for any choice of the cut-off $c_\pm$ as in the statement of lemma~\ref{lemma_state}.
	\end{rem}
	\noindent
	Using the representation provide by~\eqref{state}, we can prove the following result.
		\begin{prop}\label{prop_exists_suit_omega}
			Let $C^\infty(M) \ni \phi \mapsto \omega_\phi$ be the assignment given by the pure Hadamard $2$-point functions $\omega_\phi$ as in~\eqref{state}. This assignment is admissible in the sense of def.~\ref{def_suit_omega}.
		\end{prop}
				\begin{proof}
					To prove that $\phi \mapsto \omega_\phi$ is admissible, we compute the $\nu$-th Gateaux derivatives of $\omega_\phi$ by distributing the functional derivatives over the various factors on the right side of~\eqref{state}. The key advantage of formula~\eqref{state} is that the only places where $\phi$ occurs are in the causal propagators $E_\phi$ or $E_{\phi_-}$. Thus, we obtain that $\delta^\nu \omega_\phi / \delta \phi^\nu$ is a linear combination of terms in the form
				\begin{equation}\label{term_var_der_state}
					\begin{split}
						&\left(\frac{\delta^{|N_1|} E_\phi}{\delta \phi^{|N_1|}(\{y_{r} \}_{r \in N_1})} \circ \sigma_{c_+} \circ \frac{\delta^{|N_2|} E_{\phi_-}}{\delta \phi^{|N_2|}(\{y_{r} \}_{r \in N_2})} \circ \sigma_{c_-} \circ \omega_0 \circ \right. \\
						&\qquad \left. \circ \sigma_{c_-} \circ \frac{\delta^{|N_3|} E_{\phi_-}}{\delta \phi^{|N_3|}(\{y_{r} \}_{r \in N_3})} \circ \sigma_{c_+} \circ  \frac{\delta^{|N_4|} E_{\phi}}{\delta \phi^{|N_4|}(\{y_{r} \}_{r \in N_4})} \right)(x_1,x_2),
					\end{split}
				\end{equation}
			where $N_1, N_2, N_3, N_4$ form a partition of $\{1, \dots, \nu\}$. It is not clear a priori that~\eqref{term_var_der_state}, and so also $\delta^\nu \omega_\phi / \delta \phi^\nu$, is well-defined, since compositions of distributions are involved. To show this, we proceed using the wave-front set calculus (thm.~\ref{theo_WF_horma}).\\
			It follows from prop.~\ref{prop_var_ders_causal}\footnote{Although prop.~\ref{prop_var_ders_causal} (and prop.~\ref{prop_var_ders_A/R}, on which the proof of prop.~\ref{prop_var_ders_causal} relies) only concerns $E_\phi$, analogous results, with obvious modifications, hold for $E_{\phi_-}$.} that $\delta^{|N_i|} E_{\phi} / \delta\phi^{|N_i|}$ and $\delta^{|N_i|} E_{\phi_-} / \delta\phi^{|N_i|}$ satisfy
				\begin{equation*}
					\WF \left( \frac{\delta^{|N_i|} E_{\phi} (x_1, x_2) }{\delta\phi^{|N_i|}(\{y_r\}_{r \in N_i})}\right), \WF \left( \frac{\delta^{|N_i|} E_{\phi_-} (x_1, x_2) }{\delta\phi^{|N_i|}(\{y_r\}_{r \in N_i})} \right) \subset X_{2 + |N_i|}.
				\end{equation*}
			Furthermore, the wave-front set of $\omega_0$ is contained in $\cC^\triangleright$, by the Hadamard condition, and the wave-front sets of the compactly supported distributions $\sigma_{c_\pm}$ are given by~\eqref{WF_kernel_symp}. Then, we apply thm.~\ref{theo_WF_horma} to get that each distribution~\eqref{term_var_der_state}, and, therefore, also $\delta^\nu \omega_\phi / \delta \phi^\nu$, is well-defined.\\
			We need to show that $\delta^\nu \omega_{\phi} (x_1, x_2) /\delta\phi(y_1) \cdots \delta\phi(y_\nu)$ is compactly supported in $y_1, \dots, y_\nu$. For this purpose,  we first recall that the distributions $\delta^{|N_i|} E_{\phi}/ \delta\phi^{|N_i|}$ and $\delta^{|N_i|} E_{\phi_-}/ \delta\phi^{|N_i|}$ are compactly supported in the $y$'s variables as proved in prop.~\ref{prop_var_ders_causal}. It follows that each term~\eqref{term_var_der_state} is compactly supported in $y_1, \dots, y_\nu$, and, therefore, the same holds for $\delta^\nu \omega_{\phi}/\delta\phi^\nu$ which is precisely what we needed to show.\\
			In order to prove that $\phi \mapsto \omega_\phi$ is admissible we need to show that the conditions~\ref{o'_2},~\ref{o'_3} of def.~\ref{def_suit_omega} are fulfilled.\\
			To prove that condition~\ref{o'_2} is satisfied, we notice that thm.~\ref{theo_WF_horma} does not only ensure that $\delta^\nu \omega_\phi / \delta \phi^\nu$ is well-defined, but even provides the following upper bound for the wave-front set of $\delta^\nu \omega_\phi / \delta \phi^\nu$:
					\begin{equation}\label{better_WF_state}
							\WF \left( \frac{\delta^\nu \omega_{\phi}(x_1,x_2)}{\delta\phi(y_1) \cdots \delta\phi(y_\nu)}\right) \subset Y_{2+\nu} \subset X_{2+\nu} \subset W_{2 + \nu},
						\end{equation}
					where the set $Y_{2+\nu}$ is defined by
						\begin{equation}\label{Y}
							\begin{split}
								Y_{2+\nu} &:=  \left\{ (x_1,x_2,y_1,\dots, y_\nu; k_1, k_2, p_1, \dots,  p_\nu) \in \dot{T}^\ast M^{\nu+2} : \right. \\
								&\quad p_1=p_1' + p_1'', \dots, p_\nu=p_\nu' + p_\nu'' \mbox{ and } \exists \pi \mbox{ permutation of } \{1, \dots, \nu\} \mbox{ such that }\\
								&\quad (x_1; k_1) \sim (y_{\pi(1)}; -p_{\pi(1)}') \mbox{ or } x_1=y_{\pi(1)}, k_1= -p_{\pi(1)}' \mbox{ or } k_1,p_{\pi(1)}'=0\\
								&\quad (y_{\pi(i)}; p_{\pi(i)}'') \sim (y_{\pi(i+1)}; -p_{\pi(i+1)}') \mbox{ or } y_{\pi(i)}=y_{\pi(i+1)}, p_{\pi(i)}''= -p_{\pi(i+1)}'\\
								&\qquad \qquad \qquad \qquad \qquad \qquad \qquad \quad \mbox{ or } p_{\pi(i)}'',p_{\pi(i+1)}'=0 \\
								&\quad (y_{\pi(\nu)}; p_{\pi(\nu)}'') \sim (x_2; -k_2) \mbox{ or } y_{\pi(\nu)}=x_2, p_{\pi(\nu)}''= -k_2 \mbox{ or } p_{\pi(\nu)}'', k_2 = 0\\
								&\quad \left. \exists q \in \{k_1, p''_1, \dots, p''_\nu\} \mbox{ s.t. } q \in V^+ \mbox{ or } q=0 \right\}.
							\end{split}
						\end{equation}
					The sets $X_{2 + \nu}$ and $Y_{2 + \nu}$ differ precisely by the last condition of formula~\eqref{Y}, which is a consequence of the fact that $\omega_0$ satisfies the Hadamard condition.\\
					As a consequence of the estimate~\eqref{better_WF_state}, to prove the wave-front set estimate~\eqref{WF_better} of condition~\ref{o'_2} it is sufficient to prove
						\begin{equation*}
							Y_{2+\nu} \subset Z_{2+\nu},
						\end{equation*}
					i.e. $Y_{2+\nu} \cap C^{2;+}_{2+\nu} = \emptyset$ and $Y_{2+\nu} \cap C^{1;-}_{2+\nu} = \emptyset$, as follows from the definition of the set $Z_{2+\nu}$~\eqref{Z}.\\
					We focus on the proof of $Y_{2+\nu} \cap C^{2;+}_{2+\nu} = \emptyset$. Let $(x_1,x_2,y_1,\dots, y_\nu; k_1, k_2, p_1, \dots,  p_\nu)$ be an element in $Y_{2+\nu}$, so there are decompositions $p_r=p'_r + p''_r$ for all $r$ and a permutation $\pi$ of $\{1, \dots, \nu\}$ satisfying the requirements of~\eqref{Y}. Because of the definition of $C^{2,+}_\nu$~\eqref{aux_C_sets}, it is sufficient to consider the following two cases: (a) $p_1, \dots, p_\nu \in \overline{V}^+$, and (b) there exists $s$ such that $p_{s}$ is space-like and $p_{r} \in \overline{V}^+$ for any $r \neq s$. For this two cases separately we will show that the configuration $(x_1,x_2,y_1,\dots, y_\nu; k_1, k_2, p_1, \dots,  p_\nu)$ cannot be in $C^{2;+}_{2+\nu}$, i.e. for (a) we will prove that $k_2 \in \overline{V}^-$, while for (b) we will verify that $k_2=0$ or $k_2 \notin \overline{V}^+$.
					\begin{enumerate}[label=(\alph*), start=1]
					 \item We first assume $p_j''=0$ for a certain $j$. Because $p_r \in \overline{V}^+$ for all $r$, it follows that $k_2 \in \overline{V}^-$. This implies that $(x_1,x_2,y_1,\dots, y_\nu; k_1, k_2, p_1, \dots,  p_\nu)$ cannot belong to $C^{2;+}_{2+\nu}$.\\
					 Next, we consider $p''_r \neq 0$ for all $r$. We show that if $k_2 \notin \overline{V}^-$, which is a necessary condition for the configuration to be in $C^{2;+}_{2+\nu}$ under the assumption (a), then we contradict the hypothesis $(x_1,x_2,y_1,\dots, y_\nu; k_1, k_2, p_1, \dots,  p_\nu) \in Y_{2 + \nu}$. In fact, $k_2 \notin \overline{V}^-$ implies that $k_1 \notin \overline{V}^+$, $p''_r \notin \overline{V}^+$ for any $r$, and, furthermore, $p'_r \notin \overline{V}^-$ for any $r$. However, this configuration is not compatible with the last condition in the definition of the set $Y_{2+\nu}$. Thus, there is no element of $Y_{2 + \nu}$ satisfying condition (a) that is also contained in $C^{2;+}_{2+\nu}$, as we wanted to prove.
					\item Again we first assume that there exists a certain $j$ such that $p_j''=0$. If the permutation $\pi$ is such that $\pi^{-1}(s) \geq \pi^{-1}(j)$, then similarly as before we can conclude that $k_2 \in \overline{V}^-$. On the other hand, if $\pi^{-1}(s) < \pi^{-1}(j)$, it follows that $k_2 =0$ or $k_2 \notin \overline{V}^+$, as one can directly check. In both cases $(x_1,x_2,y_1,\dots, y_\nu; k_1, k_2, p_1, \dots,  p_\nu)$ violates the requirements to be an element of $C^{2;+}_{2+\nu}$.\\
					Next, we consider $p''_r \neq 0$ for all $r$. We show that if $k_2 \in \overline{V}^+$ and $k_2 \neq 0$, which is a necessary condition for the configuration to be in $C^{2;+}_{2+\nu}$ under the assumption (b), then we contradict the hypothesis $(x_1,x_2,y_1,\dots, y_\nu; k_1, k_2, p_1, \dots,  p_\nu) \in Y_{2 + \nu}$. In fact, $k_2 \in \overline{V}^+$ and $k_2 \neq 0$ imply that we have
						\begin{equation*}
							p''_r \in \left\{ \begin{array}{ll}
								\overline{V}^- \backslash \{0\} & \pi^{-1}(r) \geq \pi^{-1}(s) \\
								T^\ast M \backslash \overline{V}^+ & \pi^{-1}(r) < \pi^{-1}(s)
								\end{array}\right. 
							\quad p'_r \in \left\{ \begin{array}{ll}
								\overline{V}^+ & \pi^{-1}(r) > \pi^{-1}(s) \\
								T^\ast M \backslash \overline{V}^- & \pi^{-1}(r) \leq \pi^{-1}(s)
							\end{array}\right.
						\end{equation*}
					and necessarily $k_1 \notin \overline{V}^+$. However, all the configurations above violate the last condition in the definition of $Y_{2+\nu}$. Thus, there is no element of $Y_{2 + \nu}$ satisfying condition (b) that is also contained in $C^{2;+}_{2+\nu}$, as we wanted to prove.
				\end{enumerate}
				This concludes the proof of $Y_{2 +\nu} \cap C^{2;+}_{2+\nu} = \emptyset$. A similar argument implies also $Y_{2 +\nu} \cap C^{1;-}_{2+\nu} = \emptyset$, and, as discussed above, this is enough to conclude that $\WF( \delta^\nu \omega_\phi / \delta \phi^\nu) \subset Z_{2+\nu}$ as we wanted to show.\\
				Finally, to prove the condition~\ref{o'_3}, let $\bR \ni \epsilon \mapsto \phi(\epsilon) \in C^\infty(M)$ be smooth and view $\delta^\nu \omega_{\phi(\epsilon)} / \delta \phi^\nu$ as a distribution in $\bR \times M^{2+\nu}$. This distribution is again a linear combination of terms in the form~\eqref{term_var_der_state}, with the only difference that $\phi$, $\phi_-$ are replaced by $\phi(\epsilon)$, $\phi_-(\epsilon)=\chi\phi(\epsilon)$. As proved in prop.~\ref{prop_var_ders_causal}, the following upper bounds hold:
				\begin{equation*}
					\WF \left( \frac{\delta^\nu E_{\phi(\epsilon)}(x_1,x_2)}{\delta\phi(y_1) \cdots \delta\phi(y_\nu)} \right), \WF \left( \frac{\delta^\nu E_{\phi_-(\epsilon)}(x_1,x_2)}{\delta\phi(y_1) \cdots \delta\phi(y_\nu)} \right) \subset \bR \times \{0\} \times X_{2 + \nu}.
				\end{equation*}
			Using again thm.~\ref{theo_WF_horma} and the fact that $Y_{2+\nu} \subset Z_{2+\nu}$, it follows
					\begin{equation}\label{better_WF_state_s}
						\WF \left( \frac{\delta^\nu \omega_{\phi(\epsilon)}(x_1,x_2)}{\delta\phi(y_1) \cdots \delta\phi(y_\nu)}\right) \subset \bR \times \{0\} \times Y_{2+\nu} \subset \bR \times \{0\} \times Z_{2+\nu},
					\end{equation}
			which is precisely what we need to show. This concludes the proof.
				\end{proof}
				
	After establishing that the class of admissible assignments $\phi \mapsto \omega_\phi$, in the sense of def.~\ref{def_suit_omega}, is not empty, we prove that $S \ni \phi \mapsto \omega^\flat_\phi$ provides an on-shell $W$-smooth K\"{a}hler structure.
		\begin{theo}\label{theo_var_controll_state}
			Let $\phi \mapsto \omega_\phi$ be an admissible assignment in the sense of def.~\ref{def_suit_omega}. For any $\phi \in S$, let $\omega_\phi^\flat$ as in eq.~\eqref{symp_2-point} for the $2$-point function $\omega_\phi$ of the assignment chosen. The map $S \ni \phi \mapsto \omega_\phi^\flat$ is an on-shell $W$-smooth section in $\bC \otimes \boxtimes_W^2 T^* S$, denoted by $\omega^\flat$.\\
			Furthermore, the map $S \ni \phi \mapsto \mu_\phi = 2 \Rea \omega^\flat$ is an on-shell $W$-smooth symmetric covariant section of rank $2$ which is also positive definite. In other words, $\mu$ is an on-shell $W$-smooth metric satisfying eq.~\eqref{state_decomp}, i.e.  $\omega^\flat$ is an on-shell $W$-smooth K\"{a}hler structure.
		\end{theo}
		\begin{proof}
			In order to prove that $S \ni \phi \mapsto \omega_\phi^\flat$ is an on-shell $W$-smooth section, we need to define a distribution $\tilde{\omega}_\phi^\flat \in (\sigma_c \circ E_\phi)^{\otimes 2} \cE'_W(M^2)$  for a fixed cut-off $c$ as in eq.~\eqref{kernel_symp} such that: (a) it is well-defined for any $\phi \in C^\infty(M)$, (b) it is an extension of $\omega^\flat_\phi$, in the sense of eq.~\eqref{extension_on_W_tens}, and (c) it satisfies conditions~\ref{W1},~\ref{W2} of def.~\ref{def_smooth_on_tens}. We set
				\begin{equation}\label{tilde_state_flat}
					\tilde{\omega}^\flat_\phi(x_1,x_2) := \int_{M^2} \prod_{j=1}^2 (\sigma_c \circ E_\phi)(x_j,x'_j)(\sigma_c \circ \omega_\phi \circ \sigma_c)(x'_1,x'_2) dx'_1 dx'_2.
				\end{equation}
			To prove (a), we note that both distributions $\omega_\phi$ and $E_\phi$ are defined for any $\phi \in C^\infty(M)$. By definition, $\sigma_c$ is a compactly supported distribution with wave-front set contained in $W_2$, see~\eqref{WF_kernel_symp}. Since $\omega_\phi$ is a Hadamard $2$-point function, its wave-front set is contained in $\cC^\triangleright \subset W_2$ by definition. We apply lemma~\ref{lemma_W_comp} and we conclude that $\sigma_c \circ \omega_\phi \circ \sigma_c$ is a well-defined distribution in $\cE'_W(M^2)$. Thus, we have $\tilde{\omega}_\phi^\flat \in (\sigma_c \circ E_\phi)^{\otimes 2} \cE'_W(M^2)$ as we wanted to prove.\\
			To show (b), we notice that $\tilde{\omega}^\flat_\phi$ can be rewritten as
				\begin{equation}\label{tilde_state_flat_alt}
					\tilde{\omega}^\flat_\phi = \sigma_c \circ \omega_\phi \circ \sigma_c,
				\end{equation}
			as a consequence of eq.~\eqref{causal_prop_inv_symp_kernel_sol} and the fact that the $2$-point function $\omega_\phi$ is a bi-solution with respect to the operator $P_\phi$ by definition. Comparing eq.~\eqref{tilde_state_flat_alt} with the equivalent description of $\omega^\flat_\phi$ given in lemma~\ref{lemma_symp_2-pt_alt} by eq.~\eqref{symp_2-point_alt}, we conclude that necessarily $\omega^\flat_\phi(u_1, u_2) = \tilde{\omega}^\flat_\phi(u_1, u_2)$ for any $\phi \in S$ and any $u_1, u_2 \in T_\phi S$. This is precisely the condition required for $\tilde{\omega}^\flat_\phi$ to be an extension of $\omega^\flat_\phi$ and thus we have verified the requirement (b).\\
			For (c), we need to prove that for any $\nu \in \bN$ the distribution $\delta^\nu (\sigma_c \circ \omega_\phi \circ \sigma_c) / \delta \phi^\nu$ is compactly supported and
				\begin{equation}\label{var_der_tilde_state_flat}
					\WF \left( \frac{\delta^\nu (\sigma_c \circ \omega_\phi \circ \sigma_c) (x_1,x_2)}{\delta\phi(y_1) \cdots \delta\phi(y_\nu)} \right) \subset W_{2 +\nu}, \quad \WF \left( \frac{\delta^\nu  (\sigma_c \circ \omega_{\phi(\epsilon)} \circ \sigma_c) (x_1,x_2)}{\delta\phi(y_1) \cdots \delta\phi(y_\nu)} \right) \subset \bR \times \{0\} \times W_{2 + \nu},
				\end{equation}
			for any $\phi \in C^\infty(M)$ and any $\bR \ni \epsilon \mapsto \phi(\epsilon) \in C^\infty(M)$ smooth. For this purpose, it is sufficient that the assignment $C^\infty(M) \ni \phi \mapsto \omega_\phi$ is such that, for any $\nu$, $\delta^\nu \omega_{\phi} (x_1, x_2) / \delta\phi(y_1) \cdots \delta\phi(y_\nu)$ is a well-defined distribution which is compactly supported in $y_1, \dots, y_\nu$ and
				\begin{equation}\label{state_WF_cond}
					\WF \left( \frac{\delta^\nu \omega_{\phi} (x_1, x_2) }{\delta\phi(y_1) \cdots \delta\phi(y_\nu)}\right) \subset W_{2 + \nu}, \quad \WF \left( \frac{\delta^\nu \omega_{\phi(\epsilon)} (x_1, x_2) }{\delta\phi(y_1) \cdots \delta\phi(y_\nu)}\right) \subset \bR \times \{0\} \times W_{2 + \nu}.
				\end{equation}
			These conditions on $\phi \mapsto \omega_\phi$ are sufficient for our purpose: $\delta^\nu (\sigma_c \circ \omega_\phi \circ \sigma_c) / \delta \phi^\nu$ is compactly supported because of the support properties of the distributions involved, whereas the requirements~\eqref{var_der_tilde_state_flat} can be obtained from~\eqref{state_WF_cond} using the fact that the wave-front set of $\sigma_c$ is contained in $W_2$ and by applying lemma~\ref{lemma_W_comp} (for the first estimate) or the more general thm.~\ref{theo_WF_horma} (for the second estimate).\\
			As follows form eq.~\eqref{upper_bound_W}, any admissible assignment $C^\infty(M) \ni \phi \mapsto \omega_\phi$, in the sense of def.~\ref{def_suit_omega}, satisfies all the sufficient requirements above. Thus, we proved that the section $S \ni \phi \mapsto \omega_\phi^\flat$ is on-shell $W$-smooth.\\
			Finally, the on-shell $W$-smoothness of $S \ni \phi \mapsto \mu_\phi$ and the decomposition $\omega^\flat = -\frac{1}{2} \mu + \frac{i}{2} \sigma$ are straightforward consequences of eq.~\eqref{state_decomp}, eq.~\eqref{mu_phi} and the fact that both the sections $S \ni \phi \mapsto \omega^\flat_\phi$ (just proved) and $S \ni \phi \mapsto \sigma_\phi$ (thm.~\ref{theo_W_symp}) are on-shell $W$-smooth. This concludes the proof.
		\end{proof}
	
\section{The algebra structure of the on-shell $W$-smooth sections on $\cW$.}\label{subsec_algebra_W_smooth_sec}
	In the previous sections, we have introduced the notion of on-shell $W$-smooth sections. In particular, we defined the space of on-shell $W$-smooth sections on the bundle $\cW$,  $C^\infty_W(S,\cW)$, and the space of on-shell $W$-smooth forms with values in $\cW$, $\Omega_W(S,\cW)$, which replace the spaces $C^\infty(S,\cW)$ and $\Omega(S,\cW)$ in the finite-dimensional context. We provided three important concrete examples of on-shell $W$-smooth sections, namely the symplectic form $\sigma$, the almost-K\"{a}hler structure $\omega^\flat$ and the associated metric $\mu$. To guarantee the on-shell $W$-smoothness, we considered $\omega^\flat$ constructed from an admissible assignment $C^\infty(M) \ni \phi \mapsto \omega_\phi$ in the sense of def.~\ref{def_suit_omega}.\\
	Proceeding along the lines of the Fedosov quantization scheme, we now provide an algebra structure for $C_W^\infty(S, \cW)$, the formal Wick product, and then we will extend it to $\Omega_W(S,\cW)$ in a canonical way. The product is defined fiberwise, just as in the finite-dimensional case (see formula~\eqref{fiber_prod_fin}), making use of the almost-K\"{a}hler structure. However, in the infinite-dimensional case it is not evident a priori that the product of on-shell $W$-smooth sections defines again an on-shell $W$-smooth section. It is the purpose of this section to shown that.\\
	
	In the finite-dimensional case, for each $x \in S$, with $S$ an almost-K\"{a}hler manifold, the algebraic structure of $\cW_x$ is provided by the product $\bullet_x$ defined as in eq.~\eqref{star_co} using the complex matrix $\omega_x^{ij}$, i.e. the value at $x$ of the complex tensor field $\omega$. In the infinite-dimensional setting, for each non-linear solution $\phi \in S$, the product $\bullet_\phi$ on $\cW_\phi$ is defined as in eq.~\eqref{wick_prod_per} using the pure Hadamard $2$-point function $\omega_\phi(x_1,x_2)$. We assume that $\omega_\phi$ comes from an admissible assignment $C^\infty(M) \ni \phi \mapsto \omega_\phi$, which also gives an almost-K\"{a}hler structure $\omega^\flat$. The same argument we presented to prove the well-definiteness for the product of def.~\ref{defi_free_wick_per}, see in particular the discussion after lemma~\ref{lemma_W_comp}, applies in each fiber $\cW_\phi$ for the product $\bullet_\phi$. As already mentioned, we induce a product for smooth sections in $\cW$ from the product on the fibers. More precisely, for any $t,s \in C^\infty_W(S,\cW)$ and for any $\phi \in S$ we define
		\begin{equation}\label{wick_prod_inf_sec}
				\gls{t_bullet_s_phi} := t_\phi \bullet_\phi s_\phi.
		\end{equation}
	What is not immediately evident is that $S \ni \phi \mapsto (t \bullet s)_\phi$ is on-shell $W$-smooth. The proof of this claim relies on the fact that the pure Hadamard $2$-point function $\omega_\phi$ for $\phi \in S$ comes from an admissible assignment.
		\begin{prop}\label{prop_prod_smooth_on_W}
			Let $C^\infty(M) \ni \phi \mapsto \omega_\phi$ be an admissible assignment in the sense of def.~\ref{def_suit_omega}. Then, the corresponding fiberwise product $\bullet$ endows $C^\infty_W(S,\cW)$ with the structure of an associative algebra. More explicitly, let $t$ and $s$ be two on-shell $W$-smooth sections on $\cW$, then the map
				\begin{equation*}
					S \ni \phi \mapsto (t \bullet s)_\phi \in \cW_\phi,
				\end{equation*}
			is an on-shell $W$-smooth section on $\cW$.
		\end{prop}
		\begin{proof}
			Let $t$ and $s$ be two on-shell $W$-smooth sections in $C^\infty_W(S, \cW)$ homogeneous in the symmetric degree $\deg_s$ and in the formal degree $\deg_\hbar$, with $\deg_s t = n$ and $\deg_s s =m$. The product $(t \bullet s)_\phi$ is given by the sequence $((t \bullet s)^0_\phi, (t \bullet s)^1_\phi, \dots)$ where, by definition, $(t \bullet s)^j_\phi$ is
				\begin{equation}\label{bullet_prod}
					\begin{split}
						(t \bullet s)^j_\phi (x_{1}, \dots, x_j) = &\hbar^k \sC_{n,m,k} \bP^+ \int_{M^{2k}} t_\phi(z_1, \dots, z_k, x_1, \dots, x_{n-k}) \left( \prod_{\ell=1}^{k} \omega_\phi(z_\ell, z'_\ell) \right) \times \\
						&\quad \times s_\phi(z'_1, \dots, z'_k, x_{n-k+1}, \dots, x_j) \prod_{\ell=1}^k dz_\ell dz'_\ell
					\end{split}
				\end{equation}
			if $j=m+n-2k$ for some $k \leq m,n$, and $(t \bullet s)^j_\phi=0$ otherwise. By $\bP^+$ we mean that a symmetrization acts on the free variables $x_1, \dots x_j$. In the formula above, $\sC_{n,m,k} = \frac{n!m!}{k! (n-k)! (m-k)!}$ is the same combinatorial factor appearing in eq.~\eqref{star_co} and in eq.~\eqref{wick_prod_per}. Note that, by abuse of notation, we identify an equivalence class in $\vee_W^{\bullet} T_\phi^* S = \bP^+ \cE_W'(M^{\bullet})/P_\phi \bP^+ \cE_W'(M^{\bullet})$ with one of its representative in $\bP^+ \cE_W'(M^{\bullet})$. The equivalence classes corresponding to~\eqref{bullet_prod} do not depend on the choice of representative for $t$ and $s$ because $\omega_\phi$ is a bi-solution with respect to $P_\phi$.\\
			To prove that the product $\bullet$ preserves the on-shell $W$-smoothness, it is sufficient to show that for any $j$ the map $S \ni \phi \mapsto (t \bullet s)^j_\phi$ is an on-shell $W$-smooth section with rank $j$ for any $t, s$ on-shell $W$-smooth sections homogeneous in $\deg_s$ and in $\deg_\hbar$. In fact, we can extend the result to on-shell $W$-smooth sections not necessarily homogeneous exploiting the $\Deg$-filtration and the fact that each section homogeneous in $\Deg$ is a finite collection of terms homogeneous in $\deg_s$ and in $\deg_\hbar$ (see the discussion at the end of section~\ref{subsec_manifold_inf}). We consider for the rest of the proof $j= m+n-2k$, where $k \leq m,n$, because otherwise $(t \bullet s)^j_\phi$ vanishes by construction and thus it is trivially an on-shell $W$-smooth section.\\
			Let us fix a cut-off function $c$ as in eq.~\eqref{kernel_symp}. To prove that $(t \bullet s)^j_\phi$ is an on-shell $W$-smooth section, we need to provide for any $\phi \in C^\infty(M)$ a distribution $\widetilde{(t \bullet s)}{}^j_\phi$ such that (a) it belongs to $(\sigma_c \circ E_\phi)^{\otimes j} \circ \cE'_W(M^j)$, (b) it is an extension of $(t \bullet s)^j_\phi$ (up to a factor $\hbar^{k+\deg_\hbar t+ \deg_\hbar s}$) in the sense of eq.~\eqref{extension_on_W_tens}, and (c) it satisfies the requirements~\ref{W1},~\ref{W2} of def.~\ref{def_smooth_on_tens}.\\
			For any $\phi \in C^\infty(M)$, $\omega_\phi$ satisfies the conditions~\ref{o'_2},~\ref{o'_3} of def.~\ref{def_suit_omega} by hypothesis. Let $\tilde{t}_\phi \in (\sigma_c \circ E_\phi)^{\otimes n} \cE'_W(M^n)$ and $\tilde{s}_\phi \in (\sigma_c \circ E_\phi)^{\otimes m} \cE'_W(M^m)$ be two extensions of $t, s$ (up to factors $\hbar^{\deg_\hbar t}$ and $\hbar^{\deg_\hbar s}$ respectively), consequently $\tilde{t}_\phi$ and $\tilde{s}_\phi$ satisfy the requirements~\ref{W1},~\ref{W2} of def.~\ref{def_smooth_on_tens}. We define $\widetilde{(t \bullet s)}{}^j_\phi$ by
				\begin{equation}\label{tilde_prod}
					\begin{split}
						\widetilde{(t \bullet s)}{}^j_\phi := \sC_{n,m,k} \bP^+ &\int_{M^{2k}} \tilde{t}_\phi(z_1, \dots, z_k, x_1, \dots, x_{n-k}) \left( \prod_{\ell=1}^{k} \omega_\phi(z_\ell, z'_\ell) \right) \times \\
						&\qquad \times \tilde{s}_\phi(z'_1, \dots, z'_k, x_{n-k+1}, \dots, x_N) \prod_{\ell=1}^k dz_\ell dz'_\ell.
					\end{split}
				\end{equation}
			To prove (a), we first note that $\WF(\omega_\phi) = \cC^\triangleright \subset W_2$ by definition. Then, as a consequence of the hypotheses on $\tilde{t}_\phi, \tilde{s}_\phi$ we can apply lemma~\ref{lemma_W_comp} and we have that formula~\eqref{tilde_prod} is a well-defined distribution in $(\sigma_c \circ E_\phi)^{\otimes j} \circ \cE'_W(M^{j})$.\\
			In order to show (b), let $\phi \in S$ and let $u_1, \dots, u_j \in T_\phi S$. Because $u_1, \dots, u_j$ are $P_\phi$-solutions and $\omega_\phi$ is a bi-solution with respect to $P_\phi$, it holds that $\widetilde{(t \bullet s)}{}^j_\phi(u_1, \dots, u_j)$ does not depend on the choice of the extensions $\tilde{t}_\phi$, $\tilde{s}_\phi$. Furthermore, $\widetilde{(t \bullet s)}{}^j_\phi$ is an extension of $(t \bullet s)^j_\phi$ (up to a factor $\hbar^{k + \deg_\hbar t + \deg_\hbar s}$) since $\tilde{t}_\phi$ and $\tilde{s}_\phi$ are extensions of $t$ and $s$ (up to factors $\hbar^{\deg_\hbar t}$ and $\hbar^{\deg_\hbar s}$ respectively).\\
			To conclude that $(t \bullet s)^j$ is indeed an on-shell $W$-smooth section, we need to check (c), i.e. that $\widetilde{(t \bullet s)}{}^j_\phi$ satisfies the requirements~\ref{W1},~\ref{W2} of def.~\ref{def_smooth_on_tens}. We first compute the $\nu$-th Gateaux derivative of $\widetilde{(t \bullet s)}{}^j_\phi$ by distributing the variational derivatives on the factors in the right-hand side of~\eqref{tilde_prod}. It holds that $\delta^\nu \widetilde{(t \bullet s)}{}^j_\phi(x_1, \dots, x_j) / \delta \phi(y_1) \dots \delta \phi(y_\nu)$ is a finite sum of terms in the form
				\begin{equation}\label{term_var_tilde_prod_field}
					\begin{split}
						&\int_{M^{2k}} \frac{\delta^{|N_t|} \tilde{t}_\phi (z_1, \dots, z_k, \{x_{i \leq n-k}\})}{\delta \phi^{|N_t|} (\{y_{r \in N_t} \})}  \times \\
						&\qquad \qquad \qquad \times \left( \prod_{\ell =1}^k \frac{\delta^{|N_\ell|} \omega_\phi(z_\ell,z'_\ell)}{\delta \phi^{|N_\ell|} (\{y_{r\in N_\ell} \})}  \right)  \frac{\delta^{|N_s|} \tilde{s}_\phi (z'_1, \dots, z'_k, \{x_{i > n-k}\})}{\delta \phi^{|N_s|} (\{y_{r \in N_s} \})} \prod_{\ell=1}^k dz_\ell dz'_\ell,
					\end{split}
				\end{equation}
			where $N_t, N_s, N_1, \dots, N_k$ is a partition of $\{1, \dots, \nu\}$. We show that each distribution~\eqref{term_var_tilde_prod_field} is a well-defined compactly supported distribution. This is sufficient to conclude that  $\delta^\nu \widetilde{(t \bullet s)}{}^j_\phi / \delta \phi^\nu$, which is a finite sum of terms as~\eqref{term_var_tilde_prod_field}, is also a well-defined compactly supported distribution.\\
			We consider first the auxiliary distribution $\Theta_\phi$ on $M^{m + \nu - |N_t|}$ defined by
				\begin{equation*}
					\begin{split}
						&\Theta_\phi(z_1, \dots, z_k, \{x_{i > n-k}\}, \{y_{r \in N^c_t}\}) := \\
						&\quad = \int_{M^{k}} \left( \prod_{\ell =1}^k \frac{\delta^{|N_\ell|} \omega_\phi(z_\ell,z'_\ell)}{\delta \phi^{|N_\ell|} (\{y_{r \in N_\ell} \})}  \right)  \frac{\delta^{|N_s|} \tilde{s}_\phi (z'_1, \dots, z'_k, \{x_{i > n-k}\})}{\delta \phi^{|N_s|} (\{y_{r \in N_s} \})}   dz'_1 \dots dz'_k.
					\end{split}
				\end{equation*}
			By hypothesis, $\omega_\phi$ is an admissible assignment. Therefore, $\delta^{|N_\ell|} \omega_\phi(z_\ell,z'_\ell) / \delta \phi^{|N_\ell|}(y_{r \in N_\ell})$ is compactly supported in $(y_r)_{r \in N_i}$, and its wave-front set is bounded by $Z_{2+|N_\ell|}$ (this is estimate~\eqref{WF_better} of condition~\ref{o'_2}) which is contained in $W_{2+|N_\ell|}$ (see~\eqref{upper_bound_W}). By hypothesis, $\tilde{s}_\phi$ satisfies~\ref{W1}, therefore $\delta^{|N_s|} \tilde{s}_\phi / \delta \phi^{|N_s|}$ is a compactly supported distribution with wave-front set in $W_{m+ |N_s|}$.  Then, applying lemma~\ref{lemma_W_comp}, we have that $\Theta_\phi$ is a well-defined distribution which is compactly supported in $(x_{i > n-k})$ and in $(y_{r \in N^c_t})$ as follows from the support properties of the distributions involved.\\
			In order to prove that each~\eqref{term_var_tilde_prod_field} is a well-defined distribution, it is sufficient to show that the composition of $\delta^{|N_t|} \tilde{t}_\phi / \delta \phi^{|N_t|}$ with $\Theta_\phi$ exists in the sense of the wave-front set calculus (thm.~\ref{theo_WF_horma}), i.e. we need to prove that the multiplication condition~\eqref{mult_cond} and the integration condition~\eqref{int_cond} hold.\\
			The integration condition is satisfied due to the support properties of $\delta^{|N_t|} \tilde{t}_\phi / \delta \phi^{|N_t|}$ and $\Theta_\phi$.\\
			In this case, the multiplication condition reads
				\begin{equation}\label{mult_cond_der_prod}
					\begin{split}
						&\WF \left(\frac{\delta^{|N_t|} \tilde{t}_\phi (z_1, \dots, z_k, \{x_{i \leq n-k}\}) }{\delta \phi^{|N_t|} (\{y_{r \in N_t}\})}  \right)_{z_1, \dots, z_k} \cap \WF'(\Theta_\phi(z_1, \dots, z_k, \{x_{i > n-k}\}, \{y_{r \in N^c_t}\}))_{z_1, \dots, z_k}  = \\
						&\quad =\emptyset.
					\end{split}
				\end{equation}
			Using the wave-front set calculus (thm.~\ref{theo_WF_horma}) and the fact that $\WF (\delta^{|N_s|} \tilde{s}_\phi / \delta \phi^{|N_s|})$ is contained in $W_{m + |N_s|}$, we can estimate the second set appearing in the left-hand side of eq.~\eqref{mult_cond_der_prod} by
				\begin{equation}\label{theta_est}
					\begin{split}
						\WF'(\Theta_\phi)_{z_1, \dots, z_k} \subset &\left\{ (z_1, \dots, z_k; - q_1, \dots, - q_k)  \in T^\ast M^k_{\{0\}} : \exists (z'_1, \dots, z'_k;q'_1, \dots, q'_k) \in W_k \right.\\
						&\qquad \left.  (z_\ell, z'_\ell, (y_{r \in N_\ell}); q_\ell, - q'_\ell, (0, \dots, 0) ) \in \WF \left( \frac{\delta^{|N_\ell|} \omega_\phi(z_\ell, z'_\ell)}{\delta\phi^{|N_\ell|}(\{y_{r \in N_\ell} \})}\right) \right\},
					\end{split}
				\end{equation}
			Since $\delta^{|N_\ell|} \omega_\phi / \delta \phi^{|N_\ell|}$ is estimated by $Z_{2+|N_\ell|} = \dot{T}^* M^{2+ |N_\ell|} \backslash (C^{2;+}_{2+|N_\ell|} \cup C^{1;-}_{2+|N_\ell|})$, it holds that
				\begin{equation}\label{failure_of_W_omega}
					(z_\ell, z'_\ell, (y_{r \in N_\ell}); q_\ell, - q'_\ell, (0, \dots, 0) ) \in \WF \left( \frac{\delta^{|N_\ell|} \omega_\phi(z_\ell, z'_\ell)}{\delta\phi^{|N_\ell|}(\{y_{r \in N_\ell} \})}\right) \Rightarrow q_\ell \in \overline{V}^+ \backslash 0, - q'_\ell \in \overline{V}^- \backslash 0.
				\end{equation}
			By definition, the set $W_k$ does not contain elements $(z'_1, \dots, z'_k;q'_1, \dots, q'_k)$ in $T^* M^k$ with all the covectors $q'_1, \dots, q'_k$ which are future directed. Thus, we have $\WF'(\Theta_\phi)_{z_1, \dots, z_k}= \emptyset$. This implies that the multiplication condition is verified and so the distribution~\eqref{term_var_tilde_prod_field} is well-defined. Furthermore, by the support properties of $\delta^{|N_t|} \tilde{t}_\phi / \delta \phi^{|N_t|}$ and $\Theta_\phi$, it follows that each~\eqref{term_var_tilde_prod_field} is compactly supported as we wanted to prove.\\
			Next, we proceed by showing that each distribution~\eqref{term_var_tilde_prod_field} satisfies conditions~\ref{W1},~\ref{W2} given in def.~\ref{def_smooth_on_tens}. This is sufficient to ensure that $\delta^\nu \widetilde{(t \bullet s)}{}^j_\phi / \delta \phi^\nu$, which is a finite sum of terms in the form~\eqref{term_var_tilde_prod_field}, satisfies conditions~\ref{W1},~\ref{W2} as we needed to prove.\\
			To verify the condition~\ref{W1}, we need to show that the wave-front set of the distribution~\eqref{term_var_tilde_prod_field} is contained in $W_{j + \nu}$. Let $(x_1, \dots, x_j, y_1, \dots, y_\nu; k_1, \dots, k_j, p_1, \dots, p_\nu)$ be an element of the wave-front set of distribution~\eqref{term_var_tilde_prod_field}. The wave-front set calculus (thm.~\ref{theo_WF_horma}) implies that there must be 
				\begin{equation*}
					(z_1, \dots, z_k, z'_1, \dots, z'_k; q_1, \dots, q_k, q'_1, \dots, q'_k) \in T^\ast M^{2k},
				\end{equation*}
			such that
				 \begin{equation}\label{aux_covectors}
			 		\left\{ \begin{array}{l}
						(z_1, \dots, z_k, x_1, \dots, x_{n-k}, (y_{r \in N_t}); -q_1, \dots, - q_k, k_1, \dots, k_{n-k}, (p_{r \in N_t})) \in W_{n+|N_t|} \\
						\qquad \mbox{ or } q_1, \dots, q_k, k_1, \dots, k_{n-k}, p_{r \in N_t} =0\\
						(z_\ell, z'_\ell, (y_{r \in N_\ell}); q_\ell, -q'_\ell, (p_{r \in N_\ell})) \in Z_{2+|N_\ell|} \qquad \mbox{ for } \ell=1, \dots, k\\
						\qquad \mbox{ or } q_\ell, q'_\ell, p_{r \in N_\ell} =0\\
						(z'_1, \dots, z'_k, x_{n-k+1}, \dots, x_j, (y_{r \in N_s}); q'_1, \dots, q'_k, k_{n-k+1}, \dots, k_j, (p_{r \in N_s})) \in W_{m+|N_s|}\\
						\qquad \mbox{ or } q'_1, \dots, q'_k, k_{n-k+1}, \dots, k_{j}, p_{r \in N_s} =0\\
					\end{array} \right.
				\end{equation}
			We used the estimate~\eqref{WF_better} for the wave-front set of $\delta^{|N_\ell|} \omega_\phi / \delta \phi^{|N_\ell|}$ and the estimate~\eqref{WF_gateaux_t} for the wave-front sets of $\delta^{|N_t|} \tilde{t}_\phi / \delta \phi^{|N_t|}$ and $\delta^{|N_s|} \tilde{s}_\phi / \delta \phi^{|N_s|}$.\\
			Notice that $(x_1, \dots, x_j, y_1, \dots, y_\nu; k_1, \dots, k_j, p_1, \dots, p_\nu)$ is contained in $W_{j+\nu}$ as we need to prove, if we show that $k_1, \dots, k_j$, $p_1, \dots, p_\nu$ cannot be all causal future-directed or all causal past-directed except at most one covector which is space-like.\\
			We argue via reductio ad absurdum: we prove that if we assume that all the covectors $k_1, \dots, k_j$, $p_1, \dots, p_\nu$ are causal future-directed except at most one covector which is space-like, we contradict the hypotheses.\\
			We consider two cases separately: (a) all covectors $k_1, \dots, k_j, p_1, \dots, p_\nu$ belong to $\overline{V}^+$ except at most one $k_i$ or one $p_r$ with $r \in N_t \cup N_s$ which is space-like, or (b) there exists an $\ell'$ and a $r' \in N_{\ell'}$ such that $p_{r'}$ is space-like while all the remaining covectors $k_1, \dots, k_j$ and $p_r $ with $r \neq r'$ are in $\overline{V}^+$.
			\begin{enumerate}[label=\alph*), start=1]
				\item Since we assume $p_r \in \overline{V}^+$ for any $r \in N_\ell$ for any $\ell$, we have $q'_1, \dots, q'_k \in \overline{V}^+$ by definition of the sets $Z_{2 + \nu}$ (see~\eqref{Z} and~\eqref{aux_C_sets}). By the assumption (a), we have $k_{n-k+1}, \dots, k_j, p_{r \in N_s} \in \overline{V}^+$ except at most one covector which can be space-like. However, these configurations are incompatible with the conditions~\eqref{aux_covectors} because the co-vectors in $W_{m+ |N_s|}$ cannot be all causal future-directed except at most one space-like. Therefore, the assumption (a) is incompatible with the hypotheses as we wanted to prove.
				\item Since we assume $p_r \in \overline{V}^+$ for any $r \neq r'$, it follows by definition of the sets $Z_{2 + \nu}$ that $q'_{\ell}$ belongs to $\overline{V}^+$ for any $\ell \neq \ell'$, while we have $q'_{\ell'} \notin \overline{V}^-$ or $q'_{\ell'} = 0$. By the assumption (b), we have $k_{n-k+1}, \dots, k_j, p_{r \in N_s} \in \overline{V}^+$. As before, these configurations are incompatible with the conditions~\eqref{aux_covectors}, thus also the assumption (b) is incompatible with the hypotheses as we wanted to prove.
			\end{enumerate}
			Similarly, we can prove that $k_1, \dots, k_j, p_1, \dots, p_\nu$ cannot be all causal past-directed except at most one covector which is space-like. This is enough to conclude that condition~\ref{W1} holds for each distribution~\eqref{term_var_tilde_prod_field}, and consequently also for the distribution $\delta^\nu \widetilde{(t \bullet s)}{}^j_\phi / \delta \phi^\nu$.\\
			To prove~\ref{W2}, let $\bR \ni \epsilon \mapsto \phi(\epsilon) \in C^\infty(M)$ be smooth. The distribution $\delta^\nu \widetilde{(t \bullet s)}{}^j_{\phi(\epsilon)}/ \delta \phi^\nu$ can be expressed as a finite combination of terms in the form~\eqref{term_var_tilde_prod_field} with $\phi$ replaced by $\phi(\epsilon)$ everywhere. By hypothesis, $\omega_\phi$ satisfies condition~\ref{o'_3} in def.~\ref{def_suit_omega}, therefore the wave-front set $\delta^{|N_\ell|} \omega_{\phi(\epsilon)} / \delta\phi^{|N_\ell|}$ is bounded by $\bR \times \{0 \} \times  Z_{2+ |N_\ell|}$ (this is estimate~\eqref{WF_better_s} of~\ref{o'_3}). Furthermore, by hypothesis, $\tilde{t}_{\phi(\epsilon)}, \tilde{s}_{\phi(\epsilon)}$ satisfy condition~\ref{W2} in def.~\ref{def_smooth_on_tens}, therefore the wave-front sets of the distributions $\delta^{|N_t|} \tilde{t}_{\phi(\epsilon)} / \delta \phi^{|N_t|}$ and $\delta^{|N_s|} \tilde{s}_{\phi(\epsilon)} / \delta \phi^{|N_s|}$ are contained respectively in $\bR \times \{0\} \times W_{n+|N_t|}$ and $\bR \times \{0\} \times W_{m+|N_s|}$ (estimate~\eqref{WF_cont_gateaux_t}). Arguing similarly as done in the proof of~\ref{W1}, i.e. using the wave-front set calculus (thm.~\ref{theo_WF_horma}), we have that wave-front set of each term~\eqref{term_var_tilde_prod_field} in $\phi(\epsilon)$, viewed as distributions in the variables $\epsilon, x_1, \dots, x_j, y_1, \dots, y_\nu \in \bR \times M^{j+\nu}$, is contained in $\bR \times \{0 \} \times W_{j+\nu}$ which is precisely the requirement of~\ref{W2}. Consequently also $\delta^\nu \widetilde{(t \bullet s)}{}^j_{\phi(\epsilon)}/ \delta \phi^\nu$ satisfies the condition~\ref{W2}. This concludes the proof.
		\end{proof}
		
		The Fedosov construction we reviewed in sec.~\ref{subsec_Fedosov_fin} actually requires that the algebraic structure over sections on $\cW$ is extended to forms with values in $\cW$, see eq.~\eqref{graded_prod_fin}. Similarly, in our infinite-dimensional setting we present the straightforward extension of the fiberwise product~\eqref{wick_prod_inf_sec} on $\Omega_W(S,\cW)$ based on the gradings~\eqref{gradings_inf_form}.	In detail, we consider firstly $t \in \Omega^k_W(S,\cW)$ and $s \in \Omega^{k'}_W(S,\cW)$ such that both of them are homogeneous in the symmetric degree $\deg_s$ and in the formal degree $\deg_\hbar$. In particular, we set $\deg_s t =n$ and $\deg_s s = m$. For such on-shell $W$-smooth fields, we define the product $(t \bullet s)_\phi$ as the sequence $((t \bullet s)^{k+k',0}_\phi, (t \bullet s)^{k+k',1}_\phi, \dots)$, where $(t \bullet s)^{k+k',j}_\phi$ is given by
			\begin{equation}\label{prod_dega,degs-hom}
				\begin{split}
					&(t \bullet s)^{k+k',j}_\phi (y_1, \dots, y_{k+k'}, x_1, \dots, x_j) := \\
					&\quad = \hbar^\ell \sC_{n,m,\ell} \bP^+ \bP^-\int_{M^{2\ell}} t_\phi(y_1, \dots, y_k; z_1, \dots, z_\ell, x_1, \dots, x_{n-\ell}) \left( \prod_{i = 1}^\ell \omega_\phi(z_{i}, z'_{i}) \right) \times \\
					&\qquad \times s_\phi(y_{k+1}, \dots, y_{k+k'}; z'_1, \dots, z'_\ell, x_{n-\ell+1}, \dots, x_N) \prod_{i=1}^\ell dz_i dz'_i,
				\end{split}
			\end{equation}
		if $j=n+m-2\ell$ for $\ell \leq n,m$, otherwise $(t \bullet s)^{k+k',j}_\phi= 0$. In the above formula, $\sC_{n,m,\ell} = \frac{n!m!}{\ell! (n-\ell)! (m-\ell)!}$ is the same combinatorial factor appearing in eq.~\eqref{bullet_prod}. By $\bP^+, \bP^-$ we mean that a symmetrization acts on the free variables $x$ and an anti-symmetrization acts on the free variables $y$. Similarly as for eq.~\eqref{bullet_prod}, by abusing the notation, we identify an equivalence class in $\wedgevee^{k,n}_W T^*_\phi S$ (i.e. the elements in $\boxtimes_W^{k+n} T^*_\phi S$ which are anti-symmetric in the first $k$ entries and symmetric in the remaining $n$) with one of its representatives, which are distributions in $\cE_W'(M^{k+n})$ anti-symmetric in the first $k$ entries and symmetric in the remaining $n$. The equivalence classes corresponding to eq.~\eqref{prod_dega,degs-hom} do not depend on the choice of representative for $t$ and $s$.\\
			The product $\bullet$ respects the $\Deg$-grading, i.e. the total degree of the product of two factors is equal to the sum of the total degrees of the factors involved. The definition~\eqref{prod_dega,degs-hom} extends to a map $\Omega_W^k(S,\cW) \times \Omega_W^{k'} (S,\cW)\to \Omega_W^{k+k'}(S,\cW)$ making use of the $\Deg$-filtration and the fact that any form homogeneous in the total degree $\Deg$ and in the antisymmetric degree $\deg_a$ decomposes in finitely many terms homogeneous in $\deg_a$, $\deg_s$ and $\deg_\hbar$. The product extends further to $\Omega_W(S,\cW) \times \Omega_W(S,\cW) \to \Omega_W(S,\cW)$ canonically even though the $\deg_a$-grading does not not admit a maximum value as in the infinite-dimensional case. With a similar argument as the one we presented before for $C^\infty_W(S,\cW)$, we can prove that the product of two on-shell $W$-smooth forms is an on-shell $W$-smooth form. We summarize these results in the following proposition.
		\begin{prop}\label{prop_prod_smooth_on_W_form}
			The product $\bullet$ defined by formula~\eqref{prod_dega,degs-hom} defines on $\Omega_W(S,\cW)$, i.e. the space of on-shell $W$-smooth forms with values in $\cW$, the structure of an associative algebra.
		\end{prop}

\section{$W$-smooth covariant derivatives on $S$}\label{subsec_W_covariant_der}
	In sec.~\ref{subsec_W_smooth_symp_metric}, we have defined sections $\phi \mapsto \sigma_\phi$, $\phi \mapsto \mu_\phi$ and $\phi \mapsto \omega_\phi^\flat$ of the bundle $\bC \otimes \boxtimes^2_W T^* S$ that can be viewed as analogues of the tensor fields $\sigma_{ij}$, $G_{ij}$ and $\omega_{ij}$ on the bundle $\bC \otimes T^* S \otimes T^* S$ in Fedosov's construction for finite-dimensional $S$. We have shown that the sections $\phi \mapsto \sigma_\phi$, $\phi \mapsto \mu_\phi$ and $\phi \mapsto \omega_\phi^\flat$ satisfy the key property of on-shell $W$-smoothness. From these sections, we define a covariant derivative $\nabla^W$ that is well-defined on on-shell $W$-smooth sections and which preserves on-shell $W$-smoothness. This covariant derivative is also compatible with the algebraic structure we have discussed in sec.~\ref{subsec_algebra_W_smooth_sec} and it is analogous to the Yano connection $\nabla$ in the finite-dimensional case. The connection $\nabla^W$ will serve as the starting point for Fedosov's construction in the infinite-dimensional setting, just as the Yano connection did in the finite-dimensional case.\\
	Our construction will be rather pedestrian. In the finite-dimensional setting, any affine connection (and in particular the Yano-connection) can be written as 
$\nabla_i v_j = \partial_i v_j - \Gamma_{ij}^k v_k$, where $\partial$ is a flat connection such as e.g. the flat connection associated with a fixed local coordinate system. 
We are going to choose the flat derivative operator $\partial$ defined in prop.~\ref{prop_der_smooth_on} as our analogue for $\partial_i$ in the infinite-dimensional setting. The remaining task is then to show that the connection coefficients $\Gamma_{ij}^k$ have an appropriate counterpart in infinite dimensions, and that the connection $\nabla^W$ thus obtained is well defined on on-shell $W$-smooth tensor fields. This will be the case if $\phi \mapsto \omega_\phi$ is chosen to be admissible in the sense of def.~\ref{def_suit_omega}, as will be assumed throughout.\\

	First we present the general definition of covariant derivatives in our setting.
		\begin{defi}\label{def_W_cov_der}
			A {\em $W$-covariant derivative} is a linear map $\sD^W: C^\infty_W(S, \boxtimes^n_W T^* S) \to C^\infty_W(S, \boxtimes^{n+1}_W T^* S)$ such that on $C^\infty_W(S)$, i.e. for $n=0$, it equals $\partial$ defined in prop.~\ref{prop_der_smooth_on}, and such that it satisfies the Leibniz rule, i.e. for any $t,s$ on-shell $W$-smooth covariant sections with rank $n$ and, respectively, $m$, it holds
				\begin{equation}\label{cov_der_Leibniz}
					\sD^W(t \otimes s) = (\sD^W t) \otimes s + t \otimes (\sD^W s),
				\end{equation}
			where $\otimes$ is the tensor product given in prop.~\ref{prop_tensor_prod_W_sec}.
		\end{defi}
	We start defining the infinite-dimensional analogue $\mathring{\nabla}^{W}$ of the Levi-Civita connection. This is a $W$-covariant derivative which preserves the on-shell $W$-smooth covariant section $\mu$, i.e. $\mathring{\nabla}^{W} \mu = 0$, and it is torsion-free, i.e. for any $t \in C^\infty_W(S, T^* S)$
		\begin{equation}\label{inf_torsion}
			T^{(\mathring{\nabla})}(t) := \bP^- (\mathring{\nabla}^{W} t) - d t = 0,
		\end{equation}
	 where $\bP^-$ denotes anti-symmetrization. In order to define $\mathring{\nabla}^W$, we proceed as follows. For any $\phi \in C^\infty(M)$, we first construct the infinite-dimensional analogue $\mathring{\Gamma}_\phi$ of the Christoffel symbols of the finite-dimensional Levi-Civita connection. Then, we define $\mathring{\nabla} = \partial + \mathring{\Gamma}$, and we check that this maps the space of on-shell $W$-smooth covariant sections of rank $n$ into the space of on-shell $W$-smooth covariant sections of rank $n+1$, i.e. we need to construct a suitable extension $\compositeaccents{\widetilde}{\mathring{\nabla}^{W} t}_\phi$ and check~\ref{W1},~\ref{W2} of def.~\ref{def_smooth_on_tens}. Finally, we check that the proposed definition satisfies also the other requirements to be a $W$-covariant derivative. The choice of $\mathring{\Gamma}_\phi$ is made such that $\mathring{\nabla}$ is torsion-free and preserves $\mu$.
	\begin{prop}\label{prop_LC_W-smooth_on}
			Let $c$ be a cut-off function as in eq.~\eqref{kernel_symp} and let $\phi \mapsto \omega_\phi$ be an admissible assignment. For any $\phi \in C^\infty(M)$, we define the distribution $\mathring{\Gamma}_\phi \in \cD'(M^3)$ by
				\begin{equation}\label{LC_symbol_W-smooth_on}
					\begin{split}
						\mathring{\Gamma}_\phi(x_1,x_2,x_3) &:= - \frac{1}{2} \int_M G_\phi(x_1, z) \left\{ \frac{\delta (\sigma_c \circ G_\phi \circ \sigma_c)(z,x_3)}{\delta \phi(x_2)} + \frac{\delta (\sigma_c \circ G_\phi \circ \sigma_c)(x_2,z)}{\delta \phi(x_3)} \right. - \\
						&\quad - \left. \frac{\delta (\sigma_c \circ G_\phi \circ \sigma_c)(x_2,x_3)}{\delta \phi(z)} \right\} dz,
					\end{split}
				\end{equation}
			where $G_\phi$ is the symmetric part of the $2$-point function $\omega_\phi$, and where $\sigma_c$ is the distribution~\eqref{kernel_symp}.\\
			For any $t \in C^\infty_W(S, \boxtimes_W^n T^*S)$, we define
					\begin{equation}\label{tilde_LC_W-smooth_on}
						\begin{split}
							&\compositeaccents{\widetilde}{(\mathring{\nabla}^{W} t)}_\phi(x_1, \dots, x_{n+1}) :=\\
							&\quad = \widetilde{(\partial t)}_\phi(x_1, \dots, x_{n+1})  - \sum_{j =2}^{n+1} \int_{M^{3}} (\sigma_c \circ E_\phi)(x_1, x'_1) (\sigma_c \circ E_\phi)(x_j, x'_j)  \mathring{\Gamma}_\phi(z_j,x'_1,x'_j) \times \\
							&\qquad \qquad \qquad \qquad \times \tilde{t}_\phi(x_2, \dots, z_j , \dots x_{n+1}) dx'_1 dx'_j dz_j,
						\end{split}
					\end{equation}
				where $\tilde{t}_\phi \in (\sigma_c \circ E_\phi)^{\otimes n} \cE'_W(M^n)$ is an extension of $t$, and where $\widetilde{(\partial t)}_\phi \in (\sigma_c \circ E_\phi)^{\otimes n +1} \cE'_W(M^{n+1})$ is the extension of $\partial t$ defined by eq.~\eqref{tilde_diff} with respect to the fixed choice of $c$. The distribution~\eqref{tilde_LC_W-smooth_on} defines an on-shell $W$-smooth covariant section $\mathring{\nabla}^W t$ with rank $(n+1)$ by restriction to $(TS)^{\otimes n+1}$, i.e.
				\begin{equation}\label{restr_LC}
					(\mathring{\nabla}^W t)_\phi (u_1, \dots, u_{n+1}) := \compositeaccents{\widetilde}{(\mathring{\nabla}^{W} t)}_\phi(u_1, \dots, u_{n+1}) \quad \forall \phi \in S, \, u_i \in T_\phi S.
				\end{equation}
			The section $\mathring{\nabla}^W t$ does not depend on the extension $\tilde{t}$ nor the choice of the cut-off $c$.\\
			Finally, $\gls{mathring_nabla_W}$ is a $W$-covariant derivative which preserves $\mu$ and which is torsion-free.
		\end{prop}
		\begin{proof}
			We begin by proving that for any $\phi \in C^\infty(M)$ the distribution $\compositeaccents{\widetilde}{(\mathring{\nabla}^{W} t)}_\phi $ given by eq.~\eqref{tilde_LC_W-smooth_on} is well-defined and belongs to $ (\sigma_c \circ E_\phi)^{\otimes n +1} \cE'_W(M^{n+1})$. To show this, we first prove that for any $\phi \in C^\infty(M)$ the distribution $\mathring{\Gamma}_\phi(x_1, x_2, x_3)$ given by eq.~\eqref{LC_symbol_W-smooth_on} is well-defined, is compactly supported in $x_2, x_3$, and has wave-front set contained in $W_3$. By definition $G_\phi(x_1,x_2)$ is the symmetric part of the distribution $\omega_\phi(x_1,x_2)$, which is an admissible assignment in the sense of def.~\ref{def_suit_omega}. Therefore, the estimate~\eqref{WF_better} for $\omega_\phi$ implies that the wave-front set of $\delta G_\phi / \delta \phi$ is contained in $W_{3}$. The distribution $\sigma_c$ defined by eq.~\eqref{kernel_symp} is a distribution in $\cE'_W(M^2)$ which does not depend on $\phi$. Then, applying lemma~\ref{lemma_W_comp} to the right-hand side of eq.~\eqref{LC_symbol_W-smooth_on}, it follows that the distribution $\mathring{\Gamma}_\phi$ is well-defined and its wave-front set is contained in $W_3$. Furthermore, $\mathring{\Gamma}_\phi(x_1,x_2,x_3)$ is compactly supported in $x_2, x_3$ because $\sigma_c$ is compactly supported by definition, and because $\delta G_\phi(x_1,x_2) / \delta \phi(y)$ is compactly supported in $y$ as follows from the fact that $\delta \omega_\phi(x_1,x_2) / \delta \phi(y)$ is compactly supported in $y$ by hypothesis. By construction, $\widetilde{(\partial t)}_\phi$ is in $(\sigma_c \circ E_\phi)^{\otimes n+1} \cE'_W(M^{n+1})$. By what we already know about $\mathring{\Gamma}_\phi$, and because $\tilde{t}_\phi \in (\sigma_c \circ E_\phi)^{\otimes n} \cE'_W(M^n)$, it follows from lemma~\ref{lemma_W_comp} that the second term in eq.~\eqref{tilde_LC_W-smooth_on} is also a well-defined distribution in $(\sigma_c \circ E_\phi)^{\otimes n+1} \cE'_W(M^{n+1})$. Thus, we have $\compositeaccents{\widetilde}{(\mathring{\nabla}^{W} t)}_\phi \in (\sigma_c \circ E_\phi)^{\otimes n+1} \cE'_W$ as we wanted to prove.\\
			Next, we prove that $\compositeaccents{\widetilde}{(\mathring{\nabla}^{W} t)}_\phi$ defines an on-shell $W$-smooth covariant section. For this purpose we need to show that (a) for any $\phi \in S$ and any $u_i \in T_\phi S$, $\compositeaccents{\widetilde}{(\mathring{\nabla}^{W} t)}_\phi(u_1, \dots, u_{n+1})$ does not depend on the choice of the extensions $\tilde{t}_\phi \in (\sigma_c \circ E_\phi)^{\otimes n} \cE'_W(M^n)$ and $\widetilde{(\partial t)}_\phi \in (\sigma_c \circ E_\phi)^{\otimes n +1} \cE'_W(M^{n+1})$, and (b) $\compositeaccents{\widetilde}{(\mathring{\nabla}^{W} t)}_\phi$ satisfies conditions~\ref{W1},~\ref{W2} of def.~\ref{def_smooth_on_tens}. As we have already proved in prop.~\ref{prop_der_smooth_on}, $\widetilde{(\partial t)}_\phi$ satisfies both the conditions (a) and (b) above. Therefore, we need to show that the second term in~\eqref{tilde_LC_W-smooth_on} also does.
			\begin{enumerate}[label=(\alph*), start=1]
				\item By definition, $G_\phi$ is a bi-solution with respect to $P_\phi$. We have already proved that $\mathring{\Gamma}_\phi(x_1,x_2,x_3)$ is compactly supported in $x_2, x_3$ and its wave-front set is contained in $W_3$. Therefore, for any smooth functions $f, h$ we have that
					\begin{equation*}
						M \ni x_1 \mapsto \int_{M^2} \mathring{\Gamma}_\phi(x_1,x_2,x_3) f(x_2) h(x_3) dx_2 dx_3
					\end{equation*}
				is smooth because there is no element in $W_3$ in the form $(x_1,x_2,x_3; k_1, 0, 0 )$, and it is a $P_\phi$-solution by construction. Now, let $\tilde{t}_{1,\phi}, \tilde{t}_{2,\phi} \in (\sigma_c \circ E_\phi)^{\otimes n} \cE'_W(M^n)$ be two $W$-smooth extensions of $t$. We compute the difference of between the second term in~\eqref{tilde_LC_W-smooth_on} corresponding to $\tilde{t}_{1,\phi}$ and $\tilde{t}_{2,\phi}$. When evaluated at $\phi \in S$ and smeared with $u_1, \dots, u_{n+1} \in T_\phi S$, this difference reads
					\begin{equation*}
						\begin{split}
							\sum_{j =2}^{n+1} \int_{M^{3}} u_1(x'_1) u_j(x'_j)  \mathring{\Gamma}_\phi(z_j,x'_1,x'_j) (\tilde{t}_{1,\phi} - \tilde{t}_{2,\phi})(u_2, \dots, z_j , \dots u_{n+1}) dx'_1 dx'_j dz_j= 0,
						\end{split}
					\end{equation*}
				because for any $\phi \in S$ the two extension $\tilde{t}_{1,\phi}$ and $\tilde{t}_{2,\phi}$ must coincide by construction when smeared with smooth $P_\phi$-solutions. This implies that the second term in~\eqref{tilde_LC_W-smooth_on} satisfies the requirement (a) as we wanted to show.
				\item To prove that the conditions~\ref{W1},~\ref{W2} are satisfied, we compute the $\nu$-Gateaux derivative of the second term of eq.~\eqref{tilde_LC_W-smooth_on} by distributing the variational derivatives on its factors. It can be easily seen that this $\nu$-Gateaux derivative is a finite sum of appropriate compositions of $\sigma_c$, Gateaux derivatives of $E_\phi$ and Gateaux derivatives of $G_\phi$. The idea is to prove that each term in this decomposition satisfies the conditions~\ref{W1},~\ref{W2}. Rather than displaying explicitly these terms and computing their wave-front sets, we just outline the main arguments needed for this purpose and omit the tedious but entirely straightforward details, which parallel those already presented e.g. in the proof of prop.~\ref{prop_der_smooth_on}.\\
				To prove~\ref{W1}, we apply lemma~\ref{lemma_W_comp} and use the fact that $\tilde{t}$ satisfies~\ref{W1} by hypothesis, together with the estimate~\eqref{var_ders_causal_WF} of the wave-front set of $\delta^\nu E_\phi / \delta \phi^\nu$ (proved in prop.~\ref{prop_var_ders_causal}), the estimate of the wave-front set of $\delta^\nu G_\phi / \delta \phi^\nu$ induced by the estimate~\eqref{WF_better}, the wave-front set of $\sigma_c$ computed in~\eqref{WF_kernel_symp} and the support properties of the distributions involved.\\
				Let $\bR \ni \epsilon \mapsto \phi(\epsilon) \in C^\infty(M)$ be smooth. To show that~\ref{W2} holds, one shall apply thm.~\ref{theo_WF_horma} instead of lemma~\ref{lemma_W_comp} and use the fact that $\tilde{t}$ satisfy~\ref{W2} by hypothesis, together with the estimate~\eqref{cont_var_ders_causal_WF} of the wave-front set of $\delta^\nu E_{\phi(\epsilon)} / \delta \phi^\nu$ (proved in prop.~\ref{prop_var_ders_causal}) and the estimate of the wave-front set of $\delta^\nu G_{\phi(\epsilon)} / \delta \phi^\nu$ induced by the estimate~\eqref{WF_better_s}.\\
				Therefore, the second term in~\eqref{tilde_LC_W-smooth_on} satisfies the requirement (b) as we claimed.
			\end{enumerate}
			So, we have proved that $\mathring{\nabla}^W t$ is a well-defined on-shell $W$-smooth covariant section.\\
			Next, we show that $\mathring{\nabla}^W t$ does not depend on the choice of the cut-off function $c$. It is sufficient to prove that for any functions $c, c'$ as in eq.~\eqref{kernel_symp}, it holds
				\begin{equation}\label{tilde_LC_difference_c-c'}
					\compositeaccents{\widetilde}{(\mathring{\nabla}^{W} t)}_\phi -\compositeaccents{\widetilde}{(\mathring{\nabla}^{W} t)'}_\phi \approx 0,
				\end{equation}
			where the prime refers to a quantity defined with respect to the cut-off $c'$ instead of $c$, and where $\approx$ means ``equal up to distributions in $P_\phi\cE'_W(M^{n+1})$'' exactly as in lemma~\ref{lemma_approx}. The difference $\compositeaccents{\widetilde}{(\mathring{\nabla}^{W} t})_\phi - \compositeaccents{\widetilde}{(\mathring{\nabla}^{W} t)'}_\phi$ consists in two terms. One is $\widetilde{(\partial t)}_\phi - \widetilde{(\partial t)'}_\phi$, while, using lemma~\ref{lemma_approx}, the other can be written as
				\begin{equation}\label{symbols_difference_c-c'}
					\begin{split}
						&\left(\compositeaccents{\widetilde}{(\mathring{\nabla}^{W} t)}_\phi - \widetilde{(\partial t)}_\phi \right)  - \left(\compositeaccents{\widetilde}{(\mathring{\nabla}^{W} t)'}_\phi - \widetilde{(\partial t)'}_\phi \right) \approx\\
						&\quad \approx - \sum_{j =2}^{n+1} \int_{M^{3}} (\sigma_c \circ E_\phi)(x_1, x_1') (\sigma_c \circ E_\phi)(x_j, x_j') \left( \mathring{\Gamma}_\phi - \mathring{\Gamma}'_\phi \right) (z_j,x'_1,x'_j) \times \\
						&\qquad \qquad \times \tilde{t}_\phi(x_2, \dots, z_j , \dots x_{n+1}) dx'_1 dx'_{j} dz_j.
					\end{split}
				\end{equation}
			We will show that it holds
				\begin{equation}\label{tilde_LC_difference_c-c'_alt}
					\left(\compositeaccents{\widetilde}{(\mathring{\nabla}^{W} t)}_\phi - \widetilde{(\partial t)}_\phi \right)  - \left(\compositeaccents{\widetilde}{(\mathring{\nabla}^{W} t)'}_\phi - \widetilde{(\partial t)'}_\phi \right) \approx - \left( \widetilde{(\partial t)}_\phi - \widetilde{(\partial t)'}_\phi \right),
				\end{equation}
			which clearly implies the validity of formula~\eqref{tilde_LC_difference_c-c'}. For this purpose, we first need to rewrite the right-hand side of formula~\eqref{symbols_difference_c-c'}. We express the difference $\mathring{\Gamma}_\phi- \mathring{\Gamma}_\phi'$ as
				\begin{equation}\label{symbols_difference_c-c'_2}
					\begin{split}
						&\left( \mathring{\Gamma}_\phi - \mathring{\Gamma}'_\phi \right)(x_1,x_2,x_3):= \\
						&\quad = - \frac{1}{2} \int_M G_\phi(x_1, z) \left\{ \frac{\delta \Delta G_\phi(z,x_3)}{\delta \phi(x_2)} + \frac{\delta \Delta G_\phi (x_2,z)}{\delta \phi(x_3)} - \frac{\delta \Delta G_\phi (x_2,x_3)}{\delta \phi(z)} \right\} dz,
					\end{split}
				\end{equation}
			where
				\begin{equation*}
					\Delta G_\phi := \sigma_c \circ G_\phi \circ \sigma_c - \sigma_{c'} \circ G_\phi \circ \sigma_{c'} = \sigma_c \circ G_\phi \circ (\sigma_c - \sigma_{c'}) + (\sigma_c - \sigma_{c'}) \circ G_\phi \circ \sigma_{c'}.
				\end{equation*}
			As a consequence of eq.~\eqref{causal_prop_inv_symp_kernel_sol}, $\sigma_c - \sigma_{c'}$ vanishes if it is smeared with $P_\phi$-solutions. Thus, $\Delta G_\phi$ also vanishes when it is smeared with $P_\phi$-solutions, because $G_\phi$ is a bi-solution with respect to $P_\phi$. It follwos
				\begin{equation*}
					G_\phi \circ \Delta G_\phi \circ E_\phi = 0 = E_\phi \circ \Delta G_\phi \circ E_\phi.
				\end{equation*}
			Using this result and the Leibniz rule for the variational derivative, we rewrite the right-hand side of formula~\eqref{symbols_difference_c-c'} as
				\begin{equation}\label{symbols_difference_c-c'_2_bis}
					\begin{split}
						&- \frac{1}{2} \sum_{j=2}^{n+1} \int_{M^{3}} (\sigma_c \circ E_\phi)(x_1, x_1') (\sigma_c \circ E_\phi)(x_j, x_j') \left\{ \left(\Delta G_\phi \circ  \frac{\delta G_\phi}{\delta \phi(x'_1)} \right) (x'_j,z_j) + \right. \\
						&\qquad \left. + \left(\Delta G_\phi \circ  \frac{\delta G_\phi}{\delta \phi(x'_j)} \right) (x'_1,z_j) \right\} \tilde{t}_\phi(x_2, \dots, z_j, \dots, x_{n+1}) dx'_1 dx'_j dz_j- \\
						&- \frac{1}{2}\sum_{j=2}^{n+1} \int_M \left\{ \int_M (\sigma_c \circ E_\phi)(x_1, x'_1) \left( G_\phi \circ \Delta G_\phi \circ \frac{\delta E_\phi \circ \sigma_c}{\delta \phi(x'_1)} \right)(z_j,x_j) dx'_1+ \right. \\
						&\qquad \left. + \int_M (\sigma_c \circ E_\phi)(x_j, x'_j) \left(G_\phi \circ \Delta G_\phi \circ \frac{\delta E_\phi \circ \sigma_c}{\delta \phi(x'_j)}\right)(z_j,x_1) dx'_i\right\} \tilde{t}_\phi(x_2, \dots, z_j, \dots, x_{n+1}) dz_j  + \\
						&+ \frac{1}{2} \sum_{j=2}^{n+1} \int_{M^2} \left\{ \left(\frac{\delta \sigma_c \circ E_\phi}{\delta \phi(z'_j)} \circ \Delta G_\phi \circ E_\phi \circ \sigma_c \right)(x_1,x_j) + \right. \\
						&\qquad \left. + \left(\sigma_c \circ E_\phi \circ \Delta G_\phi \circ \frac{\delta E_\phi \circ \sigma_c}{\delta \phi(z'_j)}\right)(x_1,x_j) \right\} G_\phi(z'_j,z_j) \tilde{t}_\phi(x_2, \dots, z_j, \dots, x_{n+1}) dz_j dz'_j.
					\end{split}
				\end{equation}
			Since $G_\phi$ is a bi-solution, it follows from eq.~\eqref{causal_prop_inv_symp_kernel_sol} that $G_\phi \circ \sigma_c  \circ E_\phi = G_\phi$. Exploiting this result and again the fact that $\sigma_c - \sigma_{c'}$ vanishes when smeared with $P_\phi$-solutions, the first and the last term in formula~\eqref{symbols_difference_c-c'_2_bis} equal
				\begin{equation}\label{symbols_difference_c-c'_2_tris}
					\begin{split}
						& - \frac{1}{2}\sum_{j=2}^{n+1} \int_{M^{4}} \sigma_c(x_1,y_1) \left\{ E_\phi (y_1,y_3) \frac{\delta G_\phi(z_j, y_2)}{\delta \phi(y_3)} - G_\phi (z_j,y_3) \frac{\delta  E_\phi(y_1, y_2)}{\delta \phi(y_3)} \right\}   \times \\
						&\qquad \times ((\sigma_c - \sigma_{c'}) \circ G_\phi \circ \sigma_c)(y_2,x_j)  \tilde{t}_\phi(x_2, \dots, z_j, \dots, x_{n+1}) dy_1 dy_2 dy_3 dz_i + \\
						&+ \frac{1}{2} \sum_{i=2}^{n+1} \int_{M^{4}} \sigma_c(x_j,y_1) \left\{ E_\phi (y_1,y_3) \frac{\delta G_\phi(z_j, y_2)}{\delta \phi(y_3)} - G_\phi (z_j,y_3) \frac{\delta  E_\phi(y_1, y_2)}{\delta \phi(y_3)}\right\}  \times \\
						&\qquad \times ((\sigma_c - \sigma_{c'}) \circ G_\phi \circ \sigma_c)(y_2,x_1)  \tilde{t}_\phi(x_2, \dots, z_j, \dots, x_{n+1}) dy_1 dy_2 dy_3 dz_j.
					\end{split}
				\end{equation}
			The bi-distributions $E_\phi$ and $G_\phi$ satisfies the hypotheses of lemma~\ref{lemma_inf_comm} and, therefore, it follows that for any $\phi \in C^\infty(M)$ and any $f_1, f_2 \in C^\infty_0(M)$ the function
				\begin{equation}\label{E_delta_G-G_delta_E}
					M \ni x \mapsto \int_{M^3} \left( E_\phi(x_1,y) \frac{\delta G_\phi (x_2, x)}{\delta \phi(y)} - G_\phi (x_2,y) \frac{\delta E_\phi (x_1, x)}{\delta \phi(y)} \right) f_1(x_1) f_2(x_2) dydx_1 dx_2
				\end{equation}								
			is a smooth $P_\phi$-solution. Thus, the distribution given by formula~\eqref{symbols_difference_c-c'_2_tris} vanishes because $\sigma_c - \sigma_{c'}$ vanishes when smeared with $P_\phi$-solutions. On the other hand, the second term in formula~\eqref{symbols_difference_c-c'_2_bis} can be written as
				\begin{equation}\label{symbols_difference_c-c'_2_quater}
					\begin{split}
						&\frac{1}{2} \sum_{j=2}^{n+1} \int_{M^{5}} \sigma_c(x_1,y_1) \left\{ E_\phi (y_1,y_3) \frac{\delta E_\phi(y_2, z'_j)}{\delta \phi(y_3)} - E_\phi (y_2,y_3) \frac{\delta  E_\phi(y_1,z'_j)}{\delta \phi(y_3)}\right\}  \sigma_c(y_2,x_j) \times \\
						&\qquad \times ((\sigma_c - \sigma_{c'}) \circ G_\phi \circ \sigma_{c'} \circ G_\phi)(z'_j,z_j) \tilde{t}_\phi(x_2, \dots, z_j, \dots, x_{n+1}) dy_1 dy_2 dy_3 dz_j dz'_j - \\
						&- \sum_{j=2}^{n+1} \int_{M^2} (\sigma_c \circ E_\phi)(x_1, x'_1) \left(\sigma_c \circ \frac{\delta E_\phi }{\delta \phi(x'_1)} \circ (\sigma_c - \sigma_{c'}) \circ G_\phi \circ \sigma_c \circ G_\phi\right)(x_j,z_j) \times \\
						&\qquad \times \tilde{t}_\phi(x_2, \dots, z_j, \dots, x_{n+1}) dx'_1 dz_j.
					\end{split}
				\end{equation}
			The function
				\begin{equation*}
					M \ni x \mapsto \int_{M^3} \left( E_\phi(x_1,y) \frac{\delta E_\phi (x_2, x)}{\delta \phi(y)} - E_\phi (x_2,y) \frac{\delta E_\phi (x_1, x)}{\delta \phi(y)} \right) f_1(x_1) f_2(x_2) dy dx_1 dx_2
				\end{equation*}								
			is a smooth $P_\phi$-solution as follows from lemma~\ref{lemma_inf_comm}. Thus, the first term in~\eqref{symbols_difference_c-c'_2_quater} vanishes because $\sigma_c - \sigma_{c'}$ vanishes when smeared with $P_\phi$-solutions. Summing up, we obtain
				\begin{equation}\label{symbols_difference_c-c'_2_penta}
					\begin{split}
						&\left(\compositeaccents{\widetilde}{(\mathring{\nabla}^{W} t)}_\phi - \widetilde{(\partial t)}_\phi \right)  - \left(\compositeaccents{\widetilde}{(\mathring{\nabla}^{W} t)'}_\phi - \widetilde{(\partial t)'}_\phi \right) \approx \\
						&\approx - \sum_{j=2}^{n+1} \int_{M^2} (\sigma_c \circ E_\phi)(x_1, x'_1) \left(\sigma_c \circ \frac{\delta E_\phi }{\delta \phi(x'_1)} \circ (\sigma_c - \sigma_{c'}) \circ G_\phi \circ \sigma_c \circ G_\phi\right)(x_j,z_j) \times \\
						&\qquad \qquad  \times \tilde{t}_\phi(x_2, \dots, z_j, \dots, x_{n+1}) dx'_1 dz_j \\
						&\approx \sum_{j=2}^{n+1} \int_{M^2} (\sigma_c \circ E_\phi)(x_1, x'_1) \left(\sigma_c \circ \frac{\delta E_\phi }{\delta \phi(x'_1)} \circ (\sigma_c - \sigma_{c'}) \circ E_\phi \right)(x_j,z_j) \times \\
						&\qquad \qquad \times \tilde{t}_\phi(x_2, \dots, z_j, \dots, x_{n+1})  dx'_1 dz_i\\
						&\approx - \sum_{j=2}^{n+1} \int_{M^3} (\sigma_c \circ E_\phi)(x_1, x_1') (\sigma_c \circ E_\phi)(x_j, x_j') \frac{\delta (\sigma_c - \sigma_{c'}) \circ E_\phi}{\delta \phi(x'_1)}(x'_i, z_i)\times \\
						&\qquad \qquad \times \tilde{t}_\phi(x'_2, \dots, z_j, \dots, x'_{n+1}) dx'_1 dx'_j dz_j.
					\end{split}
				\end{equation}
			We used eq.~\eqref{G_tilde_G_flat}, the Leibniz rule for the variational derivative, and the fact that $E_\phi \circ (\sigma_c - \sigma_{c'}) \circ E_\phi = 0$. As follows from eq.~\eqref{rel_c_c'_partial}, the last line of eq.~\eqref{symbols_difference_c-c'_2_penta} coincides with $- (\widetilde{(\partial t)}_\phi - \widetilde{(\partial t)'}_\phi)$. This is precisely what we wanted to show. Thus, we have verified that $\mathring{\nabla} t$ is independent on the choice of the cut-off $c$.\\
			
			By construction, $\mathring{\nabla}^{W}$ reduces to $\partial$ if $n=0$ and it satisfies the Leibniz rule~\eqref{cov_der_Leibniz}. Thus, $\mathring{\nabla}^{W}$ is a $W$-covariant derivative.\\

			Finally, we need to show that $\mathring{\nabla}^W$ is torsion-free and preserves $\mu$. The torsion of $\mathring{\nabla}^{W}$ necessarily vanishes, because $\mathring{\Gamma}_\phi(x_1, x_2, x_3)$ is symmetric in $x_2, x_3$ by definition.\\
			Because we have already proved that $\mathring{\nabla}^{W} \mu$ depends neither on the choice of the extension nor on the choice of the cut-off $c$, to show that $\mathring{\nabla}^{W} \mu=0$, it is sufficient to prove that $\compositeaccents{\widetilde}{\mathring{\nabla}^{W} \mu}_\phi = 0$, where $\compositeaccents{\widetilde}{\mathring{\nabla}^{W} \mu}_\phi$ is given by eq.~\eqref{tilde_LC_W-smooth_on} for a specific $W$-smooth extension of $\mu$ and an arbitrary but fixed cut-off function $c$ as in eq.~\eqref{kernel_symp}. If we chose the distribution $\sigma_c \circ G_\phi \circ \sigma_c$ as our $W$-smooth extension of $\mu_\phi$ (we have proved in thm.~\ref{theo_var_controll_state} that this is allowed), then, using eq.~\eqref{G_tilde_G_flat}, we get
				\begin{equation}\label{LC_pres_mu}
						\begin{split}
							&\compositeaccents{\widetilde}{(\mathring{\nabla}^{W} \mu)}_\phi(x_1, x_2, x_3) = \\
							&\quad = \int_{M^3} \prod_{i=1}^3 (\sigma_c \circ E_\phi)(x_i, x'_i) \frac{ \delta (\sigma_c \circ G_\phi \circ \sigma_c)(x'_2,x'_3)}{\delta \phi(x'_1)} dx'_1 dx'_2 dx'_3 +\\
							&\qquad + \bP^+_{x_2, x_3} \int_{M^3} (\sigma_c \circ E_\phi)(x_1, x'_1) (\sigma_c \circ E_\phi)(x_2, y) \left\{  \frac{\delta (\sigma_c \circ G_\phi \circ \sigma_c)(z,y)}{\delta \phi(x'_1)} +  \right. \\
							&\qquad \qquad \left. + \frac{\delta (\sigma_c \circ G_\phi \circ \sigma_c)(x'_1,z)}{\delta \phi(y)} - \frac{\delta (\sigma_c \circ G_\phi \circ \sigma_c)(x'_1,y)}{\delta \phi(z)} \right\} (G_\phi \circ \sigma_c \circ G_\phi \circ \sigma_c)(z,x_3) dx'_1 dy dz \\
							&\quad = \int_{M^3} \prod_{i=1}^3 (\sigma_c \circ E_\phi)(x_i, x'_i) \frac{ \delta (\sigma_c \circ G_\phi \circ \sigma_c)(x'_2,x'_3)}{\delta \phi(x'_1)} dx'_1 dx'_2 dx'_3 +\\
							&\qquad - \bP^+_{x_2, x_3} \int_{M^3} (\sigma_c \circ E_\phi)(x_1, x'_1) (\sigma_c \circ E_\phi)(x_2, y) \left\{  \frac{\delta (\sigma_c \circ G_\phi \circ \sigma_c)(z,y)}{\delta \phi(x'_1)} +  \right. \\
							&\qquad \qquad \left. + \frac{\delta (\sigma_c \circ G_\phi \circ \sigma_c)(x'_1,z)}{\delta \phi(y)} - \frac{\delta (\sigma_c \circ G_\phi \circ \sigma_c)(x'_1,y)}{\delta \phi(z)} \right\} (E_\phi \circ \sigma_c)(z,x_3) dx'_1 dy dz\\
							&\quad = 0,
						\end{split}
					\end{equation}
				where $\bP^+_{x_2, x_3} $ is the symmetrization in the variables $x_2, x_3$. This concludes the proof.
		\end{proof}
		
	We point out that the $W$-covariant derivative $\mathring{\nabla}^{W}$ for any admissible assignment $\phi \mapsto \omega_\phi$ does not preserve the on-shell $W$-smooth $2$-form $\sigma$ and thus the infinite-dimensional analogue of the Levi-Civita connection is not compatible in general with the Wick product $\bullet$, just as in the finite-dimensional situation. We overcome this problem just as for finite-dimensional almost-K\"{a}hler manifolds: we define a new $W$-covariant derivative $\nabla^W$ corresponding to the finite-dimensional Yano connection. In particular, $\nabla^W$ is required to preserve both the on-shell $W$-smooth covariant fields $\mu$ and $\sigma$. The procedure to define this $W$-covariant derivative is similar to the construction we have presented for $\mathring{\nabla}^{W}$.
		\begin{prop}\label{prop_Yano_W-smooth_on}
			Let $c$ be a cut-off function as in eq.~\eqref{kernel_symp} and let $\phi \mapsto \omega_\phi$ be an admissible assignment. For any $\phi \in C^\infty(M)$, we define the distribution $\Gamma_\phi \in \cD'(M^3)$ by
				\begin{equation}\label{Yano_symbol_W-smooth_on}
					\begin{split}
						\Gamma_\phi(x_1,x_2,x_3) &:= \mathring{\Gamma}_\phi(x_1,x_2,x_3) - \frac{1}{8}\int_M (E_\phi \circ \sigma_c)(x_1, z) N_\phi(z,x_2,x_3) dz+\\
						&\qquad + \frac{1}{8} \bP^+_{x_2, x_3} \int_{M^2} G_\phi(x_1,z)  N_\phi(z',z,x_2)(\sigma_c \circ G_\phi \circ \sigma_c)(z',x_3) dz dz',
					\end{split}
				\end{equation}
			where $\bP^+_{x_2, x_3} $ is the symmetrization in the variables $x_2, x_3$, where $\mathring{\Gamma}_\phi$ is the distribution defined by formula~\eqref{LC_symbol_W-smooth_on}, where $N_\phi \in \cD'(M^3)$ is the distribution defined by
				\begin{equation}\label{Nijenhuis_inf}
					\begin{split}
						N_\phi(x_1,x_2,x_3) &:= 2 \bP^-_{x_2,x_3} \int_{M} \left\{ \frac{\delta (G_\phi \circ \sigma_c)(x_1,x_2)}{\delta \phi(z)}(G_\phi \circ \sigma_c) (z, x_3) + \right.\\
						&\qquad + \left. (G_\phi \circ \sigma_c)(x_1,z) \frac{\delta (G_\phi \circ \sigma_c)(z,x_3)}{\delta \phi(x_2)} \right\} dz,
					\end{split}
				\end{equation}
			where $\bP^-_{x_2,x_3}$ is the anti-symmetrization in the variables $x_2, x_3$, where $G_\phi$ is the symmetric parts of the $2$-point function $\omega_\phi$, and where $\sigma_c$ is the distribution~\eqref{kernel_symp}.\\
			For any $t \in C^\infty_W(S, \boxtimes_W^n T^*S)$, we define
					\begin{equation}\label{tilde_Yano_W-smooth_on}
						\begin{split}
							&\widetilde{(\nabla^W t)}_\phi(x_1, \dots, x_{n+1}) :=\\
							&\quad = \widetilde{(\partial t)}_\phi(x_1, \dots, x_{n+1}) - \sum_{j=2}^{n+1} \int_{M^{3}} (\sigma_c \circ E_\phi)(x_1, x_1') (\sigma_c \circ E_\phi)(x_j, x_j') \Gamma_\phi(z_j,x'_1,x'_j) \times \\
							&\qquad \qquad \qquad \qquad\tilde{t}_\phi(x_2, \dots, z_j , \dots x_{n+1}) dx'_1 dx'_j dz_j,
						\end{split}
					\end{equation}
				where $\tilde{t}_\phi \in (\sigma_c \circ E_\phi)^{\otimes n} \cE'_W(M^n)$ is an extension of $t$, and where $\widetilde{(\partial t)}_\phi \in (\sigma_c \circ E_\phi)^{\otimes n +1} \cE'_W(M^{n+1})$ is the extension of $\partial t$ defined by eq.~\eqref{tilde_diff} with respect to the fixed choice of $c$.\\
			The distribution~\eqref{tilde_Yano_W-smooth_on} defines an on-shell $W$-smooth covariant section $\nabla^W t$ of rank $(n+1)$ by restriction to $(TS)^{\otimes n+1}$,  i.e.
				\begin{equation}\label{restr_Yano}
					(\nabla^W t)_\phi (u_1, \dots, u_{n+1}) := \widetilde{(\nabla^{W} t)}_\phi(u_1, \dots, u_{n+1}) \quad \forall \phi \in S, \, u_i \in T_\phi S.
				\end{equation}
			The section $\nabla^W t$ does not depend on the choice of the extension $\tilde{t}$ nor the cut-off $c$.\\
			Finally, $\nabla^W$ is a $W$-covariant derivative which preserves $\mu$ and $\sigma$.
		\end{prop}
		\begin{proof}
				We begin by proving that for any $\phi \in C^\infty(M)$, the distribution $\widetilde{(\nabla^W t)}_\phi$  given by eq.~\eqref{tilde_Yano_W-smooth_on} is well-defined and belongs to $ (\sigma_c \circ E_\phi)^{\otimes n +1} \cE'_W(M^{n+1})$. To show this, we first prove that for any $\phi \in C^\infty(M)$ the distribution $N_\phi(x_1, x_2,x_3)$ given by eq.~\eqref{Nijenhuis_inf} is well-defined, is compactly supported in $x_2, x_3$, and has wave-front set contained in $W_3$. By definition $G_\phi(x_1,x_2)$ is the symmetric part of the distribution $\omega_\phi(x_1,x_2)$ which is an admissible assignment in the sense of def.~\ref{def_suit_omega}. Therefore, the estimate~\eqref{WF_better} for $\omega_\phi$ implies that the wave-front set of $\delta G_\phi / \delta \phi$ is contained in $W_{3}$. The distribution $\sigma_c$ defined by eq.~\eqref{kernel_symp} is a distribution in $\cE'_W(M^2)$ which does not depend on $\phi$. Then, applying lemma~\ref{lemma_W_comp} to the right-hand side of eq.~\eqref{Nijenhuis_inf}, it follows that the distribution $N_\phi$ is well-defined and its wave-front set is contained in $W_3$. Furthermore, $N_\phi(x_1,x_2,x_3)$ is compactly supported in $x_2, x_3$ because $\sigma_c$ is by definition compactly supported, and because $\delta G_\phi(x_1,x_2) / \delta \phi(y)$ is compactly supported in $y$ because $\delta \omega_\phi(x_1,x_2) / \delta \phi(y)$ is compactly supported in $y$ by hypothesis.\\
			We have already shown in the proof of pro p.~\ref{prop_LC_W-smooth_on} that $\mathring{\Gamma}_\phi(x_1, x_2, x_3)$ given by eq.~\eqref{LC_symbol_W-smooth_on} is well-defined, is compactly supported in $x_2, x_3$, and has wave-front set contained in $W_3$. Using the result just presented for $N_\phi$, lemma~\ref{lemma_W_comp}, and the support properties of the distributions involved, we conclude that also $\Gamma_\phi(x_1, x_2, x_3)$ given by eq.~\eqref{Yano_symbol_W-smooth_on} is a well-defined distribution which has compact support in $x_2, x_3$ and which has its wave-front set contained in $W_3$.
			By construction, $\widetilde{(\partial t)}_\phi$ is in $(\sigma_c \circ E_\phi)^{\otimes n+1} \cE'_W(M^{n+1})$. By what we already know about $\Gamma_\phi$, and because $\tilde{t}_\phi \in (\sigma_c \circ E_\phi)^{\otimes n} \cE'_W(M^n)$ by hypothesis, it follows form lemma~\ref{lemma_W_comp} that the second term in eq.~\eqref{tilde_LC_W-smooth_on} is also a well-defined distribution in $(\sigma_c \circ E_\phi)^{\otimes n+1} \cE'_W(M^{n+1})$. Thus, we have $\widetilde{(\nabla^{W} t)}_\phi \in (\sigma_c \circ E_\phi)^{\otimes n+1} \cE'_W$ as we wanted to prove.\\
			
			To prove that $\widetilde{(\nabla^{W} t)}_\phi$ defines an on-shell $W$-smooth covariant section, we need to show that (a) for any $\phi \in S$ and any $u_i \in T_\phi S$, $\widetilde{(\nabla^{W} t)}_\phi(u_1, \dots, u_{n+1})$ does not depend on the choice of the extensions $\tilde{t}_\phi \in (\sigma_c \circ E_\phi)^{\otimes n} \cE'_W(M^n)$ and $\widetilde{(\partial t)}_\phi \in (\sigma_c \circ E_\phi)^{\otimes n +1} \cE'_W(M^{n+1})$, and (b) $\widetilde{(\nabla^{W} t)}_\phi$ satisfies conditions~\ref{W1},~\ref{W2} of def.~\ref{def_smooth_on_tens}. As we have proved in prop.~\ref{prop_der_smooth_on}, $\widetilde{(\partial t)}_\phi$ satisfies both the conditions (a) and (b) above. Therefore, we need to show that the second term in ~\eqref{tilde_LC_W-smooth_on} also does.
			\begin{enumerate}[label=(\alph*), start=1]
				\item By definition, $G_\phi$ and $E_\phi$ are bi-solutions with respect to $P_\phi$. We have already proved that $\Gamma_\phi(x_1,x_2,x_3)$ is compactly supported in $x_2, x_3$ and its wave-front set is contained in $W_3$. Therefore, for any smooth functions $f, h$ we have that
					\begin{equation*}
						M \ni x_1 \mapsto \int_{M^2} \Gamma_\phi(x_1,x_2,x_3) f(x_2) h(x_3) dx_2 dx_3
					\end{equation*}
				is smooth because there is no element in $W_3$ in the form $(x_1,x_2,x_3; k_1, 0, 0 )$, and it is a $P_\phi$-solution by construction. Now, let $\tilde{t}_{1,\phi}, \tilde{t}_{2,\phi} \in (\sigma_c \circ E_\phi)^{\otimes n} \cE'_W(M^n)$ be two $W$-smooth extensions of $t$. We compute the difference of the second term in~\eqref{tilde_Yano_W-smooth_on} corresponding to $\tilde{t}_{1,\phi}$ and $\tilde{t}_{2,\phi}$. When evaluated at $\phi \in S$ and smeared with $u_1, \dots, u_{n+1} \in T_\phi S$, this difference reads
					\begin{equation*}
						\begin{split}
							\sum_{j =2}^{n+1} \int_{M^{3}} u_1(x'_1) u_j(x'_j)  \Gamma_\phi(z_j,x'_1,x'_j) (\tilde{t}_{1,\phi} - \tilde{t}_{2,\phi})(u_2, \dots, z_j , \dots u_{n+1}) dx'_1 dx'_j dz_j = 0,
						\end{split}
					\end{equation*}
				because for any $\phi \in S$ the two extension $\tilde{t}_{1,\phi}$ and $\tilde{t}_{2,\phi}$ must coincides by construction when smeared with smooth $P_\phi$-solutions. This implies that the second term in~\eqref{tilde_Yano_W-smooth_on} satisfies the requirement (a) as we wanted to show.
				\item To prove that the conditions~\ref{W1},~\ref{W2} are satisfied, we compute the $\nu$-Gateaux derivative of the second term of eq.~\eqref{tilde_Yano_W-smooth_on} by distributing the variational derivatives on the factors that compose this term. Similarly as for $\mathring{\nabla}$ (see prop.~\ref{prop_LC_W-smooth_on}), it can be easily seen that the $\nu$-Gateaux derivative of the second term of eq.~\eqref{tilde_Yano_W-smooth_on} can be decomposed into a finite sum of appropriate compositions of $\sigma_c$, Gateaux derivatives of $E_\phi$ and Gateaux derivatives of $G_\phi$. The same argument we sketched in the proof of prop.~\ref{prop_LC_W-smooth_on} implies that each term in this decomposition satisfies~\ref{W1},~\ref{W2}. Therefore, the second term in~\eqref{tilde_Yano_W-smooth_on} satisfies the requirement (b) as we claimed.
			\end{enumerate}
			Thus, $\nabla^W t$ is a well-defined on-shell $W$-smooth covariant section.\\
			
			Next, we show that $\nabla^W t$ does not depend on the choice of the cut-off function $c$. It is sufficient to prove that for any functions $c, c'$ as in eq.~\eqref{kernel_symp}, it holds
				\begin{equation}\label{tilde_Y_difference_c-c'}
					\widetilde{(\nabla^{W} t)}_\phi -\widetilde{(\nabla^{W} t)'}_\phi \approx 0
				\end{equation}
			where the prime refers to a quantity defined with respect to the cut-off $c'$ instead of $c$, and where $\approx$ means ``equal up to distributions in $P_\phi\cE'_W(M^{n+1})$'' exactly as in lemma~\ref{lemma_approx}.\\
			Because $\mathring{\nabla} t$ has already been shown to be independent of $c$ (see prop.~\ref{prop_LC_W-smooth_on}), we can equivalently write eq.~\eqref{tilde_Y_difference_c-c'} as
				\begin{equation}\label{tilde_Y_difference_c-c'_alt}
					\left( \widetilde{(\nabla^W t)}_\phi - \compositeaccents{\widetilde}{(\mathring{\nabla}^{W} t)}_\phi\right) - \left( \widetilde{(\nabla^W t)'}_\phi - \compositeaccents{\widetilde}{(\mathring{\nabla}^{W} t)'}_\phi \right) \approx 0,
				\end{equation}
			 By using lemma~\ref{lemma_approx} and the extensions $\compositeaccents{\widetilde}{(\mathring{\nabla}^{W} t)}_\phi$ and $\widetilde{(\nabla^W t)}_\phi$ provided respectively by eq.~\eqref{tilde_LC_W-smooth_on} and eq.~\eqref{tilde_Yano_W-smooth_on}, we obtain
				\begin{equation}\label{tilde_Y_difference_c-c'_alt_2}
					\begin{split}
						&\left( \widetilde{(\nabla^W t)}_\phi - \compositeaccents{\widetilde}{(\mathring{\nabla}^{W} t)}_\phi\right) - \left( \widetilde{(\nabla^W t)'}_\phi - \compositeaccents{\widetilde}{(\mathring{\nabla}^{W} t)'}_\phi \right) \approx\\
						&\quad \approx \sum_{j=2}^{n+1} \int_{M^{3}} (\sigma_c \circ E_\phi)(x_1, x_1') (\sigma_c \circ E_\phi)(x_j, x_j') \left\{ \left(\mathring{\Gamma}_\phi - \Gamma_\phi \right) - \left(\mathring{\Gamma}'_\phi - \Gamma'_\phi \right) \right\}(z_j,x'_1,x'_j) \times \\
						&\qquad \qquad  \times \tilde{t}_\phi(x_2, \dots, z_j , \dots x_{n+1}) dx'_1 dx'_j dz_j.
					\end{split}
				\end{equation}
			Next, we notice that the difference between $\mathring{\Gamma}_\phi$ and $\Gamma_\phi$ can be written as
				\begin{equation}\label{symbols_LC-Y}
					\begin{split}
						\left(\mathring{\Gamma}_\phi - \Gamma_\phi \right) (x_1,x_2,x_3) &=  \frac{1}{8}\int_M (E_\phi \circ \sigma_c)(x_1, z) N_\phi(z,x_2,x_3) dz -\\
						&\qquad - \frac{1}{8} \bP^+_{x_2, x_3} \int_{M^2} G_\phi(x_1,z) \left( N_\phi(z',z,x_2)(\sigma_c \circ G_\phi \circ \sigma_c)(z',x_3) \right) dz dz',
					\end{split}
				\end{equation}
			After a closer inspection of eq.~\eqref{tilde_Y_difference_c-c'_alt_2} and eq.~\eqref{symbols_LC-Y}, we notice that to prove eq.~\eqref{tilde_Y_difference_c-c'_alt} it is sufficient to verify
				\begin{equation}\label{sigma_N_c-c'}
					\int_{M^4} \prod_{i=1}^3 E_\phi(x_i ,x'_i) \left( \sigma_c(x'_1, z) N_{\phi}(z, x'_2, x'_3) - \sigma_{c'}(x'_1, z) N'_{\phi}(z, x'_2, x'_3) \right) dz dx'_1 dx'_2 dx'_3= 0.
				\end{equation}
			As we already mentioned in the proof of prop.~\ref{prop_LC_W-smooth_on}, it follows straightforwardly from eq.~\eqref{sympl_form/symp_kernel_sol} that $\sigma_c - \sigma_{c'}$ vanishes if smeared with $P_\phi$-solutions. Therefore, it holds $E_\phi \circ (\sigma_c - \sigma_{c'}) \circ G_\phi =0$. Using this result and the definition of the distribution $N_\phi$ (given by eq.~\eqref{Nijenhuis_inf}), we can rewrite (up to a factor $2$) the left-hand side of eq.~\eqref{sigma_N_c-c'} as
				\begin{equation}\label{sigma_N_c-c'_alt}
					\begin{split}
						&\int_{M^4} \prod_{i=1}^3 E_\phi(x_i, x'_i) \bP^{-}_{x'_2, x'_3} \left[ (\sigma_c \circ G_\phi)(x'_3,z) \left( (\sigma_c - \sigma_{c'}) \circ \frac{\delta G_\phi}{\delta \phi(z)} \circ \sigma_c \right) (x'_1 ,x'_2)\right] dz dx'_1 dx'_2 dx'_3 + \\
						&+ \int_{M^4} \prod_{i=1}^3 E_\phi(x_i ,x'_i) \bP^{-}_{x'_2, x'_3} \left[ (G_\phi \circ \sigma_c) (z, x'_3) \left( \sigma_c \circ \frac{\delta G_\phi}{\delta \phi(z)} \circ (\sigma_c - \sigma_{c'}) \right)(x'_1,x'_2) \right]  dz dx'_1 dx'_2 dx'_3  + \\
						&+ \int_{M^3} \prod_{i=1}^3 E_\phi(x_i ,x'_i) \bP^{-}_{x'_2, x'_3} \left[ \left( \sigma_c \circ G_\phi \circ (\sigma_c - \sigma_{c'}) \circ \frac{\delta G_\phi }{\delta \phi(x'_2)} \circ \sigma_c \right)(x'_1,x'_3) \right] dx'_1 dx'_2 dx'_3 + \\
						&+ \int_{M^3} \prod_{i=1}^3 E_\phi(x_i ,x'_i) \bP^{-}_{x'_2, x'_3} \left[ \sigma_c \circ \left(G_\phi \circ \sigma_{c'} \circ \frac{\delta G_\phi}{\delta \phi(x'_2)} \circ (\sigma_c - \sigma_{c'}) \right) (x'_1,x'_3) \right] dx'_1 dx'_2 dx'_3.
					\end{split}
				\end{equation}
			We now show that (1) the first term in~\eqref{sigma_N_c-c'_alt} vanishes, (2) the fourth term in~\eqref{sigma_N_c-c'_alt} vanishes, and (3) the sum of the second and the third terms in~\eqref{sigma_N_c-c'_alt} also vanishes.
				\begin{enumerate}[label=(\arabic*), start=1]
					\item Because $G_\phi$ satisfies the hypotheses of lemma~\ref{lemma_inf_comm}, it follows that the map
							\begin{equation*}
								M \ni x \mapsto \int_{M^3} \left( G_\phi(x_1,y) \frac{\delta G_\phi}{\delta \phi(y)}(x_2,x) - G_\phi(x_2,y) \frac{\delta G_\phi}{\delta \phi(y)}(x_1, x) \right) f_1(x_1) f_2(x_2) dy dx_1 dx_2
							\end{equation*}
						is a smooth $P_\phi$-solution for any test functions $f_1, f_2$. Thus, the first term in~\eqref{sigma_N_c-c'_alt} vanishes.
				\item Applying the Leibniz rule for the variational derivative, it follows from the equation $G_\phi \circ (\sigma_c - \sigma_{c'}) \circ E_\phi =0$ that it holds
							\begin{equation}\label{tech_c-c'_G-E}
								\begin{split}
									0&= \frac{\delta (G_\phi \circ (\sigma_c - \sigma_{c'}) \circ E_\phi)}{\delta \phi(y)}(x_1,x_2)\\
									&= \left( \frac{\delta G_\phi}{\delta \phi(y)} \circ (\sigma_c - \sigma_{c'}) \circ E_\phi \right)(x_1,x_2) + \left(G_\phi \circ (\sigma_c - \sigma_{c'}) \circ  \frac{\delta E_\phi}{\delta \phi(y)} \right)(x_1,x_2).
								\end{split}
						\end{equation}
					Because $G_\phi$ is a bi-solution with respect to $P_\phi$, it follows from eq.~\eqref{sympl_form/symp_kernel_sol} that $G_\phi \circ \sigma_{c'} \circ G_\phi = G_\phi \circ \sigma_c \circ G_\phi$, and so it follows from eq.~\eqref{G_tilde_G_flat} that we have
							\begin{equation}\label{tech_c-c'_G-E_2}
								- G_\phi \circ \sigma_{c'} \circ G_\phi \circ \sigma_c \circ E_\phi = E_\phi.
							\end{equation} 
					Using eq.~\eqref{tech_c-c'_G-E} and eq.~\eqref{tech_c-c'_G-E_2}, we express the last term in~\eqref{sigma_N_c-c'_alt} as
						\begin{equation*}
							 \int_{M^2} (E_\phi \circ (\sigma_c - \sigma_{c'}))(x_1, z) \left\{ E_\phi(y, x_2) \frac{\delta E_\phi(z,x_3)}{\delta \phi(y)} - E_\phi(y, x_3) \frac{\delta E_\phi(z,x_2)}{\delta \phi(y)}\right\} dz dy.
						\end{equation*}
					Since $E_\phi$ satisfies the hypotheses of lemma~\ref{lemma_inf_comm}, it follows that the map
						\begin{equation*}
							M \ni x \mapsto \int_{M} \left( E_\phi(x_1,y) \frac{\delta E_\phi}{\delta \phi(y)}(x_2,x) - E_\phi(x_2,y) \frac{\delta E_\phi}{\delta \phi(y)}(x_1, x) \right) f_1(x_1) f_2(x_2) dy dx_1 dx_2
						\end{equation*}
					is a smooth $P_\phi$-solutions for any test function $f_1,f_2$. Thus, also the last term in~\eqref{sigma_N_c-c'_alt} vanishes.
				\item Next, we focus on the second and the third terms of of eq.~\eqref{sigma_N_c-c'_alt}.
						Using  eq.~\eqref{tech_c-c'_G-E} and the fact that $G_\phi \circ \sigma_c \circ E_\phi = G_\phi$, we write the second term of eq.~\eqref{sigma_N_c-c'_alt} as
							\begin{equation*}
								\int_{M^2} (E_\phi \circ \sigma_c \circ G_\phi \circ (\sigma_c - \sigma_{c'}))(x_1, z) \bP^{-}_{x_2, x_3} \left[G_\phi (y, x_3) \frac{\delta E_\phi(z,x_2)}{\delta \phi(y)} \right]  dz dy.
							\end{equation*}
						Since $G_\phi \circ \sigma_c \circ E_\phi = G_\phi$, using the Leibniz rule for the variational derivative, we have
							\begin{equation*}
								\left( \frac{\delta G_\phi}{\delta \phi(y)} \circ \sigma_c \circ E_\phi\right)(x_1,x_2) + \left( G_\phi \circ \sigma_c \circ \frac{\delta  E_\phi} {\delta \phi(y)}\right)(x_1,x_2) = \frac{\delta G_\phi(x_1,x_2)}{\delta \phi(y)}.
							\end{equation*}
						Then, this result implies that the third term of eq.~\eqref{sigma_N_c-c'_alt} equals
							\begin{equation*}
								- \int_{M^2}(E_\phi \circ \sigma_c \circ G_\phi \circ (\sigma_c - \sigma_{c'}))(x_1, z) \bP^{-}_{x_2, x_3} \left[E_\phi(x_2, y) \frac{\delta G_\phi (z,x_3)}{\delta \phi(y)} \right] dz dy.
							\end{equation*}
						Adding the second and the third terms, we obtain
							\begin{equation*}
								\begin{split}
									&\int_{M^2}(E_\phi \circ \sigma_c \circ G_\phi \circ (\sigma_c - \sigma_{c'}))(x_1, z) \bP^{-}_{x_2, x_3} \left[ G_\phi (y, x_3) \frac{\delta E_\phi(z,x_2)}{\delta \phi(y)} - \right. \\
									&\qquad - \left. E_\phi(x_2, y) \frac{\delta G_\phi (z,x_3)}{\delta \phi(y)} \right] dz dy,
								\end{split}
							\end{equation*}
						which vanishes because, by lemma~\ref{lemma_inf_comm}, the map
							\begin{equation*}
								M \ni x \mapsto \int_{M} \left( E_\phi(x_1,y) \frac{\delta G_\phi}{\delta \phi(y)}(x_2,x) - G_\phi(x_2,y) \frac{\delta E_\phi}{\delta \phi(y)}(x_1, x) \right) f_1(x_1) f_2(x_2) dy dx_1 dx_2
							\end{equation*}
						 is a smooth $P_\phi$-solution for any test functions $f_1,f_2$.
				\end{enumerate}
			Summing up, we have that~\eqref{tilde_Y_difference_c-c'_alt_2} vanishes, i.e. we verified eq.~\eqref{sigma_N_c-c'_alt}. As already mentioned, this implies that $\nabla^W t$ is independent of the choice of the cut-off $c$.\\
		 
			By construction, the map $\nabla^W$ is a $W$-covariant derivative. Therefore, to conclude the proof, we need to show that $\nabla^W$ preserves both the covariant sections $\mu$ and $\sigma$.\\
			We have already proved that $\nabla^W \mu$ does depend neither on the choice of the extension nor on the choice of the cut-off. Then, to show that $\nabla^W \mu = 0$, it is sufficient to prove that $\widetilde{(\nabla^W \mu)}_\phi = 0$, where $\widetilde{(\nabla^W \mu)}_\phi$ is given by eq.~\eqref{tilde_Yano_W-smooth_on} for a specific $W$-smooth extension of $\mu$ and an arbitrary but fixed cut-off function $c$ as in eq.~\eqref{kernel_symp}. If we choose $\sigma_c \circ G_\phi \circ \sigma_c$ as our $W$-smooth extension of $\mu_\phi$ (we have proved in thm.~\ref{theo_var_controll_state} that this is allowed), then it follows
				\begin{equation}\label{Yano_pres_mu}
					\begin{split}
						&\widetilde{(\nabla^W \mu)}_\phi (x_1,x_2,x_3) =\\
						&\quad = \compositeaccents{\widetilde}{(\mathring{\nabla}^{W} \mu)}_\phi(x_1, x_2, x_3) - 2 \bP^+_{x_2,x_3} \int_{M^3} (\sigma_c \circ E_\phi)(x_1, x'_1) (\sigma_c \circ G_\phi \circ \sigma_c)(x_2, z)  \times \\
						&\qquad \qquad \qquad \qquad \times \left( \Gamma_\phi - \mathring{\Gamma}_\phi \right)(z, x'_1, y)  (E_\phi \circ \sigma_c)(y, x_3) dx'_1 dy dz
					\end{split}
				\end{equation}
			As we have already proved in prop.~\ref{prop_LC_W-smooth_on} (see eq.~\eqref{LC_pres_mu}), $\compositeaccents{\widetilde}{(\mathring{\nabla}^{W} \mu)}_\phi$ vanishes. We need to show that the second term in eq.~\eqref{Yano_pres_mu} also vanishes. By the definition of $\Gamma_\phi$ (see eq.~\eqref{Yano_symbol_W-smooth_on}), and since $N_\phi(x_1,x_2,x_3)$ is anti-symmetric in $x_2,x_3$, we have
				\begin{equation*}
					\begin{split}
						&\bP^+_{x_2,x_3} \int_{M^2}  (G_\phi \circ \sigma_c)(x_2, z)  \left( \Gamma_\phi - \mathring{\Gamma}_\phi \right)(z, x_1, y) E_\phi(y, x_3) dy dz =\\
						& = - \frac{1}{8} \bP^+_{x_2,x_3} \int_{M^2} (G_\phi \circ \sigma_c)(x_2, z)  N_\phi (z, x_1, y) E_\phi(y, x_3) dy dz +\\
						&\qquad + \frac{1}{8} \bP^+_{x_2,x_3} \int_{M^2} (G_\phi \circ \sigma_c \circ G_\phi)(x_2, y)  N_\phi (z, y, x_1) (\sigma_c \circ G_\phi \circ \sigma_c \circ E_\phi)(z, x_3) dy dz \\
						& = - \frac{1}{8} \bP^+_{x_2,x_3} \int_{M^2} (G_\phi \circ \sigma_c)(x_2, z)  N_\phi (z, x_1, y) E_\phi(y, x_3) dy dz -\\
						&\qquad - \frac{1}{8} \bP^+_{x_2,x_3} \int_{M^2} E_\phi(x_2, y)  N_\phi (z, y, x_1) (\sigma_c \circ G_\phi)(z, x_3) dy dz\\
						&= - \frac{1}{8} \bP^+_{x_2,x_3} \int_{M^2} (G_\phi \circ \sigma_c)(x_2, z)  \left( N_\phi (z, x_1, y)  + N_\phi (z, y, x_1) \right) E_\phi(y, x_3) dy dz = 0,
					\end{split}
				\end{equation*}
			where we used $E_\phi \circ \sigma_c \circ G_\phi = G_\phi$, which follows from eq.~\eqref{causal_prop_inv_symp_kernel_sol}, and $G_\phi \circ \sigma_c \circ G_\phi = E_\phi \circ \sigma_c \circ G_\phi \circ \sigma_c \circ G_\phi = -E_\phi$, which is a consequence of eq.~\eqref{G_tilde_G_flat}. Therefore, also the second term in eq.~\eqref{Yano_pres_mu} vanishes, as we needed to prove.\\
			To show that $\nabla^W \sigma = 0$, similarly as before, it is sufficient to prove that $\widetilde{(\nabla^W \sigma)}_\phi = 0$, where $\widetilde{(\nabla^W \sigma)}_\phi$ is given by eq.~\eqref{tilde_Yano_W-smooth_on} for a specific $W$-smooth extension of $\sigma$ and an arbitrary but fixed cut-off function $c$ as in eq.~\eqref{kernel_symp}. If we choose $\sigma_c \circ E_\phi \circ \sigma_c$ as our extension (we have proved in thm.~\ref{theo_W_symp} that this is allowed), then we obtain
				\begin{equation}\label{Yano_pres_sigma}
					\begin{split}
						&\widetilde{(\nabla^W \sigma)}_\phi (x_1, x_2, x_3) =\\
						&\quad = \widetilde{(\partial \sigma)}_\phi(x_1,x_2,x_3) + 2 \bP^-_{x_2,x_3} \int_{M^3} (\sigma_c \circ E_\phi)(x_1, x'_1) (\sigma_c \circ E_\phi \circ \sigma_c)(x_2, z)  \Gamma_\phi(z, x'_1, y) \times \\
						&\qquad \qquad \qquad \qquad \times (E_\phi \circ \sigma_c)(y, x_3) dx'_1 dy dz.
					\end{split}
				\end{equation}
			As we have already proved in thm.~\ref{theo_W_symp} (see eq.~\eqref{partial_tilde_sigma}), $\widetilde{(\partial \sigma)}_\phi$ vanishes. The second term in eq.~\eqref{Yano_pres_sigma} also vanishes as can be checked by direct calculation. The essential point is proving that it holds
				\begin{equation*}
					\begin{split}
						0 &= \int_{M^4} \prod_{j=1}^3 E_\phi(x_j,x'_j) \left[ (\sigma_c \circ G_\phi) (x'_2,z) \frac{\delta (\sigma_c \circ G_\phi \circ \sigma_c)(z,x'_3) }{\delta \phi(x'_1)} \right. -\\
						&\qquad - \left.  (\sigma_c \circ G_\phi) (x'_3,z) \frac{\delta (\sigma_c \circ G_\phi \circ \sigma_c)(z,x'_2)}{\delta \phi(x'_1)}\right] dz dx'_1 dx'_2 dx'_3,
					\end{split}
				\end{equation*}
			which is a consequence of the Leibniz rule of the variational derivative, eq.~\eqref{G_tilde_G_flat} and eq.~\eqref{causal_prop_inv_symp_kernel_sol}. This concludes the proof.
		\end{proof}
		
	As the analogy with the finite-dimensional setting suggests, the torsion of the $W$-covariant derivative $\nabla^W$ is in general non-zero. Actually, we can compute the torsion explicitly:
		\begin{equation}\label{Yano_torsion_inf}
			\begin{split}
				T^{(\nabla)} (t)_\phi &= \bP^- (\nabla^W t) _\phi - (d t)_\phi \\
				&= - \frac{1}{4} \int_{M^4} (\sigma_c \circ E_\phi)^{\otimes 2} (x_1,x_2,x_1',x_2') N_\phi(z',x_1',x_2') (\sigma_c \circ E_\phi)(z',z) t_\phi(z) dz dx'_1 dx'_2 dx'_3,
			\end{split}
		\end{equation}
	for any $t \in C^\infty_W(S,T^*S)$.\\
	
	We conclude this section by discussing the extension of the covariant derivative $\nabla^W$ to $C^\infty_W(S,\cW)$, the on-shell $W$-smooth sections on the algebra bundle $\cW$, and further to $\Omega_W(S,\cW)$, the $\cW$-valued forms.\\
	We can extend $\nabla^W$ as a map $C^\infty_W(S,\cW) \to \Omega_W^1(S,\cW)$ by the following canonical procedure. Let $t \in C^\infty(S, \cW)$ be a section homogeneous in both the degrees $\deg_s$ and $\deg_\hbar$, with $\deg_s t = n$. This means that $t$ is a complex section in $\vee_W^n T^* S$ (up to a factor $\hbar^{\deg_\hbar t}$). The on-shell $W$-smooth form $\nabla^W t$ is defined by the sequence $((\nabla^W t)^{1,0}, (\nabla^W t)^{1,1}, \dots)$, where $(\nabla^W t)^{1,\ell} = 0$ if $\ell \neq n$ and
		\begin{equation*}
			S \ni \phi \mapsto (\nabla^W t)^{1,n}_\phi(y_1,x_1, \dots, x_n) = \bP^+(\nabla^W t)(y_1, x_1, \dots, x_n).
		\end{equation*}
	By abuse of notation, we identify an equivalence classes in $\wedgevee_W^{1,n} T^*_\phi S$ (i.e. the elements in $\boxtimes_W^{n+1} T^*_\phi S$ which are symmetric in the last $n$ entries) and in $\boxtimes_W^{n+1} T^*_\phi S$ with their distributional representatives. We implicitly assumed that $\nabla^W \hbar = 0$. We note that $\nabla^W$, as a map acting on sections in $C^\infty_W(S,\cW)$ homogeneous in $\deg_s$ and $\deg_\hbar$, preserves the total degree $\Deg$. Thus, it extends as a map $C^\infty_W(S,\cW) \to \Omega_W^1(S,\cW)$ via the $\Deg$-filtration and the fact that every on-shell $W$-smooth section $C^\infty_W(S,\cW)$ which is homogeneous in $\Deg$ is a finite collection of on-shell $W$-smooth sections homogeneous in $\deg_s$ and $\deg_\hbar$. In other words, the extension is performed purely algebraically in the very same way the Yano connection in the finite-dimensional case is extended to sections on the formal Wick algebra (see sec.~\ref{subsec_Fedosov_fin}).\\
	 The extension of the covariant derivative $\nabla^W$ as an operator $\Omega_W(S,\cW) \to \Omega_W(S,\cW)$ and its properties are discussed in the following proposition:
		\begin{prop}\label{prop_Yano_W-smooth_on_form}
			Let $c$ be a cut-off function as in eq.~\eqref{kernel_symp} and let $\phi \mapsto \omega_\phi$ be an admissible assignment. For any $t_\phi \in \Omega^k_W(S, \cW)$ homogeneous in $\deg_\hbar$ and in $\deg_s$ (with $\deg_s t=n$), we define
				\begin{equation}\label{tilde_Yano_W-smooth_on_form}
					\begin{split}
						&(\widetilde{\nabla^W t})_\phi(y_1, \dots, y_{k+1}, x_1, \dots, x_{n1}) :=\\
						&\quad = \bP^+ \bP^- \widetilde{(\partial t)}_\phi(y_1, \dots, y_{k+1}, x_1, \dots, x_n)  - \bP^+ \bP^- \int_{M^3} (\sigma_c \circ E_\phi)(y_1, y'_1) (\sigma_c \circ E_\phi)(x_1, x'_1) \times \\
						&\qquad \qquad \qquad \qquad \times \Gamma_\phi(z,y'_1,x'_1) \tilde{t}_\phi(y_2, \dots , y_{k+1},z, x_2, \dots, x_n) dz dy'_1 dx'_1,
					\end{split}
				\end{equation}
			where $\Gamma_\phi$ is the distribution given by eq.~\eqref{Yano_symbol_W-smooth_on} corresponding to $\omega_\phi$, where $\tilde{t}_\phi \in (\sigma_c \circ E_\phi)^{\otimes n} \cE'_W(M^n)$ is an extension of $t$ (up to $\hbar^{\deg_\hbar t}$), and where $\widetilde{(\partial t)}_\phi \in (\sigma_c \circ E_\phi)^{\otimes n +1} \cE'_W(M^{n+1})$ is the extension (up to $\hbar^{\deg_\hbar t}$) of $\partial t$ defined by eq.~\eqref{tilde_diff} with respect to the fixed choice of $c$.\\
			The distribution~\eqref{tilde_Yano_W-smooth_on_form} defines a $\cW$-valued $k+1$-form homogeneous in $\deg_\hbar$ and in $\deg_s$ (with $\deg_s t=n$) by restriction to $(TS)^{\otimes n+k+1}$,  i.e. denoted by 
				\begin{equation}\label{restr_Yano_on_form}
					(\nabla^W t)_\phi (v_1, \dots v_{k+1}, u_1, \dots, u_{n}) := \widetilde{(\nabla^{W} t)}_\phi(v_1, \dots v_{k+1}, u_1, \dots, u_{n}) \quad \forall \phi \in S, \, v_i, u_j \in T_\phi S.
				\end{equation}
			The section $\nabla^W t$ does not depend on the choice of the cut-off $c$.\\
			$\gls{nabla_W}$ is a bilinear map $\nabla^W: \Omega_W^k(S,\cW) \to \Omega_W^{k+1}(S,\cW)$ for any $k$ which satisfies the Leibniz rule with respect to the product $\bullet$~\eqref{prod_dega,degs-hom}, i.e.
				\begin{equation}\label{Leibniz_dega,degs-hom}
					\nabla^W (t \bullet s) = (\nabla^W t) \bullet s + (-1)^k t \bullet (\nabla^W s),
				\end{equation}
			for any $t \in \Omega_W^k(S,\cW)$ and $s \in \Omega_W^{k'}(S,\cW)$. Consequently $\nabla^W$ extends to $\Omega_W(S,\cW)$ as a $\deg_a$-graded derivative which preserves the $\Deg$-grading.
		\end{prop}
		\begin{proof}
			Arguing similarly as done for prop.~\ref{prop_Yano_W-smooth_on} and exploiting the symmetry properties in eq.~\eqref{tilde_Yano_W-smooth_on_form}, we can prove that $(\widetilde{\nabla^W t})_\phi$ is a well-defined distribution which defines via eq.~\eqref{restr_Yano_on_form} a $\cW$-valued $k+1$-form homogeneous in $\deg_\hbar$ and in $\deg_s$ independently of the choice of $c$.\\
			By construction, $\nabla^W$ increases by one the degree $\deg_a$ and preserves the total degree $\Deg$. Exploiting the $\Deg$-filtration of $\Omega_W^k(S,\cW)$ and the fact that each $\cW$-valued $k$-form homogeneous in $\Deg$ is a finite collection of $k$-forms homogeneous in $\deg_s$ and in $\deg_\hbar$, the connection $\nabla^W$ extends canonically to a map $\Omega_W^{k}(S,\cW) \to \Omega_W^{k+1}(S,\cW)$. Then, $\nabla^W$ extends further to $\Omega_W(S,\cW) \to \Omega_W(S,\cW)$ in a standard way.\\
			Finally, we show that $\nabla^W$ satisfies the Leibniz rule with respect to the product $\bullet$~\eqref{prod_dega,degs-hom}. As a consequence of $\nabla^W \sigma=0$ and $\nabla^W \mu=0$, it follows that for any cut-off $c$ as in eq.~\eqref{kernel_symp} we have
				\begin{equation*}
					\begin{split}
						0 =&\int_{M^3} (\sigma_c \circ E_\phi)^{\otimes 3}(y,x'_1,x'_2,y',x_1,x_2) \times \\
						&\qquad \times \left( \frac{\delta \omega_\phi}{\delta \phi(y')}(x'_1,x'_2) + \int_{M} \left\{ \omega_\phi(z,x'_2) \Gamma_\phi(x'_1,y',z)  + \omega_\phi(x'_1,z) \Gamma_\phi(x'_2,y',z) \right\} dz\right) dy' dx'_1 dx'_2.
					\end{split}
				\end{equation*}
			As can be checked by direct computation, this result implies that $\nabla^W$ must satisfy eq.~\eqref{Leibniz_dega,degs-hom} and this concludes the proof.
		\end{proof}	
		
\section{The $W$-smooth Fedosov connection and Fedosov's theorems for QFT}\label{subsec_Fedosov_inf}
	In this section, we conclude our infinite-dimensional version of the Fedosov quantization scheme. In the previous sections,  sec.~\ref{subsec_manifold_inf}-\ref{subsec_W_covariant_der}, we defined and discussed all the geometrical notions needed, with the exception of the infinite-dimensional analogues of the Fedosov operators $\delta, \delta^{-1}$, see~\eqref{Fedosov_op},~\eqref{Fedosov_op_-1}, which are provided now. Then, we prove that Fedosov's theorems (thm.~\ref{theo_Fedosov_1_fin} and thm.~\ref{theo_Fedosov_2_fin}) extend to the infinite-dimensional framework we have set up. This result relies on two fundamental facts:
	\begin{itemize}
		\item {\em `Analytic' properties}: $\nabla^W$, $\delta$, $\delta^{-1}$, and the product $\bullet$ preserve the on-shell $W$-smoothness, i.e. these operators map on-shell $W$-smooth forms into on-shell $W$-smooth forms.
		\item {\em `Algebraic' properties}: The algebraic identities of lemma~\ref{lemma_various_res_fin}, which are used in the finite-dimensional proof, are preserved in the infinite-dimensional context.
	\end{itemize}
	Therefore, in infinite dimensions, the Fedosov's theorems can be proved repeating the same algebraic argument used in the proofs in the finite-dimensional case.\\
	
	First we define the Fedosov operators $\delta, \delta^{-1}$ in our infinite-dimensional setting. The Fedosov operator $\gls{delta}:\Omega_W^k(S,\cW) \to \Omega_W^{k+1}(S,\cW)$ (cf.~\eqref{Fedosov_op_alt}) is defined by its action on $k$-forms homogeneous in $\deg_s$ as
		\begin{equation}\label{delta_phi}
				(\delta t)^{k+1,n}_\phi (y_1, \dots, y_{k+1}; x_1, \dots, x_n) :=(n+1) \bP^- t_\phi^{k,n+1}(y_{1}, \dots, y_{k}; y_{k+1}, x_1, \dots, x_n),
		\end{equation}
	for $n+1 = \deg_s t$, while $(\delta t)^{k+1,n} = 0$ otherwise. Here $\bP^-$ acts as an anti-symmetrization on the $y$-variables. Note that, by abuse of notation, we identify an equivalence class in $\bC[[\hbar]] \otimes \wedgevee^{k,n} T^*_\phi S$ (i.e. the elements in $\bC[[\hbar]] \otimes \boxtimes_W^{k+n} T^*_\phi S$ which are anti-symmetric in the first $k$ entries and symmetric in the remaining $n$) with one of its $\bC[[\hbar]]$-valued distributional representatives in $\cE'_W(M^{k+n})$. Because $\delta t$ is a finite sum, it is clearly well-defined and it extends to $\Omega^k_W(S,\cW)$ by using the $\Deg$-filtration and the fact that each form homogeneous in the total degree $\Deg$ decomposes into a finite sum of terms homogeneous in $\deg_s$. The map $\delta$ can be extended further to $\Omega_W(S,\cW)$ in a standard way.\\
	Using the same procedure the operator $\gls{delta-1}:\Omega^k_W(S,\cW) \to \Omega_W^{k-1}(S,\cW)$ (cf.~\eqref{Fedosov_op_-1_alt}) is defined by
		\begin{equation}\label{delta_-1_phi}
				(\delta^{-1} t)^{k-1,n}_\phi (y_1, \dots, y_{k-1}; x_1, \dots, x_{n}) := \frac{k}{n+k-1} \bP^+ t_\phi^{k,n-1}(y_1, \dots, y_{k-1}, x_{1}; x_2, \dots, x_{n})
		\end{equation}
	for $n-1 =\deg_s t$ and $k \neq 0$, while $(\delta^{-1} t)^{k-1,n} = 0$ otherwise. Here $\bP^+$ acts as a symmetrization on the $x$-variables. Because $\delta^{-1}t$ involves only finitely-many terms, it is well-defined and it can be extended canonically to $\Omega_W(S,\cW)$, similarly as for $\delta$.\\
	
	Concerning the infinite-dimensional version of lemma~\ref{lemma_various_res_fin}, which collects all the necessary algebraic relations to prove Fedosov's theorems, we adopt again the pedestrian approach we have already used in the previous sections. In finite dimensions, the torsion tensor and the Riemann curvature tensor of the Yano connection, more precisely the contractions of the aforementioned tensors with the symplectic form (cf.~\eqref{R_hat_T_hat}), appear in the relations between $\delta$, $\delta^{-1}$ and $\nabla$ we are interested in. We are going to show that such tensors have appropriate infinite-dimensional counterparts.
		\begin{lemma}\label{lemma_various_res_inf}
			For any $\phi \in C^\infty(M)$ and any cut-off $c$ as in eq.~\eqref{kernel_symp}, we define the distributions $\compositeaccents{\widetilde}{\hat{T}}_\phi$ and $\compositeaccents{\widetilde}{\hat{R}}_\phi$ as
				\begin{equation}\label{symp_torsion_inf}
					\begin{split}
					\compositeaccents{\widetilde}{\hat{T}}_\phi(y_1,y_2;x) &:= - \frac{1}{8} \bP^-_{y_1, y_2} \int_{M^4} \prod_{i=1}^2 (\sigma_c \circ E_\phi)(y_i,y'_i) (\sigma_c \circ E_\phi)(x, x')  \times\\
					&\qquad \times \sigma_c(x', z) N_\phi(z,y_1',y_2') dz dx' dy'_1 dy'_2,
					\end{split}
				\end{equation}
			and
				\begin{equation}\label{symp_curv_inf}
					\begin{split}
						\compositeaccents{\widetilde}{\hat{R}}_\phi(y_1,y_2;x_1,x_2) &:= \frac{1}{4}\bP^+_{x_1, x_2} \bP^-_{y_1,y_2} \int_{M^5} \prod_{i=1}^2(\sigma_c \circ E_\phi)(y_i,y'_i) \prod_{j=1}^2(\sigma_c \circ E_\phi)(x_j,x'_j) \times\\
						&\qquad \times \sigma_c(x'_1, z)  \left\{ \frac{\delta}{\delta \phi(y'_{1})} \int_{M^2} \Gamma_\phi(z,z',z'') (E_\phi \circ \sigma_c)^{\otimes 2}(z',z'',y'_{2},x'_2) dz' dz'' \right. + \\
						&\qquad \qquad \left. + \int_{M} \Gamma_\phi(z,y'_{2},z') \Gamma_\phi(z',y'_{1},x'_2) dz' \right\} dy'_1 dy'_2 dx'_1 dx'_2 dz,
					\end{split}
				\end{equation}
			where $N_\phi$ and $\Gamma_\phi$ are defined by eq.~\eqref{Nijenhuis_inf} and eq.~\eqref{Yano_symbol_W-smooth_on}, and where $\sigma_c$ is given by eq.~\eqref{kernel_symp}.\\
			By restriction to $\phi \in S$, the distributions~\eqref{symp_torsion_inf} and~\eqref{symp_curv_inf} define two totally homogeneous on-shell $W$-smooth $\cW$-valued $2$-forms $S \ni \phi \mapsto \hat{T}_\phi$ and $S \ni \phi \mapsto \hat{R}_\phi$. In particular, $\deg_\hbar \hat{T}, \deg_\hbar \hat{R}=0$, $\deg_s \hat{T}=1$, and $\deg_s \hat{R}=2$. The two on-shell $W$-smooth $\cW$-valued $2$-forms $\hat{T}$ and $\hat{R}$ do not depend on the choice of the cut-off $c$.\\
			Moreover, the results \ref{adj_action_Fedosov_op_fin}-\ref{Hodge_fin} listed in lemma~\ref{lemma_various_res_fin} for finite dimensions translate to infinite dimensions.
		\end{lemma}
		\begin{proof}
			To prove that the distributions $\compositeaccents{\widetilde}{\hat{T}}_\phi$ and $\compositeaccents{\widetilde}{\hat{R}}_\phi$ define, by restriction to $\phi \in S$, two on-shell $W$-smooth $\cW$-valued $2$-forms, we need to show that $\compositeaccents{\widetilde}{\hat{T}}_\phi, \compositeaccents{\widetilde}{\hat{R}}_\phi$ are well-defined distributions respectively in $(\sigma_c \otimes E_\phi)^{\otimes 3} \cE'_W(M^3)$ and $(\sigma_c \otimes E_\phi)^{\otimes 4} \cE'_W(M^4)$, anti-symmetric in the $y$-variables and symmetric in the $x$-variables, and that $\compositeaccents{\widetilde}{\hat{T}}_\phi, \compositeaccents{\widetilde}{\hat{R}}_\phi$ satisfy conditions~\ref{W1},~\ref{W2} in def.~\ref{def_smooth_on_tens}.  The proof of these two facts is ultimately a consequence of the machinery of composition of distributions (thm.~\ref{theo_WF_horma} and lemma~\ref{lemma_W_comp}), the definition of $\Gamma_\phi$ (eq.~\eqref{Yano_symbol_W-smooth_on}) and $N_\phi$ (eq.~\eqref{Nijenhuis_inf}), together with the properties of the distributions $E_\phi, \sigma_c, G_\phi$ and their directional derivatives (eq.~\eqref{var_ders_causal_WF}, eq.~\eqref{kernel_symp} and eq.~\eqref{state_WF_cond}).\\
			We first show that the on-shell $W$-smooth $\cW$-valued forms $\hat{T}$ and $\hat{R}$ do not depend on the choice of the cut-off $c$. Let assume that relations~\ref{Fedosov_op_vs_Yano_fin} and~\ref{square_Yano_fin} hold in our infinite-dimensional set-up, i.e.
				\begin{equation*}
					\delta \nabla^{W} - \nabla^W \delta = \frac{i}{\hbar} \ad_\bullet \hat{T}, \quad \left( \nabla^W \right)^2 = -\frac{i}{\hbar} \ad_\bullet \hat{R}.
				\end{equation*}
			It was proved in prop.~\ref{prop_Yano_W-smooth_on_form} that $\nabla^W$ does not depend on the choice of the cut-off. The same holds for $\delta$ by definition. Therefore, if $c,c'$ are two cut-off functions as in eq.~\eqref{kernel_symp}, the difference $\hat{T} - \hat{T}'$ and $\hat{R} - \hat{R}'$ (where $\hat{T}', \hat{R}'$ are the quantities corresponding to the cut-off $c'$), are on-shell $W$-smooth $\cW$-valued forms with $\deg_s \neq 0$ such that
				\begin{equation*}
					\ad_\bullet (\hat{T} - \hat{T}') = 0 = \ad_\bullet (\hat{R} - \hat{R}').
				\end{equation*}
			As proved in lemma~\citep[prop. 2.1]{HW01}, it follows from the definition of the product $\bullet$ on $\Omega(S,\cW)$ that the center of the algebra is $\Omega_W(S)$, i.e. an element $t \in \Omega_W(S,\cW)$ satisfies $\ad_\bullet t = 0$ if and only if $t$ has $\deg_s=0$. Thus, we conclude that both $\hat{T} - \hat{T}'$ and $\hat{R} - \hat{R}'$ vanish as we needed to prove.\\
			
			The last part of the lemma, namely the fact that results \ref{adj_action_Fedosov_op_fin}-\ref{Hodge_fin} listed in lemma~\ref{lemma_various_res_fin} translate unaltered in our infinite-dimensional set-up, can be proved by tedious direct computations. Rather than displaying the details of these computations which are qualitatively similar to those presented e.g. in the proof of prop.~\ref{prop_LC_W-smooth_on} and prop.~\ref{prop_Yano_W-smooth_on}, we present to the reader the basic arguments on which the computations rely.\\
			Formula~\ref{flat_Fedosov_fin}, which means that the Fedosov operators are nilpotent, and eq.~\ref{Hodge_fin}, i.e. the Hodge-type decomposition $\delta \delta^{-1} + \delta^{-1} \delta + \tau = \id$, are simply a matter of interplay of symemtrization and anti-symmetrization operators and, therefore, these properties hold also in the infinite-dimensional case.\\
			The proof of formula~\ref{adj_action_Fedosov_op_fin}, i.e. $\delta = \frac{2i}{\hbar}\ad_\bullet(\delta^{-1} \sigma)$, follows from the fact that $(\delta^{-1} \sigma)_\phi$ can be identified with $\frac{1}{2} (\sigma_c \circ E_\phi \circ \sigma_c)(x,y)$ as a consequence of thm.~\ref{theo_W_symp} and from the fact that for any $t \in \Omega^k(S,\cW)$ and any $\phi \in S$, the $\deg_s$-homogeneous part of $\ad_\bullet (\delta^{-1} \sigma)(t)_\phi$ with $\deg_s = n$ is proportional to
				\begin{equation*}
					\int_{M^2} (\sigma_c \circ E_\phi \circ \sigma_c)(z, y_1) E_\phi(z,z') \tilde{t}^{k,n+1}_\phi (y_2, \dots, y_{k+1}, z', x_1, \dots, x_n) dz dz'.
				\end{equation*}
			Here $\tilde{t}^{k,n}$ is any $W$-smooth extension of the on-shell $W$-smooth section $t^{k,n}$ in the sequence defining the form $t = (t^{k,n})_{k,n \in \bN}$ (by abuse of notation, we identify equivalence classes with their distributional representatives).\\
			To prove formula~\ref{Fedosov_op_vs_Yano_fin}, i.e. $\delta \nabla^W + \nabla^W \delta = \frac{i}{\hbar} \ad_{\bullet} (\hat{T})$, we notice first that using $\delta = \frac{2i}{\hbar}\ad_\bullet(\delta^{-1} \sigma)$ and the fact that $\nabla^W$ is a $\deg_a$-derivation in $\Omega_W(S, \cW)$ (as proved in prop.~\ref{prop_Yano_W-smooth_on_form}), it follows that $\delta \nabla^W + \nabla^W \delta = \frac{2i}{\hbar} \ad_{\bullet} (\nabla^W(\delta^{-1} \sigma))$. Using the fact that $\partial \sigma=0$ (as shown in the proof of thm.~\ref{theo_W_symp}), it follows that $2 (\nabla^W (\delta^{-1} \sigma)) = \hat{T}$ which concludes the proof.\\
			Checking eq.~\ref{square_Yano_fin}, i.e. $(\nabla^W)^2 = -\frac{i}{\hbar} \ad_\bullet(\hat{R})$ is more involved. It can be done by computing the two sides of the equation acting on a $\deg_s$-homogeneous $k$-form. Making use of the Leibniz rule and the flatness of the exterior derivative $d$ defined in prop.~\ref{prop_der_smooth_on}, we can verify that the two sides coincides.\\
			Finally, the last three formulas~\ref{extra_fin} hold as a consequence of eq.~\ref{Fedosov_op_vs_Yano_fin}, the flatness of $d$, the Leibniz rule, the fact that $\nabla^W \sigma = 0$ and $\partial \sigma=0$.
		\end{proof}
		
		Summing up, we have constructed the following {\em dictionary} between the finite-dimensional framework of sec.~\ref{subsec_Fedosov_fin} and our infinite-dimensional framework.
	\begin{table}[H]
	\begin{center}
			\renewcommand{\arraystretch}{1.5}
			\begin{tabular}{| l | l |}
				\hline
				 	{\bf finite-dim} & {\bf infinite-dim} \\
  				\hline                       
 				$S$ finite-dim manifold	& $S$ smooth sol. of the non-lin. eq.~\eqref{eom_phi4} on $M$	\\
 				\hline
 				$T_x S$ tangent space at $x \in S$,	& $T_\phi S$ smooth sol. of lin. eq.~\eqref{leom_phi4} at $\phi \in S$,\\
 				$\otimes^n T^\ast_x S$ tensor power of the cotangent space $T^\ast_x S$, 	& $\boxtimes^n_W T^\ast_\phi S$ defined by~\eqref{cov_tens_S},\\
 				$\cW_x = \bC[[\hbar]] \otimes \bigoplus_{n \geq 0} \vee^n T^\ast_x S$	&  $\cW_\phi = \bC[[\hbar]] \otimes \bigoplus_{n \geq 0} \vee^n_W T^\ast_\phi S$\\
 				\hline
 				Smoothness, 	&	On-shell $W$-smoothness, def.~\ref{def_smooth_on_f} and def.~\ref{def_smooth_on_tens},\\
 				$d$ exterior derivative 	& $d$ defined in prop.~\ref{prop_der_smooth_on}\\
 				\hline
 				Formal Wick product $\bullet$	& Product $\bullet$ defined in prop.~\ref{prop_prod_smooth_on_W_form}\\
 				\hline
 				$\sigma$ symplectic form	& On-shell $W$-smooth $2$-form $\sigma$\\
 				 	& defined in thm.~\ref{theo_W_symp}\\
 				\hline
 				$\mu$ compatible metric	& On-shell $W$-smooth symm. section $\mu$\\
 				 	& defined in thm.~\ref{theo_var_controll_state}\\
  				\hline
  				$\nabla$ Yano connection on $\cW$-valued forms	& $\nabla^W$ defined in prop.~\ref{prop_Yano_W-smooth_on_form}\\
  				\hline
  				$\delta$ Fedosov operator,	& $\delta$ defined by~\eqref{delta_phi},\\
  				$\delta^{-1}$ ``inverse'' Fedosov operator	& $\delta^{-1}$ defined by~\eqref{delta_-1_phi}\\
  				\hline
			\end{tabular}
		\end{center}	
		\caption{The dictionary between the finite-dimensional case and our infinite-dimensional setting.}\label{table:dictionary}
		\end{table}
		
	Continuing our pedestrian approach, we keep following the finite-dimensional Fedosov method outlined in sec.~\ref{subsec_Fedosov_fin}, in particular we can make the same ansatz~\eqref{Fedosov_conn_fin} for the Fedosov connection and we can translate the results obtained in sec.~\ref{subsec_Fedosov_fin} for finite dimension in our infinite-dimensional setting.
		\begin{theo}[Fedosov's 1st and 2nd theorems in $\infty$-dim]\label{theo_Fedosov_inf}
			The Fedosov's theorems (thm.~\ref{theo_Fedosov_1_fin} and thm.~\ref{theo_Fedosov_2_fin}) hold in our infinite-dimensional setting. In particular, we can add to the dictionary the following entry:
 		\begin{equation}\label{Fedosov_conn_inf}
			D \mbox{ Fedosov flat connection } \leftrightarrow  \gls{DW} := \nabla^W - \delta + \frac{i}{\hbar} \ad_\bullet(r),
		\end{equation}
		with $r \in \Omega^1_W(S, \cW)$ denoting the unique solution for
		\begin{equation}\label{r_eq_inf}
				\nabla^W r - \delta r + \frac{i}{\hbar} r \bullet r -\hat{R} - \hat{T} = \Omega
			\end{equation}
		subject to the requirements $r = r^\dagger$, $r^{(0)}=r^{(1)}=0$, $(\delta^{-1} r)^{(k)} = s^{(k)}$, where $r^{(k)}$ and $s^{(k)}$ denote the components of the sections $r$ and $s$ homogeneous in $\Deg$ of degree $k$, where $\Omega \in \Omega^2_W(S,\cW)$ is closed $(d\Omega = 0$) and $\deg_s \Omega = 0$ (i.e. it belongs to $\bC[[\hbar]] \otimes \Omega^2_W(S)$), and where $s$ is some arbitrary self-adjoint element in $C^\infty_W(S,\cW)$ with $\Deg s \geq 3$.\\
		The infinite-dimensional translation of the Fedosov's Second Theorem provides a deformation quantization $(C_W^\infty(S)[[\hbar]], \star)$, in the sense explained in sec.~\ref{subsec_manifold_inf}. More precisely, for a given $F \in C^\infty_W(S)[[\hbar]]$, there exists a unique section $\hat{F} \in C^\infty_W(S,\cW)$ which is flat with respect to the $D^W$, i.e. $D^W \hat{F} =0$, and which satisfies $\tau \hat{F} = F$. We define by $\tau^{-1}$ the map $F \mapsto \hat{F}$. Then, we obtain a star product $\star$ by defining $F \star G = \tau( \tau^{-1}F \bullet \tau^{-1} G)$ for any $F,G \in C^\infty_W(S)[[\hbar]]$.
		\end{theo}
		\begin{proof}
			Fedosov's construction is iterative and only uses the operators $\nabla^W, \delta, \delta^{-1}, \bullet$ together with the ``auxiliary data'' $\Omega, s$ and the $\cW$-valued forms $\hat{T}, \hat{R}$. Since the former preserve on-shell $W$-smoothness (see prop.~\ref{prop_prod_smooth_on_W_form} and prop.~\ref{prop_Yano_W-smooth_on_form}), and since the latter are on-shell $W$-smooth, we never leave the space $\Omega_W(S,\cW)$ when we iteratively construct $r^{(k)}$ with $k >2$, and when we act with the projection $\tau$ onto the component with $\deg_a,\deg_s=0$ or with its iteratively defined inverse $\tau^{-1}$ (see e.g. the explicit iterative constructions of $r$ and $\tau^{-1}$ discussed in remark~\ref{rem_Fedosov}). Lemma~\ref{lemma_various_res_inf} ensures that the fundamental algebraic relations in finite dimensions extend to our infinite-dimensional framework: the core of the proofs of the finite-dimensional Fedosov theorems is the {\em fixed-point theorem} applied to the total degree $\Deg$, see~\citep{F94, neumaier2001klassifikationsergebnisse, W07}. Since we have exactly the same algebraic structure, the finite-dimensional proofs can be repeated step by step in the infinite-dimensional setting. Thus, the claims follow automatically.
		\end{proof}
		As in finite dimensions, the Fedosov connection $D^W$~\eqref{Fedosov_conn_inf} depends only on the following input data: The infinite-dimensional counterpart of the Yano-connection $\nabla^W$ (not necessarily flat), the closed on-shell $W$-smooth form $\Omega$ on $S$ taking values in $\bC[[\hbar]]$, and the datum $s$ (subject only to $\Deg s \geq 3$). As done in finite dimensions, $\Omega$ and $s$ are collectively denoted by ``auxiliary data''.
		
\chapter{The relation between perturbative quantum field theroy and Fedosov's approach in infinite dimensions}\label{sec_rel_pQFT_Fedosov}
	In this chapter, we discuss the relation between the perturbative approach to algebraic quantum field theory we reviewed in chapter~\ref{sec_pQFT} and Fedosov's approach for on-shell $W$-smooth sections we developed throughout chapter~\ref{sec_Fedosov_inf}. In particular, we would like to understand eq.~\eqref{Fedosov_per} in the light of the infinite-dimensional formalism of chapter~\ref{sec_Fedosov_inf}.\\
	At the end of sec.~\ref{subsec_int_QFT_per}, we conjectured that the derivative $\nabla^{R}- \langle \cdot, \delta /\delta \varphi \rangle$ defined in terms of the retarded connection~\eqref{R/A_der_per} is precisely the Fedosov connection associated with the assignment $\phi \mapsto \omega^R_\phi$ where $\omega^R_\phi$ is the (pure Hadamard) retarded $2$-point function defined by eq.~\eqref{in-state}. We will prove this result rigorously in sec.~\ref{subsec_gaugeeq}. In addition, we will prove that the Fedosov derivative $D^W$ constructed using the procedure of sec.~\ref{subsec_Fedosov_inf} for an admissible assignment of (pure Hadamard) $2$-point functions $\phi \mapsto \omega_\phi$ (as in def.~\ref{def_suit_omega}) is ``gauge equivalent'' to $\nabla^R - \langle \cdot, \delta /\delta \varphi \rangle$ (in the sense described below in sec.~\ref{subsec_gaugeeq}).\\
	This equivalence opens the door to understand the relation between the  way of quantising a field theory as described in chapter~\ref{sec_pQFT}, and Fedosov's method: let $S \ni \phi \mapsto \hat{F}_\phi$ be the quantum observable corresponding to the local functional $F$ given by the Haag's formula~\eqref{Haag_series}. As we have seen in thm.~\ref{theo_Fedosov_per}, this defines a flat section for $\nabla^R - \langle \cdot, \delta /\delta \varphi \rangle$. In sec.~\ref{subsec_W_smooth_ret_prod}, we will show that $S \ni \phi \mapsto \hat{F}_\phi$ is actually an on-shell $W$-smooth section. Since $D^W$ and $\nabla^R - \langle \cdot, \delta /\delta \varphi \rangle$ are gauge equivalent, we can find a gauge transformation such that the gauge-transformed section $\phi \mapsto \hat{F}'_\phi$ is $D^W$-flat. Finally, we prove that these sections $hat{F}'$ satisfies Einstein causality.\\
	To make our arguments independent of subtle ``IR-issues'', we will assume throughout this chapter that $V(\phi)=\int_M \frac{1}{4!} \lambda(x) \phi(x)^4$, where $\lambda \in C^\infty_0(M)$ is fixed.
	
\section{Gauge equivalence of perturbative quantum field theory and Fedosov's approach in infinite dimensions}\label{subsec_gaugeeq}
	As mentioned in the introduction, we focus on the derivative $\nabla^{R} - \langle \cdot, \frac{\delta}{\delta \varphi} \rangle$ defined in terms of the retarded connection~\eqref{R/A_der_per}. We will prove here that it is equal to the Fedosov connection $D^{R,W}$ associated to the assignment $\phi \mapsto \omega^R_\phi$. The construction of the Fedosov connection outlined in sec.~\ref{subsec_Fedosov_inf} requires that the assignment $\phi \mapsto \omega^R_\phi$, where $\omega^R_\phi$ is the (pure Hadamard) retarded $2$-point function defined by eq.~\eqref{in-state}, is admissible in the sense of def.~\ref{def_suit_omega}. As we will show, this is indeed true.\\
	After we have settled this point, we can apply the construction outlined throughout sec.~\ref{subsec_W_smooth_symp_metric}-\ref{subsec_Fedosov_inf} to $\phi \mapsto \omega^R_\phi$. In particular, we can define the product $\bullet^{R}$~\eqref{prod_dega,degs-hom}, the $W$-connection $\nabla^{R,W}$~\eqref{tilde_Yano_W-smooth_on_form}, and the Fedosov derivative $D^{R, W}$~\eqref{Fedosov_conn_inf} with respect to the product $\bullet^{R}$ and characterized by $\nabla^{R, W}$ together with the auxiliary data $\Omega^{R} = 0, s^{R} = 0$. Then, we will verify that the Fedosov derivative $D^{R, W}$ corresponding to the family of retarded $2$-point functions coincides, as derivative on $(C^\infty_W(S,\cW), \bullet^{R})$, with the derivative $\nabla^{R} - \langle \cdot, \frac{\delta}{\delta \varphi} \rangle$.\\
	
	First of all, we note that the Fedosov operator $\delta$ (given by eq.~\eqref{delta_phi}) when acting on on-shell $W$-smooth sections on $\cW$ equals by definition the operator $\langle \cdot, \frac{\delta}{\delta \varphi} \rangle$ (given by eq.~\eqref{variation_varphi}). In fact, we have
		\begin{equation*}
			\begin{split}
			(\delta t)_\phi (v; u_1, \dots, u_n) &= (n+1) \int_{M^{n+1}} t_\phi (y,x_1, \dots, x_{n}) v(y) u_1(x_1) \cdots u_n(x_n) dy dx_1 \dots dx_n \\
			&= \left(\langle v , \frac{\delta}{\delta \varphi} \rangle t_\phi \right) (u_1, \dots, u_n),
			\end{split}
		\end{equation*}
	 for any $t \in C_W^\infty(S, \cW)$ with $\deg_s = n+1$, for any $\phi \in S$, and for any $v, u_1, \dots, u_n \in T_\phi S$.\\
	
	Next, we prove that the assignment $\phi \mapsto \omega^R_\phi$, where $\omega^R_\phi$ is the (pure Hadamard) retarded $2$-point function defined by eq.~\eqref{in-state}, is admissible in the sense of def.~\ref{def_suit_omega}. For this purpose, it is sufficient to show that for any $\phi \in C^\infty(M)$ the $2$-point function $\omega^R_\phi$ can be written in the form~\eqref{state}, because, as already proved in lemma~\ref{lemma_state}, any $2$-point function in the form~\eqref{state} gives an admissible assignment.
		\begin{lemma}\label{lemma_ret_state_fedosov}
			For any $\phi \in C^\infty(M)$, the $2$-point function $\omega_\phi^{R}$ of the retarded state defined via~\eqref{in-state} can be written in the form~\eqref{state}.
		\end{lemma}
		\begin{proof}	
		 The argument we are going to present exploits that the coupling $\lambda$ has compact support. We fix two Cauchy surfaces $\Sigma_\pm$ such that $\Sigma_+$ does not intersect the causal future of $\supp \lambda$ and $\Sigma_- \prec \Sigma_+$, where the ordering $\prec$ is understood in terms of the causal structure. We choose an arbitrary smooth cut-off function $\chi$ which equals $1$ in the future of $\Sigma_+$ and $0$ in the past of $\Sigma_-$. As in sec.~\ref{subsubsec_kahler_inf}, we define $\phi_- := \chi \phi$. Then, we choose four further Cauchy surfaces $\Sigma_{\pm \pm}$ such that $\Sigma_{++}$ does not intersect the (causal) future of $\supp \lambda$, and $\Sigma_{--} \prec \Sigma_{-+} \prec \Sigma_- \prec \Sigma_+ \prec \Sigma_{-+} \prec \Sigma_{++}$ as in lemma~\ref{lemma_state}. We consider two smooth cut-off functions $c_\pm$ such that $c_\pm = 0$ in the future of $\Sigma_{\pm +}$ and $c_\pm = 1$ in the past of $\Sigma_{\pm -}$. We show that for these choices of $\Sigma_{\pm\pm}$ and $c_\pm$, we have
		 	\begin{equation}\label{ret_state_fedosov_form}
		 			\omega_\phi^R (x_1,x_2) = \left( E_\phi \circ \sigma_{c_+} \circ E_{\phi_-} \circ \sigma_{c_-} \circ \omega_0 \circ \sigma_{c_-} \circ E_{\phi_-} \circ \sigma_{c_+} \circ E_\phi \right)(x_1, x_2),
		 	\end{equation}
		 where $\omega_0$ is the $2$-point function of the ground state with respect to the Klein-Gordon operator $P_0=\boxempty - m^2$, i.e. the distribution~\eqref{ground_state}, and where the distributions $\sigma_{c_\pm}$ are defined by~\eqref{kernel_symp} in terms of the cut-off functions $c_\pm$. Clearly, proving this claim verifies the statement of the lemma.\\
		 As already mentioned in lemma~\ref{lemma_state}, the right-hand side of eq.~\eqref{ret_state_fedosov_form} is a pure Hadamard $2$-point function with respect to the Klein-Gordon operator $P_\phi = \boxempty - m^2  - \frac{\lambda}{2} \phi^2$. By construction, the supports of both the distributions $\sigma_{c_\pm}$ are contained in $J^-(\supp \lambda) \backslash \supp \lambda \times J^-(\supp \lambda) \backslash \supp \lambda$. Since in $ J^-(\supp \lambda) \backslash \supp \lambda$ we have $V(\phi_-)=0$, we can replace in the right-hand side of eq.~\eqref{ret_state_fedosov_form} $E_{\phi_-}$ with $E_0$. Using eq.~\eqref{causal_prop_inv_symp_kernel_sol} together with the fact that $\omega_0$ is a bi-solution for $P_0$, we can rewrite the right-hand side of eq.~\eqref{ret_state_fedosov_form} as
			\begin{equation*}
				\left( E_\phi \circ \sigma_{c_+} \circ E_{\phi_-} \circ \sigma_{c_-} \circ \omega_0 \circ \sigma_{c_-} \circ E_{\phi_-} \circ \sigma_{c_+} \circ E_\phi \right)(x_1, x_2) = \left( E_\phi \circ \sigma_{c_+} \circ \omega_0 \circ \sigma_{c_+} \circ E_\phi \right)(x_1,x_2).
			\end{equation*}
		Let $f,h$ be two test function whose support does not intersect the causal future of $\supp \lambda$. As a consequence of lemma~\ref{lemma_kernel_symp} and the support properties of $\sigma_{c_+}$, it follows
			\begin{equation*}
				\left( E_\phi \circ \sigma_{c_+} \circ \omega_0 \circ \sigma_{c_+} \circ E_\phi \right)(f,h) = \omega_0(f,h).
			\end{equation*}
		The retarded state is uniquely determined by the requirement $\omega^{R}_\phi = \omega_0$ on $M \backslash J^+ (\supp \lambda)$. Thus, we proved that the $2$-point functions $( E_\phi \circ \sigma_{c_+} \circ E_{\phi_-} \circ \sigma_{c_-} \circ \omega_0 \circ \sigma_{c_-} \circ E_{\phi_-} \circ \sigma_{c_+} \circ E_\phi )$ and $\omega^{R}_\phi$ coincide in an open region which contains a Cauchy surface. Because these two distributions obey the same hyperbolic equation, we conclude that they must coincide on the whole space-time $M$ as we wanted to prove.
		\end{proof}
		
		As corollary of the previous lemma, the whole construction exposed in sec.~\ref{subsec_W_smooth_symp_metric}-\ref{subsec_Fedosov_inf} applies to $\phi \mapsto \omega^{R}_\phi$. In particular, we can define the product $\bullet^{R}$~\eqref{prod_dega,degs-hom}, the Yano $W$-connection $\nabla^{R,W}$~\eqref{tilde_Yano_W-smooth_on_form}, and the corresponding Fedosov derivative $D^{R, W}$~\eqref{Fedosov_conn_inf} associated to $\nabla^{R, W}$ and the auxiliary data $\Omega^{R} = 0, s^{R} = 0$.\\
		We would like to show that on $C^\infty_W(S,\cW)$ the Fedosov connection $D^{R, W}$ equals the connection $\nabla^{R} - \delta$. For this purpose, we first show that $\nabla^R$ is a $W$-covariant derivative, i.e. that it is a map $C^\infty(S,\boxtimes^n_W T^*_\phi S) \to C^\infty(S,\boxtimes^{n+1}_W T^*_\phi S)$ satisfying the conditions of def.~\ref{def_W_cov_der}. Using the $\Deg$-filtration on $C^\infty_W(S,\cW)$ and the fact that any section in $C_W^\infty(S,\cW)$ homogeneous in $\Deg$ is a finite sum of sections homogeneous in $\deg_s$ and in $\deg_\hbar$, we can extend\footnote{We impose $\bC[\hbar]$-linearity.} $\nabla^R$ to a derivative $C^\infty_W(S,\cW) \to \Omega^1_W(S,\cW)$.\\
		
		The retarded connection $\nabla^R$ was defined by~\eqref{R/A_der_per} in sec.~\ref{subsec_int_QFT_per}. For any on-shell $W$-smooth covariant section $t$ of rank $n$, for any $\phi \in S$, and for any $v \in T_\phi S$, we have
			\begin{equation*}
				(\nabla^R_v t)_\phi = \left. \frac{d}{d \epsilon} \alpha^R_{\phi(\epsilon), \phi} t_{\phi(\epsilon)} \right|_{\epsilon = 0},
			\end{equation*}
		where $\bR \ni \epsilon \mapsto \phi(\epsilon) \in S$ is a smooth map such that $\phi(0) = \phi$ and $d \phi(\epsilon) / d \epsilon |_{\epsilon = 0} = v$.\\
		It is not immediately clear that the map $\nabla^R$, defined fiberwise by
			\begin{equation*}
				(\nabla^{R} t)_\phi (v,u_1, \dots, u_{n}) := (\nabla^R_{v} t)_\phi (u_1, \dots, u_{n}),
			\end{equation*}	
		for any $\phi \in S$, for any $v, u_1, \dots, u_{n} \in T_\phi S$, is a map $C^\infty_W(S, \boxtimes^n_W T^*S) \to C^\infty_W(S, \boxtimes^{n+1}_W T^*S)$ satisfying the conditions to be a $W$-derivative given in def.~\ref{def_W_cov_der}. In the following lemma, we rewrite $\nabla^R$ in an equivalent form, which resolves this issue.
			\begin{lemma}\label{lemma_nabla^R_partial}
				Let $c$ be a cut-off as in~\eqref{kernel_symp} such that $c$ vanishes in the future of a Cauchy surface $\Sigma_+$ such that $\Sigma_+ \cap J^+ (\supp \lambda) =\emptyset$ and $c$ is identically $1$ in the past of an arbitrary Cauchy surface $\Sigma_-$ in the past of $\Sigma_+$ (clearly $\Sigma_-$ is also in the (strict) past of $\supp\lambda$). Then, we have
					\begin{equation*}
						\nabla^R = \partial,
					\end{equation*}
			where $\partial$ is defined as in prop.~\ref{prop_der_smooth_on} for this specific cut-off $c$.
		\end{lemma}
		\begin{proof}
			Let $\phi \in S$ and $v, u_1, \dots, u_{n} \in T_\phi S$. Fist of all, we notice that for any $t \in C^\infty_W(S, \boxtimes_W^n T^*S)$ we can equivalently write
				\begin{equation}\label{nabla^R_on-shell}
					(\nabla^{R} t)_\phi (v,u_1, \dots, u_{n}) = \left( \left. \frac{d}{d \epsilon} \alpha^R_{\phi + \epsilon v, \phi} [\tilde{t}_{\phi + \epsilon v}] \right|_{\epsilon =0} \right)(u_1, \dots, u_{n}),
				\end{equation}	
			where $\tilde{t} \in (\sigma_c \circ E_\phi)^{\otimes n} \cE'_W(M^n)$ is an extension of $t$ satisfying the requirements of def.~\ref{def_smooth_on_tens}, and where $[\tilde{t}_\phi]$ denotes the equivalence class in $\cE'_W(M^n) / P_\phi \cE'_W(M^n)$ corresponding to $\tilde{t}_\phi$. Here, we do not require that the cut-off $c$ as in eq.~\ref{kernel_symp} satisfies also the stricter conditions of the hypothesis of the lemma. We used the fact that any smooth map $\bR \ni \epsilon \mapsto \phi(\epsilon) \in S$ such that $\phi(0) = \phi$ and $d \phi(\epsilon) / d \epsilon |_{\epsilon = 0} = v$ necessarily satisfies $\phi(\epsilon) = \phi + \epsilon v + o(\epsilon^2)$.\\
		By the definition of the isomorphism $\alpha^R$ (see~\eqref{alpha_R_wick_per},~\eqref{A_R} and~\eqref{S_R}), the right-hand side of eq.~\eqref{nabla^R_on-shell} depends neither on the choice of the extension $\tilde{t}$ nor on the choice of the cut-off appearing implicitly in $\tilde{t}$ and in $\alpha^R$ throughout $A^R$ (see~\eqref{A_R}). So we are free to use a cut-off $c$ which satisfies the requirements in the hypothesis of this lemma, since it satisfies all the conditions required by eq.~\eqref{kernel_symp} and eq.~\eqref{A_R}.\\
		 For any $\phi \in S$ and any $v, u_1, \dots, u_n \in C^\infty(M)$, we obtain
					\begin{align}
						&(\nabla^{R} t)_\phi (v, u_1, \dots, u_{n}) = \int_{M^n} \frac{d}{d \epsilon} P^{(x_1)}_\phi \dots P^{(x_{n})}_\phi \left\{ \left( c \cdot \int_{\Sigma_+}E_{\phi} \overleftrightarrow{\partial_n} E_{\phi + \epsilon v} \right)^{\otimes n} \tilde{t}_{\phi + \epsilon v}\right\}(x_1, \dots, x_n) \times \nonumber \\
						&\qquad \qquad \qquad \qquad \qquad \times u_1(x_1) \dots u_n(x_n)dx'_1 \dots dx'_{n} \nonumber\\
						&\quad=\int_{M^n} \frac{d}{d \epsilon} P^{(x_1)}_\phi \cdots P^{(x_{n})}_\phi \left\{ (c \cdot E_{\phi} \circ \sigma_c \circ E_{\phi + \epsilon v})^{\otimes n} \tilde{t}_{\phi + \epsilon v}\right\}(x_1, \dots, x_{n}) u(x_1) \dots u(x_n) dx_1 \dots dx_{n} \label{R_line1} \\
						&\quad= \int_{M^{n+1}} P^{(x_1)}_\phi \cdots P^{(x_{n})}_\phi \left( (c \cdot E_\phi)^{\otimes n} \frac{\delta \tilde{t}_{\phi}}{\delta \phi(y)} \right) (x_1, \dots, x_{n}) v(y) u_1(x_1) \dots u_n(x_n) dy dx_1 \dots dx_{n}\label{R_line2} \\
						&\quad= \int_{M^{n+1}} \frac{\delta \tilde{t}_{\phi}}{\delta \phi(y)} (x_1, \dots, x_{n})  v(y) u_1(x_1) \dots u_n(x_n) dy dx_1 \dots dx_{n}\label{R_line3}.
					\end{align}
				To get to line~\eqref{R_line1}, we used the following consequence of eq.~\eqref{kernel_symp_not_sol}:
					\begin{equation}\label{aux_formula}
						\begin{split}
							(E_\phi \circ \sigma_c \circ E_{\phi'})(x,y) &= \int_{\Sigma_-} E_\phi(x,z) \overleftrightarrow{\partial_n} E_{\phi'}(z,y) d\Sigma(z) + \int_{B} E_\phi(x,z) c(z) P^{(z)}_\phi  E_{\phi'}(z,y) dz,
						\end{split}
					\end{equation}
				where $\phi, \phi'$ are arbitrary smooth functions, and where  $B=J^+ (\Sigma_-) \cap J^- (\Sigma_+)$ is the closed space-time region bounded by $\Sigma_+$ and $\Sigma_-$. The second term in eq.~\eqref{aux_formula} vanishes because $P_\phi = P_{\phi'}$ on the support of $c$ (in fact, $\lambda=0$ on $\supp c$) and $E_{\phi'}$ is a bi-solution with respect to $P_{\phi'}$. Line~\eqref{R_line2} was obtained by recalling the definition of Gateaux derivative and by noticing that $(\sigma_c \circ E_{\phi}) \tilde{t}_\phi = \tilde{t}_\phi$ for any $\phi \in C^\infty(M)$ because $ \tilde{t}_\phi \in (\sigma_c \circ E_\phi)^{\otimes n} \cE'_W(M^n)$ by construction and $(\sigma_c \circ E_{\phi})^2 = (\sigma_c \circ E_{\phi}) $ by~\eqref{project_E_sigma}. Then, to get to line~\eqref{R_line3} we used the following equation
					\begin{equation*}
						\int_M E_\phi(x_1, z) P_\phi^{(z)} \left( c(z) E_\phi(z, x_2) \right) dz = \int_{\Sigma_-} E_\phi(x_1, z) \overleftrightarrow{\partial_n} E_\phi(z, x_2) d\Sigma(z) = E_\phi(x_1, x_2),
					\end{equation*}
				which can easily be verified using Stokes theorem and eq.~\eqref{causal_prop/symp_kernel}.\\
				Now, line~\eqref{R_line3} is precisely $(\partial t)_\phi(v, u_1, \dots, u_n)$ as one can see by comparing directly with the definition of $\partial t$ (see~\eqref{diff_on-shell}). This concludes the proof.
		\end{proof}
		\noindent
		Note that $\partial$ depends on the choice of the cut-off, and so the equivalence $\nabla^R = \partial$ holds only for the specific $c$ we choose. Nevertheless, lemma~\ref{lemma_nabla^R_partial} implies that $(\nabla^{R} t)_\phi$ is in $\boxtimes^n_W T^*_\phi(S^{n+1})$, because this is true for $(\partial t)_\phi$. Furthermore, we conclude that $S \ni \phi \mapsto (\nabla^R t)_\phi$ is on-shell $W$-smooth because $\widetilde{(\partial t)}$ defined by~\eqref{tilde_diff} for a cut-off $c$ satisfying the hypothesis of lemma~\ref{lemma_nabla^R_partial} provides an off-shell extension for $(\nabla^R t)$, and satisfies conditions~\ref{W1},~\ref{W2} in def.~\ref{def_smooth_on_tens} (as already proved in prop.~\ref{prop_der_smooth_on}). It also follows that $\nabla^R$ satisfies the conditions to be a $W$-covariant derivative listed in def.~\ref{def_W_cov_der}.\\
		The equivalence $\nabla^R = \partial$ for a cut-off $c$ as in lemma~\ref{lemma_nabla^R_partial} also implies the next result:
			\begin{theo}\label{theo_ret_conn_fedosov_conn}
				The connection $\nabla^{R} - \delta$ coincides with $D^{R, W}$ as derivative on $(C^\infty_W(S,\cW), \bullet^{R})$.
			\end{theo}
			\begin{proof}
				By definition, the Fedosov derivative $D^{R,W}= \nabla^{R,W} -\delta + \frac{i}{\hbar} \ad_\bullet(r^R)$, where $\nabla^{R,W}$ is the Yano $W$-connection associated to $\phi \mapsto \omega^R_\phi$. Therefore, to prove the theorem, we need to check that $\nabla^{R}$ coincides with $\nabla^{R,W} + \frac{i}{\hbar} \ad_\bullet(r^R)$.\\
				Remember that an on-shell $W$-smooth $\cW$-valued $1$-form is a sequence (in $n$) of $\bC[[\hbar]]$-valued on-shell $W$-smooth covariant sections of rank $n+1$ which are symmetric in the last $n$ variables. The claim is equivalent to the statement
					\begin{equation}\label{ret_=_fedo}
						\left(\nabla^{R} t\right)_\phi = \left(\nabla^{R, W} t + \frac{i}{\hbar} ( \ad_{\bullet^R}(r^R) t\right)_\phi,
					\end{equation}
				for any $\phi \in S$ and any $t \in C^\infty_W(S,\cW)$. We proceed by showing that eq.~\eqref{ret_=_fedo} holds order by order in $\deg_s$ and $\deg_\hbar$. For fixed degrees $\deg_s$ and $\\deg_\hbar$ (set $\deg_s=n$), both side of eq.~\eqref{ret_=_fedo} are equivalence classes in $\cE'_W(M^n)/ P_\phi \cE'_W(M^n)$ (up to the adequate power of $\hbar$).\\
				 We proved in prop.~\ref{prop_Yano_W-smooth_on_form} and in thm.~\ref{theo_Fedosov_inf} that $\nabla^{R, W} t$ and $\frac{i}{\hbar} ( \ad_{\bullet^R}(r^R) t)$ do not depend on the choice of the cut-off function $c$ as in~\eqref{kernel_symp} which implicitly appear in the definitions of these two on-shell $W$-smooth $1$-forms with values in $\cW$. The same holds for $(\nabla^{R} t)$ as a consequence of lemma~\ref{lemma_nabla^R_partial}. Because of the independence of the cut-off, it is sufficient to show that eq.~\eqref{ret_=_fedo} holds when both side are computed in terms of a specific cut-off $c$, i.e.
				 	\begin{equation}\label{ret_=_fedo_alt}
				 		\widetilde{(\nabla^{R} t)}_\phi(y,x_1, \dots, x_n) \approx \widetilde{(\nabla^{R, W} t)}_\phi(y,x_1, \dots, x_n) + \frac{i}{\hbar} \left(\ad_{\bullet^R_\phi} (\widetilde{r^R}_\phi) \tilde{t}_\phi\right)(y,x_1, \dots, x_n),
				 	\end{equation}
				 where $\approx$ means ``equal up to a distribution in $P_\phi \cE'_W(M^{n+1})$ symmetric in the last $n$ variables'' for each degree $\deg_s =n$, and where $\widetilde{r^R}_\phi$ is a $W$-smooth off-shell extension (in the sense of def.~\ref{def_smooth_on_tens}) of the on-shell $W$-smooth $\cW$-valued $1$-form $r^R$. Note that the individual terms in~\eqref{ret_=_fedo_alt} depend on a cut-off $c$ for general $\phi \in C^\infty(M)$.\\
				 To prove~\eqref{ret_=_fedo_alt}, we can thus use, in particular, the cut-off function $c$ we used before in lemma~\ref{lemma_nabla^R_partial}: we demand that $c \in C^\infty(M)$ vanishes in the future of a Cauchy surface $\Sigma_+$ such that $\Sigma_+ \cap J^+ (\supp \lambda) =\emptyset$ and $c$ is identically $1$ in the past of an arbitrary Cauchy surface $\Sigma_-$ in the past of $\Sigma_+$ (clearly $\Sigma_-$ is also in the past of $\supp\lambda$). We note that for a cut-off $c$ of this type, the distribution $\sigma_c$ defined in eq.~\eqref{kernel_symp} is supported in $K \times K$,  where $K$ is a compact set contained in $J^- (\supp\lambda) \backslash \supp \lambda$. It follows from the definition of the retarded $2$-point function $\omega^{R}_\phi$ that
					\begin{equation}\label{sigma_c_omega_R_sigma_c}
						\sigma_c \circ \omega^{R}_\phi \circ \sigma_c = \sigma_c \circ \omega_0 \circ \sigma_c.
					\end{equation}
				Noticing that the right-hand side of eq.~\eqref{sigma_c_omega_R_sigma_c} does not depend on $\phi$, it follows trivially that all its Gateaux derivatives vanish. This motivates our choice of $c$.\\
				We next present the extensions $\widetilde{(\nabla^{R, W} t)}_\phi$ and $\ad_{\bullet^R_\phi} (\widetilde{r^R}_\phi) \tilde{t}_\phi$ in terms such cut-off $c$. Looking at the definition~\eqref{tilde_Yano_W-smooth_on_form}, we see that in the present situation it holds
					\begin{equation}\label{nabla_R_W_=_partial}
						\widetilde{(\nabla^{R, W} t)}_\phi = \widetilde{(\partial t)}_\phi.
					\end{equation}
				This follows because $\widetilde{(\nabla^{R, W} t)}_\phi$ differs from $\widetilde{(\partial t)}_\phi$ by a finite sum of terms involving the distribution $\Gamma_\phi^{R}$ given by eq.~\eqref{Yano_symbol_W-smooth_on} in terms of the retarded $2$-point function $\omega^R_\phi$. By definition, $\Gamma_\phi^{R}$ is a finite sum of distributions involving $\delta (\sigma_c \circ \omega^{R}_\phi \circ \sigma_c) / \delta \phi$ which vanishes for the specific choice of $c$ used here. Thus, eq.~\eqref{nabla_R_W_=_partial} holds for our $c$.\\
				The fact that $\Gamma_\phi^{R}$ identically vanishes for this choice of $c$, has another consequence. The on-shell $W$-smooth $\cW$-valued forms $\hat{T}^R$ and $\hat{R}^R$ corresponding to the retarded $2$-point function $\omega^R_\phi$ are defined via the distributions~\eqref{symp_torsion_inf} and, respectively,~\eqref{symp_curv_inf}, specialized to $\omega^R_\phi$. Since we proved in lemma~\ref{lemma_various_res_inf} that $\hat{T}^R$ and $\hat{R}^R$ does not depend on the choice of the cut-off appearing in the distributions~\eqref{symp_torsion_inf} and~\eqref{symp_curv_inf}, we can choose the same cut-off $c$ we defined before. For both eq.~\eqref{symp_torsion_inf} and eq.~\eqref{symp_curv_inf}, the right-hand side depends on $\Gamma_\phi^R$. Thus, it follows straightforwardly that $\hat{T}^R$ and $\hat{R}^R$ vanish. By the Fedosov's theorem (thm.~\ref{theo_Fedosov_inf}), the on-shell $W$-smooth $\cW$-valued $1$-form $r^R$ is the unique section in $\Omega^1_W(S,\cW)$ which solves $\nabla^{R, W} r^R - \delta r^R + \frac{i}{\hbar} r^R \bullet^R r^R - \hat{R}^R - \hat{T}^R = 0$ subjected to the requirements $r^R = (r^R)^\dagger$, $(r^R)^{(0)} =0 = (r^R)^{(1)}$ and $\delta^{-1} r^R = 0$. Since $\hat{T}^R =0$ and $\hat{R}^R =0$, it follows that $r^R =0$ is a solution, and, therefore, the unique solution. Thus, we have that the part of $\ad_{\bullet^R_\phi} (\widetilde{r^R}_\phi) \tilde{t}_\phi$ with $\deg_s=n$ is simply the $0$ distribution (up to a distribution in $P_\phi \cE'_W(M^{n+1})$ symmetric in the last $n$ variables).\\
				On the other hand, we already proved in lemma~\ref{lemma_nabla^R_partial} that $\nabla^R = \partial$ for the specific cut-off $c$. Therefore, we have
					\begin{equation}\label{nabla_ret_approx_partial}
						\widetilde{\nabla^R t}_\phi \approx \widetilde{(\partial t)}_\phi \approx \widetilde{\nabla^{R, W} t}_\phi,
					\end{equation}
				where $\approx$ means ``equal up to a distribution in $P_\phi \cE'_W(M^{n+1})$'', and where our specific cut-off $c$ is chosen. This concludes the proof.
			\end{proof}
	
	In finite dimensions, we proved the existence of a gauge equivalence between two Fedosov connections corresponding to two different almost-K\"{a}hler structures (see thm.~\ref{thm1_bis}). We now investigate how this result translates in our infinite-dimensional framework. Let $\phi \mapsto \omega_\phi$ and $\phi \mapsto \omega'_\phi$ be two admissible assignments in the sense of def.~\ref{def_suit_omega} of two pure Hadamard $2$-point functions for any $\phi \in C^\infty(M)$. We will prove that the corresponding Fedosov derivatives $D^W$ and $D^{\prime W}$ are gauge equivalent. Combining this result with thm.~\ref{theo_ret_conn_fedosov_conn}, it follows that the covariant derivative $\nabla^{R} - \delta$ is gauge equivalent to the Fedosov connection corresponding to any admissible assignment $\phi \mapsto \omega_\phi$ of a pure Hadamard $2$-point function $\omega_\phi$ for any $\phi \in C^\infty(M)$. We follow the pedestrian approach we already used throughout this paper: we provide the appropriate infinite-dimensional counterpart of any object appearing in the argument presented in sec.~\ref{subsec_equivalence_fin} for finite dimensions.\\
	We proceed defining first the infinite-dimensional analogue of the isomorphism $\alpha$ between formal Wick algebras introduced in lemma~\ref{lemma_Wick_iso_fin}. In the following, we denote by $\cW_\phi$ and $\cW'_\phi$ respectively the formal Wick algebra with respect to the product $\bullet_\phi$ induced by $\omega_\phi$ and the formal Wick algebra with respect to the product $\bullet'_\phi$ induced by $\omega'_\phi$. Consistently we denote $C^\infty_W(S,\cW)$, $C^\infty_W(S, \cW')$, and more generally $\Omega_W(S,\cW)$, $\Omega_W(S, \cW)$, the algebras of the on-shell $W$-smooth sections on the corresponding bundles.\\
	Let $t$ be a element in $\cW_\phi$ homogeneous in $\deg_s$, with $\deg_s t =n$, and in $\deg_\hbar$, i.e. $t \in \bC[\hbar] \otimes \vee^n_W T^*_\phi S$. We define $\alpha_\phi(t) \in \cW'_\phi$ as the sequence $(\alpha_\phi(t)^0, \alpha_\phi(t)^1, \dots)$, where each $\alpha_\phi(t)^j$ is given by
		\begin{equation}\label{alphamap_fiber_inf}
			\begin{split}
				&\gls{alpha}_\phi(t)^{n-2\ell} (x_1, \dots, x_{n- 2\ell}) := \\ 
				&\quad =\bP^+ \frac{\hbar^\ell n!}{(n-2\ell)! (2 \ell)!} \int_{M^{2\ell}} t(z_1, \dots, z_{2\ell}, x_{1}, \dots, x_{n-2\ell} ) \prod_{i=1}^{\ell} (\omega_\phi - \omega'_\phi) (z_{2i-1}, z_{2i}) dz_1 \dots dz_{2i}.
			\end{split}
		\end{equation}
	for $0 \leq \ell \leq [n/2]$, while $\alpha_\phi(t)^j =0$ otherwise. Note that, by abuse of notation, we identify a class in $\bC[\hbar] \otimes \boxtimes_W^{\bullet} T^*_\phi S$ by one of its ($\bC[\hbar]$-valued) distributional representative in $\bC[\hbar] \otimes \bP^+ \cE'_W(M^\bullet)$. The distribution on the right-hand side of eq.~\eqref{alphamap_fiber_inf} is well defined: by definition, $t$ is a distribution in $\bP^+ \cE'_W(M^n)$ (up to a factor $\hbar^{\deg_\hbar t}$), and, by construction, the difference of the two Hadamard $2$-point functions $\omega_\phi - \omega'_\phi$ is a smooth function. We can apply thm.~\ref{theo_WF_horma} to conclude that the composition in the right-hand side is well-defined and defines a distribution in $\cE'_W(M^{n-2\ell})$ (up to a factor $\hbar^{\deg_t +k}$). Furthermore, the equivalence class in $\bC[\hbar] \otimes \boxtimes_W^{n-2\ell} T^*_\phi S$ corresponding to $\alpha(t)^{n-2 \ell}_\phi$ does not depend on the choice of the distributional representative of $t$ because both $\omega_\phi$ and $\omega'_\phi$ are bi-solution with respect to $P_\phi$.\\
	Using the filtration of $\cW_\phi$ with respect to the total degree $\Deg$ and, then, exploiting the fact each element of $\cW_\phi$ homogeneous in $\Deg$ is a finite collection of elements homogeneous in $\deg_s$ and $\deg_\hbar$, the map $\alpha_\phi: t \mapsto \alpha_\phi(t)$ defined via~\eqref{alphamap_fiber_inf} uniquely extends to a map $\cW_\phi \to \cW'_\phi$. Furthermore, as in the finite-dimensional case, it can be easily checked that $\alpha_\phi$ is an isomorphism $\cW_\phi \to \cW'_\phi$ for any $\phi \in S$.\\
	Following the finite-dimensional construction, we would next like to define an isomorphism between $C^\infty_W(S,\cW)$ and $C^\infty_W(S,\cW')$ using the maps $\alpha_\phi$ defined on each fiber $\cW_\phi$:
	\begin{prop}\label{prop_W-smooth_alpha_1}
		Let $t$ be a section in $C^\infty_W(S,\cW)$. For any $\phi \in S$, we define
			\begin{equation}\label{alphamap_inf}
				\alpha(t)_\phi := \alpha_\phi(t_\phi).
			\end{equation}
		The map $\alpha$ is an isomorphism $C^\infty_W(S,\cW) \to C^\infty_W(S,\cW')$, i.e. $\alpha(t \bullet s) = \alpha(t) \bullet' \alpha(s)$ for any $t,s \in C^\infty(S,\cW)$, and it preserves the conjugation operation $\dagger$, i.e. $\alpha(t)^\dagger = \alpha( t^\dagger)$.
	\end{prop}
	\begin{proof}
	The subtle point is proving that the proposed definition~\eqref{alphamap_inf} preserves the on-shell $W$-smoothness. Once this has been established, $\alpha$ is necessarily an isomorphism because $\alpha_\phi$ is an isomorphism which preserve the $\dagger$-operation in the fibers and both the algebraic structure and the $\dagger$-operation for the on-shell $W$-smooth sections on $\cW$ and $\cW'$ are defined fiberwise.\\
	To prove that $\alpha$ respects the on-shell $W$-smoothness, we need to provide for any on-shell $W$-smooth section $t$ an extension of $\alpha(t)_\phi$ in the sense of def.~\ref{def_smooth_on_tens}, i.e. such that the conditions~\ref{W1},~\ref{W2} hold. Actually, exploiting the filtration of the algebra $C^\infty_W(S,\cW)$ with respect to the total degree $\Deg$ and the fact that each section homogeneous in $\Deg$ is a finite collection of sections homogeneous in $\deg_s$ and $\deg_\hbar$, it is sufficient to prove the claim for an on-shell $W$-smooth section $t$ homogeneous in $\deg_s$ and $\deg_\hbar$. We assume $\deg_s t = n$.\\
	Let $c$ be an arbitrary but fixed cut-off function as in eq.~\eqref{kernel_symp} and let $C^\infty(M) \ni \phi \mapsto \tilde{t}_\phi \in (\sigma_c \circ E_\phi)^{\otimes n} \cE'_W(M^n)$ be an extension of $t$ (up to a factor $\hbar^{\deg_\hbar t}$) in the sense of def.~\ref{def_smooth_on_tens}. For any $\phi \in C^\infty(M)$, we define $\widetilde{\alpha(t)}{}^j_\phi \in (\sigma_c \circ E_\phi)^{\otimes j} \cE'_W(M^j)$ by
		\begin{equation}\label{eq_1}
			\begin{split}
				&\widetilde{\alpha(t)}{}^{n - 2\ell}_\phi (x_{1}, \dots, x_{n-2\ell} ) :=\\
				&\quad = \frac{n!}{(n-2\ell)! (2 \ell)!} \int_{M^{2\ell}} \tilde{t}_\phi(z_1, \dots, z_{2\ell}, x_{1}, \dots, x_{n-2\ell} ) \prod_{i=1}^{\ell} (\omega_\phi - \omega'_\phi) (z_{2i-1}, z_{2i}) dz_1 \dots dz_{2 \ell}.
			\end{split}
		\end{equation}
	for $\ell$ such that $0 \leq \ell \leq [n/2]$, while $\widetilde{\alpha(t)}{}^j_\phi= 0$ otherwise. It is straightforward to verify that the sequence of $\widetilde{\alpha(t)}{}^j_\phi$ (up to a suitable factor of $\hbar$) is indeed an extension of $\alpha(t)$.\\
	To conclude the proof, we need to show that $\widetilde{\alpha(t)}{}^j_\phi$ satisfies the conditions~\ref{W1},~\ref{W2} of def.~\ref{def_smooth_on_tens}. We restrict to $j = n -2 \ell$ because otherwise $\widetilde{\alpha(t)}{}^j_\phi = 0$ and the conditions are trivially satisfied. In order to verify~\ref{W1},~\ref{W2}, we first rewrite $\widetilde{\alpha(t)}{}^{n - 2\ell}_\phi$ in an equivalent form. Since both $\omega_\phi$ and $\omega'_\phi$ are bi-solutions with respect to $P_\phi$, it follows from eq.~\eqref{causal_prop_inv_symp_kernel_sol} that
		\begin{equation}\label{diff_state_trick}
			\omega_\phi - \omega'_\phi = E_\phi \circ \sigma_{c_+} \circ ( \omega_\phi - \omega'_\phi) \circ \sigma_{c_+} \circ E_\phi,
		\end{equation}
	where $c_+$ is a smooth cut-off function as in eq.~\eqref{kernel_symp} such that $c_+ = 1$ in $J^+ (\Sigma_+)$ and $c_+ = 0$ in $J^-(\Sigma_-)$ for two Cauchy surfaces $\Sigma_\pm$ such that $\Sigma_- \prec \Sigma_+ \prec \supp\lambda$, where the ordering $\prec$ is understood in terms of the causal structure. We denoted by $B$ the compact region comprised between the two Cauchy surfaces $\Sigma_{+}$ and $\Sigma_{-}$, i.e. $B= J^-(\Sigma_{+}) \cap J^+ (\Sigma_{-})$. Note that $B \cap \supp\lambda = \emptyset$. It follows
		\begin{equation}\label{eq_1_bis}
			\begin{split}
				&\widetilde{\alpha(t)}{}^{n - 2\ell}_\phi (x_{1}, \dots, x_{n-2\ell} ) =\\
				&\quad = \frac{n!}{(n -2\ell)!(2 \ell)!} \int_{M^{4\ell}} \prod_{j =1}^{2\ell} (\sigma_{c_+} \circ E_\phi)(z_j, z'_j) \tilde{t}_\phi(z'_1, \dots, z'_{2\ell}, x_{1}, \dots, x_{n-2\ell} ) \prod_{i=1}^{\ell} (\omega_\phi - \omega'_\phi) (z_{2i-1}, z_{2i}) \times\\
				&\qquad \times \prod_j dz_j dz'_j.
			\end{split}
		\end{equation}
	In other words, the distribution $\widetilde{\alpha(t)}{}^{n - 2\ell}_\phi$ is (up to a numerical factor) the composition of the compactly supported distribution
		\begin{equation}\label{partial_1}
			\begin{split}
				&\Theta_\phi(z_1, \dots, z_{2\ell}, x_{1}, \dots, x_{n-2\ell}) :=\\
				&\quad = \int_{M^{2\ell}} \prod_{j =1}^{2\ell} (\sigma_{c_+} \circ E_\phi)(z_j, z'_j) \tilde{t}_\phi(z'_1, \dots, z'_{2\ell}, x_{1}, \dots, x_{n-2\ell} ) dz_1 \dots dz_{2 \ell}
			\end{split}
		\end{equation}
	with the distributions $(\omega_\phi - \omega'_\phi)(z_{2i -1}, z_{2i})$.\\
	The distribution $\Theta_\phi$ satisfies conditions~\ref{W1},~\ref{W2} of def.~\ref{def_smooth_on_tens}. This claim follows from lemma~\ref{lemma_W_comp} and thm.~\ref{theo_WF_horma} recalling the properties of $E_\phi$ (prop.~\ref{prop_var_ders_causal}), the wave-front set $\sigma_{c_+}$ (given by~\eqref{WF_kernel_symp}) and the fact that $\tilde{t}_\phi$ satisfies conditions~\ref{W1},~\ref{W2} of def.~\ref{def_smooth_on_tens} by hypothesis. Furthermore, the support of the distribution $\Theta_\phi(z_1, \dots, z_{2\ell}, x_{1}, \dots, x_{n-2\ell})$ and its Gateaux derivatives $\delta^\nu \Theta_\phi(z_1, \dots, z_{2\ell}, x_{1}, \dots, x_{n-2\ell}) /\delta \phi^\nu$ contains only elements with $z_1, \dots, z_{2\ell} \in B$.\\
	We proceed establishing the following estimates for the variational derivatives of the difference $\omega_\phi - \omega'_\phi$ exploiting eq.~\eqref{diff_state_trick}.
			\begin{lemma}\label{lemma_diff_var_states}
			For any $\nu \in \bN$, it holds
				\begin{equation}\label{cond_exists}
					\WF \left( \frac{\delta^{\nu}  (\omega_\phi - \omega'_\phi)(x_1, x_2)}{\delta \phi(y_1) \dots \delta \phi(y_{\nu})} \right)_{x_1, x_2} = \emptyset.
				\end{equation}
			Furthermore, for $x_1, x_2 \in B$ it holds
				\begin{equation}\label{cond_for_W}
					\left.
					\begin{array}{r}
						(x_1, x_2, y_1, \dots, y_\nu; k_1, k_2, p_1, \dots, p_\nu)  \in \WF \left( \frac{\delta^{\nu}  (\omega_\phi - \omega'_\phi)(x_1, x_2)}{\delta \phi(y_1) \dots \delta \phi(y_{\nu})} \right)\\
						(y_1, \dots, y_\nu; p_1, \dots, p_\nu) \in C^\pm_\nu
					\end{array}
					\right\}  \Rightarrow k_1, k_2 \in V^\mp,
				\end{equation}
			where $C^\pm_\nu$ are the set defined by~\eqref{aux_C_sets}.
		\end{lemma}
		\begin{proof}
			By hypothesis, $\phi \mapsto \omega_\phi$ and $\phi \mapsto \omega'_\phi$ are two admissible assignments in the sense of def.~\ref{def_suit_omega}. Therefore, they satisfy the estimate~\eqref{WF_better} of condition~\ref{o'_2}, i.e.
				\begin{equation*}
					\WF \left( \frac{\delta^{\nu}  \omega_\phi}{\delta \phi^\nu} \right), \WF \left( \frac{\delta^{\nu}  \omega'_\phi}{\delta \phi^\nu} \right) \subset Z_{2 + \nu}.
				\end{equation*}
			Since $\delta^{\nu}  (\omega_{\phi} - \omega'_{\phi}) / \delta \phi^\nu$ is just the difference of $\delta^{\nu} \omega_{\phi}  / \delta \phi^\nu$ and $\delta^{\nu} \omega'_{\phi}  / \delta \phi^\nu$, the following estimate follows straightforwardly
				\begin{equation*}
					\WF \left( \frac{\delta^{\nu}  (\omega_\phi - \omega'_\phi)}{\delta \phi^\nu} \right) \subset Z_{2+\nu}.
				\end{equation*}
			By construction, $(\omega_\phi - \omega'_\phi)(x_1,x_2)$ is symmetric, and, therefore, $\delta^{\nu}  (\omega_{\phi} - \omega'_{\phi})(x_1,x_2) / \delta \phi^\nu$ is symmetric in $x_1,x_2$. Thus, we have
				\begin{equation}\label{rough}
					\WF \left( \frac{\delta^{\nu}  (\omega_\phi - \omega'_\phi)}{\delta \phi^\nu} \right) \subset \bP^+ Z_{2+\nu},
				\end{equation}
			where
				\begin{equation}\label{WF_symm_C+;2_C-;1}
					\begin{split}
						\bP^+ Z_{2+\nu} &:= \left\{ (x_1, x_2, y_1, \dots, y_n; k_1, k_2, p_1, \dots, p_\nu) \in Z_{2+\nu} : \right.\\
						&\qquad  \left. (x_2, x_1, y_1, \dots, y_n; k_2, k_1, p_1, \dots, p_\nu) \in Z_{2+\nu}  \right\}\\
						&= \dot{T}^* M^{2 +\nu} \backslash (C^{2;+}_{2+\nu} \cup C^{2; -}_{2+\nu} \cup C^{1; +}_{2+\nu} \cup C^{1; -}_{2+\nu}).
					\end{split}
				\end{equation}
			To get the last line, we used the definition of $Z_{2 + \nu}$~\eqref{Z} in terms of the sets $C^{i;\pm}_{2+\nu}$~\eqref{aux_C_sets}. It follows that if $(x_1, x_2, y_1, \dots, y_n; k_1, k_2, 0, \dots, 0)$ is in $\WF ( \delta^{\nu}  (\omega_{-} - \omega'_{-}) / \delta \phi^\nu)$, then it must have $k_1, k_2 \in \overline{V}^+ \cap \overline{V}^- = \{0\}$. However, by definition the wave-front set does not contain elements with vanishing covectors, and, therefore, condition~\eqref{cond_exists} indeed holds.\\
			Next, we notice that the estimate~\eqref{rough} is not enough to prove condition~\eqref{cond_for_W}. One can easily see from~\eqref{rough} that if $(x_1, x_2, y_1, \dots, y_\nu; k_1, k_2, p_1, \dots, p_\nu)$ belongs to $\WF ( \delta^{\nu}  (\omega_{-} - \omega'_{-}) / \delta \phi^\nu)$ with $x_1, x_2 \in B$ and $(y_1, \dots, y_\nu; p_1, \dots, p_\nu) \in C^\pm_\nu$, then it follows from~\eqref{rough} that $(k_1, k_2) \notin \overline{V}^\pm \times \overline{V}^\pm$.\\
			To strengthen this bound, we use eq.~\eqref{diff_state_trick}. We can compute the $\nu$-th Gateaux derivative of $\omega_\phi - \omega'_\phi$ by distributing the variational derivatives on the factors appearing in the right-hand side of eq.~\eqref{diff_state_trick}. By doing so, we have that $\delta^{\nu}  (\omega_{-} - \omega'_{-}) / \delta \phi^\nu$ is a finite sum of terms in the form
				\begin{equation}\label{term_var_der_state_trick}
					\begin{split}
						&\left(\frac{\delta^{|N_1|} E_\phi}{\delta \phi^{|N_1|}(\{y_{r} \}_{r \in N_1})} \circ \sigma_{c_+} \circ \frac{\delta^{|N_2|} (\omega_\phi - \omega_\phi')}{\delta \phi^{|N_2|}(\{y_{r} \}_{r \in N_2})} \circ \sigma_{c_+} \circ  \frac{\delta^{|N_3|} E_{\phi}}{\delta \phi^{|N_4|}(\{y_{r} \}_{r \in N_3})} \right)(x_1,x_2),
					\end{split}
				\end{equation}
			where $N_1, N_2, N_3$ form a partition of $\{1, \dots, \nu\}$.\\
			Because $\sigma_{c_+}$ is compactly supported, and because of the estimates of the wave-front set of $\sigma_{c_+}$ (given by~\eqref{WF_kernel_symp}), $\delta^{|N_i|} E_\phi / \delta \phi^{|N_i|}$ (given by~\eqref{var_ders_causal_WF}) and $\delta^{|N_2|} (\omega_\phi - \omega_\phi') / \delta \phi^{|N_2|}$ (given by~\eqref{rough}), it follows from thm.~\eqref{theo_WF_horma} that each term~\eqref{term_var_der_state_trick} is well-defined.\\ 
			Next, we show that if $(x_1, x_2, (y_{r \in N_i}), \dots, y_\nu; k_1, k_2, (p_{r \in N_i}))$ belongs to $\WF ( \delta^{|N_i|} E_\phi / \delta \phi^{|N_i|})$ with $x_1, x_2 \in B$, then $k_1, k_2$ must be both null covectors. For $|N_i| =0$, this claim is a consequence of the fact that the wave-front set of the causal propagator~\eqref{WF_causal} contains only null covectors. For $|N_i| >0$, we have $y_r \in \supp \lambda$ for any $r \in N_i$ as follows from prop.~\ref{prop_var_ders_causal}. Now, the estimate~\eqref{var_ders_causal_WF} implies that $(x_1, x_2, (y_{r \in N_i}), \dots, y_\nu; k_1, k_2, (p_{r \in N_i}))$ is contained in the set $X_{2 +|N_i|}$, and so, by the definition of $X_{2 +|N_i|}$ (see~\eqref{X}), there must be $(y', p'), (y'', p'') \in T^* M$ for certain points $y',y''$ among $\{y_{r \in N_i}\}$ such that
				\begin{equation*}
					\begin{split}
						&(x_1, k_1) \sim (y', -p'), \qquad x_1 = y', k_1 = - p', \qquad  k_1, p' =0 \\
						&(y'', p'') \sim (x_2, k_2), \qquad y''= x_2, p'' = -k_2, \qquad  p'', k_2 =0
					\end{split}
				\end{equation*}
			Since, by hypothesis, $x_1, x_2$ belongs to $B$ and $B$ is disjoint to $\supp \lambda$, it follows that $k_1, k_2$ must be null covectors as we wanted to prove.\\
			This result implies that any element $(x_1,x_2, y_1, \dots, y_\nu; k_1, k_2, p_1, \dots, p_\nu)$ with $x_1,x_2 \in B$ which is in the wave-front set of each distribution~\eqref{term_var_der_state_trick} must be such that $k_1, k_2$ are null covectors, as can be seen by applying the wave-front set calculus thm.~\ref{theo_WF_horma}. Thus, a similar result holds for any element $(x_1,x_2, y_1, \dots, y_\nu; k_1, k_2, p_1, \dots, p_\nu)$ of the wave-front set of $\delta^{\nu} (\omega_{-} - \omega'_{-}) / \delta \phi^\nu$ with $x_1, x_2 \in B$.\\
			Finally, combining this with the constraints imposed by estimate~\eqref{rough} we derived before, namely that each element $(x_1, x_2, y_1, \dots, y_\nu; k_1, k_2, p_1, \dots, p_\nu)$ in $\WF ( \delta^{\nu}  (\omega_{-} - \omega'_{-}) / \delta \phi^\nu)$ with $x_1, x_2 \in B$ and $(y_1, \dots, y_\nu; p_1, \dots, p_\nu) \in C^\pm_\nu$ must have $(k_1, k_2) \notin \overline{V}^\pm \times \overline{V}^\pm$. Thus, it follows that condition~\eqref{cond_for_W} holds, as we wanted to show.
		\end{proof}

		To prove that $\widetilde{\alpha(t)}{}^{n - 2\ell}_\phi$ satisfies conditions~\ref{W1},~\ref{W2} of def.~\ref{def_smooth_on_tens}, we compute its $\nu$-th Gateaux derivative distributing the variational derivatives onto each factors of the right-hand side of eq.~\eqref{eq_1_bis}. If follows that $\delta^\nu \widetilde{\alpha(t)}{}^{n - 2\ell}_\phi / \delta \phi^\nu$ is a finite sum of terms in the form
			\begin{equation}\label{term_var_eq_1_bis}
				\int_{K^{2\ell}} \frac{\delta^{|N_t|} \Theta_\phi (z_1, \dots, z_{2\ell}, x_1, \dots, x_{n- 2\ell})}{\delta \phi^{|N_t|} (\{y_{r \in N_t}\})} \frac{\delta^{|N_i|}  (\omega_{-} - \omega'_{-})(z_{2i -1}, z_{2i})}{\delta \phi ^{|N_i|} (\{y_{\nu_i \in N_i} \})} dz_1 \dots dz_{2\ell},
			\end{equation}
		where $N_t, N_1, \dots, N_\ell$ is a partition of $\{1, \dots, \nu\}$.\\
		First of all, we notice that each of the terms~\eqref{term_var_eq_1_bis} is a well-defined distribution. In fact, both the multiplication condition~\eqref{mult_cond} and the integration condition~\eqref{int_cond} of thm.~\ref{theo_WF_horma} are satisfied: the first holds because of~\eqref{cond_exists} we proved in lemma~\ref{lemma_diff_var_states}, while the second holds because $\delta^{|N_t|} \Theta_\phi / \delta \phi^{|N_t|}$ is a compactly supported distribution. Thus, by applying thm.~\ref{theo_WF_horma}, it follows that the distribution~\eqref{term_var_eq_1_bis} is well-defined.\\
		To show that $\widetilde{\alpha(t)}{}^{n - 2\ell}_\phi$ satisfies condition~\ref{W1}, it is sufficient to prove that the wave-front set of each term~\eqref{term_var_eq_1_bis} is contained in $W_{n- 2\ell +\nu}$. In other words, it is sufficient to show that there is no element $(x_1, \dots, x_{n-2\ell}, y_1, \dots, y_\nu; k_1, \dots, k_{n -2 \ell}, p_1, \dots, p_\nu)$ of the wave-front set of~\eqref{term_var_eq_1_bis} such that the all covectors $k_1, \dots, k_{n -2 \ell}, p_1, \dots, p_\nu$ belong to $\overline{V}^\pm$ except at most one which is space-like. By the wave front set calculus (thm.~\ref{theo_WF_horma}), if $(x_1, \dots, x_{n-2\ell}, y_1, \dots, y_\nu; k_1, \dots, k_{n -2 \ell}, p_1, \dots, p_\nu)$ is an element of the wave-front set of~\eqref{term_var_eq_1_bis}, then there must exist
			\begin{equation*}
					(z_1, \dots, z_{2 \ell}; q_1, \dots, q_{2\ell}) \in T^* M^{2 \ell},
				\end{equation*}
			such that $z_1, \dots, z_{2 \ell}$ are in $B$ and it holds 
				 \begin{equation}\label{aux_covectors_diff}
			 		\left\{ \begin{array}{l}
						(z_1, \dots, z_{2\ell}, x_1, \dots, x_{n-2 \ell}, (y_{r \in N_t}); -q_1, \dots, - q_{2\ell}, k_1, \dots, k_{n-2\ell}, (p_{r \in N_t})) \in W_{n +|N_t|} \\
						\qquad \mbox{ or } q_1, \dots, q_{2\ell}, k_1, \dots, k_{n-2\ell}, p_{r \in N_t} =0\\
						(z_{2i -1}, z_{2i}, (y_{r \in N_i}); q_{2i -1}, q_{2 i}, (p_{r \in N_i})) \in \WF(\delta^{|N_i|}(\omega_{-} - \omega'_{-}) / \delta \phi^{|N_i|}) \qquad \mbox{ for } i=1, \dots, \ell\\
						\qquad \mbox{ or } q_{2i-1}, q_{2i}, p_{r \in N_i} =0
					\end{array} \right.
				\end{equation}
			We used the fact that $\Theta_\phi$ satisfies~\ref{W1}, and so $\WF (\delta^{|N_t|} \Theta_\phi / \delta \phi^{|N_t|})$ is contained in $W_{n + |N_t|}$.\\
			Now, we prove the claim by reductio ad absurdum: if all covectors $k_1, \dots, k_{n -2 \ell}, p_1, \dots, p_\nu$ belong to $\overline{V}^\pm$ except at most one which is space-like, then lemma~\ref{lemma_diff_var_states} implies that the covectors $q_1, \dots, q_{2\ell}$ are in $V^\mp$. However, these configurations are incompatible with the conditions~\eqref{aux_covectors_diff}, and so we get a contradiction as we wanted to show. This concludes the proof that $\widetilde{\alpha(t)}{}^{n - 2\ell}_\phi$ satisfies condition~\ref{W1}.\\
			To prove~\ref{W2}, let $\bR \ni \epsilon \mapsto \phi(\epsilon) \in C^\infty(M)$ be smooth. The distribution $\delta^\nu \widetilde{\alpha(t)}{}^{n - 2\ell}_{\phi(\epsilon)}/ \delta \phi^\nu$ can be expressed as a finite sum of terms in the form~\eqref{term_var_eq_1_bis} with $\phi$ replaced by $\phi(\epsilon)$ everywhere. Arguing similarly as done for the proof of~\ref{W1}, but now using the fact that $\Theta_\phi$ satisfies~\ref{W2}, it follows that the wave-front set of each term~\eqref{term_var_eq_1_bis} in $\phi(\epsilon)$, viewed as distributions in the variables $\epsilon, x_1, \dots, x_{n-2\ell}, y_1, \dots, y_\nu \in \bR \times M^{n - 2\ell+\nu}$, is contained in $\bR \times \{0 \} \times W_{n -2\ell+\nu}$ which is precisely the requirement of~\ref{W2}. Consequently, also $\delta^\nu \widetilde{\alpha(t)}{}^{n - 2\ell}_{\phi(\epsilon)}/ \delta \phi^\nu$ satisfies the condition~\ref{W2}. This concludes the proof.
		\end{proof}

	To proceed, we need to the extend the isomorphism $\alpha$ to forms with values in $\cW$. In finite dimensions such extension is straightforward. In the framework of on-shell $W$-smooth sections, the desired extension is provided by the following canonical construction. We consider first $t \in \Omega^k_W(S,\cW)$ such that $t$ is a section homogeneous in $\deg_s$ and $\deg_\hbar$ with $\deg_s t = n$, we define $\alpha(t) \in \Omega^k_W(S,\cW)$ as the sequence $(\alpha(t)^{k,0}, \alpha(t)^{k,1}, \dots)$ where $\alpha(t)^{k,j}_\phi \in \bC[\hbar] \otimes \wedgevee^{k+j}_W T^*_\phi S$ is defined by the following distributional representative:
		\begin{equation}\label{alphamap_form_inf}
			\begin{split}
				&\alpha(t)^{k,n- 2 \ell}_\phi (y_1, \dots y_k; x_1, \dots, x_{n -2 \ell}) := \\
				&\quad = \frac{\hbar^\ell n!}{(n -2 \ell)! (2 \ell)!} \bP^+ \bP^- \int_{M^{2\ell}} t_\phi(y_1, \dots, y_k, z_1, \dots, z_{2\ell}, x_{1}, \dots, x_{n-2\ell} ) \prod_{i=1}^{k} (\omega_\phi - \omega'_\phi) (z_{2i-1}, z_{2i}) dz_1 \dots dz_{2\ell},
			\end{split}
		\end{equation}
	for $0 \leq \ell \leq [n/2]$, and $\alpha(t)^{k,j}_\phi =0$ otherwise. Arguing similarly as before for~\eqref{alphamap_fiber_inf}, it can be seen that the right-hand side of~\eqref{alphamap_form_inf} is well-defined.\\
	Using the filtrations with respect to $\deg_a$ and $\Deg$ of the algebra $\Omega_W(S,\cW)$, it follows that \eqref{alphamap_form_inf} gives a map acting on the whole algebra. By construction, it also preserves the total degree $\Deg$. Finally, using a similar argument as the one presented for the proof of prop.~\ref{prop_W-smooth_alpha_1}, we can verify that $\alpha$ defines an isomorphism $\Omega_W(S,\cW) \to \Omega_W(S,\cW')$. In other words, we proved the following proposition, which is the infinite-dimensional analogue of lemma~\ref{lemma_Wick_iso_fin}:
	\begin{prop}\label{prop_W-smooth_alpha}
		The map $\alpha$ defined by~\eqref{alphamap_form_inf} is an isomorphism $\Omega_W(S,\cW) \to \Omega_W(S,\cW')$, i.e. $\alpha(t \bullet s) = \alpha(t) \bullet' \alpha(s)$ for any $t,s \in \Omega(S,\cW)$, and it preserves the conjugation operation $\dagger$, i.e. $\alpha(t)^\dagger = \alpha( t^\dagger)$.
	\end{prop}
		
	Continuing our pedestrian approach, we next provide the infinite-dimensional analogue of lemma~\ref{lemma_pullback_Fedosov}.
		\begin{prop}
			$D^{\alpha, W} := \alpha D^{\prime W} \alpha^{-1}$ is a Fedosov $W$-connection. More precisely, $D^{\alpha, W}$ coincides with the derivative obtained from Fedosov's first theorem (thm.~\ref{theo_Fedosov_inf}) with respect to the product $\bullet$ and  is uniquely characterized by the following input data: the connection $\nabla^W$ and the auxiliary data $\Omega^\alpha=0$, $s^\alpha$  where $s^\alpha$ is a certain on-shell $W$-smooth section on $\cW$ with $\Deg s^\alpha \geq 3$.
		\end{prop}
		\begin{proof}
			Repeating the argument given in finite dimensions, it holds that $D^{\alpha, W}$ is a flat $\deg_a$-graded derivation of $\Omega_W(S,\cW)$. In fact, we notice that $D^{\prime W}$ preserves the on-shell $W$-smoothness because $D^{\prime W}$, $\alpha$ and $\alpha^{-1}$ do so. Moreover, the algebraic relations needed in the finite-dimensional proof persist in infinite dimensions. All that is required is that $\alpha$ is an isomorphism $\Omega_W(S,\cW) \to \Omega_W(S,\cW')$ and $D^{\prime W}$ is a flat $\deg_a$-graded derivation of $\Omega_W(S,\cW')$.\\
			What remains to be done is to establish the infinite-dimensional analogue of the elements $r^\alpha$ and $s^\alpha$ defined by~\eqref{r^alpha_fin} and by~\eqref{s_alpha}. Looking at the recursive formula~\eqref{r^alpha_fin}, we see that $r^\alpha$ is determined to all orders once is given $C$ as in~\eqref{c_fin}. To make this iterative machine work in infinite dimensions, we must show that the infinite-dimensional analogue of $C$ is on-shell $W$-smooth.\\
			The $\cW$-valued $1$-form $C$ is actually homogeneous in $\deg_s$ and $\deg_\hbar$ with $\deg_s C =2$ and $\\deg_\hbar C =0$. We provide a suitable extension for $C$ and check that this satisfies~\ref{W1},~\ref{W2} of def.~\ref{def_smooth_on_tens}.\\
			For any $\phi \in C^\infty(M)$ and for an arbitrary cut-off $c$ as in~eq.\eqref{kernel_symp}, we define the extension by:
				\begin{equation}\label{c12}
					\begin{split}
						\widetilde{C}_\phi (y, x_1, x_2) &:= \frac{1}{2} \bP^+ \int_{M^4} ( \sigma_c \circ E_\phi)(y, y') ( \sigma_c \circ E_\phi)(x_1, x'_1) ( \sigma_c \circ E_\phi) (x_2, x'_2) \times\\
						&\qquad \times \sigma(x'_1, z) (\Gamma_\phi - \Gamma'_\phi)(z,y',x'_2) dz dy' dx'_1 dx'_2,
					\end{split}
				\end{equation}
			where $\Gamma_\phi$, $\Gamma'_\phi$ are defined by~\eqref{Yano_symbol_W-smooth_on} for $\omega_\phi$, $\omega'_\phi$ respectively.\\
			It is a consequence of lemma~\ref{lemma_W_comp}, lemma~\ref{lemma_tech_X_W}, the definition of the distributions $\Gamma_\phi, \Gamma'_\phi$, and the estimates~\eqref{WF_better},~\eqref{WF_better_s},~\eqref{var_ders_causal_WF} and~\eqref{cont_var_ders_causal_WF} that $\widetilde{C}_\phi$ is a well-defined distribution in $(\sigma_c \circ E_\phi)^{\otimes 3} \circ \cE'_W(M^3)$ symmetric in the last two variables which satisfies the requirements~\ref{W1},~\ref{W2} of def.~\ref{def_smooth_on_tens}. In other words, we have that $C$ is a well-defined on-shell $W$-smooth $\cW$-valued $1$-form.\\
			Furthermore, the following the identities hold
				\begin{equation}\label{proper_c_inf}
					\alpha \nabla^{\prime W} \alpha^{-1} = \nabla^W - \frac{i}{\hbar}\ad_\bullet(C), \quad \delta C = \hat{T} - \hat{T}', \quad  \nabla^W C - \frac{i}{\hbar} C \bullet C = \alpha \hat{R}' - \hat{R}.
				\end{equation}
			These are obtained by the same algebraic manipulations as in the finite-dimensional case  (cf.~\eqref{alpha_on_Yano},~\eqref{technical_proper_c}). As in the finite-dimensional case, it then follows
				\begin{equation*}
					D^{\alpha, W} = - \delta + \alpha \nabla^{\prime W} \alpha^{-1} + \frac{i}{\hbar} \ad_{\bullet}(\alpha r') = - \delta + \nabla^W +\frac{i}{\hbar}\ad_\bullet (r^\alpha),
				\end{equation*}
			where $r^\alpha = \alpha r' - C$ is a section in $\Omega_W^1(S,\cW)$ with total degree $\Deg \geq 2$ (cf.~\eqref{D^alpha_fin}). Since the on-shell $W$-smoothness is preserved, and since the algebraic relations needed in the finite-dimensional proof persist in infinite dimensions, we can repeat the same argument already given in sec.~\ref{subsec_equivalence_fin} (cf.~\eqref{r^alpha_fin}) to show
				\begin{equation}\label{r^alpha}
					\delta r^\alpha = \nabla^W r^\alpha - \hat{R} - \hat{T} +\frac{i}{\hbar} r^\alpha \bullet r^\alpha, \quad \delta^{-1} r^\alpha = s^\alpha:= \delta^{-1} \alpha r' - \delta^{-1} C.
				\end{equation}
			Fedosov's first theorem (thm.~\ref{theo_Fedosov_inf}) ensures that $r^\alpha$ is the unique solution of the system~\eqref{r^alpha}. Thus, $D^{\alpha W}$ coincides with the Fedosov derivative with respect to the product $\bullet$, uniquely characterized by the input data $\nabla^W$, $\Omega^\alpha=0$, $s^\alpha = \delta^{-1} \alpha r' - \delta^{-1} C$.
		\end{proof}
		
	Finally, we can prove the existence of the gauge equivalence between the two Fedosov $W$-connections $D^W$ and $D^{\prime W}$, i.e. the infinite-dimensional analogue of theorem~\ref{thm1_bis}.
		\begin{theo}\label{theo_gauge_inf}
			Let $\phi \mapsto \omega_\phi$ and $\phi \mapsto \omega_\phi'$ be two admissible assignment in the sense of def.~\ref{def_suit_omega}. There exists an on-shell $W$-smooth section $H \in C^\infty_W(S,\cW)$ such that $\Deg H \geq 3$, $\tau H = 0$, $H^\dagger = H$ and
				\begin{equation}\label{gauge_inf}
					D^W = \exp \left( - \frac{i}{\hbar} \ad_\bullet (H) \right) \alpha D^{\prime W} \alpha^{-1} \exp \left( \frac{i}{\hbar} \ad_\bullet (H) \right).
				\end{equation}
			In particular, a solution $H$ for eq.~\eqref{gauge_inf} is uniquely determined by a closed on-shell $W$-smooth $1$-form $\theta \in \Omega_W(S)[[\hbar]]$.
		\end{theo}
		\begin{proof}
			The iterative construction of $H$ for $\theta = 0$ given by~\eqref{low_H_bis},~\eqref{iter_H_bis} is valid also in the infinite-dimensional case, with the obvious substitutions, because the input is on-shell $W$-smooth and all iteration steps only involve the operations $\bullet$, $\nabla^{W}$ and $\delta$ which preserves this properties.
		\end{proof}
		
		We conclude this section deriving two straightforward corollaries of thm.~\ref{theo_ret_conn_fedosov_conn} and thm.~\ref{theo_gauge_inf}. Let $\phi \mapsto \omega^R_\phi$ the admissible assignment given by the retarded $2$-point function for any $\phi$ (we proved in lemma~\ref{lemma_ret_state_fedosov} that this assignment is indeed admissible in the sense of def.~\ref{def_suit_omega}), and let $\phi \mapsto \omega'_\phi$ be any other admissible assignment. We have shown in thm.~\ref{theo_ret_conn_fedosov_conn} that $D^{W,R} = \nabla^R - \delta$. Now, if $F$ is a local functional and if $S \ni\phi \mapsto \hat{F}_\phi$ is the corresponding quantum observable (viewed as a section in the bundle $\cW$) defined by the Haag's series~\eqref{Haag_series}, then we have seen in thm.~\ref{theo_Fedosov_per} that $D^{W,R} \hat{F} = 0$. Combining with eq.~\eqref{gauge_inf}, we get the following result:
			\begin{prop}\label{corollary}
				Let $F$ be a local functional and let $S \ni \phi \mapsto \hat{F}_\phi \in \cW_\phi$ be the on-shell $W$-smooth section given by the Haag's formula~\eqref{Haag_series}, where the formal Wick algebra $\cW_\phi$ is defined for any $\phi \in S$ in terms of the retarded $2$-point function $\omega^R_\phi$. Let $\phi \mapsto \omega'_\phi$ be any admissible assignment in the sense of def.~\ref{def_suit_omega} of a pure Hadamard $2$-point function $\omega'_\phi$ for any $\phi \in C^\infty(M)$ with corresponding Fedosov connection $D^W$. Let $\hat{F}'_\phi$ be the element in $\cW_\phi'$ given by
					\begin{equation*}
						\hat{F}'_\phi := \alpha_\phi^{-1} \exp \left( \frac{i}{\hbar} \ad_{\bullet^R}(H) \right)_\phi \hat{F}_\phi.
					\end{equation*}	
				Then, the map $\hat{F}': S\ni \phi \mapsto \hat{F}'_\phi \in \cW_\phi'$ is an on-shell $W$-smooth section and 
					\begin{equation*}
						D^{\prime W} \hat{F}' = 0.
					\end{equation*}
			\end{prop}
		What we still have to prove is that $S \ni \phi \mapsto \hat{F}_\phi \in \cW_\phi$ is an on-shell $W$-smooth section. Once we have established that, the map $\hat{F}': S\ni \phi \mapsto \hat{F}'_\phi \in \cW_\phi'$ must be on-shell $W$-smooth because $\alpha^{-1}$ (as proved in prop.~\ref{prop_W-smooth_alpha}), the product $\bullet^R$, and $\exp ( \frac{i}{\hbar} \ad_{\bullet^R}(H) )$ (as a consequence of the on-shell $W$-smoothness of $H$ given by thm.~\ref{theo_gauge_inf}) preserve on-shell $W$-smoothness.\\
		Proving that $S \ni \phi \mapsto \hat{F}_\phi \in \cW_\phi$ is on-shell $W$-smooth is rather lengthy. Therefore, we devote the entire following section sec.~\ref{subsec_W_smooth_ret_prod} to it. However, before we do that, we point out the following corollary of prop.~\ref{corollary}, which shows how space-times locality (Einstein causality) can be implemented:
		\begin{prop}\label{prop_einst_loc}
			Let $F_1, F_2$ be two local functionals and let $\phi \mapsto \omega'_\phi$ an admissible assignment in the sense of def.~\ref{def_suit_omega}). Let $C^\infty_W(S,\cW')$ be the algebra of on-shell $W$-smooth sections in the local Wick algebra corresponding to the product  $\bullet'$ defined as in prop.~\ref{prop_prod_smooth_on_W} in terms of the assignment $\phi \mapsto \omega'_\phi$. We have that $\hat{F}'_1 \bullet' \hat{F}'_2 = \hat{F}'_2 \bullet' \hat{F}'_1$ if the support of $F_1$ and $F_2$ are space-like separated, i.e. $\supp F_1 \cap J(\supp F_2) = \emptyset$.
		\end{prop}
		\begin{proof}
			Let $\phi \mapsto \omega^R_\phi$ be the admissible assignment corresponding to the retarded $2$-point function and let $C^\infty_W(S,\cW^R)$ be the algebra of on-shell $W$-smooth sections corresponding to the product $\bullet^R$ which is defined as in prop.~\ref{prop_prod_smooth_on_W} in terms of $\phi \mapsto \omega^R_\phi$. By construction, $\alpha$ is an isomorphism $C^\infty_W(S,\cW^R)  \to C^\infty_W(S, \cW')$, while $\exp ( \frac{i}{\hbar} \ad_{\bullet^R}(H) )$ is an endomorphism of $C^\infty_W(S,\cW^R)$. It follows
				\begin{equation}\label{rewriting}
					\begin{split}					
							\hat{F}'_1 \bullet' \hat{F}'_2 &= \left( \alpha_\phi^{-1} \exp \left( \frac{i}{\hbar} \ad_{\bullet^R}(H) \right) \hat{F}_1 \right)  \bullet'  \left( \alpha_\phi^{-1} \exp \left( \frac{i}{\hbar} \ad_{\bullet^R}(H) \right) \hat{F}_2 \right)\\
						&= \alpha_\phi^{-1} \exp \left( \frac{i}{\hbar} \ad_{\bullet^R}(H) \right) \left( \hat{F}_1 \bullet^R  \hat{F}_2\right).
					\end{split}
				\end{equation}
			The sections $\hat{F}_1, \hat{F}_2$ are given by the Haag's formula~\eqref{Haag_series}, and, therefore, can be expressed as formal series of retarded products. Because of the GLZ formula~\ref{R11} (see sec.~\ref{subsec_int_QFT_per}), it follows that $\hat{F}_1 \bullet^R  \hat{F}_2 = \hat{F}_2 \bullet^R  \hat{F}_1$ if the support of $F_1$ and $F_2$ are space-like separated, see e.g.~\citep{DF04}. The claim then follows straightforwardly.
		\end{proof}
	
\section{On-shell $W$-smoothness of $S \ni \phi \mapsto \hat{F}_\phi \in \cW_\phi$}\label{subsec_W_smooth_ret_prod}
	In chapter~\ref{sec_pQFT}, when we discussed the perturbative approach to the quantization of interacting massive scalar theory around a classical background (sec.~\ref{subsec_int_QFT_per}) we provided an axiomatic characterization of prescriptions for retarded products $\{ R_{n,\phi}:\cF_\loc \otimes \cF_\loc^{\otimes n} \to \cW_\phi \}_{n \in \bN}$ for each $\phi \in S$. The Haag formula~\eqref{int_funct} for $\hat{F}_\phi$ expresses this quantity in terms of retarded products for each $\phi \in S$. Thus, if we can show that each retarded product $S \ni \phi \mapsto R_{n,\phi}$ is on-shell $W$-smooth, $S \ni \phi \mapsto \hat{F} \in \cW_\phi$ is also on-shell $W$-smooth. We will indeed show:
		\begin{theo}\label{thm_on_W_ret}
			For any local fuctionals $F, H_1, \dots, H_n$, there exists a prescription for retarded products such that the assignment $S \ni \phi \mapsto R_{n,\phi}(F(\phi + \varphi), \otimes_{j=1}^n H_j(\phi + \varphi)) \in \cW_\phi$ is on-shell $W$-smooth.
		\end{theo}
	\noindent
	On-shell $W$-smoothness requires that there are extensions of these maps to $C^\infty(M)$, which we have a sufficient microlocal control on their variational derivatives, more precisely the conditions~\ref{W1},~\ref{W2} in def.~\ref{def_smooth_on_tens}. It is far from  obvious that such requirements are satisfied a priori.\\
	
	We begin the proof of thm.~\ref{thm_on_W_ret} by noticing that it is sufficient to prove that there exists a prescription for time-ordered products $\{ T_{n,\phi}: \cF_\loc^{\otimes n} \to \cW_\phi \}_{n \in \bN}$ satisfying the axioms (T1)-(T10), and (T11c) defined in~\citep{HW05} ( and the axiom (T11a) necessary in the proof of the consistency of (T11c) with the other axioms, see appendix~\ref{app_proof_R12}), and such that the assignment $S \ni \phi \mapsto T_{n,\phi}[\otimes_{i=1}^n F_i(\phi + \varphi)] \in \cW_\phi$ is an on-shell $W$-smooth section for any local functionals $F_1, \dots, F_n$. Indeed, once such prescription for the time-ordered products is provided, it is well-known that a prescription for retarded products satisfying the axioms~\ref{R0}-\ref{R11} and \ref{R12} (see sec.~\ref{subsec_int_QFT_per}) can be defined by
					\begin{equation}\label{R-T}
						\begin{split}
						&R_{n, \phi} \left( F, \bigotimes_{k=1}^n H_k \right) := \sum_{I \subset \{1, \dots, n\}} (-1)^{|I|} \overline{T}_{|I|,\phi} \left[ \bigotimes_{i \in I} H_i \right] \bullet_\phi T_{|I^c|+1,\phi} \left[ F \otimes \bigotimes_{j \in I^c} H_j \right] \\
						&\quad = \sum_{I} \sum_{I_1 \sqcup \dots \sqcup I_\ell = I} (-1)^\ell T_{|I_1|,\phi}\left[ \bigotimes_{i_1 \in I_1} H_{i_1} \right] \bullet_\phi \cdots \bullet_\phi T_{|I_\ell|,\phi}\left[ \bigotimes_{i_\ell \in I_\ell} H_{i_\ell} \right] \bullet_\phi T_{|I^c|+1,\phi}\left[ F \otimes \bigotimes_{j \in I^c} H_{j} \right],
						\end{split}
					\end{equation}
	where $\overline{T}$ denotes the anti-time-ordered product, see e.g.~\citep[(T7)]{HW05}. The on-shell $W$-smoothness is clearly preserved because the product $\bullet_\phi$ preserves this notion of smoothness as we have already proved in prop.~\ref{prop_prod_smooth_on_W}.\\
	
	The strategy of our proof consists in the followings steps:
		\begin{enumerate}
			\item In sec.~\ref{subsubsec_properties_omega_H}, we present some preliminary technical results we will need later on in the proof. We first consider the case of a generic space-time $(M,g)$ (not necessarily ultra-static nor with compact Cauchy surfaces), and let $m, \phi, \lambda \in C^\infty(M)$. In this general setting, we investigate a particular distribution, the Hadamard parametrix $H_\phi$ defined by~\eqref{Hadamard_def} with respect to the linear operator $P_\phi = \boxempty - m^2 - \frac{\lambda}{2} \phi^2$. More precisely, we are interested in the variational derivatives of $H_\phi$, especially their microlocal behaviour and their scaling properties under a rescaling of $g, m , \phi, \lambda$. This will be done in lemma~\ref{lemma_H}.\\
			Then, we restrict to the more specific situation of an ultra-static space-time $(M,g)$ with compact Cauchy surfaces, a constant $m$ and  a compactly supported $\lambda$, i.e. the setting we consider throughout sec.~\ref{subsec_manifold_inf}-\ref{subsec_Fedosov_inf} and in sec.~\ref{subsec_gaugeeq}. In this situation, we discuss the properties of the difference $d_\phi := \omega^R_\phi - H_\phi$ between the retarded $2$-point function $\omega^R_\phi$ given by eq.~\eqref{in-state} and the Hadamard parametrix $H_\phi$. In particular, lemma~\ref{lemma_d} and lemma~\ref{lemma_d_coinc} provide a microlocal control on the variational derivatives of $d_\phi$ sufficient for our purposes in the following steps.
			\item With these technical results at our disposal, we begin the proof of the existence of a prescription for time-ordered products which has the desired on-shell $W$-smoothness and satisfies the axioms (T1)-(T10) and (T11c). Actually, we consider first local functionals which do not involve covariant derivatives and we consequently demand that only axioms (T1)-(T9) are fulfilled. The first step, presented in sec.~\ref{subsubsec_Wickpower}, is to provide a prescription for time-ordered products of one functional $F$. Following~\citep{HW01, HW02}, this is done considering the Wick powers defined in terms of the Hadamard parametrix $H_\phi$. We then use lemma~\ref{lemma_d_coinc} to prove that $S \ni \phi \mapsto T_{1,\phi}[F(\phi + \varphi)] \in \cW_\phi$ is on-shell $W$-smooth for any local functional $F$ which does not contain covariant derivatives.
			\item In sec.~\ref{subsubsec_suff_cond_T-prod}, we discuss time-ordered products of more factors $F_1, \dots , F_n$ (not involving covariant derivatives). We formulate sufficient conditions, collected in lemma~\ref{lemma_suff_cond}, to ensure the on-shell $W$-smoothness of any map $S \ni \phi \mapsto T_{n,\phi}[\otimes_{i=1}^n F_i(\phi +\varphi)] \in \cW_\phi$. We formulate these sufficient conditions in terms of the Wick expansion with respect to the retarded $2$-point function $\omega_\phi^R$.
			\item In sec.~\ref{subsubsec_rew_T-prod}, we review the procedure outlined in~\citep{HW02} to define the time-ordered products inductively starting from the Wick powers. The construction, which needs to be given for generic space-time $(M,g)$ and arbitrary smooth functions $m, \phi,\lambda$, relies on three fundamental concepts: the local Wick expansion (see eq.~\eqref{local_Wick_exp}), i.e. the Wick expansion in terms of the normal ordering with respect to the Hadamard parametrix $H_\phi$, the scaling expansion (see eq.~\eqref{formula_taylor_t^0}) for the distributional coefficients of the Wick expansion outside the total-diagonal, and the control of the extensions of such distributions provided by their scaling properties.
			\item In sec.~\ref{subsubsec_var}, we prove additional properties, listed in prop.~\ref{prop_t_H_W_smooth}, for the variational derivative of the distributional coefficients appearing in the local Wick expansion. The proof of these conditions relies on the fundamental properties of the Hadamard parametrix $H_\phi$ we prove in lemma~\ref{lemma_H}.
			\item The purpose of sec.~\ref{subsubsec_proof} is to prove that any prescription of time-ordered products constructed following the procedure of~\citep{HW02} satisfying the sufficient conditions of lemma~\ref{lemma_suff_cond}. This is done in prop.~\ref{prop_t_W-smooth}.
			\item The last part, sec.~\ref{subsubsec_T10,T11}, is devoted to two issues. First, we discuss the extension to local functionals which involve covariant derivatives and we prove the existence of an on-shell $W$-smooth time-ordered products prescription satisfying the Leibniz rule axiom (T10) (see~\citep{HW05}) in addition to axiom (T1)-(T9). Then, we prove the existence of a $W$-smooth prescription for time-ordered products that satisfies also the principle of perturbative agreement, i.e. we require axiom (T11c) (and axiom (T11a)) in addition to axiom (T1)-(T10).
		\end{enumerate}
		\subsection{Properties of the Hadamard parametrix and the difference of the Hadamard parametrix and the retarded $2$-point function}\label{subsubsec_properties_omega_H}		
				Let $(M,g)$ be a generic space-time and let $m, \phi, \lambda$ be generic functions in $ C^\infty(M)$. We consider the Klein-Gordon operator $P_\phi = \boxempty - m^2 - \frac{1}{2}\lambda \phi^2$.
				Let $U \subset M$ be a convex normal neighbourhood, i.e. for any two points in $U$ there is a unique geodesic connecting the two points. The {\em Hadamard parametrix} is a distribution on $U \times U$ which is a bi-solution for $P_\phi$ up to a smooth function and which is in the form
					\begin{equation}\label{Hadamard_def}
						\begin{split}
							\gls{H_phi}(x_1,x_2) &:= \frac{u_0(x_1,x_2)}{\sigma_\epsilon (x_1,x_2)} + v(x_1,x_2) \ln (\sigma_\epsilon (x_1,x_2)),
						\end{split}
					\end{equation}
				where $u_0$ and $v$ smooth functions defined in terms of the so called {\em Hadamard coefficients} described in a moment, where $\sigma_\epsilon (x_1,x_2) := \sigma (x_1,x_2) + i \epsilon(T(x_1) - T(x_2))$, where $T$ is a global time coordinate, and where $\sigma(x_1, x_2)$ denotes the (signed) square of the geodesic distance from $x_1$ to $x_2$, i.e.
					\begin{equation}\label{signed_geo_dist_square}
						\sigma(x_1, x_2) = \pm \left( \int_0^1 \left| g_{\mu \nu}(\gamma_{x_1,x_2}(t)) \frac{d \gamma^\mu_{x_1,x_2}(t)}{d t} \frac{d \gamma^\nu_{x_1,x_2}(t)}{d t} \right|^{1/2} dt \right)^{2}.
					\end{equation}
				The curve $\gamma_{x_1,x_2} :[0,1] \to M$ is a parametrization of the unique geodesic connecting $x_1, x_2$ such that $\gamma_{x_1,x_2}(0)=x_1$ and $\gamma_{x_1,x_2}(1) = x_2$. The sign in the definition of $\sigma(x_1, x_2)$ is ``$+$'' or ``$-$'' if $\gamma_{x_1,x_2}$ is time-like or space-like, respectively. The definition eq.\eqref{Hadamard_def} is given by an ``$\epsilon$-prescription'': for any $f$ test function in $U \times U$, $H_\phi(f)$ is defined by first computing it for $\epsilon >0$, and then taking the limit $\epsilon \to 0$.\\
				The first Hadamard coefficient $u_0$ is the ``Van Vleck-Morette determinant''
					\begin{equation}\label{u_0}
						u_0 (x_1,x_2) := \left( \frac{\det \left( \partial_\mu^{(x_1)} \partial_\nu^{(x_2)}  \sigma(x_1,x_2) \right)}{\sqrt{\det (g_{\mu \nu}(x)) \det(g_{\mu \nu}(y)) }} \right)^{\frac{1}{2}},
					\end{equation}
				which is a strictly positive smooth function and depends on the metric $g$ in a local and covariant way.\\
				For $k >0$ the coefficient $u_{\phi,k} = u_k[g,m,\phi,\lambda]$ (the dependence on the metric $g$, the mass $m$, the background $\phi$ and the coupling $\lambda$ is emphasized) are given by the following recursive formula:
					\begin{equation}\label{induct_u}
						u_{\phi,k+1}(x_1,x_2) = - \int_M \int_0^1 t^k \frac{u_{0}(x_1,x_2)}{u_{0}(z,x_2)} \delta (z, \gamma_{x_1,x_2}(t))  P_\phi^{(z)} u_{\phi,k}(z,x_2) dz dt.
					\end{equation}
				Since $u_0[g]$ depends locally and covariantly on the metric $g$, each $u_{\phi, k}$ depends in a local and covariant way on $g, m, \phi, \lambda$.\\
				Concerning the scaling behaviour, note that $u_0[\Lambda^{-2} g] = u_0[g]$ and $u_k[\Lambda^{-2}g, \Lambda m, \Lambda \phi, \lambda] = \Lambda^{2k} u_k[g, m, \phi, \lambda]$ for any $\Lambda >0$.\\
				On a real analytic space-time and for real analytic data $m, \phi, \lambda$, we can define $v_\phi := \sum_{k>0} u_{\phi,k+1} \sigma^k$. This series converges as shown e.g. in~\citep{friedlander1975wave}. In principle, we would like to define $v_\phi$ similarly also for space-times and data $m, \phi, \lambda$ which are only smooth. However, the series need not converge in this case. Following~\citep{HW01} (see also~\citep{BGP07, S13}), we overcome this problem defining instead
					\begin{equation}\label{v_def}
						v_\phi(x_1,x_2) := \sum_{k \geq 0} \psi( \sigma(x_1,x_2)/ \alpha_k) u_{\phi,k+1}(x_1,x_2) \sigma^k(x_1,x_2),
					\end{equation}
				where $\psi: \bR \to \bR$ is a compactly supported smooth function and $\{ \alpha_k \}_{k \in \bN}$ is a sequence of real number which are introduced to ensure the convergence of the series. More precisely, $\psi$ is chosen such that $\psi(x) = 1$ for $|x| < 1/2$ and $\psi(x) = 0$ for $|x| >1$ and $\alpha_k$ tends to zero sufficiently fast.\\
				Let us collect some properties of the Hadamard parametrix:
				\begin{itemize}
					\item The distribution $H_\phi = H[g,m,\phi,\lambda]$ is local and covariant in the following sense: let $\iota: M' \to M$ be causality-preserving isometric embedding between the space-times $(M', g')$ and $(M,g)$, i.e. $g' = \iota^* g$, and let $f$ be a test function supported in $U' \times U'$ where $\iota U' \subset U$, then it follows
						\begin{equation}\label{eq_cov}
							\left(\iota^* H[g, m, \phi, \lambda]\right) (f) = H[\iota^*g, \iota^*m, \iota^*\phi, \iota^*\lambda] (f).
						\end{equation}
					\item As proved in~\citep{radzikowski1996micro}, the wave-front set of $H_\phi$ can be estimated by
								\begin{equation}\label{H_est}
									WF( H_\phi ) \subset \left. \cC^\triangleright[g]\right|_{U \times U},
								\end{equation}
							where $\cC^\triangleright$ is the set defined by~\eqref{WF_Hadamard}. Furthermore, every Hadamard $2$-point function differs from $H_\phi$ by a smooth function, as shown also in~\citep{radzikowski1996micro}.\\
							If $\{ g^{(s)}, (m^2)^{(s)}, \phi^{(s)}, \lambda^{(s)} \}$ are smooth $1$-parameter families, then $H[g^{(s)}, m^{(s)}, \phi^{(s)}, \lambda^{(s)}]$ can be interpreted as a distribution in $\bR \times U \times U$ and it holds that
								\begin{equation}\label{H_est_s}
									\begin{split}
										\WF (H[g^{(s)}, m^{(s)}, \phi^{(s)}, \lambda^{(s)}](x_1,x_2)) \subset &\left\{ (s, x_1, x_2 ; \rho, k_1, k_2) \in \dot{T}^*(\bR \times U \times U) : \right. \\
										&\quad \left. (x_1, x_2; k_1, k_2) \in \cC^\triangleright[g^{(s)}] \right\}.
									\end{split}
								\end{equation}
						Less trivially, when the Hadamard parametrix is restricted to the total diagonal $\Delta_2$ in $U \times U$, which can be done as a consequence of the estimate~\eqref{H_est_s} and~\citep[thm. 8.2.4]{H83}, it holds in addition that
							\begin{equation}\label{H_est_s_diag}
								\left. \WF \left( H[g^{(s)}, m^{(s)}, \phi^{(s)}, \lambda^{(s)}](x_1,x_2) \right) \right|_{\bR \times \Delta_2} \perp T(\bR \times \Delta_2).
							\end{equation}
						If we vary smoothly only the background $\phi$, it i.e. for fixed $g, m, \lambda$ and for a smooth $1$-parameter family $\{\phi^{(s)}\}$, then the following estimate, stronger than~\eqref{H_est_s}, holds
							\begin{equation}\label{H_est_s_phi}
								\WF \left( H_\phi^{(s)} (x_1,x_2) \right) \subset \bR \times \{0\} \times \left. \cC^\triangleright [g] \right|_{U \times U}.
							\end{equation}
					\item As we mentioned before, on real analytic space-times and for real analytic data $m, \phi, \lambda$, the cut-off $\psi$ appearing in the series expansions of $v_\phi$, formula~\eqref{v_def}, can be omitted because the series without the cut-off already converges. Consequently, $H[g, m, \phi, \lambda]$ scales homogeneously up to logarithmic terms under the rescaling of $(g, m, \phi, \lambda)$ as before:
	  						\begin{equation*}
	  							\Lambda^{-2} H[\Lambda^{-2} g, \Lambda m, \Lambda \phi, \lambda] = H[g, m, \phi, \lambda] + \ln \Lambda^2 \left( \sum_{\ell \geq 1} u_{\ell+1}[g, m, \phi, \lambda]  \sigma^\ell\right).
	  						\end{equation*}
	  				\item Finally, for any choice of analytic $1$-parameter families $\{ g^{(s)}, m^{(s)}, \phi^{(s)}, \lambda^{(s)} \}$, estimates~\eqref{H_est_s},~\eqref{H_est_s_diag} and~\eqref{H_est_s_phi} can strengthened replacing the smooth wave-front set with the analytic wave-front set.
				\end{itemize}	
			In order to prove the $W$-smoothness of the time-ordered product, the results just outlined are not enough. We need a microlocal control also for the variational derivatives of the Hadamard parametrix. Using eq.~\eqref{Hadamard_def} and eq.~\eqref{v_def}, we can express $\delta^\nu H_\phi / \delta \phi^\nu$ for any $\nu >0$ by
				\begin{equation}\label{H_var_der}
					\begin{split}
	 					&\frac{\delta^\nu H[g,m^2,\phi,\lambda](x_1,x_2)}{\delta \phi(y_1) \dots \delta \phi(y_\nu)} = \\
	 					&\quad = \sum_{k \geq 0} \psi(\sigma(x_1,x_2)/\alpha_k) \frac{\delta^\nu u_{k+1}[g,m^2,\phi,\lambda](x_1,x_2)}{\delta \phi(y_1) \dots \delta \phi(y_\nu)} \sigma^{k}(x_1,x_2) \ln(\sigma_\epsilon(x_1,x_2)).
	 				\end{split}
				\end{equation}	
			where $x_1, x_2$ belongs to the same convex normal set $U$.\\
			Due to the presence of $\delta^\nu u_k / \delta \phi^\nu$ in the formula above, we are interested in providing estimates for the variational derivatives of the Hadamard coefficients. We present some useful properties of $\delta^\nu u_k / \delta \phi^\nu$ in the following lemma:
			\begin{lemma}\label{lemma_u}
				The distribution $\delta^\nu u_{\phi,k} /\delta \phi^\nu$ vanishes whenever $\nu > 2k$. Furthermore, if $(x_1,x_2,y_1, \dots, y_\nu)$ is in the support of $\delta^\nu u_{\phi,k}(x_1,x_2) / \delta \phi(y_1) \dots \delta \phi(y_\nu)$, then the points $y_1, \dots, y_\nu$ must belong to the unique geodesic connecting $x_1,x_2$.\\
				The distribution $\delta^\nu u_k / \delta \phi^\nu [g, m, \phi, \lambda]$ is a locally covariant distribution which scales homogeneously with degree $2k +3\nu$ under the rescaling $(g, m, \phi, \lambda) \mapsto (\Lambda^{-2} g, \Lambda^2 m^2, \Lambda \phi, \lambda)$.\\
				We have
					\begin{equation}\label{var_u}
				\WF \left( \frac{\delta^\nu u_{\phi, k}(x_1,x_2)}{\delta \phi(y_1) \dots \delta\phi(y_\nu)} \right) \subset \cC^u_{2+\nu}[g,\lambda], 
			\end{equation}
		where
			\begin{equation}\label{Cunu}
				\begin{split}
					\cC^u_{2+\nu}[g,\lambda] &:= \left\{ (x_1, x_2, y_1, \dots, y_\nu ; k_1, k_2, p_1, \dots, p_\nu) \in \dot{T}^* U^{2+\nu} : y_\ell \in \supp \lambda \, \forall \ell\right.\\
					&\qquad \left. \exists \mbox{ a partition } \{ L_m\} \mbox{ of } \{ 1, \dots, \nu\}  \mbox{ with } L_m \mbox{ proper}, |L_m| \leq 2  \right. \\
					&\qquad \left. \mbox{ for each } L_m, \, \exists T_m \in [0,1], T_{m+1} \geq T_m \right. \\
					&\qquad \left. y_\ell = \gamma_{x_1,x_2}(T_m) \, \forall \ell \in L_m, \right. \\
					&\qquad \left. -k_1 = \sum_m (1-T_m) \sum_{\ell \in L_m} \Pi_{y_\ell,x_1} p_\ell, \quad -k_2 = \sum_m T_m \sum_{\ell \in L_m} \Pi_{y_\ell,x_2} p_\ell \right\},
				\end{split}
			\end{equation}
		and where $\Pi_{x,y}$ denotes the parallel transport along the unique geodesic connecting $x,y$.\\
		For any smooth $1$-parameter families $\{ g^{(s)}, m^{(s)}, \phi^{(s)}, \lambda^{(s)} \}$, it holds that
			\begin{equation}\label{var_u_s}
				\frac{\delta^\nu u_{k}(x_1,x_2)}{\delta \phi(y_1) \dots \phi(y_\nu)} [g^{(s)}, m^{(s)}, \phi^{(s)}, \lambda^{(s)}]
			\end{equation}
		is a distribution jointly in $s$ and in $x_1,x_2, y_1, \dots, y_\nu$. Thus, we trivially have 
			\begin{equation*}
				\begin{split}
					&\WF \left( \frac{\delta^\nu u_{k}[g^{(s)}, m^{(s)}, \phi^{(s)}, \lambda^{(s)}](x_1,x_2)}{\delta \phi(y_1) \dots \delta\phi(y_\nu)} \right) \subset \\
					&\qquad \subset \left\{ (s,x_1,x_2, y_1, \dots, y_\nu; \rho, k_1, k_2, p_1, \dots, p_\nu) \in \dot{T}^* (\bR \times U^{2+\nu}) : \right. \\
					&\qquad \qquad \left. (x_1,x_2, y_1, \dots, y_\nu; k_1, k_2, p_1, \dots, p_\nu) \in \cC^u_{2+\nu}[g^{(s)}, \lambda^{(s)}]\right\}.
				\end{split}
			\end{equation*}
		Furthermore, for any smooth $\bR \ni s \mapsto \phi(s) \in C^\infty(M)$ it holds that
			\begin{equation}\label{var_u_s_phi}
				\WF \left( \frac{\delta^\nu u_{\phi(s), k} (x_1,x_2)}{\delta \phi(y_1) \dots \delta\phi(y_\nu)} \right) \subset \bR \times \{0\} \times \cC^u_{2+\nu}[g,\lambda].
			\end{equation}
	\end{lemma}
	\begin{proof}
		The proof of these properties is given by induction in $k$ exploiting the iterative definition for the Hadamard coefficients given by eq.~\eqref{induct_u} and the initial condition for $u_0$.\\
		By definition the Van Vleck-Morette determinant does not depend on $\phi$, therefore the hypotheses are trivially satisfied for $k=0$.\\
		Now, assume the results of the lemma~\ref{lemma_u} hold for all orders $\leq k$. We can compute $\delta^\nu u_{k+1,\phi} / \delta \phi^\nu$ distributing the variational derivatives on the right-hand side of eq.~\eqref{induct_u}. Since $P_\phi$ is at most quadratic in $\phi$, all derivatives $\delta^\nu P_\phi / \delta \phi^\nu$ with $\nu > 2$ vanish, so it holds
			\begin{equation}\label{induct_var_u}
				\begin{split}
					\frac{\delta^\nu u_{\phi,k+1}(x_1,x_2)}{\delta \phi(y_1) \dots \delta \phi (y_\nu)} &= - \int_U \int_0^1 t^k \frac{u_{0}(x_1,x_2)}{u_{0}(z,x_2)} \delta (z, \gamma_{x_1,x_2}(t))  \times\\
					&\qquad \times \left\{  P_\phi^{(z)}  \frac{\delta^\nu u_{\phi,k}(z,x_2)}{\delta \phi(y_1) \dots \delta \phi(y_\nu)}  + \sum_i \frac{\delta P_\phi^{(z)}}{\delta \phi(y_i)}  \frac{\delta^{\nu-1} u_{\phi,k}(z,x_2)}{\delta \phi^{\nu -1}(\{y_{r \neq i}\} )} + \right. \\
					&\qquad \qquad \left. + \sum_{i, j} \frac{\delta^2 P_\phi^{(z)}}{\delta \phi(y_i) \delta \phi (y_{j})}  \frac{\delta^{\nu-2} u_{\phi,k}(z,x_2)}{\delta \phi^{\nu -2}( \{y_{r \neq i,j} \})} \right\} dz dt.
				\end{split}
			\end{equation}
		It follows from this expression and the inductive hypothesis that $\delta^\nu u_{\phi, k+1} / \delta \phi^\nu$ vanishes if $\nu > 2k + 2$ or if the points $y_1, \dots, y_\nu$ do not belong to the unique geodesic connecting $x_1$ and $x_2$, as we wanted to prove.\\
		
		The locally covariance property and the homogeneous scaling of $\delta^\nu u_{\phi, k} / \delta \phi^\nu$ are also consequence of this expression and the inductive hypothesis.\\
		
		We come to the proof of estimate~\eqref{var_u}. The distribution in the right-hand side of eq.~\eqref{induct_var_u} is the composition in $z$ of two distribution, namely the distributions in $z,x_1,x_2$ given by
			\begin{equation}\label{delta_gamma}
				\frac{u_{0}(x_1,x_2)}{u_{0}(z,x_2)} \int_0^1 t^k \delta(z, \gamma_{x_1,x_2}(t)) dt ,
			\end{equation}
		and the distribution in $z, x_1, x_2, y_1, \dots, y_\nu$ given by
			\begin{equation}\label{P_delta_u}
				P_\phi^{(z)} \frac{\delta^\nu u_{\phi,k}(z,x_2)}{\delta \phi(y_1) \dots \delta \phi(y_\nu)}  + \sum_{I =\{i\}, \{i,j\}} \frac{\delta^{|I|} P_\phi^{(z)}}{\delta^{|I|}( \{ y_{r \neq I}\})}  \frac{\delta^{\nu-|I|} u_{\phi,k}(z,x_2)}{\delta \phi^{\nu - |I|} (\{ y_{r \notin I}\})},
			\end{equation}
		where the sum is over the subsets $I$ of $\{1, \dots, \nu\}$ containing $1$ element or $2$ distinct elements.\\
		To obtain the estimate~\eqref{var_u} for the  wave-front set of $\delta^\nu u_{\phi, k+1} / \delta \phi^\nu$, we proceed providing estimates for the wave-front set of these two distributions and then use the wave-front set calculus (thm.~\ref{theo_WF_horma}).\\
		Let us focus first on the distribution~\eqref{delta_gamma}. Consider the distribution $\delta(z, \gamma_{x_1,x_2}(t))$ in $\bR \times U^3$. Using~\citep[thm. 8.2.4]{H83} for the wave-front set of the pull-back of a distribution, it follows
			\begin{equation*}
				\begin{split}
					\WF( \delta(z, \gamma_{x_1,x_2}(t))) &\subset \left\{ (t,z,x_1,x_2; \tau, q, k_1, k_2) \in \dot{T}^* (O \times U^3) : z=\gamma_{x_1, x_2}(t), \right. \\
					&\qquad \left. -k_1 = (1-t)\Pi_{z,x_1} q, -k_2= t \Pi_{z,x_2} q, \tau = -q(\dot{\gamma}_{x_1,x_2}(t))\right\}.
				\end{split}
			\end{equation*}
		We used the fact that the unique geodesic $\gamma_{x_1,x_2}:[0,1] \to U$ connecting $x_1,x_2$ can be extended uniquely to a sufficiently small open interval $O \subset \bR$ containing $[0,1]$. By definition, $u_0$ is a strictly positive smooth function. Thus, $u_0(x_1,x_2) / u_0(z,x_2)$ is a smooth function and so does not contribute to the computation of the wave-front set. The distribution~\eqref{delta_gamma} can be equivalently written as
				\begin{equation}\label{delta_gamma_alt}
				\frac{u_{0}(x_1,x_2)}{u_{0}(z,x_2)} \int_0^1 t^k \delta(z, \gamma_{x_1,x_2}(t)) dt= \frac{u_{0}(x_1,x_2)}{u_{0}(z,x_2)} \int_\bR t^k \theta(t) \theta(1-t) \delta(z, \gamma_{x_1,x_2}(t)) dt.
			\end{equation}
		The right-hand side of eq.~\eqref{delta_gamma_alt} is the composition in the variable $t$ of the distribution $t^k \theta(t) \theta(1-t)$ with $\delta(z, \gamma_{x_1,x_2}(t))$. Since $\WF (\delta(z, \gamma_{x_1,x_2}(t)))_t = \emptyset$ (we mean the projection onto the $t$-component of the wave-front set), it follows from the wave-front set calculus (thm.~\ref{theo_WF_horma}) that the composition is well-defined and, furthermore, it holds
			\begin{equation*}
				\begin{split}
					&\WF \left( \frac{u_0(x_1, x_2)}{u_0(z,x_2)} \int_0^1 dt \, t^k \delta(z, \gamma_{x_1,x_2}(t)) \right) \subset \\
					&\qquad \subset \left\{ (z,x_1,x_2; q, k_1, k_2) \in \dot{T}^* (U^3): \right. \\
					&\qquad \qquad \exists t \in [0,1],  z=\gamma_{x_1, x_2}(t), -k_1 = (1-t)\Pi_{z,x_1} q, -k_2= t \Pi_{z,x_2} q, 0 = -q(\dot{\gamma}_{x_1,x_2}(t))\\
					&\qquad \qquad \left. \mbox{or } z=x_1, k_2=0, k_1=q, \mbox{ or } z=x_2, k_1 =0, k_2 = q \right\}.
				\end{split}
			\end{equation*}
		Let us next discuss the distribution~\eqref{P_delta_u}. By the inductive hypothesis, for any $\nu$ the wave-front set of $\delta^\nu u_{\phi, k} / \delta \phi^\nu$ is contained in $\cC^u_{2+\nu}[g,\lambda]$. Since $P_\phi^{(z)}$ is a differential operator, its action on a distribution does not enlarge the wave-front set~\citep[8.1.11]{H83}. On the other hand, the terms $\delta P_\phi^{(z)} / \delta \phi(y_i)$ and $\delta^2 P_\phi^{(z)} / \delta \phi(y_i) \delta \phi(y_j)$ are given by the distributions $- \lambda(z) \phi(z) \delta(z,y_i)$ and $- \lambda(z) \delta(z,y_i,y_j)$ respectively.\\
		Then, using the wave-front set calculus (thm.~\ref{theo_WF_horma}) we can estimate the wave-front set of the distribution~\eqref{P_delta_u}. By the definition of $\cC^u_{2+\nu}[g,\lambda]$, see~\eqref{Cunu}, $(z,x_2, y_1, \dots, y_\nu; \tau, q, k_2, p_1, \dots, p_\nu)$ is an element of the wave-front set of distribution~\eqref{P_delta_u} if there exists a subset $I \subset \{1, \dots, \nu \}$ among $\emptyset, \{i\}, \{i,j\}$, there exists a collection $\{L_m\}$ of proper subsets of $\{1, \dots, \nu \} \backslash I$ where each $L_m$ containing at most two elements, and there exists a non-decreasing collection $\{T_m\}$ of real numbers $T_m \in [0,1]$ such that
			\begin{equation*}
					y_r \in \supp \lambda \, \forall r, \quad y_{i \in I} = z, \quad  y_{\ell \in L_m} = \gamma_{z,x_2}(T_m),
			\end{equation*}
		and
		  	\begin{equation*}
		  		\begin{split} 
		  			&-q =\sum_{i \in I} p_i +\sum_{m} (1-T_m) \sum_{\ell \in L_m} \Pi_{y_\ell, z} p_\ell,\\
		  			&-k_2 =\sum_m T_m  \sum_{\ell \in L_m} \Pi_{y_\ell, x_2} p_\ell.
		  		\end{split}
		 	 \end{equation*}
		 Now, we focus on the distribution given by the right-hand side of eq.~\eqref{induct_var_u}. Using the wave-front set calculus (thm.~\ref{theo_WF_horma}) and the results just presented, we obtain the following necessary condition for $(x_1,x_2, y_1, \dots, y_\nu; k_1, k_2, p_1, \dots, p_\nu)$ to be in the wave-front set of the right-hand side of eq.~\eqref{induct_var_u}: there exists a subset $I \subset \{1, \dots, \nu \}$ among $\emptyset, \{i\}, \{i,j\}$, there exists a collection $\{L_m\}$ of proper subset of $\{1, \dots, \nu \} \backslash I$ where each $L_m$ containing at most two elements, there exists a non-decreasing collection $\{T_m\}$ of real numbers $T_m \in [0,1]$, and there exists $t \in [0,1]$ such that
			 \begin{equation*}
			 	y_r \in \supp \lambda \, \forall r, \quad y_{i \in I} = \gamma_{x_1,x_2}(t), \quad  y_{\ell \in L_m} = \gamma_{\gamma_{x_1,x_2}(t),x_2}(T_m),
			 \end{equation*}
		and
		 	\begin{equation*}
		  			\begin{split}
		  				&-k_1 = (1-t) \sum_{i \in I} \Pi_{y_i,x_1} p_i +\sum_{m} (1-t)(1-T_m) \sum_{\ell \in L_m} \Pi_{y_\ell, x_1} p_\ell,\\
		  				&-k_2 = t \sum_{i \in I} \Pi_{y_i, x_2} p_i + \sum_m (T_m +t(1-T_m))  \sum_{\ell \in L_m} \Pi_{y_\ell, x_2} p_\ell.
		  			\end{split}
		 	\end{equation*}
		 We proceed defining a new collection of subsets $\{L'_m\}$ and a corresponding collection of geodesic parameters $\{T'_m\}$ given respectively by
		 	\begin{equation*}
		 		L'_{1} = \left\{ \begin{array}{ll}
		 			L_1 & I = \emptyset \\
		 			I & \mbox{otherwise}
		 		\end{array} \right. \qquad L'_{m>1} = \left\{ \begin{array}{ll}
		 			L_m & I = \emptyset  \\
		 			L_{m-1} & \mbox{otherwise}
		 		\end{array} \right.
		 	\end{equation*}
		 and
		 \begin{equation*}
		 	T'_1 = \left\{ \begin{array}{ll}
		 			T_1(1-t)  + t & I = \emptyset  \\
		 			t & \mbox{otherwise}
		 		\end{array} \right. \qquad T'_m = \left\{ \begin{array}{ll}
		 			T_m(1-t)  + t & I = \emptyset  \\
		 			T_{m-1} (1-t)  + t & \mbox{otherwise}
		 		\end{array} \right.
		 \end{equation*}
		 It follows
		 	\begin{equation*}
		  		y_{\ell\in L'_m} = \gamma_{x_1,x_2}(T'_m), \quad -k_1 = \sum_{m} (1-T'_m) \sum_{\ell \in L'_m} \Pi_{y_\ell, x_1} p_\ell, \quad -k_2 = \sum_m T'_m  \sum_{\ell \in L'_m} \Pi_{y_\ell, x_2} p_\ell,
		 	 \end{equation*}
		 and consequently the right-hand side of eq.~\eqref{induct_var_u} defines a distribution in $x_1,x_2, y_1, \dots, y_\nu$ which has a wave-front set contained in $\cC^u_{2+\nu+1}$ as we wanted to prove.\\
		 
		A similar argument can be presented to prove the estimates~\eqref{var_u_s} and~\eqref{var_u_s_phi}. Now, in the inductive formula~\eqref{induct_var_u}, $g,m, \phi,\lambda$ depend smoothly on a parameter $s$. Note that the geodesic $\gamma$ depends on $s$. Nevertheless, one finds that the proof still goes through without non-trivial modifications. This conclude the proof.
	\end{proof}
	The following results for the variational derivatives of the Hadamard parametrix follows from formula~\eqref{H_var_der} and lemma~\ref{lemma_u} for the variational derivatives of the Hadamard coefficients.
	\begin{lemma}\label{lemma_H}
		Let $H_\phi$ be the Hadamard parametrix given by~\eqref{Hadamard_def} in the convex normal subset $U \subset M$. For any $\nu$, $\delta^\nu H_\phi (x_1,x_2) / \delta \phi(y_1) \dots \delta \phi(y_\nu)$ is a locally covariant distribution\footnote{In principle, the distribution $\delta^\nu H_\phi / \delta \phi^\nu$ is defined in $U^2 \times M^\nu$.} supported in $U^{2+\nu}$, which vanishes unless $y_1, \dots, y_\nu \in \supp\lambda$.\\
		We have
			\begin{equation}\label{var_H_est}
				\WF \left( \frac{\delta^\nu H[g,m, \phi,\lambda](x_1,x_2)}{\delta \phi(y_1) \cdots \delta \phi(y_\nu)} \right) \subset \left. Z_{2+\nu}[g]\right|_{U^{2+\nu}}
			\end{equation}
		where the set $Z_{2+\nu}$ is defined by~\eqref{Z}. On the total diagonal, a stronger bound holds:
			\begin{equation}\label{var_H_nu_diag}
				\WF \left. \left( \frac{\delta^\nu H[g, m, \phi, \lambda](x_1,x_2)}{\delta \phi(y_1) \dots \delta\phi(y_\nu)} \right) \right|_{\Delta_{2+\nu}} \perp T\Delta_{2+\nu}.
			\end{equation}
		Moreover, for any choice of smooth $1$-parameter families $\{ g^{(s)}, m^{(s)}, \phi^{(s)}, \lambda^{(s)} \}$, it holds
			\begin{equation}\label{var_H_est_s}
				\begin{split}
					&\WF \left( \frac{\delta^\nu H[g^{(s)}, m^{(s)}, \phi^{(s)}, \lambda^{(s)}](x_1,x_2)}{\delta \phi(y_1) \dots \delta\phi(y_\nu)} \right) \subset \\
					&\qquad \subset \left\{ (s,x_1,x_2, y_1, \dots, y_\nu; \rho, k_1, k_2, p_1, \dots, p_\nu) \in \dot{T}^* (\bR \times U^{2+\nu}) : \right. \\
					&\qquad \qquad \left. (x_1,x_2, y_1, \dots, y_\nu; k_1, k_2, p_1, \dots, p_\nu) \in Z_{2+\nu}[g^{(s)}]\right\},
				\end{split}
			\end{equation}
		and in addition
			\begin{equation}\label{var_H_est_s_diag}
					\WF \left. \left(  \frac{\delta^\nu H[g^{(s)}, m^{(s)}, \phi^{(s)}, \lambda^{(s)}](x_1,x_2)}{\delta \phi(y_1) \dots \delta\phi(y_\nu)} \right) \right|_{\Delta_{2 +\nu}}  \perp T(\bR \times \Delta_{2+\nu}).
			\end{equation}
		In the case of variations of only the background $\phi$, the following stronger bound is satisfied:
			\begin{equation}\label{var_H_est_s_phi}
				\WF \left( \frac{\delta^\nu H[g, m, \phi^{(s)}, \lambda](x_1,x_2)}{\delta \phi(y_1) \dots \delta\phi(y_\nu)} \right) \subset \bR \times \{0\} \times \left. Z_{2+\nu}[g]\right|_{U^{2+\nu}}.
			\end{equation}
		Finally, in any real analytic space-time and for real analytic data $m,\phi,\lambda$, $\delta^\nu H / \delta \phi^\nu [g, m, \phi, \lambda]$ scales almost homogeneously with degree $2 +3\nu$ under the rescaling $(g,m^2,\phi,\lambda) \mapsto (\Lambda^{-2}g, \Lambda m, \Lambda \phi, \lambda)$.
	\end{lemma}
	\begin{proof}
		We first note that the each term on the right-hand side of eq.~\eqref{H_var_der} is a product of distributions, and, therefore, it is not a priori well-defined. Using the estimate~\eqref{var_u}, for any $k$ we have
			\begin{equation*}
				\WF \left( \frac{\delta^\nu u_{\phi, k}(x_1,x_2)}{\delta \phi(y_1) \dots \delta \phi(y_\nu)} \right)_{x_1,x_2} = \emptyset.
			\end{equation*}
		Thus, as a consequence of the wave-front set calculus (thm.~\ref{theo_WF_horma}) each term on the right-hand side of eq.~\eqref{H_var_der} is well-defined.\\
		The fact that $\delta^\nu H_\phi / \delta \phi^\nu$ is locally covariant is a consequence of formula~\eqref{H_var_der}, the fact that $\delta^\nu u_{\phi,k} / \delta \phi^\nu$ is locally covariant for any $\nu$ as proved in lemma~\ref{lemma_u}, and the fact that $\ln(\sigma_\epsilon)$, $\psi(\sigma/\alpha_k)$ and $\sigma^k$ are clearly locally covariant. Furthermore, the support properties of $\delta^\nu u_{\phi,k} / \delta \phi^\nu$ follow from the support properties of $\delta^\nu u_{\phi,k} / \delta \phi^\nu$, see lemma~\ref{lemma_u}.\\
		We now prove estimate~\eqref{var_H_est}. Let $(x_1, x_2, y_1, \dots, y_\nu; k_1, k_2, p_1, \dots, p_\nu)$ be an element of the wave-front set of one of the terms in the right-hand side of eq.~\eqref{H_var_der}. As proved in~\citep{radzikowski1996micro}, the wave-front set of $\ln (\sigma_\epsilon)$ is $\cC^\triangleright[g]|_{U^2}$, i.e. the restriction to $U^2$ of the set defined in~\eqref{WF_Hadamard}. Furthermore, the estimate~\eqref{var_u} holds for $\delta^\nu u_{\phi,k} / \delta \phi^\nu$ as proved in lemma~\ref{lemma_u}. The wave-front set calculus (thm.~\ref{theo_WF_horma}) implies that there exist decompositions $k_1=k'_1 + k''_1$ and $k_2= k'_2 + k''_2$ such that it holds
			\begin{equation}\label{tech_WF_prod_H}
				\left\{ \begin{array}{l}
					(x_1, x_2, y_1, \dots, y_\nu; k'_1, k'_2, p_1, \dots, p_\nu) \in \cC^u_{2+\nu} \mbox{ or } k'_1, k'_2, p'_r =0,\\
					(x_1, x_2; k''_1, k''_2) \in \cC^\triangleright \mbox{ or } k''_1, k''_2=0,
				\end{array} \right.
			\end{equation}
		where $\cC^u_{2 + \nu}$ is the set~\eqref{Cunu}.
		As straightforward consequences of the definitions of $\cC^u_{2 +\nu}$ and $\cC^\triangleright$, it follows that if $p_r$ is in $\overline{V}^+$ for all $r$, then we have $k_2 \in \overline{V}^-$, while if $p_s$ is space-like and $p_r$ is in $\overline{V}^+$ for all $r \neq s$, then we have $k_2 =0$ or $k_2 \notin \overline{V}^+$. This implies precisely $(x_1, x_2, y_1, \dots, y_\nu; k_1, k_2, p_1, \dots, p_\nu) \notin C^{2;+}_{2+\nu}$, where $C^{2;+}_{2+\nu}$ is defined by~\eqref{aux_C_sets}. With a similar argument, we obtain $(x_1, x_2, y_1, \dots, y_\nu; k_1, k_2, p_1, \dots, p_\nu) \notin C^{1;-}_{2 + \nu}$, where $C^{1;-}_{2 + \nu}$ is defined by~\eqref{aux_C_sets}. By the definition of the set $Z_{2+\nu}$~\eqref{Z}, it thus follows that estimate~\eqref{var_H_est} holds, as we wanted to prove.\\
		
		We can prove estimates~\eqref{var_H_est_s} and~\eqref{var_H_est_s_phi} with similar arguments based on estimates~\eqref{var_u_s} and, respectively,~\eqref{var_u_s_phi}, instead of~\eqref{var_u}.\\
		
		To prove that the requirement~\eqref{var_H_nu_diag} is satisfied, we consider an element $(x, x, x, \dots, x; k_1, k_2, p_1, \dots, p_\nu)$ of the wave-front set of the right-hand side of eq.~\eqref{H_var_der}. Similarly as before, the wave-front calculus (thm.~\ref{theo_WF_horma}) implies that there exist decompositions $k_1=k'_1 + k''_1 = k_1$ and $k_2= k'_2 + k''_2$
		\begin{equation}\label{tech_WF_prod_H_2}
				\left\{ \begin{array}{l}
					(x, x,x, \dots, x; k'_1, k'_2, p_1, \dots, p_\nu) \in \cC^u_{2+\nu} \mbox{ or } k'_1, k'_2, p'_r =0,\\
					(x, x; k''_1, k''_2) \in \cC^\triangleright \mbox{ or } k''_1, k''_2=0.
				\end{array} \right.
			\end{equation}
		Since all the points coincide, it follows from the definitions of $\cC^u_{2 +\nu}$~\eqref{Cunu} and $\cC^\triangleright$~\eqref{Hadamard_def} that $- k'_1 - k'_2= \sum_{r} p_r$ and $k''_1 + k''_2 = 0$. This clearly implies that~\eqref{var_H_nu_diag} holds.\\
		
		We can verify the requirement~\eqref{var_H_est_s_diag} adapting, in a fairly obvious way, the argument just presented to the case of smooth families $g^{(s)}, m^{(s)}, \phi^{(s)}, \lambda^{(s)}$.\\
		
		Finally, the almost homogeneous scaling of $\delta^\nu H / \delta \phi^\nu [g, m, \phi, \lambda]$ under the rescaling of $(g, m, \phi, \lambda)$ in any real-analytic space-time and for real analytic data $m, \phi, \lambda$ is a direct consequence of the following three facts: $\delta^\nu u / \delta \phi^\nu [g, m, \phi, \lambda]$ scales homogeneously with degree $2k + 3\nu$, the factors $\psi(\sigma/ \alpha_k)$, which spoil the scaling properties, are absent if the space-time is real-analytic, and $\sigma$ scales with degree $-2$.
	\end{proof}

		So far, we have not made any assumptions on the space-time $(M,g)$ and we required only that $m, \phi ,\lambda$ are smooth functions. In the remaining part of this subsection we consider $(M,g)$ to be an ultra-static space-time with compact Cauchy surfaces, $m$ constant, and $\lambda \in C_0^\infty(M)$. We consider the retarded $2$-point function $\omega^R_\phi$ defined by~\eqref{in-state} with respect to $P_\phi = \boxempty - m^2 - \frac{\lambda}{2} \phi^2$.  As shown in lemma~\ref{lemma_ret_state_fedosov}, $\omega^R_\phi$ can be written in the form~\eqref{state} and therefore $\phi \mapsto \omega^R_\phi$ is an admissible assignment in the sense of def.~\ref{def_suit_omega}, as a consequence of lemma~\ref{lemma_state}. We want to present some properties of the difference $d_\phi = \omega^R_\phi - H_\phi$, where $H_\phi$ Hadamard parametrix with respect to $P_\phi$, in particular we want to control the wave-front set of $\delta^\nu d_\phi / \delta \phi^\nu$ and $\delta^\nu d_{\phi(\epsilon)} / \delta \phi^\nu$ for any smooth map $\bR \ni \epsilon \mapsto \phi(\epsilon) \in C^\infty(M)$.\\
		Since $H_\phi$ is defined only in $U \times U$, where $U$ is a convex normal set in $M$, also $d_\phi$ is defined only in $U \times U$ by construction. Because $\omega_R$ is a Hadamard $2$-point function, $d_\phi$ is a smooth function in $(x_1,x_2)$ (see~\citep{radzikowski1996micro}), and it is symmetric in $x_1,x_2$. Furthermore, for any smooth map $\bR \ni \epsilon \mapsto \phi(\epsilon)$, the map $(\epsilon, x_1, x_2) \mapsto d_{\phi(\epsilon)}(x_1,x_2)$ is jointly smooth. In fact, since $d_{\phi(\epsilon)}(x_1,x_2)$ is smooth in $x_1,x_2$ for any fixed $\epsilon$, it follows that $\WF( d_{\phi(\epsilon)}(x_1,x_2)) \subset \{ (s, x_1,x_2; \rho, 0 , 0) \in \dot{T}^* (\bR \times M^2) \}$. On the other hand, $d_{\phi(\epsilon)}$ is the difference of $\omega^R_{\phi(\epsilon)}$ and $H_{\phi(\epsilon)}$ which both have wave-front sets contained in $\bR \times \{0\} \times \cC^\triangleright$. Thus, it follows that $\WF( d_{\phi(\epsilon)}(x_1,x_2)) = \emptyset$, as we wanted to show.\\
		Since $\phi \mapsto \omega^R_\phi$ is an admissible assignment, $\omega^R_\phi$ must satisfy the estimate~\eqref{WF_better} and estimate~\eqref{WF_better_s} for a smooth family of backgrounds $\bR \ni \epsilon \mapsto \phi(\epsilon) \in C^\infty(M)$. In principle, we could combine these estimates for $\omega^R_\phi$ with the estimates~\eqref{var_H_est} and~\eqref{var_H_est_s} for $H_\phi$ to get bounds for the wave-front sets of $\delta^\nu d_\phi / \delta \phi^\nu$ and $\delta^\nu d_{\phi(\epsilon)} / \delta \phi^\nu$. However, such bounds are not sharp enough for the applications we are going to need in the next subsections. The following lemma gives better bounds:
	\begin{lemma}\label{lemma_d}
		Let $U$ be a convex normal set sufficiently small such that it holds $U \Subset U'$ for another convex normal set $U'$ and there exist three Cauchy surfaces $\Sigma_{++}, \Sigma_+, \Sigma_-$ which satisfies:
			\begin{itemize}
				\item $\Sigma_\pm \cap J^{+}(U) = \emptyset$ and $\Sigma_{++} \cap J^{-}(U)=\emptyset$.
				\item All the three Cauchy surfaces have non-trivial intersections with $U'$. Furthermore, $J^- (\Sigma_{++} \cap U') \cap \Sigma_- \subset \Sigma_- \cap U'$ and $U \subset J^- (\Sigma_{++} \cap U')$.
			\end{itemize}
		Consider $d_\phi$ defined on $U^2$. It holds
			\begin{equation}\label{var_d_nu}
				\WF \left( \frac{ \delta^\nu d_\phi(x_1,x_2)}{\delta \phi(y_1) \dots \delta \phi(y_\nu)} \right) \subset \cC^d_{2+\nu},
			\end{equation}
		where
			\begin{equation}\label{Cdnu}
				\begin{split}
					\cC^d_{2+\nu} := &\left\{ (x_1,x_2, y_1, \dots, y_\nu; k_1, k_2, p_1, \dots, p_\nu) \in \dot{T}^* (U^2 \times M^\nu): \right. \\
					&\quad \left. \mbox{if } p_r \in \overline{V}^\pm \, \forall r, \mbox{ then } k_1,k_2 \in \overline{V}^\mp \right. \\
					&\quad \left. \mbox{if } \exists p_s \mbox{ space-like}, p_{r} \in \overline{V}^\pm \, \forall r \neq s, \mbox{ then } k_1,k_2 \notin \overline{V}^\pm \mbox { and } k_1 + k_2 \notin \overline{V}^\pm \mbox { if } x_1 = x_2\right\}.
				\end{split}
			\end{equation}
		In addition, for any smooth map $\bR \ni \epsilon \mapsto \phi(\epsilon) \in C^\infty(M)$, the following bound is satisfied
			\begin{equation}\label{var_d_nu_s_phi}
				\WF \left( \frac{\delta^\nu d_{\phi(\epsilon)}(x_1,x_2)}{\delta \phi(y_1) \dots \delta\phi(y_\nu)} \right) \subset \bR \times \{0\} \times \cC^d_{2+\nu}. 
			\end{equation}
	\end{lemma}
	\begin{proof}
		We first prove the estimate~\eqref{var_d_nu}. We exclude two situations for which estimate~\eqref{var_d_nu} is trivially verified. Note that whenever $U \cap J^+ (\supp \lambda) = \emptyset$, we know that $\delta^\nu d_\phi / \delta \phi^\nu$ vanishes, because outside $J^+ (\supp \lambda)$ the function $d_\phi$ does not depend on $\phi$ as follows from the definitions of $H_\phi$ and $\omega^R_\phi$. From now on we can assume that $U \cap J^+ (\supp \lambda) \neq \emptyset$. By the support properties of $\delta^\nu \omega^R_\phi / \delta \phi^\nu$ (see lemma~\ref{lemma_ret_state_fedosov} and lemma~\ref{lemma_state}) and $\delta^\nu H_\phi / \delta \phi^\nu$ (see lemma~\ref{lemma_H}), the distribution $\delta^\nu d_\phi(x_1, x_2) / \delta \phi(y_1) \dots \delta \phi(y_\nu)$, which is defined on $U^2 \times M^\nu$, vanishes unless $y_1, \dots, y_\nu \in \supp \lambda$.\\
		We now discuss the remaining non-trivial possibilities by distinguishing two cases: (a) first we assume that at least one variable among $y_1, \dots, y_\nu$ does not belongs to $U$, and then (b) we assume $(x_1, x_2, y_1, \dots, y_\nu) \in U^{\nu +2}$.\\
		For case (a), since at least one of the variables $y_1, \dots, y_\nu$ does not belong to $U$, the distribution $\delta^\nu H_\phi(x_1,x_2) / \delta \phi(y_1) \dots \delta \phi(y_\nu)$ vanishes because the support of $\delta^\nu H_\phi / \delta \phi^\nu$ is contained in $U^{2 +\nu}$ as proved in lemma~\ref{lemma_H}. Thus, we can rewrite the variational derivatives of $d_\phi$ as
			\begin{equation*}
				\frac{\delta^\nu d_\phi(x_1, x_2)}{\delta \phi(y_1) \dots \delta \phi(y_\nu)} = \frac{\delta^\nu \omega^R_\phi(x_1, x_2)}{\delta \phi(y_1) \dots \delta \phi(y_\nu)}.
			\end{equation*}
		Because $\delta^\nu d_\phi(x_1,x_2) / \delta \phi^\nu$ is symmetric in $x_1, x_2$, and because of the restriction on the wave-front set of $\delta^\nu \omega^R_\phi / \delta \phi^\nu$ given by estimate~\eqref{WF_better} (see lemma~\ref{lemma_ret_state_fedosov} and lemma~\ref{lemma_state}), it follows
			\begin{equation*}
				\WF \left( \frac{\delta^\nu d_\phi(x_1, x_2)}{\delta \phi(y_1) \dots \delta \phi(y_\nu)} \right) \subset \bP^+ Z_{2+\nu},
			\end{equation*}
		where the set in the right-hand side is defined by~\eqref{WF_symm_C+;2_C-;1}. Similarly as done in lemma~\ref{lemma_diff_var_states}, we conclude that if $p_r$ is in $\overline{V}^\pm$ for all $r$, then we have $k_1, k_2 \in \overline{V}^\mp$, while if $p_s$ is space-like and $p_{r}$ is in $\overline{V}^\pm$ for any $r \neq s$, then we have$k_1, k_2 \notin \overline{V}^\pm$. This is precisely what we have to show to prove that $(x_1, x_2, y_1, \dots, y_\nu; k_1, k_2, p_1, \dots, p_\nu)$, under the assumptions (a), belongs to $\cC^d_{2+\nu}$.\\
		For the case (b), the proof is more involved. The argument we are presented is inspired by the one presented in~\citep[lemma 6.2]{HW05} (see also~\citep[Appendix A]{zahn2013locally}). Since $\omega^R_\phi$ is a bi-solution with respect to $P_\phi$, and since $H_\phi$ (defined in the convex neighbourhood $U'$) is a bi-solution with respect to $P_\phi$ modulo a smooth function, it follows that in $U' \times U'$ the functions $G^{(1,2)}_\phi(z_1, z_2)$ defined by
			\begin{equation*}
				\begin{split}
					&G^{(1)}_\phi (z_1, z_2) = P_\phi^{(z_1)} (\omega^R_\phi - H_\phi)(z_1, z_2)= - P_\phi^{(z_1)} H_\phi( z_1, z_2), \\
					&G^{(2)}_\phi (z_1, z_2) =P_\phi^{(z_2)} (\omega^R_\phi - H_\phi)(z_1, z_2)= - P_\phi^{(z_2)} H_\phi( z_1, z_2),
				\end{split}
			\end{equation*}
		are smooth. We can write $G^{(1,2)}_\phi$ explicitly in terms of the Hadamard coefficients similarly as done in~\citep[lemma 2.4.3]{BGP07} for the formal fundamental solutions of the Klein-Gordon equation:
			\begin{equation}\label{Gi}
		 		\begin{split}
		 			G^{(i)}_\phi (z_1,z_2) &= \left( \left\{ 1 - \psi ( \sigma/\alpha_{n_0}) \right\} P^{(i)}_\phi \left( u_{\phi,n_0+1} \sigma^{n_0} \ln( \sigma_\epsilon \right)\right)(z_1,z_2) + \\
		 			&\quad  + \sum_{k > n_0} \left( \left\{ \psi ( \sigma/\alpha_k) - \psi ( \sigma/\alpha_{k+1}) \right\} (P^{(i)}_\phi u_{\phi,k+1}) \sigma^{k} \ln (\sigma_\epsilon) \right) (z_1,z_2) - \\
		 			&\quad -2 \sum_{k > n_0} \left(  \nabla^{(i)} \psi ( \sigma/\alpha_k), \nabla^{(i)} (u_{\phi,k+1} \sigma^{k} \ln(\sigma_\epsilon)) \right)_g(z_1,z_2) + \\
		 			&\quad + \sum_{k > n_0} \left( \left( \boxempty^{(i)} \psi ( \sigma/\alpha_k) \right) u_{\phi,k+1} \sigma^{k} \ln(\sigma_\epsilon)\right)(z_1,z_2),
		 		\end{split}
		 	\end{equation}
		where $n_0$ is an arbitrary fixed value. To get this, one uses the recursive definition of $u_k$.\\
		Each term of eq.~\eqref{Gi} contains a cut-off which is supported where $\ln(\sigma_\epsilon)$ is smooth. Due to the properties of the Hadamard coeffiecients proved in lemma~\ref{lemma_u}, in particular estimate~\eqref{var_u}, we obtain the following estimate using the wave-front set calculus:
			\begin{equation}\label{est_G_1,2}
				\WF \left( \frac{\delta^\nu G^{(1,2)}_\phi (z_1,z_2)}{\delta \phi(y_1) \dots \delta \phi(y_\nu)} \right) \subset \cC^u_{2+\nu},
			\end{equation}
		where $\cC^u_{2+\nu}$ is the same set defined by~\eqref{Cunu}. A similar estimate holds for the smooth function $G^{(3)}_\phi(z_1,z_2)$ defined by
			\begin{equation*}
				G^{(3)}_\phi (z_1,z_2):= P_\phi^{(z_2)} G^{(1)}_\phi (z_1,  z_2) = P_\phi^{(z_1)} G^{(2)}_\phi (z_1, z_2) = - P_\phi^{(z_1)} P_\phi^{(z_2)} H_\phi( z_1, z_2).
			\end{equation*}
		Next, we exploit the hypotheses on the convex normal sets $U$, $U'$ and on the Cauchy surfaces $\Sigma_\pm, \Sigma_{++}$, see fig.~\ref{fig:2} for a sketch of the situation. 
		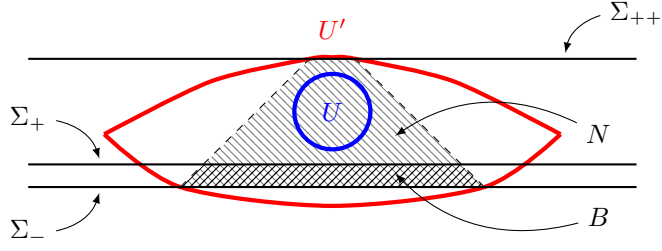
\begin{figure}
		\centering
		\begin{tikzpicture}
		 \draw [draw=red, ultra thick] plot [smooth] coordinates
			{(1,2.4) (2,1.7) (4,1.45) (6,1.7) (7,2.4)};
         \draw [draw=red, ultra  thick] plot [smooth] coordinates   
           {(1,2.4) (2.4, 3.1) (3.7,3.4) (4,3.42) (4.3,3.4) (5.6, 3.1) (7,2.4)};
        \node [red] at (4, 3.8) {${U'}$};
        \draw [draw=white, pattern=north west lines, pattern color= gray] (2,1.7) -- (3.7,3.4) -- (4.3,3.4) -- (6, 1.7) -- (2,1.7);
        \node (N1) at (4.7, 2.3) {};
		\node (N2) at (7.5,2.4) {$N$};
		\draw [-latex, bend right] (N2) to (N1);
        \draw [draw=white, pattern=north east lines] (2,1.7) -- (2.3,2) -- (5.7,2) -- (6, 1.7) -- (2,1.7);
        \node (B1) at (4.7, 2) {};
		\node (B2) at (7.5,1.3) {$B$};
		\draw [-latex, bend left] (B2) to (B1);
		\draw [draw=blue, ultra thick]  (4,2.7) circle (0.5);
		\node [blue] at (4,2.7) {${U}$};
		\draw [thick] (0,3.4) -- (8,3.4);
		\node (a1) at (7, 3.35) {};
		\node (a2) at (8,4) {$\Sigma_{++}$};
		\draw [-latex, bend right] (a2) to (a1);
		\draw [dashed] (2, 1.7) -- (3.7,3.4);
		\draw [thick] (0,2) -- (8,2);
		\node (b1) at (1, 1.95) {};
		\node (b2) at (0,2.6) {$\Sigma_+$};
		\draw [-latex, bend left] (b2) to (b1);
		\draw [dashed] (6, 1.7) -- (4.3,3.4);
		\draw [thick] (0,1.7) -- (8,1.7);
		\node (c1) at (1, 1.75) {};
		\node (c2) at (0,1.1) {$\Sigma_-$};
		\draw [-latex, bend right] (c2) to (c1);
		\end{tikzpicture}
		\caption{The set-up of $U$, $U'$.}
		\label{fig:2}
		\end{figure}
		We denote by $N$ and by $B$ the following sets:
			\begin{equation*}
				N:=J^- (U' \cap  \Sigma_{++}) \cap J^+(\Sigma_-), \quad B:= J^-( \Sigma_+) \cap N.
			\end{equation*}
		Let $c$ be a smooth cut-off function such that $c(M) \subset [0,1]$, $c = 1$ in $J^- (\Sigma_-)$ and $c=0$ in $J^+ (\Sigma_{+})$. For any $f_1, f_2 \in C^\infty_0(M)$ such that $\supp f_1, \supp f_2 \subset U$ we consider the following distribution in $U' \times U'$
			\begin{equation}\label{aux_dist}
				\beta_{\phi, \mu \nu}(x_1,x_2) : = E^A_\phi(f_1) (x_1) \overleftrightarrow{\partial_\mu} d_\phi(x_1,x_2) \overleftrightarrow{\partial_\nu} E^A_\phi(f_2)(x_2).
			\end{equation}
		It follows from the Stokes's theorem and the properties of the advanced propagator $E^A_\phi$ that it holds
			\begin{equation*}
				\begin{split}
					&\int_{\Sigma_- \times \Sigma_-} E^A_\phi (f_1)(x_1) \overleftrightarrow{\partial_n} d_\phi(x_1,x_2) \overleftrightarrow{\partial_n} E^A_\phi (f_2)(x_2) d\Sigma(x_1) d\Sigma(x_2) = \int_{N \times N} \partial^\mu \partial^\nu \beta_{\phi, \mu \nu}(x_1,x_2) dx_1 dx_2 \\
					&\quad = G^{(1)}_\phi (\chi_N E^A_\phi(f_1), f_2) + G^{(2)}_\phi(f_1, \chi_N E^A_\phi(f_2)) - d_\phi(f_1,f_2) - G^{(3)}_\phi(\chi_N E^A_\phi(f_1), \chi_N E^A_\phi(f_2)),
				\end{split}
			\end{equation*}
		where $\chi_N$ is the characteristic function of the domain $N$. Thus, the terms in the last line are integrals over $N \times N$.\\
		Using repeatedly eq.~\eqref{kernel_symp_not_sol} and the fact that $\supp c \cap \supp f_{1,2} = \emptyset$, we then obtain the following decomposition:
			\begin{equation*}
					d_\phi(f_1, f_2) = \sD_{1,\phi}(f_1,f_2) + \sD_{2, \phi}(f_1,f_2)+ \sD_{3, \phi}(f_1,f_2) + \sD_{4, \phi}(f_1,f_2),
			\end{equation*}
		where $\sD_{1,\phi}$, $\sD_{2, \phi}$, $\sD_{3, \phi}$, $\sD_{4,\phi}$ are defined respectively by
			\begin{equation}\label{sD_1}
					\sD_{1,\phi}(f_1,f_2) :=  (E^R_\phi \circ \sigma_c \circ d_\phi \circ \sigma_c \circ E^A_\phi)(f_1,f_2),
			\end{equation}
			\begin{equation}\label{sD_2}
				\sD_{2,\phi}(f_1,f_2):= ((E^R_\phi \cdot \chi_N) \circ  G^{(1)}_\phi)(f_1,f_2),
			\end{equation}
			\begin{equation}\label{sD_3}
			\sD_{3, \phi}(f_1,f_2):=(G^{(2)}_\phi \circ (\chi_N \cdot E^A_\phi))(f_1, f_2), 
			\end{equation}
			and
		 	\begin{equation}\label{sD_4}
				\begin{split}
					\sD_{4,\phi}(f_1,f_2) &:= \int_{M \times B} E^A_\phi(f_1)(x_1) (\sigma_c \circ G^{(1)}_\phi)(x_1,x_2)  c(x_2) E^A_\phi(f_2)(x_2) dx_1 dx_2 - \\
					&\quad -\int_{B \times M} E^A_\phi(f_1)(x_1) c(x_1) (G^{(2)}_\phi \circ \sigma_c)(x_1,x_2) E^A_\phi(f_2)(x_2) dx_1 dx_2 - \\
					&\quad - ((E^R_\phi \cdot \chi_N) \circ G^{(3)}_\phi \circ (\chi_N \cdot E^A_\phi))(f_1, f_2).
				\end{split}
			\end{equation}
		We treat each $\sD_{1,\phi}, \sD_{2, \phi}, \sD_{3,\phi}, \sD_{4,\phi}$ separately. We first compute their $\nu$-th variational derivatives on $U^{\nu +2}$. Then, we prove that any element $(x_1,x_2, y_1, y_\nu; k_1,k_2, p_1, \dots, p_\nu)$ of the wave-front sets of these variational derivatives must have $k_1, k_2 \in \overline{V}^\mp$ if $p_r \in \overline{V}^\pm$ for any $r$, while it must have $k_1, k_2 \notin \overline{V}^\pm$ and, when $x_1 = x_2$, $k_1 + k_2 \ni \overline{V}^\pm$ if there exists a space-like $p_s$ and if $p_{r}$ is in $\overline{V}^\pm$ for all $r \neq s$.
		\begin{enumerate}[label= $\sD_{\arabic*, \phi}$), start=1]
			\item The distribution $\sigma_c$, defined by~\eqref{kernel_symp} for the specific $c$ we chose, is such that $\supp \sigma_c \subset B \times B$. We compute $\delta^\nu \sD_{1,\phi} / \delta \phi^\nu$ by distributing the variational derivatives on each factor in the right-hand side of~\eqref{sD_1}. By the support properties of the variational derivatives of the Hadamard parametrix (see lemma~\ref{lemma_H}), we have $\delta^\nu d_{\phi} / \delta \phi^\nu = \delta^\nu \omega^R_\phi / \delta \phi^\nu$ on $B^2 \times U^\nu$. Since the wave-front set of $\sigma_c$ is given by~\eqref{WF_kernel_symp}, and since the estimates~\eqref{var_ders_A/R_WF} hold for the wave-front sets of $\delta^\nu E^{A/R}_\phi / \delta \phi^\nu$, the wave-front set calculus (thm.~\ref{theo_WF_horma}) implies the following bound:
			\begin{equation*}
				\begin{split}
					&\left. \WF\left( \frac{\delta^\nu \sD_{1,\phi}(x_1,x_2)}{\delta \phi(y_1) \dots \delta \phi (y_\nu)} \right) \right|_{U^{\nu +2}}  \subset \\
					&\quad \subset \left\{ (x_1, x_2, y_1, \dots, y_\nu; k_1, k_2, p_1, \dots, p_\nu) \in \dot{T}^* M^{2+\nu}: x_1,x_2, y_1, \dots, y_\nu \in U \right. \\
						&\qquad \left. \exists N_1, N_2, N_3 \mbox{ partition of } \{1, \dots, \nu \} \mbox{ and } (z,q), (z',q') \in T^*M \mbox{ such that }  z,z \in B \mbox{ and}\right. \\
						&\qquad \left. (x_1, z, (y_{r \in N_1}); k_1, -q, (p_{r \in N_1})) \in X_{2+ |N_1|}, \mbox{ or } k_1, q, p_{r \in N_1} = 0, \right. \\
						&\qquad \left. (z, z' (y_{r \in N_2}); q, -q', (p_{r \in N_2})) \in \WF ( \delta^{|N_2|} \omega^R_\phi/\delta \phi^{|N_2|} ), \mbox{ or } q, q', p_{r \in N_2} = 0, \right. \\
						&\qquad \left. (z', x_2, (y_{r \in N_3}); q', k_2, (p_{r \in N_3})) \in X_{2+ |N_3|}, \mbox{ or } q',k_2, p_{r \in N_3} = 0\right\}.
					\end{split}
			\end{equation*}
		Let $(x_1, x_2, y_1, \dots, y_\nu; k_1, k_2, p_1, \dots, p_\nu)$ be an element in $\WF ( \delta^\nu \sD_{1,\phi} / \delta\phi^\nu |_{U^{\nu + 2}})$. Using estimate~\eqref{WF_better} for the $\delta^\nu \omega^R_\phi / \delta \phi^\nu$ and the definition of the set $X_{2 +\nu}$ given by~\eqref{X}, it follows that if $p_r$ is in $\overline{V}^\pm$ for any $r$, then we have $k_1,k_2 \in \overline{V}^\mp$. While, if $p_s$ is space-like and if $p_{r}$ is in $\overline{V}^\pm$ for any $r \neq s$, then we have $k_1 \in \overline{V}^\mp, k_2 \notin \overline{V}^\pm$ or vice versa. These configurations satisfies the conditions we want to verify.
			\item We again compute the $\delta^\nu \sD_{2,\phi} / \delta \phi^\nu$ by distributing the variational derivatives on each factor in the right-hand side of~\eqref{sD_2}. Using the wave-front set calculus (thm.~\ref{theo_WF_horma}) together with the estimates~\eqref{est_G_1,2} and~\eqref{var_ders_A/R_WF}, we obtain
			\begin{equation*}
				\begin{split}
					&\left. \WF \left( \frac{\delta^\nu \sD_{2,\phi}(x_1,x_2)}{\delta \phi(y_1) \dots \delta \phi (y_\nu)} \right)\right|_{U^{\nu +2}}  \subset \\
					&\quad \subset \left\{ (x_1, x_2, y_1, \dots, y_\nu; k_1, k_2, p_1, \dots, p_\nu)\in \dot{T}^* M^{2+\nu} : x_1, x_2, y_1, \dots, y_\nu \in U \right. \\
					&\qquad \left. \exists N_1, N_2 \mbox{ partition of } \{1, \dots, \nu\} \mbox{ and } (z,q) \in T^* M  \mbox{ such that } z \in N \mbox{ and}\right. \\
					&\qquad \left. (x_1, z, (y_{r \in N_1}); k_1, -q, (p_{r \in N_1})) \in X_{2+ |N_1|} \mbox{ or } k_1, q, p_{r \in N_1} = 0, \right. \\
					&\qquad \left. (z, x_2, (y_{r\in N_2}); q, k_2, (p_{r \in N_2})) \in \cC^u_{2 +|N_2|} \mbox{ or }  q,k_2, p_{r \in N_2} = 0\right\}.
					\end{split}
			\end{equation*}
		Let $(x_1, x_2, y_1, \dots, y_\nu; k_1, k_2, p_1, \dots, p_\nu)$ be an element in $\WF ( \delta^\nu \sD_{2,\phi} / \delta \phi^\nu |_{U^{\nu +2}})$. As can be directly checked, it follows from the definitions of the sets $X_{2 +\nu}$ (given by~\eqref{X}) and $\cC^u_{2 +\nu}$ (given by~\eqref{Cunu}) that if $p_r$ is in $\overline{V}^\pm$ for any $r$, then we have $k_1, k_2 \in \overline{V}^\mp$. While, if there exists a space-like covector $p_s$ and if $p_{r} $ is in $\overline{V}^\pm$ for any $r\neq s$, then we have $k_1, k_2 \notin \overline{V}^\pm$ and, when $x_1 =x_2$, $k_1 + k_2 \notin \overline{V}^\pm$. This is exactly what we wanted to show.
		\item The same argument as before can be applied in this case. In particular, by distributing the variational derivatives on each factor in the right-hand side of~\eqref{sD_3},  we get
			\begin{equation*}
				\begin{split}
					&\left.\WF \left( \frac{\delta^\nu \sD_{3,\phi}(x_1,x_2)}{\delta \phi(y_1) \dots \delta \phi (y_\nu)} \right) \right|_{U^{\nu +2}}  \subset \\
					&\quad \subset \left\{ (x_1, x_2, y_1, \dots, y_\nu; k_1, k_2, p_1, \dots, p_\nu)\in \dot{T}^* M^{2+\nu} : x_1, x_2, y_1, \dots, y_\nu \in U \right. \\
					&\qquad \left. \exists N_1, N_2 \mbox{ partition of } \{1, \dots, \nu\} \mbox{ and } (z,q) \in T^* M  \mbox{ such that } z \in N \mbox{ and}\right. \\
					&\qquad \left. (x_1, z, (y_{r\in N_1}); k_1,-q, (p_{r \in N_1})) \in \cC^u_{2+|N_1|} \mbox{ or } k_1, q, p_{r \in N_1} = 0, \right. \\
					&\qquad \left. (z, x_2, (y_{r \in N_2}); q, k_2, (p_{r \in N_2})) \in X_{2+{|N_2|}} \mbox{ or } q,k_2, p_{r \in N_2} = 0 \right\}.
					\end{split}
			\end{equation*}
				Let $(x_1, x_2, y_1, \dots, y_\nu; k_1, k_2, p_1, \dots, p_\nu) $ be an element in $\WF ( \delta^\nu \sD_{3,\phi} / \delta \phi^\nu |_{U^{\nu +2}})$. We obtain again that if $p_r $ is in $\overline{V}^\pm$ for any $r$, then we have $k_1, k_2 \in \overline{V}^\mp$. While, if there exists a space-like $p_s$ and if $p_{r}$ is in $\overline{V}^\pm$ for any $r\neq s$, then we have $k_1, k_2 \notin \overline{V}^\pm$ and, when $x_1 =x_2$, $k_1 + k_2 \notin \overline{V}^\pm$. This is exactly what we wanted to prove.
		\item Since $c$ is by construction a smooth function, and since the wave-front set of $\delta^\nu G^{(3)}_\phi / \delta \phi^\nu$ is contained in $\cC^u_{2 + \nu}$, the same arguments used for $\sD_1, \sD_2, \sD_3$ allow us to conclude that the for the last term $\sD_4$ we have
			\begin{equation*}
				\begin{split}
					&\left. \WF\left( \frac{\delta^\nu \sD_{4,\phi}(x_1,x_2)}{\delta \phi(y_1) \dots \delta \phi (y_\nu)} \right)\right|_{U^{\nu +2}}  \subset \\
					&\quad \subset \left\{ (x_1, x_2, y_1, \dots, y_\nu; k_1, k_2, p_1, \dots, p_\nu) \in \dot{T}^* M^{2+\nu}: x_1,x_2, y_1, \dots, y_\nu \in U \right. \\
						&\qquad \left. \exists N_1, N_2, N_3 \mbox{ partition of } \{1, \dots, \nu \} \mbox{ and } (z,q), (z',q') \in T^*M \mbox{ such that }  z,z \in N \mbox{ and}\right. \\
						&\qquad \left. (x_1, z, (y_{r \in N_1}); k_1, -q, (p_{r \in N_1})) \in X_{2+ |N_1|}, \mbox{ or } k_1, q, p_{r \in N_1} = 0, \right. \\
						&\qquad \left. (z, z' (y_{r \in N_2}); q, -q', (p_{r \in N_2})) \in \cC^u_{2+|N_2|}, \mbox{ or } q, q', p_{r \in N_2} = 0, \right. \\
						&\qquad \left. (z', x_2, (y_{r \in N_3}); q', k_2, (p_{r \in N_3})) \in X_{2+ |N_3|}, \mbox{ or } q',k_2, p_{r \in N_3} = 0\right\}.
					\end{split}
			\end{equation*}
			Let $(x_1, x_2, y_1, \dots, y_\nu; k_1, k_2, p_1, \dots, p_\nu)$ be an element $\WF ( \delta^\nu \sD_{4,\phi} / \delta \phi^\nu |_{U^{\nu +2}})$. The estimate above implies that if $p_r$ is in $\overline{V}^\pm$, then we have $k_1, k_2 \in \overline{V}^\mp$. While, if there exists a space-like $p_s$ and if $p_{r}$ is in $\overline{V}^\pm$ for any $r\neq s$, then we have $k_1, k_2 \notin \overline{V}^\pm$ and, when $x_1 =x_2$, $k_1 + k_2 \notin \overline{V}^\pm$. This is precisely what we wanted to show.
		\end{enumerate}
		 This concludes the proof of estimate~\eqref{var_d_nu}.\\

		The estimate~\eqref{var_d_nu_s_phi} concerning the smooth variation of the background $\epsilon \mapsto \phi(\epsilon)$ is proved repeating the same argument just shown, up to some minor modifications: there is an explicit dependence on $\epsilon$ in all the distributions depending on $\phi$ and consequently we need to use estimates~\eqref{WF_better_s},~\eqref{var_u_s_phi} and~\eqref{var_ders_A/R_WF_phi} and instead of estimates~\eqref{WF_better},~\eqref{var_u} and~\eqref{var_ders_A/R_WF}. This concludes the proof of the lemma.
	\end{proof}
	
	The last result we present in this subsection is the following corollary of lemma~\ref{lemma_d}:
	\begin{lemma}\label{lemma_d_coinc}
		$\delta^\nu d_\phi(x,x) / \delta \phi(y_1) \dots \delta \phi(y_\nu)$ exists as distribution in $x, y_1, \dots, y_\nu$ and 
		\begin{equation}\label{var_est_d_coinc}
				\WF \left( \frac{\delta^\nu d_\phi(x,x)}{\delta \phi(y_1) \dots \delta \phi(y_\nu)} \right) \subset \cC^{d,\Delta}_{1+\nu}[g],
		\end{equation}
	where
		\begin{equation*}
			\begin{split}
				\cC^{d,\Delta}_{1+\nu}[g] := &\left\{ (x, y_1, \dots, y_\nu; k, p_1, \dots, p_\nu) \in \dot{T}^* M^{1 + \nu} :  \mbox{ if } p_r \in \overline{V}^\pm \, \forall r, \mbox{ then } k \in \overline{V}^\mp \right.\\
				&\qquad \qquad  \left. \mbox{ if } \exists p_s \mbox{ space-like}, \, p_r \in \overline{V}^\pm \, \forall r \neq s, \mbox{ then } k \notin \overline{V}^\pm \right\}.
			\end{split}
		\end{equation*}
	For any smooth map $\bR \ni \epsilon \mapsto \phi(\epsilon) \in C^\infty(M)$, it holds
		\begin{equation}\label{var_est_d_coinc_s_phi}
				\WF \left( \frac{\delta^\nu d_{\phi(\epsilon)} (x,x)}{\delta \phi(y_1) \dots \delta \phi(y_\nu)} \right) \subset \bR \times \{ 0 \} \times \cC^{d,\Delta}_{1+\nu}[g].
		\end{equation}
	\end{lemma}
	\begin{proof}
		Using the estimate~\eqref{var_d_nu} for the wave-front set of $\delta^\nu d_\phi/ \delta \phi^\nu$, we obtain that
			\begin{equation*}
				\begin{split}
				 &\WF \left( \frac{\delta^\nu d_\phi (x_1,x_2)}{\delta \phi(y_1) \dots \delta \phi(y_\nu)} \right)_{x_1,x_2} \subset\\
				 &\quad \subset  \{ (x_1, x_2; k_1, k_2) \in \dot{T}^*(M^2) : (x_1, x_2, y_1, \dots, y_\nu; k_1, k_2, 0, \dots, 0) \in \cC^d_{2+\nu} \} = \emptyset.
				\end{split}
			\end{equation*}
		Therefore, the wave-front set calculus (thm.~\ref{theo_WF_horma}) implies that $\delta^\nu d_\phi (x_1,x_2) / \delta \phi(y_1) \dots \delta \phi(y_\nu)$ can be contracted with the delta distribution $\delta(x,x_1,x_2)$ and, furthermore, that the wave-front set of the contraction is bounded by~\eqref{var_est_d_coinc}.\\
		
		A similar argument, based on estimate~\eqref{var_d_nu_s_phi} for the wave-front set of $\delta^\nu d_{\phi(\epsilon)}/ \delta \phi^\nu$, shows that the distribution $\delta^\nu d_{\phi(\epsilon)} (x,x) / \delta \phi(y_1) \dots \delta \phi(y_\nu)$ satisfies the bound~\eqref{var_est_d_coinc_s_phi}. This concludes the proof.
	\end{proof}

	\subsection{Proof of the on-shell $W$-smoothness of the Wick power corresponding to $\varphi^k(f)$}\label{subsubsec_Wickpower}
		Let $(M,g)$ be an ultra-static space-time and let $m$ be constant, $\lambda \in C^\infty_0(M)$, while $\phi$ is a generic function in $C^\infty(M)$. We begin by defining the ``Wick powers'' of the linear field theory corresponding to $P_\phi = \boxempty - m^2 - \frac{\lambda}{2} \phi^2$  following the prescriptions given in~\citep{HW01,HW02}. The Wick power corresponding to the classical functional $F(\varphi) =\varphi^k(f)$, where $f \in C^\infty_0(M)$, is also viewed as a time-ordered product with one factor and so it is denoted by $T_{1,\phi}[ F(\varphi)]$. It is by definition the element in $\cW_\phi = \cW[g,m,\phi,\lambda]$ given by
			\begin{equation}\label{Wick_power_fact}
				\begin{split}
					T_{1,\phi} [ \varphi^k(f)] &= \int_M f(x) : \varphi^k(x):_{H_\phi} dx \\
					&=\int_{M^{k}} f(x_1) \delta(x_1, \dots, x_{k}) :\varphi(x_1) \cdots \varphi(x_{k}):_{H_\phi} dx_1 \dots dx_k ,\\
					&=\sum_{j} (-1)^{j} \hbar^j \left({}^{2j}_k \right) \int_{M^{k-2j+1}} f(x) \delta(x, x_{1}, \dots, x_{k - 2j}) d_\phi(x,x)^j \times \\
					&\qquad \qquad \qquad \qquad \times :\varphi(x_1) \cdots \varphi(x_{k-2j}):_{\omega^R_\phi} dx dx_1 \dots dx_{k-2j},
				\end{split}
			\end{equation}
		where $d_\phi = \omega^R_\phi - H_\phi$, and where the sum is taken over $j \leq [k/2]$. Here $: \cdots:_{\omega^R_\phi}$ denotes the normal ordered Wick products with respect to the retarded $2$-point function $\omega^R_\phi$ (defined by eq.\eqref{in-state}) and $: \cdots:_{H_\phi}$ denotes the normal ordered Wick products with respect to the Hadamard parametrix $H_\phi$ (defined by eq.~\eqref{Hadamard_def}). The product of $\cW_\phi$ is defined  in terms of $\omega^R_\phi$. Then, for any $t \in \bP^+ \cE'_W(M^n)$, we identify, similarly as done in~\citep{HW03}, the normal ordering of $t$ respect to $\omega^R_\phi$ with the equivalence class of $t$ in $\cW_\phi$, i.e. we set
			\begin{equation*}
				\begin{split}
					&\int_{M^n} t(x_1,\dots, x_n) :\varphi(x_1) \cdots \varphi(x_n):_{\omega^R_\phi} dx_1 \dots dx_n = [t] \in \bP^+ \cE'_W(M^n) / P_\phi \cE'_W(M^n).
				\end{split}
			\end{equation*}
	Let $U$ be a convex normal set. In our context, for any $t \in \bP^+\cE'_W(M^n)$ with support in $U^n$, the normal ordering of $t$ with respect to the Hadamard parametrix $H_\phi$ in $U \times U$ is defined by
			\begin{equation*}
				\begin{split}
					&\int_{M^n} t(x_1,\dots, x_n) :\varphi(x_1) \cdots \varphi(x_n):_{H_\phi} dx_1 \dots dx_n := \\
					&\quad = \sum_{j \leq [n/2]} (-1)^{j} \hbar^j \left( {}^{2j}_n \right) \left[ \bP^+ \int_{M^{2 j}} t(z_1, \dots, z_j, x_{1}, \dots, x_{n -2j}) \prod_{i=1}^j d_\phi(z_{2i -1}, z_{2i}) dz_1 \dots dz_{2j} \right],
				\end{split}
			\end{equation*}
		where the right-hand side is an element in $\bC[\hbar] \otimes \oplus_{\ell \leq n} \bP^+ \cE'_W(M^{\ell}) / P_\phi \bP^+ \cE'_W(M^{\ell})$.\\
		Following the formalism we developed in sec.~\ref{subsec_manifold_inf}, each element in $\cW_\phi$ can be identified with a sequence of distributions, where the $\ell$-th entry is an element in $\bC[[\hbar]] \otimes (\sigma_c \circ E_\phi)^{\otimes \ell} \cE'_W(M^\ell)$, and where $c$ is a cut-off function as in eq.~\eqref{kernel_symp}. Going through the definitions, we find that the Wick power $T_{1,\phi} [ \varphi^k(f)] \in \cW_\phi$ corresponds to the sequence $(\tilde{t}^\ell_\phi)_{\ell \in \bN}$ given by
			\begin{equation}\label{sequence_Wick_power_fact}
			\tilde{t}^\ell_\phi = \left\{ \begin{array}{ll} (-1)^{j} \hbar^j \left({}^{2j}_k \right) (\sigma_c \circ E_\phi)^{\otimes k - 2j} \left( (f \cdot \delta^{(k -2j+1)}) \circ (d_\phi \circ \delta)^{j} \right) & \mbox{ if } \ell=k-2j\\ 0 & \mbox{ otherwise} \end{array} \right.
			\end{equation}
		where $(d_\phi \circ \delta)(x) = d_\phi(x,x)$. Note that we have $\tilde{t}^0_\phi= (-1)^{k/2} \hbar^{k/2} \int_M f(x) d_\phi(x,x)^{k/2} dx \in \bC[\hbar]$ if $k$ is even, and $\tilde{t}^0_\phi = 0$ otherwise.\\
		The Wick power $T_{1,\phi} [\varphi^k(f)]$ for $\phi \in S$ (and fixed $m$ constant and $\lambda \in C_0^\infty(M)$) corresponds to the sequence $(t^0_\phi, t^1_\phi, \dots)$, where $t^\ell_\phi$ is just $\tilde{t}^\ell_\phi$ evaluated for $\phi \in S$. In other words, we can consider the Wick powers constructed for $\phi \in C^\infty(M)$ as extensions of the corresponding the Wick powers constructed with respect to $\phi \in S$.
		\begin{prop}\label{prop_W_Wick_power}
			The section $S \ni\phi \mapsto T_{1,\phi} [\varphi^k(f)]$ is on-shell $W$-smooth.
		\end{prop}
		\begin{proof}		
			We need to prove that each $\tilde{t}^\ell_\phi$ satisfies conditions~\ref{W1},~\ref{W2} in def.~\ref{def_smooth_on_tens}. Making use of the wave-front set calculus (thm.~\ref{theo_WF_horma} and lemma~\ref{lemma_W_comp}), the properties of the causal propagator $E_\phi$ (prop.~\ref{prop_var_ders_causal}) and the definition of $\sigma_c$ (eq.~\eqref{kernel_symp}), it follows that it is sufficient to prove that for any $\ell$ the following estimates hold:
			\begin{equation}\label{est_WF_W_Wick}
				\WF \left( \frac{\delta^\nu d^\ell_{\phi}(x,x)}{\delta \phi(y_1) \dots \delta \phi(y_\nu)} \right) \subset W_{\ell + \nu},
			\end{equation}
		and
			\begin{equation}\label{est_WF_W_Wick_s}
				\WF \left( \frac{\delta^\nu d^\ell_{\phi(\epsilon)}(x,x)}{\delta \phi(y_1) \dots \delta \phi(y_\nu)} \right) \subset \bR \times \{0\} \times W_{\ell + \nu},
			\end{equation}
		for  any $\bR \ni \epsilon \mapsto \phi(\epsilon) \in C^\infty(M)$ smooth.\\
		To verify these estimates, we compute $\delta^\nu d_\phi(x,x)^{\ell} / \delta \phi(y_1) \dots \delta \phi(y_\nu)$ by distributing the Gateaux derivatives on each factor $d_\phi(x,x)$. It follows that $\delta^\nu d_\phi(x,x)^{\ell} / \delta \phi(y_1) \dots \delta \phi(y_\nu)$ is a finite sum of terms in the form
			\begin{equation}\label{var_t_Wick_prod}
				\begin{split}
					&\prod_{\ell' =1}^{\ell} \frac{\delta^{|N_{\ell'}|} d_\phi(x, x)}{\delta \phi^{|N_{\ell'}|}(\{y_{r \in N_{\ell'}}\})},
				\end{split}
			\end{equation}
		where $N_1, \dots, N_{\ell}$ form a partition of $\{1, \dots, \nu\}$. The wave-front set of $\delta^{|N_{\ell'}|} d_\phi(x,x) / \delta \phi^{|N_{\ell'}|} $ is estimated by~\eqref{var_est_d_coinc} of lemma~\ref{lemma_d_coinc} and, thus, we obtain
			\begin{equation*}
				\WF (\delta^{|N_{\ell'}|}  d_\phi(x,x) / \delta \phi^{|N_{\ell'}|} )_x =\emptyset.
			\end{equation*}
		Therefore, the product of distributions~\eqref{var_t_Wick_prod} is well-defined as a consequence of the wave-front set calculus (thm.~\ref{theo_WF_horma}).\\
		We now focus on the proof of estimate~\eqref{est_WF_W_Wick}. It follows from wave-front set calculus that whenever $(x, y_1, \dots, y_\nu; k, p_1, \dots, p_\nu)\in \dot{T}(M^{\nu +1})$ is in the wave-front set of~\eqref{var_t_Wick_prod}, there exists a decomposition $k=k^{(1)}+ \dots + k^{(\ell)}$, and for any $\ell'=1, \dots, \ell$ it must hold
			\begin{equation*}
				(x,  (y_r)_{r \in N_{\ell'}}); k^{({\ell'})}, (p_r)_{r \in N_{\ell'}})) \in \WF ( \delta^{|N_{\ell'}|} d_\phi(x, x)/ \delta \phi^{|N_{\ell'}}) \mbox{ or } k^{({\ell'})}, p_r = 0.
			\end{equation*}			 
		We prove that $(x, y_1, \dots, y_\nu; k, p_1, \dots, p_\nu)$ cannot belong to the set $C^+_{1+\nu}$ defined by~\eqref{C_set_def}. In order to do that, we split the proof in the following two cases: (a) if all covectors $p_1, \dots, p_\nu$ belongs to $\overline{V}^+$, then we get $k \in \overline{V}^-$, and (b) if there exists an $s \in N_{\ell''}$ for a certain $\ell''$ such that $p_s$ is space-like whereas $p_r$ is in $\overline{V}^+$ for any $r \neq s$, then we have $k \notin \overline{V}^+$.
		\begin{enumerate}[label=(\alph*), start=1]
			\item It follows from estimate~\eqref{var_est_d_coinc} that if $p_r$ is in $\overline{V}^+$ for any $r$, then we have $k^{(1)}, \dots, k^{(\ell)} \in \overline{V}^-$. Thus, $k$ must be in $\overline{V}^-$, as we wanted to prove.
			\item Since $p_{r}$ is in $\overline{V}^+$ for any $r\notin s$ where $s \in N_{\ell''}$, estimate~\eqref{var_est_d_coinc} implies $k^{(\ell')} \in \overline{V}^-$ with $\ell' \neq \ell''$, and $k^{(\ell'')} \notin \overline{V}^+$. Thus, putting together, we obtain $k \notin \overline{V}^+$ as we needed to prove.
		\end{enumerate}
		With a similar argument, based on estimate~\eqref{var_est_d_coinc}, we can show that $(x, y_1, \dots, y_\nu; k, p_1, \dots, p_\nu)$ does not belong to $C^-_{1+\nu}$ either. Thus, by definition, we prove $(x, y_1, \dots, y_\nu; k, p_1, \dots, p_\nu) \in W_{1+ \nu}$, which is precisely what is needed to verify estimate~\eqref{est_WF_W}.\\
		
		The proof of estimate~\eqref{est_WF_W_Wick_s} can be obtained with a similar argument as the one just presented, based on estimate~\eqref{var_est_d_coinc_s_phi} instead of estimate~\eqref{var_est_d_coinc}.
		This concludes the proof.
		\end{proof}
		
		The map $\phi \mapsto T_{1,\phi}[\varphi^k(f)]$ is still on-shell $W$-smooth if we promote the test function $f$ to be a $W$-smooth map $C^\infty(M) \ni \phi \mapsto f_\phi \in C^\infty_0(M)$. For any $F$ local functional functional not involving covariant derivatives, $F(\phi + \varphi)$ is a sum of local functionals in this generalized form. Thus, it is clear that the argument just presented implies that $S \ni \phi \mapsto T_{1,\phi}[F(\varphi + \phi)]$ is on-shell $W$-smooth.
		
	\subsection{Sufficient condition for the on-shell $W$-smoothness of the time-ordered product of $\otimes_i \varphi^{k_i}(f_i)$}\label{subsubsec_suff_cond_T-prod}
		As before, let $(M,g)$ be any ultra-static space-time and let $m$ be constant, $\lambda \in C^\infty_0(M)$, while $\phi$ is a generic function in $C^\infty(M)$. We would like to show that the $n$-fold time-ordered product defines an on-shell $W$-smooth section $\phi \mapsto T_{n,\phi} [\otimes_{i=1}^n \varphi^{k_i} (f_i)] \in \cW_\phi$. For $n=1$, we have already seen in sec.~\ref{subsubsec_Wickpower} that this is true. We now consider the case $n >1$.\\
		We make use of the Wick expansion in terms of normal ordering with respect to the retarded $2$-point function $\omega^R_\phi$, given in~\eqref{in-state}. As shown in~\citep{brunetti2000microlocal}, we obtain the following expansion:
					\begin{equation}\label{Wick_exp_state}
						\begin{split}
							T_{n,\phi} \left[ \bigotimes_{i=1}^n \varphi^{k_i} (f_i) \right] &= \int_{M^n}  f_1(x_1) \cdots f_N(x_n) T_{n,\phi} \left[ \bigotimes_{i=1}^n \varphi^{k_i} (x_i) \right] dx_1 \dots dx_n \\
							&= \sum_{{\bf j} \leq {\bf k}} \left( {}^{\bf k}_{{\bf j}} \right) \int_{M^n} f_1(x_1) \cdots f_n(x_n) \tau_{\phi, {\bf j}} (x_1, \dots, x_n) :\prod_{i=1}^n \varphi^{k_i - j_i}(x_i) :_{\omega^R_\phi} \prod_{i=1}^n  dx_i \\
							&= \sum_{{\bf j}} \left( {}^{{\bf k}}_{{\bf j}} \right) \int_{M^{n + |{\bf k} - {\bf j}|}} \left( \prod_{i} f_{i}(x_{i}) \delta(x_i, x_i^{(1)}, \dots, x_i^{(k_i - j_i)})\right)  \tau_{\phi, {\bf j}} (x_{1}, \dots, x_{n}) \times \\
							&\qquad \times :\prod_{i} \varphi(x_i^{(1)}) \cdots  \varphi(x_i^{(k_i - j_i)}):_{\omega^R_\phi} \prod_i dx_i dx_i^{(1)} \dots dx_i^{(k_i - j_i)},
						\end{split} 
					\end{equation}
				where we used the multi-index notation ${\bf k} = (k_1, \dots, k_m)$, and where the combinatorial factor appearing is just $({}^{\bf k}_{\bf j}) := ({}^{k_1}_{j_1}) \cdots  ({}^{k_m}_{j_m})$.\\
				Just as the Wick powers, the time-ordered product $T_{n,\phi}[\otimes \varphi^{k_i}(f_i)] \in \cW_\phi$ can be identified in the formalism we developed in section~\ref{subsec_manifold_inf} with a sequence of distributions $(\tilde{t}^\ell_\phi )_{\ell \in \bN}$ for any $\phi \in C^\infty(M)$. Each of these distributions $\tilde{t}^\ell_\phi$ can be expressed as the finite sum
					\begin{equation}\label{tilde_t_time}
						\tilde{t}^\ell_\phi  = \sum_{{\bf j} \leq {\bf k}, |{\bf k} - {\bf j}| = \ell} \left( {}^{{\bf k}}_{{\bf j}} \right) (\sigma_c \circ E_\phi)^{\otimes \ell} \left( (f_1 \cdot \delta \otimes  \cdots  \otimes f_n \cdot \delta) \circ \tau_{\phi, {\bf j}} \right),
					\end{equation}
			where $c$ is a cut-off function as in eq.~\eqref{kernel_symp}, and $\phi \in C^\infty(M)$.\\
			The time-ordered product $T_{n,\phi} [ \otimes_i \varphi^{k_i} (f_i)]$ for $\phi \in S$ corresponds to the sequence $(t^0_\phi, t^1_\phi, \dots)$ where $t^\ell_\phi$ is just $\tilde{t}^\ell_\phi$ evaluated for $\phi \in S$.\\
			Similarly the proof of prop.~\ref{prop_W_Wick_power} for the time-ordered products with one factor, we can formulate sufficiently conditions on the distributional coefficients of the Wick expansion to imply the on-shell $W$-smoothness of the time-ordered products.
			\begin{lemma}\label{lemma_suff_cond}
				To establish that the section $S \ni \phi \mapsto T_{n,\phi} [ \otimes_i \varphi^{k_i} (f_i)]$ is on-shell $W$-smooth it is sufficient to show that the distributional coefficients $\{\tau_{\phi, {\bf j}}\}_{\bf j}$ of the Wick expansion with respect to the state $\omega_\phi$ satisfy
			\begin{equation}\label{est_WF_W_1}
				\WF \left( \frac{\delta^\nu \tau_{\phi, {\bf j}}(x_1, \dots, x_n)}{\delta \phi(y_1) \dots \delta \phi(y_\nu)} \right) \subset W_{n+\nu},
			\end{equation}
			and
				\begin{equation}\label{est_WF_W_s_1}
					\WF \left( \frac{\delta^\nu \tau_{\phi(s), {\bf j}}(x_1, \dots, x_n)}{\delta \phi(y_1) \dots \delta \phi(y_\nu)} \right) \subset \bR \times \{0\} \times W_{n+\nu},
				\end{equation}
			for any $\bR \ni s \mapsto \phi(s) \in C^\infty(M)$ smooth.
		\end{lemma}
		\begin{proof}
				Making use of the wave-front set calculus (thm.~\ref{theo_WF_horma} and lemma~\ref{lemma_W_comp}), the properties of the causal propagator $E_\phi$ (prop.~\ref{prop_var_ders_causal}), and the definition of $\sigma_c$ (eq.~\eqref{kernel_symp}), it follows that the estimates~\eqref{est_WF_W_1} and~\eqref{est_WF_W_s_1} imply that each $\tilde{t}^\ell_\phi$ as in eq.~\eqref{tilde_t_time} satisfies the requirements~\ref{W1},~\ref{W2} of def.~\ref{def_smooth_on_tens}. This is precisely what is needed to conclude that $S \ni \phi \mapsto T_{n,\phi} [ \otimes_i \varphi^{k_i} (f_i)]$ is on-shell $W$-smooth.
		\end{proof}
		
		Once estimates~\eqref{est_WF_W_1} and~\eqref{est_WF_W_s_1} are proved, $\phi \mapsto T_{n,\phi}[\otimes_i \varphi^{k_i}(f_i)]$ will be on-shell $W$-smooth even if we promote the test functions $f_1, \dots, f_N$ to be $W$-smooth map $C^\infty(M) \ni \phi \mapsto f_{j,\phi} \in C^\infty_0(M)$. Thus, we have established that the on-shell $W$-smoothness of $S \ni \phi \mapsto T_{n,\phi}[\otimes_i F_i(\varphi + \phi)]$ for local functionals $F_i$ not containing covariant derivatives will follow from~\eqref{est_WF_W_1} and~\eqref{est_WF_W_s_1}.\\
		It is the purpose of the following subsections to prove that these sufficient conditions indeed hold.
	
	\subsection{Review of the construction of the time-ordered products of $n>1$ local functionals}\label{subsubsec_rew_T-prod}
		In order to verify the sufficient conditions of lemma~\ref{lemma_suff_cond}, we need to specify exactly how the distributions $\tau_{\phi, {\bf j}}$ are constructed. We review the procedure presented in~\citep{HW02} to define them. To do so, we need to consider a generic space-time $(M,g)$ and generic smooth functions $m, \phi, \lambda \in C^\infty(M)$. We consider a product $\bullet_\phi$ for $\cW[g,m,\phi,\lambda]$ defined in terms of an admissible assignment $\phi \mapsto \omega_\phi$ where $\omega_\phi$ is a pure Hadamard $2$-point function (such $2$-point function always exists, as proved by the deformation argument we presented in~\ref{subsubsec_kahler_inf}). Note that the retarded $2$-point function cannot be defined for such general space-times.\\
		The construction proceeds by induction on the number of the factors $n$ of the time-ordered products.  At the $n$-th induction order, we assume the a prescription for defining the time-ordered products has been constructed satisfying (T1)-(T9) for $<n$ factors. The inductive hypothesis is already know to hold for $n=1$ factor (Wick powers).\\
		The induction step relies on the causal factorization axiom (T8). Due to this property, the time-ordered product $T_{n,\phi} [ \otimes_i \varphi^{k_i}(f_i)] = \int_{M^n} f_1(x_1) \cdots f_n(x_n) T_{n,\phi}[\otimes_{i=1}^n \varphi(x_i)] dx_1 \dots dx_n$ can be expressed as a finite sum of $\bullet_\phi$-products of time-ordered products involving fewer factors whenever $f_1 \otimes \cdots \otimes f_n$ is supported outside the total diagonal. Thus, the induction hypothesis fully determines $T_{n,\phi} [ \otimes_i \varphi^{k_i}(x_i)]$ outside the total diagonal $\Delta_n$. The axioms (T1)-(T9) are satisfied in this domain.\\
		In~\citep{HW02} is provided an extension to the whole $M^n$ which is compatible (T1)-(T9). Actually, as explained in~\citep[sec. 3.1]{HW02}, it is sufficient to require that the extension satisfies axioms (T1)-(T5) and (T9), because then the axioms (T6)-(T7) can be imposed by simple redefinitions and (T8) is automatically ensured by construction.\\
		To characterize the extension, we now assume that $f_1 \otimes \cdots \otimes f_n$ is supported in a set $\cU_n$ sufficiently close to the total diagonal. In detail, let $\cU_n$ be a neighbourhood of the total diagonal in $M^n$ such that $x_1, \dots, x_n$ belong to a convex normal neighbourhood $U$ if $(x_1, \dots, x_n) \in \cU_n$. We can define the Hadamard parametrix $H_\phi$ in $U$, see eq.~\eqref{Hadamard_def}. Then, we expand $T_{n,\phi} [ \otimes_i \varphi^{k_i}(f_i)]$ in terms of the normal ordering with respect to the Hadamard parametrix, i.e.
			\begin{equation}\label{local_Wick_exp}
				\begin{split}
					&T_{n,\phi} \left[ \bigotimes_{i=1}^n \varphi^{k_i} (f_i) \right] = \sum_{{\bf j} \leq {\bf k}} \left( {}^{\bf k}_{{\bf j}} \right) \int_{M^n} f_1(x_1) \cdots f_n(x_n) \tau^H_{\phi, {\bf j}} (x_1, \dots, x_n) :\prod_{i=1}^n \varphi^{k_i - j_i}(x_i) :_{H_\phi} \prod_{i=1}^n  dx_i \\
					&\quad = \sum_{{\bf j}} \left( {}^{{\bf k}}_{{\bf j}} \right) \int_{M^{n + |{\bf k} - {\bf j}|}} \left( \prod_{i} f_{i}(x_{i}) \delta(x_i, x_i^{(1)}, \dots, x_i^{(k_i - j_i)})\right)  \tau^H_{\phi, {\bf j}} (x_{1}, \dots, x_{n}) \times \\
					&\qquad \qquad \times :\prod_{i} \varphi(x_i^{(1)}) \cdots  \varphi(x_i^{(k_i - j_i)}):_{H_\phi} \prod_i dx_i dx_i^{(1)} \dots dx_i^{(k_i - j_i)},	
				\end{split}
			\end{equation}
		where $\tau^H_{\phi, {\bf j}}$ are certain distributional coefficients. This expansion is called {\em local Wick expansion}.\\
		By comparing with formula~\eqref{Wick_exp_state}, we conclude that the distributions $\tau^H_{\phi, {\bf j}}$ are all identically $1$ for $n=1$.\\
		It is proved in~\citep[sec. 3.2]{HW02} that any definition of time-ordered products satisfying axioms (T3) and (T9) admits a local Wick expansion with coefficients $\tau^H_{\phi, {\bf j}}$ satisfying
			\begin{equation}\label{est_tau_H}
				\WF \left( \tau^H_{{\bf j}}[g,m, \phi, \lambda] \right) \subset \cC^{T}_n[g],
			\end{equation}
		where $\cC^{T}_n$ is defined in terms of decorated graphs similarly as done for $\cC^{R}_n$ in formula~\eqref{C^R_m}, namely
			\begin{equation}\label{C^T_m}
				\begin{split}
					\cC^T_n[g] := &\left\{ (x_1, \dots, x_n ; k_1, \dots, k_n) \in \dot{T}^\ast M^{n+1}: \right.\\
					&\left.\exists \mbox{ decorated graph } \sG \mbox{ with vertices } x_1, \dots, x_n \right.\\
					&\left.  k_j = \sum_{e: s(e)=x_j} p_e(x_j) - \sum_{e: t(e) =x_j} p_e(x_j) \right\},
				\end{split}
			\end{equation}
		where, in this context, a decorated graph $\sG$ is understood as an embedded graph in $M$ with vertices $x_1 \dots, x_{n}$ and with edges connecting the vertices given by oriented null-geodesic curves. The valence of a vertex $x_i$ in the graph is restricted to be less or equal to $j_i$. For an edge $e$, we denote the endpoints by $x_{s(e)}$ (called source) and $x_{t(e)}$ (called target) if $s(e) < t(e)$. We consistently impose an orientation for the null-geodesic corresponding to $e$ in such a way that the curve starts at $s(e)$. Each edge is equipped with a future-directed covector field $p_e$ which is cotangent and coparallel to the geodesic curve associated to the edge $e$. The field $p_e$ is future/past directed if $x_{t(e)}$ is in the future/past of $x_{s(e)}$.\\
		Conversely, if a prescription for time-ordered products admits a local Wick expansion with distributional coefficients satisfying the wave-front set condition above, then it satisfies the axioms (T3) and (T9).\\
		Because of the inductive hypothesis, the time-ordered product $T_{n,\phi} [ \otimes_i \varphi^{k_i}(x_i)]$ is known to satisfy all the axioms (T1)-(T9) on $\cU_n \backslash \Delta_n$, i.e. outside the total diagonal $\Delta_n$, but inside the neighbourhood $\cU_n$. Thus, it can be defined by a local Wick expansion with distributional coefficients $\tau^{H,0}_{\phi,{\bf k}}$ defined on $\cU_n \backslash \Delta_n$.\\
		As a consequence of the causal factorization axiom (T8), the distribution $\tau^{H,0}_{\phi, {\bf k}}$ is fully determined by terms $\{\tau^H_{\phi, {\bf i}}\}_{{\bf i}}$ corresponding to time-ordered products with less than $n$ factors and appropriate powers of the Hadamard parametrix. More precisely, $\tau^{H,0}_{\phi, {\bf k}}$ can be expressed as a finite sum of terms in the form
			\begin{equation}\label{gen_term_T8_H}
				f_I(x_1, \dots, x_n)  \tau^{H}_{\phi, {\bf i}} ((x_a)_{a \in I}) \tau^{H}_{\phi, {\bf i}'} ((x_b)_{b \in I^c}) \prod_{a\in I, b \in I^c} H_\phi(x_{a}, x_{b})^{n_{ab}},
			\end{equation}
		where $I$ is a proper subset of $\{1, \dots, n\}$, and where $n_{ab}$ are natural numbers such that $k_a =i_a + \sum_{b \in I^c} n_{ab}$ for $a \in I$ and $k_b =i'_b + \sum_{a \in I} n_{ab}$ for $b \in I^c$. In the formula above, $\{f_I\}_I$ is a partition of unity subordinate to the covering $\{ C_I \}_I$ of $\cU_n \backslash \Delta_n$, where $C_I$ is the open set defined by
			\begin{equation}\label{C_I_sets}
				C_I = \{ (x_1, \dots, x_n) \in \cU_n : x_a \notin J^+ (x_b) \, \forall a \in I, b \in I^c \}.
			\end{equation}
		As explained in~\citep[sec. 3.3]{HW02}, a time-ordered product $T_{n,\phi} [ \otimes_i \varphi^{k_i}(x_i)]$ which satisfies axioms (T1)-(T5) and (T9) on the whole $\cU_n$ is defined by providing an extension $\tau^H_{\phi, {\bf k}}$ on $\cU_m$ of the distribution $\tau^{H,0}_{\phi,{\bf k}}$ such that:
			\begin{enumerate}[label=(t\arabic*), start=1]
				\item\label{t_1} The distributions $\tau^H_{{\bf k}}[g, m, \phi, \lambda]$ are locally covariant: let $(M,g)$ and $(M', g')$ be two space-times, let $\iota: M' \to M$ be a causality-preserving isometric embedding, i.e. $g' = \iota^* g$, and let $f$ be a test function supported in a sufficiently small neighbourhood of the total diagonal in $(M')^n$. Then, it holds
						\begin{equation}\label{t_cov}
							\left(\iota^* \tau^H_{{\bf k}}[g, m, \phi, \lambda]\right) (f) = \tau^H_{{\bf k}}[\iota^*g, \iota^*m, \iota^*\phi, \iota^*\lambda] (f).
						\end{equation}
				\item\label{t_2} The distributions $\tau^H_{{\bf k}}[g, m, \phi, \lambda]$ scale almost homogeneously with degree $|{\bf k}|$ under the rescaling $(g, m, \phi, \lambda) \mapsto (\Lambda^{-2} g, \Lambda m, \Lambda \phi, \lambda)$, i.e. it holds
						\begin{equation*}
							\Lambda^{-d} \tau^H_{{\bf k}}[\Lambda^{-2} g, \Lambda m, \Lambda \phi, \lambda] = \tau^H_{{\bf k}}[g, m, \phi, \lambda] + \sum_{\ell = 1}^{n} \frac{\ln^\ell \Lambda}{\ell!} \beta_\ell [ g, m^2, \phi, \lambda],
						\end{equation*}
					for certain locally covariant distributions $\beta_\ell$, and for a certain $N \in \bN$.
				\item\label{t_3} It holds
						\begin{equation*}
							\WF \left( \tau^H_{{\bf k}}[g,m, \phi, \lambda] \right) \subset \cC^{T}_n[g].
						\end{equation*}
				\item\label{t_4} For any choice of smooth $1$-parameter families $\{ g^{(s)}, m^{(s)}, \phi^{(s)}, \lambda^{(s)} \}$, $\tau^H_{{\bf k}}[g^{(s)}, m^{(s)}, \phi^{(s)}, \lambda^{(s)}]$ is a distribution on $\bR \times \cU_n$ and consequently it trivially satisfies
						\begin{equation*}
							\begin{split}
								\WF \left( \tau^H_{{\bf k}}[g^{(s)},m^{(s)}, \phi^{(s)}, \lambda^{(s)}] \right) \subset &\left\{ (s, x_1, \dots, x_n; \rho, k_1, \dots, k_n) \in \dot{T}^\ast (\bR \times M^n) : \right. \\
								&\quad  \left. (x_1, \dots, x_n; k_1, \dots, k_n) \in \cC^{T}_n[g^{(s)}] \right\}.
							\end{split}
						\end{equation*}
					Less trivially, we require
						\begin{equation*}
							\left. \WF \left( \tau^H_{{\bf k}}[g^{(s)},m^{(s)}, \phi^{(s)}, \lambda^{(s)}] \right) \right|_{\bR \times \Delta_n} \perp T( \bR \times \Delta_n),
						\end{equation*}
					and, in case of a smooth variation of only the background $\phi$, we also require
						\begin{equation*}
							\WF \left( \tau^H_{{\bf k}}[g,m, \phi^{(s)}, \lambda] \right) \subset \bR \times \{0 \} \times \cC^{T}_n[g].
						\end{equation*}
				\item\label{t_5} If we assume that $\{ g^{(s)}, m^{(s)}, \phi^{(s)}, \lambda^{(s)} \}$ are analytic $1$-parameter families, then the distribution $\tau^H_{{\bf k}}[g^{(s)}, m^{(s)}, \phi^{(s)}, \lambda^{(s)}]$ varies analytically, i.e. it satisfies analogous bounds as before in~\ref{t_3}-\ref{t_4} with the smooth wave-front set replaced by the analytic wave-front set.
			\end{enumerate}
		As the notation suggest the requirements~\ref{t_1}-\ref{t_5} correspond to the axioms (T1)-(T5).\\
		
		These claims can be proved using the argument presented in~\citep{HW02}. Even though in the aforementioned paper the authors only consider the case of massless scalar field (the only dependence of the distributional coefficients $\{ \tau_{\phi,{\bf j}} \}_{\bf j}$ is on the metric $g$), the method can be adapted to the case considered here (the scalar field theory corresponding to the linear operator $\boxempty - m^2 - \frac{\lambda}{2} \phi^2$) with obvious modifications. The cornerstone of the method is the microlocal control on the Hadamard parametrix, see~\ref{subsubsec_properties_omega_H}, under smooth and analytic variation of the parameters of the theory. In the setting we are considering here, such microlocal control still holds, in the form~\eqref{H_est_s} and~\eqref{H_est_s_diag} and their analytic counterparts.\\
		For the purpose of proving prop.~\ref{prop_t_W-smooth}, we will need the following technical lemma.
		\begin{lemma}\label{lemma_tech_CT_C+_C-}
			For any $n \geq 2$, we have
				\begin{equation*}
					\cC^{T}_n[g] \subset \cC^{T;+}_n[g] \cap \cC^{T;-}_n[g],
				\end{equation*}
			where
				\begin{equation}\label{C_pm}
					\begin{split}
						\cC^{T;\pm}_n[g] := &\left\{ (x_1, \dots, x_n ; k_1, \dots, k_n) \in \dot{T}^*M^n : \mbox{ if } k_{\ell \neq i,j} \in \overline{V}^\pm \mbox{ then }\right. \\
						&\quad \left.  x_i = x_j, k_i + k_j \in \overline{V}^\mp \mbox{ or } x_i \neq x_j, k_i \in \overline{V}^\mp \mbox{ or } k_j \in  \overline{V}^\mp  \right\}.
					\end{split}
				\end{equation}
		\end{lemma}
		\begin{proof}
			We prove $\cC^{T}_n[g] \subset \cC^{T;+}_n[g]$ by induction on the number $n$ of the variables. For $n=2$ the assumption can be verified straightforwardly. We proceed assuming that the claim holds for any $n' < n$, and then we prove that the statement holds also for $n$.\\
			Let $(x_1, \dots, x_n; k_1, \dots, k_n)$ be an element of $\cC^{T}_n[g]$, corresponding to a decorated graph $\sG$. If all the points $x_1, \dots, x_n$ coincide, then the covectors $k_1, \dots, k_n$ must satisfy $\sum_i k_i = 0$ and so we have $(x_1, \dots, x_n; k_1, \dots, k_n) \in \cC^{T;+}_n[g]$.\\
			Now, if not all the points coincide, there must exists a proper subset $I \subset \{1, \dots, n\}$ such that (1) all the $x_a$ with $a \in I$ coincide and it holds $x_b \neq x_a$ for $b \in I^c$, and (2) for any $x_b$ with $b \notin I$ which is connected in the graph $\sG$ with a point $x_a$ with $a \in I$, it holds $x_b \in I^+ (x_a)$. By the definition of decorated graph, the covector $k_a$ for $a \in I$ is given by
				\begin{equation*}
					k_a =  \sum_{b \in I^c, b > a, a b \in \sG} \left. p_{a b} \right|_{x_a} - \sum_{b' \in I^c, b' <a, ab' \in \sG} \left. p_{b' a} \right|_{x_a} + \sum_{a' \in I, a' > a, aa' \in \sG} \left. p_{a a'} \right|_{x_a} - \sum_{a'' \in I, a'' <a, a''a \in \sG} \left. p_{a'' a} \right|_{x_a}.
				\end{equation*}	
		For any $b \in I^c$, we define the covector $k'_b$ by
				\begin{equation}\label{covector_new_graph}
					k'_b := k_b - \sum_{a \in I, a > b, b a \in \sG} \left. p_{b a} \right|_{x_b} + \sum_{a' \in I, a' < b, a' b \in \sG} \left. p_{a' b} \right|_{x_b}.
				\end{equation}
		Note that $k'_b - k_b \in \overline{V}^+$ because by construction the null covector field $p_{b a}$ is past directed and the null covector $p_{a' b}$ is future directed. Furthermore, we have $k'_b = k_b$ if in the graph $\sG$ the vertex $x_b$ is not connected to any vertex $x_a$ with $a \in I$.\\
		We claim that $((x_b)_{b \in I^c} ; (k'_b)_{b \in I^c})$ is in $\cC^T_{n-|I|}[g]$. In fact, it corresponds to the decorated graph obtained from $\sG$ removing the vertices $x_a$ with $a \in I$ and the edges with one of these points as source or target. This is clearly enough to prove the claim. By the inductive hypothesis, it follows that $((x_b)_{b \in I^c} ; (k'_b)_{b \in I^c})$ belongs to $\cC^{T;+}_{n-|I|}[g]$.\\
		In order to prove that $(x_1, \dots, x_n; k_1, \dots, k_n)$ belongs to $\cC^{T;+}_n[g]$, we now assume $k_{\ell} \in \overline{V}^+$ for any $\ell \neq i,j$ and we prove that if $x_i \neq x_j$, then either $k_i$ or $k_j$ is in $\overline{V}^-$, whereas we have $k_i + k_j \in \overline{V}^-$ if $x_i = x_j$. We need to consider only the following two cases: (a) $i,j \notin I$, and (b) $i \in I$, and $j \notin I$. In fact, if we assume $i,j \in I$ and $k_{\ell} \in \overline{V}^+$ for any $\ell \neq i,j$, then all covectors $k'_b$ with $b \in I^c$ belong to $\overline{V}^+$. However, this configuration violates the inductive hypothesis since $((x_b)_{b \in I^c} ; (k'_b)_{b \in I^c})$ belongs to $\cC^T_{n-|I|}[g] \subset \cC^{T;+}_{n-|I|}[g]$.
		\begin{enumerate}[label=(\alph*), start=1]
			\item The assumption $k_{\ell} \in \overline{V}^+$ for any $\ell \neq i,j$ implies that it holds $k'_{b} \in \overline{V}^+$ for any $b \in I^c$ with $b \neq i,j$ since $k'_b - k_b \in \overline{V}^+$. By the inductive hypothesis, it must hold either $k'_i \in \overline{V}^-$ or $k'_j \in \overline{V}^-$ if $x_i \neq x_j$, otherwise $k'_i + k'_j \in \overline{V}^-$ if $x_i = x_j$. Because both $k'_{i} - k_{i}$ and $k'_{j} - k_{j}$ belong to $\overline{V}^+$, it follows either $k_i \in \overline{V}^-$ or $k_j \in \overline{V}^-$ if $x_i \neq x_j$, whereas $k_i + k_j \in \overline{V}^-$ if $x_i = x_j$, as we needed to prove.
			\item By hypothesis, $x_i \neq x_j$, so we have to verify that the assumption $k_{\ell} \in \overline{V}^+$ for any $\ell \neq i,j$ implies either $k_i \in \overline{V}^-$ or $k_j \in \overline{V}^-$. It follows from this assumption that for any $b \in I^c$ with $b\neq j$ it holds $k'_{b} \in \overline{V}^+$. Because $((x_b)_{b \in I^c} ; (k'_b)_{b \in I^c}) \in \cC^{T;+}_{n-|I|}[g]$, it follows that $k'_j$ must be in $\overline{V}^-$. Because $k'_{j} - k_{j} \in \overline{V}^+$,we obtain $k_j \in \overline{V}^-$, which is enough to prove the claim.
		\end{enumerate}
		Finally, we notice that $\cC^{T}_n[g] \subset \cC^{T;-}_n[g]$ can be proved with a similar argument by time-reversal. In more detail, if $(x_1, \dots, x_n; k_1, \dots, k_n)$ is an element of $\cC^T_n[g]$ with $x_1= \dots = x_n$, then it can be proved that $(x_1, \dots, x_n; k_1, \dots, k_n) $ belongs to $\cC^{T;-}_n[g]$ arguing exactly as before. If not all the points $x_1, \dots, x_n$ coincides, we consider a subset $J \subset \{1, \dots, n\}$ such that: (1) all the points $x_a$ with $a \in J$ coincide and it holds $x_b \neq x_a$ for $b \in J^c$, and (2) for any $x_b$ with $b \notin J$ which is connected in the graph $\sG$ with a point $x_a$ with $a \in J$, it holds $x_b \in I^- (x_a)$ if $b \in J^c$. Then, we remove from the decorated graph corresponding to $(x_1, \dots, x_n; k_1, \dots, k_n)$ the vertices $x_a \in J$. Proceeding similarly as before, we obtain that $(x_1, \dots, x_n; k_1, \dots, k_n) \in \cC^{T;-}_n[g]$ as we needed to prove.
		\end{proof}
		\noindent
		As an immediate corollary of~\ref{t_3},~\ref{t_4} and this lemma, we get that
			\begin{equation}\label{WF_t_H}
				\WF \left( \tau^H_{{\bf k}}[g,m, \phi, \lambda] \right) \subset \cC^{T;+}_n[g] \cap \cC^{T;-}_n[g] \subset W_n[g],
			\end{equation}
		and for any smooth $1$-parameter family $\{\phi^{(s)}\}$
			\begin{equation}\label{WF_t_H_s}
				\WF \left( \tau^H_{{\bf k}}[g,m, \phi^{(s)}, \lambda] \right) \subset \bR \times \{0 \} \times \left( \cC^{T;+}_n[g] \cap \cC^{T;-}_n[g] \right) \subset \bR \times \{0 \} \times W_n[g].
			\end{equation}
		We used the fact $C^{T;\pm}_n \subset C^\pm_n$ which is a straightforward consequence of the definitions of the set involved~\eqref{C_pm} and, respectively,~\eqref{aux_C_sets}, and the definition of the set $W_n$~\eqref{W_set_def}.\\
		
		We continue the review of how $T_{n,\phi}(\otimes_i \varphi^{k_i}(f_i))$ is constructed. The fundamental step is to define the extension $\tau^H_{\phi, {\bf k}}$ of $\tau^{H,0}_{\phi,{\bf k}}$ to the diagonal, as also done in~\citep{HW02}.\\
		Firstly, we fix a point $x \in M$. For any sufficiently small convex normal neighbourhood $U$ of $x$ we can consider $\tau^{0,H}_{\phi, {\bf k}}$ as a distribution on the sub-manifold $\{x\} \times U^{n-1} / (x, \dots, x)$. A priori the restriction of a distribution on a sub-manifold is not well-defined. However, $\WF (\tau^{0,H}_{\phi, {\bf k}})$ does not contain elements $(x_1, \dots, x_n; k, 0, \dots, 0)$ and these elements span the co-normal bundle of $\{x\} \times U^{n-1} / (x, \dots, x)$. Then, using~\citep[thm. 8.2.4] {H83}, the restriction is well-defined.\\
		Condition~\ref{t_4}\footnote{As a matter of fact the weaker estimate where $\cC^T_n[g]$ is replaced by $\cC^{T;+}_n[g] \cap \cC^{T;-}_n[g]$ is sufficient.} implies that $\tau^{0,H}_{\phi, {\bf k}}$ can be rewritten as a Taylor expansion, the so called {\em scaling expansion} of $\tau^{0,H}_{{\bf k}}$, i.e.
			\begin{equation}\label{formula_taylor_t^0}
				\tau^{0,H}_{{\bf k}}[g,m,\phi,\lambda](x,\cdot) = \sum_{\ell =0}^{L} \frac{1}{\ell!} \theta^0_\ell [g, m, \phi, \lambda](x, \cdot)  + r^{0}_{L}[g, m, \phi, \lambda](x,\cdot),
			\end{equation}
		where
			\begin{equation}\label{taylor_t_l-th}
					\theta^0_\ell [g, m, \phi, \lambda](x, \cdot) := \left. \frac{d^\ell}{ds^\ell}\tau^{0,H}_{{\bf k}}[g^{(s)}, m^{(s)}, \phi^{(s)}, \lambda^{(s)}](x,\cdot) \right|_{s=0},
			\end{equation}
			\begin{equation}\label{taylor_t_rem}
				 	r^{0}_{L}[g, m, \phi, \lambda](x,\cdot) := \frac{1}{L!} \int_0^1 (1-s)^L  \frac{d^{L+1}}{ds^{L+1}}\tau^{0,H}_{{\bf k}}[g^{(s)}, m^{(s)}, \phi^{(s)}, \lambda^{(s)}] (x,\cdot)  ds,
			\end{equation}
		where the families $\{ g^{(s)}, m^{(s)}, \phi^{(s)}, \lambda^{(s)} \}$ are defined by
			\begin{equation}\label{family_s}
				g^{(s)} = s^{-2} \iota_s^* g, \quad  m^{(s)} = s m, \quad \phi^{(s)} = s \iota_s^* \phi, \quad \lambda^{(s)} = \iota_s^* \lambda,
			\end{equation}
		and where $\iota_s$ is the diffeomorphism in $U$ which shrinks the Riemannian coordinates with respect to $x$ by the factor $s$\footnote{For $y \in U$, described by its Riemannian coordinates $\xi_x$ with respect to $x$, $\iota_s y$ is the point corresponding to the Riemannian coordinates $s^{-1} \xi_x$ with respect to $x$.}.\\
		
		The extension is now performed using the decomposition~\eqref{formula_taylor_t^0}.  It is well-known (see e.g.~\citep[thm. 5.2]{brunetti2000microlocal}) that it is not necessary the case that the extension of a distribution to the diagonal exists nor that the extension is unique. However, in our situation, the scaling expansion guarantees the existence of an extension and, furthermore, allows a complete characterization of the non-uniqueness of such extension. The point is that for a sufficiently large $L$ we have sufficient control over the singular behaviour of $\theta^0_\ell$ with $\ell \leq L$ and of $r^{0}_{L}$ near $(x, x, \dots, x)$. More precisely, the relevant properties of $\theta^0_\ell$ and $r^0_L$ are:
		\begin{enumerate}[label=\roman*), start=1]
				\item Both $\theta^0_\ell$ and $r^0_L$ are distributions on $U^{n-1} \backslash (x, \dots, x)$ defined in terms of $g,m,\phi,\lambda$ in a local and covariant way. Namely, eq.~\eqref{t_cov} holds for any diffeomorphis $\iota$ which preserves $x$.
				\item The distribution $\theta^0_\ell$ can be expressed as
						\begin{equation}\label{formula_theo_4.1}
							\theta^0_\ell[g,m, \phi, \lambda](x, \cdot) = \sum C^{(\ell)}[g, m, \phi, \lambda, \dots](x) \cdot ((\alpha_x[g])^* u^{(\ell)0})(\cdot),
						\end{equation}
					where $\alpha_x:U \to \bR^4$ maps a point in $U$ to its Riemannian normal coordinates with respect to $x$, and where $u^{(\ell)0}$ are certain tensor valued distributions on $\bR^{4(n-1)} \backslash \{0\}$ invariant under the Lorentz transformations. Here, $C^{(\ell)}(x)$ is a sum of monomials constructed with the metric $g$, the Riemann tensor and its symmetrized covariant derivatives\footnote{We consider the covariant derivative with respect to the Levi-Civita connection of the metric $g$.}, the ``mass'' $m^2 + \lambda(x) \phi^2(x)$ and its symmetrized covariant derivatives. $C^{(\ell)}_\phi$ is required to scale homogeneously with degree $\ell$ under the rescaling of $(g, m, \phi, \lambda) \mapsto (\Lambda^{-2} g, \Lambda m, \Lambda \phi, \lambda)$. We emphasize that $\theta^0_{\ell,\phi}$ depends on the background $\phi$ only via the coefficients $C^{(\ell)}_\phi$.
				\item The distribution $u^{(\ell)0}$ scales almost homogeneously with degree $|{\bf k}| - \ell$ under rescalings of the coordinates.
				\item As distributions on $\cU_m \backslash \Delta_m$, both $\theta^0_\ell$ and $r^{0}_{L,\phi}$ scale almost homogeneously with degree $d$ under the rescaling $(g, m, \phi, \lambda) \mapsto (\Lambda^{-2} g, \Lambda m, \Lambda \phi, \lambda)$.
				\item The remainder term $r^{0}_{L}$ has a scaling degree less than $|{\bf k}|-L-1$ under rescalings of the coordinates.
			\end{enumerate}
		These properties are just the properties proved in~\citep[thm. 4.1]{HW02}, adjusted in an obvious way to reflect the presence of non-trivial $m, \phi, \lambda$ in the theory we are discussing here.\\
		
		As explained in~\citep[sec. 4.2]{HW02}, the extensions of $\theta^0$ and $r^0_L$ to the total diagonal are constructed exploiting these properties.\\
		First, we construct an extension for $\theta^0_\ell$. Making use of eq.~\eqref{formula_theo_4.1}, it will be sufficient to extend $u^{(\ell)0}$ to a Lorentz invariant distribution $u^{(\ell)}$ in $\bR^{4(n-1)}$ which scales almost homogeneously with degree $|{\bf k}| - \ell$ under rescalings of the coordinates. Such extension $u^{(\ell)}$ always exists, but it is non-unique as proved in~\citep[lemma 4.1]{HW02} making use of~\citep[thm. 5.2]{brunetti2000microlocal}. The ambiguity of the extension $u^{(\ell)}$ corresponds the {\em renormalization freedom}, which is characterized in the same reference. The extension $\theta_\ell$ of $\theta^0_\ell$ can be then constructed replacing $u^{0 (\ell)}$ with one of the possible extensions $u^{(\ell)}$ in the right-hand side of eq.~\eqref{formula_theo_4.1}.\\
		Next, we focus on the remainder term $r^0_L$. If we choose $L \geq |{\bf k}| + 4(n-1)$, then, for any $x\in M$, the distribution $r^0_{L}(x, \cdot)$ has a scaling degree less than $4(n-1) -1$ with respect to a rescaling of the coordinates. As a consequence of~\citep[thm. 5.2]{brunetti2000microlocal}, there exists a unique extension of $r^0_{L}(x, \cdot)$ with the same scaling degree for any $x \in M$. This unique extension is defined by the weak limit
			\begin{equation}\label{remainder_def}
				r_{L,\phi} (f) = \lim_{j \to \infty} r^{0}_{L,\phi} (\varrho_j f),
			\end{equation}
			where $f$ is a test function in $M^m$ with the support sufficiently close to the total diagonal $\Delta_n$, and where $\varrho_j$ is a sequence of functions with support in $U^{n} \backslash \Delta_{n}$ such that $\varrho_j$ is identically $1$ outside a neighbourhood $O_j$ of $\Delta_{n}$ with $O_j$ shrinking to $\Delta_{n}$ as $j \to \infty$. Because of the scaling properties of $r^0_{L,\phi}$, the limit exists and it does not depend on the choice of the cut-off $\varrho_j$ (see~\citep[thm. 5.2]{brunetti2000microlocal}). The desired distribution $\tau^H_{\phi, {\bf k}}$ extending $\tau^{0,H}_{\phi, {\bf k}}$ is defined as
			\begin{equation}\label{t_H_extension}
				\tau^H_{\phi,  {\bf k}} = \sum_{\ell = 0}^L \frac{1} {\ell!}\theta_{\ell, \phi} + r_{L, \phi},
			\end{equation}
		for $L \geq |{\bf k}| - 4(m-1)$. Such $\tau^H_{\phi, {\bf k}}$ satisfies the conditions~\ref{t_1}-\ref{t_5}, and, therefore, also the weaker conditions~\eqref{WF_t_H},~\eqref{WF_t_H_s}, as can be proved following the argument given in~\citep[sec. 4.3]{HW02}.
		
	\subsection{Properties of the variational derivatives of the distributional coefficients $\{\tau^H_{\phi,{\bf j}}\}_{\bf j}$}\label{subsubsec_var}
		In sec.~\ref{subsubsec_suff_cond_T-prod}, we presented two sufficient conditions for $\phi \mapsto T_{n,\phi}[\otimes_i \varphi^{k_i} (f_i)]$ to be on-shell $W$-smooth. These conditions require sufficient microlocal control on the variational derivatives of the distributions $\{ \tau_{\phi,{\bf j}} \}_{\bf j}$, i.e. the distributional coefficient of the Wick expansion with respect to the retarded $2$-point function, for $(M,g)$ ultra-static space-time with compact Cauchy surface, $m$ constant, $\phi$ a general smooth function and compactly supported coupling constant $\lambda$.\\
		As explained in the introduction, we first consider a neighbourhood $\cU_n \subset M^n$ of the total diagonal as in sec.~\ref{subsubsec_rew_T-prod}, i.e. $\cU_n$ is small enough to ensure that for any $(x_1, \dots, x_n) \in \cU_n$ it holds $x_1, \dots, x_n \in U$ for a convex normal set $U \subset M$. In $\cU_n$, we prove that the Gateaux derivatives of the distributional coefficients $\{ \tau^H_{\phi,{\bf j}} \}_{\bf j}$ of the local Wick expansion have the desired sufficient microlocal control. To do this, we enhance the inductive construction of the distribution $\tau^H_{\phi,{\bf j}}$ (governed by the conditions~\ref{t_1}-\ref{t_5}) to constrain the variational derivatives $\delta^\nu \tau^H_{\phi, {\bf j}} / \delta \phi^\nu$. In detail, we demand the additional conditions for any $\nu$:
			\begin{enumerate}[label=($\delta$t\arabic*), start=0]
					\item\label{supp_delta_t_H}
						We require
							\begin{equation*}
									\left. \supp \left( \delta^\nu \tau^H_{\phi, {\bf j}} (x_1, \dots, x_n) / \delta \phi(y_1) \dots \delta \phi (y_\nu) \right) \right|_{	(x_1, \dots, x_n) \in \Delta_n} \subset \Delta_{n+\nu},
							\end{equation*}
							and
							\begin{equation*}
									\left. \supp \left( \delta^\nu \tau^H_{\phi, {\bf j}} (x_1, \dots, x_n) / \delta \phi(y_1) \dots \delta \phi (y_\nu) \right) \right|_{	(x_1, \dots, x_n) \in U^n} \subset U^{n+\nu},
							\end{equation*}
						where $U$ is a normal convex subset of $M$ sufficiently small that we can apply lemma~\ref{lemma_H}.
					\item\label{cov_delta_t_H} We demand that $\delta^\nu \tau^H_{{\bf j}}/ \delta \phi^\nu [g, m, \phi, \lambda]$ is a locally covariant distribution. Let $(M,g)$ and $(M', g')$ be two ultra-static space-times with compact Cauchy surfaces and let $\iota: M' \to M$ be a causality-preserving isometric embedding, i.e. $g' = \iota^* g$. Then, for any test function $f$ supported in a sufficiently small neighbourhood of the total diagonal in $(M')^{n+\nu}$ it holds that
						\begin{equation*}
							\left(\iota^* \frac{\delta^\nu \tau^H_{{\bf j}}}{\delta \phi^\nu}[g, m, \phi, \lambda] \right) (f) = \frac{\delta^\nu \tau^H_{{\bf j}}}{\delta \phi^\nu}[\iota^*g, \iota^*m, \iota^*\phi, \iota^*\lambda] (f).
						\end{equation*}
					\item\label{scaling_delta_t_H} The distribution $\delta^\nu \tau^H_{{\bf j}} / \delta \phi^\nu [g,m,\phi,\lambda]$ scales almost homogeneously with degree $|{\bf j}| + 3 \nu +4$ under the rescaling $(g, m, \phi, \lambda) \mapsto (\Lambda^{-2} g, \Lambda m, \Lambda \phi, \lambda)$.
					\item\label{WF_delta_t_H} It holds
							\begin{equation}\label{est_var_delta_t_H}
								\WF \left( \delta^\nu \tau^H_{{\bf j}}/\delta \phi^\nu [g,m, \phi, \lambda] \right) \subset \cC^{\delta; +}_{n,\nu}[g] \cap \cC^{\delta; -}_{n,\nu}[g],
							\end{equation}
						where
							\begin{equation*}
								\begin{split}
									\cC^{\delta; \pm}_{n,\nu}[g]:= &\left\{ (x_1, \dots, x_n, y_1, \dots, y_\nu ; k_1, \dots, k_n, p_1, \dots, p_\nu) \in \dot{T}^*M^{n+\nu} : \right. \\
										&\quad \left. \mbox{if } p_{r \neq s} \in \overline{V}_{y_r}^\pm, p_s \mbox{ space-like}, \mbox{ then } (k_1, \dots, k_n) \notin (\overline{V}^\pm)^n \right.\\
										&\quad \left. \mbox{if } p_{r} \in \overline{V}_{y_r}^\pm, \mbox{ then } (x_1, \dots, x_n; k_1, \dots, k_n) \in \cC^{T;\pm}_n[g]\right\},
								\end{split}
							\end{equation*}
						with $\cC^{T;\pm}_n[g]$ defined by eq.~\eqref{C_pm}. Furthermore, we require
							\begin{equation*}
								\left. \WF \left( \delta^\nu \tau^H_{{\bf j}} / \delta \phi^\nu \right) \right|_{\Delta_{n+\nu}} \perp T(\Delta_{n+\nu}).
							\end{equation*}
					\item\label{WF_delta_t_H_s} For any smooth $1$-parameter families $\{ g^{(s)}, m^{(s)}, \phi^{(s)}, \lambda^{(s)} \}$, the quantity $\delta^\nu \tau^H_{{\bf j}}/\delta \phi^\nu [g^{(s)}, m^{(s)}, \phi^{(s)}, \lambda^{(s)}]$ is a distribution on $\bR \times U^{n+\nu}$ and consequently it trivially satisfies
							\begin{equation}\label{est_var_delta_t_H_s}
								\begin{split}
									&\WF \left( \delta^\nu \tau^H_{{\bf j}}/\delta \phi^\nu [g^{(s)}, m^{(s)}, \phi^{(s)}, \lambda^{(s)}] \right) \subset \\
									&\qquad \subset \left\{ (s, x_1, \dots, x_m, y_1, \dots, y_\nu ; \rho, k_1, \dots, k_m, p_1, \dots, p_\nu) \in \dot{T}^\ast (\bR \times M^{n+\nu}) : \right. \\
									&\qquad \qquad  \left. (x_1, \dots, x_m, y_1, \dots, y_\nu; k_1, \dots, k_m, p_1, \dots, p_\nu) \in \cC^{\delta; +}_{n,\nu}[g^{(s)}] \cap \cC^{\delta; -}_{n,\nu}[g^{(s)}] \right\}.
								\end{split}
							\end{equation}
						Less trivially, we require
							\begin{equation*}
								\left. \WF \left( \delta^\nu \tau^H_{{\bf j}}[g^{(s)},m^{(s)}, \phi^{(s)}, \lambda^{(s)}] / \delta \phi^\nu \right) \right|_{\bR \times \Delta_{n+\nu}} \perp T( \bR \times \Delta_{n+\nu}),
							\end{equation*}
						and, in case of a smooth variation only of the background $\phi$, we require
							\begin{equation}\label{est_var_delta_t_H_s_phi}
								\WF \left( \delta^\nu \tau^H_{{\bf j}}/\delta \phi^\nu [g,m, \phi^{(s)}, \lambda] \right) \subset \bR \times \{0 \} \times \left( \cC^{\delta; +}_{n,\nu}[g] \cap \cC^{\delta; -}_{n,\nu}[g] \right).
							\end{equation}
				\end{enumerate}
			\noindent
			As one can see, properties~\ref{WF_delta_t_H} and~\ref{WF_delta_t_H_s} give a microlocal control on the variational derivatives of $\tau^H_{\phi,{\bf j}}$. On the other hand, at this stage it does not seem clear why we have to impose also the other properties. We will see that these extra properties play an important role in the proof.\\
			
			To prove these properties, we follow the inductive construction of $\tau^H_{\phi, {\bf j}}$. The induction counter is $n$, the number $n$ of the factors $\varphi^{j_1}(f_1), \dots, \varphi^{j_n}(f_n)$ in the time-ordered product. For $n=1$, i.e. for the Wick product, a direct inspection of eq.~\eqref{Wick_exp_state} reveals that $\tau^H_{\phi, {\bf j}} =1$ and so the properties~\ref{supp_delta_t_H}-\ref{WF_delta_t_H_s} hold trivially.\\
			We assume that the variational derivatives of $\tau^H_{\phi, {\bf j}'}$ for any ${\bf j}'$ corresponding to time-ordered products with less than $n$ factors exist and satisfy properties~\ref{supp_delta_t_H}-\ref{WF_delta_t_H_s}. We prove that this is also the case for the variational derivatives of $\tau^H_{\phi, {\bf j}}$ for any ${\bf j}$ corresponding to time-ordered products with $n$ factors.\\
			As outlined in sec.~\ref{subsubsec_rew_T-prod}, for a fixed ${\bf j}$ the distribution $\tau^H_{\phi, {\bf j}}$ is an extension to the total diagonal of the distributional coefficients $\tau^{H,0}_{\phi, {\bf j}}$ of the local Wick expansion of $T_{n,\phi}[\otimes_i \varphi^{j_i} (x_i)]$ in $\cU_n \backslash \Delta_n$. In particular, after fixing a point $x \in M$, the distribution $\tau^{H,0}_{\phi, {\bf j}}(x,\cdot)$ can be expressed as in formula~\eqref{formula_taylor_t^0} for $L \geq |{\bf j}| - 4(n-1)$ in terms of the distributions $\theta_{\ell, \phi}^0(x, \cdot)$, given by~\eqref{taylor_t_l-th}, and the remainder $r^0_{\phi, L}(x, \cdot)$ given by~\eqref{taylor_t_rem}. Using eq.~\eqref{formula_theo_4.1}, we can express $\theta_{\ell, \phi}^0(x, \cdot)$ in terms of the distributions $u^{(\ell) 0} $ and the tensor fields $C^{(\ell)}_\phi$. The extension $\tau^H_{\phi, {\bf j}}$ is constructed as in eq.~\eqref{t_H_extension} providing an extension for each $u^{(\ell)0}$, for which there is not an unique choice, and for the remainder $r^0_{\phi, L}(x, \cdot)$, for which there is a unique choice since it scales with degree less than $4(n-1)$ under rescaling of the coordinates.\\
			We note that the distributions $u^{(\ell) 0}$ do not depend on the background $\phi \in C^\infty(M)$, therefore the {\em same} $u^{(\ell)0}$ appears in the factorizations of $\tau^{0,H}_{\phi, {\bf j}}$ for different background $\phi$. We can choose the extension $u^{(\ell)}$ independently of the background $\phi \in C^\infty(M)$. So, we have
				\begin{equation}\label{delta_taylor_ext}
					\begin{split}
						 &\frac{\delta^\nu \tau^H_{\phi, {\bf j}}(x, x_2, \dots, x_n)}{\delta \phi(y_1) \dots \delta \phi(y_\nu)} =\\
						 &\quad = \sum_{\ell =0}^{L} \frac{1}{\ell!} \left( \sum  \frac{\delta^\nu C_\phi^{(\ell)}(x)}{\delta \phi(y_1) \dots \delta \phi(y_\nu)} \cdot ((\alpha_x)^* u^{(\ell)})(x_2, \dots, x_n) \right)  + \frac{\delta^\nu r_{\phi, L}(x,x_2, \dots, x_n)}{\delta \phi(y_1) \dots \delta \phi(y_\nu)},
					\end{split}
				\end{equation}
			where $L$ is chosen again to be greater than $|{\bf j}| + 4(n -1)$.\\
			We aim to prove that the right-hand side of eq.~\eqref{delta_taylor_ext} is a well-defined distribution in $U^{n+\nu}$ which satisfies the properties~\ref{supp_delta_t_H}-\ref{WF_delta_t_H_s}. To do so, we proceed by the following three steps:
				\begin{enumerate}
					\item First, we investigate the properties satisfied by $\delta^\nu \tau^{H,0}_{\phi, {\bf j}}/ \delta \phi^\nu$, where $\tau^{H,0}_{\phi, {\bf j}}$ denotes $\tau^{H}_{\phi, {\bf j}}$ outside the diagonal $\Delta_n$.
					\item Then, we prove that $\delta^\nu \tau^H_{\phi, {\bf j}}(x, \cdot) / \delta \phi^\nu(\cdot)$ is an extension to the total diagonal of $\delta^\nu \tau^{H,0}_{\phi, {\bf j}}(x, \cdot) / \delta \phi^\nu(\cdot)$ once a point $x \in U$ is fixed. 
					\item Finally, we show that the desired properties~\ref{supp_delta_t_H}-\ref{WF_delta_t_H_s} hold for $\delta^\nu \tau^H_{\phi, {\bf j}}(x, \cdot) / \delta \phi^\nu(\cdot)$ exploiting the step 2.
				\end{enumerate}
			The logic behind this argument is the same adopted in~\citep{HW02}, and reviewed in sec.~\ref{subsubsec_rew_T-prod}, to prove that the extension $r_{L, \phi}$ of the remainder term $r^{0}_{L, \phi}$ exists, is unique, and satisfies the properties~\ref{t_1}-\ref{t_5}.
			\paragraph{Step 1.}
			We begin by proving the following result for $\tau^{H,0}_{\phi, {\bf j}}$:
			\begin{lemma}\label{lemma_W_t_H_0}
				For any ${\bf j}$ and any $\nu$, the distribution $\delta^\nu \tau^{H,0}_{\phi, {\bf j}} / \delta \phi^\nu$ is well-defined in $U^{n+\nu} \backslash \Delta_{n+\nu}$ and, in this domain, satisfies conditions~\ref{supp_delta_t_H}-\ref{WF_delta_t_H_s}.
			\end{lemma}
			\begin{proof}
					As discussed in sec.~\ref{subsubsec_rew_T-prod}, $\tau^{H,0}_{\phi, {\bf j}}$, as a distribution on $\cU_n \backslash \Delta_n$, can be expressed as a finite sum of products of $\tau^{H}_{\phi, {\bf i}}$ corresponding to time-ordered products with less than $n$ factors and powers of the Hadamard parametrix, see~\eqref{gen_term_T8_H}. Then, we compute $\delta^\nu \tau^{H,0}_\phi / \delta \phi^\nu$ by distributing the variational derivatives on each factor of the terms~\eqref{gen_term_T8_H}. It follows that $\delta^\nu \tau^{H,0}_{\phi, {\bf j}}(x_1, \dots, x_n) / \delta \phi(y_1) \dots \delta \phi(y_\nu)$ is a finite sum of terms in the form
					\begin{equation}\label{x_var_term}
						f_I(x_1, \dots, x_n) \frac{\delta^{|N_1|}\tau^{H}_{\phi, {\bf i}} ((x_a)_{a \in I})}{\delta \phi^{|N_1|} ((y_{r})_{r \in N_1})} \frac{\delta^{|N_2|} \tau^{H}_{\phi, {\bf i}'} ((x_b)_{bl \in I^c})}{\delta \phi^{|N_2|}((y_{r})_{r \in N_2})} \prod_{a\in I, b \in I^c} \prod_{v =1}^{n_{ab}}  \frac{\delta^{|N_{a,b,v}|} H_\phi(x_{a}, x_{b})}{\delta \phi^{|N_{a,b,v}|}((y_{r})_{r \in N_{a,b,v}})}.
					\end{equation}
				where $I$ is a proper subset of $\{1, \dots, n\}$, where $f_I$ is as in~\eqref{gen_term_T8_H}, and where $N_1$, $N_2$, $\{N_{a,b,v}\}_{a \in I, b \in I^c, v \leq n_{ab}}$ are a partition of $\{1, \dots, \nu\}$ made of disjoint subsets. Now, since $\tau^{H}_{\phi, {\bf i}} ((x_a)_{a \in I})$ and $\tau^{H}_{\phi, {\bf i}'} ((x_b)_{bl \in I^c})$ correspond to time-ordered products with less than $n$ factors, the variational derivatives of these two distributions satisfy the properties~\ref{supp_delta_t_H}-\ref{WF_delta_t_H_s} by the inductive hypothesis.\\
				We proceed proving that each terms in the form~\eqref{x_var_term} satisfies the conditions~\ref{supp_delta_t_H}-\ref{WF_delta_t_H_s}. This is clearly sufficient to imply that also $\delta^\nu \tau^{H,0}_\phi / \delta \phi^\nu$ does.
				\begin{enumerate}[label=($\delta$t\arabic*), start=0] 
					\item Outside the diagonal, this condition requires that: (1) if $(x_1, \dots, x_n, y_1, \dots, y_\nu)$ is in the support of the distribution~\eqref{x_var_term} where $(x_1, \dots, x_n)$ belongs to $U^n$, then $(x_1, \dots, x_n, y_1, \dots, y_\nu)$ belongs to $U^{n+\nu } \backslash \Delta_{n+\nu}$, and (2) there is no element $(x_1, \dots, x_n, y_1, \dots, y_\nu)$ in the support of the distribution~\eqref{x_var_term} such that $(x_1, \dots, x_n)$ belongs to $\Delta_n$.\\
					The requirement (2) holds because the functions $\{ f_I \}_I$ for a partition of unity subordinate to the covering $\{C_I\}_I$ defined by~\eqref{C_I_sets}, and so it holds $\supp f_I \cap \Delta_n = \emptyset$ by construction.\\
					To prove that requirement (1) is satisfied, we notice that $\delta^{|N_1|}\tau^{H}_{\phi, {\bf i}} / \delta \phi^{|N_1|}$ and $\delta^{|N_2|} \tau^{H}_{\phi, {\bf i}'}/\delta \phi^{|N_2|}$ satisfies condition~\ref{supp_delta_t_H} by the inductive hypothesis. As proved in lemma~\ref{lemma_H}, for $x_1, x_2 \in U$, $\delta^\nu H_\phi(x_1,x_2) / \delta \phi(y_1) \dots \delta \phi(y_\nu)$ identically vanishes if $y_1, \dots, y_\nu$ do not belongs to the unique geodesic segment connecting $x_1,x_2$. Since $U$ is a normal convex set, $y_1, \dots, y_\nu$ must belong to $U$. It follows that each of the terms in the form~\eqref{x_var_term} satisfies (1) and this concludes the proof of condition~\ref{supp_delta_t_H}.
					\item Since $\delta^{|N_1|}\tau^{H}_{{\bf i}}[g,m,\phi,\lambda] / \delta \phi^{|N_1|}$ and $\delta^{|N_2|} \tau^{H}_{{\bf i}'}[g,m,\phi,\lambda]/\delta \phi^{|N_2|}$ are locally covariant by hypothesis, the claim follows from the fact that $\delta^\nu H[g,m,\phi,\lambda] / \delta \phi^\nu$ is locally covariant as shown in lemma~\ref{lemma_H}.
				\end{enumerate}
				\begin{enumerate}[label=($\delta$t\arabic*), start=3]
					\item The wave-front set calculus implies that an element $(x_1, \dots, x_n, y_1, \dots, y_\nu; k_1, \dots, k_n, p_1, \dots, p_\nu)$ of the wave-front set of~\eqref{x_var_term} necessarily satisfies the following requirements:
				\begin{equation}\label{covectors_k}
					k_{a \in I} = k^I_a + \sum_{b \in I^c} \sum_{v \leq n_{ab}} k^L_{a,b,v}, \qquad k_{b \in I^c} = k^{I^c}_b + \sum_{a \in I} \sum_{v \leq n_{ab}} k^R_{a,b,v},
				\end{equation}
			and
				\begin{equation}\label{covectors_k_2}
					\left\{ \begin{array}{l}
						(x_a, x_b, (y_r)_{r \in N_{a,b,v}} ; k^L_{a,b,v}, k^R_{a,b,v}, (p_r)_{r \in N_{a,b,v}}) \in \WF ( \delta^{|N_{a,b,v}|} H_\phi / \delta \phi^{|N_{a,b,v}|}) \\
						\qquad \qquad \mbox{ or } k^L_{a,b,v}, k^R_{a,b,v}, p_r = 0,\\
						((x_a)_{a \in I}, (y_r)_{r \in N_1} ; (k^I_a)_{a \in I}, (p_r)_{r \in N_1}) \in \WF ( \delta^{|N_1|} \tau^H_{\phi, {\bf i}} / \delta\phi^{|N_1|} ) \\
						\qquad \qquad \mbox{ or } k^I_a, p_r =0,\\
						((x_b)_{b \in I^c}, (y_r)_{r \in N_2} ; (k^{I^c}_b)_{b \in I^c}, (p_r)_{r \in N_2}) \in \WF ( \delta^{|N_2|} \tau^H_{\phi, {\bf i}'} / \delta \phi^{|N_2|}) \\
						\qquad \qquad \mbox{ or } k^{I^c}_b, p_r = 0.
					\end{array} \right.
				\end{equation}
			We prove now that $(x_1, \dots, x_n, y_1, \dots, y_\nu; k_1, \dots, k_n, p_1, \dots, p_\nu)$ is contained in $\cC^{\delta; +}_{n,\nu}[g]$. By definition, we need to show that the following two requirements are satisfied: (a) if it holds $k_{\ell} \in \overline{V}^+$ for any $\ell \neq i,j$ and $p_r \in \overline{V}^+$ for and any $r$, then it holds $k_i \in \overline{V}^-$ or $k_j \in \overline{V}^-$ whenever $x_i \neq x_j$, while it holds $k_i + k_j \in \overline{V}^-$ in case $x_i = x_j$, and (b) if there exists $s$ such that $p_s$ is space-like and $p_{r}$ is $\overline{V}^+$ for any $r \neq s$, then not all $k_1, \dots, k_n$ are in $\overline{V}^+$.
				\begin{enumerate}[label=(\alph*), start=1]
					\item Because of estimate~\eqref{var_H_est} and because $p_r$ is in $\overline{V}^+$ for any $r$ by hypothesis, we have $k^R_{a,b,v} \in \overline{V}^-$ for any $a,b,v$. We distinguish three possibilities, namely $i,j \in I$, $i \in I$ and $j \notin I$, and $i,j \notin I$.\\
					In the first situation, it holds $k_b \in \overline{V}^+$ for all $b \in I^c$ by hypothesis. As we already mentioned, it holds $k^R_{a,b,v} \in \overline{V}^-$ for any $a,b,v$. Therefore, it follows from eq.~\eqref{covectors_k} that $k_b^{I^c}$ must be in $\overline{V}^+$ for all $b \in I^c$. However, these configurations are incompatible with the fact that $\tau^H_{\phi, {\bf i}'}$ satisfies the estimate~\eqref{est_var_delta_t_H} by hypothesis. So we cannot assume $k_{\ell},p_r \in \overline{V}^+$ for any $r$ and for any $\ell \neq i,j$ with $i,j \in I$.\\
					If $i \in I$ and $j \notin I$, then we have $k_{b} \in \overline{V}^+$ for any $b \in I^c$ such that $b \neq j$ by hypothesis. Exploiting again eq.~\eqref{covectors_k} and the fact that $k^R_{a,b,v}$ is in $\overline{V}^-$ for all possible $a,b,v$, we have $k^{I^c}_{b} \in \overline{V}^+$ for any $b \in I^c$ such that $b  \neq j$. Since the estimate~\eqref{est_var_delta_t_H} holds for $\tau^H_{\phi, {\bf i}'}$ by hypothesis, it follows that $k^{I^c}_{j}$ must belong to $\overline{V}^-$. Because $k^R_{a,j,v}$ is in $ \overline{V}^+$ for all $a$ and $v$, implies that $k_j$, defined via eq.~\eqref{covectors_k}, is in $\overline{V}^+$ and this verifies the requirements (a).\\
					Finally, if $i,j \notin I$, then we have $k_{b} \in \overline{V}^+$ for any $b \in I^c$ such that $b \neq i,j$ by hypothesis. Therefore, it must also hold $k^{I^c}_{b} \in \overline{V}^+$, defined via eq.~\eqref{covectors_k}, for any $b \in I^c$ such that $b \neq i,j$ because $k^R_{a,b,v}$ is in $\overline{V}^-$ for all possible $a,b,v$. In this case, the estimate~\eqref{est_var_delta_t_H} of $\tau^H_{\phi, {\bf i}'}$ implies that whenever $x_i \neq x_j$ it holds $k^{I^c}_i \in \overline{V}^-$ or $k^{I^c}_j \in \overline{V}^-$, while in case $x_i = x_j$ it holds $k^{I^c}_i + k^{I^c}_j \in \overline{V}^-$. Using again the fact that $k^R_{a,b,v} \in \overline{V}^-$ for all possible $a,b,v$ and eq.~\eqref{covectors_k}, we obtain $k_i \in \overline{V}^-$ or $k_j \in \overline{V}^-$ whenever $x_i \neq x_j$, and $k_i + k_j \in \overline{V}^-$ if $x_i = x_j$, which are precisely the requirements we need to prove.
					\item Let us assume $p_{r} \in \overline{V}^+$ for any $r \neq s$ and $p_s$ space-like. We distinguish the following cases: the index $s$ of the unique space-like covector $p_s$ is either in $N_2 \cup N_1$ or there exists $a',b',v'$ such that $s \in N_{a',b', v'}$. In both cases, we prove by reductio ad absurdum that we cannot assume $k_\ell \in \overline{V}^+$ for all $\ell$.\\
					Assume $s \in N_1 \cup N_2$. Because of estimate~\eqref{var_H_est} and because $p_r$ is in $\overline{V}^+$ for and any $r \neq s$, we have $k^R_{a,b,v} \in \overline{V}^-$ for all possible $a,b,v$. If we assume $k_1, \dots, k_n \in \overline{V}^+$, then eq.~\eqref{covectors_k} implies that $k^{I^c}_b$ must belong to $\overline{V}^+$ for any $b \in I^c$. However, these configurations are incompatible with the fact that $\tau^H_{\phi, {\bf i}'}$ satisfies estimate~\eqref{est_var_delta_t_H}, as we wanted to show.\\
					Next, we assume that there exist $a',b',v'$ such that $s \in N_{a',b', v'}$. Because of estimate~\eqref{var_H_est} and because we assume $p_s$ space-like whereas $p_r \in \overline{V}^+$ for any $r \neq s$, then we must have $k^R_{a,b,\lambda} \in \overline{V}^-$ for any $(a,b,v) \neq (a',b',v')$ and $k^R_{a',b',v'} \notin \overline{V}^+$.  If we assume $k_1, \dots, k_n \in \overline{V}^+$, then we get $k^{I^c}_{b} \in \overline{V}^+$ for any $b \in I^c$ such that $b \neq b'$ and $k^{I^c}_{b'} \notin \overline{V}^-$, because $k^{I^c}_{b'}$ satisfies the following equation
						\begin{equation*}
							k^{I^c}_{b'} = k_{b'} - \sum_{(a,v) \neq (a',v')} k^R_{a,b',v} - k^R_{a',b',v'}.
						\end{equation*}
					However, these configurations are incompatible with fact that $\tau^H_{\phi, {\bf i}'}$ satisfies estimate~\eqref{est_var_delta_t_H}, as we wanted to show.
				\end{enumerate}
			With a similar argument we can prove that any element $(x_1, \dots, x_n, y_1, \dots, y_\nu; k_1, \dots, k_n, p_1, \dots, p_\nu)$ of the wave-front set of distribution~\eqref{x_var_term} is contained in $\cC^{\delta; - }_{n, \nu}[g]$, i.e. that the following requirements are satisfied: (a) if $k_{\ell} \in \overline{V}^-$ for all $\ell \neq i,j$ and $p_r \in \overline{V}^-$ for all $r$, then it holds $k_i \in \overline{V}^+$ or $k_j \in \overline{V}^+$ whenever $x_i \neq x_j$, while it holds $k_i + k_j \in \overline{V}^+$ in case $x_i = x_j$, and (b) if there exists $p_s$ space-like and $p_{r}$ is in $\overline{V}^-$ for all $r \neq s$, then not all $k_1, \dots, k_n$ are in $\overline{V}^-$. For this purpose, we use the estimate~\eqref{var_H_est} to derive constraints for $k^L_{a,b,v}$ (whereas in the argument just presented we used the same estimate for $k^R_{a,b,v}$), while eq.~\eqref{covectors_k} and the conditions~\eqref{covectors_k_2} are used to derives constraints on $k^{I}_a$ and $k_a$ for $a \in I$ (whereas in the argument just presented we focused on $k^{I^c}_b$ and $k_b$ for $b \in I^c$).\\
			Putting everything together, it follows that each term~\eqref{x_var_term} satisfies condition~\ref{WF_delta_t_H}.
		\item This condition can be verified similarly as just done for condition~\ref{WF_delta_t_H}. In fact, for any smooth function $\bR \ni \epsilon \mapsto \phi(\epsilon) \in C^\infty(M)$, the distribution $\delta^\nu \tau^{H, 0}_{\phi(\epsilon) {\bf j}} / \delta \phi( y_1) \dots \delta \phi(y_\nu)$ is a finite sum of terms in the form~\eqref{x_var_term} with the only difference that $\phi$ is replaced by $\phi(\epsilon)$ everywhere. We can prove estimate~\eqref{est_var_delta_t_H_s_phi} adapting the argument we just presented to prove~\ref{WF_delta_t_H}: now we make use of estimate~\eqref{var_H_est_s_phi} of lemma~\ref{lemma_H} for $\delta^\nu H_{\phi(\epsilon)} / \delta \phi^\nu$, and the fact that for any ${\bf i}$ and any $\nu$ the quantity $\delta^\nu \tau^H_{\phi(\epsilon), {\bf i}} / \delta \phi^\nu$ corresponding to a time-ordered product less than $n$ factors satisfies estimates~\eqref{est_var_delta_t_H_s_phi} by the inductive hypothesis.
		\end{enumerate}
		\begin{enumerate}[label=($\delta$t\arabic*), start=2] 
			\item We assume first that $(M,g)$ is a real analytic space-time and $m, \phi, \lambda$ are analytic. We know from lemma~\ref{lemma_H} that for any $\nu$ the distribution $\delta^\nu H_\phi / \delta \phi^\nu$ is a locally covariant distribution which scales almost homogeneously with degree $2 + 3\nu$ under the rescaling $(g,m,\phi,\lambda) \mapsto (\Lambda^{-2} g, \Lambda m, \Lambda \phi, \lambda)$. By the inductive hypothesis, all the distributions $\delta^\nu \tau^{H}_{\phi, {\bf i}} / \delta \phi^\nu$ corresponding to time-ordered products with less than $n$ factors are locally covariant distributions which scale almost homogeneously with degree $|{\bf i}| + 3 \nu$ under the rescaling $(g,m,\phi,\lambda) \mapsto (\Lambda^{-2} g, \Lambda m, \Lambda \phi, \lambda)$. Thus, formula~\eqref{x_var_term} implies that $\delta^\nu \tau^{H,0}_{\phi, {\bf j}} / \delta \phi^\nu$ must be a locally covariant distribution which scales almost homogeneously with degree $\sum_\ell j_\ell + 3 \nu$ under the rescaling $(g,m,\phi,\lambda) \mapsto (\Lambda^{-2} g, \Lambda m, \Lambda \phi, \lambda)$.\\
					To extend the validity of the almost homogeneous scaling in the more general smooth case, we notice that we can approximate any arbitrary smooth metric in the neighbourhood $U \Subset M$ of $x$ by a sequence $\{ q^{(\ell)} \}_{\ell \in \bN}$ of real analytic metrics (see~\citep{HW02}[proof of thm. 4.1]), and similarly we approximate the smooth functions $m, \phi, \lambda$ in $U$ by sequences of real analytic functions. More precisely, we mean that
				\begin{equation*}
					\sup_{x \in U} \left| \nabla_{\alpha_1} \dots \nabla_{\alpha_\ell} g_{\mu \nu}(x) - \nabla_{\alpha_1} \dots \nabla_{\alpha_\ell} q^{(\ell)}_{\mu \nu}(x) \right|_{e} < 2^{-n}, \quad \forall \ell \leq n.
				\end{equation*}
			Similar bounds hold for $m, \phi, \lambda$ and their approximations via sequences of real analytic functions. We consider a generic symmetric smooth function $\psi: \bR \to [0,1]$ supported in $[-1,1]$ which satisfies in addition $1- \psi(x) = \psi(1-x)$ for all $x \in [0,1]$. We define a smooth family $\{h^{(s)} \}$ of metrics by setting $h^{(0)} := g$ and $h^{(s)} := \sum_n \psi( |1/s| - n) q^{(n)}$. We proceed similarly for $m, \phi, \lambda$. The almost homogeneous scaling holds for all $s \neq 0$ and, therefore, by the smoothness properties of $\delta^\nu \tau^{H,0}_{\phi, {\bf k}} / \delta \phi^\nu$, it continues to hold for $s=0$.
		\end{enumerate}
		This concludes the proof of lemma~\ref{lemma_W_t_H_0}.
		\end{proof}
		
		\paragraph{Step 2.}
		Let us fix $x \in U$. We can make use of the Taylor expansion~\eqref{formula_taylor_t^0} and the formula~\eqref{formula_theo_4.1}, to write $\delta^\nu \tau^{H,0}_{\phi, {\bf j}}(x, \cdot) / \delta \phi^\nu$ as
			\begin{equation}\label{delta_taylor}
					\begin{split}
						 &\frac{\delta^\nu \tau^{H,0}_{\phi, {\bf j}}(x, x_2, \dots, x_n)}{\delta \phi(y_1) \dots \delta \phi(y_\nu)} =\\
						 &\quad = \sum_{\ell =0}^{L} \frac{1}{\ell!} \left( \sum  \frac{\delta^\nu C_\phi^{(\ell)}(x)}{\delta \phi(y_1) \dots \delta \phi(y_\nu)} \cdot ((\alpha_x)^* u^{0 (\ell)})(x_2, \dots, x_n) \right)  + \frac{\delta^\nu r^0_{\phi, L}(x,x_2, \dots, x_n)}{\delta \phi(y_1) \dots \delta \phi(y_\nu)},
					\end{split}
				\end{equation}
			where $C_\phi^{(\ell)}(x)$, $u^{0 (\ell)}$ are as in eq.~\eqref{formula_theo_4.1}, and where $r^0_{\phi, L}(x,\cdot)$ is as in eq.~\eqref{taylor_t_rem}.\\
			By construction, $C^{(\ell)}_\phi$ is a sum of monomials constructed from the metric $g(x)$, the Riemann tensor in $x$ and its symmetrized covariant derivatives, the functions $m,\lambda, \phi$ in $x$ and their covariant derivatives. Thus, the distribution $\delta^\nu C^{(\ell)}_\phi(x) / \delta \phi^\nu$ exists and it is given by derivatives of the delta distribution $\delta(x,y_1, \dots, y_\nu)$ multiplied by sums of monomials in the same form as before.\\
			Since both the distributions $\delta^\nu \tau^{H,0}_{\phi, {\bf j}}(x, \cdot) / \delta \phi^\nu (\cdot)$ and $(C_\phi^{(\ell)}(x)/ \delta \phi^\nu(\cdot)) \cdot ((\alpha_x)^* u^{0 (\ell)})(\cdot)$ are well-defined in $U^{n+\nu -1} \backslash (x, \dots, x)$, it follows straightforwardly that this is true also for $\delta^\nu r^0_{\phi, L}(x,\cdot)/\delta \phi^\nu(\cdot)$.\\
			We consider the $1$-parameter families $\{g^{(s)}, m^{(s)}, \phi^{(s)}, \lambda^{(s)}\}$ defined by~\eqref{family_s}. As a corollary of lemma~\ref{lemma_W_t_H_0}, in particular, because of property~\ref{WF_delta_t_H_s}, we know that 
			\begin{equation*}
				\frac{\delta^\nu \tau^{H,0}_{{\bf k}}[g^{(s)}, m^{(s)}, \phi^{(s)}, \lambda^{(s)}]}{\delta \phi^\nu(\cdot)}(x, \cdot)
			\end{equation*}
		can be interpreted as a family of distributions on $U^{m+\nu-1} \backslash (x, \dots, x)$ parametrized by $(s,x)$. When smeared with a test function $f$, $\delta^\nu \tau^{H,0}_{{\bf k}}[g^{(s)}, m^{(s)}, \phi^{(s)}, \lambda^{(s)}]/ \delta \phi^\nu (x, f)$ is smooth in $(s,x)$ as a consequence of the wave-front set calculus (thm.~\ref{theo_WF_horma}) and the estimate~\eqref{est_var_delta_t_H_s} of property~\ref{WF_delta_t_H_s}. The following equations are consequences of this result:
			\begin{equation}\label{var_taylor_0}
				\begin{split}
					\sum  \frac{\delta^\nu C_\phi^{(\ell)}(x)}{\delta \phi^\nu(\cdot)} \cdot ((\alpha_x)^* u^{0 (\ell)})(\cdot) &= \frac{\delta^\nu \theta^0_\ell(x, \cdot)}{\delta \phi^\nu(\cdot)} = \left. \frac{d^\ell}{ds^\ell} \frac{\delta^\nu \tau^{0,H}_{\bf j} [g^{(s)}, m^{(s)}, \phi^{(s)}, \lambda^{(s)}](x, \cdot)}{\delta \phi^\nu(\cdot)} \right|_{s=0},
				\end{split}
			\end{equation}
		and
			\begin{equation}\label{var_rem_0}
				\begin{split}
					&\frac{\delta^\nu r^0_{L, \phi} (x,\cdot)}{\delta \phi^\nu(\cdot)} = \frac{1}{L!} \int_0^1 (1-s)^L \frac{d^{L+1}}{ds^{L+1}} \frac{\delta^\nu \tau^{0,H}_{{\bf k}}[g^{(s)}, m^{(s)}, \phi^{(s)}, \lambda^{(s)}](x, \cdot)}{\delta \phi^\nu(\cdot)} ds.
				\end{split}
			\end{equation}
		These two distributions $\delta^\nu \theta^0_\ell(x, \cdot)/ \delta \phi^\nu(\cdot)$ and $\delta^\nu r^0_{L, \phi} (x,\cdot)/ \delta \phi^\nu(\cdot)$, defined in $U^{n + \nu -1} \backslash (x, \dots, x)$, satisfy properties~\ref{supp_delta_t_H}-\ref{WF_delta_t_H_s} as can be easily seen. It follows straightforwardly from the properties of $C^{(\ell)}_\phi$ and $u^{0 (\ell)}$ that each distribution $(C_\phi^{(\ell)}(x)/ \delta \phi^\nu(\cdot)) \cdot ((\alpha_x)^* u^{0 (\ell)})(\cdot)$ satisfies~\ref{supp_delta_t_H}-\ref{WF_delta_t_H_s} and so the distribution $\delta^\nu \theta^0_\ell(x, \cdot)/ \delta \phi^\nu(\cdot)$ also does. Furthermore, we conclude that also $\delta^\nu r^0_{\phi, L} / \delta \phi^\nu$ satisfies the properties~\ref{supp_delta_t_H}-\ref{WF_delta_t_H_s}, as a simple consequence of eq.~\eqref{delta_taylor} and the fact that both $\delta^\nu \tau^0_{\phi, {\bf j}} / \delta \phi^\nu$ and $\delta^\nu \theta^0_{\ell} / \delta \phi^\nu$ satisfy~\ref{supp_delta_t_H}-\ref{WF_delta_t_H_s}.\\
		We want to show that the terms in the right-hand side of eq.~\eqref{delta_taylor_ext} are extensions of the terms that appear in the right-hand side of eq.~\eqref{delta_taylor}. Because $u^{(\ell)}$ is an extension of $u^{0 (\ell)}$ to $\bR^{4(n-1)}$, we have that $(\delta^\nu C_\phi^{(\ell)}(x) / \delta \phi^\nu(\cdot)) \cdot ((\alpha_x)^* u^{(\ell)})(\cdot)$ is trivially an extension to $U^{n+\nu-1}$ of $(\delta^\nu C_\phi^{(\ell)}(x) / \delta \phi^\nu(\cdot)) \cdot ((\alpha_x)^* u^{0 (\ell)})(\cdot)$.\\
		It remains to be shown that the distribution $\delta^\nu r_{L, \phi} / \delta \phi^\nu$ is obtained by extending to the total diagonal the distribution $\delta^\nu r^0_{L, \phi} / \delta \phi^\nu$. As we have already experienced discussing the extension of $\tau^{H,0}_{\phi, {\bf k}}$ in sec.~\ref{subsubsec_rew_T-prod}, it is not necessary that an extension of a distribution to the diagonal exists nor that such extension is unique. Nevertheless, we will show that, for any fixed $x$, $\delta^\nu r^0_{L, \phi} / \delta \phi^\nu(x, \cdot)$ scales with degree less than $4(n+\nu)$ under the rescaling of the coordinates and, therefore, it must have a unique extension.\\
		First, we prove that, for any $x$, $\delta^\nu r^0_{L, \phi}(x, \cdot) / \delta \phi^\nu (\cdot)$ has scaling degree less or equal than $|{\bf j}| +3\nu - L -1$ with respect to the rescaling of the coordinates. The idea is to adapt the argument presented in~\citep[thm. 4.1 {\em (v)}]{HW02} to this situation. Since $\delta^\nu \tau^{H,0}_{\phi, {\bf j}} / \delta \phi^\nu$ has an almost homogeneous scaling with degree $|{\bf j}| + 3 \nu$ with respect to $(g, m, \lambda, \phi) \mapsto (\Lambda^{-2} g, \Lambda m, \Lambda \phi, \lambda)$, we have
		\begin{equation}\label{rewrite_r0}
			\begin{split}
				\chi_{\Lambda}^* \frac{\delta^\nu r^{0}_{L, \phi} (x, \cdot)}{\delta \phi^\nu (\cdot)} &= \frac{\Lambda^{L+1}}{L!} \int_0^1 (1-\mu)^L \left( \partial^{L+1}_s \frac{\delta^\nu \tau^{0, H}_{{\bf j}}}{\delta \phi^\nu (\cdot)}[\Lambda^2 g^{(s)}, \Lambda^{-1} m^{(s)}, \Lambda^{-1} \phi^{(s)}, \lambda^{(s)}](x, \cdot) \right|_{s=\Lambda \mu} d\mu\\
					&= \Lambda^{L + 1 - (|{\bf j}| - 3\nu)} \sum_{\ell =0}^{n} (\ln^\ell \Lambda) \psi^0_{\ell, L, \phi}(\Lambda, x, \cdot),
				\end{split}
			\end{equation}
		where $\psi^0_{\ell, L, \phi}$ is the distribution in $\bR \times U \times U^{n-1}\backslash (x, \dots, x)$ defined by
			\begin{equation*}
				\begin{split}
					&\psi^0_{\ell, L, \phi}(\Lambda, x, \cdot) :=\\
					&\quad = \frac{1}{\ell! L!} \int_0^1 (1 - \mu)^L \left( \partial_s^{L+1} (\varepsilon\partial_\varepsilon - |{\bf j}| - 3\nu)^\ell \frac{\delta^\nu \tau^{H,0}_{{\bf j}}}{\delta \phi^\nu}[\varepsilon^2 g^{(s)}, \varepsilon^{-1} m^{(s)}, \varepsilon^{-1} \phi^{(s)}, \lambda^{(s)}] (x, \cdot) \right|_{s=\Lambda \mu, \varepsilon=1} d\mu .
				\end{split}
			\end{equation*}
		Let $f$ be a test function in $U^{n -1 + \nu}$ such that its support does not contain the point $(x, \dots, x)$. It follows from the properties~\ref{WF_delta_t_H} and~\ref{WF_delta_t_H_s} of $\delta^\nu \tau^{0, H}_{\phi, {\bf j}} / \delta \phi^\nu$ that $\psi^0_{\ell, L, \phi}(\Lambda, x, f)$ is smooth in $\Lambda$ in a sufficiently small neighbourhood of zero. Once we have established these results, it follows from eq.~\eqref{rewrite_r0} that $\delta^\nu r^0_{L, \phi}(x, \cdot) / \delta \phi^\nu (\cdot)$ has scaling degree less or equal than $|{\bf j}| +3\nu - L -1$, as we wanted to prove.\\
		Since we choose $L \geq |{\bf j}| + 4(n-1)$, this implies that $\delta^\nu r^0_{L, \phi}(x, \cdot) / \delta \phi^\nu (\cdot)$ scales with degree less that $4(n-1) +3\nu-1$ for any $x$, and so $\delta^\nu r^0_{L, \phi}(x, \cdot) / \delta \phi^\nu (\cdot)$ has a unique extension to the total diagonal. Outside $\Delta_{n+\nu}$, the distribution $\delta^\nu r_{L, \phi}(x, \cdot) / \delta \phi^\nu(\cdot)$ exists and coincides with $\delta^\nu r^0_{L, \phi} (x, \cdot)/ \delta \phi^\nu(\cdot)$. Furthermore, it must have a scaling degree less then $4(n-1) +3\nu-1$ for any $x$. Because any non-trivial distribution supported on $\Delta_{n+\nu-1}$ must be a linear combination of delta distributions and its derivative, which have scaling degree not greater than $4(n +\nu -1)$, we conclude that $\delta^\nu r_{L, \phi} / \delta \phi^\nu$ coincides with the extension of $\delta^\nu r^0_{L, \phi}(x, \cdot) / \delta \phi^\nu(\cdot)$, i.e.
		\begin{equation}\label{var_remainder_def}
			\frac{\delta^\nu r_{L, \phi}}{\delta \phi^\nu} (f)  = \lim_{j \to \infty}  \frac{\delta^\nu r^0_{L, \phi}} {\delta \phi^\nu}(\vartheta_j f),
		\end{equation}
	where $f$ is a test function in $M^{n+\nu}$ with support sufficiently close to the total diagonal $\Delta_{n+\nu}$, and where $\vartheta_j$ is sequence of functions with support in $U^{n+\nu} \backslash \Delta_{n+\nu}$ such that $\vartheta_j$ is identically $1$ outside a neighbourhood $\cO_j$ of $\Delta_{n+\nu}$ with $\cO_j$ shrinking to $\Delta_{n+\nu}$ as $j \to \infty$. The right-hand side of~\eqref{var_remainder_def} does not depend on the choice of the functions $\vartheta_j$ because of the scaling properties of $\delta^\nu r^0_{L, \phi} / \delta^\nu$ (see~\citep[thm. 5.2]{brunetti2000microlocal}).
	
	\paragraph{Step 3.}
			As we mentioned in the overview of the argument at the beginning of this subsection, to prove that $\tau^H_{\phi, {\bf j}}$ satisfies properties~\ref{supp_delta_t_H}-\ref{WF_delta_t_H_s}, we make use of the fact that the right-hand side of eq.~\eqref{delta_taylor_ext} is an extension to the total diagonal of the right-hand side of eq.~\eqref{delta_taylor} (which satisfies properties~\ref{supp_delta_t_H}-\ref{WF_delta_t_H_s} outside the total diagonal).\\
			We fist focus on the terms $\delta^\nu C_\phi^{(\ell)}(x)/ \delta \phi^\nu(\cdot) \cdot((\alpha_x)^* u^{(\ell)})(\cdot)$ and we prove the following result:
			\begin{lemma}\label{lemma_Theta}
				Each distribution
					\begin{equation}\label{ext_delta_C_alpha_u}
						\frac{\delta^\nu C_\phi^{(\ell)}(x)}{\delta \phi^\nu(\cdot)} \cdot((\alpha_x)^* u^{(\ell)})(\cdot),
					\end{equation}	
				where $C^{(\ell)}_\phi$ is as in eq.~\eqref{formula_theo_4.1}, and $u^{(\ell)}$ is a Lorentz invariant extension of $u^{0 (\ell)}$ which scales almost homogeneously with degree $|{\bf j}| - \ell$ under rescaling of the coordinates, satisfies the conditions~\ref{supp_delta_t_H}-\ref{WF_delta_t_H_s}.
			\end{lemma}
			\begin{proof}
				The distribution~\eqref{ext_delta_C_alpha_u} is an extensions to the total diagonal of the distribution $(\delta^\nu C_\phi^{(\ell)}(x) / \delta \phi^\nu(\cdot)) \cdot ((\alpha_x)^* u^{0 (\ell)})(\cdot)$ defined in $U^{n+\nu -1} \backslash (x, \dots, x)$. Before, in Step 2, we proved that each $(\delta^\nu C_\phi^{(\ell)}(x) / \delta \phi^\nu(\cdot)) \cdot ((\alpha_x)^* u^{0 (\ell)})(\cdot)$ satisfies the properties~\ref{supp_delta_t_H}-\ref{WF_delta_t_H_s} outside the total diagonal. We can adapt the argument presented in~\citep[sec. 4.3]{HW02} to prove that~\eqref{ext_delta_C_alpha_u} satisfies the conditions~\ref{supp_delta_t_H}-\ref{WF_delta_t_H_s} on $U^{n+\nu -1}$:
				\begin{enumerate}[label=($\delta$t\arabic*), start=0]
					\item Since $\delta^\nu C_\phi^{(\ell)}(x)/\delta \phi(y_1) \dots \delta \phi(y_\nu)$ is proportional to $\delta(x,y_1, \dots, y_\nu)$, it vanishes unless all the points  $x, y_1, \dots, y_\nu$ coincide. Thus, $\delta^\nu C_\phi^{(\ell)}(x) /\delta \phi^\nu(\cdot) \cdot((\alpha_x)^* u^{(\ell)})(\cdot)$ satisfies condition~\ref{supp_delta_t_H}.
					\item The Lorenz-invariance of $u^{(\ell)}$ implies the locally covariance of $\alpha^*_x u^{(\ell)}$. The term $\delta^\nu C^{(\ell)}_\phi(x) / \delta \phi^\nu$ is also locally covariant because it is given by derivatives of the delta distribution $\delta(x,y_1, \dots, y_\nu)$ multiplied by a sum of monomials constructed from the metric $g(x)$, the Riemann tensor in $x$ and its symmetrized covariant derivatives, the functions $m,\lambda, \phi$ in $x$ and their covariant derivatives. Combining these two results we obtain that $\delta^\nu C_\phi^{(\ell)}(x) /\delta \phi^\nu(\cdot) \cdot((\alpha_x)^* u^{(\ell)})(\cdot)$ satisfies property~\ref{cov_delta_t_H}.
					\item Exactly as in~\citep{HW02}, we can conclude that $\alpha^*_x u^{(\ell)}$ has an almost homogeneous scaling with degree $|{\bf j}| -\ell$ with respect to a rescaling of the metric (other rescalings do not affect $\alpha^*_x u$). Since $C^{(\ell)}_\phi(x)$ and $\delta(x, y_1, \dots, y_\nu)$ scale homogeneously respect to a rescaling $(g, m, \phi, \lambda) \mapsto (\Lambda^{-2} g, \Lambda m, \Lambda \phi, \lambda)$ respectively with degree $\ell$ and $4(\nu +1)$, it follows that $\delta^\nu C^{(\ell)}_\phi(x) / \delta \phi^\nu$ scales homogeneously with degree $|{\bf j}|+ 3 \nu +4$. Combining everything together we get precisely condition~\ref{scaling_delta_t_H}.
					\item Outside the diagonal, we already know that $(\delta^\nu C_\phi^{(\ell)}(x) /\delta \phi^\nu(\cdot)) \cdot((\alpha_x)^* u^{(\ell)})(\cdot)$ satisfies the condition~\ref{WF_delta_t_H}. We show that condition~\ref{WF_delta_t_H} holds also on the total diagonal. Similarly as done in~\citep{HW02}, making use of the wave-front set calculus, we obtain the following estimate
						\begin{equation}\label{est_tau}
							\begin{split}
								\WF \left( \frac{\delta^\nu C^{(\ell)}_\phi}{\delta \phi^\nu} \cdot (\alpha^* u) \right) &\subset \left\{ (x, x_2, \dots, x_n, y_1, \dots, y_\nu; k, k_2, \dots, k_n, p_1, \dots, p_\nu) \in \dot{T}^* M^{n+\nu} \right. \\
								&\quad x= y_1 = \dots = y_\nu, \quad x_2, \dots, x_n \in U,\\
								&\quad k = - \sum_\ell p_\ell + \sum_i \left[ \frac{\partial \alpha_x(x_i)}{\partial x} \right]^t \xi_i, \quad k_i = \left[ \frac{\partial \alpha_x(x_i)}{\partial x_i} \right]^t \xi_i,\\
								&\quad \left. \left(\alpha_x(x_2), \dots, \alpha_x(x_n); \xi_2, \dots, \xi_n\right) \in \WF( u ) \right\}.
							\end{split}
						\end{equation}
	In eq.~\eqref{est_tau}, $\partial \alpha_x(x_i) / \partial x$ denotes the matrix of partial derivatives of $\alpha_x(x_i)$ with respect $x$ at a fixed $x_i$, while $\partial \alpha_x(x_i) / \partial x_i$ denotes the matrix of partial derivatives of $\alpha_x(x_i)$ with respect to $x_i$ at a fixed $x$. When $x$ and $x_i$ coincide, such matrices satisfy $\partial \alpha_x(x_i) / \partial x = - \partial \alpha_x(x_i) / \partial x_i$.Consequently, it holds $k + \sum_\ell p_\ell + \sum_i k_i = 0$ on the total diagonal $\Delta_{n+\nu}$. In other words, we proved $\WF ( \delta^\nu C^{(\ell)}_\phi / \delta \phi^\nu \cdot \alpha^* u)|_{\Delta_{n+\nu}} \perp T \Delta_{n+\nu}$, and, thus, the wave-front set condition~\ref{WF_delta_t_H} holds for $\delta^\nu C_\phi^{(\ell)}(x) /\delta \phi^\nu(\cdot) \cdot((\alpha_x)^* u^{(\ell)})(\cdot)$ on the total diagonal.
					\item Outside the diagonal, we already know that $\delta^\nu C_\phi^{(\ell)}(x) /\delta \phi^\nu(\cdot) \cdot((\alpha_x)^* u^{(\ell)})(\cdot)$ satisfies the conditions~\ref{WF_delta_t_H_s}.  To prove that the conditions still hold on the total diagonal, we can follow a similar argument as the one given for~\ref{WF_delta_t_H}. That is indeed possible because $\alpha_x$ is the inverse of the exponential map and, therefore, it depends only on the metric in the appropriate smooth sense, $u$ is independent of $g, m, \lambda, \phi$, and $\delta^\nu C^{(\ell)}_\phi / \delta \phi^\nu$ is given by derivatives of the delta distribution multiplied by sum of monomials constructed from the metric $g$, the Riemann tensor and its symmetrized covariant derivatives, the functions $m,\lambda, \phi$ and their covariant derivatives.
				\end{enumerate}
				This concludes the proof of lemma~\ref{lemma_Theta}
			\end{proof}
			
			Now, we focus on the term $\delta^\nu r_{L, \phi} / \delta \phi^\nu$, i.e. the $\nu$-th variational derivatives of the remainder term. As we will see, proving the conditions~\ref{WF_delta_t_H} and~\ref{WF_delta_t_H_s} for the wave-front set of the extension of the remainder term is much more complicated than proving the same conditions for the extensions of the factors in the Taylor series as just done in lemma~\ref{lemma_Theta}. The reason is that we lack an explicit form of $\delta^\nu r_{L, \phi} / \delta \phi^\nu$. We proceed adapting to our situation the argument used in~\citep[prop. 4.1]{HW02}: we show that the properties of $\delta^\nu r^0_{L, \phi} / \delta \phi^\nu$ provide suitable bounds for the wave-front set of the extension $\delta^\nu r_{L, \phi} / \delta \phi^\nu$.
			\begin{lemma}\label{lemma_delta_rem}
				The distribution $\delta^\nu r_{L, \phi}/ \delta \phi^\nu$ satisfies the properties~\ref{supp_delta_t_H}-\ref{WF_delta_t_H_s}.
			\end{lemma}
			\begin{proof}
				Since $\delta^\nu r_{L, \phi}/ \delta \phi^\nu$ is the unique extension to the total diagonal of the distribution $\delta^\nu r^0_{L, \phi}(x,\cdot) / \delta \phi^\nu(\cdot)$ which satisfies the properties~\ref{supp_delta_t_H}-\ref{WF_delta_t_H_s}, we need to verify that the properties holds also on the total diagonal.
				\begin{enumerate}[label= ($\delta$t\arabic*), start=0]
					\item We argue by reductio ad absurdum: we assume that there exists an element $(x_1, \dots, x_n, y_1, \dots, y_\nu)$ in the support of $\delta^\nu r_{L, \phi}/ \delta \phi^\nu$ such that $(x_1, \dots, x_n) \in \Delta_n$, but $(x_1, \dots, x_n, y_1, \dots, y_\nu) \notin \Delta_{n+\nu}$. Since $(x_1, \dots, x_n, y_1, \dots, y_\nu)$ is not an element of the total diagonal, it must belong to the support of $\delta^\nu r^0_{L, \phi}/ \delta \phi^\nu$. However, this is incompatible with the fact that $\delta^\nu r^0_{L, \phi}/ \delta \phi^\nu$ satisfies~\ref{supp_delta_t_H} outside $\Delta_{n+\nu}$, which precisely implies that the support $\delta^\nu r^0_{L, \phi}/ \delta \phi^\nu$ does not contain any element $(x_1, \dots, x_n, y_1, \dots, y_\nu) \notin \Delta_{n+\nu}$ with $(x_1, \dots, x_n) \in \Delta_n$. This concludes the proof of property~\ref{supp_delta_t_H}.
					\item The locally covariance of the unextended distribution $\delta^\nu r^0_{L, \phi}(x,\cdot) / \delta \phi^\nu(\cdot)$ implies that the distributions $\delta^\nu r_{L, \phi} (x, \cdot)/ \delta \phi^\nu(\cdot) [\iota^* (g, m, \phi, \lambda)]$ and $\iota^* \delta^\nu r_{L, \phi}(x,\cdot) / \delta \phi^\nu(\cdot) [g, m, \phi, \lambda]$ coincide outside the total diagonal. The difference between these two distributions must be supported on the total diagonal and must have a scaling degree less than $4(n-1) +3\nu-1$ since each of the distributions has a scaling less than $4(n-1) +3\nu-1$. However, a distribution supported on the total diagonal must be a sum of the delta distribution and its covariant derivatives, which are distributions with a scaling degree not less than $4(n-1 +\nu)$. Thus, $\delta^\nu r_{L, \phi} (x, \cdot)/ \delta \phi^\nu(\cdot) [\iota^* (g, m, \phi, \lambda)]$ and $\iota^* \delta^\nu r_{L, \phi}(x,\cdot) / \delta \phi^\nu(\cdot) [g, m, \phi, \lambda]$ coincide also on the diagonal, which precisely means that $\delta^\nu r_{L, \phi} / \delta \phi^\nu$ is locally covariant.
					\item Using a similar argument as the one presented for the proof of~\ref{cov_delta_t_H}, we can prove that $\delta^\nu r_{L, \phi} / \delta \phi^\nu$ scales almost homogeneously under rescaling of $g, m, \phi, \lambda$.
				\end{enumerate}
				Proving that $\delta^\nu r_{L,\phi} / \delta \phi^\nu$ satisfies the properties~\ref{WF_delta_t_H} and~\ref{WF_delta_t_H_s} on the diagonal is more involved. We follow the argument of~\citep[prop. 4.1]{HW02}, adapted to our context:
	\begin{lemma}\label{lemma_var_remainder_diagonal}
		It holds
			\begin{equation}\label{var_remainder_diagonal}
				\left. \WF \left( \frac{\delta^\nu r_{L, \phi} (x_1, \dots, x_n)}{\delta \phi (y_1) \dots \phi(y_\nu)}\right) \right|_{\Delta_{n+\nu}} \perp T \Delta_{n+\nu}.
			\end{equation}
		Furthermore, for any smooth $1$-parameter families $\{ g^{(s)}, m^{(s)}, \phi^{(s)}, \lambda^{(s)} \}$ we have
			\begin{equation}\label{var_remainder_diagonal_s}
				\left. \WF \left( \frac{\delta^\nu r_{L, \phi}[g^{(s)}, m^{(s)}, \phi^{(s)}, \lambda^{(s)}](x_1 \dots, x_n)}{\delta \phi(y_1) \dots \delta \phi(y_\nu)} \right) \right|_{\bR \times \Delta_{n +\nu}} \perp T (\bR \times \Delta_{n+\nu}).
			\end{equation}
	\end{lemma}
	\begin{proof}
		As in~\citep[prop. 4.1]{HW02}, we proceed by induction on the number of variables $n$, i.e we assume that estimate~\eqref{var_remainder_diagonal} holds for all $\delta^\nu r_{L,\phi}(x_1, \dots, x_{n'}) / \delta \phi(y_1) \dots \delta (y_\nu)$ with $n'<n$.\\
		First, we need to show a constraint on closure in $T^* (\bR \times U^n)$ of the wave-front set of $\delta^\nu \tau^{H,0}_{\phi(s), {\bf j}} / \delta \phi^\nu$. As a consequence of lemma~\ref{lemma_Theta}, it holds
		\begin{equation}\label{var_tau_perp_diag}
			\left. \WF \left( \sum  \frac{\delta^\nu C^{(\ell)}_\phi}{\delta \phi^\nu}(s, x_1, y_1, \dots, y_\nu) \cdot ((\alpha_{x_1}[g^{(s)}])^* u^{(\ell)})(x_2, \dots, x_{n'}) \right)\right|_{\bR \times \Delta_{n'+\nu}} \perp T(\bR \times \Delta_{n'+\nu}),
		\end{equation}	
		for any $n' \leq n$. Here and in the following we use the notation $f(s,x, \dots) = f[g^{(s)}, m^{(s)}, \phi^{(s)}, \lambda^{(s)}](x, \dots)$. Using the estimate~\eqref{var_tau_perp_diag} and the inductive hypothesis on the variational derivatives of the remainder term, we obtain
		\begin{equation*}
			\left. \WF \left( \frac{\delta^{\nu} \tau^H_{\phi, {\bf i}}}{\delta \phi^{\nu}}(s, x_1, \dots, x_{n'}, y_1, \dots, y_{\nu})  \right)\right|_{\bR \times \Delta_{n' + \nu}} \perp T(\bR \times \Delta_{n' + \nu}).
		\end{equation*}
	Since $\delta^\nu \tau^{H,0}_{\phi, {\bf j}} / \delta \phi^\nu$ can be expressed as a finite sum of terms in the form~\eqref{x_var_term}, and since the estimate~\eqref{var_H_est_s_diag} holds for any variational derivatives of the Hadamard parametrix, arguing similarly as done for formulas (86),(87) of~\citep{HW02}, we conclude that it holds
		\begin{equation}\label{t_perp_diag}
			\left. \overline{ WF \left( \frac{ \delta^\nu \tau^{H,0}_{\phi, {\bf j}}}{\delta \phi^\nu} (s,x_1, \dots, x_n, y_1, \dots, y_\nu) \right) } \right|_{\bR \times \Delta_{n+\nu}} \perp T(\bR \times \Delta_{n+\nu}),
		\end{equation}
	where the overbar denote the closure in $T^* (\bR \times U^n)$.\\
	 Now, we identify $x=x_1 \in U$ with its coordinate in a smooth chart, while $\xi=(x_2, \dots, x_n, y_1, \dots, y_\nu)$ with the Riemannian normal coordinates relative to $x_1$. In this notation, the total diagonal corresponds to $\xi =0$. We identify distributions in $(x_1, \dots, x_n, y_1, \dots, y_\nu) \in M^{n+\nu}$ with distribution in $(x, \xi) \in X \times \Xi$, where $X$ is an open set in $\bR^4$, and where $\Xi$ is an open neighbourhood of the origin in $\bR^{4(n + \nu-1)}$.\\
	We chose an arbitrary $x_0$ in $X$ and a smooth function in the form $\chi(x, \xi) = \chi'(x) \chi''(\xi)$, where $\chi' \in C^\infty_0(X)$ is identically $1$ in a neighbourhood of $x_0$, where $\chi''$ vanishes in a neighbourhood of $0$ and it is identically $1$ outside a larger neighbourhood. Sufficiently close to $(x_0, 0)$, we can choose as the cut-off $\vartheta_j$ in the definition of the extension $\delta^\nu r_{L, \phi} / \delta \phi^\nu$ (see eq.~\eqref{var_remainder_def}) to be $\vartheta_j = (\chi)_{2^j}$, where $(\cdot)_{2^j}$ denotes the pull-back by the map  $(x,\xi) \mapsto (x, 2^j \xi)$.\\
	In order to prove formula~\eqref{var_remainder_diagonal}, we need to show that $(x_0, \xi_0=0; k_0, \eta_0) \in \dot{T}^* (X \times \Xi)$ is not contained in $\WF(\delta^\nu r_{L,\phi} / \delta \phi^\nu)$ if $k_0 \neq 0$. By the definition of wave-front set~\citep[def. 8.1.2]{H83} and the specific choice of the cut-off $\vartheta_j$, it is sufficient to show that there exists a conic neighbourhood $F$ of $(k_0,\eta_0)$ such that for any $(k,\eta) \in F$ and any $N,j \in \bN$ it holds that
		\begin{equation}\label{fourier_est}
			\left| \cF \left( (h)_{2^j} \cdot r^0_{L, \phi} \right) (k, \eta) \right| \leq (\const)_N 2^{-j/2} ( 1 + |k| +|\eta|)^{-N},
		\end{equation}
	where  $\cF(\cdot)$ denotes the Fourier transform, and where $h$ is the compaclty supported function defined by $h(x, \xi) := \chi (x,\xi) - \chi(x, 2\xi)$. The Fourier transform on the left-hand side of~\eqref{fourier_est} can be rewritten as
		\begin{equation}\label{fourier}
			\cF \left( (h)_{2^j} \cdot \delta^\nu r^0_{L, \phi} / \delta \phi^\nu \right) (k, \eta) = 2^{-4j( n + \nu-1)} \cF \left( h \cdot (\delta^\nu r^0_{L, \phi} / \delta \phi^\nu)_{2^{-j}} \right) (k, 2^{-j}\eta).
		\end{equation}
	Using the definition of $(\delta^\nu \psi)^0_{\ell, L, \phi}(\Lambda, x, \xi)$ given by eq.~\eqref{rewrite_r0}, it follows that the right-hand side of eq.~\eqref{fourier} can be expressed as
		\begin{equation}\label{fourier2}
			2^{-4j(m + \nu -1)} \cF \left( h \cdot (\delta^\nu r^0_{L, \phi} / \delta \phi^\nu)_{2^{-j}} \right) (k, 2^{-j}\eta) = 2^{-j} \sum_{\ell} (j \ln 2)^\ell \cF' \left( h \cdot (\delta^\nu \psi)^0_{\ell, L, \phi} \right) (2^{-j}, k, 2^{-j} \eta),
		\end{equation}
	where $\cF'$ denote the Fourier transform with respect to the variable $x$ and $\xi$.\\
	We proceed proving that for any closed conic set $F'$ in $\bR \times \bR^4 \times \bR^{4(n +\nu -1)}$ which does not contain elements in the form $(0,0,\eta)$ there is a neighbourhood $K_0 \subset \bR \times X \times \Xi$  of $(0, x_0, 0)$ such that for all $\ell$ it holds
		\begin{equation*}
			\WF( (\delta^\nu \psi)^0_{\ell, L, \phi} ) \cap (K_0 \times F') = \emptyset.
		\end{equation*}
	We consider as parameter $\overline{s}:= (\varepsilon, \mu, x) \in P_1 \times P_2 \times P_3 =: P$ where $P_1$ is a small neighbourhood of $1$ in $\bR$, where $P_2$ is a small neighbourhood of $0$ in $\bR$, and where $P_3$ is a convex normal neighbourhood of $x$ with respect to $g$, which is then identified with a subset of $\bR^4$ using the same coordinate chart used to identify $x$ with a vector in $\bR^4$. Let $\iota_{x,\mu}$ be the diffeomorphism which shrinks the Riemannian normal coordinates with respect to $x$ of a point in $P_3$ by a factor $\mu$. In terms of this family of diffeomorphisms, we construct the following smooth families, parametrized by $\overline{s} \in P$
		\begin{equation*}
			g^{(\overline{s})} = (\varepsilon \mu)^{-2} \iota^*_{x, \mu} g, \quad m^{(\overline{s})} := \varepsilon \mu m, \quad  \phi^{(\overline{s})} := (\varepsilon \mu) \iota^*_{x, \mu} \phi, \quad  \lambda^{(\overline{s})} := \iota^*_{x, \mu} \lambda.
		\end{equation*}
	The estimate~\eqref{t_perp_diag}, derived for a parameter $s \in \bR$, can be generalized to
		\begin{equation}\label{est_var_t_clos}
			\left. \overline{ \WF \left( \frac{\delta^\nu \tau^{0,H}_{\phi,{\bf j}}}{\delta{\phi^\nu}} (\overline{s}, x_1, \dots, x_n, y_1, \dots, y_\nu) \right)} \right|_{P \times \Delta_{m+\nu}} \perp T( P \times \Delta_{n+\nu}).
		\end{equation}
	We can rewrite the action of $(\delta^\nu \psi)^0_{\ell, L, \phi}$ on $f \in C^\infty_0(\bR \times (U^{n+\nu} \backslash \Delta_{n+\nu}))$ as
		\begin{equation}\label{rewrite}
			\begin{split}
				&(\delta^\nu\psi)^0_{\ell, L ,\phi} (f) =\\
				&\quad = \left( \pi^* \frac{\delta^\nu \tau^{0,H}_{{\bf k}}}{\delta \phi^\nu}[g^{(\overline{s})}, m^{(\overline{s})}, \phi^{(\overline{s})}, \lambda^{(\overline{s})}] \right) \left( \left( {}^t D^{(\ell)}_\varepsilon \delta (\cdot - 1) \right) \otimes \left( \left( {}^t R^{(L)}_\mu \otimes 1_{x_1, \dots, y_\nu} \right) f \right) \right),
			\end{split}
		\end{equation}
	where $\pi: (\varepsilon, \mu, x_1, \dots, y_\nu) \mapsto (\overline{s}=(\varepsilon, \mu, x= x_1), x_1, \dots, y_\nu) \in P \times U^{n+\nu}$, where ${}^tD^{(\ell)}$ is the transpose of the operator $D^{(\ell)}$ defined by
		\begin{equation*}
			D^{(\ell)}:= \frac{1}{\ell!}(\varepsilon \partial_\varepsilon + d - \nu)^\ell,
		\end{equation*}
	and where ${}^tR^{(L)}$ is the transpose of the map $R^{(L)}: C^\infty_0(\bR) \to C^\infty(\bR)$ defined by
		\begin{equation*}
			(R^{(L)} f)(\Lambda):= \frac{1}{L!} \int_0^1 d\mu \, (1- \mu)^L (\partial^{L+1} f)(\Lambda \mu).
		\end{equation*}
	The wave-front set of $R^{(L)}$, seen as a distribution on $\bR^2$ via the Schwartz kernel theorem, does not contain any element in the form $(\Lambda_1, \Lambda_2; 0, \rho) \in \dot{T}(\bR^2)$ as shown in the proof of~\citep[prop. 4.1]{HW02}. Using the wave-front set calculus and the estimate~\eqref{est_var_t_clos}, we obtain from eq.~\eqref{rewrite} the following constraint
		\begin{equation*}
			\left. \overline{ \WF ((\delta^\nu \psi)^0_{\ell, L, \phi})} \right|_{\bR \times \Delta_{n+\nu}} \perp T( \bR \times \Delta_{n+\nu}).
		\end{equation*}
	Once $\bR \times (U^{n+\nu} \backslash \Delta_{n+\nu})$ is identified with a subset of $\bR \times X \times (Y \backslash 0)$, the previous result reads
		\begin{equation*}
			\left. \overline{ \WF ((\delta^\nu \psi)^0_{\ell, L, \phi})} \right|_{\bR \times X \times \{0\}} \perp T( \bR \times X \times \{0\}).
		\end{equation*}
	We conclude that the open set $T^* (\bR \times X \times \Xi) \backslash \overline{\WF ((\delta^\nu \psi)^0_{\ell,L,\phi})}$ contains a set of the form $K_0 \times F'$ where $F'$ can be any closed cone which do not contain elements in the form $(0,0,\eta)$ and $K_0$ is a sufficiently small neighbourhood of $(0, x_0, 0)$.\\
	With a suitable redefinition of the cut-off, we can consider $h \in C^\infty_0(K_0)$ and, then, for all $(\rho, k ,\eta) \in F'$ it holds
		\begin{equation*}
			\left| \cF\left( h \cdot (\delta^\nu \psi)^0_{\ell, L, \phi} \right)(\rho, k ,\eta) \right| \leq (\const)_N (1 + |\rho| + |k| + |\eta|)^{-N}.
		\end{equation*}
	We match the choice of the cones $F$ and $F'$ such that $(\rho, k ,2^{-j} \eta) \in F'$ for all $(k,\eta) \in F$, for all $j$, and for all $\rho \neq 0$. In particular, there must exist $C >0$ such that for any $(k,\eta) \in F$ it holds $|k| >C |\eta|$ and $C \neq 0$. In fact, if these requirements did not hold, then $F'$ would contain elements in the form $(0,0,\eta)$ contradicting the hypotheses on $F'$.\\
		Finally, we can rewrite eq.~\eqref{fourier2} as
		\begin{equation*}
			\cF \left( h_{2^j} \cdot \delta^\nu r^0_{L, \phi} / \delta \phi^\nu \right) (k, \eta) = (2\pi)^{-1/2} 2^{-j} \sum_{\ell} (j \ln 2)^\ell \int  e^{-i 2^{-j} \rho} \cF \left( h \cdot (\delta^\nu \psi)^0_{\ell, L, \phi} \right)(\rho, k , 2^{-j} \eta) d\rho,
		\end{equation*}
		and then it holds
		\begin{equation*}
			\left| \cF \left( h_{2^j} \cdot \delta^\nu r^0_{L, \phi} / \delta \phi^\nu \right) (k, \eta) \right| \leq (\const)_N 2^{-j/2} (1 + |k|)^{-N} \leq (\const)'_N 2^{-j/2} (1 + |k| + |\eta|)^{-N},
		\end{equation*}	
	for all $(k, \eta) \in F$, and for all natural numbers $N$ and $j$. This concludes the proof of estimate~\eqref{var_remainder_diagonal}\\
	
	The argument just presented can be generalized to the case of smooth variations of $g, m, \phi, \lambda$, and so it follows that also estimate~\eqref{var_remainder_diagonal_s} also holds. This concludes the proof of the lemma~\ref{lemma_var_remainder_diagonal}.
	\end{proof}
	\noindent
	The results of lemma~\ref{lemma_var_remainder_diagonal} are precisely what we need to show to conclude that the wave-front conditions~\ref{WF_delta_t_H}, \ref{WF_delta_t_H_s} hold for $\delta^\nu r_{L, \phi} / \delta \phi^\nu$, because we have already proved in Step 2 that $\delta^\nu r_{L, \phi} / \delta \phi^\nu$ satisfies the conditions~\ref{WF_delta_t_H}, \ref{WF_delta_t_H_s} outside the total diagonal.
	\end{proof}
			
		Summing up the results of lemma~\ref{lemma_Theta} and lemma~\ref{lemma_delta_rem}, we have:
		\begin{prop}\label{prop_t_H_W_smooth}
				Let $\cU_m$ be a neighbourhood of the total diagonal $\Delta_n$ such that for any $(x_1, \dots, x_n) \in \cU_n$ the point $x_1, \dots, x_n$ are contained in a normal convex subset $U \subset M$ sufficiently small that we can apply lemma~\ref{lemma_H}.  For any ${\bf j}$, the distributional coefficient $\tau^H_{\phi,{\bf j}}$ of the local Wick expansion of $T_{n,\phi}[\otimes_i \varphi^{k_i} (x_i)]$ in $\cU_n$ is such that for any $\nu \in \bN$ its Gateaux derivative $\delta^\nu \tau^H_{\phi,{\bf j}}(x_1, \dots, x_n)/\delta \phi(y_1) \cdots \delta \phi(y_\nu)$ is a well-defined distribution and satisfies the properties~\ref{supp_delta_t_H}-\ref{WF_delta_t_H_s}.
			\end{prop}

	\subsection{The distribution coefficients $\tau_{\phi, {\bf j}}$ satisfy the sufficient conditions for the on-shell $W$-smoothness of $\phi \mapsto T_{n,\phi}[\otimes_i \varphi^{k_i}(f_i)]$}\label{subsubsec_proof}
		In this subsection, we prove that the sufficient conditions for an on-shell $W$-smooth $\phi \mapsto T_{n,\phi}[\otimes_i \varphi^{k_i}(f_i)]$ we introduced in sec.~\ref{subsubsec_suff_cond_T-prod} are satisfied:
		\begin{prop}\label{prop_t_W-smooth}
			Let $(M,g)$ be an ultra-static space-time with compact Cauchy surfaces, let $m$ be a constant, let $\phi$ be a general smooth function, and let $\lambda$ be a compactly supported coupling constant. We have
				\begin{equation}\label{est_WF_W}
					\WF \left( \frac{\delta^\nu \tau_{\phi, {\bf j}}(x_1, \dots, x_n)}{\delta \phi(y_1) \dots \delta \phi(y_\nu)} \right) \subset W_{n+\nu}.
				\end{equation}
			Furthermore, for any $\bR \ni \epsilon \mapsto \phi(\epsilon) \in C^\infty(M)$ smooth, we have
				\begin{equation}\label{est_WF_W_s}
					\WF \left( \frac{\delta^\nu \tau_{\phi(\epsilon), {\bf j}}(x_1, \dots, x_n)}{\delta \phi(y_1) \dots \delta \phi(y_\nu)} \right) \subset \bR \times \{0\} \times W_{n+\nu}.
				\end{equation}
		\end{prop}
		\begin{proof}
			We first prove the estimates~\eqref{est_WF_W} and~\eqref{est_WF_W_s} in a neighbourhood $\cU_n$ of the total diagonal sufficiently small that the hypotheses of prop.~\ref{prop_t_H_W_smooth} are satisfied, then we prove that these estimates hold also outside $\cU_n$.\\
			Let us consider $T_{n,\phi} [\otimes_i \varphi^{k_i}(x_i)]$ inside $\cU_n$. In this space-time domain, we can define its Wick expansion with respect to the retarded $2$-point function $\omega^R_\phi$, given by~\eqref{Wick_exp_state} in terms of the distributions $\{\tau_{\phi,{\bf j}} \}_{{\bf j} \leq {\bf k}}$, and its local Wick expansion with respect to the Hadamard parametrix $H_\phi$, given by eq.~\eqref{local_Wick_exp} in terms of the distributions $\{\tau^H_{\phi,{\bf j}} \}_{{\bf j} \leq {\bf k}}$. Since both these two expansions must give $T_{n,\phi} [\otimes_i \varphi^{k_i}(x_i)]$, there is a relation between the collections of distributions $\{\tau_{\phi,{\bf j}} \}_{{\bf j} \leq {\bf k}}$ and $\{\tau^H_{\phi,{\bf j}} \}_{{\bf j} \leq {\bf k}}$. More precisely, this relation is given by the following formula:
				\begin{equation}\label{t-t^H}
					\tau_{\phi, {\bf j}}(x_1, \dots, x_n) = \sum_{{\bf j}' \leq {\bf j}} \sum_{\{n_{ab}\}} \sC_{n_{ab}, {\bf j}, {\bf j}'} \hbar^{\sum_a n_{ab}}  \tau^H_{\phi, {\bf j}'}(x_1, \dots, x_n) \prod_{a \leq b} d_\phi^{n_{ab}}(x_a,x_b),
				\end{equation}
			where $d_\phi = \omega^R_\phi - H_\phi$, where the sum $\sum_{\{n_{ab}\}}$ is taken over all possible family of natural numbers $\{ n_{ab} \}_{a \leq b}$ such that $j_a = j'_a + \sum_c (n_{ac} + n_{ca})$, and where $\sC$ are certain combinatorial factors.\\
			The right-hand side of eq.~\eqref{t-t^H} is well-defined because $d_\phi$ is a smooth function.\\
			For any smooth map $\bR \ni \epsilon \mapsto \phi(\epsilon) \in C^\infty(M)$ a similar decomposition holds, namely we just need to replace $\phi$ by $\phi(\epsilon)$ everywhere in eq.~\eqref{t-t^H}. Because $d_{\phi(\epsilon)}(x_1, x_2)$ is jointly smooth in $\epsilon, x_1, x_2$ as proved in sec.~\ref{subsubsec_properties_omega_H}, the decomposition is again well-defined.
			Using formula~\eqref{t-t^H} and its analogue for smooth variations of the background $\phi$, we can prove the following lemma:
			\begin{lemma}\label{lemma_t_W_neigh}
				The distributions $\{ \tau_{\phi, {\bf j}} \}_{\bf j}$ satisfy estimates~\eqref{est_WF_W} and~\eqref{est_WF_W_s} in $\cU_n$.
			\end{lemma}
			\begin{proof}
				As seen in sec.~\ref{subsubsec_rew_T-prod}, the distributions $\{ \tau^H_{\phi, {\bf j}} \} _{\bf j}$ satisfy condition~\ref{t_3} and~\ref{t_4}. Using lemma~\ref{lemma_tech_CT_C+_C-} we can conclude that for any ${\bf j}$ we have
				\begin{equation*}
					\WF (\tau^H_{\phi, {\bf j}}) \subset \cC^{T;+}_n \cap \cC^{T;-}_n \subset W_n, \qquad \WF (\tau^H_{\phi(\epsilon), {\bf j}}) \subset \bR \times \{0\} \times \cC^{T;+}_n \cap \cC^{T;-}_n \subset \bR \times \{0\} \times W_n,
				\end{equation*}
			for any smooth map $\bR \ni \epsilon \mapsto \phi(\epsilon) \in C^\infty(M)$. Since $d_\phi \in C^\infty(M^2)$ and $d_{\phi(\epsilon)} \in C^\infty(\bR \times M^2)$, using the wave-front set calculus (thm.~\ref{theo_WF_horma}), we conclude that estimates~\eqref{est_WF_W} and~\eqref{est_WF_W_s} hold for $\nu =0$.\\
			To prove the estimate~\eqref{est_WF_W} for $\nu > 0$, we need to compute $\delta^\nu \tau_{\phi, {\bf j}} / \delta \phi^\nu$ by distributing the variational derivatives on each factor in the right-hand side of eq.~\eqref{t-t^H}. It follows that the distribution $\delta^\nu \tau_{\phi, {\bf j}}(x_1, \dots, x_n) / \delta \phi(y_1) \dots \delta \phi(y_\nu)$ is a finite sum of terms in the form (up to a constant factor)
				\begin{equation}\label{var_term_t-t^H}
					\frac{\delta^{|N|}\tau^{H}_{\phi, {\bf j}'} (x_1, \dots, x_n)}{\delta \phi^{|N|} (\{ y_{r}\}_{r \in N} )} \prod_{a \leq b} \prod_{v = 1}^{n_{ab}}  \frac{\delta^{|N_{a,b,v}|} d_\phi(x_{a}, x_{b})}{\delta \phi^{|N_{a,b,v}|}(\{y_r\}_{r \in N_{a,b,v}})}.
				\end{equation}
			where $N$, $\{ N_{a,b,v} \}_{a \leq b, v \leq n_{ab}}$ gives a disjoint partition of $\{1, \dots, \nu\}$. To prove estimate~\eqref{est_WF_W}, it is sufficient to show that each distribution~\eqref{var_term_t-t^H} has a wave-front set contained in $W_{n+\nu}$.\\
			Let $(x_1, \dots, x_n, y_1, \dots, y_\nu; k_1, \dots, k_n, p_1, \dots, p_\nu)$ be an element of the wave-front set of the distribution~\eqref{var_term_t-t^H}. By the wave-front set calculus (thm.~\ref{theo_WF_horma}), it must hold
				\begin{equation}\label{formula_k_t-t^H}
					k_{i} = k_{i}' +\sum_{b > i} \sum_{v\leq n_{ib}} k^L_{i,b,v} + \sum_{a <i} \sum_{v \leq n_{ai}} k^R_{a,i,v} + \sum_{v=1}^{n_{ii}} k_{i,i, v},
				\end{equation}
			where $k^L_{a,b \neq a,v}$, $k^R_{a,b \neq a,v}$ and $k_{a,a, v}$ satisfy
				\begin{equation*}
					\left\{ \begin{array}{l}
						(x_a, x_b, (y_r)_{r \in N_{a,b,v}} ;k^L_{a,b,v}, k^R_{a,b,v}, (p_r)_{r \in N_{a,b,v}}) \in \WF ( \delta^{|N_{a,b,v}|} d_\phi / \delta \phi^{|N_{a,b,v}|}) \\
						\qquad \qquad \qquad \qquad \mbox{ or } k^R_{a,b,v}, k^L_{a,b,v}, p_r = 0\\
						(x_a, (y_r)_{r \in N_{a,a,v}} ;k_{a, a, v}, (p_r)_{r \in N_{a,a,v}}) \in \WF( \delta^{|N_{a,a,v}|} d_\phi(x_a, x_a) / \delta \phi^{|N_{a,a,v}|})\\
						\qquad \qquad \qquad \qquad \mbox{ or } k_{a,a,v}, p_r = 0\\
						(x_1, \dots, x_n, (y_r)_{r \in N}; k'_1 \dots, k'_n, (p_r)_{r \in N}) \in \WF ( \delta^{|N|} \tau^H_{\phi, {\bf j}'} / \delta\phi^{|N|} )\\
						\qquad \qquad \qquad \qquad \mbox{ or } k'_1, \dots, k'_n, p_r =0. \\
					\end{array} \right.
				\end{equation*}
			We prove that $(x_1, \dots, x_n, y_1, \dots, y_\nu; k_1, \dots, k_n, p_1, \dots, p_\nu)$ cannot belong to the set $C^+_{n + \nu}$ defined by~\eqref{C_set_def}. By definition of $C^+_{n + \nu}$, we need to consider just the following two cases: (a) all the covectors $k_1, \dots, k_n$, $p_1, \dots, p_\nu$ are causal future-directed except at most one among $k_1,\dots, k_n, p_{r \in N}$ which can be space-like, and (b) there exist $a' \in I$, $b' \in I^c$, $v' \leq n_{a' b'}$ such that one and only one covector $p_s$ with $s \in N_{a',b',v'}$ is space-like whereas the other covectors are all causal future-directed. We prove that neither of these two cases can be realized:
			\begin{enumerate}[label=(\alph*), start=1]
				\item Since, we assume $p_r \in \overline{V}^+$ for any $r \in N_{a,b,v}$ and for any $a,b,v$, the estimates~\eqref{var_d_nu} and~\eqref{var_est_d_coinc} for the wave-front set of $\delta^\nu d_\phi(x_1,x_2) / \delta \phi^\nu$ and $\delta^\nu d_\phi(x,x) / \delta \phi^\nu$ (see lemma~\ref{lemma_d} and in lemma~\ref{lemma_d_coinc}) imply that $k^L_{a,b,v}$, $k^R_{a,b,v}$ and $k_{a,a,v}$ belong to $\overline{V}^-$ for any $a,b, v$. Because we also assume $k_1, \dots, k_m,p_1, \dots, p_\nu \in \overline{V}^+$ except at most one covector among $\{k_1, \dots, k_m\}$ or $\{ p_r\}_{r \in N}$ which can be space-like, eq.~\eqref{formula_k_t-t^H} implies that all $k'_1, \dots, k'_m$ must be in $\overline{V}^+$ except for at most one $k'_\ell$ which can be space-like. Actually, $k'_\ell$ can be space-like only if $k_\ell$ is space-like. However, these configurations are incompatible with condition~\ref{WF_delta_t_H} for the variational derivatives of $\tau^H_{\phi, {\bf j}}$. This concludes the proof that the case (a) cannot be realized.
				\item We first assume that there exist $a', b',v'$ with $b' \neq a'$ such that the unique space-like covector $p_s$ among $p_{1}, \dots p_\nu$ has $s \in N_{a', b' \neq a', v'}$. Since we assume $p_r \in \overline{V}^+$ for all $r \neq s$, using estimates~\eqref{var_d_nu} and~\eqref{var_est_d_coinc} (see lemma~\ref{lemma_d} and in lemma~\ref{lemma_d_coinc}), we obtain $k^L_{a',b',v'}, k^R_{a',b',v'} \notin \overline{V}^+$ and $k^L_{a,b,v}$, $k^R_{a,b,v}$, $k_{a,a,v} \in \overline{V}^-$ for any $(a,b,v) \neq (a',b',v')$. Since, by hypothesis, $k_\ell$ is in $\overline{V}^+$ for all $\ell$, eq.~\eqref{formula_k_t-t^H} implies
				\begin{equation}\label{req}
					k'_{a'} + k^L_{a', b', v'} \in \overline{V}^+, \quad k'_{b'} + k^R_{a', b', v'} \in \overline{V}^+, \quad k'_{i \neq a', b'} \in \overline{V}^+.
				\end{equation}
			If $x_{a'} \neq x_{b'}$, condition~\ref{WF_delta_t_H} requires that either $k'_{a'}$ or $k'_{b'}$ is in $\overline{V}^-$. Thus, we conclude that the requirement~\eqref{req} cannot be realized. On the other hand, if $x_{a'} = x_{b'}$, then estimate~\eqref{var_d_nu} implies $k^L_{a', b', v'} + k^R_{a', b', v'} \notin \overline{V}^+$, whereas condition~\ref{WF_delta_t_H} for $\delta^\nu \tau^H_{\phi, {\bf j}}/ \delta \phi^\nu$ requires $k'_{a} + k'_{b'} \in \overline{V}^-$. It follows that the requirements~\eqref{req} cannot be fulfilled also in this case.\\
			Now, we assume that there exist $a',v'$ such that the unique space-like covector $p_s$ has $s \in N_{a', a', v'}$. By hypothesis, $k_\ell, p_r$ are in $\overline{V}^+$ for any $\ell$ and any $r \neq s$. Thus, we obtain $k_{a', a', v'} \notin \overline{V}^+$ using the estimates~\eqref{var_d_nu},~\eqref{var_est_d_coinc} (see lemma~\ref{lemma_d} and in lemma~\ref{lemma_d_coinc}) and the hypotheses on $k_\ell, p_r$. Similarly as just done for the case $b' \neq a'$, it follows from the assumptions chosen that all $k^L_{a,b,v}$, $k^R_{a,b,v}$ and all $k_{a,v}$ with $(a,v) \neq (a',v')$ are in $\overline{V}^-$. Then, eq.~\eqref{formula_k_t-t^H} implies the following conditions which replace~\eqref{req} in this case:
				\begin{equation}\label{req_2}
					k'_{a'} + k_{a', a', v'} \in \overline{V}^+, \quad k'_{i \neq a'} \in \overline{V}^+.
				\end{equation}
			These conditions~\eqref{req_2} are incompatible with property~\ref{WF_delta_t_H} for the variational derivatives of $\tau^H_{\phi, {\bf j}}$. This concludes the proof that the case (b) cannot be realize.
			\end{enumerate}	
		With a similar argument we can prove that $(x_1, \dots, x_n, y_1, \dots, y_\nu; k_1, \dots, k_n, p_1, \dots, p_\nu)$ cannot belong to the set $C^-_{n + \nu}$ defined by~\eqref{C_set_def}. By the definition of the set $W_{n+\nu}$ as the complement of $C^+_{n +\nu} \cup C^-_{n+\nu}$, see eq.~\eqref{W_set_def}, we have proven that each distribution~\eqref{var_term_t-t^H} has wave-front set contained in $W_{n + \nu}$. Thus, we proved estimate~\eqref{est_WF_W} in $\cU_m$.\\
		
		To prove estimate~\eqref{est_WF_W_s} for $\nu >0$, let be $\bR \ni \epsilon \mapsto \phi(\epsilon) \in C^\infty(M)$ smooth. The distribution $\tau_{\phi(\epsilon), {\bf j}}$ is defined as in eq.~\eqref{t-t^H} with $\phi$ replaced by $\phi(\epsilon)$ everywhere, i.e. $\tau_{\phi(\epsilon), {\bf j}}$ is a finite sum of products of $\tau^H_{\phi(\epsilon), {\bf j}' \leq {\bf j}}$ and $d_{\phi(\epsilon)}^{n}$ with appropriate coefficients. For any $\nu$, the variational derivatives $\delta^\nu \tau_{\phi(\epsilon), {\bf j}}(x_1, \dots, x_n) / \delta \phi(y_1) \dots \delta \phi(y_\nu)$ is a finite sum of terms in the form~\eqref{var_term_t-t^H} with the only difference that $\phi$ is replaced by $\phi(\epsilon)$ everywhere. We can show that estimate~\eqref{est_WF_W_s} holds, adapting the argument we used to prove estimate~\eqref{est_WF_W}: instead of using estimates~\eqref{var_d_nu},~\eqref{var_est_d_coinc} and condition~\ref{WF_delta_t_H}, we use estimates~\eqref{var_d_nu_s_phi},~\eqref{var_est_d_coinc_s_phi} and estimate~\eqref{est_var_delta_t_H_s_phi} of condition~\ref{WF_delta_t_H_s}. This conclude the proof of lemma~\ref{lemma_t_W_neigh}.
		\end{proof}
		To conclude the proof of prop.~\ref{prop_t_W-smooth}, we still need to prove the claims outside the neighbourhood $\cU_m$ of the total diagonal:
		\begin{lemma}\label{lemma_t_W_out}
			The distributions $\{ \tau_{\phi, {\bf j}} \}_{\bf j}$ satisfy estimates~\eqref{est_WF_W} and~\eqref{est_WF_W_s} outside $\cU_n$.
		\end{lemma}
		\begin{proof}
		We proceed by induction on the number $n$ of the factors involved in the time-ordered product $T_{n,\phi}[\otimes_{i=1}^n \varphi^{k_i}(x_i)]$ corresponding to the distributions $\{\tau_{\phi, {\bf j}}\}_{{\bf j} \leq {\bf k}}$. For $n=1$, a direct inspection of formula~\eqref{Wick_power_fact} defining the Wick product shows that
			\begin{equation}\label{t_Wick_prod}
				\tau_{\phi, j}(x) = \left\{ \begin{array}{ll} 
				d_\phi(x,x)^{j'} & \mbox{ if } j = 2j'\\
				0 & \mbox{ otherwise.} 
				\end{array} \right.
			\end{equation}
		In the proof of prop.~\ref{prop_W_Wick_power}, we have already proved that $d_\phi(x,x)^{j'}$ satisfies estimates~\eqref{est_WF_W_Wick} and~\eqref{est_WF_W_Wick_s}, which are precisely estimates~\eqref{est_WF_W} and~\eqref{est_WF_W_s} for $n=1$.\\
		We now assume that $\tau_{\phi, {\bf j}}$ corresponding to time-ordered products with less than $n$ factor satisfies both estimates~\eqref{est_WF_W} and~\eqref{est_WF_W_s}. Since $(x_1, \dots, x_n)$ does not belong to the neighbourhood $\cU_n$ of the total diagonal, we use the causal factorization axiom (T8) to express $\tau_{\phi, {\bf j}}(x_1, \dots, x_n)$ as finite sum of terms in the form
				\begin{equation}\label{t_outside_factor}
					f_{I}(x_1, \dots, x_n) \tau_{\phi, {\bf i}}((x_{a})_{a \in I}) \tau_{\phi, {\bf i}'}((x_{b})_{b \in I^c}) \prod_{a\in I, b\in I^c} \omega^R_\phi(x_a, x_b)^{n_{ab}}
				\end{equation}
			where $I \subset \{1, \dots, n\}$ proper, where $\{f_{I}\}_I$ is a partition of unity subordinate to the covering $\{ C_I\}_I$ of $M^n \backslash \Delta_n$ defined by~\eqref{C_I_sets}, where ${\bf i}, {\bf i}'$ and $\{n_{ab}\}$ satisfy $j_a =i_a + \sum_c (n_{a c} + n_{c a})$ for $a \in I$ and $j_b =i'_b + \sum_c (n_{b c} + n_{c b})$ for $b \in I^c$.
			We compute $\delta^\nu \tau_{\phi, {\bf j}}/ \delta \phi^\nu$ by distributing the Gateaux derivatives among the factors of~\eqref{t_outside_factor}. In detail, $\delta^\nu \tau_{\phi, {\bf j}}(x_1, \dots, x_n)/ \delta \phi(y_1) \dots \delta \phi(y_\nu)$ is given by a finite sum of terms in the form
				\begin{equation}\label{var_t_outside_factor}
					\begin{split}
						&f_{I}(x_1, \dots, x_n) \frac{\delta^{|N_1|} \tau_{\phi, {\bf i}}(\{x_{a \in I}\})}{\delta \phi^{|N_1|}(\{y_{r \in N_1}\})} \frac{\delta^{|N_2|} \tau_{\phi, {\bf i}'}(\{x_{b \in I^c}\})}{\delta^{|N_2|}(\{y_{r \in N_2}\})} \prod_{a \in I, b\in I^c} \prod_{v \leq n_{ab}} \frac{\delta^{|N_{a, b, v}|} \omega^R_\phi(x_a, x_b)}{\delta \phi^{|N_{a, b, v}|}(\{y_{r \in N_{a,b,v}}\})},
					\end{split}
				\end{equation}
			where $N_1, N_2, \{N_{a, b, v}\}$ form a disjoint partition of $\{1, \dots, \nu\}$.\\
			We now prove the estimate~\eqref{est_WF_W} by verifying it on each term in the form~\eqref{var_t_outside_factor}. Let $(x_1, \dots, x_n, y_1, \dots, y_\nu; k_1, \dots, k_n, p_1, \dots, p_\nu)$ be an element in the wave-front set of the distribution~\eqref{var_t_outside_factor}. The wave-front set calculus (thm.~\ref{theo_WF_horma}) implies that there exist the following decompositions
				\begin{equation}\label{condition}
					k_{a} = k_{a}'+\sum_{b \in I^c, v \leq n_{a b}} k^L_{a,b,v}, \qquad k_{b} = k_{b}''+\sum_{a \in I, v \leq n_{a b}} k^R_{a,b,v},
				\end{equation}
			and it holds
				\begin{equation*}
					\left\{ \begin{array}{l}
						(x_a, x_b, (y_r)_{r \in N_{a,b,v}} ; k^L_{a,b,v}, k^R_{a,b,v}, (p_r)_{r \in N_{a,b,v}}) \in \WF ( \delta^{|N_{a,b,v}|} \omega^R_\phi / \delta \phi^{|N_{a,b,v}|})\\
						\qquad \qquad \qquad \qquad \mbox{ or } k^L_{a,b,v}, k^R_{a,b,v}, p_r = 0\\
						((x_{a})_{a \in I}, (y_r)_{r \in N_1} ; (k'_a)_{a \in I}, (p_r)_{r \in N_1}) \in \WF ( \delta^{|N_1|} \tau^H_{\phi, {\bf i}} / \delta\phi^{|N_1|} )\\
						\qquad \qquad \qquad \qquad \mbox{ or } k'_{a}, p_r =0 \\
						((x_{b})_{b \in I^c}, (y_r)_{r \in N_2}); (k''_b)_{b \in I^c}, (p_r)_{r \in N_2}) \in \WF ( \delta^{|N_2|} \tau^H_{\phi, {\bf i}'} / \delta\phi^{|N_2|} )\\
						\qquad \qquad \qquad \qquad \mbox{ or } k''_{b}, p_r =0 \\
					\end{array} \right.
				\end{equation*}
			Remember that the estimate~\eqref{WF_better} holds for $\omega_\phi^R$ because $\phi \mapsto \omega_\phi^R$ is an admissible assignment as follows lemma~\ref{lemma_ret_state_fedosov} and lemma~\ref{lemma_state}.\\
			We prove by reductio ad absurdum that $(x_1, \dots, x_n, y_1, \dots, y_\nu; k_1, \dots, k_n, p_1, \dots, p_\nu)$ cannot belong to the set $C^+_{m+\nu}$ defined by~\eqref{C_set_def}. We consider the following two cases separately: (a) all the covectors $k_1, \dots, k_n, p_1, \dots, p_\nu$ are causal future-directed except at most one among $k_1,\dots, k_n, p_{r \in N_1 \cup N_2}$ which can be space-like, and (b) there exist $a' \in I$, $b' \in I^c$, $v' \leq n_{a' b'}$ such that one covector $p_s$ with $s \in N_{a',b',v'}$ is space-like whereas the other covectors are all causal future-directed. We show that both of the two cases contradict the inductive hypothesis. 
		\begin{enumerate}[label=(\alph*), start=1]
			\item Since we assume $p_r \in \overline{V}^+$ for any $r \in N_{a,b,v}$ and any $a,b,v$, the estimate~\eqref{WF_better} implies $k^R_{a,b,v} \in \overline{V}^-$ for all $a,b,v$. Furthermore, using eq.~\eqref{condition}, we obtain
				\begin{equation}\label{condition_2}
					k''_b = k_{b} - \sum_{a \in I, v \leq n_{ab}} k^R_{a,b,v},
				\end{equation}
			By hypothesis, the covectors $k_b, p_r$ with $b \in I^c$ and $r \in N_2$ belong to $\overline{V}^+$ except at most one which is space-like. It follows that all the covectors $k''_{b}, p_{r}$ with $b \in I^c$ and $r \in N_2$ must belong to $\overline{V}^+$ except at most one which is space-like. However, these configurations are incompatible with the inductive hypothesis on $\tau^H_{\phi, {\bf i}'}$, more precisely they violate the requirement $\WF ( \delta^{|N_2|} \tau^H_{\phi, {\bf i}'} / \delta\phi^{|N_2|} ) \subset W_{|I^c| + |N_1|}$. This is precisely what we wanted to show.
			\item By hypothesis, the unique space-like covector is $p_s$ for a certain $s \in N_{a',b',v'}$. Since we assume $p_r \in \overline{V}^+$ for any $r \neq s$ and $p_s$ space-like, estimate~\eqref{WF_better} implies $k^R_{a, b,v} \in \overline{V}^-$ for all $(a, b,v) \neq (a',b',v')$, whereas $k^R_{a',b',v'} \notin \overline{V}^+$. Combining these results with the assumption $k_1, \dots, k_n \in \overline{V}^+$ and using eq.~\eqref{condition_2}, we obtain $k''_{b} \in \overline{V}^+$ for any $b \neq b'$ and $k''_{b'} \notin \overline{V}^-$. By hypothesis, $p_{r} $ is in $\overline{V}^+$ for any $r \in N_2$. Therefore, we obtain again that all the covectors $k''_{b}, p_{r}$ with $b \in I^c$ and $r \in N_2$ must belong to $\overline{V}^+$ except at most one which is space-like. Thus, the assumptions (b) contradicts the inductive hypothesis on $\tau^H_{\phi, {\bf i}'}$ as we wanted to show.
		\end{enumerate}
		With a similar argument we can prove that $(x_1, \dots x_n, y_1, \dots, y_\nu; k_1, \dots, k_n, p_1, \dots, p_\nu)$ does not belong to $C^-_{n+\nu}$, defined by~\eqref{C_set_def}. Thus, by definition of the sets $W_{\nu}$, see~\eqref{W_set_def}, the wave-front set of each term in~\eqref{var_t_outside_factor} is contained in $W_{n+\nu}$, which is precisely what is needed to prove estimate~\eqref{est_WF_W}.\\
		The proof of estimate~\eqref{est_WF_W_s} can be obtained with a similar argument as the one just presented, based on estimates~\eqref{est_WF_W_s} and~\eqref{WF_better_s} (the latter is a consequence of the fact that $\phi \mapsto \omega^R_\phi$ is an admissible assignment as proved in lemma~\ref{lemma_ret_state_fedosov} and lemma~\ref{lemma_state}) instead of estimates~\eqref{est_WF_W} and~\eqref{WF_better}.
		\end{proof}
		This concludes the proof of prop.~\ref{prop_t_W-smooth}.
		\end{proof}
		
	\subsection{Local functionals involving covariant derivatives and fulfilment of axioms (T10) and (T11c)}\label{subsubsec_T10,T11}
		In this subsection, we first extend the previous construction to the case of local functional containing covariant derivatives. In particular, we verify that there is a prescription which in addition satisfies the Leibniz-rule axiom (T10) and which still satisfies the $W$-smoothness requirement. In this context, we consider more general local functionals in the form
			\begin{equation}\label{der_loc_func}
				F(\varphi) = \int_M f(x) \cdot C(x) \cdot (\nabla)^{r_1}\varphi(x) \cdots (\nabla)^{r_k} \varphi(x) dx = \int_M f(x) \cdot C(x) \cdot \prod_j \left( (\nabla)^{j} \varphi(x) \right)^{\kappa_{ij}} dx,
			\end{equation}
		where $(\nabla)^{r}$ is a short-hand notation for the symmetrized $r$-th covariant derivative (namely, the Levi-Civita connection of $g$), where $\boldsymbol{\kappa} = (\kappa_0, \kappa_1, \dots)$ is a multi-index, where $C$ is an arbitrary curvature tensor, where $f$ is a smooth compactly supported tensor field, and where  ``$\cdot$'' means ``contractions of space-time indices'' (note that there are no free space-time indices in $F(\varphi)$). Any possible local functional can be written as a finite sum of terms in form~\eqref{der_loc_func}. To simplify the notation in the following, we denote $\prod_j ( (\nabla)^{j} \varphi )^{\kappa_{ij}}$ by $\nabla^{\boldsymbol{\kappa}} \varphi$ and $C \cdot \nabla^{\boldsymbol{\kappa}} \varphi$ by $\Phi$.\\
		As proved in~\citep[prop. 3.1]{HW05}, there exists a prescription for time-ordered products satisfying also the Leibniz-rule axiom (T10). The construction is given inductively on the number of factors involved in the time-ordered product similarly as the case discussed in sec.~\ref{subsubsec_rew_T-prod}.\\
		First of all, we note that the Leibniz-rule axiom (T10) can be imposed consistently with the axiom (T1)-(T9) on time-ordered products involving only one factor. This is done by defining the Wick monomial $T_{1,\phi}[f \cdot \Phi]$ as in eq.~\eqref{Wick_power_fact}, but the distribution $f(x) \delta(x, x_1, \dots, x_{k-2n})$ is now replaced by a more general distribution in the form $f(x) \cdot C(x) \cdot (\nabla_{x_1})^{r_1} \cdots (\nabla_{x_{k-2n}})^{r_{k-2n}} \delta(x, x_1, \dots, x_{k-2n})$, and $d_\phi(x_1,x_2)$ is replaced by an appropriate product of its symmetrized covariant derivatives $(\nabla_{x_1})^{r} (\nabla_{x_2})^{r'} d_\phi(x_1,x_2)$. This definition is the same presented in~\citep[eq. (60)-(61)]{HW05}. The modification just outlined does not affect the proof of the on-shell $W$-smoothness because the operator $\nabla^{(x_\ell)}$ commutes with the variational derivative $\delta / \delta \phi(y)$ and differential operators do not enlarge the wave-front set (see~\citep[8.1.11]{H83}).\\
		Next, we consider time-ordered products $T_{n,\phi}[ \otimes_{i=1}^n f_i \cdot \Phi_i]$ with $n >1$. As already mentioned the construction is given by induction on the number of factors involved. Assuming that the time-ordered products satisfying the axioms (T1)-(T10) and involving less than $n$ factors are already given, one constructs the time-ordered products involving $n$ factors by proceeding similarly as done in sec.~\ref{subsubsec_rew_T-prod}: exploiting the causal factorization axiom (T8), the inductive hypothesis fixes the time-ordered products for local functionals $\otimes_{i=1}^n f_i \cdot \Phi_i$ supported outside the total diagonal. Then, in a sufficiently small neighbourhood of the total diagonal one performs the local Wick expansion. By the causal factorization axioms (T8) and the inductive hypothesis, the distributional coefficients of the local Wick expansion are known outside the total diagonal. The time-ordered products with $n$ factors are obtained by constructing extensions of these distributional coefficients on the total diagonal  which implement axiom (T1)-(T5) and (T10). As already pointed out in sec.~\ref{subsubsec_rew_T-prod}, it is proved in~\citep{HW02}[sec. 3.1] that axioms (T6)-(T7) can be enforced by simple redefinitions and axiom (T8) holds by construction.\\
		We need to show what is the constraint imposed by axiom (T10) on the distributional coefficients of the local Wick expansion. In this context, the local Wick expansion of $T_{n,\phi}[ \otimes_{i=1}^n f_i \cdot \Phi_i]$, for $f_1 \otimes \cdots \otimes f_n$ supported in a sufficiently small neighbourhood of the total diagonal, is given by (cf.~\eqref{local_Wick_exp})
			\begin{equation}\label{local_wick_exp_der}
		 		\begin{split}
		 			&T_{n,\phi} \left[\bigotimes_{i=1}^n f_i \cdot \Phi_i \right] = T_{n,\phi} \left[\bigotimes_{i=1}^n f_i C_i \cdot \nabla^{\boldsymbol{\kappa}_i} \varphi \right]\\
		 			&\quad = \sum \sC_{\boldsymbol{\alpha}_1 \dots \boldsymbol{\alpha}_n} \int_{M^n} \prod_i f_i(x_i) \cdot \tau^H_\phi [\otimes_i C_i \cdot \nabla^{\boldsymbol{\alpha}_i} \varphi ] (x_1, \dots, x_n) : \prod_{i=1}^n \nabla^{\boldsymbol{\kappa}_{i} - \boldsymbol{\alpha}_{i}} \varphi(x_i):_{H_\phi} dx_1 \dots dx_n\\
		 			&\quad = \sum \sC_{\boldsymbol{\alpha}_1 \dots \boldsymbol{\alpha}_n} \int_{M^{n + \sum_i \ell_i}}  \prod_{i} f_i(x_i) \cdot (\nabla_{x^{(1)}_i})^{r_{i,1}} \cdots (\nabla_{x^{(\ell_i)}_{i}})^{r_{i, \ell_i}} \delta(x_i, x^{(1)}_i, \dots, x^{(\ell_i)}_{i})) \cdot \\
		 			&\qquad \cdot \tau^H_\phi [\otimes_i C_i \cdot \nabla^{\boldsymbol{\alpha}_i} \varphi] (x_1, \dots, x_n) : \prod_{i=1}^m  \varphi(x_i^{(1)}) \cdots \varphi(x_i^{(\ell_i)}):_{H_\phi} \prod_i dx_i dx_i^{(1)} \dots dx_i^{(\ell_i)},
		 		\end{split}
		 	\end{equation}
		 where the sum is over the multi-indices such that $\boldsymbol{\alpha}_i \leq \boldsymbol{\kappa}_i$, where $\sC_{\boldsymbol{\alpha}_1 \dots \boldsymbol{\alpha}_n} = \frac{1}{\boldsymbol{\alpha}_1 ! \cdots \boldsymbol{\alpha}_n!}$, where $\boldsymbol{\alpha} ! = \alpha_1! \dots \alpha_n!$, and where $\ell_i = |\boldsymbol{\kappa}_i - \boldsymbol{\alpha}_i|$. In this setting, the distributional coefficients are the distributions $\tau^H_{\phi}[\otimes_i C_i \cdot \nabla^{\boldsymbol{\alpha}_i} \varphi] = C_1 \otimes \cdots \otimes C_m \cdot \tau^H_{\phi}[\otimes_i \nabla^{\boldsymbol{\alpha}_i} \varphi]$. Note that this is consistent with the situation discussed in sec.~\ref{subsubsec_rew_T-prod}. In fact, $\tau^H_{\phi} [ \otimes_i C_i \cdot (\nabla)^{\boldsymbol{\alpha}_i} \varphi]$ with $\boldsymbol{\alpha}_i = (j_i, 0, \dots)$ and $C_i =1$, i.e. $\tau^H_{\phi} [ \otimes_i \varphi^{j_i}]$, is the distributional coefficient $\tau^H_{\phi, {\bf j}}$ of the local Wick expansion defined by~\eqref{local_Wick_exp} for a time-ordered product of local functionals not involving covariant derivatives.\\
		As proved in~\citep[prop. 3.1]{HW05}, the Leibniz-rule axiom (T10) imposes the following additional constraint for the distributional coefficient of the local Wick expansion
		 	\begin{equation}\label{t_10}
		 		\nabla_{x_j} \tau^H_{\phi}[\otimes_i \Phi_i](x_1, \dots, x_n) = \tau^H_{\phi}[\Phi_1 \otimes \cdots \otimes \nabla_{x_j} \Phi_j \otimes \cdots \Phi_n] (x_1, \dots, x_m).
		 	\end{equation}
		As explained in the proof of~\citep[prop. 3.1]{HW05}, the suitable extensions are obtained by induction on the number of covariant derivatives acting on $\varphi$. Ultimately, the extension is provided combining the procedure based on the scaling expansion, as in sec.~\ref{subsubsec_rew_T-prod}, and using~\eqref{t_10} to define the right-hand side for the so-called ``Leibniz depended'' part (for more details see~\citep[prop. 3.1]{HW05}).\\
		Because the operator $\nabla_{x_j}$ commutes with the variational derivative $\delta / \delta \phi(y)$, and because differential operators do not enlarge the wave-front set (see~\citep[8.1.11]{H83}), it follows that we can adapt the argument given in sec.~\ref{subsubsec_proof} and we have that the variational derivatives $\delta^\nu \tau^H_{\phi}[\otimes_i \Phi_i] / \delta \phi^\nu$ satisfy the conditions~\ref{supp_delta_t_H}-\ref{WF_delta_t_H_s} we defined in sec.~\ref{subsubsec_var}.\\
		Using the analogue of eq.~\eqref{t-t^H} in this context, in a sufficiently small neighbourhood of the diagonal, we express the distributional coefficients $\tau_{\phi}[\otimes_i C_i \cdot \nabla^{\boldsymbol{\alpha}_i}\varphi]$ of the Wick expansion with respect to the retarded $2$-point function in terms of $(\nabla_{x_a})^{r_a} (\nabla_{x_b})^{r_b} d_\phi (x_1,x_b)$, for appropriate $r_a, r_b$, and $\tau^H_{\phi}[\otimes_i C_i \cdot \nabla^{\boldsymbol{\alpha}'_i} \varphi]$, with $\boldsymbol{\alpha}'_i \leq \boldsymbol{\alpha}_i$. We can extend the argument used in lemma~\ref{lemma_t_W_neigh} to the more general case of local functionals containing covariant derivatives proving that the distributional coefficient of the Wick expansion with respect to the retarded $2$-point function satisfies estimates~\eqref{est_WF_W} and~\eqref{est_WF_W_s} in a sufficiently small neighbourhood of the total diagonal. The causal factorization axiom (T8) implies that this result must hold also outside this neighbourhood, as can be checked similarly as we did in lemma~\ref{lemma_t_W_out}. Thus, there exists a prescription for time-ordered product satisfying axioms (T1)-(T10) and the desired on-shell $W$-smoothness, in the sense of thm.~\ref{thm_on_W_ret}.\\
		 
		To conclude the proof of thm.~\ref{thm_on_W_ret}, we need to show that if we start with a time-ordered prescription satisfying axioms (T1)-(T10) and the $W$-smooth condition, then the changes of prescription required to impose the axiom (T11c) (see app.~\ref{app_proof_R12}) preserve the $W$-smooth condition. Actually, in the proof of the consistency of axiom (T11c) with (T1)-(T10), we need axiom (T11a), which is enforced also using a change of prescription for time-ordered products, see~\citep{HW05}.\\
		Let $\{T_{n,\phi} \}_{n \in \bN}$ and $\{T'_{n, \phi}\}_{n \in \bN}$ be two prescription for time-ordered products satisfying axioms (T1)-(T10) for a fixed background $\phi \in C^\infty(M)$. These two prescriptions must be related by a hierarchy of maps $\{D_{n,\phi} \}_{n \in \bN}$ in the following way:
			\begin{equation}\label{ren_freedom_T_prod}
				T'_{n,\phi} \left[ \bigotimes_{i+1}^n F_i \right] = T_{n,\phi} \left[ \bigotimes_{i=1}^n F_i \right] + \sum_{I_0 \sqcup \dots \sqcup I_{k}=\{1, \dots, n\}} T_{|I_0|+1,\phi} \left[ \bigotimes_{\ell = 1}^k D_{|I_\ell|,\phi} \left( \bigotimes_{j \in I_\ell} F_j \right) \otimes \bigotimes_{i \in I_0} F_i \right],
			\end{equation}
		where the linear maps $D_{n,\phi} : \otimes^n \cF_\loc \to \cF_\loc$ satisfy the properties listed in thm.~\ref{theo_R12}. In particular, $D_{n,\phi}$ can be written as
			\begin{equation*}
				\begin{split}
					&D_{n,\phi} \left( \bigotimes_{i=1}^n f_i \cdot \Phi_i \right)(x) = D_{n,\phi} \left( \bigotimes_{i=1}^n f_i C_i \cdot \nabla^{\boldsymbol{\kappa}_i} \varphi \right) \\
					&\quad =  \sum \sC_{\boldsymbol{\alpha}_1 \dots \boldsymbol{\alpha}_n} \int f_1(x_1) \cdots f_n(x_n) \cdot c_\phi [\otimes_i C_i \cdot \nabla^{\boldsymbol{\alpha}_i} \varphi ] (x_1, \dots, x_n) \prod_{i=1}^n \nabla^{\boldsymbol{\kappa}_{i} - \boldsymbol{\alpha}_{i}} \varphi(x_i) dx_1 \dots dx_n,
				\end{split}
			\end{equation*}
		where the sum is over the multi-indices such that $\boldsymbol{\alpha}_i \leq \boldsymbol{\kappa}_i$, and where $\sC_{\boldsymbol{\alpha}_1 \dots \boldsymbol{\alpha}_n}$ is the same factor that appears in~\eqref{local_wick_exp_der}. Each distribution $c_\phi [\otimes_i C_i \cdot \nabla^{\boldsymbol{\alpha}_i} \varphi ] = C_1 \cdots C_n \cdot c_\phi [\otimes_i \nabla^{\boldsymbol{\alpha}_i} \varphi ]$ can be expressed as a sum of terms which are products of derivatives of the delta distribution and polynomials in the Riemann tensor, $m$, $\lambda$, $\phi$ and their covariant derivatives. Because of this polynomial behaviour in $\phi$ and in its covariant derivatives, a redefinition of the time-ordered prescription cannot spoil the on-shell $W$-smoothness. Therefore, we can conclude that any time-ordered prescription satisfying axioms (T1)-(T10), (T11a) and (T11c) (under the assumption that for any $\phi \in C^\infty(M)$ the product $\bullet_\phi$ in $\cW_\phi$ is given in terms of the retarded $2$-point function $\omega^R_\phi$)\footnote{As a matter of fact, it suffices that the algebra structure is given in terms of an admissible assignment $\phi \mapsto \omega_\phi$ in the sense of def.~\ref{def_suit_omega}.} gives on-shell $W$-smooth maps $S \ni  \phi  \mapsto T_{n,\phi}[\otimes_{i=1}^n F_i(\phi + \varphi)] \in \cW_\phi$ for any local functional $F_i$.
		
\chapter*{Conclusions and outlook}
\addcontentsline{toc}{chapter}{Conclusions and outlook}
	We conclude by presenting an overview of the results obtained in this work and indicate some open issues and possible directions for future investigations. Our main 
	result is that we succeeded in constructing a deformation quantization for a class functionals on the smooth solutions to the non-linear Klein-Gordon equation on the space-time $M$ which parallels a construction of Fedosov (devised originally for finite-dimensional phase spaces). We then  compared this approach to the causal approach to perturbative quantum field theory.\\
	
	We started by constructing a geometrical framework for the set $S$ of these solutions. The cornerstones of our set-up are the definition of the formal Wick algebra bundle $\cW = \sqcup_{\phi \in S} \cW_\phi$ over $S$, and the notion ``on-shell $W$-smoothness'' for functions on $S$, sections in $\cW$, or, more generally, forms with values in $\cW$. The elements of the formal Wick algebra $\cW_\phi$ are ultimately identified (up to formal power series in the formal parameter $\hbar$) with sequences of distributions on $M^n$ which satisfy a certain restriction of the wave-front set given by the collection $\{W_n\}$ of sets $W_n \in T^*M^n$. The product $\bullet_\phi$ in $\cW_\phi$ is constructed in terms of a pure Hadamard $2$-point function $\omega_\phi$. We imposed a further constraint considering only a particular class of assignments $\phi \mapsto \omega_\phi$, named ``admissible assignments'', which have a specific dependence on $\phi$. Using the methods of microlocal analysis, we were able to show that the fiberwise product endows the space of on-shell $W$-smooth sections in $\cW$ and, more generally, on-shell $W$-smooth forms with values in $\cW$ with a well-defined algebra structure, the product $\bullet$.\\
	Then, we proved that the recursion procedure to define the flat Fedosov connection in finite dimensions can be performed also in our infinite-dimensional set-up . The resulting connection $D^W$ is flat, preserves the on-shell $W$-smoothness, and is determined by the choice of the assignment $\phi \mapsto \omega_\phi$ (and some auxiliary data). We obtained a deformation quantization of the set of on-shell $W$-smooth functions on $S$ using the product $\bullet$ and inverting the map $\tau$ that projects a flat on-shell $W$-smooth section in $\cW$ into its $S \to \bC[[\hbar]]$ part, similarly as done in finite dimensions.\\
	
	We showed that different choices of (admissible) assignments $\phi \mapsto \omega_\phi$ and $\phi \mapsto \omega'_\phi$ give ``gauge equivalent'' Fedosov connections $D^W, D^{\prime W}$. This reflects in our set-up the equivalence of the Fedosov connections in finite dimensions corresponding to two different almost-K\"{a}hler structures which are both compatible with the same symplectic form. The gauge transformation is determined by the same recursion procedure as in the finite-dimensional case. The new result is that this recursion process remains well-defined in the infinite-dimensional setting.\\
	We then investigated the relation of Fedosov's approach to quantum field theory with the method of ``causal perturbation theory''.  In the latter method, for each classical local polynomial function $F$ on $S$, one constructs using Haag's formula its corresponding quantum observable $\hat{F}_\phi \in \cW_\phi$ for each classical background $\phi \in S$. We proved that the map $\phi \mapsto \hat F_\phi$ is on-shell $W$-smooth and it is ``gauge equivalent'' to a section $\hat{F}'$ in $\cW$ which is flat with respect to the Fedosov connection $D^{\prime W}$ corresponding to a generic admissible assignment $\phi \mapsto \omega'_\phi$. The flat sections of the form $\hat{F}'$ generate an algebra, with respect to the product $\bullet$, and we proved that Einstein causality holds in this algebra.\\
	
	Our results leave plenty of room for further investigations. First of all, we point out that we have constructed the infinite-dimensional set-up only for scalar field theories on ultra-static space-times with compact Cauchy surface with interactions given by a potential in the form $V(\phi) = \int \lambda(x) \phi^4(x)$, where $\lambda$ is a smooth compactly supported ``coupling constant''. Considering more general space-times and potentials could affect the infinite-dimensional manifold structure we have assigned to $S$. In our construction, in fact, the manifold structure of $S$ is related to the initial value problem of the non-linear equation of motion, which, in general, is not a priori globally well-defined for smooth (compactly supported) Cauchy data in arbitrary $M$.\\
	
	Another prospect is investigating further the relations between the sections $\tau^{\prime -1} F$ and  $\hat{F}'$ for a local functional $F$. Both of them are flat sections with respect to the same Fedosov connection $D^{\prime W}$ corresponding to an admissible assignment $\phi \mapsto \omega_\phi'$. A priori these two sections differ since their components proportional to the section $1$ are not equal. This fact is not surprising because to define $\hat{F}'$ we have implicitly chosen one of the many admissible (due to renormalization freedom) prescriptions for the retarded products, while that is not the case for $\tau^{\prime -1} F$, which is essentially unique. Nevertheless, there is a remaining freedom in defining the on-shell $W$-smooth section $H$ that appears in $\hat{F}' = \alpha^{-1} \exp (\frac{i}{\hbar} \ad_{\bullet_R}(H)) \hat{F}$. This freedom is characterized by the choice of a closed $1$-form $\theta$ with values in $\bC[[\hbar]]$. It would be interesting to see if it is possible to choose $\theta$ such that $\hat{F}'$ and $\tau^{\prime -1}F$ would coincide.\\
	
	It would also be interesting to analyse the problem of convergence in $\hbar$ of the star product. In our set-up, the star product for the algebra of functionals on the solutions $S$ is constructed in terms of the Wick product $\bullet$ for the algebra (flat) sections on formal Wick algebra bundle $\cW$. Roughly speaking, for a fixed $\phi \in S$, the formal Wick algebra $\cW_\phi$ can be interpreted as the algebra of formal polynomials on $T_\phi S$, i.e. it is not just a formal series in the parameter $\hbar$, but also in the degree of the polynomials. For increasing order in $\hbar$, also the polynomial order required for our constructions increases. Therefore, to even start talking about convergence, one has to replace ``polynomials'' by some class of more general ``functions''. The aim would then be to find a suitable topology to get hopefully convergence of the various series in $\hbar$. It is not obvious, however, how this could be done in practice, even in the finite-dimensional case. One approach in this direction has been suggested in~\citep{waldmann2014nuclear}. It would be interesting to see if it is possible to adapt this construction to our set-up. This must be left for a future work.\\
	
	Another open direction  is to extend our construction to quantum field theories with fermionic fields and/or gauge fields. For fermionic fields, two rather different approaches come to mind. On the one hand, one could consider ``classical'' fermions, i.e. solutions to a Dirac-type equation,  possibly non-linear. Whenever the Cauchy problem is well-posed (which it is clearly a non-trivial question), one can provide an infinite-dimensional manifold structure for the set $S$ of solutions to the non-linear equation. The tangent space $T_\phi S$ at a fixed solution $\phi$ of the non-linear equation would be again identified with the solutions of the linearised equation around $\phi$. Similarly as for the scalar case we discussed here, the causal  approach to the quantization of fermionic field theories (see~\citep{dappiaggi2009extended,sanders2010locally,rejzner2011fermionic,zahn2014renormalized}) might be expected to provide guidelines how to define $T^*_\phi S$, $\boxtimes_W^n T^*_\phi S$ and $\cW_\phi$ in the infinite-dimensional setting for the fermionic case.  For the linear Dirac equation, one still has the notion of the causal propagator and Hadamard $2$-point functions (see~\citep{hollands2001hadamard}). The main differences with the scalar field seem to be the following: (1) the algebra of classical observable has a graded structure, which should be also incorporated in Fedosov's method, and (2) the fundamental notions in our infinite-dimensional set-up, in particular the notion of on-shell $W$-smoothness, need to be extended to vector-valued distributions.\\
	There is, on the other hand, also a different possible approach. One could avoid introducing ``classical'' fermions --which seems, after all, physically questionable-- and introduce them only at the quantum level. This idea could be realized as follows: one considers for instance a supersymmetric theory containing both bosonic and fermionic degrees of freedom. Instead of considering all the possible solutions to the  equations of motions, one defines the classical solutions $S$ as those with vanishing fermionic components. Nevertheless, the fermionic degrees of freedom will appear at the linearised level, i.e. in the tangent space $T_\phi S$ at a classical (bosonic) solution $\phi$, and, therefore, also in the formal Wick algebra. It is not clear to us if and how Fedosov's method can be implemented in this situation.\\
	
	For gauge fields, the equations of motion are not hyperbolic, so one cannot directly proceed constructing perturbatively the quantum field. It is, however, well understood how to circumvent this problem by adding further fields (the ghost, anti-ghost, and auxiliary fields) to the theory in order to make the equations of motion hyperbolic. At the classical level, the unphysical fields can be removed by a symmetry, called ``BRST-symmetry'', which restores the gauge invariance: the gauge invariant classical observables are obtained as the cohomology (with respect to the BRST-operator) of the auxiliary algebra containing also the unphysical fields. This can be viewed as a symplectic-reduction of the unphysical phase space. To quantize this theory, one proceeds by defining the deformation quantization of the auxiliary algebra and a suitable deformed extension of the BRST-operator. This sophisticated and complex procedure is described in~\citep{hollands2008renormalizedYM, fredenhagen2013batalin} in the framework of the algebraic approach to quantum field theory. It is not obvious to us how our approach can be adapted to this case. There are some results in the literature, e.g.~\citep{bordemann2000brst}, for the finite-dimensional case, and maybe this could be used as a guideline for the case of field theory. We must leave this, too, to a future investigation.
	
\appendix

\chapter{Proof of theorem~\ref{theo_R12}}\label{app_proof_R12}
	In this appendix, we present the proof for the consistency of axioms~\ref{R0}-\ref{R12}, thm.~\ref{theo_R12}. As already mentioned, it is well-known that there exists a prescription for retarded products satisfying axioms~\ref{R0}-\ref{R11}. The assertion follows e.g. from~\citep{HW05}. Here, the authors proved the existence of a prescription for {\em time-ordered products} satisfying the corresponding axioms (T1)-(T10) (and also axiom (T11a), which will be required later) of~\citep{HW05}. The retarded products are obtained from the time-ordered products by
		\begin{equation}\label{retarded_vs_time-ordered}
			R_{n,m,\phi}\left(\bigotimes_{i=1}^n F_i;\bigotimes_{j=1}^m H_j\right) := \sum_{I \subset \{1, \dots, m\}} (-1)^{|I|} \overline{T}_{|I|,\phi} \left[ \bigotimes_{\ell \in I} H_\ell \right] \bullet_\phi T_{|I^c| + n,\phi} \left[ \bigotimes_{i} F_i \otimes \bigotimes_{j \in I^c} H_j \right],
		\end{equation}
	where $\overline{T}$ denotes the anti-time-ordered product, see e.g.~\citep[(T7)]{HW05}. Note that in sec.~\ref{subsec_int_QFT_per} we used the alternative notation $R_{m,\phi}=R_{1,m,\phi}$.\\
	We claim that it indeed follows from (T1)-(T10) that~\ref{R1}-\ref{R11} hold: the proof is not complicated and one can see that one by one the axioms (T1)-(T10) imply their counterparts~\ref{R1}-\ref{R10}. The requirement $R_{0,\phi}(\varphi(f)) = \varphi(f)$, which is the second ``initial condition'' in axiom~\ref{R0}, is a consequence of the implicit assumption $T_{1,\phi}(\varphi(f)) = \varphi(f)$. The GLZ formula~\ref{R11} and the requirement $R_{n,\phi}(A, \otimes_i H_i) = A \delta_{n,0}1$, which is the first ``initial condition'' in axiom~\ref{R0}, are consequence of the definition~\eqref{retarded_vs_time-ordered}.\\
	
	We want to construct a prescription for the retarded products which satisfies also~\ref{R12}. This will follow if the time-ordered products satisfies the following condition:
		\begin{equation}\label{T11c}
			\frac{\partial}{\partial s} \alpha^R_{\phi_s,\phi} T_{n,\phi_s} \left[ \bigotimes_{i=1}^n F_{i,\phi_s} \right] = \frac{i}{\hbar} R_{n,1,\phi} \left( \bigotimes_{i} F_{i,\phi} ; \frac{\partial I^{(2)}_{\phi_s}}{\partial s} \right) + \sum_{\ell =1}^n T_{n,\phi} \left[ \bigotimes_{i \neq \ell} F_{i,\phi} \otimes \frac{\partial F_{\ell,\phi_s}}{\partial s} \right],
		\end{equation}
	where $\phi_s$ is a smooth $1$-parameter family of backgrounds such that $\phi_0 = \phi$. As already stated in sec.~\ref{subsec_int_QFT_per}, the derivative $\partial/\partial s$ is always evaluated in $s=0$. This additional condition on the time-ordered product corresponds in~\citep{HW05} to the formulation of the principle of perturbative agreement for an external potential variation (T11c), i.e. for a variation in the $\varphi^2$-term of the Lagrangian. However, it was not demonstrated in~\citep{HW05} that condition (T11c) can actually be imposed. We now fill this gap following an analogous argument as given in~\citep{HW05} for the proof of condition (T11b).\\
	Consider local fuctionals in the form
		\begin{equation*}
			F_{i,\phi} = \int_M f_i(x) \cdot \Phi_{i,\phi}(x) dx = \int_M f_{i}(x) \cdot C_{i,\phi}(x) \cdot (\nabla)^{\kappa_{i1}} \varphi(x) \cdots (\nabla)^{\kappa_{ij}} \varphi(x) dx,
		\end{equation*}
	where $f_{i}$ is a generic compactly supported tensor field, and where $C_\phi$ is a generic tensor depending polynomially on the metric, the curvature tensors, $m^2$, $\phi$ and their derivatives. Then, we define
			\begin{equation}\label{deviation}
				\begin{split}
					D_{n,\phi}(h_\phi; f_{1}, \dots, f_{n}) &:= \frac{\partial}{\partial s} \alpha^R_{\phi_s,\phi} T_{n, \phi_s} \left[ \bigotimes_{i} F_{i,\phi_s} \right]  - \frac{i}{\hbar} R_{n,1,\phi} \left( \bigotimes_{i} F_{i,\phi} ; \varphi^2(h_\phi) \right) -\\
					&\quad -  \sum_{\ell} T_{n,\phi} \left[ \bigotimes_{i \neq \ell} F_{i,\phi} \otimes \frac{\partial F_{\ell,\phi_s}}{\partial s} \right],
				\end{split}
			\end{equation}
		where $h_\phi$ is the compactly supported smooth function defined by
			\begin{equation}\label{h_var}
				h_\phi(x) := \frac{\partial v_{\phi_{s}}(x)}{\partial s},
			\end{equation}
		which implies
			\begin{equation}\label{U_var}
				\varphi^2(h_\phi) = \frac{\partial I_{\phi_s}^{(2)}}{\partial s}.
			\end{equation}
		Thus, eq.~\eqref{T11c} holds if $D_{n,\phi} = 0$.\\
		In order to simplify the notation, in the following we do not explicitly write the dependence on $\phi$ and we just denote by the subscript $s$ the dependence on $\phi_s$ in local functionals, time-ordered products or retarded products.\\  			
		We first note that trivially $D_0=0$. Arguing as in~\citep{HW05}, we proceed by induction in the number $N$ of factors of $\varphi$ and its derivatives that appear in the collection $F_{1}, \dots, F_{n}$, and prove that a prescription $\{T_n\}_{n \in \bN}$ for the time-ordered products, which gives $D_n(<N) = 0$, can be adjusted to a new prescription $\{T'_n\}_{n \in \bN}$ such that $D'_n(\leq N) = 0$. More precisely, the new prescription is defined by subtracting $D_n(N)$ from the corresponding time-ordered products. We must show that the replacement we just described is admissible, i.e. consistent with the renormalization freedom characterizing the non-uniqueness of the time-ordered products prescription given originally in~\citep{HW01, HW05} or, more concisely, in~\citep[thm. 2]{hollands2015quantum}. To show this, we must prove the following conditions (see~\citep{HW05, zahn2013locally}):
			\begin{enumerate}[label=(d\arabic*)]
				\item\label{d_1} $D_n$ is a functional of $h, f_{1}, \dots, f_{n}$ supported on the total diagonal $\Delta_{n+1}$.
				\item\label{d_2} $D_n$ is a $c$-number, i.e. $D_n = c 1 \in \cW$.
				\item\label{d_3} $D_n$ is local and covariant and scales almost homogeneously with scaling degree equal to the sum of the engineering dimensions of the classical functionals $F_1, \dots, F_n$.
				\item\label{d_4} $D_n$ vanishes if one of the entries is in the form $\varphi(f) =\int_M f(x) \varphi(x)$.
				\item\label{d_5} $D_n$ is a distribution with smooth dependence upon the metric and the background $\phi$. 
				\item\label{d_6} $D_n$ has the appropriate symmetry.
			\end{enumerate}
		As we have already mentioned, the new prescription $\{ T'_{n} \}_{n \in \bN}$ is defined by
			\begin{equation*}
				T'_{n+1} \left[ \varphi^2(h) \otimes \bigotimes_{i=1}^n F_i\right] := T_{n+1} \left[ \varphi^2(h) \otimes \bigotimes_{i=1}^n F_i\right] + 2i D_{n}(h; f_1, \dots, f_n),
			\end{equation*}
		if one of the factor in the time-ordered product is $\varphi^2(h)$ for a function $h$ as in~\eqref{h_var}, and simply by $T'_{n} \left[ \otimes_{i=1}^n F_i\right] = T_{n} \left[ \otimes_{i=1}^n F_i\right]$ otherwise. This new prescription $\{ T'_{n} \}_{n \in \bN}$ satisfies $D'_n(h; f_1, \dots, f_n)=0$. We now prove the conditions~\ref{d_1}-\ref{d_6}.
		
		\paragraph*{Proof of~\ref{d_1}.}
			For a given $h$ as in~\eqref{h_var}, choose $f_{1}, \dots, f_{n}$ such that the support of $h \otimes f_{1} \otimes \cdots \otimes f_{n}$ does not intersect the total diagonal $\Delta_{n+1}$. We must be in one of the following cases:
				\begin{enumerate}[label=(\alph*)]
					\item There is a Cauchy surface $\Sigma$ such that $\supp h \subset J^+(\Sigma)$ and $\supp f_{i} \subset J^-(\Sigma)$ for all $i$.
					\item The same as (a), but with ``$+$'' and ``$-$'' interchanged.
					\item There is a Cauchy surface $\Sigma$ and a proper subset $I \subset \{1, \dots n\}$ such that $\supp h \subset J^+(\Sigma)$, $\supp f_{i} \subset J^+(\Sigma)$ for $i \in I$, and $\supp f_{j} \subset J^-(\Sigma)$ for $j \notin I$.
					\item The same as (c), but with ``$+$'' and ``$-$'' interchanged.
				\end{enumerate}
			In case (a), the infinitesimal variation $\varphi^2(h)$ of the quadratic term $I^{(2)}$ of the action occurs in the future of the support of all the functionals $F_{i}$ and, therefore,
				\begin{equation}\label{infinitesimal_separated_R}
						\frac{\partial}{\partial s} \alpha^R_{\phi_s,\phi} T_{n,s} \left[ \bigotimes_{i} F_{i,s} \right] -  \sum_{\ell} T_{n,\phi} \left[ \bigotimes_{i \neq \ell} F_{i,\phi} \otimes \frac{\partial F_{\ell,\phi_s}}{\partial s} \right] =0,
				\end{equation}	
			by the definition of the isomorphism $\alpha^R$ (see~\eqref{alpha_R_wick_per}). The remaining term in \eqref{deviation} also vanish because of the support properties of the retarded product. Thus, necessarily $D_n = 0$.\\
			In case (b), because of the separation of the support of the infinitesimal variation $h$ and the supports of the functionals $F_{1} \dots F_{n}$, the third term in \eqref{deviation} must vanish and  the ``time-reversed'' version of \eqref{infinitesimal_separated_R} holds, i.e.
				\begin{equation*}
				\frac{\partial}{\partial s} \alpha^A_{\phi_s,\phi} T_{n,s} \left[ \bigotimes_{i} F_{i,s} \right] -  \sum_{\ell} T_{n,\phi} \left[ \bigotimes_{i \neq \ell} F_{i,\phi} \otimes \frac{\partial F_{\ell,\phi_s}}{\partial s} \right] = 0,
				\end{equation*}
			where $\alpha^A_{\phi,\phi'}$ is the isomorphism of $\cW_\phi \to \cW_{\phi'}$ constructed similarly as done for $\alpha^R_{\phi, \phi'}$ identifying those algebras in a neighbourhood of a Cauchy surface not intersecting the past of the support of the interaction $V$, i.e. via the so-called ``advanced state'' (or ``out-state''). Using the explicit formula~\eqref{alpha_R_wick_per} for $\alpha^R$ (and its analogue for $\alpha^A$), we get that for any $t \in \cW_\phi$ it holds
				\begin{equation*}
					\frac{\partial}{\partial s} \alpha^R_{\phi_s,\phi} \circ (\alpha^A_{\phi_s,\phi})^{-1} t = \frac{i}{\hbar} \left[ \frac{\partial I^{(2)}_{s}}{\partial s}, t \right]_{\bullet}.
				\end{equation*}
			Under the hypothesis of (b), it follows
				\begin{equation*}
					\begin{split}
						&\frac{\partial}{\partial s} \alpha^R_{\phi_s,\phi} T_{n,s} \left[ \bigotimes_{i} F_{i,s} \right] -  \sum_{\ell} T_{n,\phi} \left[ \bigotimes_{i \neq \ell} F_{i,\phi} \otimes \frac{\partial F_{\ell,\phi_s}}{\partial s} \right] =\\
						&= \frac{\partial}{\partial s} \alpha^R_{\phi_s,\phi} \circ (\alpha^A_{\phi_s,\phi})^{-1} \circ \alpha^A_{\phi_s,\phi} T_n \left[ \bigotimes_{i} F_{i} \right] -  \sum_{\ell} T_{n,\phi} \left[ \bigotimes_{i \neq \ell} F_{i,\phi} \otimes \frac{\partial F_{\ell,\phi_s}}{\partial s} \right]\\
						&= \frac{i}{\hbar} \left[ \frac{\partial I^{(2)}_{s}}{\partial s}, T_n \left[ \bigotimes_{i} F_{i} \right] \right]_{\bullet} + 0\\
						&= \frac{i}{\hbar} R_{n,1} \left( \bigotimes_{i} F_{i} ; \frac{\partial I^{(2)}_{s}}{\partial s} \right),
					\end{split}
				\end{equation*}
			where we also used the causal factorization property (T8) for the time-ordered products and formula~\eqref{retarded_vs_time-ordered}. Therefore, we have $D_n = 0$.\\
			Finally, it is similarly seen in cases (c) and (d) that $D_n=0$ holds as consequence of the causal factorization properties for the time-ordered products, the inductive hypothesis, and the fact that $\alpha^{R/A}$ are $*$-isomorphisms. The details are similar as in~\citep[sec. 6.2.2]{HW05}, so we omit.\\
			We have proved that $D_n(h, f_{1}, \cdots, f_{n}) = 0$ if the support of $h \otimes f_{1} \otimes \cdots \otimes f_{n}$ does not intersect the total diagonal $\Delta_{n+1}$. This clearly implies that the functional $D_n$ must be supported on the total diagonal as we wanted to prove.
			
			\paragraph*{Proof of~\ref{d_2}.}
				We proceed first by giving an equivalent characterization of $c$-numbers in $\cW_\phi$, which corresponds to \citep[prop. 2.1]{HW01} in our framework.
				\begin{lemma}\label{lemma_c-num}
					Let $t$ be an element of $\cW_\phi$ such that $[t, \varphi(f)]_{\bullet_\phi} = 0$ for any $f \in C^\infty_0(M)$, where $\varphi(f)$ is the equivalence class in $C^\infty_0(M)/(\boxempty - m^2 - v_\phi) C^\infty_0(M)$ corresponding to $f$. It holds $t=c 1$ with $c \in \bC[[\hbar]]$.
				\end{lemma}
				\begin{proof}
					Due to the $\Deg$-filtration, any element $t \in \cW_\phi$ is identified with the series\footnote{We used an informal notation here. The series should be interpreted as a sequence of distributions in $\cE'_W$.} $\sum_k \sum_{n =0}^k t^{(k),n}$, where each $t^{(k),n}$ is homogeneous in the total degree $\Deg$ and in the symmetric degree $\deg_s$, i.e. $t^{(k),n}$ is a symmetric distribution in $\cE'_W(M^n)$ (modulo elements of the ideal defined by $(\boxempty - m^2 - v_\phi) \cE'_W(M^n)$) up to a factor which is a power of $\hbar$. Since $[\cdot, \varphi(f)]_{\bullet_\phi}$, seen as a map on $\cW_\phi$, preserves the $\Deg$-grading and reduces by $1$ the $\deg_s$-grading, $[t, \varphi(f)]_{\bullet_\phi}$ vanishes if and only if $[t^{(k),n}, \varphi(f)]_{\bullet_\phi}$ vanishes for all $k,n >0$. By definition of the product $\bullet_\phi$, it is equivalent to require 
						\begin{equation*}
							\int_{M} E_\phi(x_1,z') t^{(k),n}(z',x_2, \dots, x_{n}) dz' = 0. 
						\end{equation*}												
					By the definition of the causal propagator, this implies
						\begin{equation*}
							\int_{M} E^A_\phi(x_1,z') t^{(k),n}(z',x_2, \dots, x_{n}) dz' = \int_{M} E^R_\phi(x_1,z') t^{(k),n}(z',x_2, \dots, x_{n}) dz'.
						\end{equation*}	
					Therefore, the distribution $s$ defined by
						\begin{equation*}
							s(x_1, \dots, x_n) := \int_{M} E^A_\phi(x_1,z') t^{(k),n}(z',x_2, \dots, x_{n}) dz'
						\end{equation*}
					is compactly supported by the support properties of $E^{A/R}_\phi$, and by the hypotheses on $t^{(k),n}$. Moreover, applying lemma~\ref{lemma_W_comp}, we have that $s$ is a distribution in $\cE'_W(M^n)$ because $\WF (E^{A/R}_\phi) = \cC^{A/R} \subset W_2$, and because $t^{(k),n} \in \cE'_W(M^n)$ by hypothesis. It follows
						\begin{equation*}
							t^{(k),n} (x_1, \dots, x_n)= \bP^+ (\boxempty - m^2 - v_\phi)_{x_1} s(x_1, \dots, x_n),
						\end{equation*}
					which means that $t^{(k),n}$ belongs to the ideal defined by $\boxempty - m^2 - v_\phi$, i.e. $t^{(k),n}$ must be the zero element in the quotient space $\cW_\phi$. So we conclude that the non-trivial elements $t \in \cW_\phi$ such that $[t, \varphi(f)]_{\bullet_\phi} = 0$ for any $f \in C^\infty_0(M)$ must have $\deg_s t =0$, i.e. $t=c1$ for a $c \in \bC[[\hbar]]$, which is precisely what we wanted to show.
				\end{proof}
			\noindent
			In the light of the previous lemma, we need to prove that for any $f \in C_0^\infty(M)$, the quantity $D_n$ commutes with $\varphi(f)$. The fact that $\alpha^R_{\phi_s,\phi}$ is an algebra homomorphism implies that
				\begin{equation}\label{c-number_eq}
					\begin{split}
						\left[ \frac{\partial}{\partial s} \alpha^R_{\phi_s,\phi} T_{n,s} \left[ \bigotimes_{i} F_{i,s} \right], \varphi(f) \right]_{\bullet} &= \frac{\partial}{\partial s} \alpha^R_{\phi_s,\phi} \left[ T_{n,s} \left[ \bigotimes_{i} F_{i,s} \right], \varphi_s(f)  \right]_{\bullet_{s}}  -\\
						&\quad -  \frac{\partial}{\partial s} \left[ T_n \left[ \bigotimes_{i} F_{i} \right], \alpha^R_{\phi_s,\phi} (\varphi_s(f)) \right]_{\bullet}.
					\end{split}
				\end{equation}
			Because of the explicit definition of $\alpha^R$ given by~\eqref{alpha_R_wick_per}, it holds
				\begin{equation}\label{alpha^R_f}
					\alpha^R_{\phi_s,\phi} (\varphi_{\phi_s}(f)) = (\boxempty - m^2 - v_\phi) E^A_{\phi_s}(f),
				\end{equation}
			up to a compactly supported smooth function in $(\boxempty - m^2 - v_\phi)C^\infty_0(M)$. We can pull the derivative $\partial/\partial s$ inside the commutator in the second term of the right-hand side of eq.~\eqref{c-number_eq} above. Using the formula~\eqref{var_der_A} for the variations the advanced propagator $E^A_\phi$ with respect to the background $\phi$, we then rewrite such term as
				\begin{equation*}
					\frac{\partial}{\partial s} \left[ T_n \left[ \bigotimes_{i} F_{i} \right], \alpha^R_{\phi_s,\phi} (\varphi_s(f))  \right]_{\bullet}  = \left[ T_n \left[ \bigotimes_{i} F_{i} \right], h E^A(f)  \right]_{\bullet}.
				\end{equation*}
			The commutator property (T9) of the time-ordered products allows us to rewrite eq.~\eqref{c-number_eq} as
				\begin{equation}\label{d_2_part_1}
					\begin{split}
						\left[ \frac{\partial}{\partial s} \alpha^R_{\phi_s,\phi} T_{n,s} \left[ \bigotimes_{i} F_{i,s} \right], \varphi(f) \right]_{\bullet} &= i \hbar \sum_{\ell =1}^n \frac{\partial}{\partial s} \alpha^R_{\phi_s,\phi} T_{n,s} \left[ \langle E_{s}(f), \frac{\delta}{\delta \varphi}\rangle F_{\ell,s} \otimes \bigotimes_{i \neq \ell} F_{i,s} \right] +\\
						&\quad +  i \hbar \sum_{\ell =1}^n T_{n} \left[ \langle E(h E^A(f)), \frac{\delta}{\delta \varphi}\rangle F_{\ell} \otimes \bigotimes_{i \neq \ell} F_{i} \right].
					\end{split}
				\end{equation}
			The first term in the right-hand side of eq.~\eqref{d_2_part_1} contains less than $N$ factors of the field $\varphi$. Therefore, by inductive hypothesis, it follows
				\begin{equation}\label{c-number_eq_2}
					\begin{split}
						&\left[ \frac{\partial}{\partial s} \alpha^R_{\phi_s,\phi} T_{n,s} \left[ \bigotimes_{i} F_{i,s} \right], \varphi(f) \right]_{\bullet} =  - \sum_{\ell} R_{n,1} \left( \langle E(f), \frac{\delta}{\delta \varphi}\rangle F_{\ell} \otimes \bigotimes_{i \neq \ell} F_{i}  ; \frac{\partial I^{(2)}_{s}}{\partial s} \right)  - \\
						&\qquad - i \hbar  \sum_{\ell, \ell' \neq \ell} T_n \left[ \langle E(f), \frac{\delta}{\delta \varphi}\rangle F_{\ell} \otimes \bigotimes_{i \neq \ell,\ell'} F_{i} \otimes \frac{\partial F_{\ell',s}}{\partial s} \right] -\\
						&\qquad - i \hbar  \sum_\ell T_n \left[ \bigotimes_{i \neq \ell} F_{i} \otimes \frac{\partial}{\partial s} \langle E_{s}(f), \frac{\delta}{\delta \varphi}\rangle F_{\ell, s} \right] +  i \hbar \sum_{\ell =1}^n T_{n} \left[ \langle E(h E^A(f)), \frac{\delta}{\delta \varphi}\rangle F_{\ell} \otimes \bigotimes_{i \neq \ell} F_{i} \right].
					\end{split}	
				\end{equation}
			Using the commutator property (T9) and the explicit expression for $\partial E_s / \partial s$ given by~\eqref{var_ders_causal_WF}, it follows that the second, the third, and the last terms in the right-hand side of eq.~\eqref{c-number_eq_2} can be rewritten as
				\begin{equation*}
					\sum_\ell \left[ T_n \left[ \bigotimes_{i \neq \ell} F_{i} \otimes \frac{\partial F_{\ell,s}}{\partial s} \right], \varphi(f)  \right]_{\bullet} - i \hbar \sum_\ell  T_n \left[ \bigotimes_{i \neq \ell} F_{i} \otimes \langle E^R(h  E(f)), \frac{\delta}{\delta \varphi}\rangle F_{\ell} \right].
				\end{equation*}
			Using the commutator property (T9) and the formula~\eqref{retarded_vs_time-ordered} which give $R_{n,m}$ in terms of the time-ordered products, we have
				\begin{equation*}
					\begin{split}
						\left[ R_{n,m} \left( \bigotimes_{i=1}^n F_{i}  ; \bigotimes_{j=1}^m H_{j} \right), \varphi(f) \right]_{\bullet} &= i\hbar \sum_{\ell=1}^n R_{n,m} \left( \langle E(f), \frac{\delta}{\delta \varphi} \rangle F_{\ell} \otimes \bigotimes_{i \neq \ell} F_{i}  ; \bigotimes_{j} H_{j} \right) + \\
						&\quad + i \hbar \sum_{r=1}^m R_{n,m} \left( \bigotimes_{i} F_{i}  ;  \langle E(f), \frac{\delta}{\delta \varphi} \rangle H_{r} \otimes\bigotimes_{j \neq r} H_{j} \right),
					\end{split}
				\end{equation*}
			for any local functionals $F_i, H_j$. Applying this result, we can rewrite the first term of the right-hand side of eq.~\eqref{c-number_eq_2} as
				\begin{equation*}
					\begin{split}
						- \sum_{\ell} R_{n,1} \left( \langle E(f), \frac{\delta}{\delta \varphi}\rangle F_{\ell} \otimes \bigotimes_{i \neq \ell} F_{i}  ; \frac{\partial I^{(2)}_{s}}{\partial s} \right) &=  \left[ \frac{i}{\hbar} R_{n,1} \left( \bigotimes_{i} F_{i}  ; \frac{\partial I^{(2)}_{s}}{\partial s} \right), \varphi(f) \right]_{\bullet} \\
						&\quad  + R_{n,1} \left( \bigotimes_{i} F_{i}  ; \langle E(f), \frac{\delta}{\delta \varphi}\rangle \frac{\partial I^{(2)}_{s}}{\partial s} \right).
					\end{split}
				\end{equation*}
			Since $\langle E(f), \delta /\delta \varphi \rangle \partial I^{(2)}_{s}/\partial s = \varphi(2 h E(f))$ and since $h E(f) \in C^\infty_0(M)$, axiom (T11a) implies
				\begin{equation*}
					R_{n,1} \left( \bigotimes_{i} F_{i}  ; \langle E(f), \frac{\delta}{\delta \varphi}\rangle \frac{\partial I^{(2)}_{s}}{\partial s} \right) = i \hbar \sum_\ell T_n \left[ \bigotimes_{i \neq \ell} F_{i} \otimes \langle E^R( h E(f)), \frac{\delta}{\delta \varphi}\rangle F_{\ell} \right].
				\end{equation*}
			Putting together, we have obtained 
				\begin{equation*}
					\begin{split}
						\left[ \frac{\partial}{\partial s} \alpha^R_{\phi_s,\phi} T_{n,s} \left[ \bigotimes_{i} F_{i,s} \right], \varphi(f)  \right]_{\bullet} &= \frac{i}{\hbar}  \left[ R_{n,1} \left( \bigotimes_{i} F_{i}  ; \frac{\partial I^{(2)}_{s}}{\partial s} \right), \varphi(f)  \right]_{\bullet} + \\
						&\quad  +  \sum_\ell \left[ T_n \left[ \bigotimes_{i \neq \ell} F_{i} \otimes \frac{\partial F_{\ell,s}}{\partial s} \right], \varphi(f)  \right]_{\bullet},
					\end{split}
				\end{equation*}
			i.e. that $[D_n, \varphi(f)]_\bullet = 0$ for any $f \in C^\infty_0(M)$. Therefore, by lemma~\ref{lemma_c-num}, $D_n$ is a $c$-number as we needed to prove.
			
	\paragraph*{Proof of~\ref{d_3}.}
		The proof is based on the locality/covariance property (T1), the scaling property (T2) for the time-ordered products, and the fact that the map $\alpha^R$ is well-behaving under isometric embeddings and rescalings. The details of the proof are the same as in~\citep[sec. 6.2.3]{HW05}, so we omit.
		
	\paragraph*{Proof of~\ref{d_4}.}
		Let $F_{n}$ be the functional $\varphi(f)$ where $f \in C^\infty_0(M)$ and let $F_1, \dots, F_{n-1}$ be arbitrary local functionals. Under this hypothesis, eq.~\eqref{deviation} reads
		\begin{equation}\label{deviation2}
				\begin{split}
					D_{n}(h; f_{1}, \dots, f_{n-1}, f) &= \frac{\partial}{\partial s} \alpha^R_{\phi_s,\phi} T_{n,s} \left[ \bigotimes_{j=1}^{n-1} F_{j,s} \otimes \varphi(f) \right]  - \frac{i}{\hbar} R_{n,1} \left( \bigotimes_{j=1}^{n-1} F_{j} \otimes \varphi(f); \varphi^2(h) \right) -\\
					&\quad -  \sum_{\ell} T_{n} \left[ \bigotimes_{j \neq \ell,n} F_{j} \otimes \frac{\partial F_{\ell,s}}{\partial s} \otimes \varphi(f) \right],
				\end{split}
			\end{equation}
		Note that $\varphi(f)$, as functional $C^\infty(M) \ni \phi \mapsto \int_M f(x) \phi(x) dx$, does not depend on $\phi$, i.e. $\partial \varphi(f) / \partial s = 0$. Using axiom (T11a), we rewrite each term appearing in the right-hand side of eq.~\eqref{deviation2}. For the last term, we obtain
			\begin{equation*}
				\begin{split}
					&\sum_{\ell =1}^{n-1} T_n \left[ \bigotimes_{j \neq \ell,n} F_{j} \otimes \frac{\partial F_{\ell,s}}{\partial s}  \otimes \varphi(f) \right] = \\
					&\quad = \sum_{\ell =1}^{n-1} R_{n-1,1} \left(\bigotimes_{j \neq \ell,n} F_{j} \otimes \frac{\partial F_{\ell,s}}{\partial s} ; \varphi(f) \right) - \sum_{\ell =1}^{n-1}  \varphi(f) \bullet T_{n-1} \left[ \bigotimes_{j \neq \ell,n} F_{j} \otimes \frac{\partial F_{\ell,s}}{\partial s} \right] \\
					&\quad = i \hbar \sum_{\ell =1}^{n-1} \sum_{r \neq \ell,n} T_{n-1} \left[ \langle E^R(f), \frac{\delta}{\delta \varphi} \rangle F_{r} \otimes \bigotimes_{j \neq \ell,r,n} F_{i} \otimes \frac{\partial F_{\ell,s}}{\partial s} \right] +\\
					&\qquad + i \hbar \sum_{\ell =1}^{n-1}  T_{n-1} \left[ \bigotimes_{j \neq \ell,n} F_{j} \otimes \langle E^R(f), \frac{\delta}{\delta \varphi} \rangle \frac{\partial F_{\ell,s}}{\partial s} \right] - \sum_{\ell =1}^{n-1}  \varphi(f) \bullet T_{n-1} \left[ \bigotimes_{j \neq \ell,n} F_{j} \otimes \frac{\partial F_{\ell,s}}{\partial s} \right].
				\end{split}
			\end{equation*}
		We rewrite the second term in the right-hand side of eq.~\eqref{deviation2} as
			\begin{equation*}
				\begin{split}
					&-\frac{i}{\hbar} R_{n,1} \left( \bigotimes_{j=1}^{n-1} F_{j} \otimes \varphi(f) ; \frac{\partial I^{(2)}_{s}}{\partial s} \right) =\sum_{r=1}^{n-1} R_{n-1,1} \left( \langle E^R(f), \frac{\delta}{\delta \varphi} \rangle F_{r} \otimes \bigotimes_{j \neq r,n} F_{j} ; \frac{\partial I^{(2)}_{s}}{\partial s} \right) - \\
					&\qquad + i\hbar \sum_{r=1}^{n-1} T_{n-1} \left[ \langle E^R( h E^R(f)), \frac{\delta}{\delta \varphi} \rangle F_{r} \otimes \bigotimes_{j \neq r, n} F_{j} \right] -\frac{i}{\hbar} \varphi(f) \bullet R_{n-1,1} \left( \bigotimes_{j \neq n} F_{j} ; \frac{\partial I^{(2)}_{s}}{\partial s} \right) - \\
					&\qquad + \varphi(h E^A(f)) \bullet T_{n-1} \left[ \bigotimes_{j \neq n} F_{j} \right].
				\end{split}
			\end{equation*}
		We used the definition of $h_\phi$ and the definition of $R_{n,1}$ in terms of time-ordered products.\\
		Finally, the first term in the right-hand side of~\eqref{deviation2} reads
			\begin{equation*}
				\begin{split}
					&\frac{\partial}{\partial s} \alpha^R_{\phi_s,\phi} T_{n,s} \left[ \bigotimes_{j =1}^{n-1} F_{i,s} \otimes \varphi(f) \right] = i \hbar \sum_{r=1}^{n-1} \frac{\partial}{\partial s} \alpha^R_{\phi_s,\phi} T_{n-1,s} \left[ \langle E^R_{s}(f), \frac{\delta}{\delta \varphi} \rangle F_{r,s} \otimes \bigotimes_{j \neq r,n} F_{j,s} \right] - \\
					&\qquad - \varphi(f) \bullet \frac{\partial}{\partial s} \alpha^R_{\phi_s,\phi} T_{n-1,s} \left[ \bigotimes_{j \neq n} F_{j,s} \right]  - \varphi(h E^A(f)) \bullet T_{n-1} \left[ \bigotimes_{j \neq n} F_{j} \right],
				\end{split}
			\end{equation*}
		where we used formula~\eqref{alpha^R_f} and the fact that the map $\alpha^R_{\phi_s, \phi}$ is an isomorphism of algebras.\\
		Putting together, we have that the quantity $D_n$ corresponding to the functionals $F_1, \dots, F_{n-1}, \varphi(f)$ can be written in terms of $D_{n-1}$, which vanishes by the inductive hypothesis, in detail
			\begin{equation*}
				\begin{split}
					D_n (F_1, \dots, F_{n-1}, \varphi(f))  &= i\hbar \sum_{r=1}^{n-1}  D_{n-1} \left( F_1, \dots, \langle E^R(f), \frac{\delta}{\delta \varphi} \rangle F_r, \dots, F_{n-1} \right) \\
					&\quad -\varphi(f) \bullet D_{n-1} (F_1, \dots, F_{n-1})\\
					&=0.
				\end{split}
			\end{equation*}
		Here we used the notation $D_n (F_1, \dots, F_{n})$ for the element in $\cW$ given by the right-hand side of eq.~\eqref{deviation}. This is precisely condition~\ref{d_4}.

	\paragraph*{Proof of~\ref{d_5}.}
		We need to show that $D_n$ is a distribution on $M^{n+1}$ and it satisfies the following conditions:
			\begin{itemize}
				\item It holds
						\begin{equation}\label{WF_deviation}
							\left. \WF(D_{n,\phi}) \right|_{\Delta_{n+1}} \perp T\Delta_{n+1}.
						\end{equation}
				\item Let $\bR \ni \epsilon \mapsto \phi_\epsilon$ be smooth (respectively analytic). It holds
						\begin{equation}\label{WF_deviation_smooth}
							\left. \WF(D_{n,\phi_\epsilon}) \right|_{\bR \times \Delta_{n+1}} \perp T(\bR \times \Delta_{n+1}),
						\end{equation}
					(where the smooth wave-front set is replaced with the analytic wave-front set in the analytic case).
			\end{itemize}
		Since $D_{n,\phi}$ is a $c$-number, it is equal to its expectation value in any state of $\cW_\phi$. To simplify, we consider a quasi-free state $\omega_\phi$ such that its $2$-point function coincides in $M \backslash J^+ (\supp v)$ with a fixed pure Hadamard $2$-point function $\omega_0$ with respect to $P_0=\boxempty - m^2$. We can still conclude, following the same argument as in lemma~\ref{lemma_ret_state_fedosov}, that it must hold $\omega_\phi = E_\phi \circ \sigma_{c_+} \circ \omega_0 \circ \sigma_{c_+} \circ E_\phi$. We write $D_{n,\phi}$ as
			\begin{equation}\label{alt_deviation}
					D_{n,\phi}(h_\phi; f_1, \dots, f_n) = r_{n,\phi} (h_\phi, f_1, \dots, f_n) - \frac{1}{\hbar} \omega_\phi \left( R_{n,1,\phi} \left( \bigotimes_{i} F_{i,\phi} ; \varphi^2(h_\phi) \right) \right),
			\end{equation}
		where $r_{n,\phi} (h_\phi, f_1, \dots, f_n)$ is defined by
			\begin{equation}\label{r_n_def}
				r_{n,\phi} (h_\phi, f_1, \dots, f_n) := \omega_\phi \left( \frac{\partial}{\partial s} \alpha^R_{\phi_s,\phi} T_{n,\phi_s} \left[ \bigotimes_{i} F_{i,\phi_s} \right] \right) - \sum_{\ell} \omega_\phi \left( T_{n,\phi} \left[ \bigotimes_{i \neq \ell} F_{i,\phi} \otimes \frac{\partial F_{\ell,\phi_s}}{\partial s} \right] \right),
			\end{equation}
		To prove the properties~\eqref{WF_deviation} and~\eqref{WF_deviation_smooth}, we show that each term in the right-hand side of eq.~\eqref{alt_deviation} satisfies the desired properties.\\
		As a straightforward consequence of the microlocal spectrum condition~\ref{R3}, the last term in eq.~\eqref{alt_deviation} is a well-defined distribution which satisfies the wave-front set condition~\eqref{WF_deviation}. The condition~\eqref{WF_deviation_smooth} can be treated similarly: if we consider a background $\phi_\epsilon$ depending smoothly (respectively analytically) on $\epsilon$ and a corresponding family of quasi-free states $\{ \omega^{(\epsilon)} \}$ depending smoothly (respectively analytically) on $\epsilon$ in the sense of~\ref{R4} (respectively~\ref{R5}), then the smoothness property~\ref{R4} (respectively the analyticity property~\ref{R5}) imply that the last term in $D_{n,\phi_\epsilon}$, given by eq.~\eqref{alt_deviation} with the obvious changes due to the dependence upon $\epsilon$, satisfy condition~\eqref{WF_deviation_smooth} (respectively its analytic counterpart).\\
		To prove that $D_{n}$ satisfies the conditions~\eqref{WF_deviation} and~\eqref{WF_deviation_smooth}, we then need to show that $r_{n,\phi} (h_\phi, f_1, \dots, f_n)$ does. We notice first that whenever $h_\phi \otimes f_1 \otimes \cdots\otimes f_n$ are supported outside the diagonal, then $D_{n}(h_\phi; f_1, \dots, f_n)$ vanishes. Therefore, $r_{n}(h_\phi, f_1, \dotsm f_n)$ equals minus the second and the third terms in~\eqref{alt_deviation}. Consequently, it must be a well-defined distribution since both the second and the third terms are already known to be well-defined distributions. So we need to investigate $r_{n}$ near the total diagonal $\Delta_{n+1}$.\\
				 For this purpose, we assume that $h_\phi \otimes f_1 \otimes \cdots \otimes f_n$ is supported in neighbourhood of the total diagonal $\Delta_{n+1}$ sufficiently small that $\supp(h_\phi \otimes f_1 \otimes \cdots \otimes f_n) \subset U^{n+1}$, where $U$ is a convex normal subset of $M$. Actually, we require that $U$ is sufficiently small to satisfy the hypotheses of lemma~\ref{lemma_d}. Under this assumption we can express the time-ordered product using the local Wick expansion (see~\eqref{local_wick_exp_der}):
			\begin{equation}\label{local_wick_R12}
				\begin{split}
					&T_{n,\phi} \left[ \bigotimes_{i} F_{i,\phi} \right]  =\\
					&\quad = \sum \sC_{\boldsymbol{\alpha}_1 \dots \boldsymbol{\alpha}_n}  \int_{M^n} \prod_{i=1}^n f_i(z_i) \tau^H_\phi [\otimes_i C_{i,\phi} \cdot \nabla^{\boldsymbol{\alpha}_i} \varphi ] (z_1, \dots, z_n) : \prod_{i=1}^n \nabla^{\boldsymbol{\kappa}_{i} - \boldsymbol{\alpha}_{i}} \varphi(z_i):_{H_\phi} dz_1 \dots dz_n\\
					&\quad = \sum_\ell \int_{M^{n+\ell}} \prod_{i=1}^n C_{i,\phi}(z_i) f_i(z_i) w^\ell_\phi (z_1, \dots, z_n, x_1, \dots, x_\ell) : \prod_{j=1}^\ell \varphi(x_j) :_{H_\phi} dz_1 \dots dz_n dx_1 \dots dx_\ell,
		 		\end{split}
			\end{equation}
		where $H_\phi$ is the Hadamard parametrix~\eqref{Hadamard_def}, where $: \cdots:_{H_\phi}$ is the ordering with respect to the Hadamard parametrix, and where $w^\ell_\phi$ are suitable distributions locally and covariantly constructed from the metric, depending on $\phi$ via $v_\phi$. Note, in particular, that each $w^\ell_\phi$ is a finite sum of appropriate products of distributions $\tau^H_\phi$ and derivatives of the delta distribution. Note that the dependence of $F_{i,\phi}$ on $\phi$ is encoded in the dependence of $C_{i,\phi}$ on $\phi$, so it does not affect $w^\ell_\phi$.\\
		Inserting eq.~\eqref{local_wick_R12} into the definition of $r_n$~\eqref{r_n_def}, we find
			\begin{equation*}
				\begin{split}
					&r_{n,\phi}(h_\phi, f_1, \dots, f_n) = \sR_{1,\phi} + \sR_{2,\phi}\\
					&\quad = \sum_\ell \int_{M^{n+\ell}} \prod_{i=1}^n C_{i,\phi}(z_i) f_i(z_i) w^\ell_\phi (z_1, \dots, z_n, x_1, \dots, x_\ell) \, \omega_\phi \left( \frac{\partial}{\partial s} \alpha^R_{\phi_s,\phi} : \prod_{j=1}^\ell \varphi(x_j) :_{H_{\phi_s}} \right) \prod_i dz_i \prod_j dx_j+ \\
					&\qquad  +  \sum_\ell \int_{M^{n+\ell}} \prod_{i=1}^n C_{i,\phi}(z_i) f_i(z_i)  \frac{\partial}{\partial s} w^\ell_{\phi_s} (z_1, \dots, z_n, x_1, \dots, x_\ell)  \, \omega_\phi \left(  : \prod_{j=1}^\ell \varphi(x_j) :_{H_\phi} \right) \prod_i dz_i \prod_j dx_j.
				\end{split}
			\end{equation*}
		We analyse the two terms $\sR_{1,\phi}, \sR_{2,\phi}$ separately and we prove that each of them satisfies the requirements~\eqref{WF_deviation} and~\eqref{WF_deviation_smooth}.\\
		To prove the claim for $\sR_{1,\phi}$, we first need to rewrite it. Let $\omega_{\phi_s}$ be the unique quasi-free Hadamard state with respect to $P_{\phi_s}$ such that $\omega_{\phi_s}$ coincides with $\omega_\phi$ on $M\setminus J^+(K)$, where $K$ denotes the region in which $v_{\phi_s}$ and $v_\phi$ differ (at most the whole support of the interaction, which is compact by hypothesis). Necessarily, it holds $\omega_{\phi_s} = E_{\phi_s} \circ \sigma_{c_+} \circ \omega_0 \circ \sigma_{c_+} \circ E_{\phi_s}$. Inside the convex normal set $U$, we define the difference $d_{\phi_s} (x_1,y) = \omega_{\phi_s}(x,y) - H_{\phi_s}(x,y)$. Then, $\sR_{1,\phi}$ reads
			\begin{equation}\label{term_R_1}
				\sR_{1,\phi} = \sum_\ell \sum_{\{ab\}} \int_{M^{n+\ell}} \prod_{i=1}^n C_{i,\phi}(z_i) f(z_i) w^\ell_\phi (z_1, \dots, z_n, x_1, \dots, x_\ell)  \frac{\partial}{\partial s} \prod_{ab} d_{\phi_s}(x_a,x_b) \prod_{i} dz_i \prod_{j} dx_j.
			\end{equation}
		We what to express $\sR_{1, \phi}$ as a distributional kernel in $U^{n+1}$ evaluated on $h_\phi \otimes f_1 \otimes \cdots \otimes f_n$. To do so, we need to prove
				\begin{equation}\label{delta_d}
					\frac{\partial}{\partial s} d_{\phi_s} (x_1,x_2) = \int_{M} \frac{\delta d_\phi(x_1, x_2)}{\delta v_\phi(y)} h_\phi(y) dy,
				\end{equation}
			for a well-defined distribution $\delta d_\phi/\delta v_\phi$ for which we have sufficient microlocal control.\\
			Making use of the results of appendix~\ref{app_bkgr_dep_propa}, in particular eq.~\eqref{var_der_A}, one can see that the advanced/retarded propagators satisfies
				\begin{equation*}
					\frac{\partial}{\partial s} E^{A/R}_{\phi_s} (x_1,x_2) = \int_M E^{A/R}_{\phi} (x_1,y) \frac{\partial v_{\phi_s}(y)}{\partial s} E^{A/R}_{\phi} (y,x_2) dy = \int_M \frac{\delta E^{A/R}_{\phi}(x_1,x_2)}{\delta v_\phi(y)} h_\phi(y) dy,
				\end{equation*}
			and, then, we have
				\begin{equation*}
					\WF \left( \frac{\delta E^{A/R}_{\phi}}{\delta v_\phi} \right) \subset X_{2+1},
				\end{equation*}
			where the set $X_{2+1}$ is given by~\eqref{X}. Since we chose $\omega_\phi = E_\phi \circ \sigma_{c} \circ \omega_0 \circ \sigma_{c} \circ E_\phi$, where $\omega_0$ is a pure Hadamard $2$-point function with respect to $P_0=\boxempty - m^2$, it follows
				\begin{equation*}
					\frac{\partial}{\partial s} \omega_{\phi_s} (x_1,x_2) = \int_{M} \frac{\delta \omega_\phi(x_1, x_2)}{\delta v_\phi(y)} h_\phi(y) dy,
				\end{equation*}
			and, similarly as in lemma~\ref{lemma_state}, we also have
				\begin{equation*}
					\WF \left( \frac{\delta \omega_{\phi}}{\delta v_\phi} \right) \subset Z_{2+1},
				\end{equation*}
			where the set $Z_{2+1}$ is defined by~\eqref{Z}.\\
			Furthermore, we can modify the argument of lemma~\ref{lemma_u} to obtain
				\begin{equation*}
					\frac{\partial}{\partial s} u_{\phi_s,k} (x_1,x_2) = \int_{M} \frac{\delta u_{\phi,k}(x_1, x_2)}{\delta v_\phi(y)} h_\phi(y) dy,
				\end{equation*}
			where $\delta u_{\phi,k}/\delta v_\phi$ is a well-defined distribution which satisfies
				\begin{equation*}
					\WF \left( \frac{\delta u_{\phi,k}}{\delta v_\phi} \right) \subset \cC^u_{2+1},
				\end{equation*}
			where the set $\cC^u_{2+1}$ is given by~\eqref{Cunu}. Then, following the argument of lemma~\ref{lemma_H}, we have
				\begin{equation*}
					\frac{\partial}{\partial s} H_{\phi_s} (x_1,x_2) = \int_{M} \frac{\delta H_{\phi}(x_1, x_2)}{\delta v_\phi(y)} h_\phi(y) dy,
				\end{equation*}
			where $\delta H_{\phi}/\delta v_\phi$ is a well-defined distribution which satisfies
				\begin{equation*}
					\WF \left( \frac{\delta H_{\phi,k}}{\delta v_\phi} \right) \subset Z_{2+1}, \qquad \left. \WF \left( \frac{\delta H_{\phi,k}}{\delta v_\phi} \right)\right|_{\Delta_3} \perp T\Delta_3.
				\end{equation*}
			By the results just proved for $\delta \omega_{\phi}/\delta v_\phi$ and $\delta H_{\phi}/\delta v_\phi$, we conclude that $\delta d_{\phi}/\delta v_\phi$ is a well-defined distribution. We can adapt the argument used in lemma~\ref{lemma_d} to obtain the following upper bound for its wave-front set:
				\begin{equation}\label{WF_delta_d}
					\left. \WF \left( \frac{\delta d_\phi}{\delta v_\phi} \right) \right|_{\Delta_3} \subset \{ (y,x_1,x_2; p, k_1, k_2) : y=x_1=x_2, \, p+k_1+k_2=0\}.
				\end{equation}
		This microlocal condition corresponds to~\citep[lemma 3.6]{zahn2013locally} in our framework.\\
		Next, we note that by construction $w^\ell_\phi$ is a finite sum of products of the distributional coefficients $\tau^H_\phi$ and derivatives of the delta distribution. The wave-front sets of the distributional coefficients $\tau^H_\phi$ of the local Wick expansion are estimated by the set $\cC^T$ defined in~\eqref{C^T_m} (see~\eqref{est_tau_H} in sec.~\ref{subsubsec_rew_T-prod} based on~\citep{HW02, HW05}). Therefore, using the wave-front set calculus (thm.~\ref{theo_WF_horma}), we have
			\begin{equation*}
				\begin{split}
					\WF( w^\ell_\phi) \subset &\left\{ (z_1, \dots, z_n, x_1, \dots, x_\ell; q_1, \dots, q_n, k_1, \dots, k_\ell) \in \dot{T}^* (M^{n+\ell}) : \right.\\
					&\quad \exists \mbox{ partition } I_1 \sqcup \dots \sqcup I_n= \{1, \dots, \ell \} \mbox{ such that } x_i = z_j \; \forall i \in I_j \mbox{ and }\\
					&\quad \left. (z_1, \dots, z_n; q'_1, \dots, q'_n) \in \cC^T_n \mbox{ where } q'_j = q_j + \sum_{i \in I_j} k_i \right\}
				\end{split}
			\end{equation*}
		We then estimate the wave-front set of the distributional kernel $\sR_{1,\phi}(y;z_1, \dots, z_n)$ corresponding to~\eqref{term_R_1} by using the wave-front set calculus (thm.~\ref{theo_WF_horma}). It follows from the estimates we provided for $\WF(w^\ell_\phi)$ and for $\WF( \delta d_\phi / \delta v_\phi)$, that $\WF(\sR_1)|_{\Delta_{n+1}} \perp T \Delta_{n+1}$, i.e. that condition~\eqref{WF_deviation} holds. By considering the background $\phi$ depending smoothly (or analytically) on a further parameter, we can show that also condition~\eqref{WF_deviation_smooth} holds.\\
		
		To prove the claim for $\sR_{2,\phi}$, we notice that, inside $U^n$, $\sR_{2,\phi}$ reads 
			\begin{equation}\label{term_R_2}
				\sR_{2,\phi} = \sum_{\ell} \sum_{\{ab\}} \int_{M^{\ell+n}} \prod_{i=1}^n C_{i,\phi}(z_i)f(z_i) \frac{\partial}{\partial s} w^\ell_{\phi_s} (z_1, \dots, z_n, x_1, \dots, x_\ell) \prod_{ab} d_{\phi}(x_a,x_b) \prod_{i} dz_i \prod_{j} dx_j.
			\end{equation}
		Since the distributional coefficients of the local Wick expansion $\tau^H_\phi$ depend on $\phi$ only via $v_\phi$, it follows that each $w^\ell_\phi$, which is a finite sum of products of $\tau^H_\phi$ and derivatives of the delta distribution, satisfies
			\begin{equation*}
				\frac{\partial}{\partial s} w^\ell_{\phi_s}(z_1, \dots, z_n, x_1, \dots, x_\ell) = \int_M \frac{\delta w^\ell_\phi(z_1, \dots, z_n, x_1, \dots, x_\ell)}{\delta v_\phi(y)} h_\phi(y) dy.
			\end{equation*}
		Following the argument presented in sec.~\ref{subsubsec_var} (for time-ordered products of functionals which do not involve covariant derivatives) and the generalization discussed in sec.~\ref{subsubsec_T10,T11}, we have the following restrictions on wave-front set of $\delta \tau^H_\phi / \delta v_\phi$:
			\begin{equation*}
				\WF \left( \delta w^\ell_\phi/\delta v_\phi \right) \subset \cC^{\delta; +}_{n,1} \cap \cC^{\delta; -}_{n,1}, \qquad \left. \WF \left( \delta w^\ell_\phi/\delta v_\phi \right) \right|_{\Delta_3} \perp T \Delta_3,
			\end{equation*}
		where the sets $\cC^{\delta; \pm}_{n,1}$ are defined in~\eqref{Cdnu}. Using the wave-front set calculus (thm.~\ref{theo_WF_horma}) and the fact that $d_\phi$ is a smooth function, we conclude that the distributional kernel $\sR_{2,\phi}(y,z_1, \dots, z_n)$ corresponding to~\eqref{term_R_2} is such that $\WF(\sR_{2,\phi})|_{\Delta_{n+1}} \perp T \Delta_{n+1}$, as we wanted to prove.\\
		By considering the background $\phi$ depending smoothly (or analytically) on a further parameter, we can show that also condition~\eqref{WF_deviation_smooth} holds for $\sR_{2,\phi}$. This concludes the proof of~\ref{d_5}.
		
	\paragraph*{Proof of~\ref{d_6}.}
		It follows form the definition of $D_{n,\phi}$ together with the symmetry properties of the time-ordered products (T6) that $D_{n,\phi}(h_\phi; f_1, \dots, f_n)$ is symmetric in $f_1, \dots, f_n$. Because $h_\phi$ appears on a completely different footing than $f_i$, the non-trivial question is about the behaviour of $D_{n,\phi}(h_\phi; f_1, \dots, f_n)$ when $h_\phi$ is exchanged with $f_1$. More precisely, consider two smooth families $\{\phi^{(1)}_s\}$ and $\{\phi^{(2)}_s\}$ such that
			\begin{equation*}
				\phi^{(1)}_{s=0} = \phi^{(2)}_{s=0} =\phi, \quad h^{(i)}_\phi := \frac{\partial v(\phi^{(i)}_s)}{\partial s}.
			\end{equation*}
		We want to prove
			\begin{equation}\label{symm_prop_devi}
				\Delta D_{1}(h^{(1)}, h^{(2)}):= D_{1}(h^{(1)}; h^{(2)}) - D_{1}(h^{(2)}; h^{(1)})=0,
			\end{equation}
		and, for $n\geq 2$, also
			\begin{equation}\label{symm_prop_devi_great}
				\Delta D_{n}(h^{(1)}, h^{(2)} ; f_2, \dots, f_n):= D_{n}(h^{(1)}; h^{(2)}, f_2, \dots, f_n) - D_{n}(h^{(2)}; h^{(1)}, f_2, \dots f_n)=0,
			\end{equation}
		 where $D_{n}(h^{(1)}; h^{(2)}, f_2, \dots, f_n)$ is understood as the $c$-number given by eq.~\eqref{deviation} for the local functional $F_1 = \varphi^2(h^{(2)})$.\\
		 Following the same argument presented in~\citep[eq. (248)]{HW05} or~\citep[prop. 3.7]{zahn2013locally}, once it is proved eq.~\eqref{symm_prop_devi}, then eq.~\eqref{symm_prop_devi_great} would be a consequence of the flatness of $\nabla^R$. Therefore, we focus on the case $n=1$. More explicitly, $\Delta D_1 (h^{(1)}, h^{(2)})$ is given by
		 	\begin{equation*}
		 		\begin{split}
		 		\Delta D_1 (h^{(1)}, h^{(2)}) &= \frac{\partial}{\partial s} \alpha^R_{\phi^{(1)}_s, \phi} T_{1,\phi^{(1)}_{s}} \left[\varphi^2(h^{(2)}_{\phi^{(1)}_s}) \right] - \frac{\partial}{\partial s} \alpha^R_{\phi^{(2)}_s, \phi} T_{1,\phi^{(2)}_{s}} \left[ \varphi^2(h^{(1)}_{\phi^{(2)}_s}) \right] - \\
		 		&\quad - \frac{i}{\hbar} R_{1,1} \left( \varphi^2(h^{(2)}) ; \varphi^2(h^{(1)}) \right) + \frac{i}{\hbar} R_{1,1} \left( \varphi^2(h^{(1)}) ; \varphi^2(h^{(2)}) \right) - \\
		 		&\quad - T_1 \left[ \frac{\partial}{\partial s} \varphi^2(h^{(2)}_{\phi^{(1)}_s}) \right] + T_1 \left[ \frac{\partial}{\partial s} \varphi^2(h^{(1)}_{\phi^{(2)}_s}) \right].
		 		\end{split}
		 	\end{equation*}
		  It follows from the properties~\ref{d_2} and~\ref{d_5} of $D_{1}$ that $\Delta D_1$ is a c-number distribution supported on the total diagonal in $M \times M$ which satisfies the wave-front set constraints~\eqref{WF_deviation} and~\eqref{WF_deviation_smooth}. Moreover, the properties~\ref{d_1} and~\ref{d_3} of $D_{1}$ imply that $\Delta D_1$ is covariantly constructed out of $g$, $m^2$ and $v_\phi$, and scales almost homogeneously with degree $4$ under the rescaling of $g,m, \phi, \lambda$. Therefore, $\Delta D_1$ is necessarily in the form
		 	\begin{equation*}
		 		\Delta D_{1} (h^{(1)}, h^{(2)}) = \sum_{r \geq 1}^{n_f} \int_M h^{(1)}(x) \left( \nabla_{(\mu_1 \dots \mu_r)} h^{(2)} \right)(x) C^{\mu_1 \dots \mu_r}(x) - (1 \leftrightarrow 2),
		 	\end{equation*}
		where $n_f$ is a finite number, where $C^{\mu_1 \dots \mu_r}$ are polynomials of scaling dimension $4$ constructed from the metric, the Riemann tensor, $m^2$, $v_\phi$ and their derivatives. However, there are no tensors $C$ with the correct dimension that give a non-vanishing $\Delta D_1$. This concludes the proof of condition~\ref{d_6} and the consistency of the principle of perturbative agreement for variations of the background $\phi$.
		
\chapter{Continuity properties of the non-linear and the linearized Cauchy problems}\label{app_nnlin}
	In the first part of this appendix, we discuss some aspects of the initial value problem for the non-linear equation of motion~\eqref{eom_phi4} corresponding to the $\lambda \phi^4$-theory for the ultra-static space-time
		\begin{equation*}
			M=\bR \times \Sigma, \qquad g = -dt^2 + h_{ij} dx^i dx^j,
		\end{equation*}
	where $\Sigma$ is a $3$-dimensional compact Riemannian manifold. We will prove that the initial value problem for smooth (global in time) solutions with smooth Cauchy data on $\Sigma$ is well-posed. Then, we will show that the map which takes smooth Cauchy data $q,p$ and gives their corresponding unique smooth solution $\phi$ is continuous.\\
	In the second part of this appendix, we fix a solution $\phi$ for the non-linear equation, and we consider the linearized equation~\eqref{leom_phi4}. For this linear equation, it is known that the initial value problem for smooth (global in time) solutions with smooth Cauchy data is well-posed. We will show that the unique smooth solution $u_\phi(q,p)$ of the linearised equation in $\phi$ corresponding to a fixed pair of smooth Cauchy data $q,p$, depends continuously on $\phi$.
	
	\section{Continuity properties of the non-linear Cauchy problem}
	We first prove by well-known methods the existence and the uniqueness of a global smooth solution $\phi \in C^\infty(\bR \times \Sigma)$ to the initial value problem
		\begin{equation}\label{IVP_phi4}
			\left\{\begin{array}{l}
				\phi_{tt}(t,x) - (\Delta^{(h)} \phi)(t,x) + m^2 \phi(t,x) =- \frac{1}{3!} \lambda(t,x) \phi^3(t,x), \\
				\phi|_{t=0}=q, \\
				\phi_t|_{t=0}=p,
			\end{array}\right.
		\end{equation}
	where $q, p \in C^\infty(\Sigma)$, where $\Delta^{(h)}$ denotes the Laplace operator for $(\Sigma, h)$, and where $(\cdot)_t$ denotes the partial derivative with respect to the coordinate $t$. We assume $m>0$ and that $\lambda$ is a positive constant or a non-negative cutoff function in $C^\infty_0(\bR \times \Sigma)$ such that
		\begin{equation*}
			\sup_{(t,x) \in \bR \times \Sigma} \left| \frac{\lambda_t(t,x)}{\lambda(t,x)} \right| < \infty.
		\end{equation*}
	It is easy to see that this class of cut-off functions is not empty.\\
	Then, we prove that the map that associates a pair of smooth Cauchy data to the corresponding solution is continuous with respect to the natural Fr\'{e}chet topology on $\sE = C^\infty(\Sigma) \oplus C^\infty(\Sigma)$ and the compact-open topology on $C^\infty(\bR \times \Sigma)$.
		
	\paragraph{Global well-posedness of the non-linear Cauchy problem.} The second order partial differential equation in~\eqref{IVP_phi4} is hyperbolic and quasi-linear, therefore it is well-known that the initial value problem is well-posed locally~\citep[prop. 3.1 in chapter 16]{T96b}. More precisely, it is known that there exists a closed interval $I \subset \bR$ around $0$ such that the system~\eqref{IVP_phi4} for $q \in H^{s+1}(\Sigma)$ and $p \in H^{s}(\Sigma)$ with $s >3/2 +1$ admits a unique local solution
		\begin{equation}\label{sol_sob}
			\phi \in C(I, H^{s+1}(\Sigma)) \cap C^1(I,H^{s}(\Sigma)),
		\end{equation}
	where $H^{s}(\Sigma)$ refers to the $L^2$-Sobolev spaces~\citep[sec. 1 and 6 in chapter 13]{T96b},~\citep{emmanuel1996sobolev} for the compact Riemannian manifold $(\Sigma, h)$, i.e. $H^s(\Sigma)$ it is the completion of $C^\infty(\Sigma)$ with respect to the norm
		\begin{equation*}
			 \Vert f \Vert_s := \sum_{j=0}^k \left( \int_\Sigma \vert (\partial^{(h)})^j f \vert_h^2 d\vol^h \right)^{\frac{1}{2}},
		\end{equation*}
	where $\partial^{(h)}$ denotes the Levi-Civita covariant derivative on $(\Sigma, h)$, and where $\vert \cdot \vert_h$ is the natural norm for tensor fields on $\Sigma$ defined via the Riemannian metric $h$\footnote{\label{footnote} More explicitly, $\vert \cdot \vert_h$ is defined for $k$-rank covariant tensor field $t$ by
			\begin{equation*}
				\vert t \vert_h^2(x) := h^{\mu_1 \nu_1}(x) \cdots ^{\mu_k \nu_k}(x) t_{\mu_1 \cdots \mu_k}(x)t_{\nu_1 \cdots \nu_k}(x).
			\end{equation*}}. Here and in the following, we always consider integer Sobolev orders $s$.\\
	We remind the reader about the following well-known properties (see e.g.~\citep[chapter 4]{T96a},~\citep[chapter 13 and 16]{T96b}, and~\citep{emmanuel1996sobolev}) of the Sobolev spaces on a compact manifold:
		\begin{itemize}
			\item The Sobolev norm $\Vert \cdot \Vert_{H^s}$ is equivalent to the norm given by $\Vert A^s \cdot \Vert_{L^{2}}$, where $A$ is the square root of the unique self-adjoint extension of the operator $m^2-\Delta^{(h)}$ on $\Sigma$.
			\item We have
					\begin{equation*}
						\Vert f h\Vert_{H^s} \leq \sC \Vert f \Vert_{L^\infty} \Vert h \Vert_{H^s} + \sC' \Vert h \Vert_{L^\infty} \Vert f \Vert_{H^s}
					\end{equation*}
				where $\Vert \cdot \Vert_{L^\infty}$ is the usual supremum norm, and where $\sC$, $\sC'$ are constants depending on $s$.
			\item It holds $H^{s+1}(\Sigma) \subset H^{s}(\Sigma)$, i.e. for any $f \in H^s(\Sigma)$ it holds $\Vert f \Vert_{H^s} < \sC \Vert f \Vert_{H^{s+1}}$, for a constant $\sC$ depending on $s$.
			\item For $s> \frac{3}{2}$, we have $H^s(\Sigma) \subset L^\infty(\Sigma)$, i.e. for any $f \in H^s(\Sigma)$, it holds $\Vert f \Vert_{L^\infty} < \sC_s \Vert f \Vert_{H^s}$, for a constant $\sC$ depending on $s$.
			\item For any $k$, it holds $H^s(\Sigma) \subset C^{k}(\Sigma)$ if $s >3/2 + k$.
			\item For any $s > 3/2$,  the space $H^s(\Sigma)$ is an algebra if equipped with the pointwise product.
		\end{itemize}
	The size of the time interval in which the local solution exists is controlled by the Sobolov norms, see~\citep[prop. 8.5 in ch. 13 and thm. 3.5 in ch. 16]{T96b}. Namely, there must exist a maximal $T^* \in (0,+\infty]$ such that any other local existence interval $I$ is contained in $(-T^*,T^*)$ and
		\begin{equation}\label{fin_time_blowup}
			\lim_{t \to \pm T^*} \left( \Vert \phi(t,\cdot) \Vert_{H^{s+1}} + \Vert \phi_t(t, \cdot) \Vert_{H^{s}} \right) = \infty.
		\end{equation}
	Indeed, whenever the limit is finite, the local initial value problem can be posed again for the initial data $\phi(T^*,\cdot), \phi_t(T^*,\cdot)$ and so $T^*$ is not maximal. We now show that the initial value problem we are considering is well-posed:
		\begin{prop}\label{prop_nn-lin_Cauchy_prob}
			There exists a global, unique, smooth solution $\phi \in C^\infty(\bR \times \Sigma)$ of the system~\eqref{IVP_phi4} for smooth Cauchy data $q, p \in C^\infty(\Sigma)$.
		\end{prop}
		\begin{proof}
			The proof is based on standard results, we give the argument for completeness. Let $\phi$ be a local solution of~\eqref{IVP_phi4}. We choose $t$ in the interval of local existence. Because $A$ does not depend on $t$ by definition, it follows that
				\begin{equation}\label{d_A_dt}
					\frac{d}{dt} \left( \Vert A^{s+1} \phi (t, \cdot) \Vert_{L^{2}}^2 + \Vert A^{s} \phi_t (t, \cdot) \Vert_{L^{2}}^2 \right) =- \frac{1}{3} \left( A^{s} (\lambda (t, \cdot) \phi^3 (t, \cdot)), A^{s} \phi_t (t, \cdot) \right)_{L^{2}}.
				\end{equation}
			Using the Cauchy-Schwartz inequality for the $L^2$-norm and the properties of the Sobolev spaces on compactly supported manifolds, for $s > \frac{1}{2}$ we obtain
				\begin{equation*}
					\begin{split}
					&\left\vert \frac{d}{dt} \left( \Vert \phi (t,\cdot) \Vert_{H^{s+1}}^2 + \Vert \phi_t (t,\cdot) \Vert_{H^{s}}^2 \right) \right\vert \leq \\
					&\qquad \leq (\const) \Vert A^{s} (\lambda(t,\cdot)  \phi^3(t,\cdot) ) \Vert_{L^{2}} \Vert A^{s} \phi_t (t,\cdot) \Vert_{L^{2}} \\
					&\qquad \leq (\const) \Vert \lambda(t,\cdot) \phi^3  (t,\cdot)  \Vert_{H^{s+1}} \Vert \phi_t  (t,\cdot)  \Vert_{H^{s}} \\
					&\qquad \leq (\const) \left(\supp_{t \in \bR} \Vert \lambda(t,\cdot) \Vert_{H^{s+1}} \right) \Vert \phi^3  (t,\cdot)  \Vert_{H^{s+1}} \Vert \phi_t  (t,\cdot)  \Vert_{H^{s}} \\
					&\qquad \leq (\const)  \Vert \phi^3  (t,\cdot)  \Vert_{H^{s+1}} \Vert \phi_t  (t,\cdot)  \Vert_{H^{s}} \\
					&\qquad \leq (\const) \; \Vert \phi  (t,\cdot)  \Vert^2_{L^{\infty}} \Vert \phi  (t,\cdot) \Vert_{H^{s+1}} \Vert \phi_t  (t,\cdot) \Vert_{H^{s}} \\
					&\qquad \leq \sC^{G} \; \Vert \phi  (t,\cdot)  \Vert^2_{L^{\infty}} \left( \Vert \phi (t,\cdot) \Vert_{H^{s+1}}^2 + \Vert \phi_t (t,\cdot) \Vert_{H^{s}}^2 \right),
				\end{split}
			\end{equation*}
		where $(\const)$ and $\sC^{G}$ are appropriate constants depending on $s$ and, eventually, on $\lambda$ via $\supp_{t \in \bR} \Vert \lambda(t,\cdot) \Vert_{H^{s+1}}$, which is necessarily finite because $\lambda \in C^\infty_0(\bR \times \Sigma)$. The Bellman-Gronwall inequality (see e.g.~\citep{bellman1943} or~\citep[thm.3 in ch. XII]{mitrinovic1991inequalities}) implies that
			\begin{equation}\label{est_gronwall}
				\Vert \phi (t,\cdot) \Vert_{H^{s+1}}^2 + \Vert \phi_t (t,\cdot) \Vert_{H^{s}}^2 \leq \left(  \Vert q \Vert_{H^{s+1}}^2 + \Vert p \Vert_{H^{s}}^2 \right) \exp \left( \sC \int_0^t   \Vert \phi  (\tau,\cdot)  \Vert^2_{L^{\infty}} d\tau\right).
			\end{equation}
		We emphasize that the constant $\sC$ is independent of $t$. If we can show that for any time $t$ in the domain of local existence the $L^\infty$-norm is finite, then the global existence of a solution with Sobolev regularity as in~\eqref{sol_sob} for the initial value problem follows as a consequence of~\eqref{est_gronwall} and the condition on the maximal time $T^*$ of local existence given by~\eqref{fin_time_blowup}.\\
		The first step is to get a bound on a suitably defined energy. It follows from~\eqref{d_A_dt} for $s=0$ that
			\begin{equation*}
				\frac{d}{dt} \left( \Vert \phi(t, \cdot) \Vert_{H^{1}}^2 + \Vert \phi_t (t, \cdot) \Vert_{L^{2}}^2 + \frac{1}{3}  \left( \lambda(t, \cdot), \phi^4(t, \cdot) \right)_{L^2} \right) = \frac{1}{3} \left( \lambda_t(t, \cdot), \phi^4(t, \cdot) \right)_{L^2} . 
			\end{equation*}
We define the energy as:
			\begin{equation*}
				e(\phi,t) := \Vert \phi(t, \cdot) \Vert_{H^{1}}^2 + \Vert \phi_t (t, \cdot) \Vert_{L^{2}}^2 + \frac{1}{3} \left( \lambda(t, \cdot), \phi^4(t, \cdot) \right)_{L^2}
			\end{equation*}
We can therefore bound the growth of $e(\phi,t)$ by 
			\begin{equation*}
				\begin{split}
					\left| \frac{d}{dt}  e(\phi,t) \right| &= \frac{1}{3} \left| \left( \frac{\lambda_t (t, \cdot) }{\lambda (t, \cdot) }\lambda(t, \cdot),\phi^4(t, \cdot) \right)_{L^2} \right| \leq \frac{1}{3} \sup_{(t,x) \in \bR \times \Sigma} \left| \frac{\lambda_t(t, x)}{\lambda(t, x)} \right| \left( \lambda(t, \cdot), \phi^4(t, \cdot) \right)_{L^2} \\
					&\leq  \sC^{s=0} e(\phi, t),
				\end{split}
			\end{equation*}
		where the constant $\sC^{s=0}$ is independent of $t$ and is proportional to $\sup_{(t,x) \in \bR \times \Sigma} | \lambda_t / \lambda |$ which is finite by hypothesis. We can apply again the Bellman-Gronwall inequality to obtain
			\begin{equation*}
				e(\phi,t) = e(\phi,0) \exp \left( \sC^{s=0} t \right),
			\end{equation*}						
		where $e(\phi,0)$ is the ``energy'' of the initial data, namely
			\begin{equation*}
				e(\phi,0) = \Vert  q \Vert_{H^{1}}^2 + \Vert p \Vert_{L^{2}}^2 + \frac{1}{3} \left( \lambda(0, \cdot), q^4(\cdot) \right)_{L^2}.
			\end{equation*}
		Starting form~\eqref{d_A_dt} with $s=1$ and exploiting the properties of the Sobolev norms, we obtain
			\begin{equation*}
				\begin{split}
				&\left\vert \frac{d}{dt} \left( \Vert A^2 \phi(t,\cdot) \Vert^2_{L^{2}} + \Vert A \phi_t(t, \cdot) \Vert^2_{L^{2}} \right) \right\vert = \frac{1}{3} \left\vert \left(A (\lambda(t, \cdot) \phi^3(t,\cdot)), A \phi_t(t, \cdot) \right)_{L^{2}} \right\vert \leq\\
				&\qquad \leq (\const)  \Vert \lambda(t, \cdot) \phi^3(t,\cdot) \Vert_{H^{1}} \Vert \phi_t(t,\cdot) \Vert_{H^{1}} \\
				&\qquad \leq (\const) \sup_{t \in \bR} \left( \sum_{ j=0,1} \sup_{x \in \Sigma} \vert (\partial^{(h)})^j \lambda(t, x) \vert_{h} \right) \Vert \phi^3(t,\cdot) \Vert_{H^{1}} \Vert \phi_t(t,\cdot) \Vert_{H^{1}}\\
				&\qquad \leq (\const) \left( \Vert A \phi^3(t,\cdot) \Vert_{L^{2}}^2 + \Vert \phi_t(t,\cdot) \Vert_{H^{1}}^2 \right)\\
				&\qquad \leq (\const) \left( \Vert \vert \partial^{(h)} \phi^3 \vert_h (t,\cdot) \Vert_{L^{2}}^2 + m^2 \Vert \phi^3(t, \cdot) \Vert_{L^{2}}^2 + \Vert \phi_t (t,\cdot) \Vert_{H^{1}}^2 \right)\\
				&\qquad \leq (\const) \left( \Vert \phi(t,\cdot) \Vert_{L^{6}}^4 \Vert \vert \partial^{(h)} \phi \vert_h (t,\cdot) \Vert_{L^{6}}^2 + \Vert \phi (t, \cdot) \Vert_{L^{6}}^6 + \Vert \phi_t (t,\cdot) \Vert_{H^{1}}^2 \right),
				\end{split}
			\end{equation*}
		where $(\const)$ are appropriate constants which does not depend on $t$ (remember that $\lambda \in C^\infty_0(\bR \times \Sigma)$). In the last step we used the H\"{o}lder inequality\footnote{In detail, for $p,q >0$ are such that $1/p + 1/q = 1$ it holds
				\begin{align}
					&\Vert \vert \partial^{(h)} \phi^3 \vert_h (t,\cdot) \Vert_{L^{2}}^2 = \int_\Sigma \left( \partial^{(h)} \phi^3 \cdot \partial^{(h)} \phi^3 \right)(t, x) d\Sigma(x) = 9 \int_\Sigma \phi^4(t,x) \left( \partial^{(h)} \phi \cdot \partial^{(h)} \phi\right) (t,x) d\Sigma(x) \\
					&\quad \leq 9 \left( \int_\Sigma \phi^{4p}(t,x) d\Sigma(x)\right)^{\frac{1}{p}}\left(\int_\Sigma \vert \partial^{(h)}\phi\vert_h^{2q}(t,x) d\Sigma(x)\right)^{\frac{1}{q}}.
			\end{align}
		The claimed result is obtained choosing $p=3/2$ and $q=3$.}. As a consequence of the Sobolev embedding $H^{1}(\Sigma) \subset L^{6}(\Sigma)$ and the Kato inequality\footnote{In detail, let $E \to \Sigma$ be a vector bundle and assume it is equipped with a Riemannian metric $e$. Then, for any section $\xi$ on $E$, the Kato's Inequality reads
				\begin{equation*}
					\vert d \vert \xi \vert_{e} \vert_{h}^2 \leq \vert \partial^{(e)} \xi \vert_{h^* \otimes e}^2.
				\end{equation*}
			Specializing it for $\xi = \partial^{(h)}\phi$, which is a section on the cotangent bundle $T^*\Sigma \to \Sigma$, we get
				\begin{equation*}
					\Vert \vert \partial^{(h)} \phi \vert_h \Vert_{H^1}^2 \leq 2 \int_\Sigma \left( \vert \partial^{(h)} \phi  \vert_h^2 + \vert \partial^{(h)} \vert \partial^{(h)} \phi \vert_h  \vert_h^2 \right) d\Sigma \leq 2 \int_\Sigma \left( \vert \partial^{(h)} \phi  \vert_h^2 + \vert (\partial^{(h)})^2 \phi \vert_h^2 \right) d\Sigma \leq 2  \Vert \phi \Vert_{H^2}^2.
\end{equation*}} we get
			\begin{align}
				\left\vert \frac{d}{dt} \left( \Vert \phi(t,\cdot) \Vert^2_{H^{2}} + \Vert \phi_t(t, \cdot) \Vert^2_{H^{1}} \right) \right\vert &\leq \sC^{s=1} \Vert \phi(t,\cdot) \Vert^4_{H^{1}} \left( \Vert \phi(t,\cdot) \Vert^2_{H^{2}} + \Vert\phi_t(t,\cdot) \Vert^2_{H^{1}} \right) \\
				 &\leq \sC^{s=1} e(\phi,0)^2 \exp \left( \sC^{s=0} t \right) \left( \Vert \phi(t,\cdot) \Vert^2_{H^{2}} + \Vert \phi_t(t,\cdot) \Vert^2_{H^{1}} \right),
			\end{align}
		where $\sC^{s=0}$ and $\sC^{s=1}$ are appropriate constants independent of $t$. Applying again the Bellmann-Gronwall's inequality, we obtain
			\begin{equation*}
				\Vert \phi(t,\cdot) \Vert^2_{H^{2}} + \Vert \phi_t(t, \cdot) \Vert^2_{H^{1}} \leq \left( \Vert q \Vert^2_{H^{2}} + \Vert p \Vert^2_{H^{1}}\right) \exp \left( \sC^{s=1} e(\phi,0)^2 \left( \exp( \sC^{s=0} t) - 1 \right) \right).
			\end{equation*}
		It follows from the result just proved that the $L^\infty$-norm of $\phi(t,\cdot)$ is finite for any finite time $t$. In fact, using the Sobolev embedding theorem $H^2(\Sigma) \subset L^\infty(\Sigma)$ we obtain
			\begin{equation}\label{estimate_Linfty}
				\Vert \phi(t, \cdot) \Vert_{L^\infty}^2 \leq \sC \Vert \phi(t, \cdot) \Vert_{H^2}^2 \leq \sC \left( \Vert q \Vert^2_{H^{2}} + \Vert p \Vert^2_{H^{1}}\right) \exp \left( \sC^{s=1} e(\phi,0)^2 \left( \exp( \sC^{s=0} t) - 1 \right) \right).
			\end{equation}
		Since the right-hand side of the inequality~\eqref{estimate_Linfty} is finite for any finite value of $t$, we conclude that for any $s >0$ there exists a global solution $u \in C^0(\bR, H^{s+1}(\Sigma)) \cap C^1(\bR, H^s(\Sigma))$ for initial value problem
			\begin{equation}\label{IVP_Sobolev}
				\left\{\begin{array}{l}
					\phi_{tt} - \Delta \phi + m^2 \phi =- \frac{1}{3!} \lambda \phi^3 \\
					\phi|_{t=0} = q \in H^{s+1}(\Sigma) \cap H^2(\Sigma)\\
					\phi_t|_{t=0} = p \in H^s(\Sigma) \cap H^1(\Sigma)
				\end{array}\right.
			\end{equation}
		Note that $H^{s+1}(\Sigma) \cap H^2(\Sigma) \subset H^{s+1}(\Sigma)$ and $H^s(\Sigma) \cap H^1(\Sigma) \subset H^s(\Sigma)$ for any $s>1$.\\
		
		We are interested in the initial value problem~\eqref{IVP_phi4} corresponding to smooth data $q,p$, and, for such data, we now establish that the solution is globally defined and smooth. Since $\Sigma$ is compact, it follows that the smooth data $q,p$ must satisfy $q \in H^{s+1}(\Sigma)$ and $p \in H^s(\Sigma)$ for any $s >0$. It follows that there exists a unique solution $\phi$ for the initial value problem~\eqref{IVP_Sobolev} in $C^0(\bR, H^{s+1}(\Sigma)) \cap C^1(\bR, H^s(\Sigma))$ for any $s>1$. For any $s> 3/2$ the space $H^{s}(\Sigma)$ equipped with the point-wise product forms an algebra, it holds
			\begin{equation*}
				\phi_{tt}=\Delta \phi - m^2 \phi - \frac{1}{3!} \lambda \phi^3 \in C^0(\bR, H^{s-1}(\Sigma)).
			\end{equation*}
		Taking an increasing number of derivatives of our partial differential equation and arguing in a similar way, we get
			\begin{equation*}
				\phi \in \bigcap_{s > \frac{3}{2}} \bigcap_{\ell < s+1} C^\ell(\bR, H^{s+1-\ell}(\Sigma)).
			\end{equation*}
		If we rearrange the intersections defining the set above choosing $s=2k+1$ and $\ell=k$ for any $k \in \bN$, then we get
			\begin{align}
				\phi & \in \bigcap_{s > \frac{3}{2}} \bigcap_{\ell < s+1} C^\ell(\bR, H^{s+1-\ell}(\Sigma)) \subset \bigcap_{k} C^k(\bR,H^{k+2}(\Sigma))\\
				&\qquad \subset \bigcap_k C^k(\bR, C^m(\Sigma)) = \bigcap_k C^k(\bR \times \Sigma) = C^\infty(\bR \times \Sigma),
			\end{align}
		where we used $H^s(\Sigma) \subset C^{k}(\Sigma)$ if $s >3/2 + k$. This concludes the proof.
	\end{proof}
		
	\paragraph{Continuous dependence of solutions on initial data:} Having established that the initial value problem~\eqref{IVP_phi4} is well-posed, we now want to prove the continuity of the map
		\begin{equation*}
			\sE = C^\infty(\Sigma) \oplus C^\infty(\Sigma) \ni (q,p) \mapsto U(q,p) \in S \subset C^\infty(M),
		\end{equation*}
	i.e. the map that assigns to each pair of Cauchy data the corresponding unique smooth global solution. We recall the construction of the topologies involved. First consider $C^\infty(X)$ with $X$ a finite-dimensional manifold. The {\em compact-open topology} on $C^\infty(X)$ is the topology of uniform convergence of functions and all their derivatives on any compact set $K \subset X$. More precisely, this topology is induced by the {\em supremum seminorms} defined as
		\begin{equation*}
			p^{(C^\infty(X))}_{\infty,n,K}(f):= \sup_{x \in K} \left( \sum_{j=0}^n \vert (\partial^{(e)})^j f \vert_e^2(x) \right)^{1/2},
		\end{equation*}
	where $e$ is some Rimannian metric on $X$, where $(\partial^{(e)})^j$ denotes the $j$-th covariant derivative defined in terms of the Levi-Civita connection of $e$, and where $\vert \cdot \vert_e$ is the natural norm for tensors defined via the metric $e$ (see footnote~\ref{footnote}). By the Sobolev embedding theorem, the compact-open topology can be generated by another family of seminorms, the local {\em Sobolev seminorms} defined as
		\begin{equation}\label{Sobolev_semi}
			p^{(C^\infty(X))}_{H,n,K}(f):= \left( \sum_{j=0}^n \int_K \vert (\partial^{(e)})^j f \vert_e^2 d\vol^{(e)} \right)^{1/2},
		\end{equation}
	where $d\vol^{(e)}$ is the volume form with respect to the Riemannian metric $e$. Both the supremum seminorms and the Sobolev seminorms are separating, i.e. if $f \neq 0$, then there exist $n,K$ such that $p_{\infty/H,n,K}(f) \neq 0$. If the manifold $X$ is locally compact, then there exists a countable family of compact sets $\{K_n\}_{n \in \bN}$, such that $K_n$ is contained in the interior of $K_{n+1}$ and $\cup_n K_n = X$. Evidently both $\Sigma$ and $M=\bR \times \Sigma$ satisfy this condition. We can extract from each of the families of seminorms defined before a countable family, i.e. $p^{(C^\infty(X))}_{\infty/H, n} := p^{(C^\infty(X))}_{\infty/H, K_n, n}$. The compact-open topology does not depend on the choice of $e$ and $\{K_n\}_{n \in \bN}$, and, furthermore, gives on $C^\infty(X)$ the structure of a Fr\'{e}chet space.\\
	Since $\Sigma$ is compact, $p^{(C^\infty(\Sigma))}_{H,n} (\cdot)$ is equivalent to $\Vert \cdot \Vert_{H^n}$. Therefore, the Fr\'{e}chet  topology on $\sE=C^\infty(\Sigma) \oplus C^\infty(\Sigma)$ is given by the family of seminorms
			\begin{equation*}
				p^{(\sE)}_n (q,p) := \Vert q \Vert_{H^{n+1}} + \Vert p \Vert_{H^n}.
			\end{equation*}
	On $C^\infty(M)$, one is free to choose $e= dt^2 + h_{ij} dx^idx^j$ and $K_n = I_n\times \Sigma$, where $I_n$ is the interval $(-(T+n), T+n)$ for a fixed $T$. The compact-open topology on $C^\infty(M)$ is defined in terms of the seminorns
		\begin{equation}\label{M_Sob_Semi}
			\begin{split}
				p^{(C^\infty(M))}_{n} (\phi) := p^{(C^\infty(M))}_{H,n} (\phi) &= \left( \sum_{j=0}^n \int_{I_n \times \Sigma} \vert (\partial^{(e)})^j \phi \vert_e^2 d\vol^{(e)} \right)^{1/2}\\
				&= \left( \sum_{k,j \leq n} \left( {}^{k}_{j} \right) \int_{I_n \times \Sigma} \vert (\partial^{(h)})^{j-k} (\partial_t)^k \phi \vert_e^2 d\vol^{(e)} \right)^{1/2}.
			\end{split}
		\end{equation}
	We evidently have
		\begin{equation*}
			\begin{split}
				p^{(C^\infty(M))}_{n} (\phi) &\leq (\const)  \; \sup_{t \in I_n} \left( \sum_{k,j \leq n} \Vert ((\partial_t)^k\phi)(t, \cdot) \Vert^2_{H^{j-k}} \right)^{1/2} \\
				&\leq (\const)  \; \sup_{t \in I_n} \left( \sum_{k,s \leq n} \Vert ((\partial_t)^k\phi)(t, \cdot) \Vert^2_{H^s} \right)^{1/2},
			\end{split}
		\end{equation*}
	where $(\const)$ is a constant which depend on $n$ but not on $t$.\\
	The compact-open topology on $C^\infty(M)$ induces a topology on the set $S \subset C^\infty(M)$ of the smooth solutions of the initial value problem~\eqref{IVP_phi4}.\\
	It follows immediately that if $p^{(C^\infty(M))}_n (\phi) \to 0$ for any $n$, then necessarily $\Vert \phi(t,\cdot) \Vert_{H^s} \to 0$ for any $s$ and any fixed $t$. In other words, the restriction to the surface $\{t\} \times \Sigma$ is a continuous map $C^\infty(M) \to C^\infty(\Sigma)$.
		\begin{prop}\label{prop_U_cont}
			The map $U:\sE \to S \subset C^\infty(M)$ is continuous with respect to the topologies we introduced before.
		\end{prop}
		\begin{proof}
			We need to verify the following implication
				\begin{equation*}
					p^{(\sE)}_k ((q,p) - (q', p')) \rightarrow 0 \; \forall k \quad \Rightarrow \quad p^{(C^\infty(M))}_\ell (\phi - \phi') \rightarrow 0 \; \forall \ell,
				\end{equation*}
			where $\phi$ and $\phi'$ are the unique solutions corresponding respectively to initial data $(q,p)$ and $(q', p')$. Since $\phi$, $\phi'$ are solutions to the same non-linear equation, it follows that $d:=(\phi - \phi')$ is a smooth solution for the following initial value problem
				\begin{equation}\label{IVP_diff}
					\left\{\begin{array}{l}
						d_{tt} + (m^2 - \Delta^{(h)}) d = - \frac{1}{2}\lambda \phi \phi' d - \frac{1}{3!} \lambda d^3 \\
						d|_{t=0}=q - q' \in C^\infty(\Sigma) \\
						d_t|_{t=0}=p - p' \in C^\infty(\Sigma)
					\end{array}\right.
				\end{equation}
			We can adapt the same argument based on the Gronwall-Bellmann inequality we used in~\eqref{est_gronwall} to obtain
				\begin{equation*}
					\begin{split}
						&\Vert d (t,\cdot) \Vert_{H^{s+1}}^2 + \Vert d_t (t,\cdot) \Vert_{H^{s}}^2 \leq \\
						&\quad \leq \left(  \Vert q - q' \Vert_{H^{s+1}}^2 + \Vert p - p' \Vert_{H^{s}}^2 \right) \exp \left( \sC \int_0^t (1 + \Vert \phi (\tau,\cdot) \Vert^4_{H^{s+1}}  + \Vert \phi' (\tau,\cdot) \Vert^4_{H^{s+1}} ) d\tau \right),
							\end{split}
				\end{equation*}
			for any $s > 1/2$, and for a positive constant $\sC$ independent of $t$ (depending, however, on $s$ and on $\lambda$). We extensively used the properties of the Sobolev norms. Since $\phi$ and $\phi'$ are smooth solutions, both $\phi(t,\cdot)$ and $\phi'(t,\cdot)$ are $H^{s+1}$-bounded for any finite time $t$. It follows that we have the following bound:
				\begin{equation*}
					\Vert d (t,\cdot) \Vert_{H^{s+1}}^2 + \Vert d_t (t,\cdot) \Vert_{H^{s}}^2 \leq D_s^2 F_s(t),
				\end{equation*}
			where $D_s$ simply denotes $p^{(\sE)}_s(q -q', p - p')$, while $F_s$ denotes the continuous and positive function
				\begin{equation*}
					F_s(t) := \exp \left( \sC \int_0^t (1 + \Vert \phi (\tau,\cdot) \Vert^4_{H^{s+1}}  + \Vert \phi' (\tau,\cdot) \Vert^4_{H^{s+1}} ) d\tau \right).
				\end{equation*}
			Thus, we have proved that for $s > 3/2$ the following bounds hold
				\begin{equation}\label{first_lines}
					\begin{split}
						\Vert d (t,\cdot) \Vert_{H^s}^2 &\leq D_{s-1}^2 F_{s-1}(t), \\
						\Vert d_t (t,\cdot) \Vert_{H^s}^2 &\leq D_s^2 F_s(t). \\
					\end{split}
				\end{equation}
			Using the fact that $d$ is a solution for~\eqref{IVP_diff} and the properties of the Sobolev norms, we obtain the following bound for $s > 3/2$:
				\begin{equation}\label{last_line}
					\begin{split}
					&\Vert d_{tt}(t, \cdot) \Vert_{H^{s}}^2 \leq\\
					&\quad \leq (\const) \left( \Vert (m^2 - \Delta^{(h)})d (t, \cdot) \Vert_{H^s} + \Vert \lambda(t,\cdot) d^3 (t,\cdot) \Vert_{H^s} + \Vert \lambda(t, \cdot) (\phi \phi' d) (t, \cdot) \Vert_{H^s} \right)^2\\
					&\quad \leq (\const) \left( 1 + \Vert \phi (t, \cdot) \Vert^4_{H^s}  + \Vert \phi' (t,\cdot)\Vert_{H^s})^4  \right) \Vert d (t, \cdot) \Vert^2_{H^{s+2}},
					\end{split}
				\end{equation}
			where $(\const)$ are constants depending on $s$ and on $\lambda$. Since $\phi$ and $\phi'$ are smooth solutions with smooth Cauchy data, $1 + \Vert \phi (t, \cdot) \Vert^4_{H^s} + \Vert \phi' (t,\cdot)\Vert^4_{H^s}$ is a continuous function in $t$. Furthermore, using the estimates~\eqref{first_lines}, we can rewrite the inequality~\eqref{last_line}, as 
				\begin{equation}\label{bound_tt}
					\Vert d'_{tt}(t, \cdot) \Vert_{H^{s}}^2 \leq D_{s+1}^2 F_{2,s}(t),
				\end{equation}
			where the function $F_{2,s}$ is given by
				\begin{equation*}
					F_{2,s}(t) := \sC^{2} \left( 1 + \Vert \phi (t, \cdot) \Vert^4_{H^s}  + \Vert \phi' (t,\cdot)\Vert_{H^s}^4  \right) F_{s + 1}(t),
				\end{equation*}
			To obtain bounds of the form~\eqref{bound_tt} for higher order time-derivatives of $d$, one takes further derivatives in $t$ of the partial differential equation $d_{tt} = -(m^2 - \Delta^{(h)}) d - \lambda \phi \phi'd - \frac{\lambda}{3!} d^3$. Repeating these kinds of arguments, one can show inductively that for any order $k$ (and for any Sobolev order $s >3/2$) there is a continuous function $F_{k, s}(t)$ which depends continuously on $\Vert (\partial_t)^n \phi (t, \cdot) \Vert_{H^s}$ and $\Vert (\partial_t)^n \phi' (t,\cdot)\Vert_{H^s}$ with $n \leq k -2$, such that it holds
				\begin{equation*}
					\Vert (\partial_t)^k d(t,\cdot) \Vert_{H^s}^2 \leq D_{s + k -1}^2 F_{k, s}(t).
				\end{equation*}
			This last result implies the following bound:
				\begin{equation*}
					p^{(C^\infty(M))}_\ell (d) \leq \sC  \; \sup_{t \in I_\ell} \left( \sum_{k,s \leq \ell}  D^2_{s+ k-1} F_{k,s}(t) \right)^{1/2},
				\end{equation*}
			where $\sC$ is an appropriate constant independent of $t$. Now, if $D_k \to 0$ for any $k$, then necessarily $p^{(C^\infty(M))}_\ell (d) \to 0$, as we wanted to prove.
		\end{proof}

	\section{Continuity in $\phi$ of the Cauchy problem for the linearized equation around $\phi$}
		Let now $\phi \in S$, i.e. $\phi$ is a smooth solution of the initial value problem~\eqref{IVP_phi4}. It is well-know that the Cauchy problem for the linearized equation, i.e.
			\begin{equation}\label{IVP_lin}
				\left\{ \begin{array}{ll}
					u_{tt}(t,x) - (\Delta^{(h)} u)(t,x) + (m^2 + \frac{1}{2} \lambda(t,x) \phi^2(t,x)) u(t,x) = 0,\\
					u|_{t=0} = q, \\
					u_t |_{t=0} = p,
				\end{array} \right.
			\end{equation}
		is globally well-posed for $u \in C^\infty(\bR \times \Sigma)$ and $q, p \in C^\infty(\Sigma)$. We denote by $u_{\phi}(q,p)$ the unique solution of~\eqref{IVP_lin} and we investigate its dependence on $\phi \in S$:
			\begin{prop}\label{prop_cont_causal}
				For any $p,q \in C^\infty(\Sigma)$, the map
					\begin{equation*}
						S \ni \phi \mapsto u_\phi(q,p) \in T_\phi S
					\end{equation*}
				is continuous in the topologies for $S$ and $T_\phi S$ induced by the compact-open topology of $C^\infty(M)$.
			\end{prop}
			\begin{proof}
				We consider the more general case of a background $\phi \in C^\infty(M)$ not necessarily a solution of the initial value problem~\eqref{IVP_phi4}. In particular, let $\phi$ and $\phi'$ be in $C^\infty(M)$, and let $q,p \in C^\infty(\Sigma)$. We define the smooth function $d:= u_\phi(q,p) - u_{\phi'}(q,p)$. We need to prove the following implication:
					\begin{equation*}
						p^{(C^\infty(M))}_k (\phi - \phi')\rightarrow 0 \; \forall k \quad \Rightarrow \quad p^{(C^\infty(M))}_\ell (d) \rightarrow 0 \; \forall \ell.
					\end{equation*}
				Since $u_\phi(q,p)$ and $u_{\phi'}(q,p)$ are the unique solutions of the initial value problem~\eqref{IVP_lin} with data $q,p$ respectively for $\phi$ and $\phi'$, we deduce that $d$ is a global smooth solution for the following initial value problem
					\begin{equation}\label{IVP_lin_diff}
						\left\{\begin{array}{l}
							d_{tt}(t,x) + (m^2 - \Delta^{(h)}) d(t,x) = - \lambda(t,x) \phi^2(t,x) d(t,x) - \lambda(t,x) (\phi(t,x) - \phi'(t,x)) j_{\phi,\phi'}(t,x)\\
							d|_{t=0}=0 \\
							d_t|_{t=0}=0
						\end{array}\right.
					\end{equation}
				where $j_{\phi,\phi'}:=(\phi + \phi') u_{\phi'}(q, p)$. Exploiting the properties of the Sobolev norms, we obtain from~\eqref{IVP_lin_diff} the following inequality for $s > 3/2$:
					\begin{equation}\label{ineq_lin_diff}
						\left\vert \frac{d}{dt}  \left( \Vert d(t,\cdot) \Vert^2_{H^{s+1}} + \Vert d_t(t, \cdot) \Vert^2_{H^{s}} \right) \right\vert \leq A(t) \left( \Vert d(t,\cdot) \Vert^2_{H^{s+1}} + \Vert d_t(t, \cdot) \Vert^2_{H^{s}} \right) + B(t),
					\end{equation}
				where $A(t), B(t)$ are defined by
					\begin{equation*}
						A(t) \leq \sC^A \left( 1 + \Vert \phi(t,\cdot) \Vert^2_{H^{s}} \right),
					\end{equation*}
				and
					\begin{equation*}
						B(t) \leq \sC^B \Vert (\phi - \phi')(t,\cdot) \Vert_{H^s} \Vert j_{\phi,\phi'}(t,\cdot) \Vert_{H^s},
					\end{equation*}
				where $\sC^A, \sC^B$ are appropriate constants. Since $\phi$, $\phi'$ and $j_{\phi,\phi'}$ are smooth functions, it follows that $A(t)$ and $B(t)$ are finite for all $t$. Now we fix a time $t$. As a consequence of inequality~\eqref{ineq_lin_diff}, for any $\tau \in [0,t+1]$ we obtain
					\begin{equation*}
						\Vert d'(\tau,\cdot) \Vert^2_{H^{s+1}} + \Vert d_t(\tau, \cdot) \Vert^2_{H^{s}} \leq M_B \tau + M_A \int_0^\tau \left( \Vert d(\tau',\cdot) \Vert^2_{H^{s+1}} + \Vert d_t(\tau', \cdot) \Vert^2_{H^{s}} \right) d\tau',
					\end{equation*}
				with $M_A$ and $M_B$ respectively the maximum of $A$ and $B$ in $[0,t+2]$. We apply a slight generalization of the Gronwall-Bellman inequality (see~\citep[thm. 3 ch. XII]{mitrinovic1991inequalities}) and we get
					\begin{equation*}
						\Vert d(t,\cdot) \Vert^2_{H^{s+1}} + \Vert d_t(t, \cdot) \Vert^2_{H^{s}} \leq M_B \left( t + M_A \int_0^t \tau' \exp(M_A (t-\tau')) \right) d\tau'.
					\end{equation*}
				The fact that $p^{(C^\infty(M))}_k(\phi - \phi') \to 0$ for any $k$ implies that $\Vert (\phi - \phi')(t,\cdot) \Vert_{H^k} \to 0$ for any finite $t$, and then $B(t) \to 0$. This means that both $\Vert d(t,\cdot) \Vert^2_{H^{s}}$ and $\Vert d_t(t,\cdot) \Vert^2_{H^{s}}$ (for a sufficiently large order $s$) must vanish in the limit. Arguing as in prop.~\ref{prop_U_cont}, we can prove similar results for $\Vert (\partial_t)^k d(t,\cdot) \Vert^2_{H^{s}}$ for higher order $k$.\\
				Finally, making use of formula~\eqref{M_Sob_Semi}, we can conclude that $p_\ell (d)$ is bounded by a quantity which vanishes if $p^{(C^\infty(M))}_k(\phi - \phi') \to 0$ for any $k$.
			\end{proof}

\chapter{Background dependence of the propagators $E^{A/R}_\phi$}\label{app_bkgr_dep_propa}
	In this appendix, we investigate the behaviour of $E^{A/R}_\phi$, the advanced/retarded propagators with respect to the operator $P_\phi=\boxempty - m^2 -V''(\phi)$, under variations of the background $\phi \in C^\infty(M)$. We assume that $V$ is a local functional, e.g. $V(\phi) = \int_M \frac{1}{4!} \lambda(x) \phi^4(x) dx$ with $\lambda \in C^\infty_0(M)$.
	First of all we recall the defining relations of $E^{A/R}_\phi$:
		\begin{equation*}
			P_\phi^{(x_1)} E^{A/R}_\phi(x_1,x_2) = \delta(x_1,x_2) = P_\phi^{(x_2)} E^{A/R}_\phi(x_1,x_2) \qquad \supp( E^{A/R}_\phi(f)) \subset J^{\mp}(\supp f) \quad  \forall  f \in C^\infty_0(M).
		\end{equation*}
	Let $\phi$ be a fixed smooth function. Consider a smooth map $\bR \ni s \mapsto \phi(s) \in C^\infty(M)$ such that $\phi(0) = \phi$. We regard $E^{A/R}_{\phi(s)}(x_1,x_2)$ as distributions in $\cD^\prime(\bR \times M^2)$, i.e. in the variables $s, x_1,x_2$. From the wave-front set of the advanced/retarded propagators, see~\eqref{A/R_WF}, it follows
		\begin{equation}\label{rough_adv_WF}
			\WF(E^{A/R}_{\phi(\cdot)}) \subset \left\{ (s, x_1, x_2; \rho, k_1, k_2) \in \dot{T}^\ast (\bR \times M^2)| (x_1, x_2; k_1, k_2) \in \cC^{A/R}(M)\right\},
		\end{equation}
	where the set $\cC^{A/R}(M)$ is defined by eq.~\eqref{C^A/R}.\\
		The estimate~\eqref{rough_adv_WF} does not put any restriction on the $s$-part of the wave-front set, i.e. it does not impose any control on the dependence under variations of the background. We prove a stronger bound:
		\begin{prop}\label{prop_WF_A_bkgr_dep}
			It holds
				\begin{equation}\label{WF_A_bkgr_dep}
					\WF(E^{A/R}_{\phi(s)} (x_1,x_2)) \subset \bR\times\{0\} \times \cC^{A/R}.
				\end{equation}
		\end{prop}
		\begin{proof}
			We present explicitly the proof of the claim for the advanced propagator $E^A_\phi$. The proof relies on the propagation of singularities for hyperbolic partial differential equations and the Hadamard expansion for the advanced propagator. Except for some obvious adjustments, the argument we are presenting holds also for the retarded propagator $E^R_\phi$. \\
			First of all, we prove that it would be sufficient to prove the claim for $x_1,x_2$ both contained in the same convex normal set $U$. Let $(s, x_1, x_2; \rho, k_1, k_2)$ be an element of $\WF(E^{A/R}_{\phi(\cdot)})$. Assume that $x_1, x_2$ do not belong to the same convex normal neighbourhood $U$. By the estimate~\eqref{rough_adv_WF}, we have $(x_1, k_1) \sim (x_2, -k_2)$. By definition, $E^{A}_{\phi(s)}(x_1,x_2)$ satisfies the following equation
				\begin{equation*}
					P_{\phi(s)}^{(x_1)} E^{A}_{\phi(s)}(x_1,x_2)=\delta(x_1,x_2)= P_{\phi(s)}^{(x_2)} E^{A}_{\phi(s)}(x_1,x_2).
				\end{equation*}
			The propagation of singularities (see~\citep[thm. 8.3.3']{H83} and~\citep[thm. 6.1.1]{duistermaat1972fourier}) implies that $\WF(E^{A/R}_{\phi(\cdot)}) \backslash (\bR \times \{0\} \times \WF(\delta))$ is invariant under the action of the Hamiltonian vector field $\hami$ associated to the principal symbol of the differential operator $P_{\phi(s)}^{(x_2)}$, namely
				\begin{equation*}
					\hami(s, x_1, x_2; \rho, k_1, k_2):=\frac{\partial g^{\lambda \nu}_{x_2}}{\partial x_2^\mu}(k_2)_\lambda (k_2)_\nu \frac{\partial}{\partial (k_2)_\mu} - 2g_{x_2}^{\mu \nu} (k_2)_\nu \frac{\partial}{\partial x_2^\mu}.
				\end{equation*}
			Since $\hami$ depends only on $x_2, k_2$, it follows that if $(s, x_1, x_2; \rho, k_1, k_2)$ is contained in $\WF(E^{A/R}_{\phi(\cdot)})$, then there is $(s, x_1, x'_2; \rho, k_1, k'_2)$ in $\WF(E^{A/R}_{\phi(\cdot)})$ such that $(x'_2, k'_2) \sim (x_2,k_2)$. For any convex normal neighbourhood $U$ of $x_1$, we are free to choose $x'_2 \in U$.\\
			Next, we prove the estimate~\eqref{WF_A_bkgr_dep} for $x_1, x_2$ in the same convex normal set $U$. In $U^2$, we can construct the advanced Hadamard parametrix, i.e. a bi-distribution $H^{A}_\phi(x,y)$ on $U \times U$ such that:
			 \begin{itemize}
			 	\item It is a fundamental $P_\phi$-solution modulo smooth functions, i.e.
						\begin{equation*}
							P_\phi^{(x_1)} H^{A}_\phi(x_1,x_2) = \delta(x_1,x_2)+G_\phi^{A(1)}(x_1,x_2), \qquad P_\phi^{(x_2)} H^{A}_\phi(x_1,x_2) =\delta(x_1,x_2)+G_\phi^{A(2)}(x_1,x_2),
						\end{equation*}
					for some $G^{A(1,2)}_\phi \in C^\infty(U^2)$.
				\item Its support satisfies
						\begin{equation*}
							\supp (H^{A}_\phi(f)) \subset J^-(\supp f) \cap U
						\end{equation*}
					for any test function $f$ supported in $U$.
			\end{itemize}
			Inside $U \times U$, we can trivially decompose $E^A_{\phi}$ in $H^A_{\phi} + d^A_{\phi}$, where $d^A_{\phi} := E^A_{\phi} - H^A_{\phi}$. The advantage of this decomposition is that $d^A_\phi(x_1,x_2)$ is smooth in $(x_1,x_2)$ (it is shown in~\citep[proof of prop. 2.5.1]{BGP07} that $d^A$ must be $C^k$ for any $k$), and $H^A_{\phi}$ is locally and covariantly constructed in terms of the metric and $m^2 + V''(\phi)$. We proceed by showing that both $H^A_{\phi(s)}(x_1,x_2)$ and $d^A_{\phi(s)}(x_1,x_2)$ separately satisfy the estimate~\eqref{WF_A_bkgr_dep}.\\
			Following~\citep{BGP07, S13}, the advanced Hadamard parametrix is
				\begin{equation*}
					\begin{split}
						H_\phi^{A}(x_1, x_2) &:= u_0(x_1, x_2) \delta(\sigma(x_1, x_2)) \theta(-(x_1^0 - x_2^0)) +\\
						&\quad + \sum_{k \geq0} \psi\left( \frac{\sigma(x_1, x_2)}{\alpha_k} \right) u_{\phi, k+1} (x_1, x_2) \sigma(x_1,x_2)^k \theta(-\sigma(x_1, x_2)) \theta(-(x_1^0 - x_2^0)),
						\end{split}
				\end{equation*}
			where $\sigma$ is the signed squared geodesic distance~\eqref{signed_geo_dist_square}, and where $u_{k,\phi}$ are the Hadamard coefficients defined recursively by formula~\eqref{induct_u} starting from $u_0$ defined by~\eqref{u_0}. Here, $\psi: \bR \to \bR$ is a compactly supported smooth function and $\{ \alpha_k \}_{k \in \bN}$ is a sequence of real number which are introduced to ensure the convergence of the series in case $(M, g)$ is not a real analytic space-time. More precisely, $\psi$ is chosen such that $\psi(x) = 1$ for $|x| < 1/2$ and $\psi(x) = 0$ for $|x| >1$ and, for increasing $k$, $\alpha_k$ tends to zero sufficiently fast such that the series converges in the sense of~\citep[lemma 2.4.2]{BGP07}.\\
			For any $k$, the function $u_{k, \phi(s)}(x_1, x_2)$ is jointly smooth in $s, x_1 ,x_2$ as can be proved by induction on $k$ using the recursive definition~\eqref{induct_u}. The wave-front sets of the distributions $\delta(\sigma(x_1, x_2))$, $\theta(-(x_1^0 - x_2^0))$ and $\theta(-\sigma(x_1, x_2))$ can be explicitly computed, and, using the wave-front set calculus (thm.~\ref{theo_WF_horma}), it follows that $H^{A}_{\phi(s)} (x_1, x_2)$ satisfies the following wave-front set condition:
				\begin{equation*}
					\WF(H^{A}_{\phi(s)}(x_1,x_2)) \subset \bR\times\{0\} \times \cC^A|_{U^2}.
				\end{equation*}
			To conclude the proof, it is sufficient to prove that $d_{\phi(s)}(x_1,x_2)$ is jointly smooth in $s,x_1,x_2$. First, we note that by construction
				\begin{equation*}
					P_{\phi(s)}^{(x_1)} d^{A}_{\phi(s)}(x_1,x_2)=G_{\phi(s)}^{A(1)}(x_1,x_2), \qquad P_{\phi(s)}^{(x_2)}d^{A}_{\phi(s)}(x_1,x_2)=G_{\phi(s)}^{A(2)}(x_1,x_2).
				\end{equation*}
			The remainders $G_\phi^{A(1,2)}$ can be calculated explicitly (see~\citep[lemma 2.4.3]{BGP07}) and, furthermore, one can show that the $G_{\phi(s)}^{A(1,2)}(x, y)$ are jointly smooth in $s, x, y$.\\
			We are free to choose $U$ sufficiently small such that $U \Subset U'$ for another normal convex set $U'$ and such that there must be two Cauchy surfaces $\Sigma_+$ and $\Sigma_-$ which satisfy $\Sigma_\pm \cap U' \neq \emptyset$ and $\Sigma_\pm \cap J^\mp( U)= \emptyset$. We consider a cut-off function $\chi \in C^\infty_0(M)$ such that $\chi(M) \subset [0,1]$, $\chi =1$ in $U$, and $\Sigma_\pm \cap J^\mp (\supp \chi)= \emptyset$.\\
			In this set-up, consider the advanced propagator $E^{A}_{\chi \phi(\cdot)}$, the advanced Hadamard parametrix $H^{A}_{\chi \phi(\cdot)}$, the remainder terms $G^{A(1,2)}_{\chi \phi(\cdot)}$ and the difference $d^{A}_{\chi \phi(\cdot)}$ defined in $\bR \times U' \times U'$ with respect to the operator $P_{\chi \phi} = \boxempty - m^2 - V''(\chi \phi)$. These distributions fulfil similar properties of their counterparts $E^A_{\phi(\cdot)}$, $H^A_{\phi(\cdot)}$, $G^{A(1,2)}_{\phi(\cdot)}$ and $d^A_{\phi(\cdot)}$ (defined in $\bR \times U \times U$) corresponding to $P_\phi$. Since $\chi$ is identically $1$ in $U$, we claim that in $\bR \times U \times U$ the distributions $E^A_{\chi \phi(\cdot)}$ and $H^A_{\chi \phi(\cdot)}$ coincide with their counterparts $E^A_{\phi(\cdot)}$ and $H^A_{\phi(\cdot)}$. For $E^A_{\chi \phi(\cdot)}$, the claim follows from the uniqueness of the advanced propagator in $U \times U$ and the fact that $P_\phi = P_{\chi \phi}$ in $U$. For $H^A_{\chi \phi(\cdot)}$, the claim is a consequence of the fact that the Hadamard coefficients $u_{k,\phi}(x_1,x_2)$ depend on $\phi$ only along the geodesic connecting $x_1$ and $x_2$, which is contained in $U$ if $x_1, x_2 \in U$ because $U$ is a convex normal set. It clearly follows that $d^A_{\chi \phi(\cdot)}$ and $d^A_{\phi(\cdot)}$ must coincide in $\bR \times U \times U$.\\
			Next, we proceed arguing similarly as in~\citep[lemma 6.2]{HW01}. For any test functions $f_1,f_2$ supported in $U$, we have
				\begin{equation}\label{rewrite_d^A}
					\begin{split}
						d^A_{\phi(s)}(f_1, f_2) &= d^A_{\chi \phi(s)}(f_1, f_2)\\
						&= - \int_{\Sigma_- \times \Sigma_-} (E^A_{\chi \phi(s)} f_1)(z_1) \overleftrightarrow{\partial_n} d^A_{\chi \phi(s)}(z_1, z_2) \overleftrightarrow{\partial_n} (E^A_{\chi \phi(s)} f_2)(z_2) d\Sigma(z_1) d\Sigma(z_2) -\\
						&\quad - G^{A(2)}_{\chi \phi(s)}(E^A_{\chi \phi(s)} f_1, f_2) - G^{A(1)}_{\chi \phi(s)}(f_1, E^A_{\chi \phi(s)} f_2) - \\
						&\quad - \frac{1}{2} G^{A(2)}_{\chi \phi(s)}(P_{\chi \phi(s)} f_1, f_2) - \frac{1}{2} G^{A(1)}_{\chi \phi(s)}(f_1, P_{\chi \phi(s)} f_2).
					\end{split}
				\end{equation}
			By construction, there must be a neighbourhood of $\Sigma_-$ sufficiently small such that $\chi =0$ in this neighbourhood. Therefore, for $x_1, x_2$ sufficiently close to $\Sigma_-$ the function $d^A_{\chi \phi(s)}(x_1,x_2)$ coincides with $d^A_0(x_1,x_2)$, and so it does not depend on $s$. As we already discussed $G^{A(1,2)}_{\chi \phi(s)}(x_1,x_2)$ are jointly smooth in $s,x_1,x_2$. We use the estimate~\eqref{WF_A_bkgr_dep} for the wave-front set of $E^A_{\chi \phi(s)}(x_1,x_2)$. In particular, we note that it cannot contain elements $(s,x_1,x_2; \rho, k_1, k_2)$ with $k_1=0$ or $k_2=0$. Then, using the wave-front set calculus (thm.~\ref{theo_WF_horma}) and eq.~\eqref{rewrite_d^A}, we find that $d^A_{\phi(s)}(x_1,x_2)$ is jointly smooth in $s, x_1, x_2$, for $x_1, x_2 \in U$.\\
			Putting together the results we derived for $H^A_{\phi(\cdot)}$ and $d^A_{\phi(\cdot)}$, it follows that $\WF( E^A_{\phi(\cdot)} ) \subset \bR \times \{0\} \times \cC^{A}|_{U^2}$ as we wanted to prove.\\
			A similar argument holds for the retarded product.
		\end{proof}
		
		We next want to compute the directional derivative of $E^{A}_\phi$ in $\phi \in C^\infty(M)$ along the direction $h \in C^\infty(M)$, i.e.
			\begin{equation*}
				\left. \frac{d}{d\epsilon} E^{A}_{\phi +\epsilon h} (f_1,f_2) \right|_{\epsilon=0}
			\end{equation*}
		for any $f _1,f_2 \in C^\infty_0(M)$. Because $\epsilon \mapsto \phi +\epsilon h$ is clearly a smooth function, the estimate.~\eqref{WF_A_bkgr_dep} of prop.~\ref{prop_WF_A_bkgr_dep} holds and, thus, we can apply the Leibniz rule for the directional derivative to obtain
			\begin{equation*}
				0 = \left. \frac{d}{d\epsilon} E^{A}_{\phi +\epsilon h} (f_1,P_{\phi + \epsilon h}f_2) \right|_{\epsilon=0} = E^{A}_\phi \left(f_1, \left. \frac{d}{d\epsilon} P_{\phi +\epsilon h} f_2 \right|_{\epsilon=0} \right) + \left. \frac{d}{d\epsilon} E^{A}_{\phi +\epsilon h} (f_1, P_\phi f_2) \right|_{\epsilon=0}.
			\end{equation*}
		Then, it follows
			\begin{equation*}
				\left. \frac{d}{d\epsilon} E^{A}_{\phi +\epsilon h} (f_1, P_\phi f_2)\right|_{\epsilon=0} = - E^{A}_\phi \left(f_1, \left. \frac{d}{d\epsilon} P_{\phi +\epsilon h} f_2 \right|_{\epsilon=0} \right) = E^{A}_\phi (f_1, h V'''(\phi) f_2).
			\end{equation*}
		We would like to replace $f_2$ by $E^A_\phi(f_2)$ in the formula above. However, this cannot be done since the smooth function $E^A_\phi(f_2)$ is not compactly supported. To circumvent this issue we choose a partition of unity $\{\psi_n\}$ for $M$. In detail, we get
			\begin{equation*}
				\begin{split}
				\left. \frac{d}{d\epsilon} E^{A}_{\phi +\epsilon h} (f_1,f_2) \right|_{\epsilon = 0} &= \left. \frac{d}{d\epsilon} E^{A}_{\phi +\epsilon h} (f_1, P_\phi E^{A}_\phi (f_2)) \right|_{0} = \left. \frac{d}{d\epsilon} E^{A}_{\phi +\epsilon h} (f_1, P_\phi \sum_{n \in \bN} \psi_n E^{A}_\phi (f_2))\right|_{0}\\
				&= \sum_{n} \left. \frac{d}{d\epsilon} E^{A}_{\phi +\epsilon h} (f_1,P_\phi \psi_n E^{A}_\phi (f_2)) \right|_{0} = \sum_{n} E^{A}_\phi (f_1, h V'''(\phi) \psi_n E^{A}_\phi (f_2)) \\
				&=E^{A}_\phi (f_1, h V'''(\phi) E^{A}_\phi(f_2)).
				\end{split}
			\end{equation*}
 		We used the fact that the sum in the equation above contains only a finite number of non-vanishing terms because $E^{A}_{\phi +\epsilon h} (f_1, P_\phi \psi_n E^{A}_\phi (f_2))$ vanishes if $\supp \psi_n$ does not intersect the compact set $J^-(\supp f_2) \cap J^+(\supp f_1)$. Summing up we have proved that
			\begin{equation*}
				\left. \frac{d}{d\epsilon} E^{A}_{\phi +\epsilon h} (f_1,f_2) \right|_{\epsilon=0} \equiv \int_{M^3} h(y) \frac{\delta E^{A}_\phi (x_1,x_2)}{\delta \phi(y)} f_1(x_1) f_2(x_2) dy dx_1 dx_2,
			\end{equation*}
		where
			\begin{equation}\label{var_der_A}
				\frac{\delta E^{A}_\phi (x_1,x_2)}{\delta \phi(y)} := E^{A}_\phi(x_1,y) V'''(\phi)(y) E^{A}_\phi(y,x_2)
			\end{equation}
		which is a well-defined distribution in $\cD^\prime(M^3)$. Since $V$ is a compactly supported functional, it follows that the distribution $\delta E^A(x_1, x_2) / \delta \phi(y)$ is compactly supported in $y$.
		An analogous result (with $A \leftrightarrow R$) holds for the retarded propagator.\\
		From the estimate~\eqref{WF_A_bkgr_dep} and the wave-front set calculus (thm.~\ref{theo_WF_horma}), it follows that
			\begin{equation*}
				\WF\left( \frac{\delta E^{A/R}_\phi (x_1,x_2)}{\delta \phi(y)} \right) \subset X_{2+1}, \quad \WF \left(\frac{\delta E^{A/R}_{\phi (\epsilon)} (x_1,x_2)}{\delta \phi(y)}\right) \subset \bR \times \{0\} \times X_{2+1},
			\end{equation*}
		where the set $X_{2 + 1}$ is defined by~\eqref{X}, where $\bR \ni \epsilon \mapsto \phi(\epsilon) \in C^\infty(M)$ is a  smooth map, and where $\delta E^{A/R}_{\phi(\epsilon)}(x_1,x_2)/\delta \phi(y)$ is viewed as a distribution in the variables $\epsilon, x_1,x_2,y$.\\
		We can compute Gateaux derivatives of higher orders by simply distributing the variational derivatives onto each factor appearing on the right-hand side of eq.~\eqref{var_der_A}. Again from the support properties of the interaction $V$ and the wave-front set calculus (thm.~\ref{theo_WF_horma}) we obtain the following results:
		\begin{prop}\label{prop_var_ders_A/R}
				For any $\phi \in C^\infty(M)$, and for any $\nu \in \bN$, the $\nu$-th Gateaux derivative
					\begin{equation*}
						\frac{\delta^\nu E^{A/R}_\phi (x_1,x_2)}{\delta \phi(y_1) \dots \delta \phi(y_\nu)}
					\end{equation*}
				is a well-defined distribution which satisfies the following properties:
				\begin{enumerate}
					\item The distribution $\delta^\nu E^{A/R}_\phi (x_1,x_2) / \delta \phi(y_1) \dots \delta \phi(y_\nu)$ is compactly supported in $y_1, \dots, y_\nu$\footnote{If $V(\phi) = \int_M \frac{1}{4!}\lambda(x) \phi^4(x) dx$ for $\lambda \in C^\infty_0(M)$, it holds that $y_1, \dots, y_\nu$ must belong to $\supp \lambda$.}.
					\item It holds
							\begin{equation}\label{var_ders_A/R_WF}
								\WF\left( \frac{\delta^\nu E^{A/R}_{\phi}(x_1,x_2)}{\delta \phi(y_1) \cdots \delta \phi(y_\nu)} \right) \subset X_{2+\nu},
							\end{equation}
							where $X_{2 + \nu}$ is defined by~\eqref{X}.
					\item Let $\bR \ni \epsilon \mapsto \phi(\epsilon) \in C^\infty(M)$ be smooth and view $\delta^\nu E^{A/R}_{\phi(\epsilon)}(x_1, x_2)/\delta \phi(y_1) \cdots \delta \phi(y_\nu)$ as a distribution in $\bR \times M^{2+\nu}$. It holds
							\begin{equation}\label{var_ders_A/R_WF_phi}
								\WF\left( \frac{\delta^\nu E^{A/R}_{\phi(\epsilon)}(x_1,x_2)}{\delta \phi(y_1) \cdots \delta \phi(y_\nu)} \right) \subset \bR \times \{0\} \times X_{2+\nu}.
							\end{equation}
				\end{enumerate}
			\end{prop}
			
	\newpage
	\printglossary[title={List of symbols}]
	
	\bibliographystyle{abbrv}	
	\bibliography{biblio}
\end{document}